%% file: UPC_iop_final.tex
\documentclass[12pt]{iopart}
\usepackage{graphicx,axodraw,color}
\usepackage{epsfig}
\usepackage{here,citesort}
\usepackage{tabularx,rotating}
\include{abrev}

\begin{document}
\title[The Physics of Ultraperipheral Collisions at the LHC]{The Physics of
Ultraperipheral Collisions at the LHC}
\author{
{\bf Editors and Conveners}: K.~Hencken$^{7,8}$, M.~Strikman$^{18}$, 
                             R.~Vogt$^{11,19,20}$, P.~Yepes$^{22}$\\
{\bf Contributors}: 
   A.~J.~Baltz$^{1}$,
   G.~Baur$^{2}$,
   D.~d'Enterria$^{3}$, 
   L.~Frankfurt$^{4}$, 
   F.~Gelis$^5$,
   V.~Guzey$^{6}$, 
   K.~Hencken$^{7,8}$,
   Yu.~Kharlov$^9$,
   M.~Klasen$^{10}$,
   S.~R.~Klein$^{11}$,
   V.~Nikulin$^{12}$,
   J.~Nystrand$^{13}$,
   I. A. Pshenichnov$^{14,15}$,
   S.~Sadovsky$^{9}$,
   E.~Scapparone$^{16}$,
   J.~Seger$^{17}$,
   M.~Strikman$^{18}$,
   M. Tverskoy$^{12}$,
   R.~Vogt$^{11,19,20}$, 
   S.~N.~White$^1$, 
   U.~A.~Wiedemann$^{21}$, 
   P.~Yepes$^{22}$,
   M.~Zhalov$^{12}$
}
\address{$^{1}$Physics Department, Brookhaven National Laboratory, Upton, 
NY, USA}
\address{$^{2}$Institut fuer Kernphysik, Forschungszentrum Juelich, Juelich, 
Germany}
\address{$^{3}$Experimental Physics Division, CERN, Geneva, Switzerland}
\address{$^{4}$Nuclear Physics Department, Tel Aviv University, Tel Aviv,
Israel}
\address{$^5$CEA/DSM/SPhT, Saclay, France}
\address{$^{6}$Institut f\"ur Theoretische Physik II, Ruhr-Universit\"at 
Bochum, Bochum, Germany}
\address{$^7$University of Basel, Basel, Switzerland}
\address{$^8$ABB Corporate Research, Baden-Daettwil, Switzerland}
\address{$^9$Institute for High Energy Physics, Protvino, Russia}
\address{$^{10}$Laboratoire de Physique Subatomique et de Cosmologie, 
Universit\'e Joseph Fourier/CNRS-IN2P3, Grenoble, France}
\address{$^{11}$Nuclear Science Division, Lawrence Berkeley National 
Laboratory, Berkeley, USA}
\address{$^{12}$Petersburg Nuclear Physics Institute, Gatchina, Russia}
\address{$^{13}$Department of Physics and Technology, University of Bergen, 
Bergen, Norway}
\address{$^{14}$Frankfurt Institute for Advanced Studies, Frankfurt am Main,
Germany}
\address{$^{15}$Institute for Nuclear Research, Russian Academy of Sciences,
Moscow, Russia}
\address{$^{16}$INFN, Sezione di Bologna, Bologna, Italy}
\address{$^{17}$Physics Department, Creighton University, Omaha, NE, USA}
\address{$^{18}$Physics Department, Pennsylvania State University, State 
College, PA, USA}
\address{$^{19}$Physics Department, University of California at Davis, Davis,
CA, USA} 
\address{$^{20}$Lawrence Livermore National Laboratory, Livermore, CA, USA}
\address{$^{21}$Theory Division, CERN, Geneva, Switzerland}
\address{$^{22}$Physics and Astronomy Department, Rice University, 
Houston, TX, USA}

\begin{abstract}
We discuss the physics
of large impact parameter interactions at the LHC: ultraperipheral
collisions (UPCs).  The dominant processes in UPCs are
photon-nucleon (nucleus) interactions.  
The current LHC detector configurations can explore small $x$ hard 
phenomena with 
nuclei and nucleons at photon-nucleon center-of-mass energies above 1 TeV, 
extending the $x$ range of HERA by a factor of ten. 
In particular, it will be possible to probe 
diffractive and inclusive parton densities in nuclei using several processes.
The interaction of small dipoles with protons and nuclei
can be investigated in elastic and quasi-elastic $J/\psi$ and
$\Upsilon$ production as well as in high $t$ $\rho^0$ production 
accompanied by a rapidity gap. 
Several of these phenomena provide clean signatures of the onset 
of the new high gluon density QCD regime.
The LHC is in the kinematic range where 
nonlinear effects are several times larger than at HERA.
Two-photon processes in UPCs are also studied.  In addition, while
UPCs play a role in limiting the maximum beam luminosity, they can
also be used a luminosity monitor by measuring
mutual electromagnetic dissociation of the beam nuclei.  We also
review similar studies at HERA and RHIC as well as describe the
potential use of the LHC detectors for UPC measurements.
\end{abstract}

\maketitle

\pagebreak 

\pagenumbering{arabic}

\section{Introduction}
{\it Contributed by: K. Hencken, M. Strikman, R. Vogt and P. Yepes}\\

\begin{figure}[ht]
 \centering
 \includegraphics[totalheight=2.5 in]{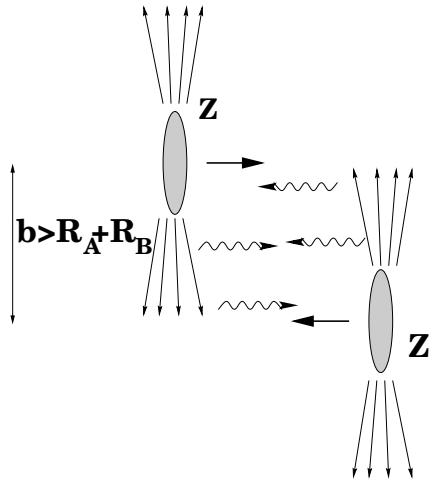}
\caption[]{Schematic diagram of an ultraperipheral collision of two ions.
The impact parameter, $b$, is larger than the sum of the two radii, $R_A + 
R_B$.  Reprinted from Ref.~\protect\cite{Baur:2001jj} with permission from
Elsevier.}
  \label{fig:basicdiagram}
\end{figure}

In 1924 Enrico Fermi, 23 at the time, proposed 
the equivalent photon method
\cite{fermi25} which treated the
moving electromagnetic
fields of a charged particle as a flux of virtual photons.
A decade later, Weizs{\"a}cker and Williams applied
the method~\cite{ww1934} to relativistic ions. 
Ultraperipheral collisions, UPCs, are those reactions in which two
ions interact via their cloud of virtual photons. 
The intensity of the electromagnetic field, and therefore the number of
photons in the {\it cloud} surrounding the nucleus, is proportional to 
$Z^2$. Thus these types of interactions
are highly favored when heavy ions collide.
Figure~\ref{fig:basicdiagram} shows a schematic view of an
ultraperipheral heavy-ion collision. The pancake shape of the nuclei is
due to Lorentz contraction.

\begin{figure}[!ht]
\centering
 \includegraphics[width=1.5 in]{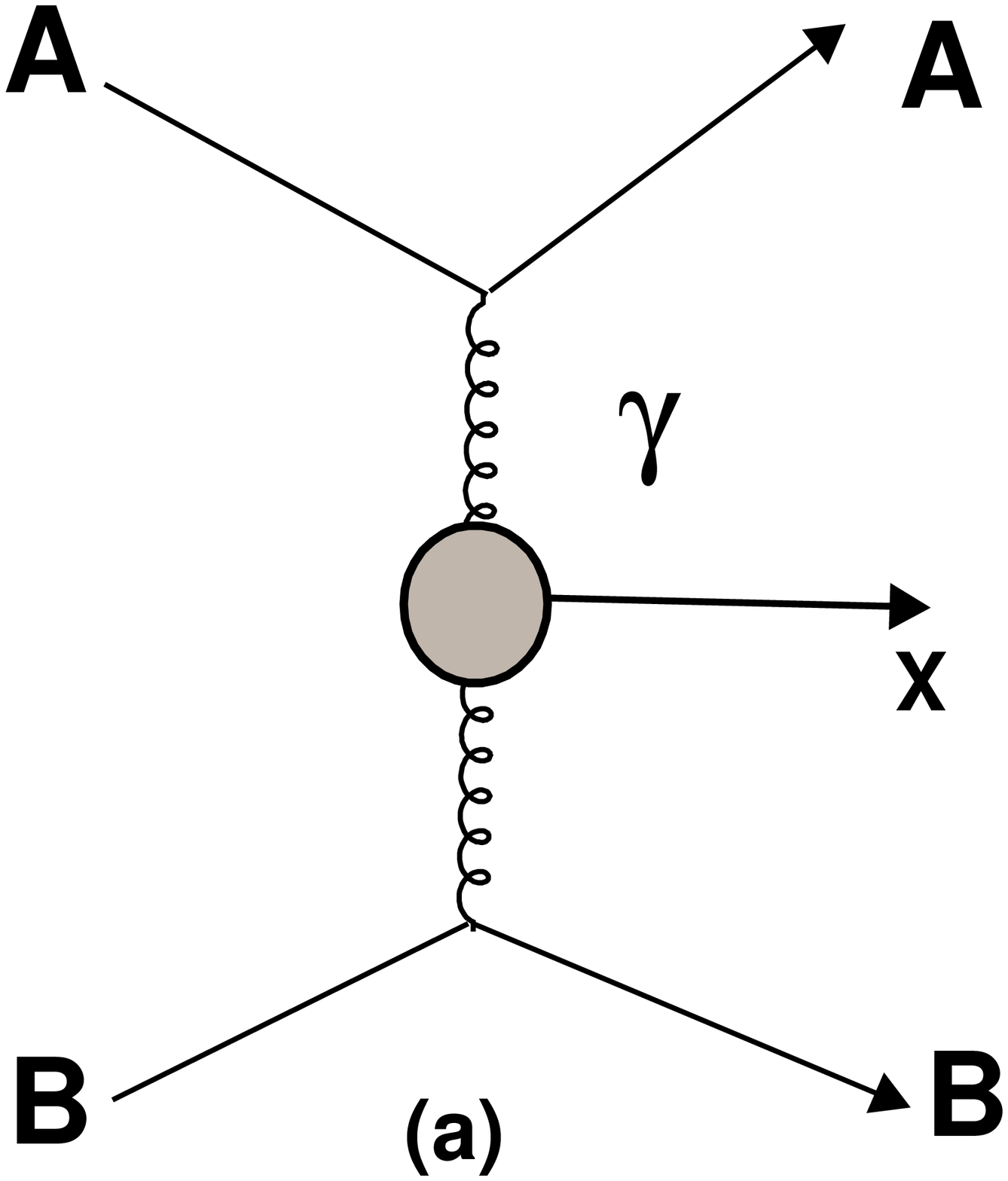}
 \includegraphics[width=1.5 in]{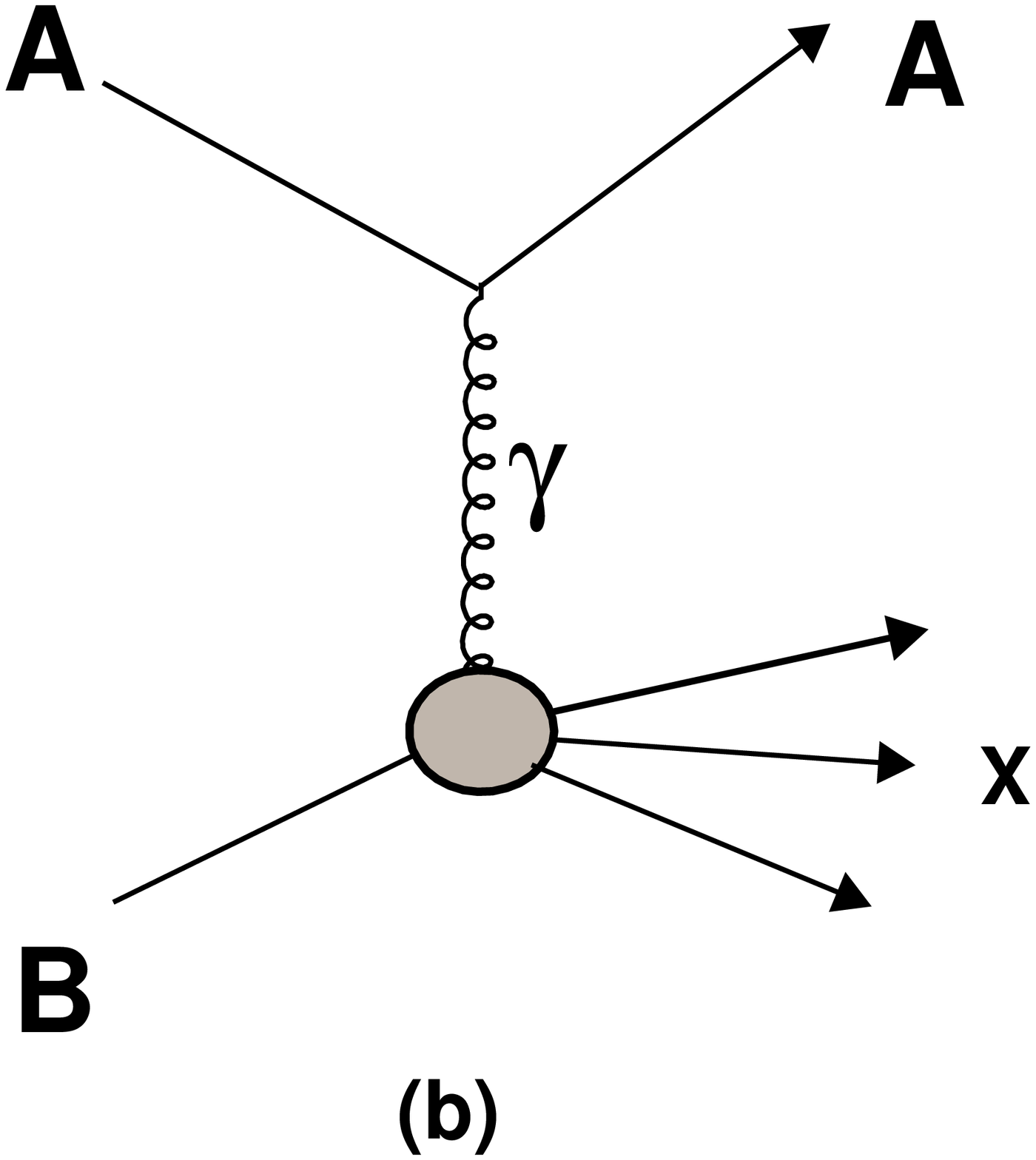}
 \includegraphics[width=1.5 in]{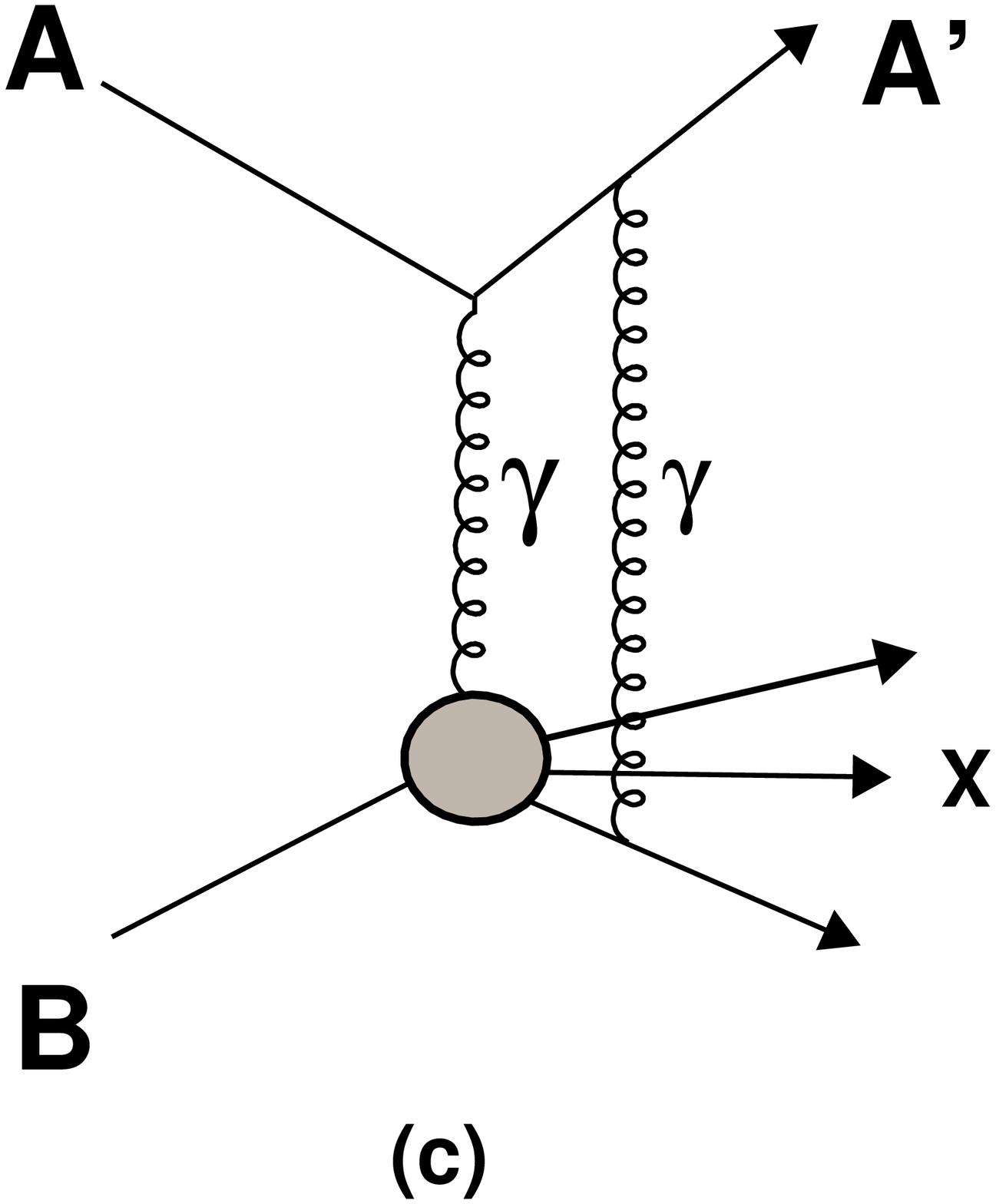}
  \caption[]{A schematic view of (a) an electromagnetic interaction where
photons emitted by the ions interact with each other, (b) a photon-nuclear   
reaction in which a photon emitted by an ion interacts with the other nucleus,
(c) photonuclear reaction with nuclear breakup due to photon exchange.}
\label{fig:upc_diagrams}
\end{figure}

Ultraperipheral photon-photon collisions are 
interactions 
where the radiated photons interact with each other.
In addition, photonuclear
collisions, where one radiated photon interacts with a constituent of the
other nucleus, are also possible. The two processes
are illustrated in Fig.~\ref{fig:upc_diagrams}(a) and (b).
In these diagrams the nucleus that emits the photon remains intact 
after the collision. However, it is possible to have an ultraperipheral
interaction in which one or both nuclei break up. The breakup may occur through
the exchange of an additional photon, as illustrated in 
Fig.~\ref{fig:upc_diagrams}(c).  

In calculations of ultraperipheral $AB$ collisions, 
the impact parameter is usually required to
be larger than the sum of the two nuclear radii, $b>R_A+R_B$. 
Strictly speaking, an ultraperipheral electromagnetic interaction could 
occur simultaneously with a hadronic collision. However, since 
it is not possible to separate the hadronic and electromagnetic components
in such collisions, the hadronic components are excluded by 
the impact parameter cut.

Photons emitted by ions are coherently radiated by the whole nucleus, 
imposing a limit on the minimum photon wavelength of greater than the 
nuclear radius. In the transverse
plane, where there is no Lorentz contraction, the uncertainty principle sets
an upper limit on the
transverse momentum of the photon emitted by ion $A$ of
$p_T \lesssim \hbar c/R_A \approx$ 28~(330) MeV$/c$ for Pb~($p$) beams.
In the longitudinal direction, the maximum possible momentum is
multiplied by a Lorentz factor, $\gamma_L$, due to the Lorentz contraction of
the ions in that direction: $k \lesssim \hbar c \gamma_L/R_A $.
Therefore the maximum $\gamma\gamma$ collision energy in a symmetric $AA$
collision is $2\hbar c\gamma_L/R_A$, about 6 GeV at the Relativistic Heavy 
Ion Collider (RHIC) and 200 GeV at the Large Hadron Collider (LHC).

The cross section for two-photon processes is \cite{BB88}
\begin{equation}
\sigma_{X}=\int dk_1 dk_2 \frac{dL_{\gamma \gamma}}{dk_1 dk_2} \sigma
_{X}^{\gamma\gamma}\left(k_1,k_2\right)
\;,\label{eq:two-photon}%
\end{equation}
where $\sigma_{X}^{\gamma\gamma}\left(k_1,k_2\right)$ is the
two-photon production cross section of final state $X$ and $dL_{\gamma 
\gamma}/dk_1 dk_2$ is the two-photon luminosity,
\begin{equation}
{dL_{\gamma\gamma} \over dk_1dk_2} = 
\int_{b>R_A}\int_{r>R_A} d^2b d^2r
\frac{d^3N_\gamma}{dk_1 d^2b}\frac{d^3N_\gamma}{dk_2 d^2r} \ ,\label{lumin}%
\end{equation}
where $d^3N_\gamma/dk d^2r$ is the photon flux from a charge $Z$ nucleus at a
distance $r$.  The two-photon cross section can also be written in terms of
the two-photon center-of-mass energy, $W_{\gamma \gamma} =
\sqrt{s_{\gamma \gamma}} = \sqrt{4k_1 k_2}$ 
by introducing the delta function $\delta(s_{\gamma \gamma} - 4 k_1 k_2)$ 
to integrate over $k_1$ and changing the integration variable from $k_2$ 
to $W_{\gamma \gamma}$ so that
\begin{eqnarray}
\sigma_X = \int 
\frac{dL_{\gamma \gamma}}{dW_{\gamma \gamma}} W_{\gamma \gamma}
\sigma_X^{\gamma 
\gamma}(W_{\gamma \gamma}) \, \, .
\end{eqnarray}
(Note that we use $W$ and $\sqrt{s}$ for the center-of-mass energy 
interchangeably throughout the text.

\begin{figure}[!ht]
\centering
 \includegraphics[width=3.0 in]{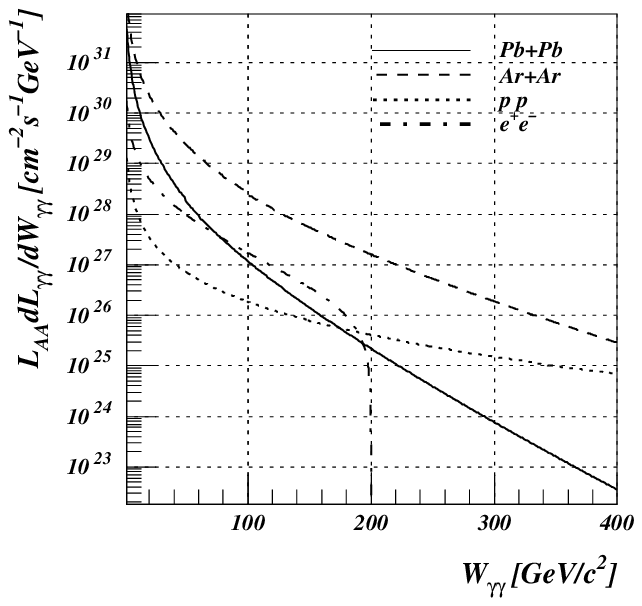}
 \includegraphics[width=3.0 in]{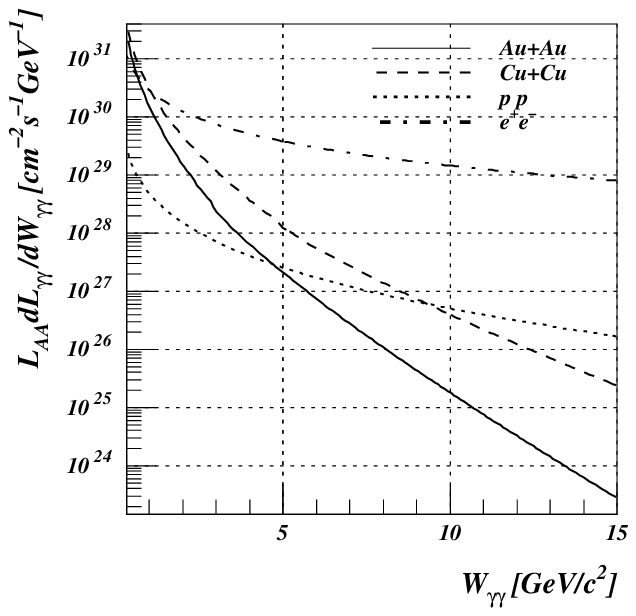}
\caption[]{Effective $\GG$ luminosity at LHC (left) and RHIC (right) 
for different ion species and protons as well as at LEP II.
In $pp$ and $e^+ e^-$ collisions, $L_{AA}$ corresponds to the $pp$ or $e^+e^-$
luminosity.  Reprinted from Ref.~\protect\cite{Baur:2001jj} with permission
from Elsevier.}
\label{fig:gglum}
\end{figure}

The two-photon luminosity in Eq.~(\ref{lumin}) can be multiplied 
by the ion-ion luminosity, $L_{AA}$, yielding an effective 
two-photon luminosity,
$dL_{\gamma \gamma}^{\rm eff}/dW_{\gamma \gamma}$, 
which can be directly compared to two-photon
luminosities at other facilities such as $e^{+}e^{-}$ or $pp$ colliders 
\cite{Khoze02}. 
Figure \ref{fig:gglum} shows the
two-photon effective luminosities for various ion species and protons as a
function of $W_{\gamma \gamma}$ 
for the LHC (left) and for RHIC (right) \cite{Baur:2001jj}. 
Note the difference in energy scales between the LHC and RHIC.
The ion collider luminosities are also
compared to the $\GG$ luminosity at LEP II.  The LHC will have 
significant energy and luminosity reach beyond LEP II and could be a
bridge to $\gamma\gamma$ collisions at a future linear $e^+e^-$
collider.  Indeed, the LHC two-photon luminosities for light ion beams are 
higher than available elsewhere for energies up to $W_{\gamma \gamma}
\approx 500$ GeV/$c^2$.

\begin{figure}
\center{\psfig{figure=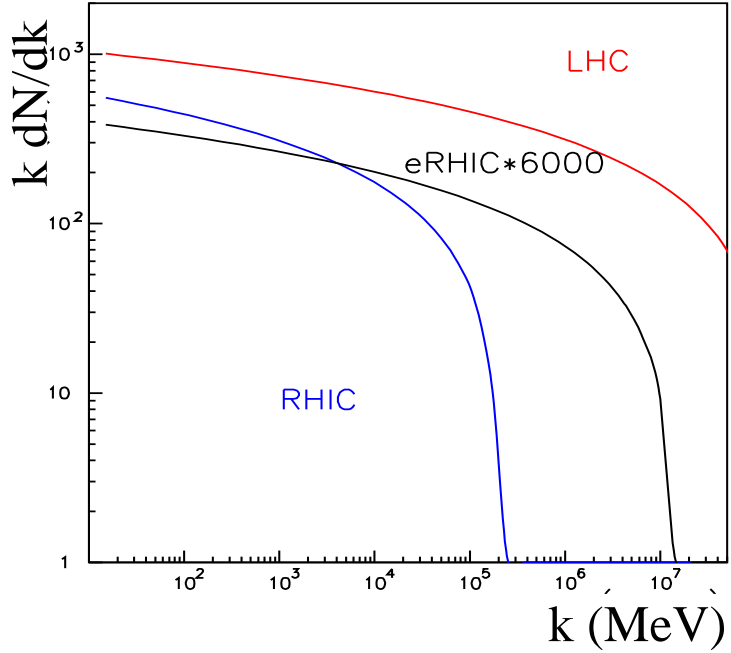,height=3.5in}}
\caption[]{The photon flux from $\sqrt{s_{NN}} = 200$ GeV Au+Au collisions at 
RHIC and $\sqrt{s_{_{NN}}} = 5.5$ TeV Pb+Pb collisions at the LHC,
compared with that expected for 10 GeV + 100 GeV $e$Au collisions at the 
proposed eRHIC~\protect\cite{EIC1,EIC2}. The
eRHIC curve has been multiplied by 6000 to account for improved gold beam
parameters at eRHIC. 
$k$ is given in the rest frame of the target nucleus in all three cases.
Modified from Ref.~\protect\cite{Erice_full} with permission from World
Scientific.}
\label{fig:gammaa}
\end{figure}

\begin{figure}[htb]
    \begin{center} \includegraphics[width=10cm]{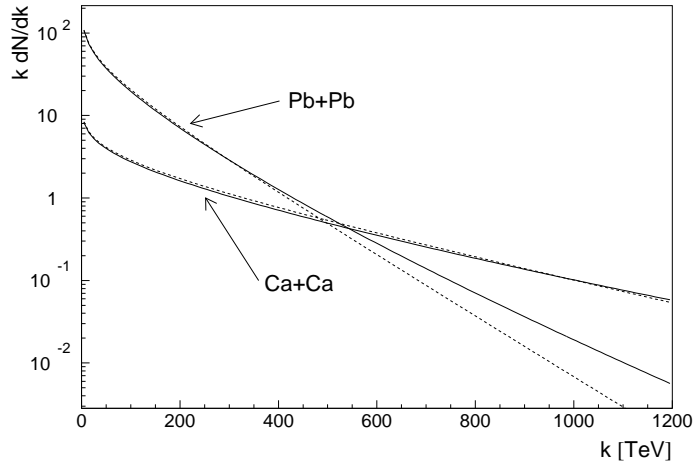}
      \end{center} 
\caption[]{The equivalent photon spectrum in Pb+Pb and Ca+Ca interactions 
at the LHC, evaluated in the rest frame of the target nucleus. The solid curves
correspond to the numerical result of Eq.~(\protect\ref{eq:3:nofe}) while
the dashed curves are the analytical result, Eq.~(\protect\ref{analflux}).
}
\label{fig:3:dndk} 
\end{figure}

The photoproduction cross section can also be factorized into
the product of the photonuclear cross section and the photon flux, $dN_\gamma
/dk$,
\begin{equation}
\sigma_{X}=\int dk \frac{dN_\gamma}{dk} \sigma_{X}^{\gamma}(k)  
\;,\label{eq:photonuclear}%
\end{equation}
where $\sigma_{X}^{\gamma}(k)$ is the photonuclear cross
section. 

The photon flux used to calculate the two-photon luminosity in 
Eq.~(\ref{lumin}) and the photoproduction cross section in 
Eq.~(\ref{eq:photonuclear}) is given by the Weizs\"acker-Williams method
\cite{Jackson}.  The flux is evaluated in impact
parameter space, as is appropriate for heavy-ion
interactions \cite{Cahn:1990jk,Baur:1990fx}.
The flux at distance $r$ away from a charge $Z$ nucleus is
\begin{equation}
{d^3N_\gamma \over dkd^2r} = 
{Z^2\alpha w^2\over \pi^2kr^2} \left[ K_1^2(w) + {1\over
\gamma_L^2} K_0^2(w) \right] \, \, 
\label{wwr}
\end{equation}
where $w=kr/\gamma_L$ and $K_0(w)$ and $K_1(w)$ are modified Bessel
functions.  The photon flux decreases exponentially above a cutoff
energy determined by the size of the nucleus.  In the laboratory frame, the
cutoff is $k_{\rm max} \approx \gamma_L \hbar c/R_A$.  In the rest frame of the
target nucleus, the cutoff is boosted to 
$E_{\rm max}=(2\gamma_L^2-1)\hbar c/R_A$,
about 500 GeV at RHIC and 1 PeV (1000 TeV) at the LHC.  The photon flux  
for heavy ions at RHIC and the
LHC is depicted in Fig. \ref{fig:gammaa}. Also shown, for
comparison, is the flux for the proposed electron-ion collider at RHIC, 
eRHIC\footnote{We give estimates for the 10 GeV + 100 GeV version of the
proposed electron-ion collider eRHIC.}.
The $eA$ flux has been multiplied by 6000 to include the
expected luminosity increase for eRHIC relative to RHIC. Although both
RHIC and eRHIC are high luminosity $\gamma A$ colliders, the 
LHC has an energy reach far
beyond other existing or planned machines.

In these collisions, the accelerated ion is surrounded
by a cloud of almost real photons of virtuality $|q^2| < (\hbar c/R_A)^2$
where $R_A$ is the nuclear radius.  The virtuality, less than $(60 \, {\rm
MeV})^2$ for nuclei with $A> 16$, can be neglected.  Since the photon
interaction is long range, photons can interact with partons in the opposite
nucleus even when the nuclei themselves do not interpenetrate.  Because the
photon energies are less than those of the nucleons, these photonuclear
interactions have a smaller average center-of-mass energy than hadronic
parton-parton collisions.  However, even though the energy is smaller, 
coherent photon beams have a flux proportional to the square of the nuclear 
charge, $Z$, enhancing the rates relative to those of photoproduction in
$pp$ collisions.  Although the photons are nearly real, their high energy
allows interactions at high virtualities, $Q^2$, in the photon-parton center of
mass.  Thus, massive vector mesons, heavy quarks and jets can be produced
with high rates in UPCs.

Table~\ref{gamfacs} shows the
nucleon-nucleon center-of-mass energies, $\sqrt{s_{_{NN}}}$, the
beam energies in the center-of-mass frame, $E_{\rm beam}$, 
Lorentz factors, $\gamma_L$, $k_{\rm
max}$, and $E_{\rm max}$, as well as the corresponding maximum $\gamma A$ 
center-of-mass energy per nucleon, $\sqrt{s_{\gamma N}} = W_{\gamma N} =
[2k_{\rm max} \sqrt{s_{_{NN}}}]^{1/2} = \sqrt{2E_{\rm max}m_p}$.  We give the
appropriate default 
kinematics for $AA$, $pA$ and $pp$ collisions at the LHC.  The resulting
values are compared to the fixed-target kinematics of the SPS as well as the 
proton and gold beams at the RHIC collider.  In fixed-target kinematics,
$E_{\rm max}$ is obtained from $\gamma_L \hbar c/R_A$ with the Lorentz boost
of the beam while $k_{\rm max}$ is calculated with $\gamma_L = 
\sqrt{s_{_{NN}}}/2m_p$.  In $pA$ collisions,
the photon field of the nucleus is stronger so that the interacting photon
almost always comes from the nucleus.  Note also that the LHC $pA$ results
are calculated in the center-of-mass kinematics although the different $Z/A$
ratios in asymmetric collisions mean that the beams have different velocities.
In $pp$ collisions, we use $r_p = 0.6$ fm to calculate $E_{\rm max}$ and
$k_{\rm max}$.  Note that, at high energy, the maximum photon energy is 25\% 
of the proton energy for this choice of $r_p$, significantly increasing the 
probability of proton breakup.  More
work is required to understand the usable $pp$ luminosity in this case.

We have also included the best available estimates 
\cite{lhcdesign,Brandttalk,BrandtEM94} of the beam-beam luminosities for $AA$
and $pp$ collisions in Table~\ref{gamfacs} to aid rate
calculations.  No beam-beam luminosity is given for the fixed-target
kinematics of the SPS.  
Only an estimate of the initial LHC $pA$ luminosities are 
given \cite{Brandttalk}.  
The maximum machine luminosities are applicable to CMS and ATLAS.
Unfortunately the interaction rate in ALICE is limited to 200 kHz.  Therefore
its maximum $pp$ luminosities are significantly lower. 
The luminosities for collision modes other than $pp$
and Pb+Pb are unofficial and, as such, are subject to revision.

\begin{table}[htpb]
\begin{center}
\caption[]{Pertinent parameters and kinematic limits for some projectile-target
combinations at several accelerators.  
We first give the luminosities and the $NN$ collision
kinematics,  the nucleon-nucleon center-of-mass energies, $\sqrt{s_{_{NN}}}$, 
the corresponding beam energies, $E_{\rm beam}$, and the
Lorentz factors, $\gamma_L$.  We then present the
photon cutoff energies in the center-of-mass frame, $k_{\rm max}$, and
in the nuclear rest frame, $E_{\rm max}$, as well as the equivalent maximum
photon-nucleon and photon-photon center-of-mass energies, 
$\sqrt{s_{\gamma N}^{\rm max}}$ and $\sqrt{s_{\gamma \gamma}^{\rm max}}$ 
respectively.}
\label{gamfacs}
\vspace{0.4cm}
\begin{tabular}{|c|c|c|c|c|c|c|c|c|} \hline
$AB$       & $L_{AB}$ & $\sqrt{s_{_{NN}}}$ & $E_{\rm beam}$ 
& $\gamma_L$ & $k_{\rm max}$ & $E_{\rm max}$
& $\sqrt{s_{\gamma N}^{\rm max}}$ & $\sqrt{s_{\gamma \gamma}^{\rm max}}$ \\ 
& (mb$^{-1}$s$^{-1}$) & (TeV) & (TeV) &  & (GeV) & (TeV) & (GeV) & (GeV)
\\ \hline
\multicolumn{9}{|c|}{SPS} \\ \hline
In+In & - & 0.017 & 0.16 & 168 & 0.30 & 5.71 $\times 10^{-3}$ & 3.4  & 0.7 \\ 
Pb+Pb & - & 0.017 & 0.16 & 168 & 0.25 & 4.66 $\times 10^{-3}$ & 2.96 & 0.5\\ 
\hline
\multicolumn{9}{|c|}{RHIC} \\ \hline
Au+Au & 0.4 & 0.2 & 0.1 & 106 & 3.0 & 0.64 & 34.7 & 6.0 \\
$pp$  & 6000 & 0.5 & 0.25 & 266 & 87 & 46.6 & 296 & 196 \\ \hline 
\multicolumn{9}{|c|}{LHC} \\ \hline
O+O   & 160  & 7 & 3.5 & 3730 & 243 & 1820 & 1850 & 486 \\
Ar+Ar & 43   & 6.3 & 3.15 & 3360 & 161 & 1080 & 1430 & 322 \\
Pb+Pb & 0.42 & 5.5 & 2.75 & 2930 &  81 &  480 &  950 & 162 \\ \hline
$p$O  & 10000 & 9.9 & 4.95 & 5270 & 343 & 3620 & 2610 & 686 \\
$p$Ar & 5800  & 9.39 & 4.7 & 5000 & 240 & 2400 & 2130 & 480 \\
$p$Pb & 420   & 8.8 & 4.4 & 4690 & 130 & 1220 & 1500 & 260 \\ \hline
$pp$  & $10^7$  & 14 & 7 & 7455 & 2452 & 36500 & 8390 & 4504 \\ 
\hline 
\end{tabular}
\end{center}
\end{table}

The total photon flux striking the target nucleus is the integral of
Eq.~(\ref{wwr}) over the transverse area of the target for all impact
parameters subject to the constraint that the two nuclei do not
interact hadronically.  A reasonable analytic approximation for $AB$
collisions is given by the photon flux integrated over radii larger than
$R_A + R_B$.  The analytic photon flux is
\begin{equation}
{dN_\gamma\over dk} = 
{2Z^2 \alpha \over\pi k} \left[ w_R^{iA}K_0(w_R^{iA})
K_1(w_R^{iA})- {(w_R^{iA})^2\over 2} \big(K_1^2(w_R^{iA})-K_0^2(w_R^{iA})
\big) \right] \, \, 
\label{analflux}
\end{equation}
where $w_R^{AA}=2kR_A/\gamma_L$ and $w_R^{pA} = k(r_p + R_A)/\gamma_L$.  
This analytic flux is compared to the full numerical 
result, Eq.~(\ref{eq:3:nofe}), in Fig.~\ref{fig:3:dndk} for Pb+Pb and Ca+Ca
collisions at the LHC.  The numerical result gives a harder photon spectrum
for Pb+Pb collisions at the same $k$.  On the other hand, there is little 
difference between the two results for Ca+Ca collisions.  (Note that there is
some discussion of Ca+Ca interactions in the text since some initial UPC
studies were done before argon was chosen over calcium beams.  While $A=40$
for both, the $Z$ is different, changing both the flux and the energy range
for Ar relative to Ca.)

\begin{figure}[tbhp]
  \centering
  \includegraphics[height=6.5cm]{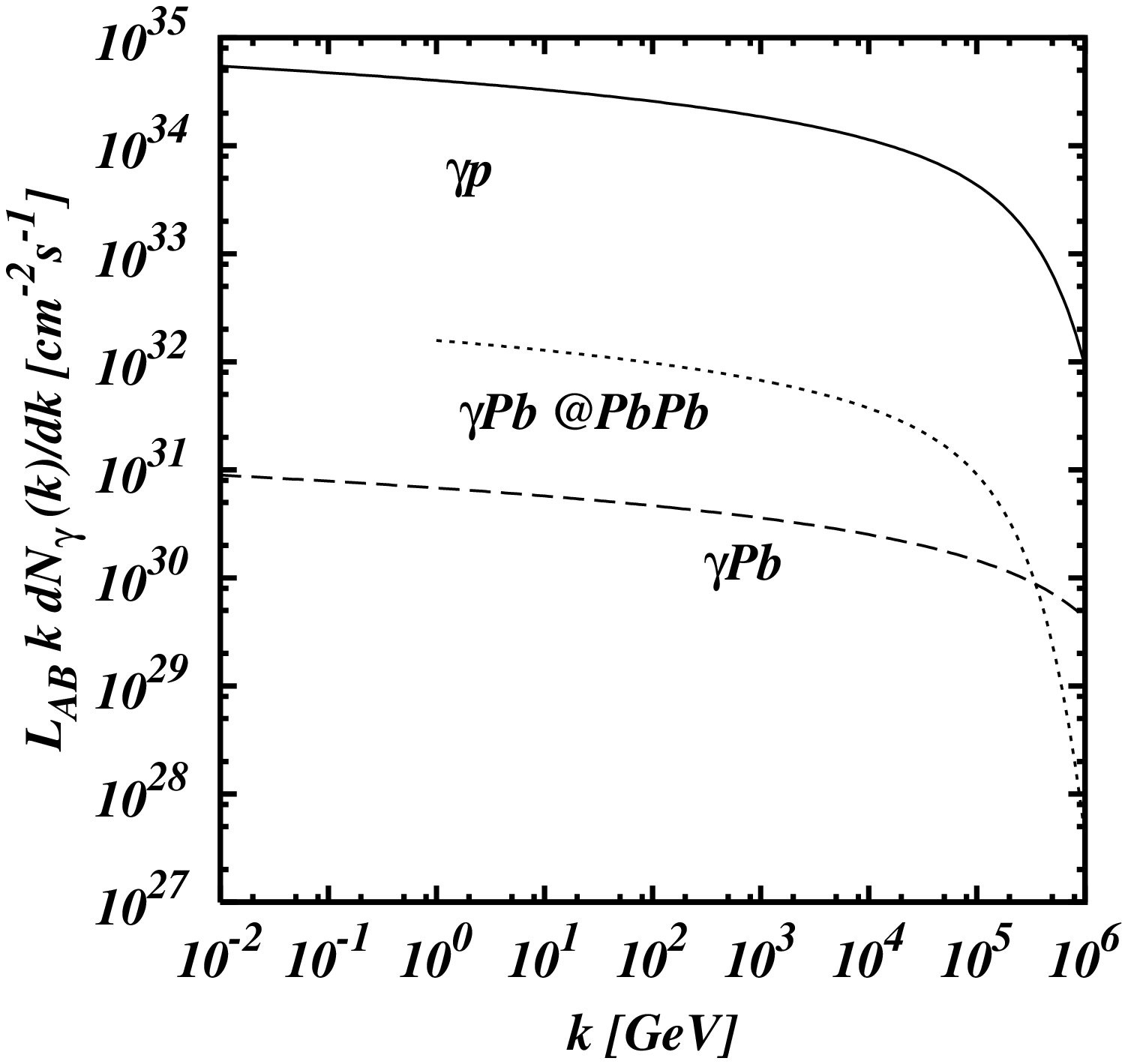}
  \includegraphics[height=6.5cm]{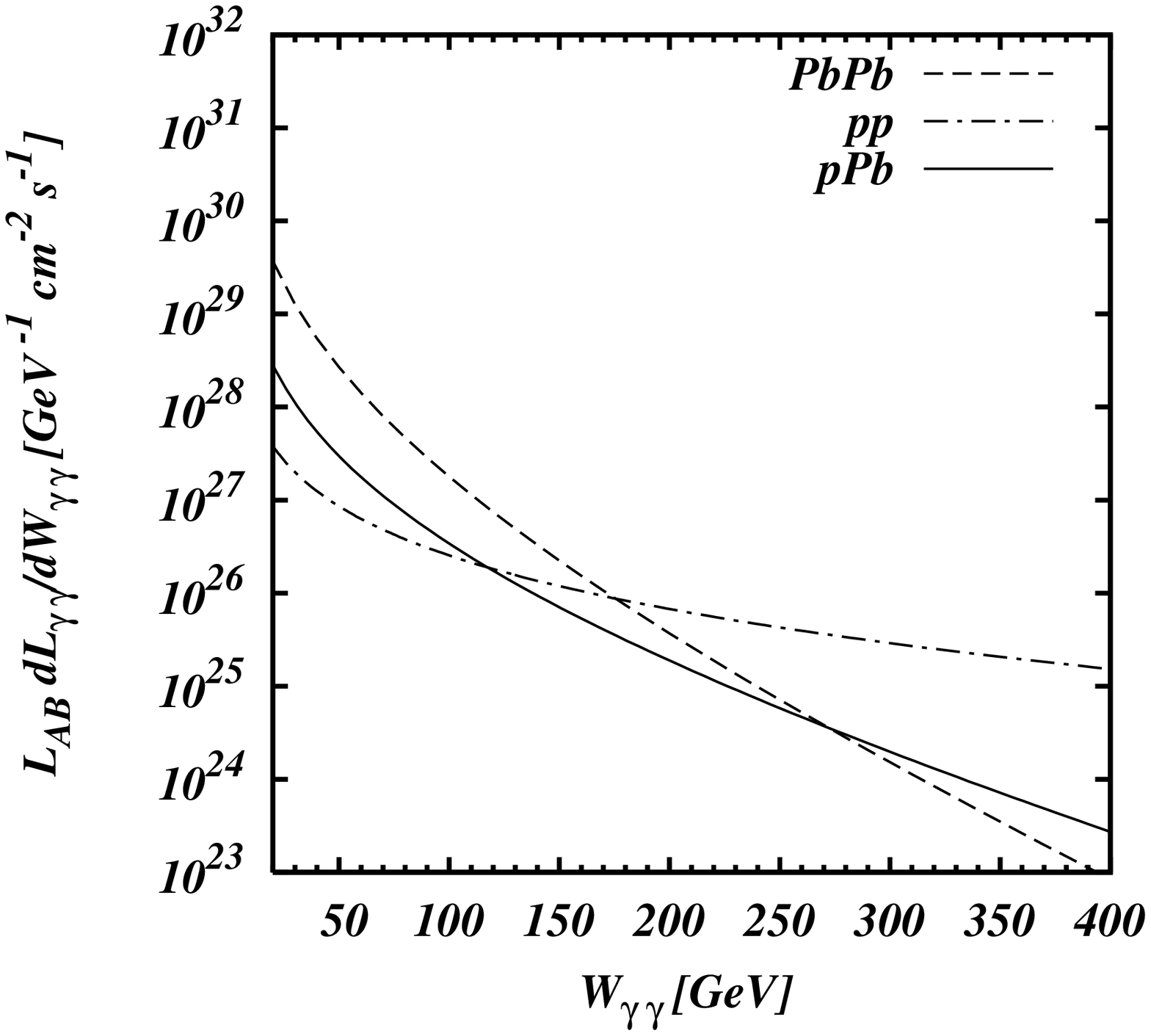}
  \caption[]{(left) The effective $\gamma A$ luminosity, $L_{AB}(k 
    dN_\gamma/dk)$, is shown for the cases where the photon is emitted from
    the proton ($\gamma$Pb) and the ion ($\gamma p$) as well as when the photon
    is emitted from the ion in a Pb+Pb collision ($\gamma$Pb@Pb+Pb).  
    (right) The photon-photon 
    luminosities, $L_{AB}(dL_{\gamma\gamma}/dW_{\gamma \gamma})$, 
    are compared for $pp$, $p$Pb and Pb+Pb collisions
    at the LHC. }
  \label{fig:luminosities}
\end{figure}

Since photonuclear
rates increase more slowly than $A^2$, there may be advantages in
$pA$ relative to $AA$ collisions. As presented above, 
event rates in ultraperipheral $AA$ collisions depend 
both on the photon flux, $dN_\gamma/dk$, which scales as $Z^2$ 
in photoproduction and $Z^4$ in two-photon processes, and on
the beam-beam luminosity, $L_{AB}$.  
Lighter ions are favored for many UPCs since the higher luminosities
\cite{BrandtEM94} compensate for the larger $Z$ in lower luminosity
Pb+Pb collisions. 
In the case of $p$Pb collisions, $L_{p{\rm Pb}}$ 
is two orders of magnitude higher than $L_{\rm PbPb}$. While it is more 
probable for the photon to be emitted by the ion and interact with
the proton ($\gamma p$), it could also be emitted by the proton 
and interact with the ion
($\gamma$Pb).  The relevant figure of merit is the effective photon-nucleus
luminosity, $L_{AB}(k dN_\gamma/dk)$. The left-hand side of 
Fig.~\ref{fig:luminosities} compares $L_{AB}(kdN_\gamma/dk)$ for 
$\gamma p$ (solid) and $\gamma$Pb (dashed) collisions
in $p$Pb interactions to the case where the photon is emitted from the ion
in lower energy and lower luminosity Pb+Pb collisions.  The effective
$\gamma p$ luminosities are enhanced by the larger $p$Pb luminosity. 
Thus photonuclear processes on protons can be studied at energies beyond the
HERA range so that {\it e.g.} the energy dependence of $\Upsilon$ production
can be measured.  

As shown on the right-hand side of Fig.~\ref{fig:luminosities}, 
the two-photon luminosities 
for $p$Pb collisions at the LHC are only
slightly lower than those for Pb+Pb collisions at low $W_{\gamma \gamma}$, 
and even become higher for $W_{\gamma \gamma}>250$ GeV due to the larger 
$p$Pb energy. While these luminosities are lower than the 
$pp$ luminosities, heavy ions suppress
the diffractive background. The potential for the discovery of new
physics in $pA$ is rather limited at low $W_{\gamma \gamma}$ but there are 
again some advantages at higher $W_{\gamma \gamma}$.  Thus two-photon studies 
are still possible, as are electroweak studies. 
When the photon is emitted from the proton, the luminosity could be further
enhanced by allowing for inelastic processes such as proton breakup
\cite{OhnemusWZ94,DreesEZ89}. 

The physics of UPCs has been reviewed by a number of groups.  
The first comprehensive
study was by Baur and Bertulani in 1988 \cite{BB88}. 
More recent reviews are by Krauss, Greiner and Soff \cite{Krauss:1997vr},
Baur and collaborators \cite{Baur:2001jj}, and by Bertulani, Klein and Nystrand
\cite{BKN}.  The LHC FELIX proposal also did much to advance
UPCs \cite{FELIX}, as did a UPC workshop in Erice, Italy 
\cite{Erice,Erice_full}.
Useful related material is discussed in a recent photoproduction review
by Butterworth and Wing \cite{BW2005}.  

The remainder of this introduction will address some of the physics issues 
that can be studied with UPCs. A few of these will be described
in more detail in the body of the report.

\subsection{Physics of photonuclear reactions}

Data from HERA show that the gluon and sea quark distributions rise
quickly as their momentum fraction $x$ drops.  At small enough $x$,
the growth of the proton parton densities may decrease proportionally with
$\ln (1/x)$.  The 
increase of the parton densities is regulated by phenomena such as
shadowing, recombination reactions, {\it e.g.} $g g\rightarrow g$,
as well as possible tunneling between different QCD vacua that are suppressed 
at large $x$.   These phenomena are most
significant in the central core of a nucleon. Scattering off the periphery
of the nucleon will dominate at small $x$, causing the cross section to
increase asymptotically  as fast as $\propto \ln^3(1/x)$ \cite{annual}.  
The large diffractive gluon 
densities observed at HERA demonstrate nonlinear effects for squared momentum
transfer of the virtual photon of up to 
$Q^2\sim 4$ GeV$^2$ at the smallest $x$ values studied, $x \sim 10^{-4}$.  
At the LHC, these QCD phenomena should
be visible at larger $x$ in central collisions of both protons and heavy ions.

Studies of small $x$ deep inelastic scattering (DIS) at HERA substantially 
improved our understanding of  strong interactions at high energies.
There are several key findings of HERA in this field.
Rapid growth of the small $x$ parton densities was observed over a wide 
range of $Q^2$.
A significant probability for hard diffraction was seen,
consistent with approximate scaling and a logarithmic $Q^2$ 
dependence  (``leading-twist" dominance).
HERA also found a new 
class of hard exclusive processes --   light vector meson production
at large $Q^2$ and heavy $Q \overline Q$ vector mesons 
at all $Q^2$. These processes are described by the QCD factorization theorem 
\cite{Brodsky:1994kf,Collins:1996fb} and 
related to the generalized parton distributions in 
the target.  In the small $x$ limit, they can be calculated for zero 
squared momentum transfer, $t$, using standard parton 
distributions. This new class of interactions probes small  
$q\bar q$ dipole interactions with hadrons. The $t$-dependence 
provides direct information on the gluon distribution of hadrons in the
transverse plane as a function of $x$.

Combined analyses of inclusive DIS and hard vector 
meson production suggest that the strength of the interactions, especially in 
channels where a hard probe directly couples to low $x$ gluons, approaches the
maximum possible strength -- the black disk regime (BDR) -- for $Q^2 \leq 4$ 
GeV$^2$. This conclusion is confirmed by studies of hard
inclusive diffraction \cite{annual}. 

However, the $Q^2$ range over which the black disk regime holds
is relatively small, with even smaller 
values for processes where a hard probe couples to a $q\overline q$ dipole
with $Q^2 \sim  1\,$ GeV$^2$,
making it difficult 
to separate perturbative from nonperturbative effects and draw 
unambiguous conclusions. 

The interaction regime where hard probes of small target $x$ occur with
high probability should be a 
generic feature of strong interactions
at high energies.  This feature is related to high gluon densities,
reached for any target at sufficiently small $x$.  Extended targets are 
expected to reach this high density regime at substantially higher $x$. 
At very high gluon density, even the notion of inclusive 
parton densities is ill-defined.

The onset of the BDR corresponds to a drastic departure from the linear 
regime of QCD.  Observing the onset of nonlinear QCD dynamics at 
small $x$ would be of great importance.
The problems which emerge in the BDR kinematics can be visualized by 
considering DIS interactions and exclusive diffractive processes 
in the language of small dipoles interacting with the target. In the 
leading-log approximation,
the inelastic quark-antiquark (gluon-gluon) dipole-hadron
cross section for a dipole of size $d$ has the form 
\cite{BBFS93,Frankfurt:1993it,Radyush}
\begin{eqnarray}
\sigma_{{\rm dip}\, h} (s_{{\rm dip}\, h}, d^2) = 
\frac{\pi^2}{4}  C_F^2  d^2  \alpha_s (Q^2_{\rm eff}) 
x g (x, Q^2_{\rm eff}) \, \,  
\label{sigma_d_DGLAP}
\end{eqnarray}
where $x = Q^2_{\rm eff}/s_{{\rm dip} \, h}$ and $s_{{\rm dip} \, h}$
is the square of the dipole-hadron center-of-mass energy.
Here $C_F^2 $ is the Casimir operator, equal to 4/3 for $q\bar q$ and $3$ for 
$gg$, $\alpha_s (Q_{\rm eff}^2)$ is the leading order (LO) strong 
coupling constant and $g (x, Q_{\rm eff}^2 )$ is the LO gluon density in 
the target.  The coupling constant and the gluon density 
are evaluated at $Q_{\rm eff}^2 \propto d^{-2}$.
Since the gluon densities increase at small $x$, 
the cross section in Eq.~(\ref{sigma_d_DGLAP})
ultimately becomes larger than allowed by the
unitarity constraint, $\pi r_h^2$, where $r_h$ is the transverse 
radius of the gluon distribution in the hadron
at the corresponding $x$. Since the unitarity bound 
corresponds to complete absorption at impact parameters $b \leq r_h$, 
the resulting diffractive cross section reflects
absorption at small $b$.
If the regime of complete absorption at $b \leq r_h$ is 
reached, the diffractive absorption  cross section  becomes nearly equal to 
the inelastic scattering cross section.
At sufficiently high energies, the small $x$ gluon fields resolved by the
small color dipole become so strong 
that the dipole cannot propagate through extended nuclear media 
without absorption, signaling the breakdown of the linear scaling regime
of Eq.~(\ref{sigma_d_DGLAP}) and the onset of the BDR.

In the dipole picture, a high energy photon can be considered to be a 
superposition of large and small size dipoles.  Smaller and
smaller dipoles begin to interact in the BDR with increasing energy.  
Photons contain more
small dipoles than hadrons such as pions, leading to faster growth
of $\sigma_{\rm tot}(\gamma p)$ than given by the Froissart bound
for hadrons.  Thus real photon interactions are sensitive to these 
small dipoles.  As a result,
a number of theoretical issues concerning the onset of the BDR
can be studied using UPCs. The energy scale at which the dipole-target cross 
section in Eq.~(\ref{sigma_d_DGLAP}) is tamed by 
the unitarity constraint near the BDR and no longer undergoes rapid growth
is unknown, as is the energy dependence 
of the cross section. The energy at which the dipole cross section makes the
transition from color transparency (no screening) to color opacity (strong 
screening) and, ultimately, the BDR also needs to be determined. 
Answers may be found by selecting processes where gluons interact directly. 
High gluon densities may be achieved at lower energies using nuclei, as we now 
discuss. 

To reach the regime where Eq.~(\ref{sigma_d_DGLAP}) breaks down, 
measurements need to be extended to higher energies, smaller $x$, and to 
higher gluon densities, at the same energy and $x$, using nuclei. 
Nuclear beams were discussed for HERA
\cite{Arneodo:1996qa} but will not be implemented. 
Studies of small $x$ physics at the LHC using hadronic $pp$ or $pA$ collisions
will be rather difficult because, at central rapidities, the backgrounds 
due to multiple hard collisions will likely 
prevent measurements at virtualities less than $Q^2_{\rm eff} \sim 100-200$ 
GeV$^2$. Although the fragmentation region at forward rapidity, with smaller 
backgrounds, is likely beyond the acceptance of the currently planned 
detectors, some small $x$ studies using the CMS forward hadron calorimeter, HF,
or CASTOR have been performed \cite{CMS-TOTEM,dde_lowx06}.
Thus, instead of using 
$eA$ collisions to reach the small $x$ regime, many of the approaches used
at HERA could be implemented at the LHC using UPCs in both 
$AA$ and $pA$ collisions.

\begin{figure}
\center{\psfig{figure=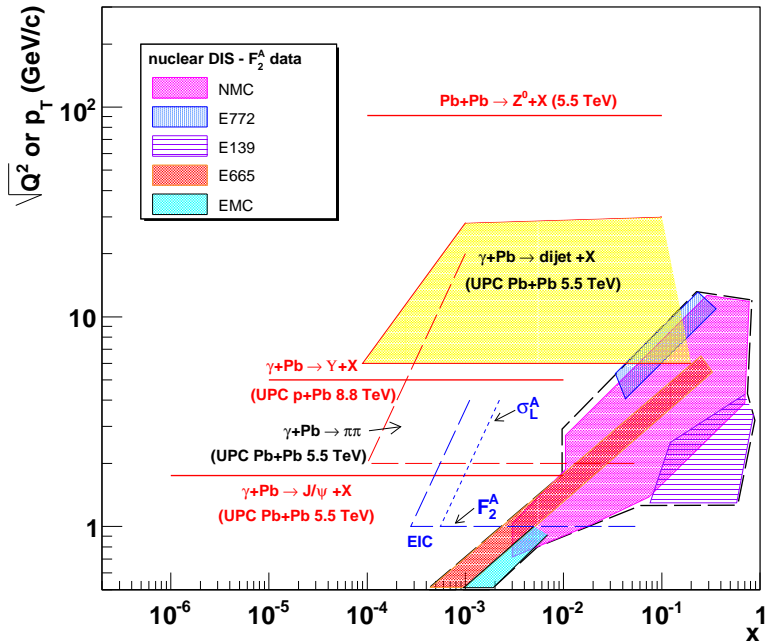,height=3.5in}}
\caption[]{The kinematic range in which UPCs at the LHC can 
probe gluons in protons and nuclei in quarkonium production,
dijet and dihadron  production. The $Q$ value for typical gluon virtuality in 
exclusive quarkonium photoproduction is shown for $J/\psi$ and $\Upsilon$. 
The transverse momentum of the jet or leading pion sets the scale for dijet 
and $\pi \pi$ production respectively.  For comparison, the kinematic 
ranges for $J/\psi$ at RHIC, $F_2^A$ and $\sigma_L^A$ at eRHIC and $Z^0$
hadroproduction at the LHC are also shown.
}
\label{xpablo}
\end{figure}

A primary focus of UPC studies in $AB$ and $pA$ collisions is on hard 
interactions in the kinematics which probe high gluon
densities in nucleons and nuclei.
Hard scatterings on nuclear targets will extend the low $x$ range of
previous studies by nearly three orders of magnitude. In $pA$ 
collisions, the HERA $x$ range
could be extended to an order of magnitude smaller $x$. Thus all three 
HERA highlights: gluon density measurements, gluon-induced hard diffraction, 
and exclusive $J/\psi$ and $\Upsilon$ production can be studied in
ultraperipheral $pA$ and $AB$ collisions.  Figure~\ref{xpablo} shows the
$x$ and $Q^2$ ranges covered by UPCs at the LHC.  For comparison, the kinematic
range of both $Z^0$ production in $pp$ collisions at the LHC and the nuclear
structure function at eRHIC are also shown.  The
$x$ range of $ep$ collisions at eRHIC is a factor of $\sim 30$ lower than at
HERA for the same $p_T$.
In the remainder of the introduction, we summarize some of 
the possible UPC measurements that could
further our understanding of small $x$ dynamics. \\

{\it Measurements of parton distributions in nuclei/nucleons} \\

The studies in Section~\ref{section-pdf} will demonstrate that 
hard ultraperipheral collisions investigate hard photon-nucleus 
(proton) collisions at significantly higher energies than at HERA.
The dominant process is photon-gluon fusion to two jets with leading light 
or heavy quarks, $\gamma g \rightarrow {\rm jet}_1 \, + \, {\rm jet}_2$,
fixing the gluon densities in protons/nuclei.
The LHC rates will be high enough to 
measure dijets and $c$ and $b$ quarks, probing the gluon distribution at 
$x \sim 5 \times 10^{-5}$ for $p_T\ge 6$ GeV/$c$ \cite{strikman06}.

\begin{figure}[tbh]
\centering
\includegraphics[width=0.5\textwidth,totalheight=0.35\textheight]{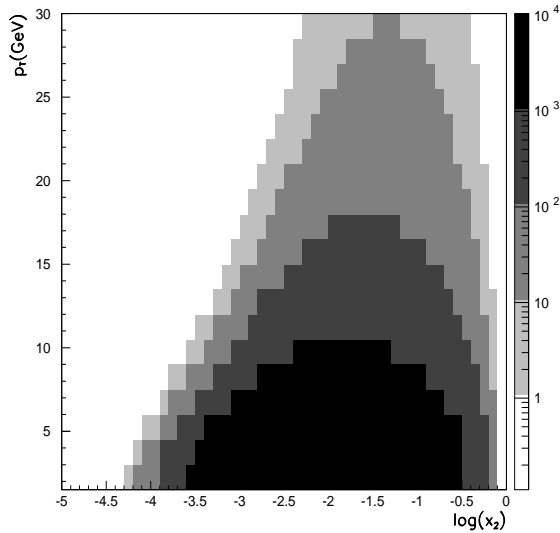}
\caption{The rate for inclusive $b\bar{b}$ photoproduction for a one month
LHC Pb+Pb run
at $0.42 \times 10^{27}$cm$^{-2}$s$^{-1}$.
Rates are in counts per bin of $\pm 0.25  x_2$ and $\pm 0.75$ GeV in 
$p_T$.  From Ref.~\protect\cite{strikman06}.  Copyright 2006 by the American
Physical Society (http://link.aps.org/abstract/PRL/v96/e082001).
}
\label{XXX}
\end{figure}

The virtualities that can be probed in UPCs will be 
much higher than those reached in lepton-nucleon/nucleus interactions. 
The larger $x$ range and direct gluon couplings will make these measurements 
competitive with those at HERA and the planned eRHIC as a way to probe 
nonlinear effects.  Indeed if it is possible to go down to $p_T\sim 5$ GeV/$c$,
the nonlinear effects in UPCs would be a factor of six higher than at HERA 
and a factor of two larger than at eRHIC \cite{strikman06}. An example of the
$b$ quark rate in the ATLAS detector \cite{strikman06} is 
presented in Fig.~\ref{XXX}.

{\it Hard diffraction} 

One of the cleanest signals of the proximity of the BDR is the ratio of the 
diffractive to total cross sections.  In the cases we discuss, 
rapidity gap measurements will be straightforward in 
both ATLAS and CMS. If the 
diffractive rates are $\sim 20$\% of the total rate, as expected 
in current models, the statistics will be sufficient 
for inclusive measurements over most of the $x$ range. 
(Note that a 20\% diffractive probability at $p_T\ge 5$ GeV$/c$ 
suggests a $\sim 40$\% diffractive probability at $p_T\sim 2$ GeV$/c$.)
Production of two pions with $p_T \geq 2$ GeV/$c$ will probe still further
into the low $x$ regime, albeit at slightly higher $x$, see Fig.~\ref{xpablo}.

{\it Exclusive quarkonium production}

Although calculations of the absolute cross section do 
involve significant higher-twist corrections,
the strong increase in the $J/\psi$ photoproduction cross section at HERA
clearly indicates that heavy quarkonia are produced via coupling to small 
$x$ gluon fields. 
Thus $J/\psi$ and $\Upsilon$ photoproduction provide one of the cleanest 
tests of small $q\bar q$ dipole interactions with gluon fields. In the case 
of nuclear targets, several channels will be accessible:  
coherent processes, $\gamma A\to V A$;  quasi-elastic processes,
$\gamma A\to V A'$; and rapidity gap processes such as large-$t$
light vector meson production, $\gamma A\to V X$.
 
A highly nontrivial prediction of QCD is that, at sufficiently high energies,  
even small dipoles should be strongly absorbed by extended targets both due 
to leading-twist gluon shadowing and higher-twist multiple dipole rescattering.
The $A$ dependence of the coherent and quasi-elastic reactions, 
both change by $A^{-2/3}$ when going from weak absorption to the regime
of strong absorption, as we now illustrate.  The coherent dipole scattering 
cross section is $\propto A^{4/3}$ in the weak absorption impulse approximation
(a combination of $A^2$ from coherence at $t=0$ and $A^{-2/3}$ from the 
integral over $t$) and $\propto A^{2/3}$ for strong absorption over the 
surface area of the target.  Likewise, the quasi-elastic $A$ dependence 
varies between $A$ (weak absorption: volume emission) and 
$A^{1/3}$ (strong absorption: edge emission). 

Dipole absorption is expected to reveal itself through strong 
suppression of coherent quarkonium production at 
$x_{\rm eff}\equiv m_V^2/s_{\gamma N}\leq 10^{-3} $ and at midrapidity for 
$x_{\rm eff}\leq 5\times 10^{-3}$.  The $AA$ measurements 
probe $x_{\rm eff} =m_V/2E_N$ since $s_{\gamma N} = 2E_N m_V$ when 
$E_N \gg m_V, m_N$, corresponding to $x_{\rm eff} \equiv 2.5 \times 
10^{-3}$ for $\Upsilon$ and $7.5 \times 10^{-4}$ for $J/\psi$.
Measurements at lower $x_{\rm eff}$ (higher effective energy) would require 
identifying which nucleus emitted the photon.
An advantage of studying quasi-elastic reactions is the dissociation  
of the nucleus that absorbed the photon.
As a result, the quasi-elastic $x_{\rm eff}$ range is a 
factor of 10 higher than coherent processes because the measurement is not
restricted to midrapidity.
Measurements of low $p_T$ $J/\psi$ production away from $y=0$
appear to be easier for several of the detectors.  At forward rapidity,
the difference between the minimum $x$ reached in breakup processes
and coherent production is even larger.

Processes with rapidity gaps are most interesting 
for sufficiently large vector meson $p_T$ since they probe 
whether the elementary reaction $\gamma j \to V \, + {\rm jet}$
where $j$ is a parton, leading to
$\gamma A(N) \to V \,  + \, {\rm rapidity \, gap} +X$, is 
dominated by elastic scattering of small $q\overline q$ dipole
components of the photon wavefunction with partons in the
nucleon.  Light vector mesons, including the $\rho^0$, are then also
effective probes.  Such reactions are an effective 
way of studying the properties of perturbative colorless 
interactions in the vacuum (the ``perturbative Pomeron") at 
finite $t$.  The LHC kinematics and detector acceptances would
greatly increase the energy range covered by HERA.  Nuclear scattering
would provide a complementary method of studying the dynamics
of small dipole propagation through the nuclear medium.  UPCs
at the LHC are expected to reach both the large $t$ and moderate
$W$ regime where the onset of the perturbative color transparency
limit, $\sigma\propto A$, is expected as well as the onset of the BDR 
at large $W$ where $\sigma \propto A^{1/3}$.

{\it UPCs in pA interactions}

Proton-nucleus collisions are also an important 
part of the LHC program.  Ultraperipheral $pA$ studies will further
extend the HERA range for several 
important processes.  The small $x$ gluon densities can be studied through
heavy quark production by photon-gluon fusion when the gluon comes from the
nucleus and, in the diffractive case, when the gluon comes from the Pomeron.

Exclusive $J/\psi$ production should be able to determine whether the growth 
of the $J/\psi$ cross section with $W$ decreases as the BDR is approached. 
If the proposed forward proton counters at 420 m downstream are
approved \cite{loi}, accurate measurements of the 
$t$-dependences of these reactions 
could determine the transverse gluon distribution over a wide $x$ range.  In
contrast, HERA could not directly detect protons and had to rely on vetoing.
Measurements of the $\Upsilon$ photoproduction cross section
could verify the prediction that the cross section should increase
as $W_{\gamma p}^{1.7}$ \cite{Frankfurt:1998yf,Martin:1999rn}.  

The ATLAS and CMS detectors can study vector meson production both as functions
of the vector meson rapidity and the rapidity gap, $\Delta y$, between
the vector meson and other produced particles.
While $\Delta y_{\rm max} \sim 2$ at HERA, at the LHC
$\Delta y_{\rm max} \sim 8$, making studies of Pomeron
dynamics much more effective.

In summary, UPC studies in $pA$ interactions will probe the small $x$ dynamics 
for $x\ge 10^{-4}$ in a number of complementary ways. They will address 
the high density regime, a primary motivation 
for the proposals to extend HERA running beyond 2007 \cite{LOIH1}, 
with the added advantage of much higher densities than accessible
in $ep$ collisions.  Since these measurements will cover the $x$ range probed
in $AA$ collisions at the LHC,  
these studies are also important for understanding the $AA$ collision
dynamics.

\subsection{Overview of interesting $\gamma\gamma$ processes}

Two-photon collisions are fundamental processes that have previously 
been studied at every lepton collider,
particularly in $e^+ e^-$ at the CERN LEP and also in $ep$ at HERA.  There 
are three
areas of two-photon physics that may be studied using UPCs at the LHC: QED
processes in strong electromagnetic fields; QCD processes; and new physics
searches.

At low photon energies, QED processes in strong
electromagnetic fields can be studied. The photon-ion coupling constant is
$Z\alpha\approx 0.6$.  Therefore Coulomb corrections,
processes beyond leading order, can become
important. In the case of $e^+ e^-$ pair production, higher-order 
processes can be studied either as unitarity corrections,
resulting in multiple pair production in single collisions, or as
Coulomb corrections, giving a reduction relative to the Born cross 
section. Together with the possibility of tagging additional nuclear
excitations, these processes can be studied at small impact parameter where 
the effects may be enhanced. 

An important beam-physics effect is ``bound-free pair
production'' or ``electron capture from pair production'', a
pair production process where the electron is produced in a bound
state with one of the ions. As the $Z/A$ ratio changes, the
ion is no longer kept in the beam. These ions then hit the wall of the
beam pipe, leading to large heating
and potentially quenching the superconducting magnets. This
is the dominant process restricting the maximum Pb+Pb luminosity at the
LHC. They also cause approximately half of the beam losses
and therefore shorten the heavy-ion beam lifetime.  Bound-free pair production
was observed during the 2005 RHIC Cu+Cu run \cite{BFPP}.

At higher photon energies, QCD two-photon
processes may be of interest. The large photon flux allows more detailed
studies of processes that are separable from diffractive
$\gamma A \to X A$ processes.  In double vector meson production, 
not only light mesons like $\rho^0 \rho^0$ but also $J/\psi J/\psi$ or
pairs of two different vector mesons could be studied. Vector meson pair
production can be distinguished from production of two independent vector
mesons in coherent $\gamma A$ scattering since the transverse momenta
of two vector mesons produced in $\gamma \gamma$ processes are much larger and
back-to-back. 

The high photon energies and the correspondingly large available two-photon 
invariant mass, together with the large photon flux, motivated previous
new physics searches such as Higgs and supersymmetric particle production 
in two-photon interactions.
However, experimental limits on the masses of many new particles have 
increased in recent years, making their discovery in $\gamma \gamma$ processes
at the LHC unlikely.  The parameter space for production beyond the Standard
Model may still be explored. In $pp$ collisions, it is possible to tag
the photons if they have lost more than 10\% of their energy, making
electroweak studies of $\gamma \gamma$ or $\gamma W$ processes possible.
Although the cross section are not large, the higher energies, longer runs
and high beam luminosities in $pp$ collisions offer some advantages.

\section{Exclusive photonuclear processes}

\subsection{Introduction}
{\it Contributed by: L.~Frankfurt, V.~Guzey, M.~Strikman, 
R.~Vogt, and M.~Zhalov} \\

During the last decade, studies of small $x$ phenomena at HERA have revealed
that, at the highest energies available in $ep$ collisions, 
the interaction strength becomes comparable to the
maximum allowed by unitarity over a wide range of $Q^2$.
An increase in interaction energies and/or the extension to ion beams
is needed to reach higher interaction
strengths. 

The most practical way to carry out such a program 
in the next decade appears to be investigation of photon-nucleus interactions
at the LHC \cite{Baur:2001jj,FELIX,alice}. 
Though it is not possible to vary the virtuality of the photon in photonuclear 
interactions, as in lepton-nucleus scattering, the isolated nature of 
direct photon events provides an effective means of determining the
virtuality of the probe. An important advantage of ultraperipheral heavy-ion
collisions relative to the HERA program is the ability to simultaneously
study $\gamma N$ and $\gamma A$ scattering, making it possible
to investigate the onset of a variety of hard QCD phenomena leading to a new 
strong interaction regime including:
color transparency and color opacity; leading-twist nuclear shadowing
and the breakdown of linear QCD evolution, 
These phenomena will be clearer in hard scattering with nuclear beams since
the onset should occur at larger $x$ than in nucleons.
In general, nuclear 
targets are ideal probes of the space-time evolution of small dipoles of size 
$d$ which can be selected in high energy $\gamma N$
scattering either by considering small $x$ processes with 
$Q^2 \propto 1/ d^2$, or by studying special diffractive processes such 
as quarkonium or dijet production.  Understanding the 
space-time evolution has consequences for other branches of physics,
including the early universe since the emergence of color-singlet clusters
may play a role in the quark-hadron transition.  
This program makes it possible to
study coherent (and some incoherent) photonuclear interactions 
at energies which exceed those at HERA by at least an order of magnitude.
Thus coherent UPC studies at the LHC will answer a number of fundamental 
questions in QCD.  They will identify and investigate a new regime of strong 
interactions
by probing the dependence on the projectile, the final state, and
the nuclear size and thickness.

Several QCD regimes may be accessible, depending on the incident energy, 
the $Q^2$ of the process and the nuclear thickness. High-energy interactions 
of hadrons with nuclei  rapidly approach the black-disk regime (BDR) where 
the total interaction cross section is $\approx 2\pi R_A^2$ where 
$R_A \simeq 1.2 A^{1/3}$.  At another extreme, the photon interacts like a 
small color singlet dipole. In this case, the system remains 
small over a wide energy range while traversing the nucleus, known as
color transparency. In this regime, small dipole interactions with nuclei 
are rather weak and proportional to $A$.  Color 
transparency predicts that the forward scattering cross section in $\gamma A$
collisions should be proportional to $A^2$ since the amplitude is proportional
to $A$.  Color transparency 
has recently been observed in exclusive dijet production by coherent 
diffraction in $\pi A$ interactions \cite{E791}.  A similar $A$ 
dependence has also been observed in coherent $J/\psi$ 
production in fixed-target $\gamma A$ interactions at FNAL \cite{Sokoloff}.
At higher energies, the interactions of small color dipoles may be
described in the perturbative color opacity regime. Here, the 
dipole still couples to the gluon field of the nucleus through the nuclear 
gluon density, $g_{A}(x,Q^2)$, as in the color 
transparency regime. However, the scattering amplitude is not $\propto A$ 
due to leading-twist (LT) shadowing, resulting in $g_A(x,Q^2)/Ag_N(x,Q^2) < 1$.
The onset of LT gluon shadowing partially tames the increase of 
$g_A(x,Q^2)$ for $10^{-4} < x < 10^{-2}$, slowing the increase of the 
dipole-nucleus cross section with energy.  However, the reduction of 
$g_A(x,Q^2)$ at small $x$
is insufficient to prevent the LT
approximation of the total inelastic cross section from reaching 
and exceeding its maximum value, violating unitarity, an unambiguous 
signal of the breakdown of the LT approximation at small $x$.
We will discuss how to unambiguously distinguish between 
leading-twist nuclear shadowing and the blackening of hard 
interactions.

It is important to determine 
whether dipole-nuclear interactions are strongly modified by LT shadowing 
at small $x$ \cite{Frankfurt:1998ym}. Some models neglect this effect
\cite{McLerran:1994vd} and focus on higher-twist effects, often modeled using 
the impact-parameter space eikonal approach
\cite{Mueller:1988xy,Kovchegov:1997pc}. If LT shadowing was small and
only higher-twist effects reduced the increase of
the dipole-nucleus cross section, the DGLAP
approximation of parton evolution would break down at rather large $x$. 
On the other hand, the DGLAP breakdown may be due to the onset of the BDR, 
taming the dipole-nucleus cross section at smaller $x$.  We argue 
that the relative importance of leading and higher-twist contributions could 
be experimentally resolved using coherent quarkonium photoproduction.

If LT gluon shadowing effects are small, 
$q\bar q$ dipoles with 
$d \geq 0.3 - 0.4$ fm could be in the BDR in central $AA$ collisions 
at ${\it x}\geq 10^{-3}$, the kinematic regime where 
$\ln x$ effects on the parton evolution are also small.
In any case, the limiting behavior of the dipole-nuclear interaction is of 
great theoretical interest since it represents a new regime of strong 
interactions where the LT QCD approximation, and 
therefore the notion of parton distributions, 
becomes inapplicable at small 
$x$ even though $\alpha_s$ is small.  
We emphasize that, besides higher parton densities in nuclei,
the dependence of the scattering amplitude on impact parameter is rather
weak over a wide range of $b$.  Thus the dependence of the amplitudes on
the nuclear thickness can be studied by employing both heavy and light
nuclear targets.  On the other hand, nucleon scattering at large $b$ is
important at small $x$, making the change of interaction regime at small $b$
and leading to different energy dependencies of the deep-inelastic scattering
cross sections for nucleons $(\propto \ln^3 s)$ and nuclei $(\propto \ln s)$.
In hard diffraction, the forward cross sections and the $t$
dependence of the slope parameter $B$ also increase 
rapidly with energy: $\sigma \propto \ln^{4}s$ and $B \approx B_0 + B_1 
\ln^2 s$ respectively.

Theoretical studies of the limiting behavior of the dipole-nucleus cross 
sections have so far not produced any definitive results.  QCD dynamics may 
slow the increase of the dipole-nucleus cross section at central impact 
parameters ($b \sim 0$) at significantly
larger $x$ than allowed by the BDR. In the following discussion, we assume
that the BDR is reached at small $b$ to emphasize the distinguishing features
of the new regime where the elastic and inelastic cross sections are equal.

In many processes, the projectile wavefunction may be described as 
a superposition of different size configurations 
($q \overline q$, $q \overline q g$, {\it etc.})
leading  to fluctuations in the interaction strength. 
Interactions of real and virtual photons with heavy nuclei can therefore
provide unique information since the 
photon wavefunction contains both ``hadron-like'' configurations (vector 
meson dominance) and ``photon-like'' configurations (light $q\bar q$ 
components and heavy $Q \overline Q$ components).
In high-energy photonuclear interactions, the BDR is manifested by inelastic
diffraction of the photon into a multitude of hadronic final states
while elastic diffraction, $\gamma \rightarrow \gamma$, is negligible.
On the other hand, only elastic hadron diffraction survives in the BDR, hiding
the detailed dynamics.
Moreover, it is possible to post-select a small or large
configuration of the photon wavefunction by choosing a particular
final state. 
Such post-selection is more difficult for hadrons since the configuration
size distribution is wider for photons.

Spectacular manifestations of the BDR in (virtual) photon diffraction include
strong enhancement of the high mass tail of the diffractive spectrum
relative to the triple Pomeron limit and large dijet production cross
sections at high $p_T$ \cite{Frankfurt:2001nt}.
We emphasize that the study of diffractive channels can distinguish between 
the two scenarios of strong cross section suppression: leading-twist shadowing 
and the black-disk regime.
Studies of coherent diffraction in the BDR will
uniquely measure components of the light-cone photon wavefunction, providing 
more detailed information than similar 
measurements where leading-twist dominates.

\subsection{Color transparency, nuclear shadowing and quarkonium production}
{\it Contributed by: L.~Frankfurt, V.~Guzey, M.~Strikman, 
R.~Vogt, and M.~Zhalov} \\

The interaction of small color singlets with hadrons is one
of the most actively studied issues in high-energy QCD. 
In exclusive electroproduction of mesons at high $Q^2$ as well as 
$J/\psi$ and $\Upsilon$ photoproduction, the QCD factorization theorem 
separates the vector meson wave function at zero transverse separation into
the hard scattering amplitude and the generalized parton densities, making
evaluation of the vector meson production amplitude possible
\cite{Collins:1996fb,Brodsky:1994kf}\footnote{The proportionality of hard
diffractive amplitudes to the nucleon gluon density
was discussed for hard $pp$ diffraction \cite{FS89}, 
$J/\psi$ production \cite{Ryskin:1992ui} in the BFKL approximation,
and pion diffraction into two jets \cite{Frankfurt:1993it} in the leading
log $Q^2$ approximation \cite{Frankfurt:1993it}.}. 
The leading-twist approximation differs
strongly from predictions based on the Glauber model and 
two-gluon exchange models.  The LT approximation accounts for the
dominance of the space-time evolution of small quark-gluon wave packets in 
electroproduction, leading to the formation of a softer gluon field which 
effectively increases the dipole size with energy.

In perturbative QCD, similar to QED, the total cross section for the 
interaction of small systems with hadrons is proportional to the area
occupied by the color charge in the projectile hadron \cite{Low}, predicting
color transparency hard interactions with nuclei.
Incoherent cross sections are expected to be proportional 
to the nuclear mass number, $A$, while  the  coherent 
amplitude is proportional to $A$ times the 
nuclear form factor, $F$. The approximation of 
a quarkonium projectile as a colorless $Q \overline Q$ dipole can be formally 
derived from QCD within 
the limit $m_{Q}\to \infty$ and 
a fixed, finite momentum fraction, $x=4m_{Q}^2/s$ \cite{FKS}. In these
kinematics, the quarkonium radius is sufficiently small to justify
the applicability of pQCD.

It is important to determine the $Q^2$ in vector meson production where 
squeezing becomes effective and the dipole size decreases as $1/Q$. 
Perhaps the most sensitive indicator of small dipole size 
is the $t$-dependence of vector meson production. The current HERA data
are consistent with the prediction \cite{Brodsky:1994kf,FKS}
that the slopes of the $\rho^0$ and $J/\psi$ production amplitudes 
should converge to the same value.
Thus configurations much smaller than average, $d \sim 0.6$ fm
in light mesons, dominate small $x$ $\rho^0$
production at  $Q^2\geq 5$ GeV$^2$.  However, at all $Q^2$, $J/\psi$ 
production is dominated by small size configurations.
Therefore, color transparency is expected for $x \geq 0.03$ where gluon 
shadowing is either very small or absent. 

Color transparency (CT) was observed
at Fermilab~\cite{E791} with coherent dissociation in $\pi A \rightarrow {\rm
jet}_1 \, + \, {\rm jet}_2 +A$ 
interactions at 500 GeV. Diffractive masses of up to 5 GeV were
observed, consistent with two jets.  The results confirmed the
$A$ dependence and the $p_T$ and longitudinal jet momentum distributions
predicted in Ref.~\cite{Frankfurt:1993it}.
Color transparency was also
previously observed in coherent $J/\psi$ photoproduction
at $\left<E_{\gamma}\right>=120$ GeV \cite{Sokoloff}.

It is not clear whether CT will hold
at arbitrarily high energies since two phenomena are expected to counter 
it at high energies: leading-twist gluon shadowing and the increase of the 
dipole-nucleon cross section with energy. 

Leading-twist gluon shadowing predicts that the gluon distribution in a 
nucleus will be depleted at low
$x$ relative to the nucleon, $g_A(x,Q^2)/A g_N(x,Q^2) < 1$.
Such expectations are tentatively supported by the current analyzes of nuclear 
DIS although the data does not
extend deep enough into the shadowing region for confirmation.
Shadowing should lead to a gradual but calculable
disappearance of color transparency \cite{Frankfurt:1993it,Brodsky:1994kf} and 
the onset of a new regime, the color opacity regime.  It is possible to
consider color opacity to be generalized color transparency 
since the small $q\bar q$ dipole still couples to the gluon
field of the target by two gluons with an 
amplitude  proportional to the generalized nuclear gluon
density. 

The small dipole-nucleon cross section is expected to increase with energy 
as $xg_{N}(x,Q^2)$ where $x \propto 1/s_{(q \overline q)N}$.
For sufficiently large energies the cross section becomes comparable to 
the meson-nucleon cross sections which may result in significant suppression  
of hard exclusive diffraction
relative to the leading-twist approximation. 

While this suppression may be beyond the kinematics achievable
for $J/\psi$ photoproduction in UPCs at RHIC \cite{dde_qm05},
$x\approx 0.015$ and $Q^2_{\rm eff} \approx 4$ GeV$^2$, it could be important
in UPCs at the LHC. 
Thus systematic studies of coherent quarkonium production in ultraperipheral 
$AA$ interactions at collider energies should be very interesting.
We emphasize that the eikonal (higher-twist) contributions 
die out quickly with decreasing quarkonium size for fixed $x$. 
In particular, for the $\Upsilon$, nuclear gluon fields at transverse scale
$\sim 0.1$ fm ($Q^2_{\rm eff} \sim 40$ GeV$^2$) are probed.  
The $J/\psi$ is  closer to the border
between the perturbative and nonperturbative domains. As a result, the
nonperturbative region appears to give a significant contribution to
the production amplitude \cite{Frankfurt:2000ty}.

\begin{figure}
\centering
  \includegraphics[totalheight=0.45\textheight]{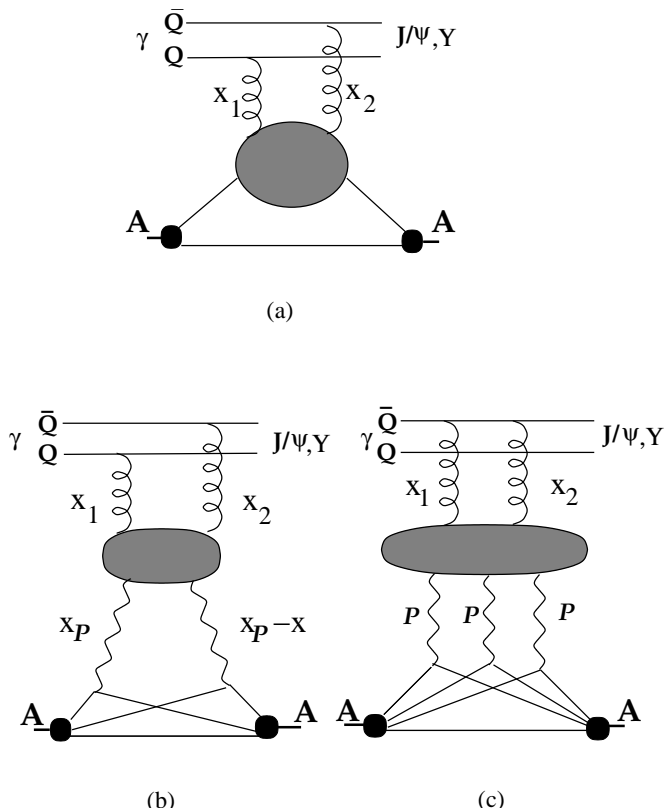}
\caption[]{Leading-twist diagrams for quarkonium production from nuclear
targets.}
\label{ltdiag}
\end{figure}

We now discuss the quarkonium photoproduction amplitude,
$\gamma  A\to J/\psi \, (\Upsilon) A$, in greater detail.
The $W_{\gamma p}$ range
probed at the LHC corresponds to rather small $x$.
The key theoretical issue is how to properly incorporate nuclear shadowing.  
A number of coherent mechanisms have been
suggested.  Here leading-twist shadowing, shown in the diagrams of 
Fig.~\ref{ltdiag}, is employed.
There is a qualitative difference between the interaction of a small dipole 
with several nucleons and a similar interaction with a single hadron.
For example, we consider an interaction with two nucleons.
The leading-twist contribution is described by diagrams
where two gluons attach to the
dipole. To ensure that the nucleus remains intact, 
color singlet lines should be attached to both nucleons.
These diagrams, especially Fig.~\ref{ltdiag}(b), are closely related
to those describing diffractive gluon densities measured
at HERA and thus also to
similar diagrams for nuclear gluon shadowing \cite{Frankfurt:1998ym}.

The amplitude for coherent 
quarkonium photoproduction is proportional to the generalized gluon 
density of the target,
$G_A(x_1,x_2,t,Q_{\rm eff}^2)$,
which depends on the light-cone fractions $x_1$ and $x_2$ of the two gluons 
attached to the quark loop, as shown in the top parts of the diagrams in
Fig.~\ref{ltdiag}. The momentum fractions satisfy the relation
\begin{equation}
x_1-x_2={m^2_V\over s_{(q \overline q)N}}\equiv x \, \, .
\label{skewed}
\end{equation}
If Fermi motion and binding effects are negligible,
$x_2\ll x_1$. The resolution scale, $Q_{\rm eff}$, is large,
$Q_{\rm eff}^2\geq m_Q^2$ where $m_Q$ is
the heavy quark mass.
Numerical estimates of $J/\psi$ photoproduction give
$Q^2_{\rm eff} \sim 3- 4$ GeV$^2$ \cite{FKS,Frankfurt:2000ty},
reflecting the relatively small charm quark mass and 
indicating that this process bridges the nonperturbative 
and perturbative regimes.
On the other hand, the bottom quark mass
is very large on the scale of soft QCD. In this case, hard physics 
dominates and the effect of attaching more than two gluons to the $b\overline 
b$ is negligible.  The QCD factorization theorem then provides a reliable  
description of $\Upsilon$ production.
Higher-twist effects due to the 
overlap of the $b\overline b$ component of 
the photon and the $\Upsilon$ cancel in the ratio of $\Upsilon$ production
on different targets.  As a result, in the leading-twist shadowing 
approximation, the $\gamma A\to \Upsilon A$ cross section is proportional to 
the square of the generalized nuclear gluon density so that
\begin{eqnarray}
\sigma_{\gamma A\to V A}(s_{\gamma N}) & = &  
{d \sigma_{\gamma N \to V N}(s_{\gamma N})\over dt}\bigg|_{t=t_{\rm min}}
 \Biggl [\frac {G_{A}(x_1,x_2,t=0,Q_{\rm eff}^2)}
{AG_{N}(x_1,x_2,t=0,Q_{\rm eff}^2)}\Biggr ]^2 \nonumber \\
&  & \mbox{} \times
\int \limits_{-\infty}^{t_{\rm min}} dt
{\left|
\int d^2bdz e^{i{\vec q_T}\cdot {\vec b}}
e^{-i q_{l} z}\rho_A ({\vec b},z)
\right|
}^2 \, \, .
\label{phocs}
\end{eqnarray}

Numerical estimates using realistic potential model wave
functions indicate that for  $J/\psi$, 
$x_2/x_1 \sim 0.33$ \cite{Frankfurt:2000ty} while for the $\Upsilon$, 
$x_2/x_1\sim 0.1$ \cite{Frankfurt:1998yf}.
Models of generalized parton distributions (GPDs) at moderate $Q^2$ 
suggest that, for any hadron or nucleus, $G(x_1,x_2,t=0,Q^2)$ can be 
approximated by the inclusive gluon density, $g(x,Q^2)$, at $x=(x_1+x_2)/2$ 
\cite{Brodsky:1994kf,Radyushkin:2000uy}. 
At large $Q^2$ and small $x$, the GPDs
are dominated by evolution from $x_i^{\rm init} \gg
x_i$. Since evolution on the gluon ladder conserves $x_1-x_2$,
the effect of skewedness ($x_2/x_1 < 1$) is
determined primarily by evolution from nearly diagonal ($x_1 \sim x_2$) 
distributions \cite{MF}.

Skewedness
increases the $\Upsilon$ cross section by a factor of $\sim 2 $ 
\cite{Frankfurt:1998yf,Martin:1999rn}, potentially obscuring
the connection between the suppression of the cross section discussed above 
and nuclear gluon shadowing. 
However, Ref.~\cite{Frankfurt:2000ty} showed that 
the ratio $G_A(x_1,x_2,t,Q_{\rm eff}^2)/AG_N(x_1,x_2,t,Q_{\rm eff}^2)$ 
is a weak function of $x_2$ at $t=0$,
slowly dropping from the diagonal value, $x_2=x_1$, for decreasing $x_2$.  
This observation suggests that it is more appropriate to compare the
diagonal and non-diagonal (skewed) ratios at $x=(x_1+x_2)/2$.

In the following, the ratio of generalized nuclear to nucleon 
gluon densities is approximated by the ratio of  
gluon densities at $x=m_V^2/s_{(q \overline q)N}$,
\begin{eqnarray}
\frac{G_A(x_1,x_2,t=0,Q_{eff}^2)}{A G_N(x_1,x_2,t=0,Q_{eff}^2)} \approx
\frac{g_A(x, Q_{eff}^2)}{A g_N(x, Q_{eff}^2)} \, \, . 
\label{gpdrat}
\end{eqnarray} 
 For the 
$\Upsilon$, $x/2$ may be more appropriate, leading to slightly 
larger shadowing effects than with $x$ alone. 

Reference~\cite{Frankfurt:1998ym} showed that nuclear shadowing
may be expressed in a model-independent
way, through the corresponding diffractive parton densities, using
the Gribov theory of inelastic shadowing~\cite{Gribov:1969,Gribovinel} and
the QCD factorization theorem for the hard diffraction
\cite{Collins:1997sr}.  HERA demonstrated that
hard diffraction is dominated by the leading-twist contribution with
gluons playing an important role in diffraction,
referred to as ``gluon dominance of the Pomeron''. Analysis of diffractive 
HERA data indicates that the probability of diffraction
in gluon-induced processes is significantly 
larger than in quark-induced processes
\cite{Frankfurt:1998ym}. The recent H1 data on diffractive dijet production
\cite{H1} provide an additional confirmation of this observation.
The large probability of diffraction in gluon-induced hard scattering
can be understood in the $s$-channel language as the formation of large
color-octet dipoles which can diffractively scatter with a correspondingly
large cross section. The interaction strength can be quantified using
the optical theorem, introducing the effective cross section, 
\begin{eqnarray}
\sigma_{{\rm eff}}^g(x,Q_0^2)&=&\frac{16 \pi}{\sigma_{{\rm tot}}(x,Q_0^2)}
\frac{d \sigma_{{\rm diff}}(x,Q_0^2, t_{{\rm min}})}{dt} \nonumber\\
&=&\frac{16 \pi}{xg_N(x,Q_0^2)}
\int_x^{x_{\Pomeron}^0} d x_{\Pomeron} \,\beta 
g_N^D(\frac{x}{x_{\Pomeron}},x_{\Pomeron},Q_0^2,t_{{\rm min}}) \, \, 
\label{sigef}
\end{eqnarray}
for hard scattering of a virtual photon off the gluon field
of the nucleon. Here $Q_0^2 = 4$ GeV$^2$ is the resolution scale for the
gluons; $x_{\Pomeron}$ is the longitudinal momentum fraction
of the Pomeron; $x_{\Pomeron}^0=0.1$; and
$g^D_N$ is the diffractive gluon density of the nucleon, known from the H1
Fit B diffractive analysis \cite{unknown:2006hy,:2006hx}.
While this coherent mechanism may effectively be absent for
$x\geq 0.02 - 0.03$, it may quickly become important at smaller $x$. 

\begin{figure}[h]
\begin{center}
\epsfig{file=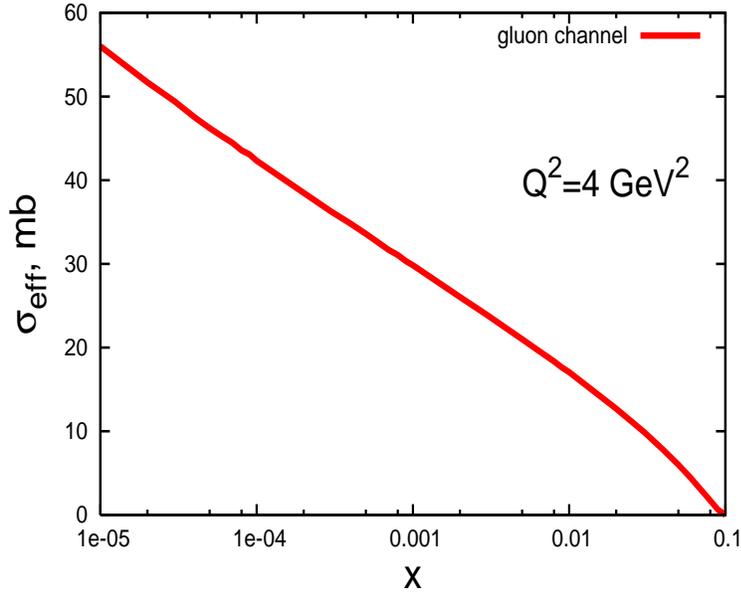,width=10cm,height=8cm}
\caption[]{
The effective gluon shadowing cross section,
$\sigma^g_{\rm eff}(x)$, at $Q^2=4$ GeV$^2$ as a function of $x$ for the 
H1 parameterizations of the diffractive gluon density.}
\label{figsigef}
\end{center}
\end{figure}

The ratio of the inclusive gluon densities in 
Eq.~(\ref{gpdrat}) is calculated 
using leading-twist shadowing~\cite{Frankfurt:1998ym}, see 
Ref.~\cite{Frankfurt:2003zd} for details. 
First,
the nuclear gluon density, including leading-twist shadowing is calculated at 
the minimum scale, $Q_{0}^2=4$ GeV$^2$ 
\begin{eqnarray}
g_A(x,Q_0^2) &=& A g_N(x,Q_0^2) - 8 \pi {\rm Re} \Bigg[
\frac{(1-i\eta)^2}{1+\eta^2} \label{ga} \\ 
\mbox{} &   & \times \int d^2 b \int^{\infty}_{-\infty} dz_1
 \int^{\infty}_{z_1} dz_2 \int_x^{x_{\Pomeron}^0} d x_{\Pomeron}
\beta g_N^D(\frac{x}{x_{\Pomeron}},x_{\Pomeron},Q_0^2,t_{{\rm min}}) 
\nonumber \\
& & \times 
\rho_A(b,z_1)\rho_A(b,z_2) \,e^{i x_{\Pomeron} m_N(z_1-z_2)}\,
e^{-\frac{1-i\eta}{2} \sigma_{\rm eff}^g(x,Q_0^2) \int^{z_2}_{z_1}
dz^{\prime} \rho_A(b,z^{\prime})} \Bigg] \,, \nonumber
\end{eqnarray}
where $\eta$ is the ratio of the real to imaginary parts of the elementary
diffractive amplitude.
The H1 parametrization of
$g^{D}_{N}(x/x_{\Pomeron},x_{\Pomeron},Q_{0}^2,t_{\rm min})$
is used as input.
The effective cross section, $\sigma^g_{\rm eff}(x,Q_{0}^2)$, determined by 
Eq.~(\ref{sigef}), accounts for elastic rescattering of the produced 
diffractive state with a nucleon.  Numerically, $\sigma_{\rm eff}^g$ is
very large at $Q_0^2$, see
Fig.~\ref{figsigef}, and corresponds to a probability for gluon-induced 
diffraction of close to $\sim 50$\% at $x \sim 10^{-5}$ (see 
Fig.~\ref{fig:Pdiff_nucleon}).  Consequently at $Q_0^2$, gluon interactions 
with nucleons approach the BDR at $x\sim 10^{-4} - 10^{-5}$ while, for nuclei, 
a similar regime should hold for $x \leq 10^{-3}$ over a large range of impact 
parameters.

The double scattering term in Eq.~(\ref{ga}), proportional to 
$\sigma_{g \, {\rm eff}}$, for the nuclear parton densities 
satisfies QCD evolution, while higher-order terms (higher powers of 
$\sigma_{g \, {\rm eff}}$) do  not. 
Thus if a different $Q_0^2$ is used, a
different $g(x,Q^2)$ would be obtained since the higher-order terms, 
$\propto (\sigma^g_{\rm eff})^n$, $n\geq 2$ are
sensitive to the $Q^2$-dependent fluctuations in the diffractive cross 
sections. The $Q^2$ dependence of the fluctuations are 
included in the QCD evolution, violating the Glauber-like 
structure of shadowing for $Q^2>Q_0^2$.
The approximation for $n\geq 3$ in Eq.~(\ref{ga})
corresponds to the assumption that the
fluctuations are small at $Q_0^2$ since this scale is close to the soft
interaction scale \cite{Frankfurt:1998ym}.
Thus we use NLO QCD evolution
to calculate shadowing at larger $Q^2$ using the $Q^2_0$ result as a boundary 
condition.  We also include gluon enhancement at $x\sim 0.1$ which 
influences shadowing at larger $Q^2$.  The proximity to the BDR, reflected in 
large $\sigma_{g \, {\rm eff}}$, may result in
corrections to the LT evolution.

\begin{figure}[h]
\begin{center}
\epsfig{file=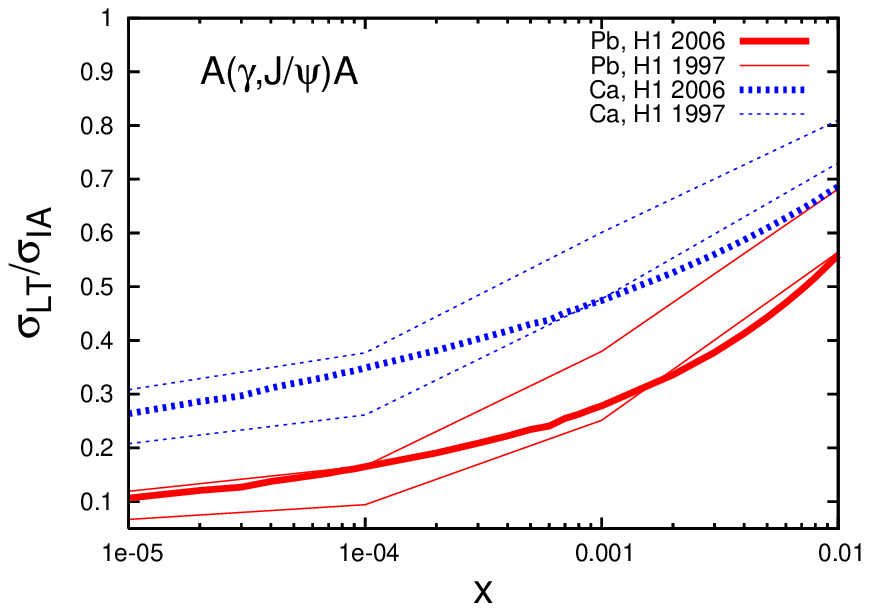,width=7cm,height=7cm}
\epsfig{file=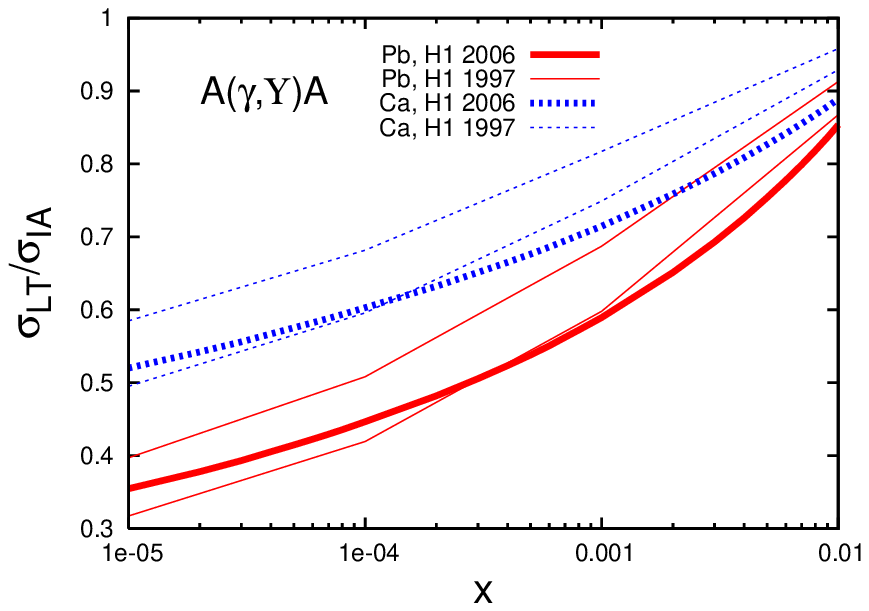,width=7cm,height=7cm}
\caption[]{The $x$ dependence of the ratio 
of $J/\psi$ and $\Upsilon$ production in Ca+Ca and Pb+Pb interactions in the 
leading-twist Glauber model (LT) to that
in the Impulse Approximation (IA), without shadowing.
Calculations employing the new H1
parametrization~\protect\cite{unknown:2006hy,:2006hx} of the diffractive
PDFs are compared to those of Ref.~\protect\cite{Frankfurt:2003zd}.}
\label{fshad}
\end{center}
\end{figure}

We first present the coherent $J/\psi$ and $\Upsilon$ photoproduction cross
section ratios in ultraperipheral $AA$ collisions, shown in Fig.~\ref{fshad}.
The thick curves show the leading-twist shadowing results using the recent H1
Fit B to the hard diffraction DIS data~\cite{unknown:2006hy,:2006hx}.
The thin curves are calculations~\cite{Frankfurt:2003zd} using the older H1
fits from 1995~\cite{FP}. The spread between the two thin curves
corresponds to the difference between the ``low'' and ``high" gluon shadowing
scenarios introduced in Ref.~\cite{Frankfurt:2003zd}.

We point out that, as shown in Fig.~\ref{fshad},
the leading-twist
shadowing predictions using the 2006 H1 fits are consistent with the
theoretical band determined from the 1997 H1 diffractive fits.
Therefore, predictions made using the results of Ref.~\cite{Frankfurt:2003zd}
elsewhere in this report should not change much when the more recent
fits are used.

Finally, we comment on the difference between the present and earlier
leading-twist shadowing predictions using the H1 fits to hard diffraction in
DIS with protons.
First, in the analysis of Ref.~\cite{Frankfurt:2003zd}, the 1997 H1
diffractive gluon PDF was multiplied by 0.75 to
phenomenologically account for the difference between the 1994 and 1997 H1
hard diffractive gluon data.
Second, in earlier analyses \cite{Frankfurt:2003zd1}, 
a somewhat larger slope for the
$t$-dependence of the gluon diffractive PDF was used.
Thus, predictions of leading-twist nuclear gluon shadowing made with the
unmodified 1997 H1 fits and the larger slope result in larger nuclear
shadowing relative to Ref.~\cite{Frankfurt:2003zd}.
Thus the FGS calculations presented in Section~\ref{section-pdf}, based on 
the earlier predictions \cite{Frankfurt:2003zd1},
somewhat overestimate the shadowing effect compared to predictions based on
the most recent fits~\cite{unknown:2006hy,:2006hx}, see Fig.~\ref{fshad}.

The ratios in Fig.~\ref{fshad} are independent of uncertainties in the 
elementary cross sections, providing a sensitive test of LT shadowing effects.
In the case of $J/\psi$ photoproduction, $Q^2 \sim 4$ GeV$^2$.  A significant 
fraction of the amplitude comes from smaller virtualities 
\cite{FKS,Frankfurt:2000ty} which may result in a larger shadowing 
effect.  We use $Q^2 =40$ GeV$^2$ to calculate $\Upsilon$ photoproduction.
However, the result is not very sensitive to
the precise $Q^2$, since the scale dependence of
gluon shadowing at higher $Q^2$ is rather small. Despite 
the small $\Upsilon$ size, which precludes higher-twist
shadowing at very small $x$, the perturbative color opacity effect
is important.  The effective rescattering cross
section in the Glauber model is determined by the
dipole-nucleon cross section, $\sigma_{\rm in}^{(Q \overline Q) N}$, with
$d\sim 0.25 - 0.3$ fm for $J/\psi$ and $\sim 0.1$ fm for
the $\Upsilon$.  These distances correspond to cross sections of $\sim 10-15$ 
mb for $J/\psi$ and $\sim 3$ mb for $\Upsilon$ at $x\sim 10^{-4}$,
reducing the cross section by a factor of $\sim 1.5-2$ for $x\sim 10^{-3}$, 
see Fig.~13 in Ref.~\cite{Frankfurt:2000ty}. The cross sections are not reduced
as much as those calculated using LT gluon shadowing.

\begin{figure}
\centering
  \includegraphics[totalheight=0.45\textheight]{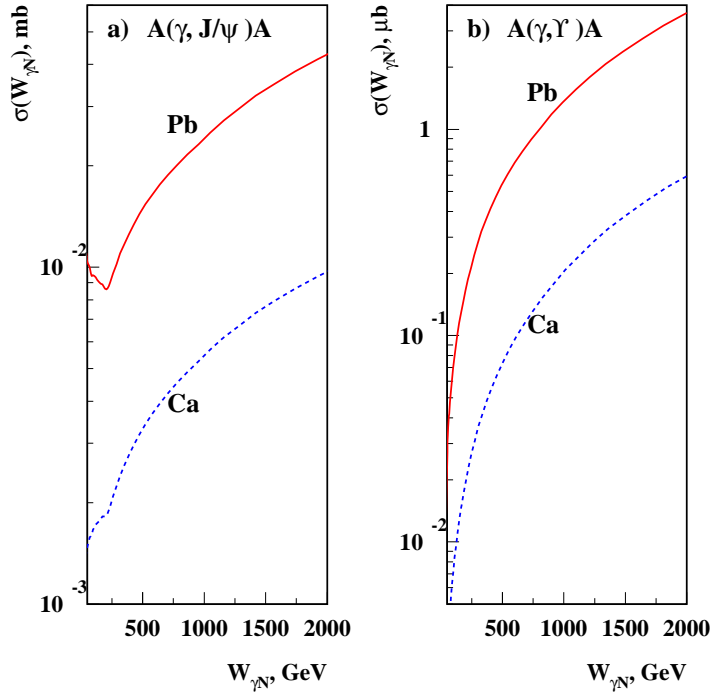}
\caption[]{The energy dependence of coherent $J/\psi$ and $\Upsilon$
photoproduction in ultraperipheral Ca+Ca and Pb+Pb collisions 
in the LT approximation. Reprinted from Ref.~\protect\cite{Frankfurt:2003wv}
with permission from Acta Physica Polonica.}
\label{onics}
\end{figure}

The absolute quarkonium photoproduction cross sections were estimated over a 
wide range of photon energies. 
The energy dependence of the momentum-integrated cross sections, 
$\sigma(W_{\gamma N})$ where
$W_{\gamma N}=\sqrt{s_{\gamma N}}$, is presented in Fig.~\ref{onics}. 
At low $W_{\gamma N}$, there is a dip in the $\gamma \, {\rm Pb}
\rightarrow J/\psi$ cross section due to the fast onset of
shadowing at $x \sim 10^{-3}$ in the leading-twist parameterization employed
in the calculation.
The $J/\psi$ calculations are straightforward since accurate HERA data are 
available. However, the $\gamma N\to \Upsilon N$ data are very limited,
with only ZEUS and H1 total cross section data for $\sqrt{s_{\gamma N}} 
\approx 100$ GeV, complicating the $\Upsilon$ predictions. 
A simple parametrization is used 
to calculate the photoproduction cross section,
\begin{equation}
 {d \sigma_{\gamma N\to VN}(s_{\gamma N},t)\over dt}=10^{-4}
B_{\Upsilon}\left({s_{\gamma N}\over s_0}\right)^{0.85}
\exp(B_{\Upsilon}t) \, \, 
\mu{\rm b/GeV}^2 \, \, ,
\label{eq:cs}
\end{equation}
where $s_0=6400$ GeV$^2$ and $B_{\Upsilon}=3.5$ GeV$^{-2}$ were fixed from
the analysis of the two-gluon form factor in 
Ref.~\cite{Frankfurt:2002ka}. The energy dependence follows from the 
$\Upsilon$ photoproduction calculations of Ref.~\cite{Frankfurt:1998yf}
in the leading $\log Q^2$ approximation, including the skewedness of the PDFs.

\subsection{Vector meson production}
{\it Contributed by: L.~Frankfurt, V.~Guzey, S.~R.~Klein, 
J.~Nystrand, M.~Strikman, R.~Vogt, and M.~Zhalov}
\label{vmaa}

Exclusive photonuclear vector meson production in relativistic heavy-ion 
interactions are interactions of the type 
\begin{equation}
A+A \rightarrow A+A+V
\end{equation}
where the nuclei normally remain intact. These interactions typically
occur for impact parameters much larger than the sum of the
nuclear radii, $b\gg 2R_A$, and proceed through an interaction between the
electromagnetic field of one of the nuclei with the nuclear field of
the other.  The experimental feasibility of studying these
interactions at heavy-ion colliders has been demonstrated by STAR
\cite{Adler:2002sc} and PHENIX \cite{dde_qm05} at RHIC.    

\subsubsection{Vector Meson Dominance Model} 
\label{vector}

\bigskip

Coherent production of light vector mesons, $V$, off nucleons and nuclei,
$\gamma A \rightarrow V A$, at high energies can be 
described within the framework of the Vector Meson
Dominance Model (VDM) or the Generalized Vector Meson Dominance Model
(GVDM) \cite{Gribov,Brodsky} reviewed in 
Refs.~\cite{DonnachieShaw,GVDM,Baur:iq}.

At low and moderate energies, the hadronic interaction of a
low-virtuality photon is dominated by quantum mechanical fluctuations
into a strongly interacting state, preferentially a vector meson,
with the quantum numbers of the photon, $J^{PC} = 1^{- -}$.

The photon wavefunction can be written as a sum of Fock states, 
\begin{equation}
|\gamma\rangle = C_{\rm pure} |\gamma_{\rm pure}\rangle + C_{\rho^0} 
|\rho^0\rangle + C_{\omega} |\omega\rangle + C_{\phi} |\phi\rangle 
+ C_{J/\psi} |J/\psi\rangle + \cdots + C_{q\overline{q}} |q\overline{q}\rangle 
\,\, ,
\label{decomp}
\end{equation}
where $|\gamma_{\rm pure}\rangle$ corresponds to a bare photon which may
interact with a parton in the target, $C_{\rm pure} \approx 1$.  The
amplitude, $C_V$, for the photon to fluctuate into vector meson $V$ is
proportional to the inverse of the photon-vector meson coupling,
$f_V$. This coupling can be related to the measured dilepton
decay width, $\Gamma_{V \rightarrow e^+e^-}$,
\begin{equation}
\left| C_V \right|^2 = \frac{4 \pi \alpha}{f_V^2} = 
\frac{3 \, \Gamma_{V \rightarrow e^+e^-}}{\alpha^{2} M_V } \, \, ,
\label{csubv}
\end{equation}
where $\alpha$ is the electromagnetic coupling constant and $M_V$ the 
vector meson mass. 

The VDM neglects contributions from non-diagonal transitions, {\it i.e.}
$\langle \rho^0 |\omega \rangle = 0$.  The GVDM includes these non-diagonal
transitions.  In such transitions, the photon fluctuates
into a different hadronic state from the observed final-state vector meson. 
The observed final state is produced by hadronic rescattering, $V' A 
\rightarrow VA$ where $V'$ is the initially-produced vector meson and $V$
the final-state meson.

Squaring Eq.~(\ref{decomp}) and assuming the diagonal approximation of the VDM,
the differential photoproduction cross section, $d\sigma_{\gamma A \to 
VA}/dt$, calculated using the Glauber scattering model, is
 \begin{eqnarray}
{d \sigma_{\gamma A\to VA}\over dt} & = &
\left. {d \sigma_{\gamma N\to VN}\over dt}\right|_{t=0} \nonumber \\
\mbox{} &  & \times \left|
\int d^2b \, dz \, e^{i{\vec q_T} \cdot {\vec b}} e^{iq_L z} \rho_A(b,z)
e^{-\frac {1} {2} \sigma^{VN}_{\rm tot}(1-i\epsilon)
\int \limits ^{\infty}_{z} dz' \, \rho_A(b,z')} 
\right|^2 \,  .
\label{dsig}
\end{eqnarray}
The square of the transverse momentum transfer in the $\gamma \to V$ 
transition, $|{\vec q}_T|^{~2}= |t_T| = |{t_{\rm min}-t}|$, depends on the
photon energy, $\omega$, through $t_{\rm min}$ since $-t_{\rm min}=
M_{V }^4/4\omega^2$.  The ratio of the real to imaginary parts of the
vector meson scattering amplitude is denoted $\epsilon$ in Eq.~(\ref{dsig}). 

The longitudinal momentum transfer, $q_L$, reflects the large longitudinal 
distances over which the transition $\gamma  \to V$ occurs.
The hadronic fluctuation extends over distance $l_c$, the
coherence length, restricted by the uncertainty principle so that
\begin{equation}
l_c =1/q_L= \Delta t c = \frac{\hbar c}{\Delta E} = \frac{\hbar c}{E_V -
E_{\gamma}} = \frac{2 E_{\gamma}}{M_V^2 + Q^2} \hbar c \, \, ,
\end{equation}
where $E_V$ is the vector meson energy while $E_{\gamma}$ and $Q$ are the 
energy and virtuality of the photon, respectively. In the limit where the 
coherence length is much larger than the nuclear radius, $l_c\gg R_A$, 
Eq.~(\ref{dsig}) is reduced to the usual Glauber expression for elastic 
hadron-nucleus scattering by making the substitutions
$(d \sigma_{\gamma N\to VN}/ dt)|_{t=0} \rightarrow (d \sigma_{V N\to 
VN}/dt)|_{t=0}$ and
$(d \sigma_{\gamma A\to VA}/ dt) \rightarrow (d \sigma_{V A\to 
VA}/dt)$.

In the nuclear rest frame, for light vector meson production at midrapidity
the limit $l_c \gg R_A$ holds  
at RHIC and LHC so that
\begin{equation}
\frac{d \sigma_{\gamma A \rightarrow V A}}{dt} = 
\left| C_V \right|^2 \frac{d \sigma_{VA \rightarrow VA}}{dt} \, \, .
\end{equation}
The exclusive photo-nuclear scattering amplitude is thus proportional to 
the amplitude for elastic vector meson scattering. If two vector meson 
states, $V$ and $V'$, contribute then non-diagonal transitions,  
$V' A \rightarrow V A$, have to be considered in GVDM \cite{Pautz:qm}. 
The more general expression for the scattering amplitude,
\begin{equation}
{\mathcal M}_{\gamma A \rightarrow V A} =
C_V \, {\mathcal M}_{VA \rightarrow VA} + 
C_{V'} \, {\mathcal M}_{V' A \rightarrow VA} \, \, ,
\end{equation}
is then needed.

The $t$-dependence of the differential cross section for coherent 
elastic scattering off a heavy nucleus is primarily 
determined by the nuclear form factor, $F(t)$, 
\begin{equation}
\frac{d \sigma_{\gamma A \rightarrow V A}}{dt} = \left. | F(t) |^2
\frac{d \sigma_{\gamma A \rightarrow V A}}{dt} \right|_{t=0} \, \, ,
\end{equation}
where $F(t)$ is the Fourier transform of the nuclear density distribution
The elastic cross section at $t=0$ is related to the total cross 
section, $\sigma_{\rm tot}$, by the optical theorem,
\begin{equation}
\left. \frac{d \sigma_{V A \rightarrow V A}}{dt} \right|_{t=0} = 
\frac{\sigma_{\rm tot}^2}{16 \pi} \left( 1 + \epsilon^2 \right) \, \, .
\end{equation}

\begin{figure}
\centering
    \epsfysize=0.4\vsize
     \epsffile{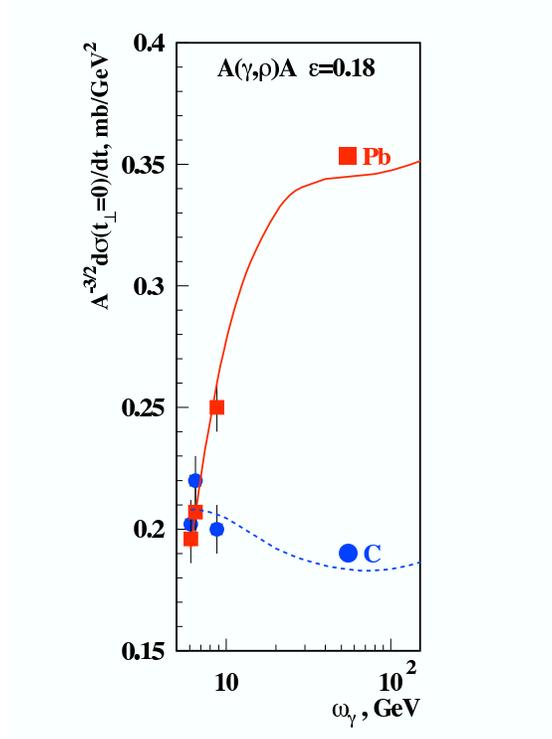}
\caption[]{The energy dependence of the $\rho^0$ photoproduction cross section
calculated in the GVDM with Glauber scattering.  The data are from
Ref.~\protect\cite{rhodata}.  Reprinted from 
Ref.~\protect\cite{Frankfurt:2002wc} with permission from Elsevier.}
\label{rhoen}
\end{figure}

The GVDM describes all available data at intermediate energies, see {\it e.g.} 
Fig.~\ref{rhoen} from Ref.~\cite{Frankfurt:2003wv}. Hence vector meson 
production is very useful for checking the basic approximations of UPC theory.

\subsubsection{Cross sections in heavy-ion colliders}
\label{section-aa-xsections} \bigskip

The first calculations of exclusive vector meson production at
heavy-ion colliders were made in Ref.~\cite{Klein:1999qj}. The model
is briefly described here.
The total photo-nuclear cross section is the convolution of the photon flux 
with the differential photo-nuclear cross section, integrated over 
the photon energy,
\begin{equation}
\sigma_{AA\rightarrow AAV} = 
\int_0^\infty dk \, {dN_\gamma(k)\over dk} \,
{d\sigma_{\gamma A\rightarrow VA}\over dt}\bigg|_{t=0}
\int_{-t_{\rm min}}^\infty dt \, |F(t)|^2 \, \, .
\label{eq:3:vmcross}
\end{equation}
Here $-t_{\rm min} = (M_V^2/2k)^2$ is 
the minimum momentum transfer squared needed to produce a vector meson of
mass $M_V$. The nuclear form factor, $F(t)$, is significant only for 
$|t| < (\hbar c/R_A)^2$.  Thus only photons with $k>M_V^2 R_A/2 \hbar c$ can
contribute to coherent production. 

The expression for $dN_\gamma/dk$ in Eq.~(\ref{analflux}) corresponds 
to the photon flux at the center of the target nucleus, $r = b$. 
The flux on the target surface 
will be higher near the photon-emitting projectile, $b-R_A < r < b$ and lower 
further away, $b < r < b+R_A$. In coherent interactions, where the fields 
couple to the entire nucleus or at least to the entire nuclear surface, a 
better estimate of the total flux is obtained by taking the average over the
target surface
\begin{equation}
\frac{dN_\gamma(k)}{dk} =  
2 \pi \int_{2R_A}^{\infty} db \, b 
\int_0^R  {dr \, r \over \pi R_A^2} 
\int_0^{2\pi} d\phi 
\ \ {d^3N_\gamma(k,b+r\cos \phi)\over dkd^2r} \, \, .
\label{eq:3:nofeave} 
\end{equation} 
The $r$ and $\phi$ integrals, over the surface of the target nucleus for 
a given $b$, are evaluated 
numerically.  A sharp cutoff at $b=2R_A$ in the lower limit of the integral
over $b$ treats the nuclei as hard 
spheres. In a more realistic model, accounting for the diffuseness of the 
nuclear surface, all impact parameters are included and the integrand is 
weighted by the probability for no hadronic interaction, $1 - P_H(b)$,
\begin{equation}
\frac{dN_\gamma(k)}{dk} =  2 \pi
\int_{0}^{\infty} \! db \, b \, [1 - P_H(b)]
\int_0^{R_A}  {dr \, r \over \pi R_A^2} 
\int_0^{2\pi} d\phi \!
{d^3N_\gamma(k,b+r\cos \phi)\over dkd^2r}  \; .
\label{eq:3:nofe} 
\end{equation} 
Here the probability of a hadronic interaction, $P_H(b)$, is often taken to
be a step function, $P_H(b) = 1$ for $b>2R_A$ and 0 otherwise.  Other, more
sophisticated approaches, make a $(10-15)$\% difference in the flux.
This expression, used for the photon flux in the following calculations, is 
compared to the analytical approximation, Eq.~(\ref{analflux}), 
in Fig.~\ref{fig:3:dndk} for Pb+Pb and Ca+Ca interactions at the LHC. 

As discussed previously, the optical theorem relates the forward scattering 
amplitude to the total interaction cross section, leading to the scaling
\begin{equation}
\frac{\left. d \sigma_{\gamma A \rightarrow V A} / dt \right|_{t=0} }
{\left. \, \, d \sigma_{\gamma N \rightarrow V N} / dt \right|_{t=0} } =
\bigg( \frac{\sigma_{\rm tot}^{VA}}{\sigma_{\rm tot}^{VN}} \bigg)^2 = A^\beta
\end{equation}
for $\gamma A$ relative to $\gamma N$ ($\gamma p$).
The total interaction cross section in nuclei is a function of the total
cross section on a nucleon and the absorption in nuclear medium.  Two
limits for the $A$ scaling can be obtained. First,
if $\rho_A R_A \sigma_{\rm tot}^{VN} \ll 1$, one expects scaling with target 
volume, $A$, and $\beta =2$.
When $\rho_A R_A \sigma_{\rm tot}^{VN} \gg 1$, 
the amplitude is proportional to 
the surface area of the target, $A^{2/3}$, and $\beta = 4/3$.

A more accurate estimate of the effect of absorption on
$\sigma_{\rm tot}^{VA}$ is obtained by a Glauber
calculation. In Refs.~\cite{Klein:1999qj,DrellTrefil}, 
the total cross section was calculated from the classical Glauber formula
\begin{equation}
\sigma_{\rm tot}^{VA} = \int d^2b \, [1 -
\exp(-\sigma_{\rm tot}^{VN} T_{A}(b)) ] \, \, .
\label{eqabs}
\end{equation}
where $T_{A}(b)$ is the nuclear profile function, normalized so that $\int d^2b
T_A(b) = A$.
Equation~(\ref{eqabs}) gives $\sigma_{\rm tot}^{VA} \approx 
\pi R_A^2$ for $\rho^0$ and $\omega$ production. The model input is based on
parameterizations of exclusive vector meson production data from HERA
and lower energy, fixed-target experiments.  We take $\sigma_{\rm tot}^{J/\psi 
N}(W_{\gamma p}) = 1.5 W_{\gamma p}^{0.8}$ nb from HERA data. 
We use Eq.~(\ref{eq:cs}) for the $\Upsilon$, in
agreement with the limited $\Upsilon$ HERA data 
\cite{Klein:2003vd,Klein:2003qc}. The total production cross sections in
different systems at RHIC and the LHC are given in
Table~\ref{tab:3:sigmavm}.

References \cite{Frankfurt:2002wc,Frankfurt:2002sv} compare the
classical, Eq.~(\ref{eqabs}), and quantum mechanical,
\begin{eqnarray}
\sigma^{VA}_{\rm tot} = 2 \int d^2b \left[1-\exp(-\sigma_{\rm 
tot}^{VN}T_A(b)/2)
\right] \, \, ,
\end{eqnarray} 
Glauber formulas. They also
include contributions from the cross term $\rho^{0 \prime} N \rightarrow
\rho^0 N$ and the finite coherence length, both of
which are neglected above.

\subsubsection{Comparison to RHIC data} \bigskip

\begin{figure}[h]
    \centering
        \leavevmode
        \epsfxsize=0.8\hsize
        \epsffile{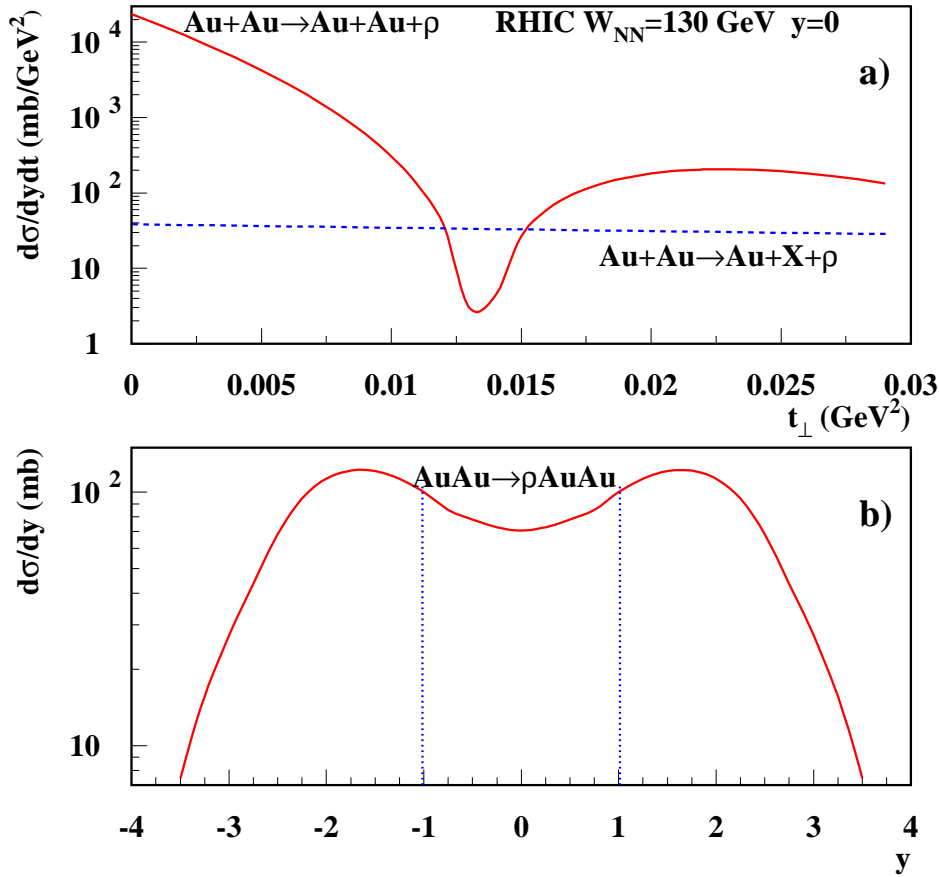}
\caption[]{The $t$ dependence of coherent (solid) and
incoherent (dashed) (a) and the coherent rapidity distribution (b)
of $\rho^0$ production in Au+Au UPCs at $\sqrt{s_{_{NN}}} = 130$ 
GeV, calculated in the GVDM ~\protect\cite{Frankfurt:2002sv}.  The photon
$p_T$ is neglected.  Copyright 2003 by the American Physical Society
(http://link.aps.org/abstract/PRC/v67/e034901).}
\label{figstar1}
\end{figure}

The STAR collaboration has measured 
the coherent $\rho^0$ production cross section in ultraperipheral Au+Au
collisions at $W_{NN}= \sqrt{s_{_{NN}}} = 130$ GeV \cite{Adler:2002sc},  the 
first opportunity to check the basic model features.  The primary 
assumptions include the Weizs\"{a}cker-Williams
approximation of the equivalent photon spectrum and the vector meson 
production model in $\gamma A$ interactions. 
The basic process is better understood for $\rho^0$ production than
other vector mesons. Hence, the $\rho^0$ study can prove that
UPCs provide new information about photonuclear interactions.
Inelastic shadowing effects remain a few percent  
correction at energies less than 100 GeV, relevant for the STAR kinematics.
In the LHC energy range, the blackening of nuclear
interactions should be taken into account. In this limit, inelastic 
diffraction in hadron-nucleus collisions should approach zero.  
Therefore the $\rho^{0 \prime}$ contribution to diffractive $\rho^0$ 
photoproduction is negligible \cite{Frankfurt:2002wc}. 
The $t$ distributions at $y=0$ and 
the $t$-integrated rapidity distribution for ultraperipheral Au+Au collisions 
at $\sqrt{s_{_{NN}}} = 130$ GeV  are presented in 
Fig.~\ref{figstar1} \cite{Frankfurt:2002sv}.  The photon $p_T$ spread, which
would smear the minimum in the $t$ distribution, and interference
are neglected.

\begin{figure}[h]
    \centering
        \leavevmode
        \epsfxsize=0.8\hsize
        \epsffile{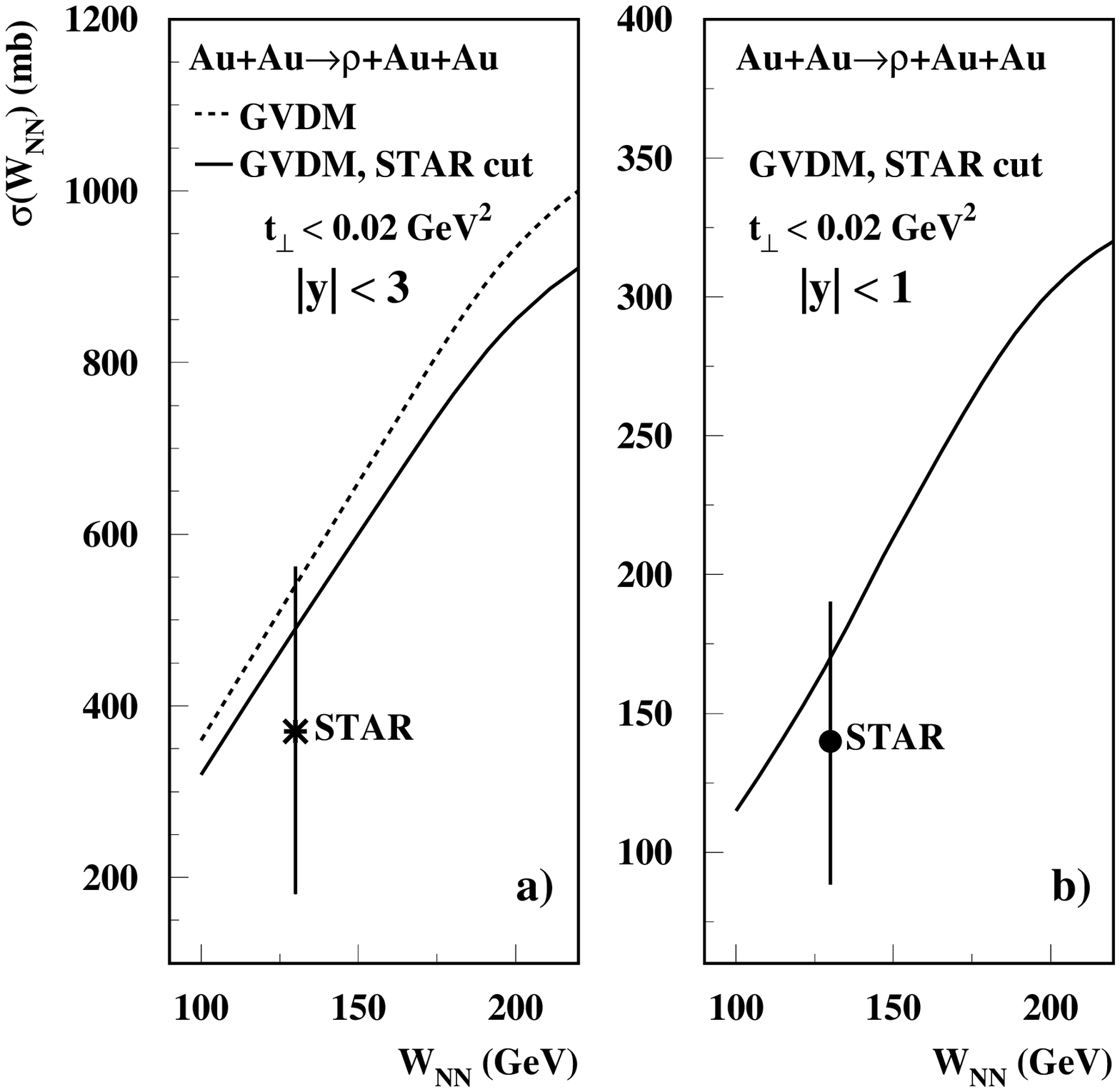}
\caption[]{The energy dependence of the total coherent $\rho^0$ production 
cross section in ultraperipheral Au+Au collisions, calculated in the GVDM
\protect\cite{Frankfurt:2002sv}.  Copyright 2003 by the American Physical 
Society (http://link.aps.org/abstract/PRC/v67/e034901).}
\label{figstar2}
\end{figure}

The total coherent $\rho^0$ production cross section at RHIC,
calculated in the GVDM, is shown in Fig.~\ref{figstar2} 
\cite{Frankfurt:2002sv}.  The cross section is $\sigma_{\rm coh}
= 540$ mb at $\sqrt{s_{_{NN}}} = 130$ GeV. STAR measured $\sigma_{\rm coh} =
370 \pm 170 \pm 80$ mb for $t_\perp \leq 0.02$ GeV$^2$. 
This $t_\perp$ cut, reducing the cross section by $\sim 10$\%, shown 
in the dashed curve in Fig.~\ref{figstar2}, should be included before comparing
to the data.  The $t_{\perp}$-dependence
of the elementary amplitudes was not taken into account since it is relatively
independent of energy in the RHIC regime compared to the nuclear form factor.
If included, it would further reduce the cross section slightly.  Smearing 
due to the photon $p_T$ and interference of the production amplitudes of the
two nuclei are also neglected~\cite{Klein:2000aa}. 

Interference produces the narrow dip in the coherent 
$t_{\perp}$-distribution at 
$t_{\perp}\leq 5\times 10^{-4}$ GeV$^2$, in addition to the Glauber
diffractive minimum at $_\perp \sim 0.013$.  While these effects
do not strongly influence the $t_{\perp}$-integrated
cross section, they can easily be taken into account, giving 
$\sigma_{\rm coh} = 490$ mb, closer to the STAR value.
Since the calculation does not have any free parameters, the cross section is
in reasonable agreement with the STAR data.

\subsubsection{LHC Estimates} \bigskip

\begin{figure}[htb]
\centering
\epsfxsize=0.8\hsize
    \epsffile{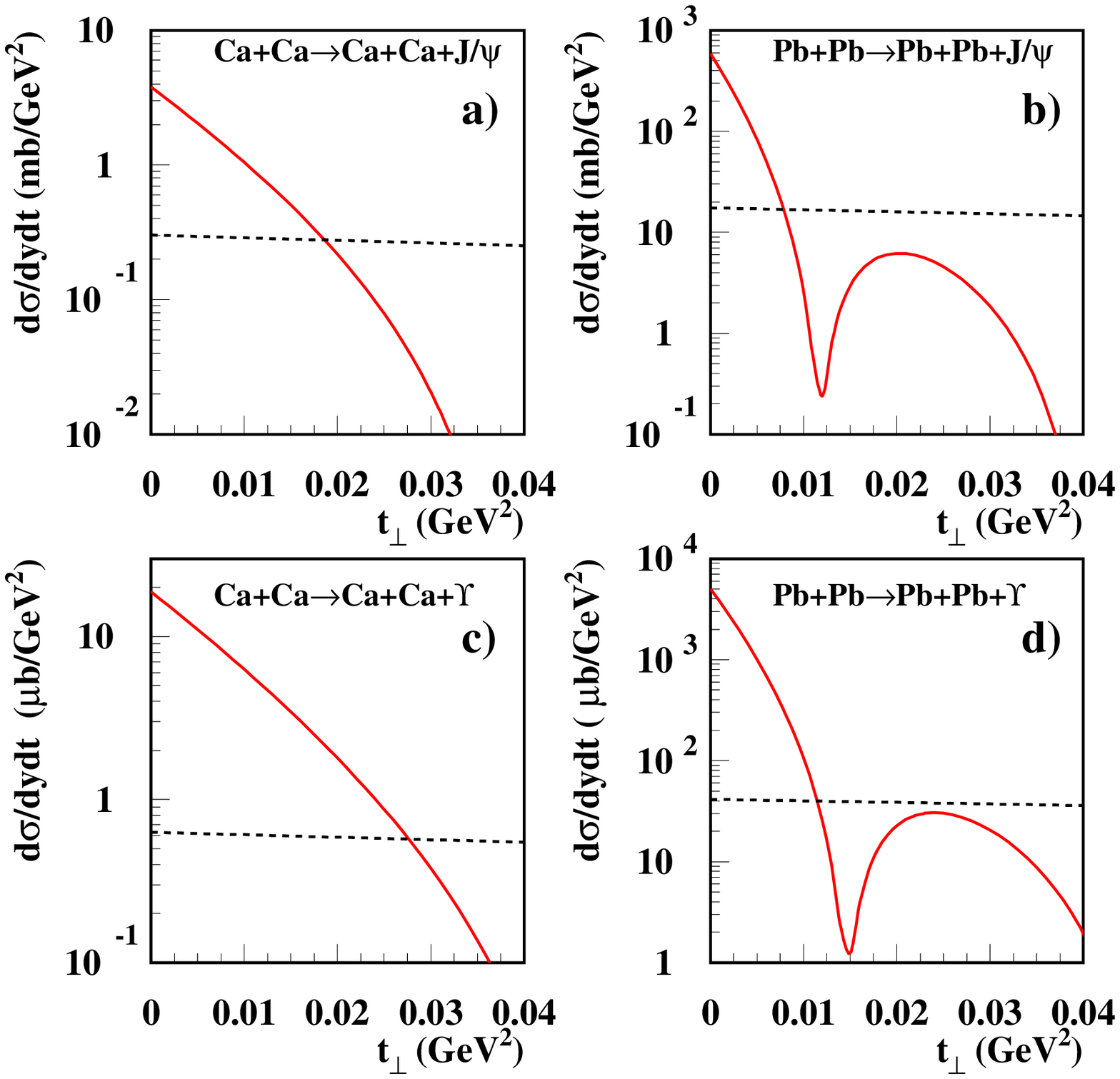}
\caption[]{The $t_\perp$ distribution of coherent
$J/\psi$ and $\Upsilon$ production in Ca+Ca and Pb+Pb UPCs at the LHC, 
including leading-twist shadowing but neglecting the photon $p_T$ spread.  
The dashed curves show the incoherent distributions.  Reprinted from 
Ref.~\protect\cite{Frankfurt:2003wv} with permission from Acta Physica 
Polonica.}
\label{lhcdst}
\end{figure}

\begin{figure}[htb]
\centering
        \leavevmode
        \epsfxsize=0.8\hsize
       \epsffile{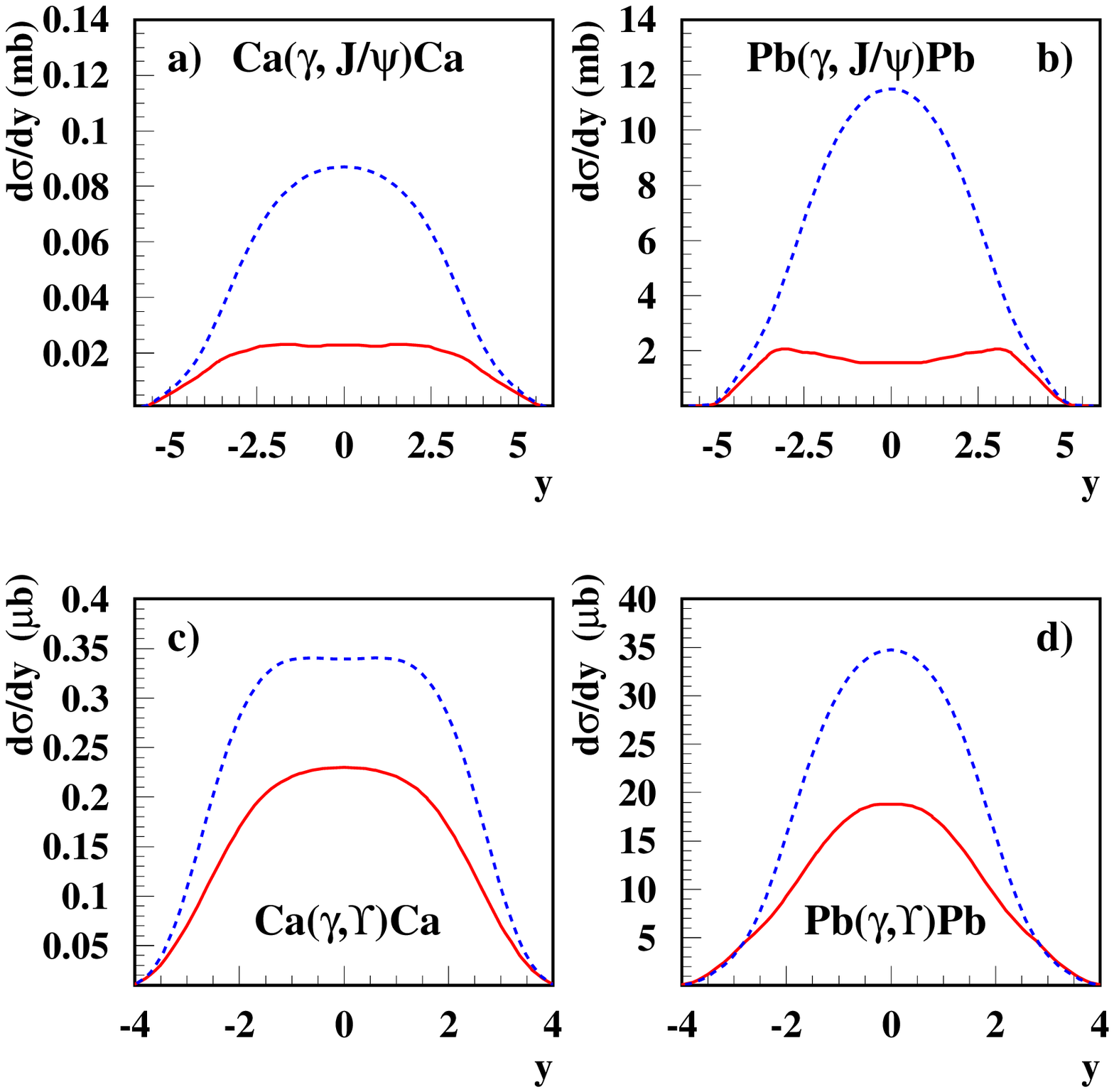}
\caption[]{The coherent $J/\psi$ and $\Upsilon$ rapidity distributions 
in Ca+Ca and Pb+Pb UPCs at the LHC calculated in the impulse approximation
(dashed) and including leading-twist shadowing based on the 
H1 gluon density parametrization (solid).  Reprinted from 
Ref.~\protect\cite{Frankfurt:2003wv} with permission from Acta Physica 
Polonica.}
\label{figlhc1}
\end{figure}

References~\cite{FKS,Frankfurt:1998ym} suggested using $J/\psi$ 
(electro)photoproduction to search for color opacity.  However, this requires 
energies much larger than those available at fixed-target facilities, such
as electron-nucleus colliders. FELIX rate estimates \cite{FELIX} 
demonstrated that the effective photon
luminosities  generated in peripheral heavy-ion collisions
at the LHC would lead to significant coherent vector meson
photoproduction rates, including $\Upsilon$.
It is thus possible to study vector meson photoproduction in Pb+Pb and Ca+Ca 
collisions at the LHC with much higher energies than $W_{\gamma p}\leq 17.3$ 
GeV, the range of fixed-target experiments at FNAL \cite{Sokoloff}. 
Even current experiments at RHIC, with $W_{\gamma p}\leq 25$ GeV, also exceed 
the fixed-target limit.  As indicated by the STAR study, coherent 
photoproduction, leaving both nuclei intact, can be reliably identified 
using veto triggering from the zero degree calorimeters (ZDCs).  
Selecting low $p_T$ quarkonia removes 
incoherent events where the residual nucleus is in the ground state.

Hadronic absorption should be moderate or small for heavy vector mesons.
The production cross sections
are, however, sensitive to gluon shadowing in the parton
distribution functions. If two-gluon exchange is the dominant
production mechanism \cite{Ryskin:1992ui,Ryskin:1995hz},
\begin{equation}
   \frac{\left. d \sigma_{\gamma A \rightarrow V A} / dt \right|_{t=0}}
        {\left. \, \, d \sigma_{\gamma N \rightarrow V N} / dt \right|_{t=0}} =
   \left[ \frac{g_A(x,Q^2)}{g_N(x,Q^2)} \right]^2
\label{gratios}
\end{equation}
where $g_A$ and $g_N$ are the gluon distributions in the nucleus and nucleon, 
respectively.

The sensitivity of heavy quarkonia to the gluon
distribution functions can be further illustrated by a model comparison.
In Fig.~\ref{lhcdst}, the $t_\perp$ distributions of coherent $J/\psi$ and 
$\Upsilon$ photoproduction, calculated with leading-twist shadowing,
are compared to incoherent photoproduction.  The spread in photon 
$p_T$ is again neglected.  The maximum incoherent cross section 
is estimated to be the elementary cross section on a nucleon target scaled by
$A$.
\begin{figure}[htb]
    \begin{center}
      \includegraphics[width=0.85\textwidth]{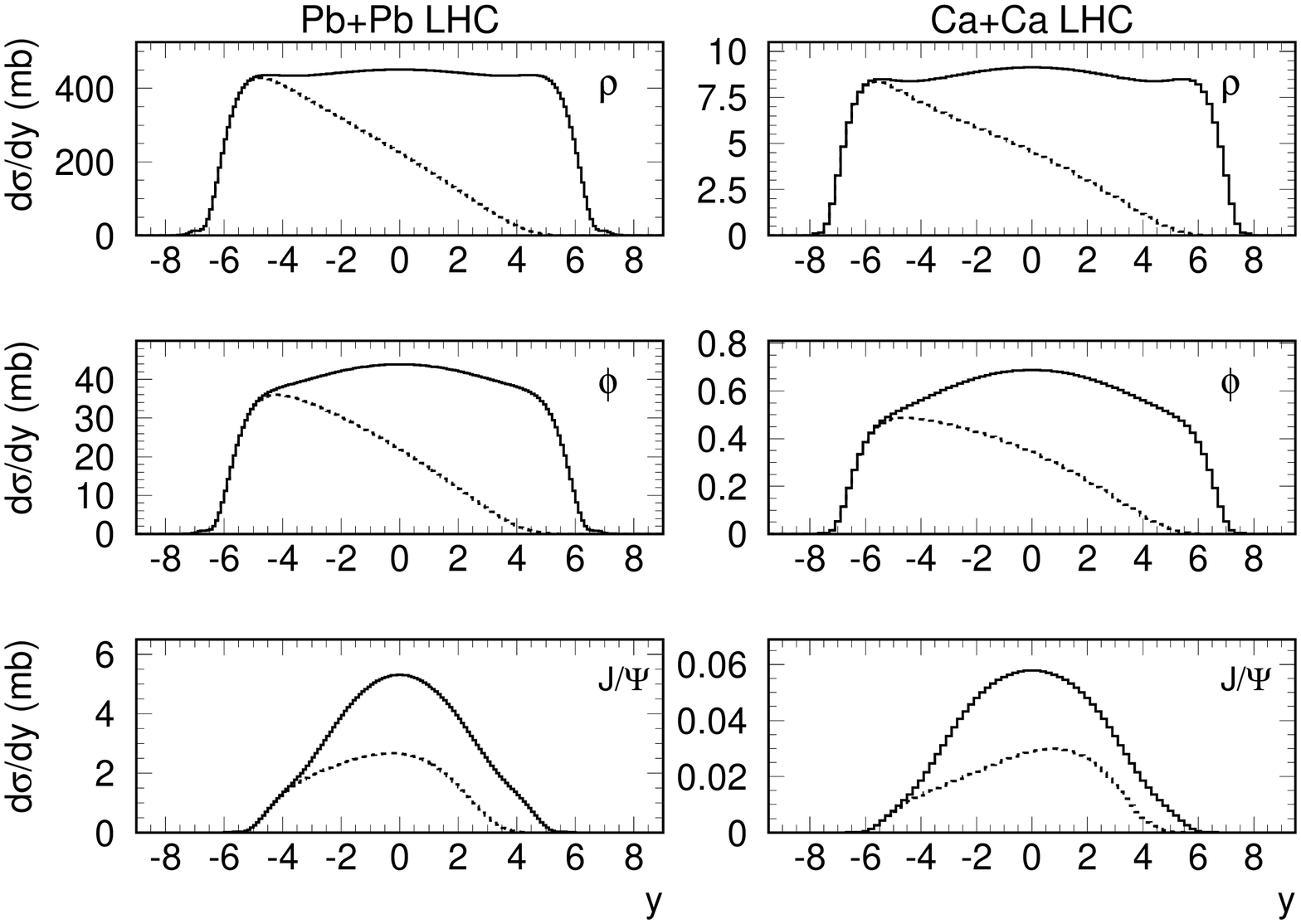}
    \end{center}
    \caption[]{The rapidity distributions of coherent $\rho^0$, $\phi$ and 
$J/\psi$ production in Pb+Pb and Ca+Ca UPCs at the LHC. From 
Ref.~\protect\cite{Klein:1999qj}.  The solid lines correspond to the total
rate while the dashed lines are the contribution due to photon emission
by the nucleus with negative rapidity. Copyright 1999 by the American Physical
Society (http://link.aps.org/abstract/PRC/v60/e014903).}
    \label{fig:3:dndy}
\end{figure}

Figure~\ref{figlhc1} shows the coherent $J/\psi$ and $\Upsilon$ rapidity 
distributions calculated in the impulse approximation and with nuclear
gluon shadowing.  At central rapidities, $J/\psi$ production is suppressed 
by a factor of four (six) for Ca+Ca (Pb+Pb).  For comparison, the $\rho^0$,
$\phi$ and $J/\psi$ rapidity distributions calculated with the parametrization
of Section~\ref{section-aa-xsections} are presented in Fig.~\ref{fig:3:dndy}.
While at RHIC energies the rapidity distributions 
have two peaks, corresponding to production off each of the two nuclei 
\cite{Klein:1999qj}, the higher LHC energies largely remove the two-peak
structure, as shown in Figs.~\ref{figlhc1} and \ref{fig:3:dndy}.

\begin{table}[htpb]
\begin{center}
\caption[]{Vector meson production cross sections in ultraperipheral Au+Au
interactions at RHIC and Pb+Pb and Ca+Ca interactions at the LHC.
The results are shown with the cross section parametrization of 
Ref.~\protect\cite{Klein:1999qj} (CP), the impulse approximation (IA) 
and the IA including leading-twist (LT)
shadowing \protect\cite{fguzsz,Frankfurt:2001db}.}
\vspace{0.4cm}
\begin{tabular}{|c|c||c|c|c||c|c|c|}
\hline
&  Au+Au & \multicolumn{3}{|c||}{Pb+Pb}& \multicolumn{3}{|c|}{Ca+Ca} \\ 
\hline{2-8}
VM & $\sigma_{\rm CP}$ (mb) 
& $\sigma_{\rm CP}$ (mb) & $\sigma_{\rm IA}$ (mb) & 
$\sigma_{\rm LT}$ (mb) & $\sigma_{\rm CP}$ (mb) & $\sigma_{\rm IA}$ (mb) 
& $\sigma_{\rm LT}$ (mb) \\ \hline
$\rho^0$        & 590  & 5200  &    &    & 120  & &     \\ 
$\omega$        & 59   &  490  &    &    & 12   & &     \\ 
$\phi$          & 39   &  460  &    &    & 7.6  & &     \\ 
$J/\psi$        & 0.29 &   32  & 70 & 15 & 0.39 & 0.6 & 0.2 \\ 
$\Upsilon(1S)$  & $5.7 \times 10^{-6}$ &  0.17  & 0.133 & 0.078 & 0.0027 & 
0.0018 & 0.0012 \\ \hline
\end{tabular}
\end{center}
\label{tab:3:sigmavm}
\end{table}

The $J/\psi$ and $\Upsilon$ total cross sections are given in 
Table~\ref{tab:3:sigmavm} for the impulse approximation (IA) and including
leading-twist shadowing (LT).  Comparison of the LT and IA calculations
shows that the $\Upsilon$ yield is predicted to be suppressed by a factor of
$\sim 2$ due to leading-twist shadowing.  The suppression factor is higher for
the $J/\psi$.  Hence, coherent quarkonium photoproduction
at the LHC can probe shadowing effects on the nuclear gluon distributions 
in a kinematic regime that would be hard to probe at other facilities.
For comparison, the cross sections calculated with the parametrization in 
Ref.~\cite{Klein:1999qj}
are also given in Table~\ref{tab:3:sigmavm}.

\subsubsection{Cross sections at $pp$ colliders} \bigskip

The strong electromagnetic fields generated by high-energy protons may 
also lead to exclusive vector meson production in $pp$ and 
$\overline{p}p$ collisions \cite{Klein:2003vd,Klein:2003qc}. 

Although there is no coherent enhancement of the photon spectrum or
the photon-nucleon cross section, the LHC $pp$ luminosity is about seven orders
of magnitude larger than in \PbPb\ collisions. In addition, since protons are
smaller than ions, photoproduction can occur at smaller
impact parameters.  Together, these factors more than compensate
for the coherent enhancement in \PbPb\ collisions.  Also, due to the smaller
proton size, the photon spectrum extends to higher energies, increasing 
the kinematic reach.

\begin{figure}[htb]
    \begin{center}
      \includegraphics[width=10cm]{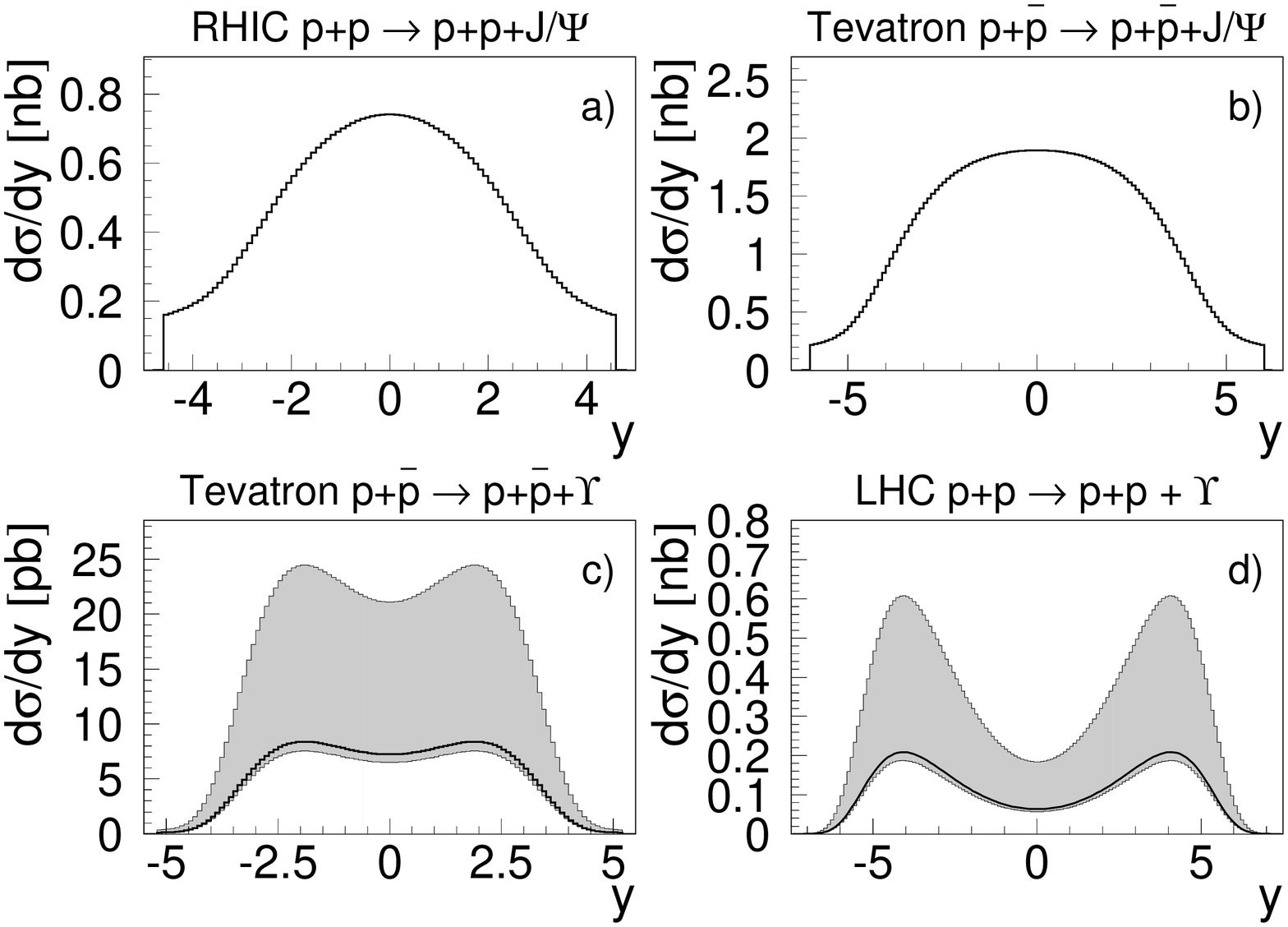}
    \end{center}
    \caption[]{The $J/\psi$ (RHIC, Tevatron) and $\Upsilon$ (Tevatron, LHC) 
rapidity distributions are shown in $pp$ and $\overline p p$ interactions.
The cross sections are calculated using $b_{\rm min}
= 1.4$ fm.  The shaded area in the lower $\Upsilon$ plots due to the poorly 
known $\gamma p \to \Upsilon p$ cross section. From 
Refs.~\protect\cite{Klein:2003vd,Klein:2003qc}.  Copyright 2004 by the American
Physical Society (http://link.aps.org/abstract/PRL/v92/e142003).}
\label{fig:3:ppdndy}
\end{figure}

The calculations are similar to those discussed above for nuclei. The 
photon spectrum for relativistic protons is, however, different, since 
the impact parameter is not always well defined for $pp$ collisions. 
Interference between production sources also differs in $\overline{p}p$ 
relative to $pp$ and $AA$ collisions due to the different $CP$ 
symmetry. 

The proton photon spectrum, calculated using the dipole formula for the 
electric form factor, is \cite{Drees:1988pp} 
\begin{equation}
\frac{dN_\gamma}{dk} = \frac{\alpha}{2\pi z} \big[1+ (1-z)^2\big]
\bigg(\ln{D} - \frac{11}{6} + \frac{3}{D} - \frac{3}{2D^2}
+\frac{1}{3D^3}\bigg) \, \, 
\label{eq:ffluxp}
\end{equation}
where
\begin{equation}
D = 1 + \frac{\rm 0.71 \, GeV^2}{Q_{\rm min}^2} \, \, ,
\label{eq:ddef}
\end{equation}
$z=W_{\gamma p}^2/s_{pp}$ and $Q_{\rm min}$ is the
minimum momentum transfer needed to produce the vector meson.

The $J/\psi$ and $\Upsilon$ rapidity distributions in $pp$ collisions at RHIC 
and the LHC and in $\overline{p}p$ collisions at the Tevatron are shown in
Fig.~\ref{fig:3:ppdndy}. The corresponding total cross sections are
listed in Table~\ref{tab:sigmapp}.

\begin{table}[htpb]
\begin{center}
\caption[]{The vector meson production cross sections in $pp$ and 
$p\overline{p}$ collisions at RHIC, the Tevatron and the LHC.}
\vspace{0.4cm}
\begin{tabular}{|c|c|c|c|}
\hline
& $pp$ RHIC   & $\overline p p$ Tevatron  & $pp$ LHC \\ \hline{2-4}
VM     & $\sigma(500 \, {\rm GeV})$ ($\mu$b) & $\sigma(1.96 \, {\rm TeV})$ 
($\mu$b)  & $\sigma(14 \, {\rm TeV})$ ($\mu$b)     \\ \hline
$\rho^0$  & 5.1  & 8.6  & 17  \\
$J/\psi$  & 0.0043      & 0.15      & 0.075     \\ 
$\Upsilon$& $5.2 \times 10^{-6}$ & $5.5 \times 10^{-5}$ & 0.0016  \\ \hline
\end{tabular}
\end{center}
\label{tab:sigmapp}
\end{table}

\subsubsection{Multiple vector meson production} \bigskip

A unique feature of heavy-ion colliders, not accessible in $ep$ or $eA$
interactions, is multiple vector meson production in a single
event. This is a consequence of the extreme electromagnetic 
fields present in grazing ultrarelativistic heavy-ion collisions.  
since the photon spectrum in Eq.~(\ref{wwr}) is
proportional to $Z^2$ and inversely proportional to $k$ and $b^2$.  Thus
the low energy photon density is large at small $b$.

\begin{figure}[htb]
    \begin{center}
      \includegraphics[width=8cm]{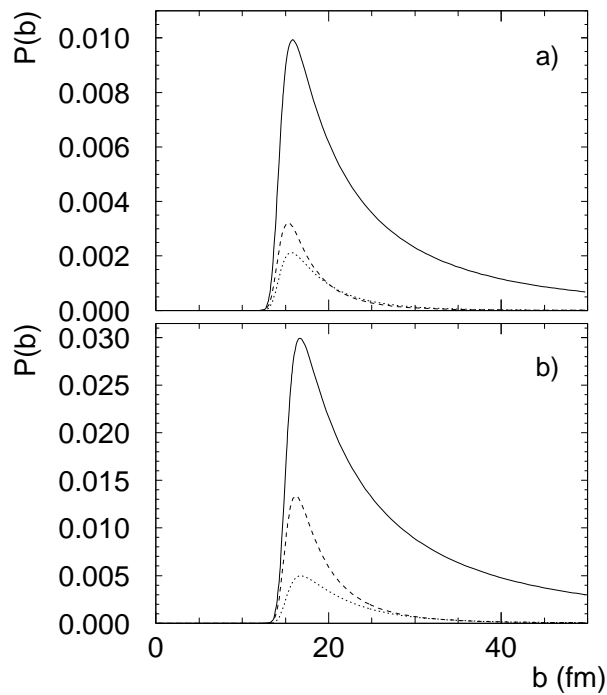}
    \end{center}
    \caption[]{The interaction probability for $\rho^0$ production as a 
function of $b$ in a) Au+Au interactions at RHIC and b) Pb+Pb interactions 
at the LHC. The solid curve is exclusive 
production, the dashed and dotted curves are for $(Xn,Xn)$ and $(1n,1n)$ 
neutron emission due to Coulomb breakup, as described 
in the text. The dotted curves have been scaled up by a factor of 
10 \protect\cite{Baltz:2002pp}.  Copyright 2002 by the American Physical 
Society (http://link.aps.org/abstract/PRL/v89/e012301).}
    \label{fig:3:prob}
\end{figure}

By changing the order of integration in Eqs.~(\ref{eq:3:nofe}) and
(\ref{eq:3:vmcross}), we can write the unnormalized interaction probability 
for single vector meson photoproduction as a function of
impact parameter as
\begin{equation}
P_V^{(1)} (b) = \frac{d \sigma} {d^2b} = \int  dk \frac{d^2N_\gamma}{dk d^2b}
{d\sigma_{\gamma A\rightarrow VA}\over dt}\bigg|_{t=0} \int_{t_{\rm 
min}}^\infty dt \, |F(t)|^2 \; 
\label{eq:p1ofb}
\end{equation}
where $d^2N_\gamma/dk d^2b$ is the photon density of Eq.~(\ref{wwr}). The
superscript `(1)' indicates a single photon exchange. 
At small $b$, $P_V^{(1)}(b) > 1$ for 
photonuclear processes with low thresholds and/or
high cross sections.  Thus $P_V^{(1)} (b)$ cannot be interpreted as an ordinary
probability but should be interpreted as a first-order amplitude. Unitarity
can be restored by accounting for multiple exchanges where the
probability of having exactly $N$ exchanges is given by a Poisson
distribution,
\begin{equation}
P_V^{(N)} (b) = \frac{ [ P_V^{(1)} (b) ]^N \, \exp[- P_V^{(1)} (b)] }{N!} \; .
\label{eq:pnofb}
\end{equation}
The $\rho^0$ production probability, $P_{\rho^0}^{(1)}(b)$, 
in Au+Au
interactions at RHIC and Pb+Pb interactions at the LHC is shown in
Fig.~\ref{fig:3:prob}. Since $P^{(1)}_V(b) \ll 1$, $P_V(b)
\approx P^{(1)}_V(b)$).

Neglecting correlations, the cross section for producing a pair of
identical mesons is then
\begin{equation}
\sigma_{VV} = \frac{1}{2} \, \int d^2b \, [P_V (b)]^2 \, \, .
\end{equation}
If the mesons are not identical,
\begin{equation}
\sigma_{V_1 V_2} = \frac{1}{2} \, \int d^2b \, [P_{V_1}(b)][P_{V_2}(b)] \, \, .
\end{equation}
This gives $\sigma_{\rho^0 \rho^0} \sim 9$ mb, $\sigma_{\phi\phi} \approx
\sigma_{\omega \omega} \approx 70 $ $\mu$b and $\sigma_{\rho^0 J/\psi} \sim
0.2$ mb in ultraperipheral Pb+Pb collisions at the LHC \cite{Klein:1999qj}.  

Production of two vector mesons in a single $AA$ collision 
introduces angular correlations among the decay products.  The linear
polarization of the photon follows the electric field vector of the 
emitting ion so that, at the target, the
photon polarization is parallel to $\vec{b}$ \cite{Baur:2003ar}.  In the
case of multiple vector meson production, the photon polarizations are 
either parallel or anti-parallel.   

The photon polarizations affect the angular distribution of the decay 
products.  In the rest frame of a vector meson making a
two-body decay such as $\rho^0\rightarrow\pi\pi$ or $J/\psi\rightarrow ee$, 
the final-state decay particle angular distribution with respect to the photon 
polarization goes as $\cos \phi$  where
$\phi$ is the azimuthal angle, perpendicular to the direction of motion,
between the photon polarization and the decay particle direction.  
Although the polarization is
not directly observable, a correlation function can be defined for double 
vector meson production,
\begin{equation}
C(\Delta\phi) = 1 + {1\over 2}\cos{(2\Delta\phi)} \, \, ,
\label{eq:corr}
\end{equation}
where $\Delta\phi = \phi_1-\phi_2$ is the azimuthal difference 
between the two positively (or negatively) charged decay particles.

Similar neutron correlations are expected in neutron emission in 
giant dipole resonances (GDRs) which typically decay by single 
neutron emission. The direction of the neutron $p_T$ 
follows the same azimuthal angle distribution as vector meson decays. 
For mutual Coulomb excitation to a double giant dipole resonance, the
azimuthal separation between the two emitted neutrons should follow 
Eq.~(\ref{eq:corr}).

These angular correlations make UPC studies possible with linearly
polarized photons.  If the direction of the neutron $p_T$ can be measured,  
a single or mutual GDR excitation tag
can be used to determine the polarization direction of any additional 
photons in the event, allowing studies of polarized proton collisions.  
The RHIC ZDCs have been upgraded to
include position-sensitive shower-maximum detectors which can make
directional measurements \cite{ZDC}.
Similar detectors could be useful at the LHC.

These calculations neglect quantum statistics, addressed
in the next section.  The cross sections are large enough for
multiple vector meson production to be observable, making correlation
studies possible.

\subsubsection{Vector meson production in coincidence with nuclear breakup}
\label{vmabreakup} \bigskip

As discussed in the previous section, strong fields in heavy-ion collisions 
may lead to large cross sections for interactions involving multiple photon
exchanges. The additional photons 
may interact with the target nucleus in a number of ways.
In particular, they may lead to the breakup of one or both of the interacting 
nuclei. The largest contribution comes from the nuclear excitation to a GDR
\cite{Baltz:as}.  About 83\% of GDR decays are via
single neutron emission~\cite{veyssiere}. 

\begin{figure}[htb]
    \begin{center}
      \includegraphics[width=0.5\textwidth]{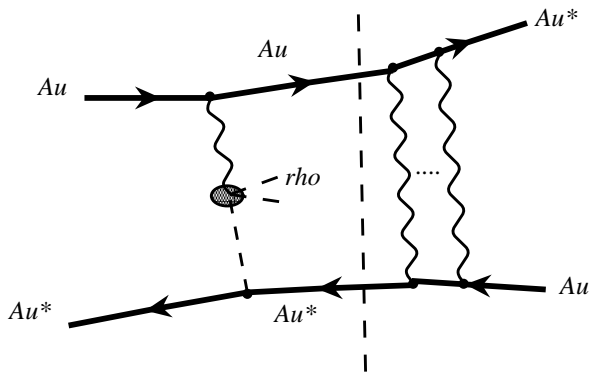}
    \end{center}
    \caption[]{Diagram of nuclear excitation accompanied by $\rho^0$
production. From Ref.~\protect\cite{Baltz:2002pp}.  Copyright 2002 by the
American Physical Society (http://link.aps.org/abstract/PRL/v89/e012301).}
    \label{fig:3:feynman}
\end{figure}

There is a $\sim 35$\% probability for mutual excitation of both ions 
in the same event by two-photon exchange in Au+Au collisions with $b \sim 2R_A$
at RHIC.  The cross section for vector meson production in coincidence with 
mutual Coulomb breakup of the beam nuclei, see Fig.~\ref{fig:3:feynman}),
was calculated in Ref.~\cite{Baltz:2002pp}
based on parameterizations of measured
photo-nuclear cross sections, $\sigma_{\gamma A \rightarrow A^*}$
\cite{Baltz:1998ex}. The probability
for Coulomb breakup of a single nucleus is
\begin{equation}
P_{Xn} (b) = \int dk \frac{dN_\gamma}{dk db^2} \,
\sigma_{\gamma A \rightarrow A^*} (k) \; .
\end{equation}
Assuming that Coulomb excitation and vector meson production
are independent, the probabilities factorize so that
\begin{equation}
\sigma_{AA \rightarrow A^* A^* V} = 2 \int db^2 P_V(b) P_{(Xn,Xn)} (b) 
\exp( -P_H(b) )
\end{equation}
where $P_{(Xn,Xn)} = P_{Xn} P_{Xn}$ 
is the probability for mutual Coulomb excitation
followed by the emission of an arbitrary number of neutrons in either
direction, and $\exp( -P_H(b) )$ is the probability of no
hadronic interaction. The cross sections for Pb+Pb interactions are
shown in Table~\ref{tab:3:breakup} for the Coulomb breakup of a single
nucleus with multiple neutron emission, $Xn$, and for single and multiple 
neutron emission and Coulomb breakup of both nuclei, $(1n,1n)$ and $(Xn,Xn)$ 
respectively, in addition to the total cross section.  

\begin{table}[htpb]
\begin{center}
\caption[]{The cross sections and average impact parameters, $\langle b 
\rangle$, for vector meson production in Pb+Pb interactions at the LHC.}
\vspace{0.4cm}
\begin{tabular}{|c|c|c|c|c|c|c|c|c|} \hline
& \multicolumn{2}{c|}{total}  & \multicolumn{2}{c|}{$Xn$}  
& \multicolumn{2}{c|}{$(Xn,Xn)$} & \multicolumn{2}{c|}{$(1n,1n)$}  \\ 
\cline{2-9}
VM      & $\sigma$ (mb) &$\langle b \rangle$ (fm) & $\sigma$ (mb)  
& $\langle b \rangle$ (fm) & $\sigma$ (mb)  & $\langle b \rangle$ (fm) & 
$\sigma$ (mb)  & $\langle b \rangle$ (fm) \\
\hline
$\rho^0$& 5200          & 280   & 790   & 24 & 210  & 19    & 12    & 22    \\
$\omega$& 490           & 290   &  73   & 24 &  19  & 19    & 1.1   & 22    \\
$\phi$  & 460           & 220   &  74   & 24 &  20  & 19    & 1.1   & 22    \\
$J/\psi$& 32            &  68   & 8.7   & 23 & 2.5  & 19    & 0.14  & 21   \\ 
$\Upsilon (1S)$ & 0.17  &  31   & 0.078 & 21 & 0.025& 18    & 0.0013& 20    \\
\hline
\end{tabular}
\end{center}
\label{tab:3:breakup}
\end{table}

Vector meson production in coincidence with nuclear breakup is of
experimental as well as theoretical interest.
All the RHIC and LHC experiments are equipped with ZDCs. These detectors
measure neutrons emitted at forward angles by the fragmenting
nuclei and may be used
for triggering and thus provide a UPC trigger for experiments that 
lack a low multiplicity trigger at midrapidity.

Requiring the vector mesons to be produced in coincidence with Coulomb
excitation alters the impact parameter distribution compared with
exclusive production. The probability for $\rho^0$ production for single
and multiple neutron emission by both nuclei is shown in 
the dashed and dotted curves of Fig.~\ref{fig:3:prob}.  As seen in the figure 
and in Table~\ref{tab:3:breakup}, the mean impact parameter, 
$\langle b \rangle$,
is dramatically reduced in interactions with Coulomb dissociation. The 
decreased average impact parameter changes the photon energy spectrum, 
enhancing the relative hard photon yield and modifying the rapidity
distributions in interactions with breakup. As discussed in the next section,
since both single and multiple neutron emission usually
involves at least 3 photons,
$\langle b \rangle$ is essentially independent of the photon energy.  Thus
$\langle b \rangle$ is similar for $(1n,1n)$ and $(Xn,Xn)$ and is independent
of vector meson mass.  The change in $\langle b \rangle$
also affects interference between vector meson photoproduction on two
nuclear targets, discussed in 
the next section.

\subsubsection{Interference effects on the $p_T$ distribution}
\label{section-vm-interference} \bigskip

One important feature of $AA$ and $pp$ collisions is that
the two incoming particles are identical, {\it i.e.}\
indistinguishable.  The initial state is completely symmetric
and, because of the small momentum transfers in vector meson
photoproduction, it is not possible to tell which nucleus emitted the photon 
and which was the target.  Since the two processes (nucleus one as the photon
emitter, nucleus two as the target and nucleus two as the emitter, one the
target) are indistinguishable, the amplitudes must be added rather than the
cross sections.  Changing nucleus one from the photon target to the
photon emitter is the equivalent of a parity transformation. Since vector
mesons have negative parity, the amplitudes are subtracted.
If the amplitudes for the two processes are ${\mathcal A}_1$ and ${\mathcal 
A}_2$, then
\begin{equation}
\sigma = |{\mathcal A}_1 - {\mathcal A}_2\exp[i(\vec{b}\cdot\vec{p}_T + 
\delta)]|^2
\end{equation}
where $\delta$ accounts for possible changes in $\rho^0$ photoproduction
with photon energy.
The exponential is a propagator accounting for the phase shift
when nucleus one becomes the target.  At midrapidity, ${\mathcal A}_1= 
{\mathcal A}_2 = {\mathcal A}$
and the expression simplifies to
\begin{equation}
\sigma = {\mathcal A}^2 [1-\cos(\vec{b}\cdot\vec{p}_T)] \, \, .
\end{equation}
Since the impact parameter is unknown, it is necessary
to integrate over all $b$.  There is significant
suppression when
\begin{equation}
p_T < \frac{\hbar}{\langle b\rangle} \, \, .
\label{ptestim}
\end{equation}
Using Eq.~(\ref{eq:pnofb}), $\langle b\rangle = 46$ fm for $\rho^0$ 
production in Au+Au UPCs at RHIC, rising to 290 fm in Pb+Pb interactions 
at the LHC.  There is thus significant suppression for $p_T < 5$ MeV/$c$ at
RHIC and $p_T < 1$ MeV/$c$ at the LHC.  At the LHC, observing this suppression
may be challenging.  A detailed study of the transverse momentum distribution
within the framework of the semi-classical and Glauber approximations is done
in Ref.~\cite{HBT}.
The numerical results are rather similar to those in Ref.~\cite{Klein:2000aa}.

Multiple interactions in heavy-ion collisions makes observation of the 
interference effect easier since the more photons that are exchanged, the 
smaller the impact parameter.  The
average impact parameter is \cite{Baur:2003ar}
\begin{equation}
\langle b \rangle = \frac{\int d^2b \, b \, P(b)}{\int d^2b \, P(b)}
\end{equation}
for any probability $P(b)$.  In the case of vector meson production,
as long as $\gamma_L/k > b > 2R_A$, $P_V(b)
\approx 1/b^2$.  For single photon exchange, $P_V(b) = P_V^{(1)}(b)$, and
\begin{equation}
\langle b \rangle = {b_{\rm max}-b_{\rm min}\over \ln(b_{\rm max}/b_{\rm 
min})} \, \, 
\end{equation}
where $b_{\rm min}=2R_A$ and $b_{\rm max} = \gamma_L/k$.  For $N$ photon 
exchanges, $N \geq 2$, and 
$b_{\rm max} >> b_{\rm min}$, the result is
\begin{equation}
\langle b_N \rangle = {2N-2\over 2N-3} \, b_{\rm min}
\end{equation}
so that, for example, for vector meson production accompanied by
mutual nuclear excitation ($N=3$), $\langle b \rangle = 1.33 \, b_{\rm min}
\approx 18$ fm, 
almost independent of the details of the interaction.  Since this independence
does not change with beam energy, interference should significantly affect
the cross section of mutual nuclear excitation
for $p_T < 11$ MeV/$c$ at both RHIC and the LHC.  
Interference has already been observed at RHIC
\cite{Klein:2004kq,Klein:2003rx} for both single and multiple photon exchange. 
It should thus be equally observable at the LHC.

\begin{figure}[htb]
    \begin{center} \includegraphics[width=10cm]{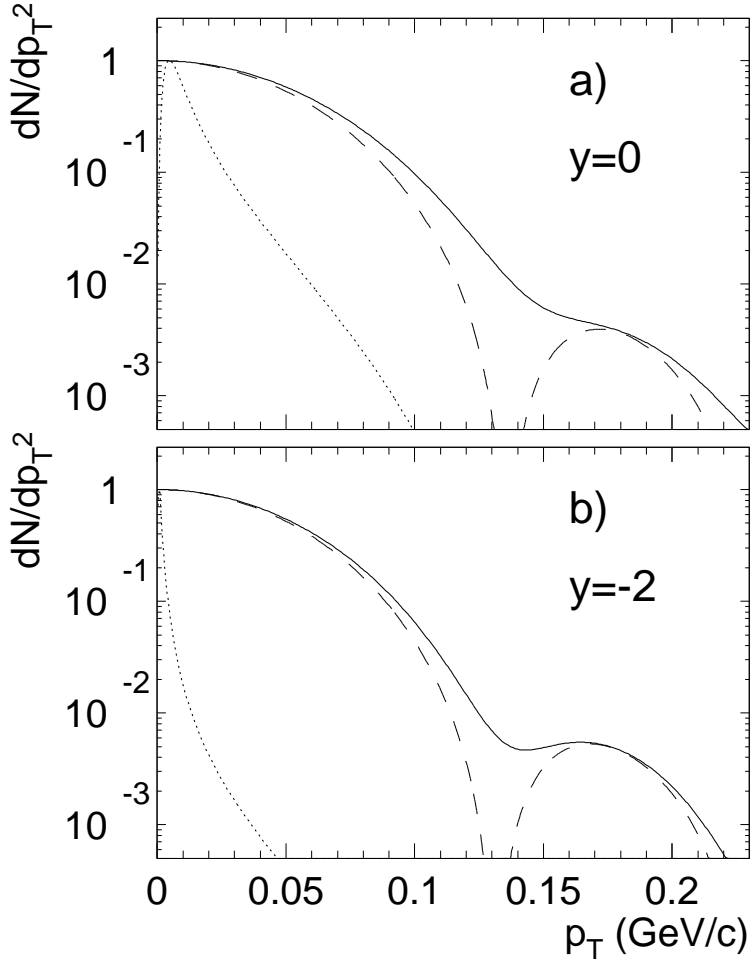}
      \end{center}
\caption[]{The $p_T$ distributions of exchanged photons (dotted) and 
Pomerons (dashed) as well as final state $\phi$ mesons (solid) in 
$\sqrt{s_{_{NN}}} = 200$ GeV Au+Au collisions
at RHIC, at (a) $y=0$ and (b) $y=-2$. The curves are normalized to unity
at $p_T=0$.  Clear diffraction minima are visible in the Pomeron
spectra but are largely washed in the final state.  From 
Ref.~\protect\cite{Klein:2000aa}.  Copyright 2000 by the
American Physical Society (http://link.aps.org/abstract/PRL/v84/p2330).}
\label{phipt}
\end{figure}

A detailed calculation of the $p_T$ spectrum requires consideration
of both the photon $p_T$ and the Pomeron momentum transferred to the vector 
meson during scattering \cite{Klein:2000aa}.  These two sources, shown in
Fig.~\ref{phipt} for $\phi$ production, neglecting interference, must be added
in quadrature.  The photon $p_T$ is approximately peaked around
$p_T \sim k/\gamma_L$.  At midrapidity, $k=M_V/2$, resulting in a peak at
$M_V/2\gamma_L \sim 5$~MeV/$c$ while the photon flux goes to zero as
$p_T \to 0$.  The Pomeron $p_T$ is peaked at zero with a
width of $\sim \hbar c/R_A \approx 30$ MeV/$c$ for heavy ions.

\begin{figure}[htb]
    \begin{center} 
\includegraphics[width=10cm]{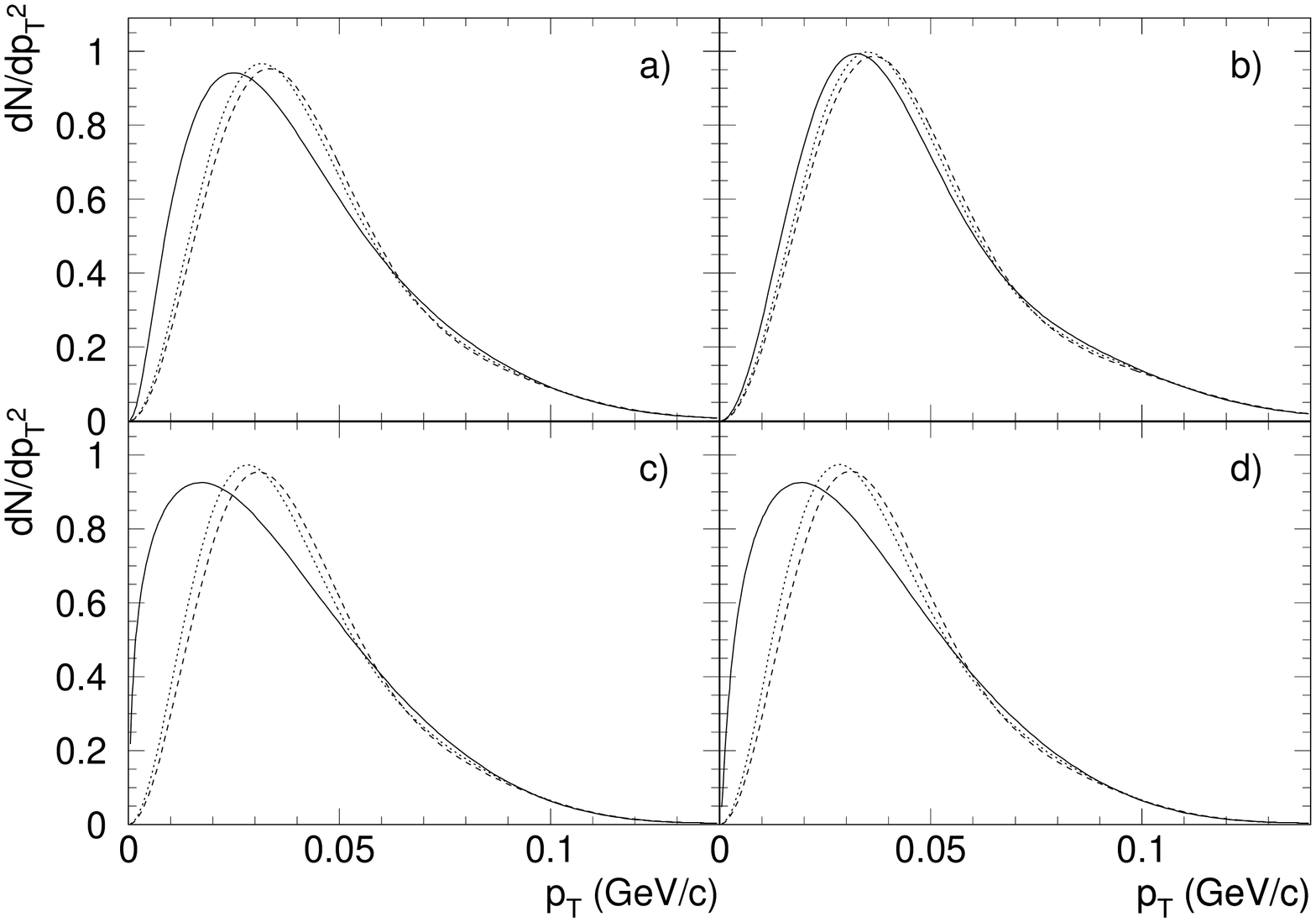}
      \end{center} 
\caption[]{The $p_T$ distributions, $dN/dp_T^2$, for $\rho^0$ production
at (a) RHIC and (c) the LHC and $J/\psi$ production at (b) RHIC and (d)
the LHC.  The solid curve shows exclusive vector meson production, the dashed,
mutual Coulomb dissociation and the dotted, mutual
Coulomb dissociation with single neutron emission in both directions $(1n,1n)$.
All the calculations include interference.  The results in 
Fig.~\protect\ref{phipt} 
are comparable $\phi$ calculations without interference.  From
Ref.~\protect\cite{Baltz:2002pp}.  Copyright 2002 by the
American Physical Society (http://link.aps.org/abstract/PRL/v89/e012301).}
\label{ptinterfere}
\end{figure}

If $P_V(b)$ is known, the interference may be
calculated.  Figure~\ref{ptinterfere} shows $dN/dp_T^2$ for $\rho^0$ and
$J/\psi$ production at RHIC and the LHC, both for exclusive vector meson
production and vector mesons accompanied by mutual Coulomb
excitation.  The RHIC and LHC results are
very different for exclusive production, and, at the LHC, production is 
only reduced for $p_T < 5$ MeV/$c$.  While small, this $p_T$ is much
larger than estimates using Eq.~(\ref{ptestim}) because
significant production occurs for $b\approx 2R_A$.  For
vector mesons accompanied by mutual Coulomb dissociation, the RHIC
and LHC curves are quite similar.
Interference has been studied by the STAR Collaboration at RHIC.
The results are presented in Section~\ref{star-interference}.

Interference effects are also present in $pp$
and $\overline pp$ collisions.  One complication for
these smaller systems is that $b_{\rm min}$ is not easily defined.  However, 
$b_{\rm min} = 0.7$ fm is reasonable choice.  Thus $\langle b \rangle$ 
is smaller than in $AA$ collisions, extending 
the $p_T$ scale of the interference to $p_T \sim 200$ MeV/$c$.  

In $pp$ and $\overline pp$ collisions the
interference depends on the symmetry of the system.
For $pp$ or $AA$ collisions, there is a parity transformation from the 
situation where hadron one emits a photon while hadron two is the photon target
to the opposite situation where hadron two is the photon
emitter.  However, a charge-parity transformation is required for
$p\overline p$ collisions.  Since vector mesons have $CP= --$, the
amplitudes from the two sources add so that
\begin{equation}
\sigma = |{\mathcal A}_1 + {\mathcal A}_2\exp[i(\vec{b}\cdot\vec{p}_T)]|^2 
\, \, .
\end{equation} 
Thus the $p\overline p$ and $pp$ interference patterns are of opposite sign.

\begin{figure}[htb]
    \begin{center} 
\includegraphics[width=10cm]{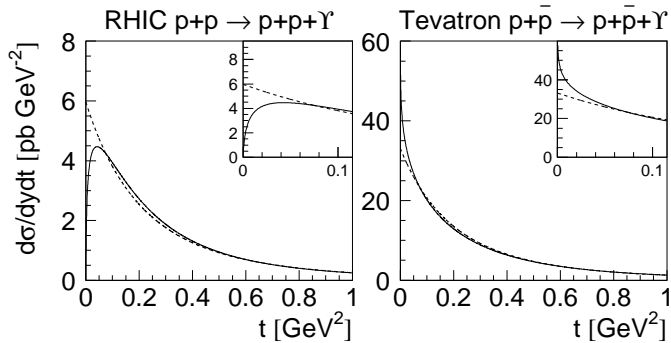}
      \end{center} 
\caption[]{The cross sections, $d\sigma/dydt$, at $y=0$ for $\Upsilon$ 
production in (a) $\sqrt{s}=500$ GeV $pp$ collisions at RHIC and (b)
$\sqrt{s}=1.96$ TeV $\overline pp$ collisions at the Tevatron.  The sign of 
the interference differs in the two
cases.  From Refs.~\protect\cite{Klein:2003vd,Klein:2003qc}.  Copyright 2004 
by the American Physical Society 
(http://link.aps.org/abstract/PRL/v92/e142003).}
\label{pppt}
\end{figure}

Figure \ref{pppt} compares the $t$ distributions for
$\Upsilon$ production at RHIC ($pp$ collisions at $\sqrt{s}=500$ GeV)
and the Fermilab Tevatron ($\overline pp$ collisions at
$\sqrt{s}=1.96$ TeV).  Interference makes the
two spectra very different.
At the LHC, the average impact parameters will be somewhat larger.
Since the vector meson rates will also be much larger, it
should be possible to study interference with the $\Upsilon$.

\subsubsection{Interferometry with short-lived particles} \bigskip

Interference is especially interesting because the vector
mesons have very short lifetimes.  For example, the $\rho^0$ decays before 
it travels 1 fm while interference typically occurs at  $20<b<40$ fm.  
Thus the $\rho^0$ photoproduction amplitudes
cannot overlap before the $\rho^0$ decay.  The conventional
interpretation \cite{Klein:2003aa} is that the interference must
occur at a later time, after the wavefunctions of the two sources
overlap, thus involving the vector meson decay products.  Since 
interference depends on the sum of the final-state momenta, it is a global 
quantity and involves non-local effects.
In other words, the global final-state wavefunction, $|\Psi \rangle$, is not
the product of the individual final-state particle wavefunctions, $|\Psi_n
\rangle$,
\begin{equation}
|\Psi \rangle \ne |\Psi_1 \rangle |\Psi_2 \rangle \cdots |\Psi_n \rangle \,\, ,
\end{equation}
an example of the Einstein-Podolsky-Rosen paradox.

\subsection{Coherent vector meson production in ultraperipheral $pA$ 
collisions} 
\label{paonium}
{\it Contributed by: L. ~Frankfurt, M. ~Strikman and  M.~Zhalov}

\subsubsection{Introduction} \bigskip

The $pA$ runs at the LHC will provide another means for studying photonuclear
processes.  Ultraperipheral vector meson production in $pA$ interactions 
originate predominantly from protons scattering off the photon field of the
nucleus.  Interactions where the nucleus scatters with a photon emitted by
the proton give a smaller contribution, see Fig.~\ref{fig:luminosities} of the
introduction.

The elementary reaction, $\gamma p\to V p$, is the only high-energy 
two-body reaction dominated by vacuum exchange which can readily be compared
to elastic $pp$ scattering.
Moreover, studying production of light to heavy vector mesons probes 
increasingly hard interactions.

UPC studies in $AA$ collisions have two limitations.  In heavy-ion
collisions, the photon can be emitted by either of the two nuclei, making it 
difficult to study coherent quarkonium production at $x<M_V/2E_N$ where $E_N$ 
is the ion energy per nucleon in the center-of-mass frame.  
Different systematic errors can also hinder the
comparison of data taken at more than one facility such as $\gamma p$ 
data at HERA and $\gamma A$ data at the LHC.
 
Studies of UPCs in $pA$ collisions (or d$A$ collisions, studied at RHIC for
technical reasons) can circumvent these problems by measuring
vector meson production over a much larger energy and momentum transfer range
than at HERA \cite{FSZ06_01}.  Effective Pomeron trajectories for light 
vector meson production and elastic $pp$ scattering can be compared  at 
similar energies, complementing the planned elastic $pp$ scattering studies
by TOTEM \cite{CMS-TOTEM} and possibly also ATLAS \cite{DetectorTDRs}. 

Calculations of the reaction
\begin{equation}
p + A \rightarrow p + A + V
\end{equation}
are performed within the same formalism as vector meson production in 
ultraperipheral $AA$ collisions. The $k$ integrand in
Eq.~(\ref{eq:3:vmcross}) for $AA$ collisions can be replaced by
\begin{equation}
{d \sigma_{pA\to pAV}\over dydt}=
{dN_{\gamma}^{Z}(y)\over dk}  {d\sigma_{\gamma p\rightarrow V p}(y)\over dt}+
{dN_{\gamma}^{p}(-y)\over dk}  {d\sigma_{\gamma A\rightarrow V A}(-y)\over dt}
\label{base}
\end{equation}
where the rapidity of the produced vector meson is
\begin{equation}
y={1\over 2}\ln{E_V-p_{z \,V} \over E_V +p_{z\, V}} \, \, .
\end{equation}
For large rapidities, the suppression of the finite-$t$ cross section is
negligible.
We have arbitrarily chosen the nuclear beam to be at positive rapidity and
the proton to be at negative rapidity.  The equivalent photon flux from the
nucleus, ${N_{\gamma}(y)}$, corrected for absorption 
at small impact parameters is given by Eq.~(\ref{analflux}).
The condition that the nucleus remains intact restricts scattering to 
large impact parameters, $b\ge R_A+r_N$, quantified by the nuclear thickness 
function, similar to $AA$ interactions.
At LHC energies, the total nucleon-nucleon cross section is $\sim 100$ mb. 
Therefore, interactions at $b<R_A+r_N$ give a negligible contribution. 
The transition region, $b\sim R_A+r_N$, where absorption,
while incomplete, is still significant, gives a very small contribution.
Thus inelastic screening corrections can be neglected.  The photon flux from
the proton is given in Eqs.~(\ref{eq:ffluxp}) and (\ref{eq:ddef}). 
The squared vector meson-nucleon center-of-mass energy, $s_{\gamma p}$, is 
$s_{\gamma p} = 2E_N (E_V + p_{z\, V}) = 2E_N M_V \exp(y)$.  Recall that
we generally refer to the squared energy as $s_{\gamma p}$ and the energy 
itself as $W_{\gamma p}$.  In Eq.~(\ref{eq:ffluxp}),
$z = W_{\gamma p}/\sqrt{s_{_{NN}}}$
and $s_{_{NN}}=4\gamma_{L}^p \gamma_{L}^A m_{N}^2$ is the nucleon-nucleon
center-of-mass energy while
$\gamma_{L}^p$ and $\gamma_{L}^A$ are the proton and nuclear
Lorentz factors in the lab frame of each beam.

\subsubsection{Heavy quarkonium production} \bigskip

We begin with heavy quarkonium photoproduction.  A fit \cite{Strikman} to the 
data was used for $J/\psi$ photoproduction while for $\Upsilon$ 
production, we approximate the cross section as in Eq.~(\ref{eq:cs}),
consistent 
with the limited HERA data.
The energy dependence follows from the calculations
in the leading $\log Q^2$ approximation \cite{Frankfurt:1998yf}, taking into  
account the inequality of the light-cone momentum fractions on the gluon
ladder, see Eq.~(\ref{skewed}).

\begin{figure}[htbp]
\begin{center}
\epsfig{file=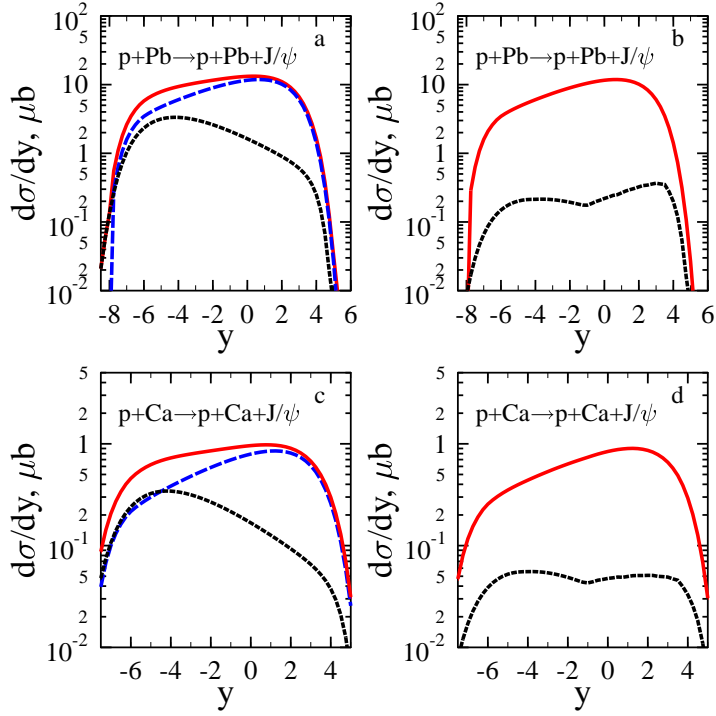, height=0.4\textheight}
 \caption[]{The $J/\psi$ rapidity distribution 
in $p$Pb, (a) and (b), and $p$Ca, (c) and (d), UPCs at the LHC. 
The long-dashed curve is the $\gamma p$ contribution and the short-dashed 
curve is the $\gamma A$ contribution. The solid curve is the sum.  
Leading-twist nuclear shadowing is included in (b) and (d).  Here the $\gamma 
p$ contribution is indistinguishable from the sum. Reprinted from 
Ref.~\protect\cite{FSZ06} with permission from Elsevier.}
 \label{psilhc}
\end{center}
\end{figure}

The coherent quarkonium photoproduction cross section is calculated 
with leading-twist nuclear shadowing, see Ref.~\cite{Frankfurt:2003wv} 
and Section~\ref{vmaa}. 
The QCD factorization theorem for  exclusive meson 
photoproduction~\cite{Brodsky:1994kf,Collins:1996fb} expresses the 
imaginary part of the forward amplitude for $\gamma A \to V A$
by convolution of the meson wavefunction at zero $q \overline q$
transverse separation, the hard scattering amplitude, and the generalized 
parton distribution of the target,
$G(x_1,x_2,Q^2,t_{\rm min})$, where $t_{\rm min}\approx -x^2m_N^2$.
To a good approximation, 
$G_A(x_1,x_2,Q^2,t=0) \approx g_A(x,Q^2)$ where $x=(x_1+x_2)/2$ 
\cite{Brodsky:1994kf,Radyushkin:2000uy}. Hence, the
amplitude for $\Upsilon$ photoproduction at $k_T^2=0$ is \cite{FKS}
\begin{equation}
{\mathcal M}(\gamma A \to \Upsilon A) = {\mathcal M}(\gamma N \to 
\Upsilon N) {\frac{g_A(x,Q^2_{\rm eff})} {Ag_N(x,Q^2_{\rm eff})}} 
F_A(t_{\rm min})\, \, 
\label{amplitude}
\end{equation}
since the meson wavefunction cancels in the ratio.  The nuclear form factor,
$F_A$, is normalized so that $F_A(0)=A$, giving Eq.~(\ref{gratios})
for the cross section ratio at $t=0$. We use the same model of gluon shadowing 
as in Section 4.2.  Current uncertainties in leading-twist gluon shadowing  
will be reduced after the recent H1 data on hard diffraction \cite{Newman} is 
incorporated in the analysis.

In our calculations, we neglect quasi-elastic nuclear scattering since the 
probability is relatively small and is easily separated from 
coherent production using information from the ZDCs, see Section~\ref{neutron} 
in Ref.~\cite{Strikman}.

\begin{figure}[htbp]
\begin{center}
\epsfig{file=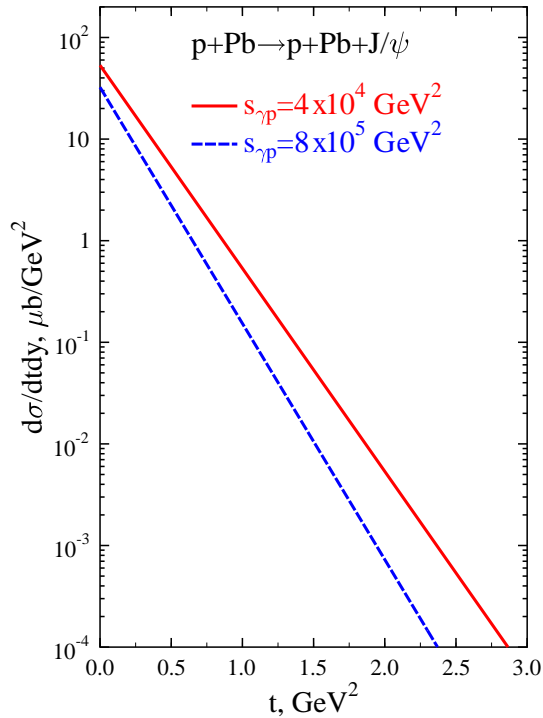, height=0.4\textheight}
\caption[]{The $J/\psi$ $t$ distribution in $p$Pb UPCs. Reprinted from 
Ref.~\protect\cite{FSZ06} with permission from Elsevier.}
 \label{psidsdt}
\end{center}
\end{figure}

The $J/\psi$ results are presented in Figs.~\ref{psilhc} and \ref{psidsdt}. 
The direction of the incoming nucleus corresponds to positive rapidities. 
One can see from Fig.~\ref{psilhc} that the $\gamma p \to J/\psi p$ cross 
section in ultraperipheral $pA$ collisions is large enough to be measured in 
the interval $20 < W_{\gamma p} < 2\times 10^3$ GeV. 
The minimum $W_{\gamma p}$ reflects the estimated maximum rapidity at 
which $J/\psi$'s could be detected\footnote{In the case when the photon 
comes from the left, the ALICE detector is expected to
have good $J/\psi$ acceptance to $y \sim 3.5$ and $\Upsilon$ acceptance to 
$y\sim 2$ as well for several smaller $|y|$ intervals \cite{note}.}.
The maximum $W_{\gamma p}$ corresponds to
$x_{\rm eff} \sim M_{J/\psi}^2/W_{\gamma p}^2 \sim 2\times 10^{-6}$, low 
enough to reach the domain where  
interaction of small dipoles contributing to the $J/\psi$ photoproduction  
amplitude already requires significant taming (see {\it e.g.} \cite{ted}).

For large $W_{\gamma p}$ (positive $y$),  the coherent $\gamma A \to J/\psi A$
contribution to $d\sigma /dy$ is negligible.  At 
negative $y$, $W_{\gamma p}$ is small and $W_{\gamma A}$ is large so that
$\gamma A$ contribution becomes relevant.  
Nevertheless, it remains a correction to the total even without 
nuclear shadowing. 
The $t$-dependence of the coherent $\gamma A$ contribution is
determined primarily by the nuclear matter form factor.  On the other
hand, the $t$-dependence of the coherent $\gamma p$ contribution is due to the
gluon transverse momentum distribution at the relevant $x$ value and
is a much weaker function of $t$.  Both these $t$-dependencies can be
approximated by exponentials.  Accordingly, the $\gamma A$ contribution
can be determined by fitting $d\sigma/dt$ to a sum of two exponentials.
The $\gamma A$ contribution to $J/\psi$ production could be effectively
enhanced or reduced by introducing a $p_T$ cut, {\it e.g.} $p_T < 300$ MeV/$c$ 
to enhance the $\gamma A$ contribution or 
$p_T \geq$ 300 MeV/$c$ to reduce it. 
An observation of the $\gamma A$ contribution with the low $p_T$ cut
would probe the small dipole interaction with nuclei at $x_A\sim 10^{-5} - 
10^{-6}$. If gluon shadowing is large, observing the $\gamma A$ contribution 
for such $x_A$ would require very good $p_T$ resolution, 
$p_T\leq 150$ MeV/$c$ or better.  It may be possible to eliminate/estimate 
the $\gamma p$ contribution by measuring recoil protons produced with 
$x_{\Pomeron}=M^2_{J/\psi}/W^2_{\gamma p}$. 
In the kinematic range $0.1 \leq x_{\Pomeron} \leq 0.001$, the
proton could be detected, for example, by the TOTEM T1 and T2 trackers or by  
the Roman pot system proposed in Ref.~\cite{loi}. 

\begin{figure}[htbp]
\begin{center}
\epsfig{file=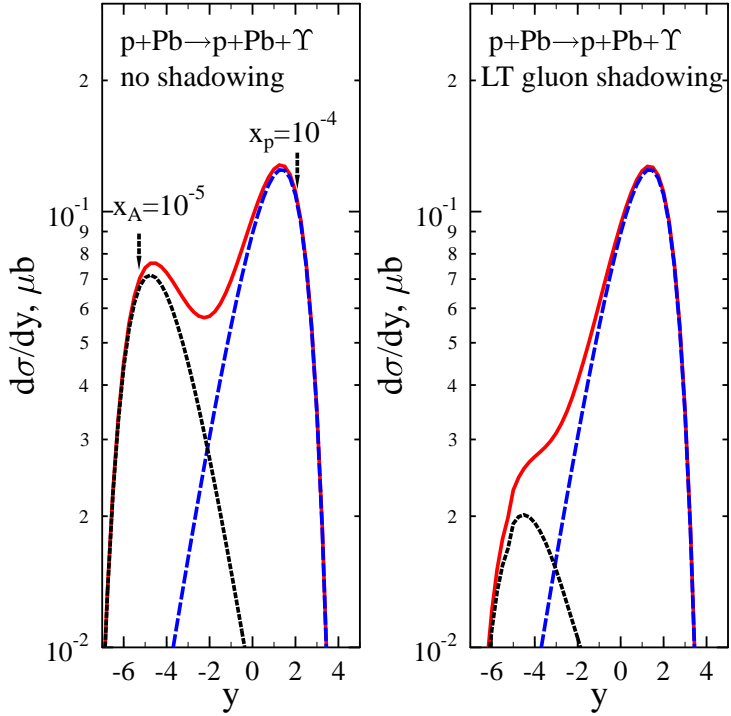, height=0.4\textheight}
 \caption[]{The $\Upsilon$ rapidity distribution in $p$Pb UPCs at the LHC 
without (left-hand side) and with (right-hand side) leading-twist nuclear 
shadowing.  The $\gamma p$ contribution is given by the long-dashed lines, 
$\gamma A$ the short-dashed lines and the solid curves are the sum.
Reprinted from 
Ref.~\protect\cite{FSZ06} with permission from Elsevier.}
 \label{upslhc}
\end{center}
\end{figure}
  
The  \jp production cross section will be sufficiently 
large to  measure the $t$ dependence of the $\gamma p$ contribution, shown
in Fig.~\ref{psidsdt}, up 
to $-t \sim 2$ GeV$^2$, if the contribution from proton dissociation
can be suppressed.  This measurement provides a unique opportunity to study
the $t$ dependence of the hard Pomeron trajectory since the power of
$s_{\gamma p}/s_0$ in {\it e.g.} Eq.~(\ref{eq:cs}) can be written as 
$2(\alpha_{\Pomeron}^{\rm hard}(t) -1)$.  
This study will complement measurements
of $\alpha_{\Pomeron}^{\rm hard}(t)$ in vector meson production with rapidity
gaps, discussed in Section~\ref{deltay}

\begin{figure}[htbp]
\begin{center}
\epsfig{file=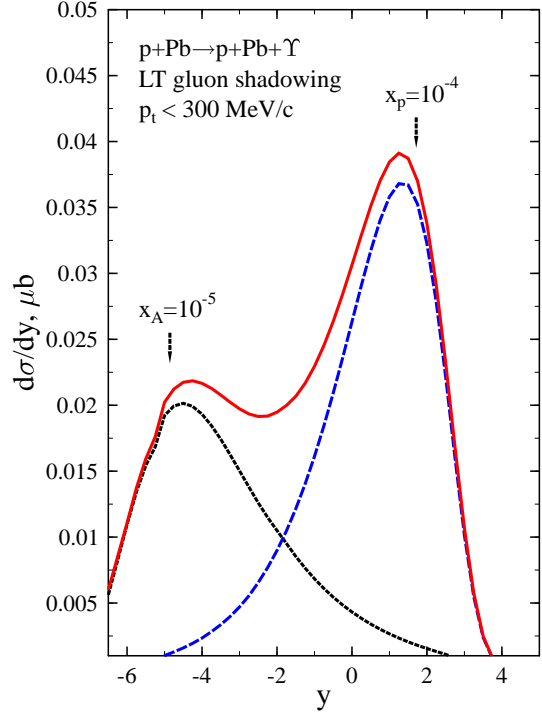, height=0.4\textheight}
 \caption{The $\Upsilon$ rapidity distribution in $p$Pb UPCs including 
leading-twist shadowing and a $p_T$ cut on the $\Upsilon$, $p_T < 300$ MeV/$c$.
The curves are the same as in Fig.~\protect\ref{upslhc}. Reprinted from 
Ref.~\protect\cite{FSZ06} with permission from Elsevier.
}
 \label{upslhccut}
\end{center}
\end{figure}
   
For $\Upsilon$ production, Fig.~\ref{upslhc}, 
the $\gamma p$ contribution can be studied 
for  $10^2 \leq W_{\gamma p} \leq 10^3$ GeV.  The $W_{\gamma p}$ interval is 
smaller due to the strong drop 
of the cross section with $W_{\gamma p}$, not compensated by the
larger photon flux at small $W_{\gamma p}$. 
Still, this interval is sufficient for a study of the $\Upsilon$ energy 
dependence since the cross section is expected 
to increase with $W_{\gamma p}$ by a factor of $\sim 30$ as $W_{\gamma p}$ 
increases from 100 GeV to 1 TeV.
The statistics will also be sufficient to determine the slope of the 
$t$-dependence. Since the $\Upsilon$ is the smallest dipole, this would 
provide a valuable 
addition to the measurement of the transverse gluon 
distribution.

The relative $\gamma A$ contribution is much larger for $\Upsilon$ 
production.
Without nuclear shadowing, $\gamma A$ would dominate at 
rapidities corresponding to $x_A\sim 10^{-5}$, as shown in 
Fig.~\ref{upslhc}(a). Even if nuclear shadowing reduces the $\gamma A$ 
cross section by a factor of $3-4$, Fig.~\ref{upslhc}(b), the cross section is 
still dominated by the nuclear contribution at negative rapidities. If 
the $p_T < 300$ MeV/$c$ cut is applied,  Fig.~\ref{upslhccut}, the background 
could be suppressed further. Hence $pA$ scattering can probe the interaction 
of $\sim 0.1$ fm dipoles with nuclei at very small $x$, virtually impossible 
in any other process available in the next decade.

The estimates of Ref.~\cite{FSZ06_01} indicate that
$J/\psi$ photoproduction measurements
would also be feasible in high luminosity $pA/$d$A$ runs at RHIC.
Thus we find that the LHC $pA$ runs will significantly add to 
quarkonium photoproduction studies in $AA$ collisions by 
new studies of the elementary reaction $\gamma p \rightarrow V p$ in an energy
substantially exceeding that of HERA.
In addition, ultraperipheral $pA$ collisions provide independent $\Upsilon$
and possibly also $J/\psi$ photoproduction at very small $x_A$ in $\gamma A$
interactions.

\subsubsection{Light vector meson production} \bigskip

The Pomeron hypothesis of universal strong interactions has provided
a good description of $pp$ and $\overline p p$ interactions at collider 
energies \cite{Land1}. 
The total and elastic hadron scattering cross sections are hypothesized to
proceed via single Pomeron exchange.  The cross section can be written as
\begin{equation}
{d \sigma_{h_1  h_2 \to h_1  h_2}\over dt } = f_{h_1 h_2}(t) 
\left({s_{h_1 h_2}\over s_0}\right)^{2\alpha_{\Pomeron}(t) -2}
\end{equation} 
where $f_{h_1 h_2}(t)$ parametrizes the $t$ dependence of the cross section,
$s_{h_1 h_2}$ is the square of the $h_1 h_2$ center-of-mass energy 
and $s_0 \sim 1$ GeV$^2$.  Here $\alpha_{\Pomeron}(t) $ is the Pomeron 
trajectory.  At small $t$, as in coherent production,
\begin{equation}
\alpha_{\Pomeron}(t)=\alpha_0 + \alpha^{\prime}t \, \, .
\label{pomeq}
\end{equation}
The $pp$ and $\overline p p$ total and elastic cross sections can be well 
described with \cite{Land1}
\begin{equation}
\alpha_0=1.0808 \, \, , \,\,\,\,\,\,\, \alpha^{\prime}=0.25 \, \mbox{GeV}^{-2}.
\label{univ}
\end{equation}
Checking the universality hypothesis at fixed-target energies is
hampered by exchanges from non-Pomeron sources that die out at 
high energies.  However, significant deviations from universality cannot be 
ruled out. For example,
studies of the total $\Sigma^- N$ cross section \cite{SELEX} 
are consistent with Lipkin's prediction of  
$\alpha_0=1.13$ for this reaction \cite{Lipkin}.

Vector meson photo/electroproduction plays a unique 
role in strong interaction studies.  Light vector meson photoproduction is
the only practical way to check the accuracy of the universality hypothesis
for soft interactions at collider energies.  The exclusive photoproduction
is predicted to be
\begin{equation}
{d \sigma (\gamma p \to V p)\over dt } = f_{\gamma p}(t) 
\left({s_{\gamma p}\over s_0}\right)^{2\alpha_{\Pomeron}(t) -2} \, \, .
\label{vm}
\end{equation} 

There are several mechanisms which could cause the predicted 
universality to break down.  In soft interactions there are non-universal
multi-Pomeron exchanges which are generally more important at large $t$. 
The $\rho^0$ data are consistent with 
Eq.~(\ref{vm}) for the universal Pomeron trajectory in Eq.~(\ref{pomeq}) with
the parameters of Eq.~(\ref{univ}) \cite{LD}. 
The very recent H1 results \cite{Olsson} 
for $\alpha_{\Pomeron}(t) $ using
Eq.~(\ref{vm}), assuming a linear trajectory lead, to  
\begin{equation}
\alpha_{\Pomeron}(t)= 1.093\pm 0.003^{+0.008}_{-0.007} +
(0.116\pm0.027^{+0.036}_{-0.046}\,\,\mbox{GeV}^{-2}) t \, \, .
\end{equation} 
This result agrees well with the previous ZEUS analysis based on a comparison 
with fixed-target data \cite{ZEUSrho}, seemingly contradicting the universality
of $\alpha'$. However, the data allow another interpretation: a significant 
$t$-dependence with $\alpha'(t) \sim 0.25$ GeV$^{-2}$ for $-t \leq 0.2$ 
GeV$^2$.
 
Thus new questions about soft dynamics arise from the HERA studies of light 
vector meson photoproduction:
\begin{itemize}
\item To what accuracy is the Pomeron trajectory linear?  
\item Is $\phi$ production purely soft or will a larger $\alpha_0$ be observed,
as in $J/\psi$ photoproduction?
\item Does $\alpha'$  decrease with increasing vector meson mass as expected 
in pQCD or it is the same for $M\leq M_{J/\psi}$ 
as the current HERA data may suggest?
\item Are nonlinearities in the effective Pomeron trajectories,
where $\alpha^\prime$ is not constant, the same 
for all vector mesons?
\end{itemize}

To address these questions, it is necessary to measure $\rho^0$ and
$\phi$ photoproduction over the largest possible interval of 
$W_{\gamma p}$ and $t$.  To determine the feasibility of this program in
ultraperipheral $pA$ interactions, we used the Donnachie-Landshoff 
parametrization of the elementary cross section \cite{LD},
\begin{equation}
{\frac {d\sigma_{\gamma p\rightarrow Vp}} {dt}}=
{\vert T_{S}(s_{\gamma p},t)+T_{H}(s_{\gamma p},t)\vert}^2 \, \, ,
\label{ldparam}
\end{equation}
where $T_S(s_{\gamma p},t)$ is the amplitude for soft Pomeron and Reggeon
exchange and $T_H(s_{\gamma p},t)$ is the hard Pomeron amplitude.
Two Regge trajectories were used \cite{LD} to parameterize $T_S$:
$\alpha_{\Pomeron_1} (t)=1.08+\alpha_{\Pomeron_1}^\prime t$,  
$\alpha_{\Pomeron_1}^\prime =0.25 \,\mbox{GeV}^{-2}$ for soft Pomeron
exchange and
$\alpha_{R}(t)=0.55+\alpha_{R}^\prime t$,
$\alpha_{R}^\prime =0.93\, \mbox{GeV}^{-2}$ for Reggeon exchange.
The Regge trajectory for the hard Pomeron also uses a linear parametrization:
$\alpha_{\Pomeron_0}=1.44+\alpha_{\Pomeron_0}^\prime t$,
$\alpha_{\Pomeron_0}^\prime =0.1\, \mbox{GeV}^{-2}$.

\begin{figure}[htbp]
\begin{center}
\epsfig{file=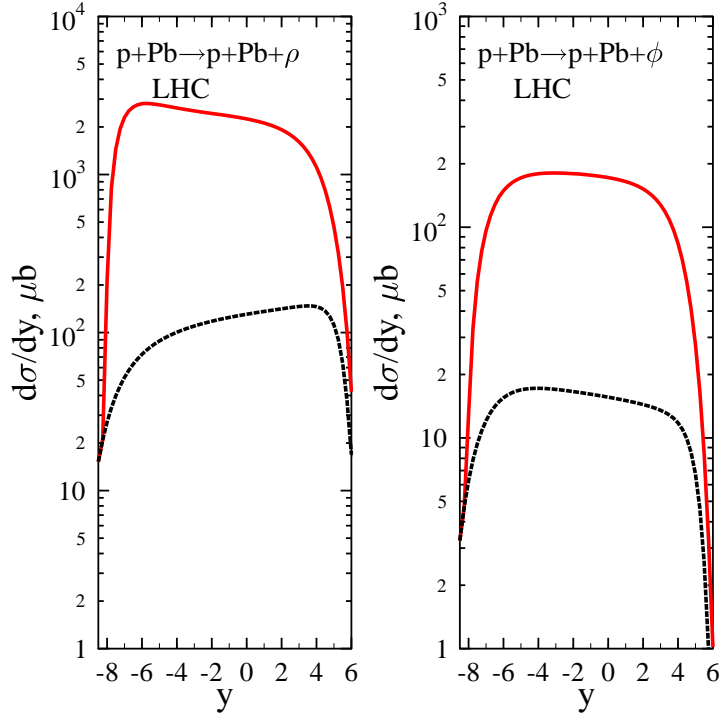, height=0.4\textheight}
\caption{The $\rho^0$ and $\phi$ rapidity distributions in $p$Pb UPCs at the 
LHC. The short-dashed lines are the $\gamma A$ contribution while the solid
curves are the total, indistinguishable from the $\gamma p$ contribution.
Reprinted from 
Ref.~\protect\cite{FSZ06} with permission from Elsevier.}
 \label{lightlhc}
\end{center}
\end{figure}

\begin{figure}[htbp]
\begin{center}
\epsfig{file=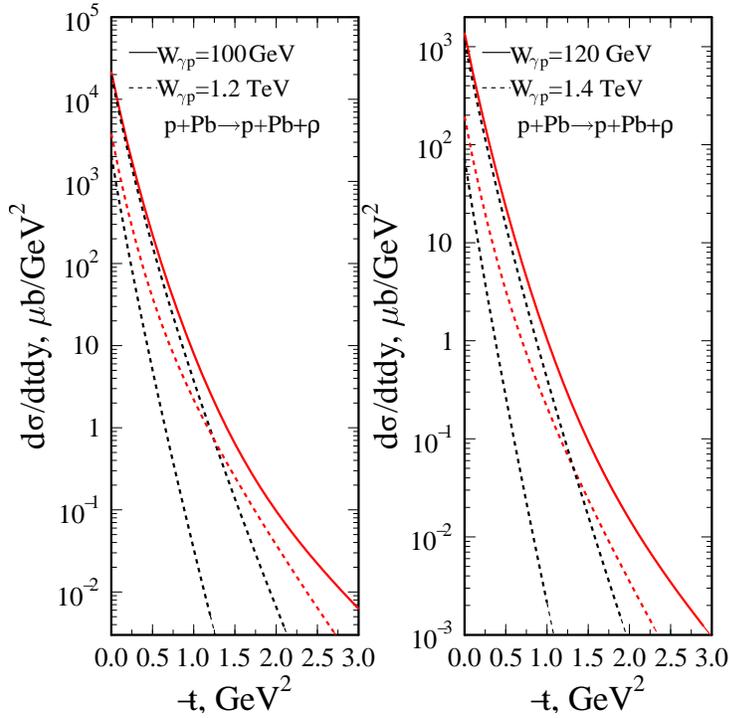, height=0.4\textheight}
\caption{The $\rho^0$ and $\phi$ $t$ distributions in $p$Pb UPCs.  The solid
and long-dashed lines are the results of Eq.~(\protect\ref{ldparam}) 
for two different 
values of $W_{\gamma p}$.  The short-dashed lines are the same results without
the contribution from $T_H$. Reprinted from 
Ref.~\protect\cite{FSZ06} with permission from Elsevier.}
\label{dtlhc}
\end{center}
\end{figure}

The coherent light vector meson production cross section for $\gamma A$
interactions was calculated using the  vector dominance model combined with 
Glauber-Gribov multiple scattering. The final-state interaction is determined 
by the total $VN$ cross sections.  The $\rho^0$ cross section was calculated
using vector dominance with the Donnachie-Landshoff
parameterizations for the $\gamma p\rightarrow \rho^0 p$ amplitude.
The energy dependence of the $\phi N$ total cross section was assumed to be
$\sigma_{\phi N}=9.5 (s_{\phi N}/1 \, {\rm GeV}^2)^{0.11}$ mb,
taken from a fit to the data. 
The $t$-integrated results are presented in Fig.~\ref{lightlhc}. 
The rates at the expected LHC $pA$ luminosities are very large, even for 
$W_{\gamma p}=2$ TeV.  The $t$-dependence is shown in Fig.~\ref{dtlhc},  
demonstrating that the rates at $L_{p {\rm Pb}} \approx 1.4\times 10^{30}$ 
cm$^{-2}$s$^{-1}$ are sufficient for studying the differential cross sections 
from  $|t | \ge  2$ GeV$^2$ up to $\sqrt {s_{\gamma N}}\approx 1$ TeV.

Measurements of the $t$-dependence over two orders of magnitude in 
$W_{\gamma p}$ in the same experiment would allow precision measurements of 
$\alpha'$ for $\rho^0$ and $\phi$ production.  For example, if the $t$
dependence of $f_{h_1 h_2}(t)$ is parametrized as $\exp[B_0 t]$ and
$(s_{\gamma p}/s_0)^{[2(\alpha(t) - 1)]} = (s_{\gamma p}/s_0)^{[2\alpha^\prime
t]}$ is reformulated as $\exp[2 \alpha^\prime t \ln(s_{\gamma p}/s_0)]$, then,
in general, the cross section is proportional to $\exp[Bt]$ where
$B = B_0 + 2 \alpha^\prime \ln (s_{\gamma p}/s_0)$.  Thus, if $\alpha^\prime
= 0.25$ GeV$^{-2}$, the change in slope is $\Delta B = B - B_0 \sim 4.6$ 
GeV$^{-2}$, a $\sim 50\%$ change.  The data should then be sensitive 
to any nonlinearities in the Pomeron trajectory.
It therefore appears that light meson production studies will substantially 
contribute to the understanding of the interplay between soft and hard 
dynamics.

Thus UPC studies in $pA$ collisions at the LHC will provide unique new 
information about diffractive $\gamma p$ collisions, both in the hard regime, 
down to $x\sim 10^{-6}$, and in the soft regime. 

\subsection{Neutron tagging of quasi-elastic \jp\ and 
$\Upsilon$ photoproduction}
{\it Contributed by: M.~Strikman, M.~G.~Tverskoy and M.~B.~Zhalov}
\label{neutron}

In Section~\ref{paonium} we argued that ultraperipheral heavy-ion 
collisions could study coherent vector meson production up to 
$s_{\gamma N}= 2M_VE_N$. Although coherent events can be easily identified by 
selecting vector mesons with sufficiently small transverse  momentum, $p_T \leq
\sqrt{3}/R_A$, it is very difficult to determine whether the left or right 
moving nucleus was the source of the photon that converted into a vector meson.
Since the photon flux strongly decreases with increasing photon energy, lower
energy photons are the dominant contribution at $y\neq 0$.  
 
Another vector meson production process is governed by similar dynamics 
and comparable cross section: quasi-elastic production,
$\gamma +A \to V +A'$. It is as sensitive to the dynamics of the vector meson
interaction in the nuclear medium as coherent processes.
The $A$ dependence of this process varies from $\propto 
A$ for weak absorption to $A^{1/3}$ for strong absorption since
only scattering off the nuclear rim contributes.   
Thus, the sensitivity to the change of the interaction regime from color 
transparency to the black disk regime is up to 
$\propto A^{2/3}$, as is the case for coherent processes 
where the $t$-integrated cross section
is $\propto \sigma^{2}_{\rm tot}(VA)/R_A^2$ and changes from $\propto A^{4/3}$
to $\propto A^{2/3}$.  Thus the ratio of quasi-elastic to coherent cross 
sections should be a weak function of the $Q \overline Q$ dipole interaction
strength in the medium.  This expectation is consistent with Glauber-model
estimates where the ratio of quasi-elastic to coherent
$J/\psi$ and $\Upsilon$ production cross sections is $0.3 - 0.2$ over the
entire energy range from color transparency (impulse approximation)
to the BDR.  The $\mu^+\mu^-$ continuum, an important background,
for coherent vector meson production,
see Sections~\ref{alice-vm} and \ref{cms-vm}, is reduced 
in incoherent production.
    
The QCD factorization theorem for quarkonium leads to color transparency for 
moderate energies, $s_{\gamma N} \leq M_V^2/x_0$ where $x_0\sim 0.01$ is a
minimum scale where little absorption is expected.  As $x$ decreases and 
$s_{\gamma N}$ increases, color transparency gives way to leading-twist 
nuclear shadowing and, ultimately, the BDR, violating the factorization 
theorem. 

In most of the LHC detectors, it is much easier 
to trigger on vector meson production if it is accompanied by the breakup of 
at least one of the nuclei, resulting in one or more neutrons with
the per nucleon beam energy, 
$\sim E_N\approx 0.5\sqrt{s_{_{NN}}}$,
hitting one of the ZDCs. 
Current measurements and numerical estimates indicate that,
at RHIC, given coherent $J/\psi$ production, there is a $50-60$\% probability
for excitation.
The probability will be somewhat larger at the LHC
\cite{Baltz:2002pp}.  The removal of a nucleon from a heavy
nucleus in the quasi-elastic process should also lead to significant nuclear 
breakup, resulting in the emission of several neutrons
with a probability of order one.  
Hence, the detection rates for quasi-elastic and coherent processes
in UPCs at RHIC and the LHC  should be comparable.
 
Here we summarize the first study \cite{Strikman} of the 
characteristics of quasi-elastic processes relevant for their identification 
in UPCs. As a starting point, we use \jp\ photoproduction at 
RHIC and $\Upsilon$
production at the LHC. In both cases, the effective cross sections for the 
$Q\overline Q$ pair interaction with the medium are rather small. 

We then use the impulse approximation to model the neutron yields with
quarkonium production.  Data shows that the $t$ 
dependence of the $\gamma +N\to J/\psi +N$ cross section is rather flat,
$B_{J/\psi} \sim 4 - 5$ GeV$^{-2}$, in the RHIC and LHC energy range.  
The $\Upsilon$ slope, $B_{\Upsilon}$, is expected to be even smaller, 
$\sim 3.5$ GeV$^{-2}$.  The effective $t$ range in quasi-elastic production
can be rather large, up to $\sim 1$ GeV$^2$, relative to coherent quarkonium 
photoproduction where $|t| \leq 0.015$ GeV$^2$ since higher $t$ is suppressed 
by the nuclear form factor. 
The ejected nucleons have average momenta $p_{N}\approx \sqrt{|t|} \approx
1/B_V \sim 0.3$ GeV,
large enough for strong reinteraction in the nucleus, making the probability
for the nucleus to break up when a nucleon is emitted of order one.

\begin{figure}[htbp]
\begin{center}
\epsfig{file=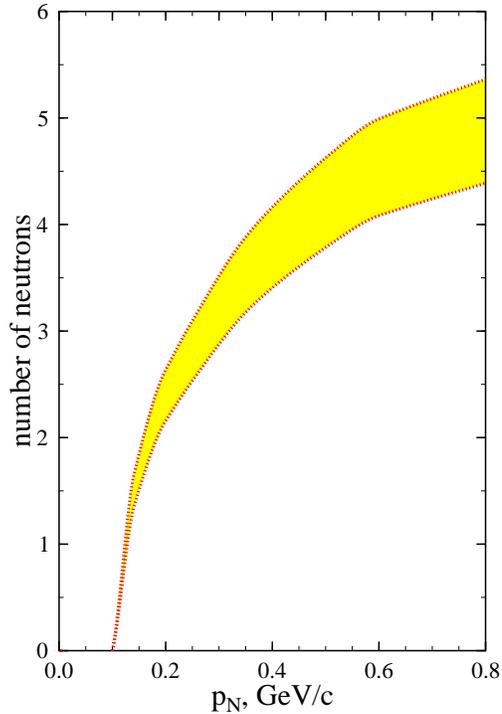, height=0.4\textheight}
\caption[]{
The average number of neutrons emitted in incoherent $J/\psi$ production in
Au+Au UPCs at RHIC and $\Upsilon$ production in Pb+Pb UPCs at the LHC
as a function of the recoil nucleon momentum, 
$p_{N}=\sqrt{|t|}$.
The band indicates the estimated uncertainties of the Monte Carlo.
Reprinted from 
Ref.~\protect\cite{Strikman} with permission from Elsevier.}
\label{neutmom}
\end{center}
\end{figure}
 
To characterize the interaction of the recoil nucleon with the
residual nucleus in the reaction, $N+(A-1)\to C_{i}+kn$, we introduce the 
excitation function, $\Phi_{C_{i},kn}(p_N)$, the probability to emit $k$ 
neutrons with $C_i$ charged fragments. 
The excitation function was calculated including the nucleon cascade 
within the nuclear medium followed by
the evaporation of nucleons and nuclear fragments from the nucleus. 
In Ref.~\cite{Strikman:1998cc}, the same Monte Carlo 
was used to analyze neutron production in the E665 fixed-target 
experiment at Fermilab which studied soft neutron production in $\mu$Pb DIS. A
a good description of these data \cite{E665}, as well as other intermediate
energy neutron
production data in $pA$ interactions, was obtained.
The dependence of the average number of emitted neutrons on the recoil
nucleon momentum is shown in Fig.~\ref{neutmom}.
For typical quasi-elastic \jp\ or $\Upsilon$ production, $p_T \sim 
B^{-1/2}_{J/\psi}\sim 0.5$ GeV/$c$, 
about four neutrons are emitted per event.

In Ref.~\cite{Strikman}, a more realistic estimate of the absolute $J/\psi$
production rate at RHIC was obtained, including
absorption of the $c\overline c$ in the nuclear medium. An effective  
$c \overline c $ interaction cross section, $\sigma_{\rm eff}(x\ge 0.015)=3$ 
mb, was used, based on Ref.~\cite{oscil}.
In these kinematics, the contribution of double elastic scattering
can be neglected since $\sigma_{\rm el}/\sigma_{\rm in}$ is very small for 
quarkonium interactions. Thus a simple Glauber-type model 
approximation can be used to obtain the 
probability for exactly one elastic rescattering and no inelastic interactions,
\begin{eqnarray}
\sigma^{\rm incoh}_{\gamma A\to J/\psi A'} 
= 2 \pi \sigma_{\gamma N\to J/\psi N} 
\int \limits_{0}^{\infty} db \, b \int \limits_{-\infty}^{\infty} dz 
\rho_A(b,z) \exp [-\sigma_{\rm tot}^{J/\psi N} T_A(b)] \, \, .
\end{eqnarray}
Here $\sigma_{\rm tot}^{J/\psi N}$ is the effective quarkonium-nucleon total 
cross section, $\sim 3$ mb. 

The coherent and incoherent \jp\ photoproduction cross sections in UPCs, 
integrated over rapidity and momentum transfer in the RHIC kinematics, are 
given in Table~\ref{tcrsec}.  The table also shows the quasi-elastic 
\jp\ partial cross sections without any emitted neutrons,  
$(0n,0n)$, and with the breakup of one nucleus,
$(0n,Xn)$, where $X \geq 1$.
The ratios $(0n,0n)$/total and $(0n,Xn)$/total should be similar for $\Upsilon$
production at $y=0$ at the LHC.

\begin{table}[htbp]
\begin{center}
\caption[]{The total coherent and incoherent $J/\psi$ photoproduction
cross sections calculated in the impulse approximation (IA) and the Glauber
approach in Au+Au UPCs at RHIC.}
\vspace{0.4cm}
\begin{tabular} {|c|c|c|c|c|}\hline 
 &    $\sigma_{\rm coh}$ ($\mu$b) & $\sigma_{\rm incoh}$ ($\mu$b) 
& $\sigma_{\rm incoh}^{(0n,0n)}$ ($\mu$b) &
 $\sigma_{\rm incoh}^{(0n,Xn)}$ ($\mu$b) \\ \hline
 IA &   212  & 264  &  38 & 215  \\ 
 Glauber &   168  & 177  & 25.5 & 144 \\ \hline
\end{tabular}
\label{tcrsec}
\end{center}
\end{table}

\begin{figure}[htbp]
\begin{center}
\epsfig{file=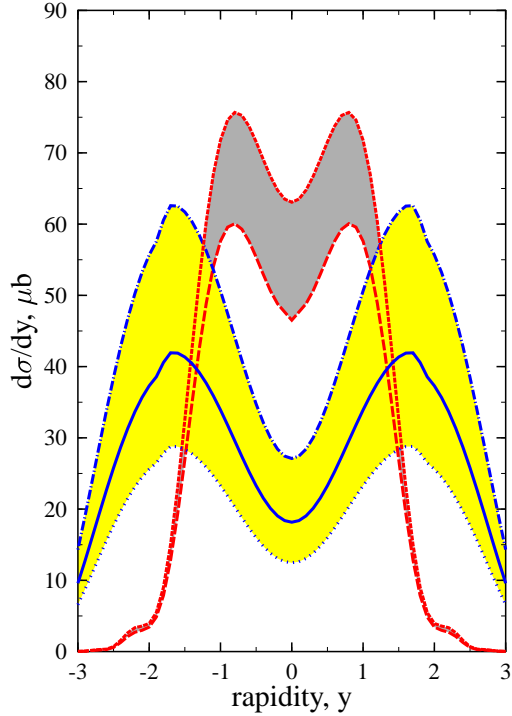, height=0.4\textheight}
\caption[]{
The $t$-integrated rapidity distributions for coherent 
$J/\psi$ photoproduction 
in Au+Au UPCs at RHIC calculated in the impulse approximation (short-dashed
line) and with $\sigma_{\rm tot}^{J/\psi N} = 3$ mb (long-dashed line).
The incoherent \jp\ cross section calculated in the Glauber model for
$\sigma_{\rm tot}^{J/\psi N} = 0$ (dot-dashed line), 3 (solid) and 6 (dotted) 
mb. Reprinted from 
Ref.~\protect\cite{Strikman} with permission from Elsevier.}
\label{rap}
\end{center}
\end{figure}

The coherent and quasi-elastic \jp\ rapidity distributions, 
integrated over $t$,
are shown in Fig.~\ref{rap} for several values of $\sigma_{\rm tot}^{J/\psi N}$
to illustrate their sensitivity to the $J/\psi N$ interaction 
strength. The coherent distribution is narrower because it is suppressed
by the nuclear form factor in the region where the longitudinal momentum
transfer, $p_{z}= M_{J/\psi}^{2}m_{N}/{s_{\gamma N}}$, is still significant.
The predictions of Ref.~\cite{Strikman} at $y=0 $ agrees with the preliminary 
PHENIX data, see Section~\ref{phenix_res}.

\begin{figure}[htbp]
\begin{center}
\epsfig{file=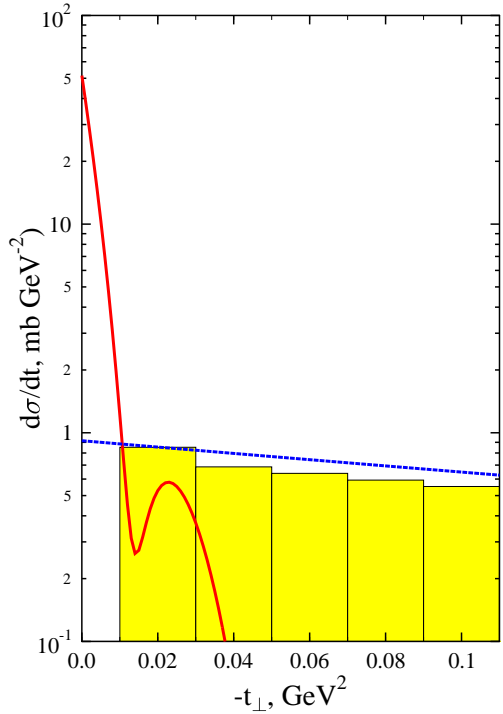, height=0.4\textheight}
\caption[]{The $t$ distribution integrated over $-3 \leq y\leq 3$
for coherent (solid line) and incoherent (dashed line) $J/\psi$
photoproduction in UPCs at RHIC.  The shaded histogram shows the incoherent
cross section with neutron emission. Reprinted from 
Ref.~\protect\cite{Strikman} with permission from Elsevier.}
\label{dst}
\end{center}
\end{figure}

\begin{figure}[htbp]
\begin{center}
\epsfig{file=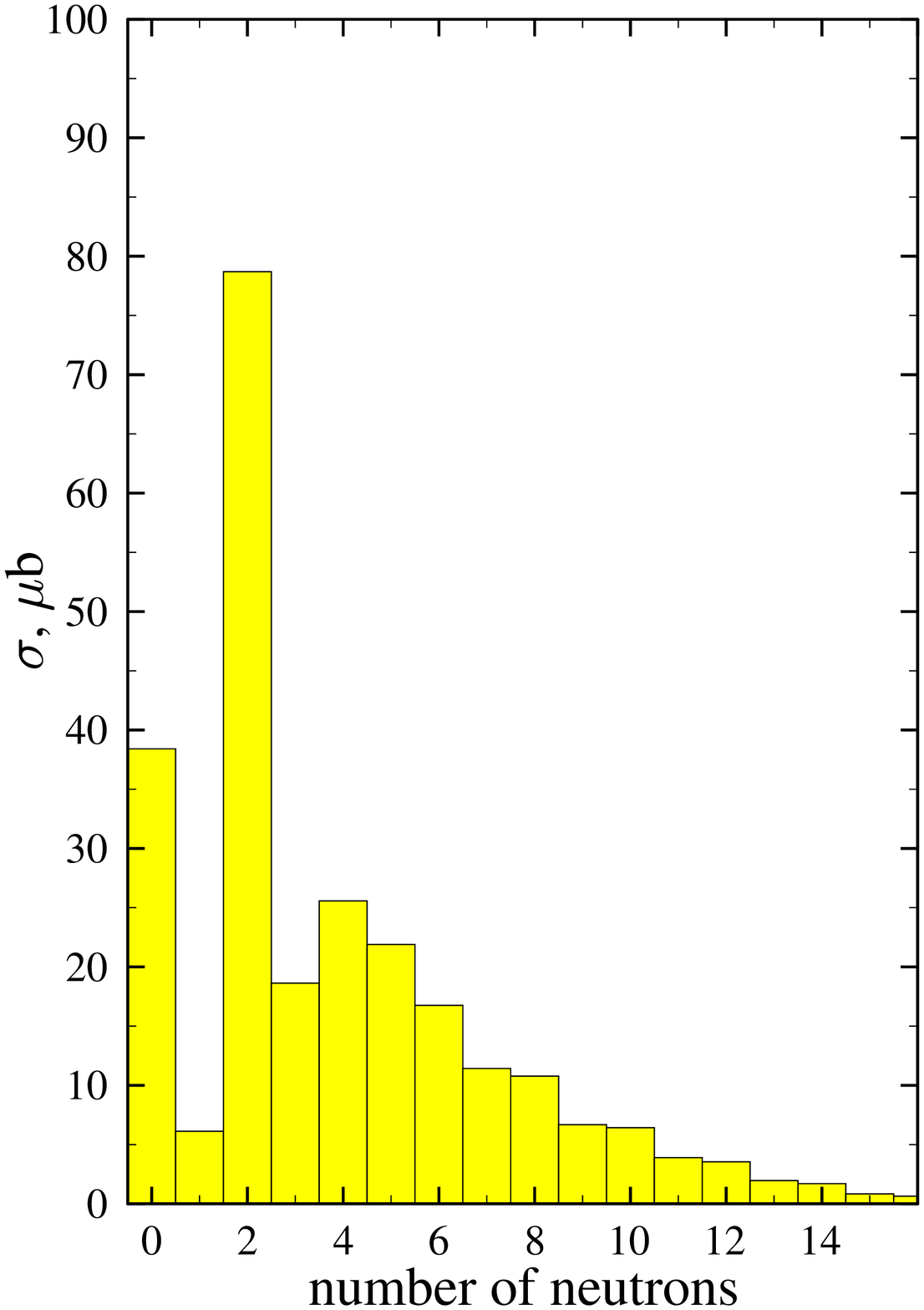, height=0.4\textheight}
\caption[]{
The incoherent $J/\psi$ cross section in Au+Au UPCs at RHIC as a function of 
the number of emitted neutrons.  A similar dependence is expected for 
$\Upsilon$ production at the LHC. Reprinted from 
Ref.~\protect\cite{Strikman} with permission from Elsevier.}
\label{neutdist}
\end{center}
\end{figure}

The $t$ dependence of the rapidity-integrated cross sections is shown in 
Fig.~\ref{dst}. It is easy to discriminate between the coherent and 
quasi-elastic events by selecting different $t$.
At $t\leq 0.01$ GeV$^2$ the quasi-elastic contribution
(dashed line) is small while it is dominant at higher $t$.
The shaded histogram shows incoherent \jp\ photoproduction accompanied by 
neutron emission due to final-state interactions with the recoil nucleon. 
Quasi-elastic \jp\ production accompanied by neutron emission has 
a probability of almost unity.
The only exception is 
the  region of very small $t$ where the recoil energy  
is insufficient for nucleon removal. In gold, the minimum separation energy is 
$\sim 5$ MeV. 
Generally, the ratio of the incoherent cross section with emission of 
one or more neutrons is about 80\% of the total incoherent
cross section. The dependence of the incoherent cross section, 
integrated over rapidity and $t$, on the number of emitted neutrons is 
presented in Fig.~\ref{neutdist}.
The distribution has a pronounced peak for $k=2$ with a long tail up 
to $k=14$. The average number of emitted neutrons is $\langle k \rangle
\approx 4.5$ with a standard deviation of $\approx 2.5$.  
Single neutron emission 
is strongly suppressed due to the low probability of the decay of the hole 
produced by knock-out nucleon into a single neutron.  The probability for the 
knock-out nucleon to emit a neutron while propagating through the nucleus is
greater than 50\%.
  
Neutron tagging of incoherent quarkonium photoproduction can determine which
nucleus was the photon target since the neutrons are emitted by the target.
It is then possible to resolve the ambiguity between photon-emitter and 
photon-target for a given rapidity, not possible for coherent production on
an event-by-event basis.
  
To a first approximation, neutron emission due to electromagnetic
dissociation does not depend on the quarkonium $p_T$. Hence, this
mechanism can be quantified in coherent production at small $t_{\bot }$ and   
folded into quasi-elastic $J/\psi$ production at larger $t$.
 
The pattern of neutron emission we find in quasi-elastic \jp\ production
is qualitatively different from electromagnetic excitation. 
Reference~\cite{Baltz:2002pp} predicts that $\sim 50-70$\% of RHIC
collisions occur without electromagnetic excitation.
The largest partial channel is one-neutron emission, $1n$, followed by 
two-neutron emission, $2n$, about 35\%
of $1n$ events, and a long tail with a broad and falling distribution 
\cite{vidovic,Pshenichnov:2001qd,Chiu:2001ij,Golubeva}. 
On the other hand, two-neutron emission is most probable for the quasi-elastic
mechanism.  In addition, the correlation between emitted neutrons
in the quasi-elastic and electromagnetic mechanisms is different. In the 
quasi-elastic case, neutrons are emitted in only one 
of two directions while simultaneous emission in both directions is possible
in the electromagnetic case. 

\begin{figure}[htpb]
\begin{center}
\epsfig{file=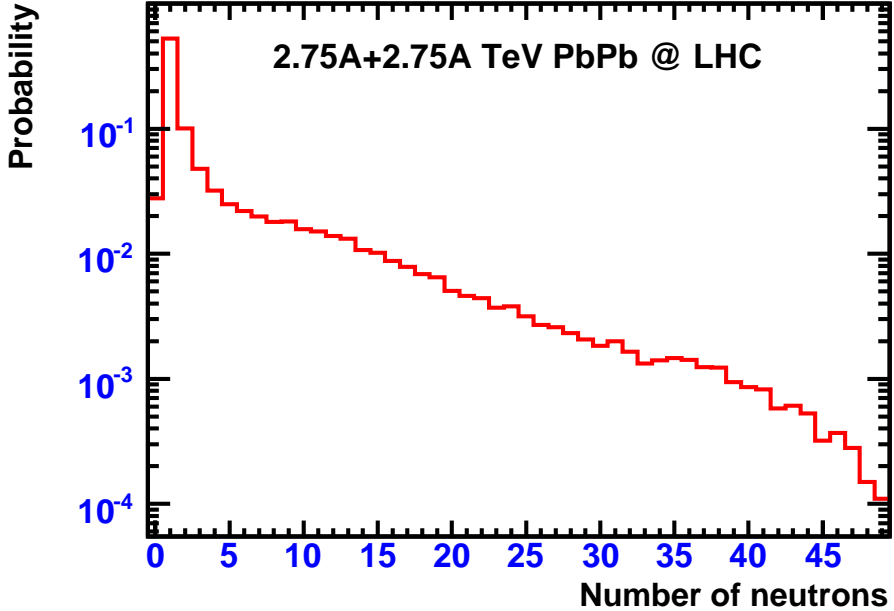, height=0.4\textheight}
\caption[]{The neutron production probability in mutual electromagnetic
dissociation in Pb+Pb collisions
at the LHC.}
\label{emneutrons}
\end{center}
\end{figure}

At the LHC, electromagnetic neutron emission is more important than at RHIC. 
The probability of nuclear dissociation is close to 50\% 
\cite{Pshenichnov:2003aq}.  Most likely, only one neutron is emitted, see 
Fig.~\ref{emneutrons}, calculated with 
{\sc reldis} \cite{Pshenichnov:2001qd,Pshenichnov:ALICE}.  It is possible
to either select only events where one nucleus did not dissociate or use
a deconvolution procedure to separate events where neutron emission is due
to electromagnetic excitation rather than nuclear dissociation.
For example, the difference between the number of neutrons emitted by two
nuclear decays could be studied.  A more detailed analysis, including 
both electromagnetic and quasi-elastic neutron emission in quarkonium 
photoproduction in UPCs will be presented elsewhere.

We have neglected diffractive quarkonium production with nucleon breakup,
$\gamma + p \to J/\psi (\Upsilon) + M_X$. For relatively small $M_X$, the 
dissociation products will be not detected in the 
 central detector and the process would be identified as quasi-elastic. 
At HERA, the ratio of quasi-elastic to elastic channels at $t=0$,
$(d\sigma_{\gamma p \to J/\psi M_X}/ dt)/(d\sigma_{\gamma p \to J/\psi p}/ dt)
\approx  0.2$ and increases with $t$.  Hence, although this process will be a 
small correction to low $t$ quasi-elastic scattering, it will dominate at $|t|
\ge 0.5$ GeV$^2$.  It will thus further enhance the quasi-elastic signal.
In principle, diffractive production could be separated using the 
$t$-dependence of quasi-elastic quarkonium production with the neutron signal. 

In summary, neutron tagging of incoherent quarkonium photoproduction
in ultraperipheral heavy-ion collisions may provide reliable event selection
for quarkonium production by high energy photons.
Precision measurements of quasi-elastic processes, combined with 
improved $\gamma p$ measurements, described in Section~\ref{paonium},
could improve the understanding of $Q \overline Q$ propagation through the
nuclear medium.

\subsection{Quarkonium photoproduction in ALICE} 
{\it Contributed by: V.~Nikulin and M.~Zhalov}
\label{alice-vm} 

The quarkonium cross sections have been calculated in both the impulse and 
leading-twist approximations, IA and LTA respectively, see
Refs.~\cite{fguzsz,Frankfurt:2001db}.  A more detailed discussion and further
references can be found in Ref.~\cite{note}.
Estimates of the total cross sections are presented in Table~\ref{tab:tCrSec}.
  \begin{table}[htbp]
    \begin{center}
\caption[]{The $J/\psi$ and $\Upsilon$ total cross sections in \UPC
at the LHC.}
\vspace{0.4cm}
      \begin{tabular}{|c|c|c||c|c|} \hline
 &  \multicolumn{2}{|c||}{Ca+Ca} & \multicolumn{2}{|c|}{Pb+Pb} \\ \cline{2-5}
 &  $\sigma^{J/\psi}_{\rm tot}$ (mb) & $\sigma^\Upsilon_{\rm tot}$ ($\mu$b)
&  $\sigma^{J/\psi}_{\rm tot}$ (mb) & $\sigma^\Upsilon_{\rm tot}$ ($\mu$b)
\\ \hline
IA  & 0.6  & 1.8  & 70  & 133  \\ 
LTA  & 0.2  & 1.2  & 15  &  78  \\ \hline
      \end{tabular}
    \end{center}
     \label{tab:tCrSec}
  \end{table}
    
The coherent $J/\psi$ and $\Upsilon$  $t$ distributions  for Ca and 
Pb beams at $y=0$ are shown in Fig.~\ref{lhcdst} while the rapidity 
distributions are presented in Fig.~\ref{figlhc1}. 
The incoherent contribution in the impulse approximation, the upper 
limit of the expected cross section, was estimated in 
Refs.~\cite{fguzsz,Frankfurt:2001db}.

Comparison of the LTA and IA results shows that leading-twist shadowing
suppresses the $\Upsilon$ yield by a factor of two at central rapidity. 
The $J/\psi$ suppression is a factor of $4 - 6$ larger. 
In principle, multiple eikonal-type rescatterings due to 
gluon exchanges could also suppress vector meson production.
This mechanism predicts up to a factor of two less 
suppression than leading-twist shadowing, at least for $x \leq 0.001$. 

Coherent quarkonium photoproduction in ultraperipheral $AA$
collisions has a clear signature: a single muon pair in the detector. 
Since the ions remain in the ground state, there should be
no ZDC signal. Any hadronic interaction would show ZDC activity.
The ZDC inefficiency is expected to be very low.
Further improvement could be achieved, if necessary, with
a veto from the outer rings of the V0 detectors, see 
Section~\ref{section-alice}
  
The standard ALICE  Level-0 trigger, L0, was not well suited for 
ultraperipheral studies since it did not cover the barrel rapidity region,
$|y|<1$.  However, the recently proposed inner tracking system (ITS) 
pixel L0 trigger~\cite{pixelL0} 
might improve the situation.  Additional studies are still required to 
determine its utility for UPCs. Therefore
events with two muons in the barrel are not considered here.
The photon detector (PHOS) L0 trigger has recently been introduced. 
It covers a relatively small area,
about 10\% of the barrel solid angle. 
A fast veto from PHOS or the future electromagnetic calorimeter
can suppress more central events.

The dimuon trigger~\cite{dimuonTDR}, covering $-4<\eta<-2.5$, together with 
the PHOS veto could select very low multiplicity events accompanied by fast 
muons.  In addition, low multiplicity selection could be applied at Level 1,
L1. The dimuon L0 processor can produce three kinds of triggers.
The minimal-$p_T$ trigger, initially intended for monitoring and testing,
fires when a single muon passes a loose cut of $p_T>0.5$ GeV/$c$. 
The low-$p_T$ trigger, used predominantly to select two-muon events 
with $p_T>1$ GeV/$c$, is designed to tag $J/\psi$ decays.
The high-$p_T$ trigger selects heavy resonances 
($\Upsilon,\ \Upsilon'$) by tagging muon pairs with $p_T>2$ GeV/$c$.
The minimal trigger rate is expected to be at the level of 8 kHz 
for Pb+Pb interactions.
The two last, tighter, triggers are intended to reduce the dimuon rate 
to 1 kHz or less.

We have studied which trigger configurations may be most useful for studies
of quarkonium photoproduction in ALICE.  At L0, only the muon trigger is 
used.  The minimal-$p_T$ muon trigger is vetoed by activity in PHOS. It 
selects events with at least one muon in the muon spectrometer, including 
events with one muon in the spectrometer and a second muon in the barrel.
If the trigger rates are too high, the low-$p_T$ and/or the high-$p_T$ triggers
will be utilized. Here, only events with two muons in the spectrometer 
can be triggered.
At L1, ZDC information will be used to perform additional 
selection of very peripheral events.
The Level-2 trigger performs standard TPC past-future protection while the
high level trigger checks that the event contains only a few tracks.
 
Thus the proposed trigger enables the study of a class of events 
with ``abnormally" low multiplicity tagged by a muon.
Among the reactions that could be measured are coherent and incoherent
quarkonium photoproduction and lepton pair production in $\gamma \gamma$ 
interactions.  Such a trigger could be integrated into the standard ALICE 
running conditions. 

The expected rates were estimated using 
the ALICE simulation code {\tt AliRoot}~\cite{ALIROOT}.  
In the simulation, a muon which traverses
ten tracking and four trigger chambers of
the muon spectrometer or produces hits in both the ITS and TPC is considered
detected. 
The LTA distributions were used for the analysis, resulting in a $\sim 5$\% 
acceptance for the $J/\psi$ and $\sim 2$\% for the $\Upsilon$. These 
acceptances correspond to $\sim 1000$ muon pairs/day from $J/\psi$ decays
and about 3 pairs/day from $\Upsilon$ decays detected in 
the muon spectrometer.  The machine-induced (beam-gas) background is 
expected to be negligible.
  
The physical background due to coherent quarkonium production in 
coherent and incoherent diffractive (Pomeron-Pomeron) interactions 
is expected to be small.  This contribution still needs to be evaluated.
Another source of physical background is 
muon pair production in $\gamma\gamma$ interactions.
The total number of triggers could be significant since the background was 
underestimated in Ref.~\cite{note}. 
The degradation of the mass resolution due to the 
uncertainty in the interaction point and the far forward peaked muon angular
distribution should be taken into account.
The ratio of the coherent signal, $S$, to background events below the
signal peak, $B$, the signal-to-background ratio, $S/B$, is of order unity
for the $J/\psi$ (Table~\ref{photoproductionRatesPbJ})
and $\sim 0.5$ for the $\Upsilon$ (Table~\ref{photoproductionRatesPbU}).

The statistical significance, $S/\sqrt{S + B}$, 
of data collected during a $10^6$ s run is estimated to be $\sim 100$ for 
the $J/\psi$ (Table~\ref{photoproductionRatesPbJ}) and $3-4$ for the 
$\Upsilon$ (Table~\ref{photoproductionRatesPbU}).  A significance of $\sim 100$
is sufficient for study of the differential distributions.
The LTA $J/\psi$ and $\Upsilon$ rates expected in a $10^6$ s Pb+Pb run
are given in Table~\ref{photoproductionRatesPbJ} 
and \ref{photoproductionRatesPbU}, respectively, along with the 
signal-to-background ratios and the significance.  In 
Table~\ref{photoproductionRatesPbJ}, the suppression of the rate due to LT
shadowing is given by the ratio IA/LTA.
The corresponding Ar+Ar rates are also shown.  The mass bin, $\Delta M$, 
used is approximately three times the detector mass resolution at the 
quarkonium mass.  The interaction point resolution is also taken into account.
The resolution is better if one muon goes to the barrel and the other to the
muon spectrometer than if both muons go to the spectrometer.  Since the 
$J/\psi$ resolution is not noticeably affected, the mass bin $\Delta M = 
0.2$~GeV has been used in both cases. For the $\Upsilon$,
$\Delta M = 0.3$~GeV is used when one muon is in the spectrometer and the other
in the barrel while $\Delta M = 0.4$~GeV is used when both muons are 
accepted in the spectrometer.

\begin{table}[htbp]
    \begin{center}
     \caption[]{The expected $J/\psi$ photoproduction rates in a $10^6$ s
run for Pb+Pb and Ar+Ar collisions.
The Ar+Ar luminosity assumed is $4\times 10^{28}$ cm$^{-2}$s$^{-1}$.} 
\vspace{0.4cm}
      \label{photoproductionRatesPbJ}
       \begin{tabular}{ | l| c| c| c| c || c | } \hline
            & \multicolumn{4}{c||}{Pb+Pb} & Ar+Ar \\ \cline{2-6}
            &   LTA  & IA/LTA &  $S/B$   & Significance & LTA \\ \hline 
Muon Arm    & 25,000 & 2.28  &  6        & 150          & 25,000   \\ 
Barrel      & 21,400 & 6.19  &  0.7      &  90          & 13,000  \\ \hline
        \end{tabular}
      \end{center}
\end{table}

\begin{table}[htbp]
    \begin{center}
     \caption[]{The expected $\Upsilon$ photoproduction rates in a $10^6$ s
run for Pb+Pb and Ar+Ar collisions.} 
\vspace{0.4cm}
      \label{photoproductionRatesPbU}
       \begin{tabular}{ | l| c| c| c || c | } \hline
            & \multicolumn{3}{c||}{Pb+Pb} & Ar+Ar \\ \cline{2-5}
            & LTA  &  $S/B$  & Significance & LTA \\ \hline 
Muon Arm    & 25   &  0.7    & 3          & 33   \\ 
Barrel      & 60   &  0.26   & 4          & 72  \\ \hline
        \end{tabular}
      \end{center}
\end{table}

The dimuon invariant mass and $p_T$ will be reconstructed offline.
Since coherent events are peaked at $p_T \sim 0$, it is possible to
estimate the incoherent contribution and reconstruct the coherent cross 
section.  The Monte Carlo acceptance will be used for reconstruction of 
the coherent cross section.
  
The muon spectrometer and barrel measurements are complementary in their
rapidity coverage.  When one muon is detected in the barrel and the other in 
the spectrometer, $-2.5 < y < -1$ for the vector meson, 
the measurement is more central.  Effects related
to the reaction mechanism are dominant and IA/LTA $\sim 6.2$.  When both muons
are detected in the spectrometer, $-4 < y < -2.5$ for the vector meson, IA/LTA 
$\sim 2.2$ and the forward
cross sections are more sensitive to the gluon density.
    
Comparison of $J/\psi$ and $\psi'$ yields from
different collision systems (Pb+Pb, Ar+Ar and $pA$) may provide further 
information about the gluon density at $x$ values as yet unexplored.

\subsection{Detection and reconstruction of vector mesons in CMS}
{\it Contributed by: D. d'Enterria and P. Yepes}
\label{cms-vm}

In this section, we present the CMS capabilities for diffractive 
photoproduction measurements of light ($\rho^0$) and heavy ($\Upsilon$) vector 
mesons as well as two photon production
of high-mass dileptons ($M_{l^+l^-} >  5$ GeV/$c^2$), part of the 
$\Upsilon$ photoproduction background. On one hand, $\rho^0$ photoproduction  
studies extend the HERA measurements \cite{Aktas:2006qs} and provide new 
information about the interplay of soft and hard physics in diffraction
~\cite{FSZ06_01,FSZ06}. 
A clean signature with a low $\pi^+\pi^-$ invariant mass background makes this 
measurement relatively straightforward in UPCs, as demonstrated in Au+Au 
collisions at RHIC~\cite{Adler:2002sc}.
On the other hand, heavy quarkonium ($J/\psi$, $\Upsilon$) production provides
valuable information on the nuclear gluon density, $xg_A(x,Q^2)$
\cite{annual}, and extends studies at RHIC 
\cite{dde_qm05} into a previously unexplored $x$ and $Q^2$ range,
see Fig.~\ref{fig:x_Q2}.

Table~\ref{tab:cms1} lists the expected $\rho^0$, $J/\psi$ and $\Upsilon$
photoproduction cross sections in UPCs at the LHC, as given by {\sc 
starlight} \cite{Klein:1999qj,Baltz:2002pp,Klein:2003vd,starlight}.
which satisfactorily reproduces the present RHIC UPC $\rho^0$
\cite{Adler:2002sc} and $J/\psi$ \cite{dde_qm05} data as well as the 
low \cite{Adams:2004rz} and high mass \cite{dde_qm05}
dielectron data.  For comparison, we note that the calculated
$\Upsilon$ cross section in inelastic $pp$ collisions at 5.5 TeV is $\sim 600$ 
times smaller, $\sigma_{pp\rightarrow \Upsilon X}\approx 0.3$ 
$\mu$b~\cite{vogt02}, while the inelastic minimum bias Pb+Pb $\Upsilon$ 
cross section is $\sim 100$ times larger, $\sigma_{{\rm PbPb} \rightarrow 
\Upsilon X} = A^2 \sigma_{pp\rightarrow \Upsilon X}\approx 13$ mb.

\begin{table}[htbp]
\begin{center}
\caption[]{Exclusive vector meson photoproduction cross sections predicted by 
{\sc starlight} \protect\cite{Klein:1999qj,starlight,Baltz:2002pp}
in ultraperipheral Pb+Pb interactions at 5.5 TeV accompanied by neutron 
emission in single ($Xn$) or double ($Xn|Xn$) dissociation of the lead
nuclei, shown on the left-hand side of Fig.~\protect\ref{fig:diag_gg_gA}.
(Note that $\sigma_{Xn}$ includes $\sigma_{Xn|Xn}$).}
\vspace{0.4cm}
\label{tab:cms1}
\begin{tabular}{|c|c|c|c|}\hline
Vector Meson & $\sigma_{\rm tot}$ (mb) & $\sigma_{Xn}$ (mb) 
& $\sigma_{Xn|Xn}$ (mb)
\\ \hline
$\rho^0$   & 5200 &  790 & 210 \\
$J/\psi$    & 32   &  8.7 & 2.5 \\
$\Upsilon(1S)$ & 0.173 & 0.078 & 0.025 \\
\hline
\end{tabular}
\end{center}
\end{table}

The most significant physical background for these measurements is coherent
lepton pair production in two-photon processes, shown on the right-hand side of
Fig.~\ref{fig:diag_gg_gA}.  Table~\ref{tab:cms2} lists the expected dilepton
cross sections in the mass range relevant for quarkonium measurements. The 
fraction of the continuum cross sections accompanied by nuclear breakup with 
neutron emission is expected to be the same as for quarkonia
photoproduction, on the order of $\sim 50$\% for high-mass dileptons.

\begin{table}[htbp]
\begin{center}
\caption{Dilepton production cross sections
predicted by {\sc starlight} \protect\cite{Klein:1999qj,starlight,Baltz:2002pp}
for two-photon interactions in ultraperipheral Pb+Pb interactions at 5.5 TeV,
see the right-hand side of Fig.~\protect\ref{fig:diag_gg_gA}. The results
are given in 
the mass regions of interest for $J/\psi$ and $\Upsilon$ production, $M>1.5$ 
GeV and $M>6$ GeV respectively.}
\vspace{0.4cm}
\label{tab:cms2}
\begin{tabular}{|l|c|c|}\hline
Mass                 & $\sigma_{\gaga\rightarrow e^+e^-}$ (mb) 
& $\sigma_{\gaga\rightarrow\mu^+\mu^-} $ (mb) \\ \hline
$M>1.5$ GeV/$c^2$        & 139    & 45  \\
$M>6.0$ GeV/$c^2$        & 2.8    & 1.2 \\
\hline
\end{tabular}
\end{center}
\end{table}

\subsubsection{Trigger considerations}
\label{sec:upc_trigg} \bigskip

Ultraperipheral collisions are mediated by photon exchange 
with small momentum transfer and are characterized by a large rapidity gap 
between 
the produced system and the beam rapidity.  After the interaction, the nuclei 
either remain essentially intact or in a low excited state. Thus UPCs can be 
considered `photon-diffractive' processes sharing many
characteristics with `hadron-diffractive' (Pomeron-mediated) collisions. An 
optimum UPC trigger is thus usually defined based on these typical signatures.

UPCs are characterized by a large rapidity gap
between the produced state and the interacting 
nuclei accompanied by forward neutron emission 
from the de-excitation of one or
both nuclei. Single or mutual Coulomb excitation, indicated by the soft 
photon exchange in Fig.~\ref{fig:diag_gg_gA}, occurs in about 50\% of UPCs. 
The Coulomb excitation generates a Giant-Dipole Resonance (GDR) in the nucleus 
which subsequently decays via neutron emission.  Since the global multiplicity
is very low, the central detector is virtually empty apart from
the few tracks/clusters 
originating from the produced system.  The resulting rapidity distribution is 
relatively narrow, becoming narrower with increasing mass of the produced
system, $M_X$, and centered at midrapidity.
Note that although the energies of the $\gamma$ and the
``target'' nucleus are very different and the produced final state is 
boosted in the direction of the latter, since each of the nuclei can act as
both ``emitter'' and ``target'', the sum of their rapidity distributions
is symmetric around $y = 0$.

Given these general properties of UPC events and based upon our previous 
experience with the $J/\psi$ in Au+Au UPCs at RHIC~\cite{dde_qm05}, 
we devised the following CMS Level-1 primitives for the ultraperipheral 
trigger.

To ensure a large rapidity gap in one or in both hemispheres, we
reject events with signals in the forward hadron calorimeters towers,
$3 < |\eta|< 5$, above the default energy threshold for triggering on 
minimum-bias nuclear interactions 
($\mathtt{\overline{HF+}.OR.\overline{HF-}}$).  Although pure $\gamma$Pb 
coherent events have rapidity gaps in both hemispheres,
we are also interested in triggering on ``incoherent'' $\gamma\,N$ 
photoproduction which usually breaks the target nucleus, partially 
filling one of the hemispheres with particles.

To tag Pb$^*$ Coulomb breakup by GDR neutron de-excitation, we
require energy deposition in the Zero Degree Calorimeters~\cite{ZDC-CMS} 
($\mathtt{ZDC+.OR. ZDC-}$) above
the default threshold in normal Pb+Pb running. 
The availability of the ZDC signals in the L1 trigger decision is an
advantage of CMS.

\subsubsection{Light meson reconstruction}
\label{section-cms-lightmesons} \bigskip
{\it Contributed by: P. Yepes} \bigskip

Here we present a feasibility study of light meson analysis in UPCs
with CMS.  Triggering on reactions 
without nuclear breakup in CMS is difficult because the detector is 
designed to trigger on transverse energy rather than multiplicity. The 
mesons considered here, with masses less than a 
few GeV/$c^2$, will deposit little energy in the calorimeters. 
However, even for low mass particles, triggering on reactions with nuclear
breakup should be feasible using the CMS ZDCs.
The $\rho^0$ is used as a test case.  We show that, despite the 4 T magnetic
field of CMS and a tracker designed for high $p_T$ particles, acceptable
reconstruction efficiencies are achieved.

A set of 1000 $\rho^0$s produced in ultraperipheral Pb+Pb collisions
was generated \cite{Baltz:2002pp,Klein:1999qj} and run through the detailed 
{\sc geant}-3 based CMS simulation package, CMSIM 125, using a silicon pixel 
detector with three layers.  Events were then passed through 
the digitization packages using version 7.1.1 of the ORCA reconstruction 
program.  Only information from the silicon pixels was used. 
The performance of the reconstruction algorithm does not 
significantly improve with one or two additional silicon layers. 
The $\rho^0$ candidates are reconstructed by combining opposite-sign tracks.
The same-sign background was negligible. 
The overall reconstruction efficiency is $\epsilon
=35$\%. For central rapidities, $|\eta| < 1$, $\epsilon =42$\%, while for 
more forward rapidities, $1 < \eta < 1.8$, $\epsilon =16$\%.
Therefore, we conclude that light mesons produced in UPCs with nuclear breakup
can be reconstructed in CMS if they are 
triggered with the ZDCs.

\subsubsection{$\upsi$ Detection in CMS} \bigskip
{\it Contributed by: D. d'Enterria} \bigskip

\begin{figure}[htbp]
\begin{center}
\includegraphics[height=4.8cm]{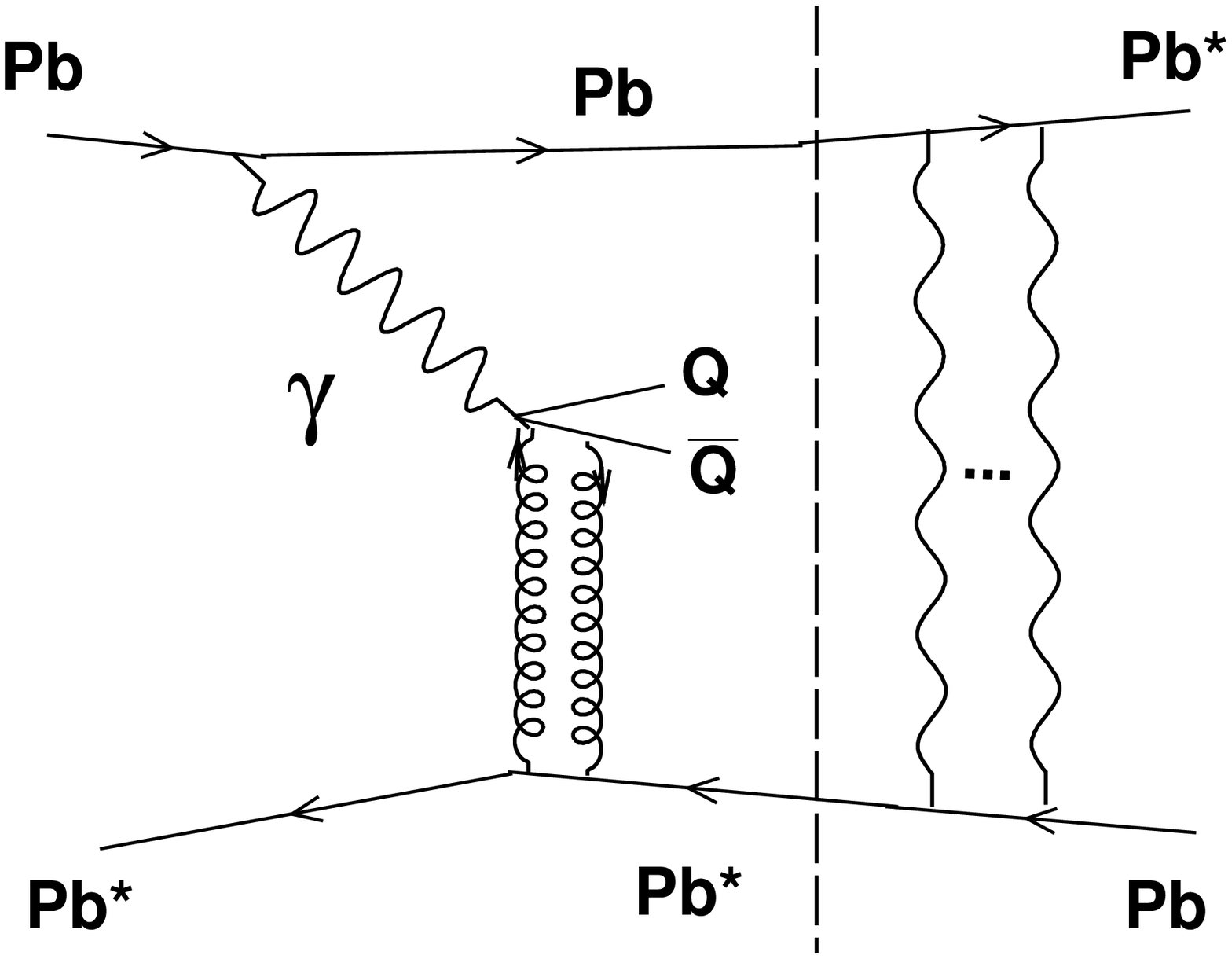}
\includegraphics[height=5.1cm]{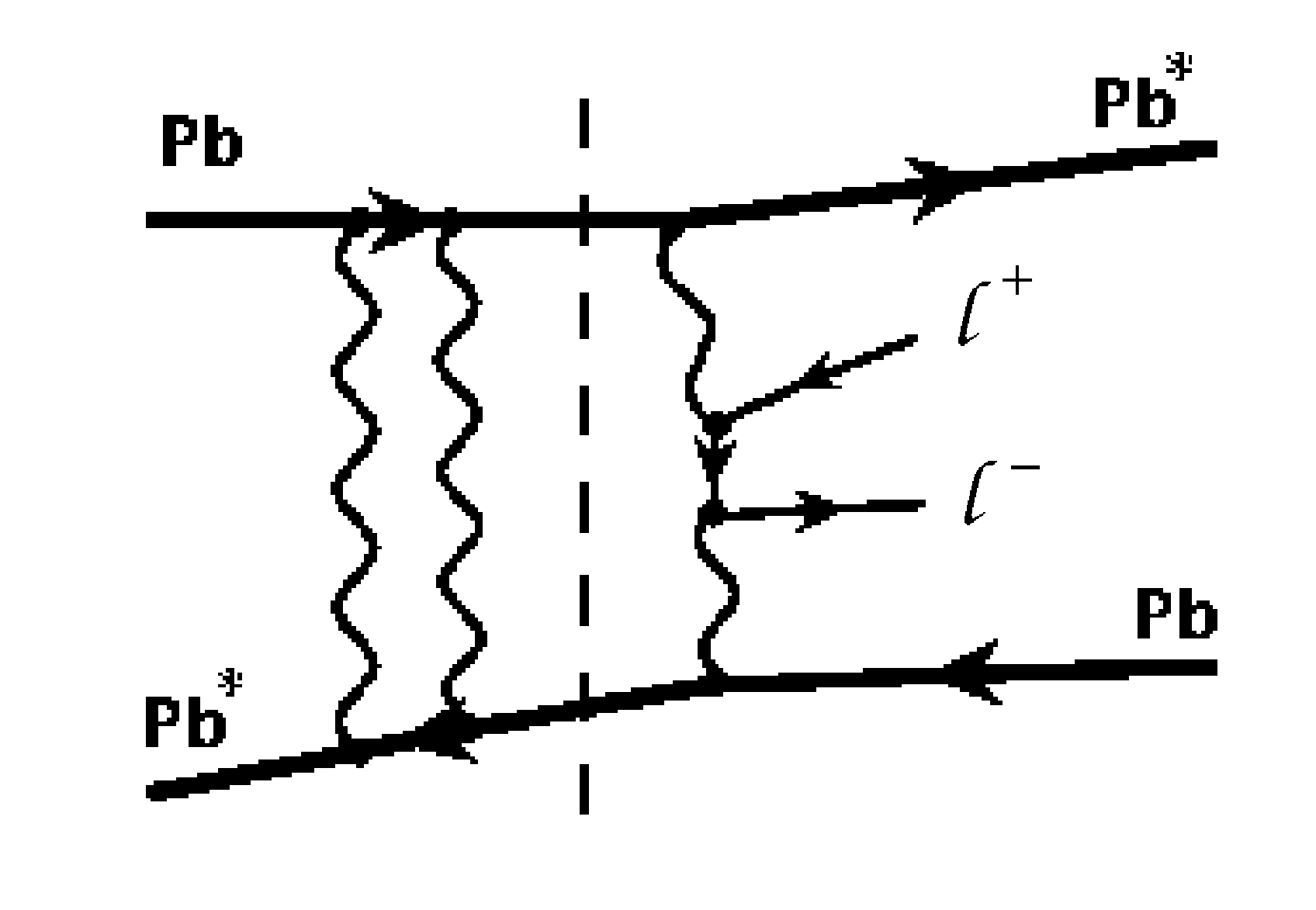}
\end{center}
\vspace{-0.2cm}
\caption[]{The leading order diagrams for $\Upsilon$ (left) and lepton 
pair \protect\cite{CMS-PTDR1} (right) 
production in $\gA$ and $\gaga$ processes accompanied by Coulomb 
excitation in ultraperipheral Pb+Pb collisions.}
\label{fig:diag_gg_gA}
\end{figure}

\begin{figure}[htbp]
\begin{center}
\includegraphics[height=6.cm,width=8.cm]{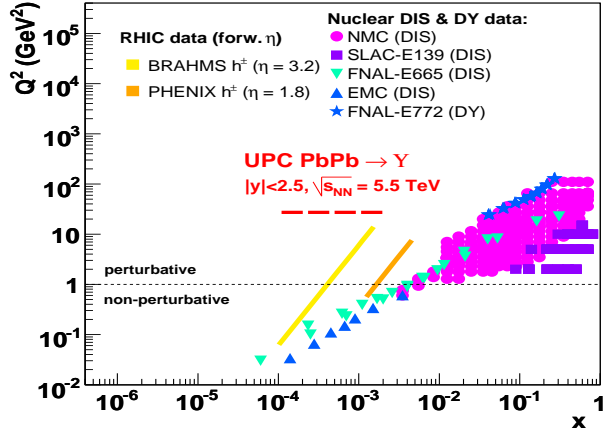}
\end{center}
\vspace{-0.4cm}
\caption[]{Measurements in the  $(x,Q^2)$ plane used to constrain the 
nuclear PDFs. The approximate $(x,Q^2)$ range covered 
by ultraperipheral $\upsi$ photoproduction in Pb+Pb collisions at 
$\sqrt{s_{_{NN}}} = 5.5$ TeV in $|\eta|<2.5$ is indicated.  Reprinted from
Ref.~\protect\cite{dde_qm04} with permission from the Institute of Physics.}
\label{fig:x_Q2}
\end{figure}

At leading order, diffractive $\gamma A \rightarrow 
J/\psi\,(\Upsilon)$ proceeds 
through a colorless two-gluon (Pomeron) exchange, see the left-hand side of
Fig.~\ref{fig:diag_gg_gA}. After the
scattering, both nuclei remain intact, or at a low level of excitation, 
and separated from the produced state by a rapidity gap.  Such hard 
diffractive processes are thus valuable probes of the gluon 
density since their cross sections are proportional to the square of the 
gluon density,
$(d\sigma_{\gamma p,A\rightarrow V\,p,A}/dt)|_{t=0} \propto 
[xg(x,Q^2)]^2$ where $Q^2\approx M_V^2/4$ and $x=M_V^2/W_{\gamma p,A}^2$,
see Eq.~(\ref{phocs}).
At $y=0$, $x \sim 2\times 10^{-3}$ in $\gamma A \rightarrow \Upsilon A$ 
interactions at the LHC.  The $x$ values
can vary by an order of magnitude in the range $|y| \leq 2.5$, thus probing 
the nuclear PDFs in an $x$ and $Q^2$ 
range so far unexplored in nuclear DIS or in lower energy $AA$
collisions, see Fig.~\ref{fig:x_Q2}.  Photoproduction measurements thus help 
constrain the low-$x$ behavior of the nuclear gluon distribution in a range
where saturation effects due to nonlinear evolution of the PDFs are expected 
to set in~\cite{dde_lowx06,strikman06}.  

\paragraph{Expected cross sections} \bigskip

\noindent
The expected $\jps$ and $\upsi$ photoproduction cross sections in
ultraperipheral Pb+Pb collisions at the LHC given by 
{\sc starlight}~\cite{Klein:1999qj,starlight,Baltz:2002pp} 
are listed in Table~\ref{tab:cms1}. The $\gpb$ cross sections 
do not include the
$\sim 10-20$\% feeddown contributions from excited $S$ states. 
They also do not include contributions from incoherent $\gamma N$ processes 
which should increase the $J/\psi$ and $\Upsilon$ yields by $\sim 50$\%
~\cite{Strikman}. 
Other $\gpb \rightarrow \Upsilon$ predictions for LHC energies,
{\it e.g.}\ $\sigma_{\upsi} = 135$ $\mu$b ~\cite{fguzsz}, give cross sections 
comparable to Table~\ref{tab:cms1}. Including leading-twist 
shadowing reduces the $\Upsilon$ yield by up to 
a factor of $\sim 2$ to 78 $\mu$b~\cite{fguzsz}, see 
Table~\ref{tab:3:sigmavm}. Even larger reductions are expected when saturation 
effects, see Section~\ref{sec:cgc-desc}, are included
\cite{goncalves06}.  Our motivation is to precisely pin down the differences
between the lead and proton PDFs at low $x$ and relatively large 
$Q^2$, $\approx 40$ GeV$^2$.

Roughly 50\% of the UPCs resulting in $\upsi$ production are accompanied by 
Coulomb excitation of one or both nuclei due to soft photon exchange, as
shown in Fig.~\ref{fig:diag_gg_gA}.  The excitations can lead to nuclear
breakup with neutron emission at very forward rapidities, covered by the
ZDCs.  This dissociation, 
primarily due to the excitation and decay of giant dipole resonances, 
provides a crucial UPC trigger, as discussed in the next section. 

The coherent photon fields generated by the ultrarelativistic nuclei have
very small virtualities, $p_T < 2\hbar c/R_A \approx$ 50 MeV/$c$.  
Coherent production thus results in very low $p_T$ $J/\psi$s
so that the $p_T$ of the decay 
leptons, $\sim m_{J/\psi}/2$, is too low to 
reach the detectors due to the large CMS magnetic field.
We thus concentrate on the $\upsi$ since the decay lepton energies are
$\sim 5$ GeV and can, therefore, reach the electromagnetic calorimeter (ECAL) 
and the muon chambers to be
detected. In particular, this analysis focuses on $\Upsilon$ measurements 
in the CMS barrel and endcap regions, $|\eta|<2.5$, for:
\begin{description}
\item (1) $\gpb \rightarrow  \upsi + {\rm Pb}^\star {\rm Pb}^{(\star)}$,
$\upsi\rightarrow e^+ e^-$ measured in the ECAL;
\item (2) $\gpb \rightarrow  \upsi + {\rm Pb}^\star {\rm Pb}^{(\star)}$, 
$\upsi\rightarrow \mu^+ \mu^-$ measured in the muon chambers.
\end{description}
The $\star$ superscript indicates that one or both lead nuclei may
be excited. Here and below, the presence of the lead nucleus that emits the
photon is implied but not explicitly shown.

The most significant background source is coherent dilepton production in 
two-photon processes, shown on the right-hand side of 
Fig.~\ref{fig:diag_gg_gA}:
\begin{description}
\item (1) $\gaga \rightarrow {\rm Pb}^\star {\rm Pb}^{(\star)} + e^+ e^-$, 
measured in the ECAL;
\item (2) $\gaga \rightarrow {\rm Pb}^\star {\rm Pb}^{(\star)} + \mu^+\mu^-$, 
measured in the muon chambers.
\end{description}
These are interesting pure QED processes and have been proposed as a 
luminosity monitor in $pp$ and $AA$ 
collisions at the LHC~\cite{Bocian04,lumi2}.  As such, they
may be used to normalize the absolute cross section of this and other 
heavy-ion measurements. Table~\ref{tab:cms2} lists the expected dilepton
cross sections in the mass ranges relevant for the quarkonia measurements. 
The fraction of the continuum cross sections accompanied by nuclear breakup 
with neutron emission is of the order of $\sim 50$\%, as is the case for 
quarkonium photoproduction.

\paragraph{Level-1 and High-Level Triggers}
\label{sec:cms_upc_trigg}

\subsection*{Level-1 trigger}

Given the general considerations discussed in Section~\ref{sec:upc_trigg} and 
experience 
with $\jps$ photoproduction studies in ultraperipheral Au+Au collisions at 
RHIC~\cite{dde_qm05}, we propose to use several CMS Level-1 (L1)
primitives as part of the ultraperipheral trigger.

We require a large rapidity gap in one or both hemispheres (forward/backward
of midrapidity).  Thus we veto signals in the forward hadron (HF) calorimeters,
$3 < |\eta|< 5$, above the default minimum bias energy threshold, 
\ttt{$\mathtt{\overline{HF+}.OR.\overline{HF-}}$},
where the $\pm$ refers to the forward/backward region and the bar over the
HF signifies veto.  We do not make an $\mathtt{AND}$ veto to require the 
absence of a signal in {\it both} HF towers, a signal of coherent 
$\gamma$Pb events
with gaps in both hemispheres, because we also want to trigger on incoherent
photoproduction ($\gamma N$) where the target nucleus breaks up,
partially populating the hemisphere on that side.
There should be one or more neutrons, $Xn$, in at least one ZDC, 
$\mathtt{ZDC+.OR.ZDC-}$, to tag Coulomb breakup of the excited lead nucleus
due to de-excitation of the GDR by neutron emission.  

Leptons from $\Upsilon$ decays have energy $E_l \gtrsim m_\Upsilon/2 \sim 4.6$
GeV.  Electrons and muons from these decays are triggered in two different
ways.  Electrons from $\Upsilon$ decays are selected by energy deposition in
an isolated ECAL trigger tower with threshold energy greater than 3 GeV.
Muons can be selected by hits in the muon resistive plate chambers (RPCs),
$|\eta|< 2.1$, or cathode strip chambers (CSCs), $0.8 <|\eta|< 2.4$. 
No track momentum threshold is required since the material budget in front of 
the chambers effectively reduces any muon background below
$\sim 4$ GeV.

The following two dedicated L1 UPC triggers are thus proposed:
\begin{description}
\item \ttt{UPC-mu-L1=(ZDC+.OR.ZDC-).AND.$(\overline{\mathtt{HF+}}$.OR.$\overline{\mathtt{HF-}})$.AND.(muonRPC.OR.muonCSC)} \, \, ;
\item \ttt{UPC-e-L1=(ZDC+.OR.ZDC-).AND.$(\overline{\mathtt{HF+}}$.OR.$\overline{\mathtt{HF-}}$).AND.ECALtower($E>2.5$ GeV)} \, \, .
\end{description}

\subsection*{Expected L1 trigger rates}

The coherent $\upsi\rightarrow \lele$ photoproduction rate, $N_\Upsilon$, 
assuming a perfect trigger, full acceptance and no efficiency losses at the 
nominal Pb+Pb luminosity,
$L_{\rm PbPb} = 0.5$ mb$^{-1}$s$^{-1}$, is
\begin{eqnarray}
N_{\upsi} & = & L_{\rm PbPb} \, B(\upsi\rightarrow \lele) \, \sigma_{\upsi} \\ \nonumber
& = & 0.5 \mbox{ mb}^{-1}\mbox{s}^{-1} \times 0.024 \times 0.078 \mbox{ mb} 
= 0.001 \mbox{ Hz}
\label{eq:upsrates}
\end{eqnarray}
or 1000 $\Upsilon (1S)$ dilepton decays in a $10^6$ s run.

There will be several sources of background that will also satisfy the UPC-L1
triggers defined above.  For the purpose of estimating the trigger rates, we 
consider 
sources of physical and ``non-physical'' backgrounds which have characteristics
similar to a UPC event and, therefore, can potentially fulfill UPC-L1 trigger
criteria.  All these processes give a ZDC signal.

Beam-gas and beam-halo collisions do not have a good vertex.  They have a
comparatively large multiplicity with an asymmetric $dN/dy$ and relatively
low transverse energy, $E_T$. However, this process will be suppressed by the
rapidity gap requirement, 
$(\overline{\mathtt{HF+}}$.OR.$\overline{\mathtt{HF-}})$, 
and will not be discussed further.

High L1 background rates may be generated by the coincidence of cosmic-ray 
muons with electromagnetic nuclear dissociation (ED) and peripheral nuclear 
collisions.  Cosmic rays in coincidence with ED, $\gamma A \rightarrow 
{\rm Pb}^\star \, + \, {\rm Pb}^{(\star)}$ with $Xn$ neutrons hitting the ZDC, 
will have large net $p_T$ tracks in the muon chambers alone but no collision 
vertex.  The ZDC
signal is from lead dissociation.  Peripheral nuclear collisions, $AA 
\rightarrow X$ also have a ZDC signal but have relatively large hadron 
multiplicities with large $p_T$.

A background process with an almost indistinguishable signal at L1 arises from 
two-photon production of dileptons, $\gamma \gamma \rightarrow l^+ l^-$.
We thus discuss this process further as a possible reference process in the
remainder of this section.  This background can be significantly reduced by an
asymmetry cut on the lepton pair while the residual contribution below the
$\Upsilon$ mass can be statistically subtracted in the offline analysis, see
Section~\ref{sec:minv}.

Finally, some background arises from interesting low rate processes that can
be studied offline as a byproduct of the UPC trigger.  These include hadronic
diffraction, hard diffractive photoproduction and two-photon hadronic
production.  Hadronic diffractive collisions, $\Pomeron \, {\rm Pb}, \,
\Pomeron\Pomeron \rightarrow X$, have larger multiplicities than diffractive
photoproduction and predominantly produce pions rather than vector states.  
A like-sign subtraction can remove the pion background.
The $p_T$ is also larger for diffractive processes, 
$p_T(\Pomeron\Pomeron) > p_T(\gamma\Pomeron)> 
p_T(\gaga)$.  Hard diffraction, {\it e.g.} dijet and $Q \overline Q$ 
production, is also characterized by larger multiplicities.  These background
events can be removed offline using standard subtraction techniques.  

The single and mutual electromagnetic lead dissociation cross section 
at the crossing point of the two beams is $\sigma^{\rm S(M)ED} = 215$ b
\cite{Pshenichnov:2001qd},
the main limitation on the Pb+Pb luminosity achievable 
at the LHC. Such large cross sections translate into very large rates,
$N^{\rm S(M)ED} = L_{\rm PbPb} \sigma^{\rm S(M)ED} = 10^5$ Hz.  Thus 
accidental 
coincidences with cosmic ray muons traversing the muon chamber and activating 
the \ttt{UPC-mu-L1} trigger are possible. 
The typical cosmic ray muon rate at ground level is about 60 Hz/m$^2$ 
with $\mean{E_\mu}\approx 4$ GeV~\cite{pdg}. At the IP5 cavern, $\sim 80$ m 
underground, the rate is reduced to $\sim 6$ Hz/m$^2$. Muons which traverse
the rock overburden above CMS typically have an energy at the surface of
at least 10 GeV.
Since the surface area of the muon chambers is $\sim 20 \times 15$ m$^2$, 
the total rate of cosmic ray muons entering 
the chambers is $N_{\rm cosmic}\approx  2$ kHz. The accidental coincidence 
rate for two detectors with counting rates $N_1$ and $N_2$ in a trigger time 
window $\Delta t_{\rm trig}$ is $N_{\rm acc} = 2 N_1\,N_2\,\Delta t_{\rm 
trig}$.  If $\Delta t_{\rm trig} = 10$ ns around the nominal bunch 
crossing time of 25 ns, we have
\begin{eqnarray}
N_{\rm cosmic}^{\rm S(M)ED} & = & 2 \, N^{\rm S(M)ED}\, N_{\rm cosmic} \, 
\Delta t_{\rm trig} \\ \nonumber
& \approx & 2 \times 10^{5} \mbox{ Hz} \times 2000 \mbox{ Hz} \times 10^{-8} 
\mbox{ Hz}^{-1} \approx 4 \mbox{ Hz} \, \, .
\label{eq:edrates}
\end{eqnarray}
However, very few cosmic ray muons pass the trigger if we require the tracks 
to be pointing to the vertex.  There is a factor of 2500 reduction when
we require $z_{\rm hit}<  60$ cm and $R_{\rm hit} < 20$ cm.
In the high-level trigger (HLT), this background can be
reduced by requiring vertex reconstruction.
 
At RHIC energies, usually $\epsilon_{\rm periph} \approx 5$\% of the 
most peripheral
nuclear $AA$ interactions ($95-100$\% of the total $AA$ cross section) 
do not generate activity within $3<|\eta|<4$ but still produce a signal in 
the ZDC~\cite{ppg014}.  Assuming that the same fraction of Pb+Pb collisions
at the LHC will be accepted by the virtually identical L1 condition
\ttt{(ZDC+.OR.ZDC-).AND.$(\overline{\mathtt{HF+}}$.OR.
$\overline{\mathtt{HF-}})$}.  Such Pb+Pb collisions will fire 
\ttt{UPC-e-L1} and/or \ttt{UPC-mu-L1} 
provided that these reactions also produce a lepton of sufficient energy.
An analysis of a few hundred thousand 
minimum-bias $pp$ events at 5.5 TeV generated using {\sc pythia} 6.4 
\cite{Pythia}, 
$\epsilon_{l}\approx 1$\% 
of the collisions generate at least one lepton within $|\eta|< 2.5$ 
above the UPC-L1 energy thresholds. 
We assume the same relative fraction
will hold for peripheral Pb+Pb interactions. 
The corresponding rate for this background is
\begin{eqnarray}
N_{\rm had} & = &  L_{\rm PbPb}\,\sigma_{\rm tot} \, \epsilon_{\rm periph} \, 
\epsilon_{l}  \\ \nonumber
& = & 0.5 \mbox{ mb}^{-1}\mbox{s}^{-1} \times 8000\mbox{ mb} \times 0.05 
\times 0.01 \approx 2 \mbox{ Hz} \, \, .
\label{eq:periphrates}
\end{eqnarray}

If the HF veto is insufficient to reduce these background rates at L1, 
an additional L1 primitive may be considered such as the total energy in the 
ECAL-HCAL system, requiring an energy deposition only a few GeV above the 
calorimeter noise level. This will suppress peripheral hadronic interactions
since their multiplicity is much larger than ultraperipheral events.

We now discuss the $\gamma \gamma \rightarrow e^+ e^-$ background in more 
detail.  The known $\gamma \gamma \rightarrow e^+ e^-$ QED cross sections 
are given in Table~\ref{tab:cms2}.  Simulations have shown that the
fraction of these events producing an electron within $|\eta|< 2.5$ and
above the 3 GeV ECAL trigger threshold and potentially firing \ttt{UPC-e-L1}, 
is only $\epsilon_e \sim 5$\%.  Since the corresponding fraction of muons 
triggering \ttt{UPC-mu-L1} is even lower, we do not include it here.  
There is a 50\% probability for neutron emission, $P_n$, incorporated
into the rate.  Thus, 
the expected rate for this background is
\begin{eqnarray}
N_{\gaga} & = & L_{\rm PbPb} \, 
\sigma_{\gaga\rightarrow \eepair} \, P_n \, \epsilon_e  
\\ \nonumber
& = & 0.5 \mbox{ mb}^{-1}\mbox{s}^{-1} \times 139\mbox{ mb} \times 0.5
\times 0.05 = 1.7 \mbox{ Hz} \, \, .
\label{eq:gagarates}
\end{eqnarray}

The conservative background sum,
$N_{L1} = N_{\rm cosmic}^{\rm 
S(M)ED} + N_{\rm had} + N_{\gaga} +$ others $\sim 5-7$ Hz,
is a factor of $\sim 5000-7000$ smaller than the $\Upsilon$ rate in
Eq.~(\ref{eq:upsrates}). It is therefore important to not have any significant 
trigger dead-time and not to remove good events in the high-level trigger
selection.

\subsection*{High Level Trigger}

The CMS L1 trigger can pass {\it all} selected Pb+Pb events, $\sim 3$ kHz on 
average, and send them to the HLT without reduction~\cite{gunther,ferenc07}. 
The UPC trigger bandwidth allocated in the HLT is 2.25 MByte/s (1\% of the 
total rate) or $\sim 1-2$ Hz for an ultraperipheral event of $1 - 2$ MB. 
The estimated event size of a very peripheral Pb+Pb hadronic interaction with 
$b> 12$ fm is 0.3 MB plus a conservative 1 MB ``noise'' overhead. Since events 
triggering the UPC-L1 trigger have, by design, 
very low multiplicities, they will be below 2 MB already at L1. 

Recording
UPC-HLT rates at the allocated $1-2$ Hz rate requires a factor of $2.5-7$ 
reduction relative to the expected UPC-L1 rates.  To do so, we will need to 
apply one or more simple algorithms at the HLT level to match the allocated 
bandwidth.
First, the L1 electron/muon candidates should be verified with a
L1-improved software check to remove fake triggers.  Next, the
event vertex should be within $z< 15$ cm of (0,0,0). The inherent low 
track/cluster multiplicity of UPC events results in a rather wide vertex 
distribution.  An even looser cut, $z < 60$ cm, is expected to reduce the 
cosmic ray background by a factor of 2500, as well as any remaining beam-gas or
beam-halo events.  Finally, two $p_T$ cuts can be applied.  The total $p_T$
of all particles should be low.  This can be checked by making a rough 
determination of the net $p_T$ of all muon/electron HLT candidates in the 
event.  Hadrons emitted in peripheral hadronic events at 
$\sqrt{s_{_{NN}}}= 5.5$ TeV have 
$\mean{p_T}\approx 600$ MeV/$c$, much larger than the $\mean{p_T}\approx 70$ 
MeV/$c$ expected for coherent photoproduction events.
Thus this cut should significantly reduce the peripheral $AA$ background. 
However, we may also want to study other hard photoproduction events with 
larger $p_T$ which satisfy the UPC-L1 trigger.  Therefore it is probably 
more appropriate to select back-to-back dileptons, part of the global
calorimeter and muon triggers. All these considerations can be taken into
account when setting the final L1 thresholds and 
HLT algorithms and do not affect the quantitative conclusions 
about the $\upsi$ measurement described here.

\paragraph{Input Monte Carlo}
\label{sec:MC} \bigskip

\begin{figure}[htbp]
\begin{center}
\hspace{-0.5cm}
\includegraphics[width=0.5\textwidth,height=7.cm]{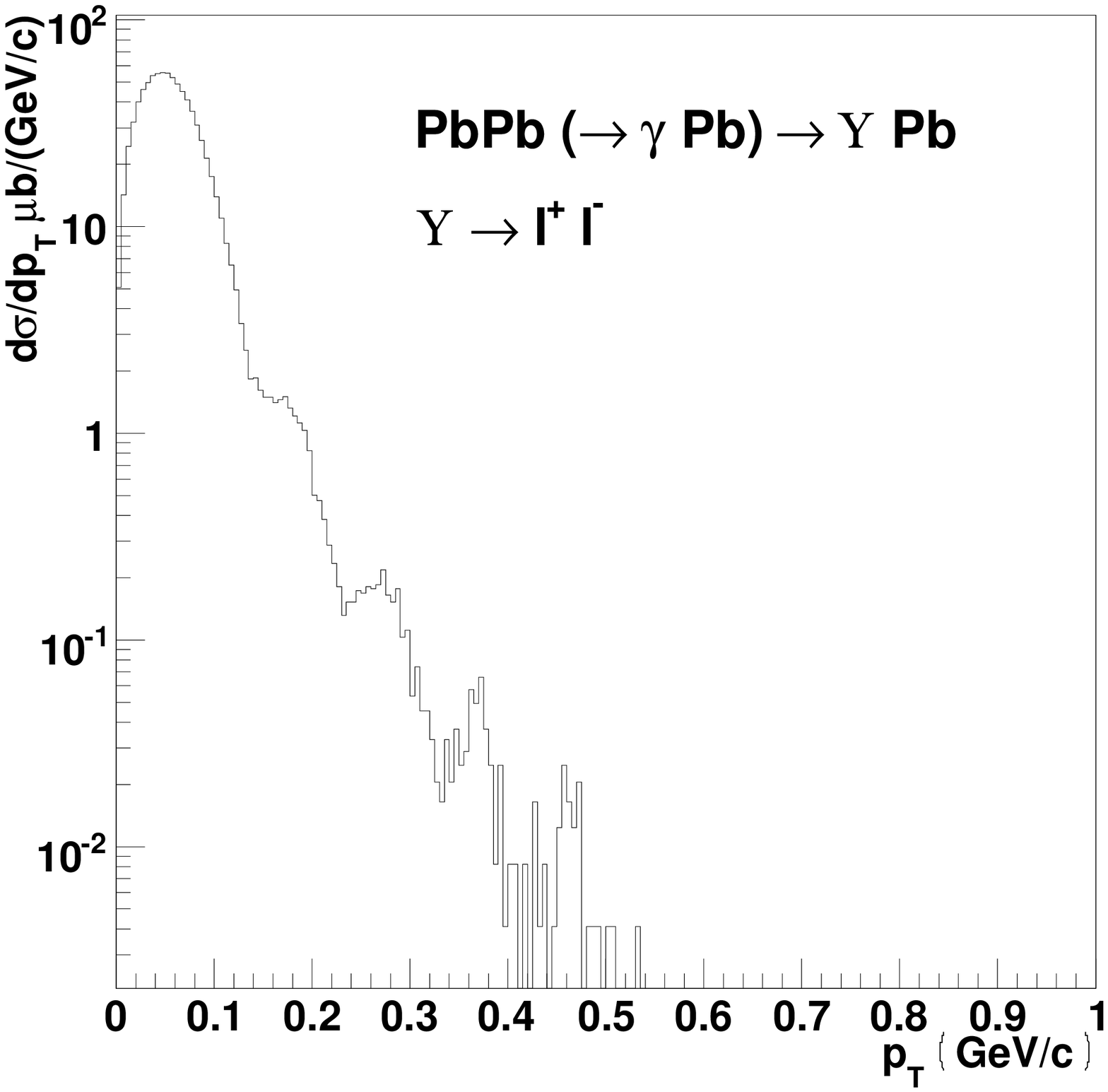}
\includegraphics[width=0.5\textwidth,height=6.8cm]{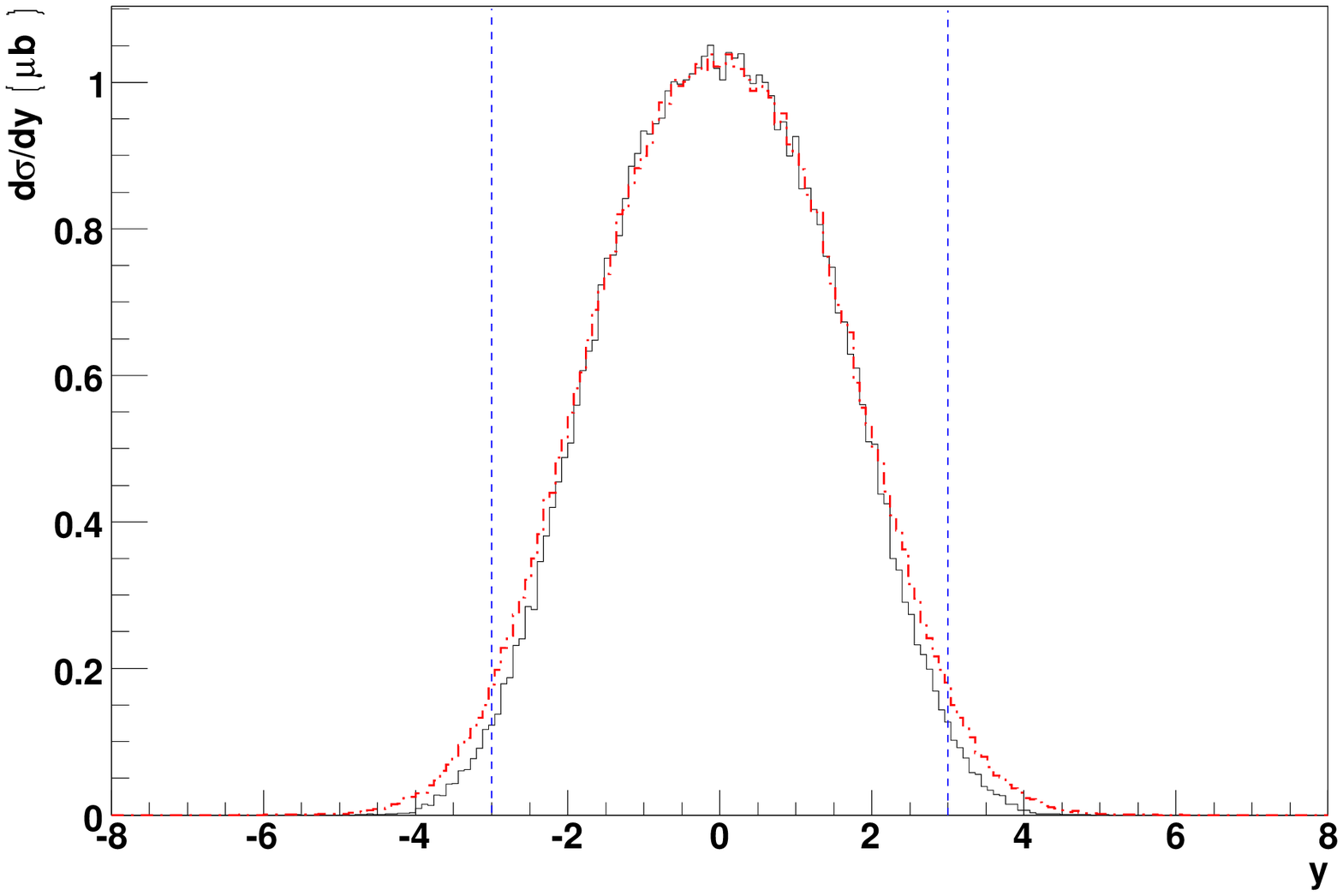}
\hspace{+0.5cm}
  \vspace{-0.6cm}
\end{center}
\caption[]{The {\sc starlight} $p_T$ (left-hand side) and $y$ (right-hand side)
distributions for coherent $\Upsilon$ photoproduction in ultraperipheral Pb+Pb 
collisions at $\sqrt{s_{_{NN}}} = 5.5$ TeV \protect\cite{dde_ahees}.  Note the
diffractive-like peaks in the $p_T$ distribution.  The rapidity distribution 
of single leptons from $\Upsilon$ decays, dot-dashed curve, is also shown on
the right-hand side.  The vertical dashed lines indicate the approximate CMS 
acceptance.}
\label{fig:sim_ups}
\end{figure}

\begin{figure}[htbp]
\begin{center}
\hspace{-0.5cm}
\includegraphics[width=0.5\textwidth,height=6.8cm]{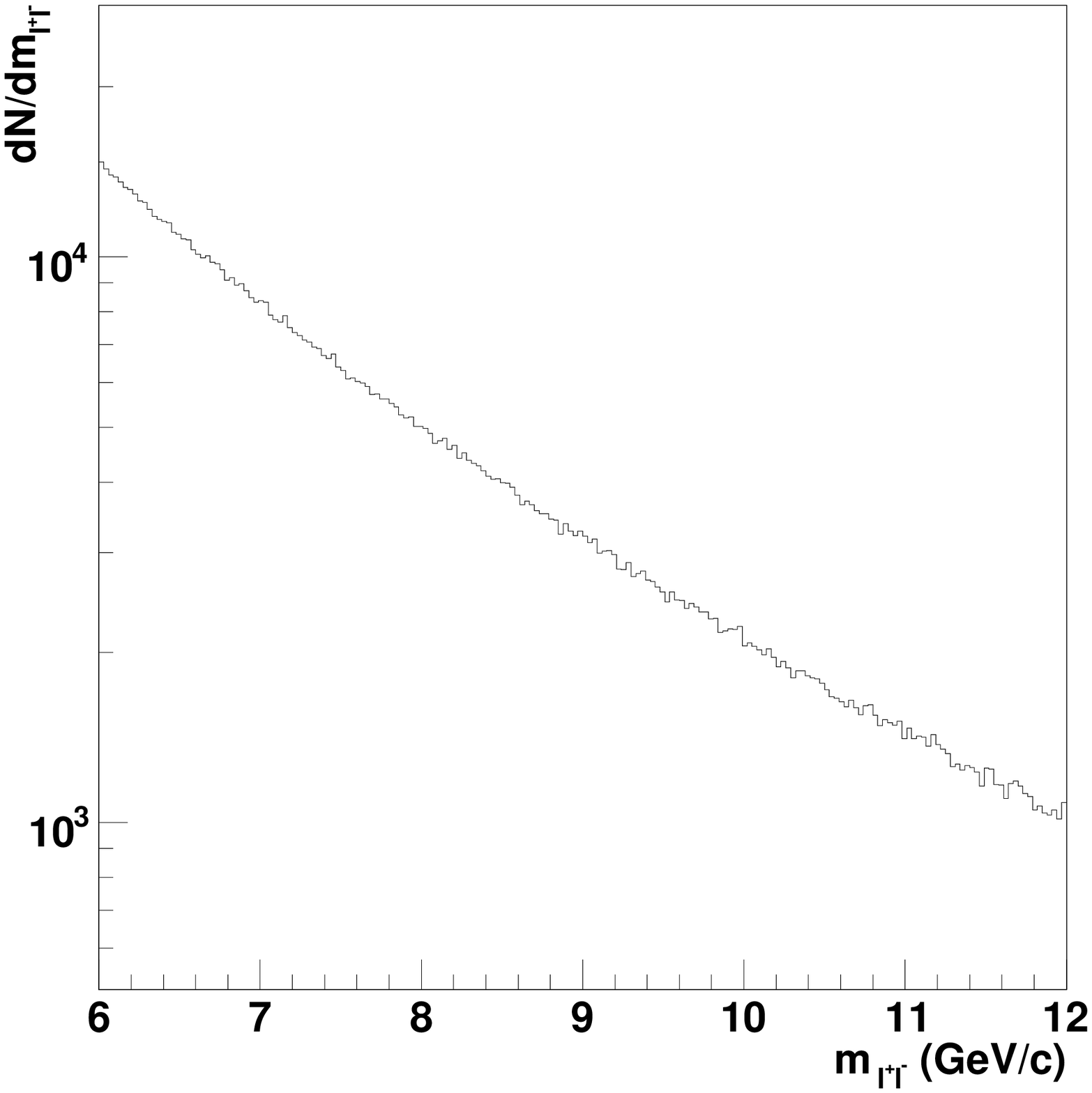}
\includegraphics[width=0.5\textwidth,height=6.8cm]{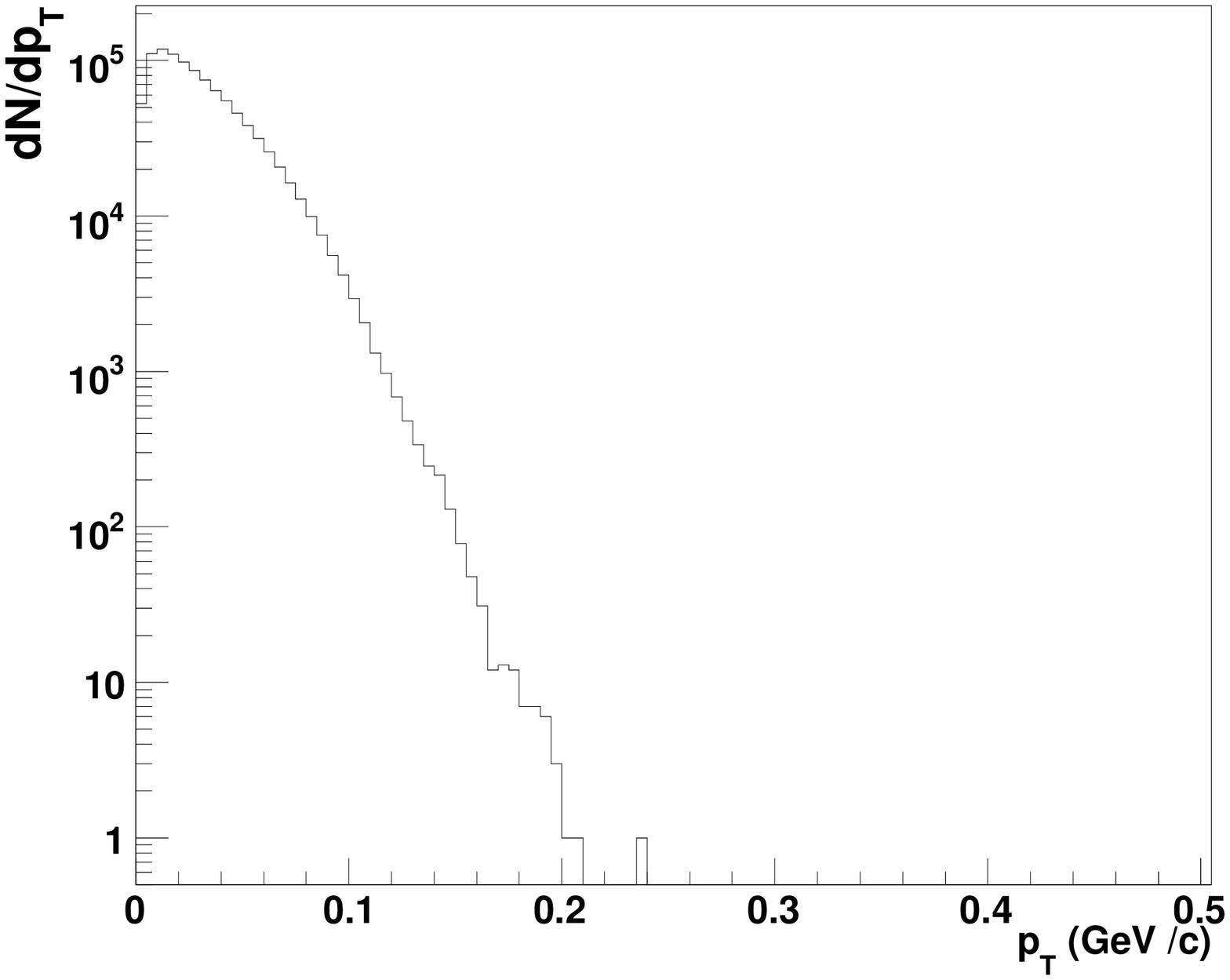}
\hspace{-0.5cm}
\includegraphics[width=0.5\textwidth,height=6.8cm]{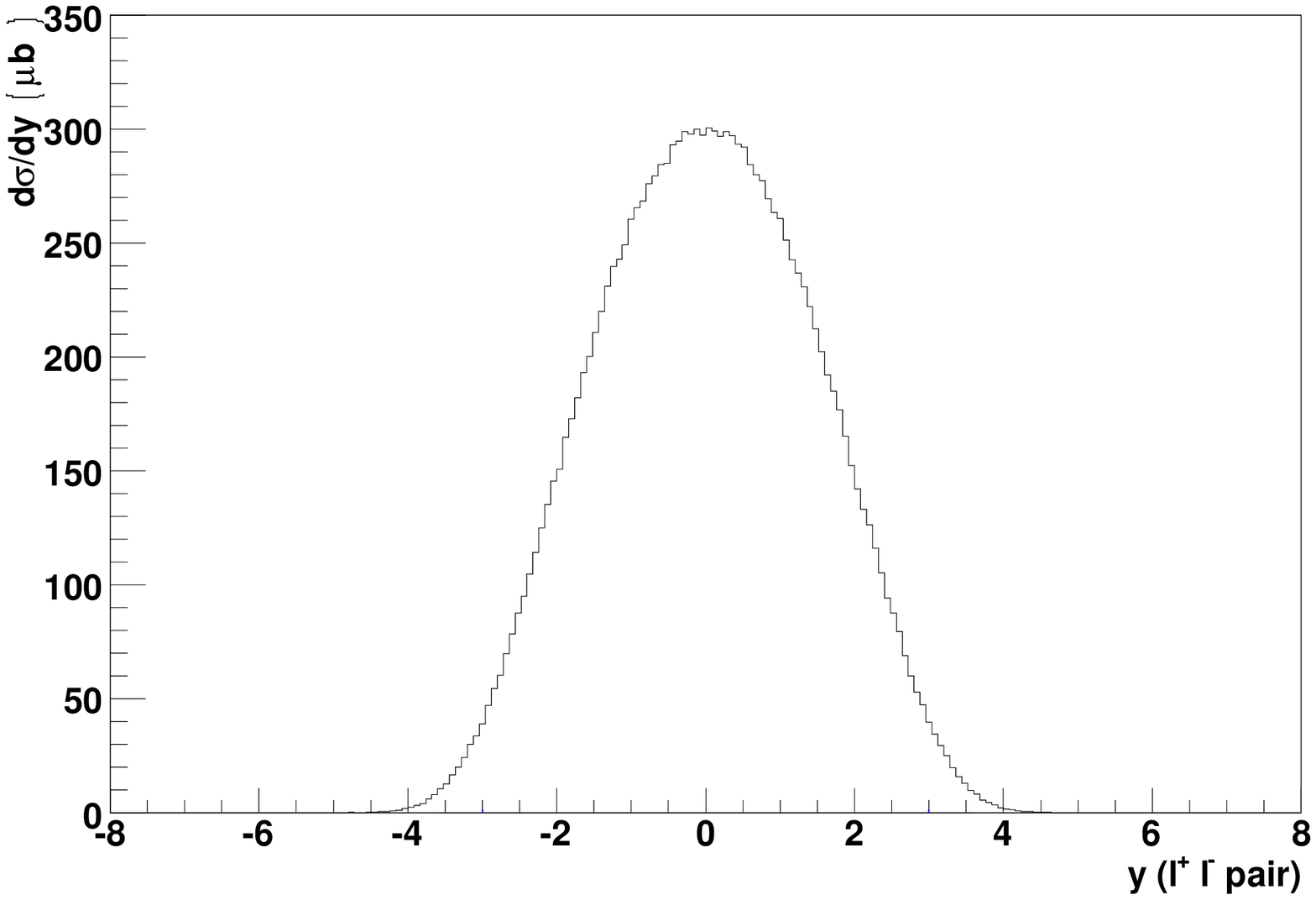}
\includegraphics[width=0.48\textwidth,height=6.8cm]{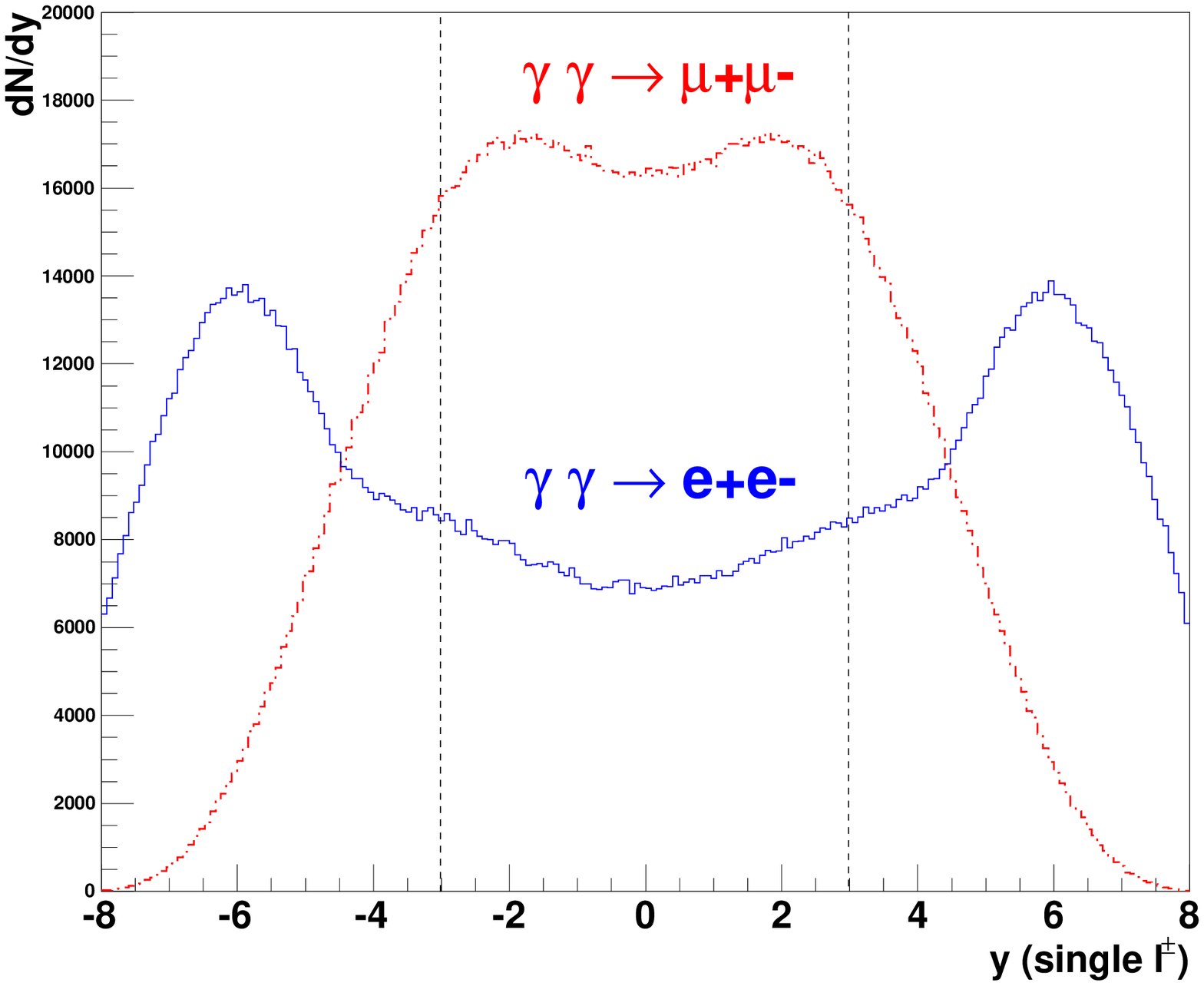}
 \vspace{-0.2cm}
\end{center}
\caption[]{The {\sc starlight} dilepton ($e^\pm,\mu^\pm$) invariant mass 
(top left), pair $p_T$ (top right), pair rapidity (bottom left) and single 
lepton rapidity (bottom right) distributions in ultraperipheral Pb+Pb 
collisions at $\sqrt{s_{_{NN}}} = 5.5$ TeV \protect\cite{dde_ahees}.  
The single muon (dashed) and electron (solid)
rapidity distributions are shown separately in the bottom right plot.  The
vertical dashed lines indicate the CMS acceptance.}
\label{fig:sim_ll}
\end{figure}

Event samples for the $\upsi\rightarrow l^+l^-$ signal and the dilepton 
continuum are generated with the {\sc starlight} Monte Carlo
\cite{Klein:1999qj,starlight,Baltz:2002pp}.  
The input Monte Carlo $p_T$, rapidity and lepton pair invariant mass 
distributions for the signal and background are shown in 
Figs.~\ref{fig:sim_ups} and ~\ref{fig:sim_ll}. 

The most significant characteristic of coherent particle production in UPCs 
is the extremely soft $p_T$ distribution. The $\upsi$ and the lepton pairs 
are produced almost at rest. The $\Upsilon$ 
$p_T$ distribution is also sensitive to the nuclear form factor for lead.
Figure~\ref{fig:sim_ups} shows a diffractive pattern with several diminishing 
local maxima. The dilepton mass distribution decreases locally like an 
exponential or power law, shown in the top left plot of Fig.~\ref{fig:sim_ll}.
The signal and background rapidity distributions are peaked at $y =0$. 
The continuum distribution is broader because it also includes lower mass
pairs. Interestingly, the rapidity distributions of the single {\it decay}  
leptons are much narrower for the $\upsi$ (Fig.~\ref{fig:sim_ups}, right) than
the $\lele$ continuum (Fig.~\ref{fig:sim_ll}, bottom right).  One or both 
leptons from the continuum is often emitted outside the CMS rapidity coverage 
and, therefore, will not affect the $\upsi$ invariant mass reconstruction.

\paragraph{$\upsi\rightarrow l^+ l^-$ acceptance and reconstruction 
efficiency} \bigskip

\begin{figure}[htbp]
\begin{center}
    \includegraphics[width=0.95\textwidth,height=7.5cm]{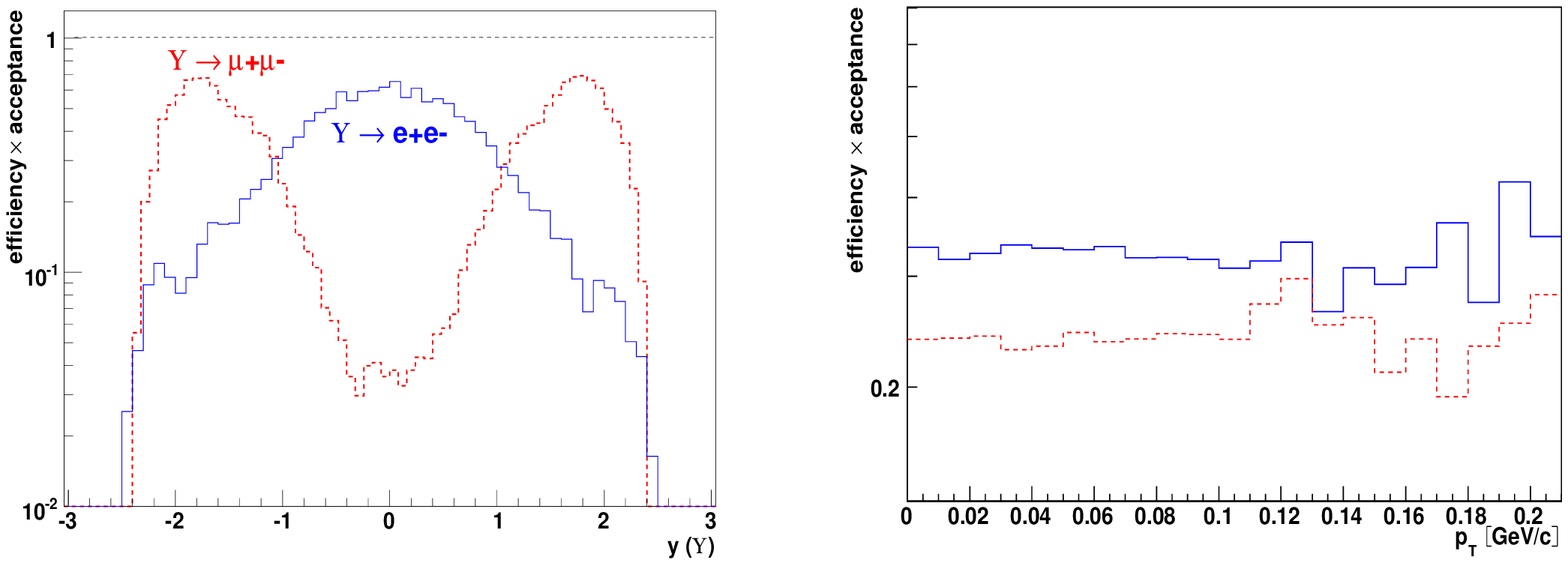}
  \vspace{-0.5cm}
\end{center}
\caption[]{The combined efficiency and acceptance for $\upsi$ decays to 
$\mumupair$ (dashed) and $\eepair$ (solid) obtained from a full CMS simulation
and reconstruction of the input {\sc starlight} distributions as a function 
of the $\upsi$ rapidity (left-hand side) and $p_T$ 
(right-hand side)~\protect\cite{dde_ahees}.}
\label{fig:acceff_ups_reco}
\end{figure}

Figure~\ref{fig:acceff_ups_reco} shows the 
convolution of efficiency with acceptance for CMS as a function of the 
$\upsi$ rapidity and transverse momentum 
respectively in the $\mumupair$ (dashed) and $\eepair$ (solid) analyses, 
obtained by taking the ratio of reconstructed relative to input spectra. 
Note that although the rapidity acceptances of 
both analyses are very different and complementary -- the muon efficiency 
is peaked around $|y| = 2$ and the electron efficiency at $|y|< 1$ -- the 
$p_T$ efficiencies are very similar. The efficiency is about 8\% 
for $\upsi$ produced at rest. 
At the expected coherent production peak, 
$p_T\approx 40-80$ MeV/$c$, the average efficiency is $\sim 10$\%, increasing 
with $p_T$ thereafter. 
The reconstructed spectrum is higher than the generated one for $p_T \geq 130$
MeV/$c$. This `artifact' is due to the combination 
of a steeply-falling spectrum and a reconstruction yielding larger $p_T$ 
$\Upsilon$ than the inputs. 

The integrated combination of the geometric acceptance with the reconstruction 
efficiency in both analyses is 26\% for $\eepair$ and 21\% for $\mumupair$.

\paragraph{Invariant mass distributions and continuum subtraction}
\label{sec:minv} \bigskip

To determine the $\Upsilon$ invariant mass distribution,
it is necessary to include the lepton pair continuum in the mass background.
Any residual combinatorial background 
can be removed from the measured $dN/dM$ distributions by subtracting 
the like-sign, $l^\pm l^\pm$, background from the opposite-sign, $l^\pm l^\mp$,
signal.  In this simulation, the like-sign background is negligible because
we reconstruct only the opposite-sign pairs.

\begin{figure}[htbp]
\begin{center}
\includegraphics[width=0.48\textwidth]{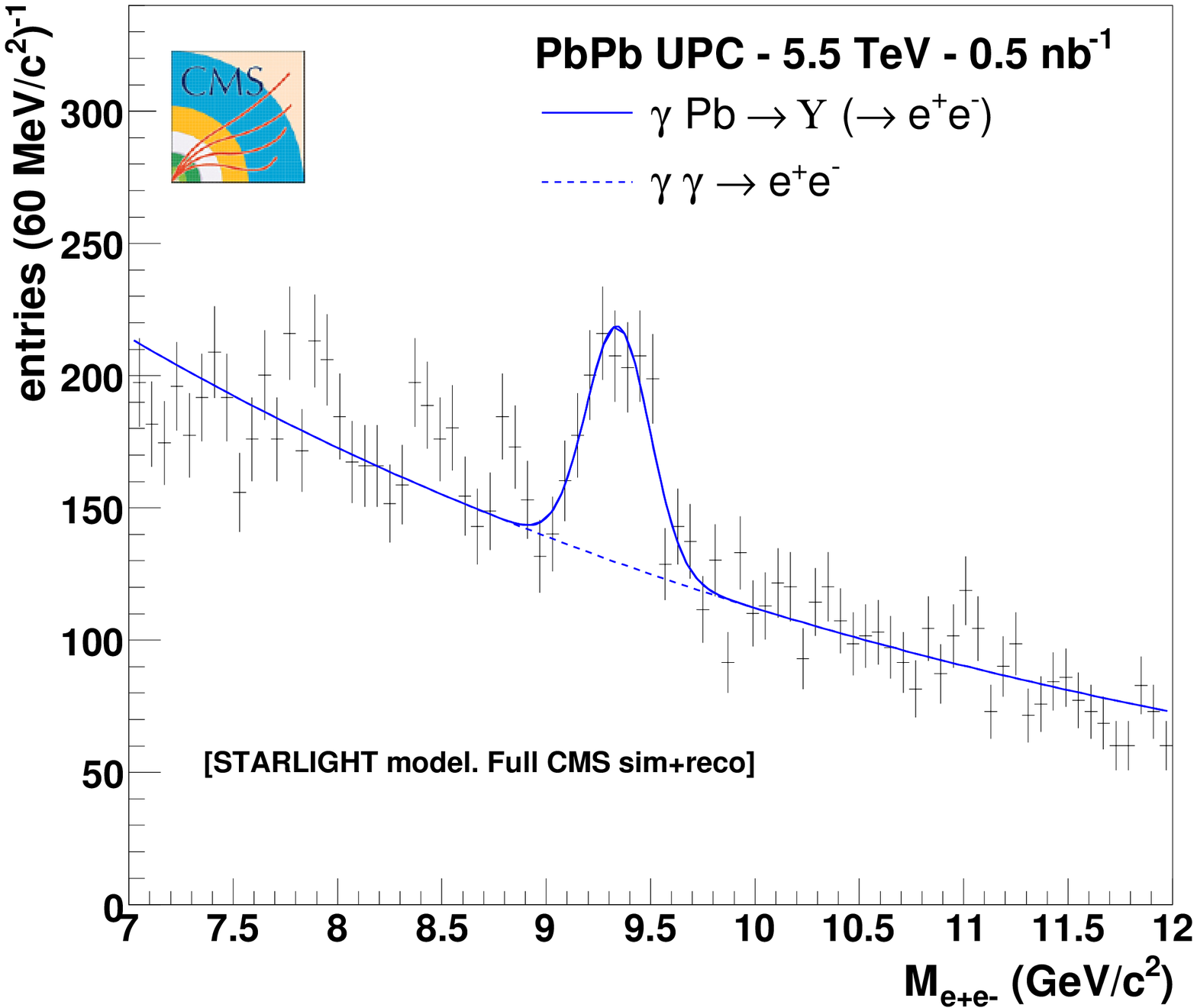}
\includegraphics[width=0.48\textwidth]{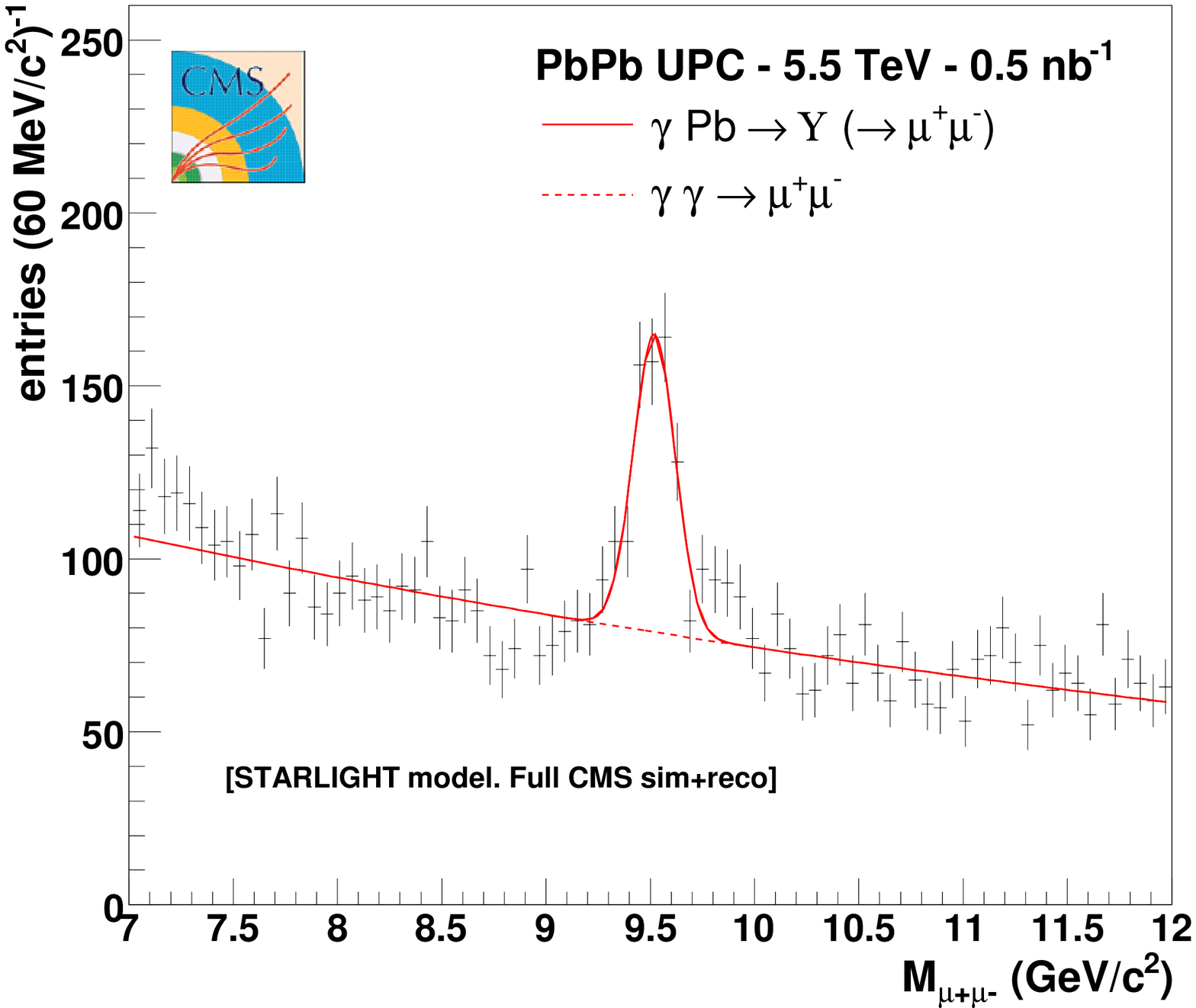}
  \vspace{-0.2cm}
\end{center}
\caption[]{The $\eepair$ (left) and $\mumupair$ (right) mass distributions for 
the $\Upsilon$ signal and dilepton continuum.  Reprinted from 
Ref.~\protect\cite{dde_ahees2} with permission from the Institute of Physics.}
\label{fig:dNdminv}
\end{figure}

The generated $\Upsilon$ signal and lepton pair continuum, 
$6 < M< 12$ GeV/$c^2$, events are mixed according to their 
relative cross sections in Tables~\ref{tab:cms1} and~\ref{tab:cms2}, taking 
the $\upsi$ branching ratio, $B(\Upsilon \rightarrow l^+ l^-) \sim 2.4$\%,
into account.  The input signal-to-background ratio integrated over all 
phase space is rather low,
\begin{equation}
\frac{N_{S}}{N_{B}} = \frac{B(\upsi\rightarrow \lele) 
\sigma_{\upsi}}{\sigma_{\lele}(6 < M < 12\mbox{ GeV/}c^2)} 
\approx 0.35\% \; (\mu^+ \mu^-); \, \, 0.15\%\; (\eepair) \, \, .
\end{equation}
However, coherent lepton pair production is asymmetric and more forward 
than $\upsi \rightarrow l^+ l^-$ so that single leptons from continuum pairs
often fall outside the CMS acceptance,  $|\eta|<2.5$.
In practice, more electrons than muons miss the central CMS region, 
see Fig.~\ref{fig:sim_ll}, making the ratio $N_{S}/N_B$ very similar for 
the $\eepair$ and $\mumupair$ analyses if the different detector
responses are not included.

\begin{figure}[htbp]
\begin{center}
\includegraphics[width=0.48\textwidth]{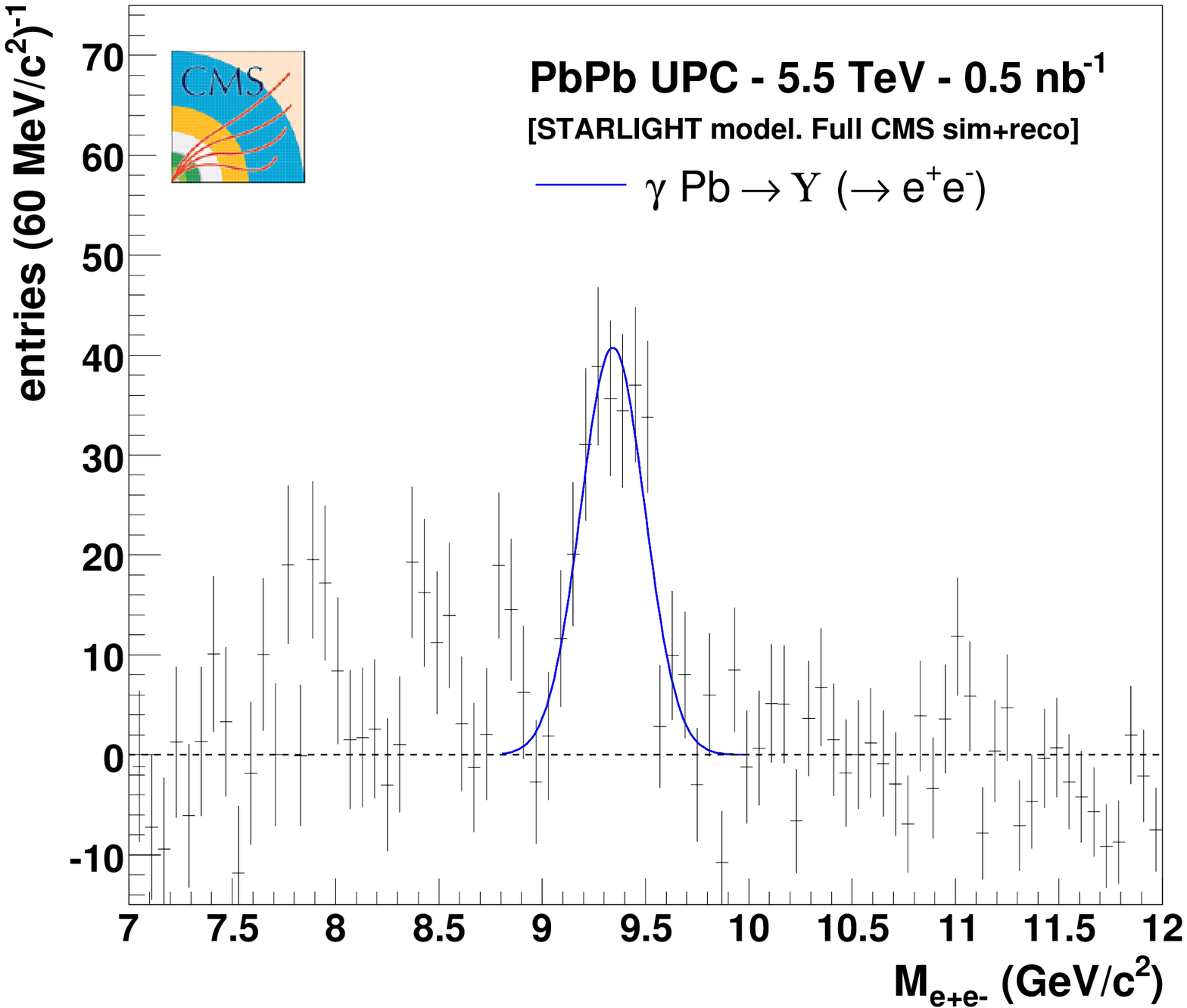}
\includegraphics[width=0.48\textwidth]{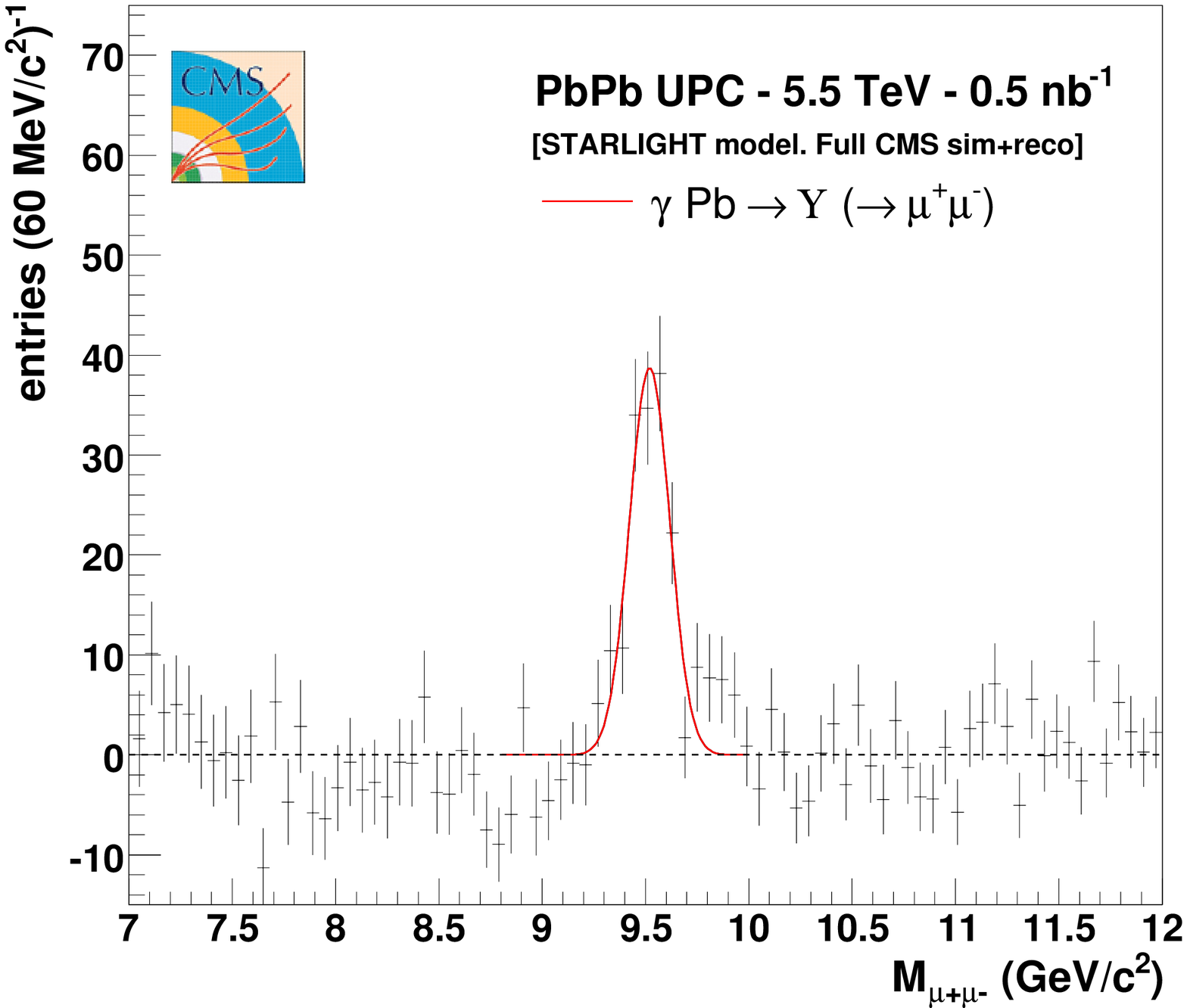}
  \vspace{-0.2cm}
\end{center}
\caption[]{The $\eepair$ (left) and $\mumupair$ (right) mass distributions for 
the $\Upsilon$ signal after background subtraction.  Reprinted from 
Ref.~\protect\cite{dde_ahees2} with permission from the Institute of Physics.}
\label{fig:dNdminv2}
\end{figure}

Figure~\ref{fig:dNdminv} shows the combined signal+background mass spectra in 
the dielectron and dimuon channels.  We find $N_S/N_B \sim 1$ for both cases. 
The combined reconstructed mass spectra are fitted to a Gaussian for the 
$\upsi$ peak plus an exponential for the continuum. The exponential fit to 
the continuum is subtracted from the signal+background entries. The resulting 
background-subtracted $\upsi$ mass distributions fitted to a Gaussian 
alone are shown in Fig.~\ref{fig:dNdminv2}. The final $\Upsilon$ masses and 
widths are $M = 9.52$ GeV/$c^2$ and $\sigma_{\rm res} = 0.090$ GeV/$c^2$ 
for the $\mu^+ \mu^-$ channel and $M = 9.35$ GeV/$c^2$ and $\sigma_{\rm res} 
= 0.16$ GeV/$c^2$
for the $e^+ e^-$ channel, very close to the nominal $\Upsilon$ mass,
$M = 9.46$ GeV/$c^2$ \cite{pdg}. In the dimuon channel, the mass resolution 
is sufficient to allow clean separation of the higher $\Upsilon$ $S$ states,
$\upsi'$ (10.02 GeV/$c^2$) and $\upsi''$ (10.36 GeV/$c^2$), which can also
be produced coherently but were not included in the current simulation.

\paragraph{Total rates} \bigskip

The extracted yields, integrating the counts within 
$3\sigma_{l^+ l^-}$ around the $\upsi$ peak after continuum
background subtraction, are computed for both decay modes.
The efficiency of the yield extraction procedure is
$\epsilon_{\rm extract} = 85$\% for the $e^+ e^-$ and 90\% for the
$\mu^+\mu^-$ analysis. The efficiency is lower in the dielectron channel
due to the larger background. The total $\Upsilon$ production yields
expected with an integrated design Pb+Pb luminosity of 0.5 nb$^{-1}$
are $N_{\upsi\rightarrow \eepair}\approx  220 \pm 15$ (stat) and 
$N_{\upsi\rightarrow \mumupair}\approx 180 \pm 14$ (stat).
Systematic uncertainties are estimated to be $\sim 10$\%  by using different 
functional forms for the continuum and the method of $\Upsilon$ yield
extraction. The uncertainty in the luminosity normalization will be $\sim 5$\% 
since the concurrent continuum measurement provides a direct calibration
scale for the QED calculations~\cite{piotrzk,lumi2}. Combining 
the statistics from 
both channels, the $\upsi$ $y$ and $p_T$ spectra 
will test theoretical predictions of low-$x$ saturation in the 
nuclear PDFs. Even reducing the $\upsi$ yields by a factor of 4, as predicted 
by calculations of nonlinear parton evolution at small $x$, would still
provide a statistically significant sample to compare with theory.

\section{Inclusive photonuclear processes}

\subsection{Large mass diffraction in photon-induced processes} 
{\it Contributed by L. Frankfurt, V. Guzey, M. Strikman and M. Zhalov}

Studies of inelastic diffraction at small $t$ through the $A$ dependence of
hadron-nucleus scattering provide information 
about fluctuations in the interaction strength
\cite{FP,GW}. The total cross section of inelastic diffraction
has been calculated and used to study the $A$ dependence in two ways.  First,
assuming that the $A$ dependence of a particular diffractive channel is the
same as the $A$ dependence of the total cross section, the calculations were 
compared to diffractive $pA \rightarrow p \pi A$ and $\pi A \rightarrow 
\pi \pi \pi A$ scattering.  Second, the total cross section of inelastic 
diffraction has been measured in $pA$ interactions for $A = ^4$He and emulsion
at $p_{\rm lab} = 200$ and 400 GeV \cite{Frankfurt:2000ty}.  Since the $NN$ 
cross section increases with energy, 
fluctuations in the elementary amplitudes lead to much smaller 
fluctuations in absorption in scattering off heavy nuclei.  As a result, 
a much weaker $A$ dependence is expected for the diffractive cross 
section~\cite{Frankfurt:1993qi} at colliders.  In particular,
$\sigma^{\rm diff}_{p A\to X A}\propto A^{0.25}$ at LHC energies
\cite{FELIX} relative to $\sigma^{\rm diff}_{p A\to X A}\propto
A^{0.7}$ at fixed-target energies. 
For high-mass diffraction, this suppression can also be understood by using 
the $t$ channel picture of Pomeron exchange due to
the stronger screening of the triple Pomeron exchange \cite{Kaidalov:2003vg}.

Diffraction in deep-inelastic scattering
corresponds to the transition of the (virtual) photon
into its hadronic components, leaving the nucleus intact.
Hence diffractive DIS has more in common with elastic 
hadron-nucleus scattering than inelastic diffractive
hadron-nucleus scattering. The approach of the elastic cross section
to half the total cross section is a direct indication of the proximity of the 
interaction regime to the BDR.
Correspondingly,  the most direct information on the proximity of hard
interactions, such as $c \overline c$ photoproduction, to the BDR can be
obtained if the diffractive fraction of the total cross section 
can be measured.

In the following, leading-twist diffraction and diffraction in 
the BDR will be discussed and applied to the analysis of diffractive
UPCs.

\subsubsection{Nuclear diffractive parton densities}

The key ingredient in calculations of hard diffractive processes
in photon-nucleus scattering is nuclear diffractive PDFs (NDPDFs).
In the photon case, the NDPDFs can be determined from direct photon studies,
such as photon-gluon fusion or large angle Compton scattering, $\gamma q 
\to \gamma q$. 
Since the leading-twist NDPDFs satisfy the factorization 
theorem, they can be analyzed on the basis of diffraction in DIS.

There is a deep connection between shadowing and
diffractive scattering off nuclei.
The simplest way to investigate this connection is to apply 
the AGK cutting rules~\cite{AGK}. Several processes contribute to nuclear
diffraction: coherent diffraction where the nucleus 
remains intact; nuclear breakup without hadron production in
the nuclear fragmentation region; and 
rapidity gap events with hadron production in the 
nuclear fragmentation region. For 
$x\leq 3\times 10^{-3}$ and $Q^2\sim 4$ GeV$^2$, the fraction of DIS events 
with rapidity gaps reaches $\sim 30-40$\% for heavy 
nuclei, rapidly decreasing with $A$ \cite{FS96}.

\begin{figure}[htb]
\begin{center}
\includegraphics[width=0.6\textwidth]{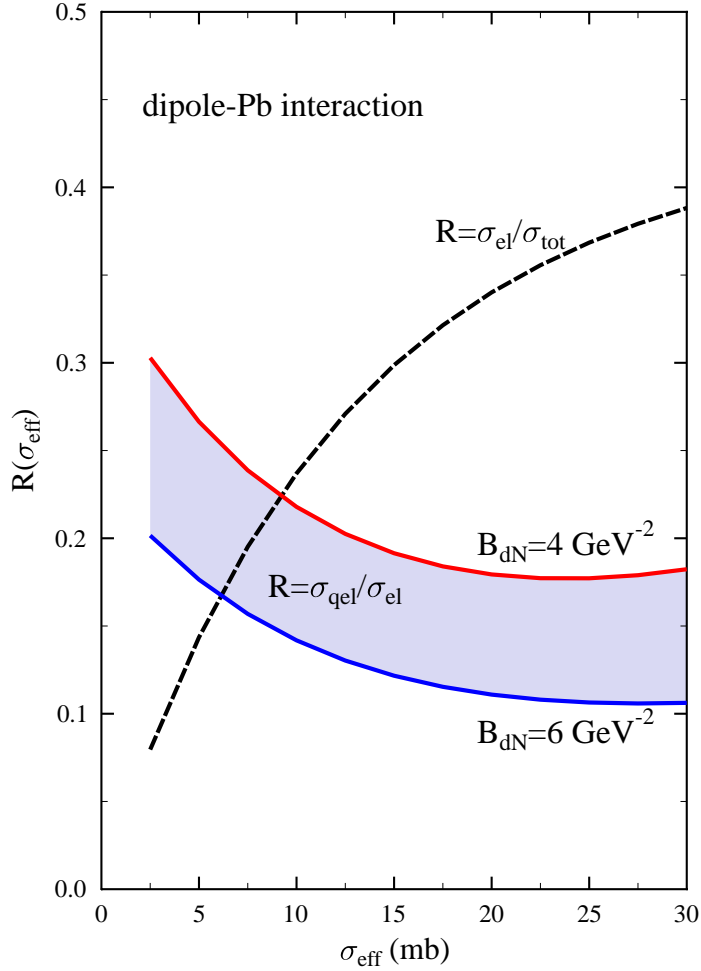}
\caption[]{The dashed curve is the ratio of the coherent to total dipole-lead  
cross sections as a function of the effective dipole-nucleon cross section. 
The solid lines are the quasi-elastic to coherent dipole-nucleus cross section
ratios for two different values of the slope, $B$, of the elastic 
dipole-nucleon $t$ distribution.}
\label{figcsratio}
\end{center}
\end{figure}

The effective cross section, $\sigma_{\rm eff}$, in Eq.~(\ref{sigef})
which describes diffractive hard interactions of quark-gluon configurations
with a nucleon can be used to estimate the probability of diffractive
interactions in nuclei for a number of hard triggers beginning at resolution
scale $Q_0^2$.  The $\sigma_{\rm eff}$ dependence of the fraction of events 
attributable to coherent diffraction and diffraction with 
nuclear breakup was considered,
neglecting fluctuations in the interaction strength.  For realistic values of 
$\sigma_{\rm eff}$, the probability of coherent diffraction 
is quite large.  The probability increases slowly with $\sigma_{\rm eff}$ and 
does not approach 50\% even for very large $\sigma_{\rm eff}$, reflecting  
a significant diffuse nuclear surface, even for large $A$, see 
Fig.~\ref{figcsratio}.  Thus, the probability is not sensitive to fluctuations 
in $\sigma_{\rm eff}$.  In the quasi-eikonal approximation, the ratios
$R = \sigma_{\rm qel}/\sigma_{\rm el}$ describe the dependence of the
quasi-elastic to elastic scattering ratio on $\sigma_{\rm eff}$.

The diffractive parton densities were calculated by extending
the leading-twist theory of nuclear shadowing on the total cross 
sections to the case of diffractive scattering \cite{Frankfurt:2003gx},
\begin{eqnarray}
x f_{j/A}^{D(3)}(x,Q_0^2,x_{\Pomeron})&=& 4\,\pi \beta\,
f_{j/N}^{D(4)}(x,Q_0^2,x_{\Pomeron},t_{{\rm min}}) \int d^2 b \nonumber\\
&\!\!\!\!\!\!\!\!\!\!\times &\!\!\!\!\!\!\!\!\!\!
\left| \int^{\infty}_{-\infty} dz \, \rho_A(b,z)
\,e^{i x_{\Pomeron} m_N z} e^{-\frac{1-i \eta}{2}\sigma_{{\rm eff}}^j(x,Q_0^2) 
\int_{z}^{\infty} d z^{\prime}\rho_A(b,z^{\prime})} \right|^2 \,\, . 
\label{eq:masterD}
\end{eqnarray}
The 2006 H1 Fit B~\cite{unknown:2006hy,:2006hx} to the nucleon diffractive PDFs
was used in the analysis of Eq.~(\ref{eq:masterD}).
The superscripts $(3)$ and $(4)$ denote the dependence of the diffractive
PDFs on three and four variables, respectively. 
Equation (\ref{eq:masterD}) is presented for the $t$-integrated nuclear DPDFs
since it is more compact and since it is not feasible to 
measure $t$ in diffraction off nuclei in colliders.
In deriving Eq.~(\ref{eq:masterD}) any possible dependence of 
$\sigma_{\rm eff}^j(x,Q^2)$ on $\beta = x/x_{\Pomeron}$ 
in the exponential factor was neglected and an average value of
$\sigma_{\rm eff}^j$ was employed.
Note that any suppression of small $\beta$ diffraction
in interactions with nuclei in the soft regime is neglected since there
are only elastic components for heavy nuclei (inelastic diffraction is zero). 
Hence the
soft contribution at $Q^2_0$ due to triple Pomeron exchange is strongly 
suppressed \cite{FS96}. As a result, the small $\beta$
nuclear DPDFs are suppressed by a factor $\propto A^{1/3}$
at  $Q^2_0$.  This suppression will be less pronounced
at large $Q^2$ due to QCD evolution.

The nucleon DPDFs are well approximated by the factorized product of two
functions, one dependent on $x_{\Pomeron}$ and $t$ and the other dependent on
$\beta$ and $Q^2$.  However, it is clear from Eq.~(\ref{eq:masterD}) that  
the factorization approximation is not valid for the 
nuclear DPDFs.  At fixed $x_{\Pomeron}$, the right-hand side of
Eq.~(\ref{eq:masterD}) depends not only on $\beta$ but also on
Bjorken $x$ since the screening factor depends on $\sigma_{{\rm eff}}^j$, 
a function of $x$. Equation (\ref{eq:masterD}) also depends on $A$ since   
nuclear shadowing increases with $A$. 
The breakdown of factorization results from the increase of the nuclear 
shadowing effects with incident energy and $A$.
The resulting nuclear DPDFs are evolved to higher $Q^2$ using the NLO 
leading-twist (DGLAP) evolution equations.

\subsubsection{Numerical results}
\label{numer}

It is convenient for our discussion to 
quantify the nucleon and nuclear diffractive PDFs
by introducing $P_{\rm diff}^j$, the
probability of diffraction for a given parton flavor $j$,
\begin{eqnarray}
P^j_{{\rm diff}}=\frac{\int_x^{x_{\Pomeron}^0} dx_{\Pomeron}\,
x f_{j}^{D(3)}(x,Q^2,x_{\Pomeron})}{x f_{j}(x,Q^2)} \,.
\label{eq:probability}
\end{eqnarray}

\begin{figure}[h]
\begin{center}
\epsfig{file=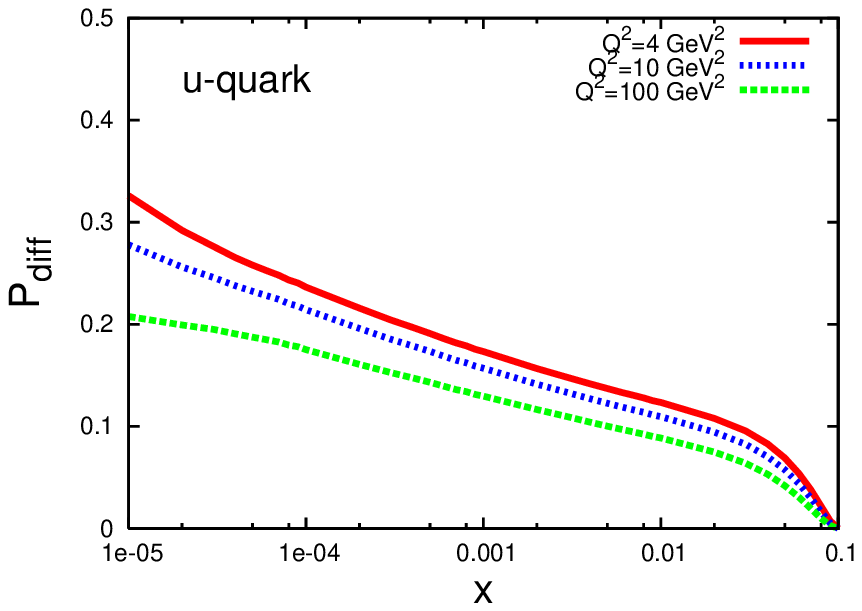,width=7cm,height=7cm}
\epsfig{file=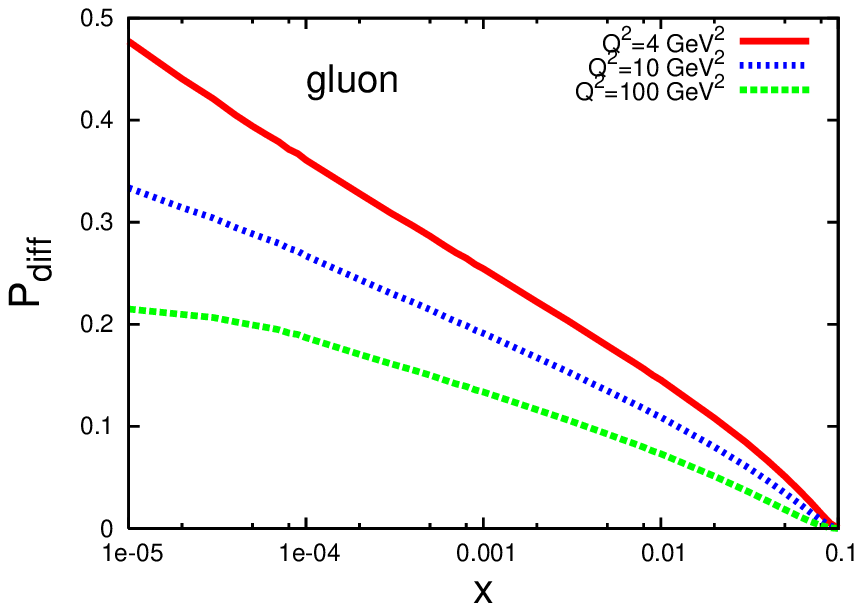,width=7cm,height=7cm}
\caption{The probability of hard diffraction on the nucleon,
$P^j_{{\rm diff}}$, defined in Eq.~(\protect\ref{eq:probability}),
as a function of $x$ and $Q^2$ for $u$ quarks (left)
and gluons (right).}
\label{fig:Pdiff_nucleon}
\end{center}
\end{figure}

First we discuss nucleon diffractive PDFs.
Figure~\ref{fig:Pdiff_nucleon} presents the nucleon $P^j_{{\rm diff}}$ as a
function of $x$ for $Q^2=4$, 10 and 100 GeV$^2$ for $u$ quarks and gluons.
At low $Q^2$, $P_{\rm diff}^g > P_{\rm diff}^u$.  Note also that
$P_{\rm eff}^g$ is very close to the unitarity
limit, $P^j_{{\rm diff, max}}=1/2$.
The larger probability of diffraction for gluons is related
to the larger gluon color dipole cross section in the ${\bf 8 \times 8}$
representation relative to the triplet $q {\bar q}$ dipole.

\begin{figure}[h]
\begin{center}
\epsfig{file=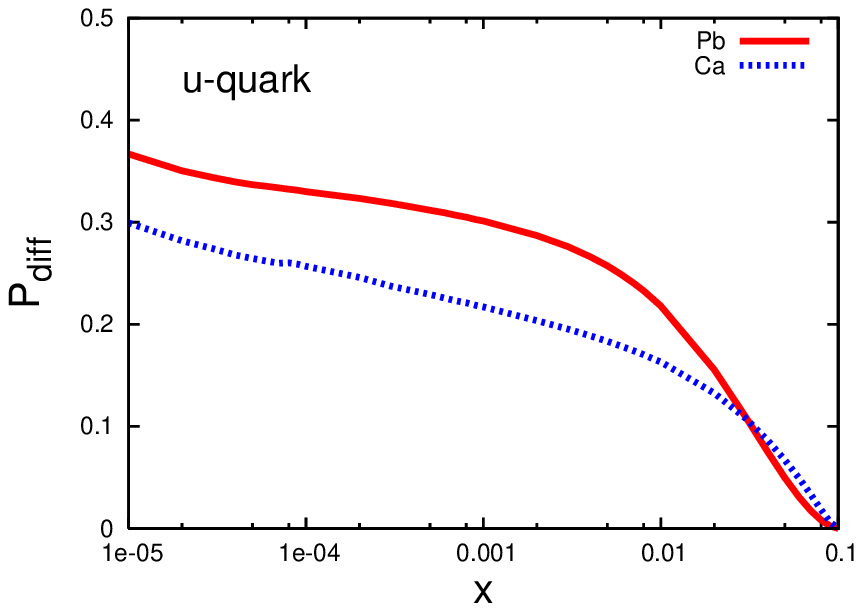,width=7cm,height=7cm}
\epsfig{file=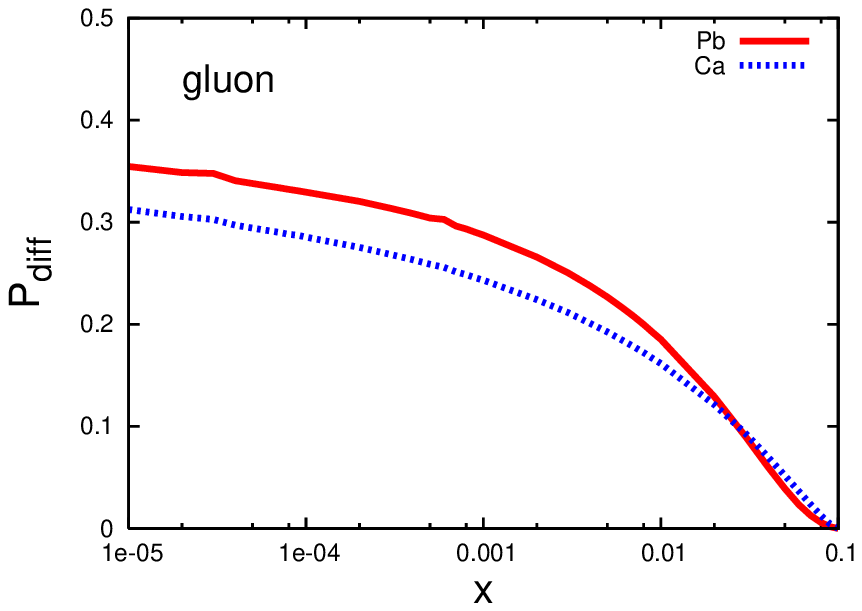,width=7cm,height=7cm}
\caption{The probability of hard diffraction, $P^j_{{\rm diff}}$, on
$^{40}$Ca and $^{208}$Pb, at $Q^2=4$ GeV$^2$
as a function of $x$ for $u$ quarks (left) and gluons (right).}
\label{fig:Pdiff_nuclei}
\end{center}
\end{figure}

Next, we turn to hard diffraction with nuclear targets.
Figure~\ref{fig:Pdiff_nuclei} presents $P^j_{{\rm diff}}$ for $^{40}$Ca and
$^{208}$Pb at $Q^2=4$ GeV$^2$ as a function of $x$ for $u$ quarks and gluons.
The $A$ dependence of
$P^j_{{\rm diff}}$ is rather weak for $A \geq 40$ because at large $A$ and
small $b$, the interaction is almost completely absorptive (black) with a
small contribution from the opaque nuclear edge. The $A$ dependence for gluons
is somewhat weaker since gluon
interactions are closer to the black disk regime.

At small $x$, the $A$ dependence of $P_{\rm diff}^j$ is qualitatively
different for quarks and gluons.  While the $A$ dependence of
$P^g_{{\rm diff}}$ is expected to be very weak\footnote{The probability
$P^g_{{\rm diff}}$ for nuclei increases faster than for nucleons with 
decreasing $x$ since the nuclear center is like a black disk.  However, 
scattering off nucleons near the edge of the nucleus slows the increase of
$P^g_{{\rm diff}}$ for nuclei as the ratio
$\sigma_{\rm diff}/\sigma_{\rm tot}$ approaches 0.5.},
$P^q_{{\rm diff}}$ is expected to grow with
$A$ since the diffractive probability for quarks, shown in
Fig.~\ref{fig:Pdiff_nucleon}, is rather far from the BDR and thus can increase.

We now turn to the $Q^2$ dependence of
$P^j_{{\rm diff}}$.  For both nucleons and nuclei, $P^q_{{\rm diff}}$ changes
weakly with $Q^2$ and is $\sim 20-30$\% at small $x$, in good agreement with
early estimates \cite{FS96}.
While $P_{\rm diff}^g$ decreases faster with increasing $Q^2$, the probability
is still $\sim 15-20$\%
at $Q^2=100$ GeV$^2$, making {\it e.g.} heavy flavor studies feasible in UPCs
at the LHC, similar to inclusive production, considered
in Ref.~\cite{Klein:2002wm}.
Dijet production is another alternative, studied by ZEUS \cite{Derrick:1995tw}
and H1 \cite{Newman} using protons\footnote{The recent HERA data seem to 
indicate that the factorization theorem for direct photoproduction 
holds at lower transverse momentum for charm
production than typical dijet production.}.

The discussion presented here is relevant for hard processes produced in
direct proton interactions. Spectator parton interactions will suppress
the probability of diffraction for resolved photons. Estimates
\cite{Frankfurt:1996bh} indicate that spectator interactions will decrease 
the probability of nuclear diffraction by at least a
factor of two for $A \sim 200$. Thus, the $A$ dependence of diffraction
with resolved photons will also be interesting since it will measure the
interaction strength of the spectator system with the media, providing another
handle on the photon wavefunction.

\subsubsection{Large mass diffraction in the black disk regime}

One striking feature of the BDR is the orthogonality of the Fock
components of the photon wavefunction \cite {Gribov}.  Thus there can be
no transitions between non-diagonal components, {\it e.g.} 
$\langle q \overline q|
q \overline q g \rangle \equiv 0$.  Since the dominant contribution to 
coherent diffraction in the BDR
originates from a `shadow' of fully-absorptive interactions
for $b\leq R_A$, the orthogonality argument 
is applicable. The orthonormality condition is used to derive 
the BDR expression for the differential cross section of the process
$\gamma A \rightarrow X A$ where $X$ is a final state of 
invariant mass $M$ \cite{Frankfurt:2001nt}. In the real photon case,
\begin{equation}
{{d\sigma_{\gamma A\to X A}}
\over dt dM^2} ={\alpha \over 3 \pi}{(2\pi R_A^2)^2\over 16\pi}
{\rho(M^2)\over M^2} {4\left|J_1(\sqrt{-t}R_A)\right|^2
\over -t R_A^2} \, \, 
\label{ccsb}
\end{equation}
where $\rho(M^2) = \sigma_{e^+ e^-\to {\rm hadrons}}/ \sigma_{e^+ e^-\to 
\mu^+ \mu^- }$ at $s = M^2$.
Diffractive measurements of states with a range of masses would determine
the blackness of the photon wavefunction as a function of mass by comparing to
the BDR results in Eq.~(\ref{ccsb}).
A similar equation for production of specific final states
is valid in the BDR in the case of coherent nuclear recoil. It is then possible
to determine the components of the photon wavefunction which interacts 
with the BDR strength in the coherent processes. 

The onset of the BDR limit for hard processes should also reveal itself in a 
faster increase of the photoproduction cross sections of radial
excited states with energy relative to the ground state cross section.
Utilizing both an intermediate and a heavy nuclear beam, such as Ar and Pb,
would make it possible to remove edge effects as well as maximize the path
length through nuclear matter, about 10 fm in a large nucleus.

One especially interesting process is exclusive 
diffractive dijet production by real photons.
For the $\gamma A$ energies available at the Electron Ion Collider 
\cite{EIC1,EIC2} 
or in UPCs at the LHC, the BDR would be a good approximation for $M \sim$ few 
GeV in the photon wavefunction, the domain described by perturbative QCD for
${\it x}\sim 10^{-3}$ with proton targets,
and larger $x$ for nuclei. The condition of large longitudinal 
distances, a small longitudinal momentum 
transfer,  will be applicable up to quite large values of the
produced diffractive mass $M$.
In the BDR, the dominant channel for large mass diffraction is
dijet production with a total cross section given by
Eq.~(\ref{ccsb}) and characteristic center-of-mass angular distribution
$(1+ \cos^2 \theta)$ \cite{Frankfurt:2001nt}. In contrast, except for charm,
diffractive dijet production is strongly suppressed in the perturbative QCD 
limit \cite{Brodsky:1968,diehl}. The suppression
is due to the coupling of the  $q\overline q$ component of the 
photon wavefunction to two gluons, calculated to lowest order in 
$\alpha_s$. As a result, for real photons, hard diffraction
with light quarks is connected to the production of $q\overline q g$ and 
higher states. The mass distribution of diffractively-produced 
jets thus provides an important  test of 
the onset of the BDR. In the DGLAP/color transparency regime, 
forward diffractive dijet production cross sections should
should be $\propto 1/M^8$ and dominated by charm and bottom jet production,  
strikingly different from the BDR expressions of Ref.~\cite{Frankfurt:2001nt}.

Thus, dijet photoproduction should be very sensitive to the
onset of the BDR.  The $q \overline q$ component
of the  photon light-cone wavefunction can be measured using three
independent diffractive phenomena:
in the BDR off protons and heavy nuclei and in the 
color transparency regime where the wavefunction  can be measured as a 
function of the  inter-quark distance~\cite{Frankfurt:1993it}.
A competing process for dijet photoproduction off heavy nuclei is
the process $\gamma \gamma \rightarrow {\rm dijets}$ where the second photon
comes from the Coulomb field of the opposite nucleus.  Dijets produced
in $\gamma \gamma$ collisions have positive
$C$ parity.  Thus this amplitude does not interfere
with dijet production in $\gamma \Pomeron$ interactions with negative 
$C$-parity.  Therefore $\gamma \gamma \rightarrow 
{\rm dijets}$ are a small background over a 
wide range of energies \cite{Frankfurt:2002wc}.

\subsubsection{High mass diffraction in UPCs}
\label{upcdiff} \bigskip

The large predicted hard diffraction probability
can be checked in UPCs at the LHC.
For example, $\gamma A \to {\rm jet}_1 + {\rm jet}_2 +X + A $
can be studied in the kinematics where the direct photon process,
$\gamma g \to q\overline q$, dominates.
In this case, for $p_T \sim 10$ GeV/$c$ and
$Q^2\sim 100$ GeV$^2$, $\sim 20$\% of the events will be diffractive.
The hadroproduction background originates from glancing collisions
where two  nucleons interact through the double diffractive process 
$pp\to pp X$ where $X$ contains jets.
The probability of hard processes with two gaps is very small 
at collider energies, even smaller than the probability of
single diffractive hard processes \cite{Goulianos:2002hp}. 
Therefore, the relative backgrounds in the diffractive case are expected to
be at least as good as in the
inclusive case \cite{Klein:2002wm}.
Thus, it would  be rather straightforward to extract coherent diffraction 
by simply using  anti-coincidence with the forward neutron  detector,
especially for heavy nuclei \cite{Strikman:1998cc}.
As a result, it would be possible to measure the nuclear DPDFs 
with high statistical 
accuracy. In contrast to diffractive vector meson production, it would be 
possible to determine the energy of the photon which induced 
the reaction on an event-by-event basis since the photon rapidity
is close to the rapidities of the two jets.
It would be possible to measure large rapidities by selecting photoproduction
events with the highest kinematically allowed
energies of the produced particles in the rapidity interval $ y_1 < y < y_2$
and determine the DPDFs for rather small $x$.

There are two contributions to dijet photoproduction, direct and resolved.
In the direct process, the entire photon energy contributes to the hard
process.  In the resolved process, only a fraction of the photon energy,
$z_\gamma$, is involved.  HERA studies indicate that the requirement 
$z_\gamma \geq 0.8$ eliminates the resolved photon contribution.  However,
at higher $Q^2$, DGLAP evolution increases the relative
importance of the resolved component.

In $AA$ collisions, there are two possible contributions since the photon can
come from either nucleus.  It is thus more convenient to refer to the $x$
of the photon and the Pomeron.  
The values of $x$ can be reconstructed from
the kinematics of the diffractive state, $X$, with mass $M_X$, produced in 
the reaction $AA \rightarrow AA X$.  The light cone momentum fractions of
the two nuclei, $x_1$ and $x_2$ are normalized to $A$ and satisfy the
kinematic relation
\begin{equation}
x_1x_2s_{_{NN}}=M_X^2.
\end{equation}
One $x$ is carried by the photon and the other by the Pomeron. (Here Pomeron
is used to define the kinematics of the process without specifying a 
particular dynamical mechanism).
When no high $p_T$ jets are produced, the values of $x$ are related to the
rapidity range of the produced system. In a symmetric $AA$ interaction, the
convention is to define $y_1 = y_A - \ln (1/x_1)$ and $y_2 = -y_A + 
\ln(1/x_2)$.)
The cross section for the production of state $X$ is
\begin{eqnarray}
{d \sigma_{AA\to X AA}\over dx_1dx_2} & = & \frac{dN_{\gamma}(x_1)}{dk} \,
{d\sigma_{\gamma A\to X A}(\tilde{s}=x_1s_{_{NN}},x_{\Pomeron}=x_2)\over 
d x_{\Pomeron}} \nonumber \\ &  & \mbox{} + \frac{dN_{\gamma}(x_2)}{dk} \,
{d\sigma_{\gamma A\to X A}(\tilde{s}=x_2s_{_{NN}},
x_{\Pomeron}=x_1)\over d x_{\Pomeron}} \, \, .
\label{basediff}
\end{eqnarray}
The direct photoproduction cross section for a hard process such as dijet
or heavy quark production is given by the standard pQCD convolution formulas 
over the nuclear DPDFs and the photon flux. In the resolved case,
$z_{\gamma} \ll 1$, diffraction should
be suppressed by interaction of spectators in the photon wavefunction
with the target, increasing the multiplicity and reducing the rapidity gap. 
Though these processes appear to be negligible for protons, they are likely
to reduce the diffractive cross section considerably, see Section \ref{numer}.
 
\begin{figure}[t]
\begin{center}
\includegraphics[width=10cm]{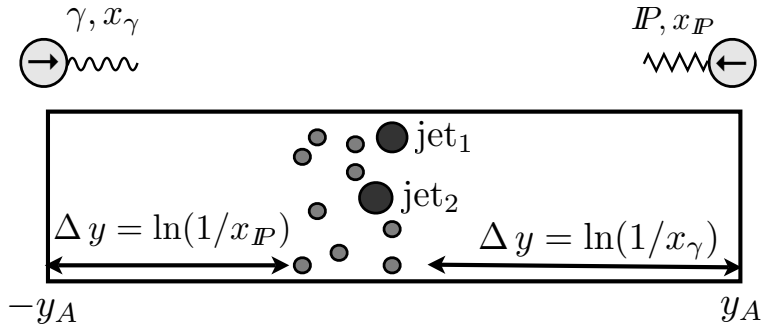}
\end{center}
\caption[]{A schematic lego plot of a diffractive photoproduction event
showing the gap between the photon-emitter nucleus and the produced dijet
system on the right-hand side and the additional gap between the
Pomeron-emitter nucleus and the dijet system on the left-hand side.  The
dijet is accompanied by fewer soft hadrons than in inclusive photoproduction
where the nucleus that emits the parton breaks up.  From 
Ref.~\protect\cite{strikman06}.  Copyright 2006 by the American Physical
Society (http://link.aps.org/abstract/PRL/v96/e082001).
}
\label{4}
\end{figure}

There is a potential problem specific
to diffractive events: determining which nucleus emitted the photon and which
emitted the ``Pomeron''.  Such an event is shown schematically in Fig.~\ref{4}.
The photon source can generally be identified by comparing the invariant mass
of the entire produced system, the dijet and the accompanying soft hadrons,
the diffractive mass $M_X$, to that of the dijet alone, $M_{\rm dijet}$.  
For most events, the
diffractive mass is much larger than the dijet mass, $M_X \gg M_{\rm dijet}$, 
and the gap between the
dijet and the photon-emitting nucleus is larger than that on the
Pomeron-emitter side, making identification of the photon source possible.
In the rare cases where $M_X \sim M_{\rm dijet}$, fewer
accompanying hadrons are produced in a limited rapidity range and the gaps on
both sides of the produced system are nearly the same, making identification
of the photon source impossible.  In this case, the $x$ range is more
restricted.

\subsection{Large $t$ diffractive $\rho^0$ photoproduction with target 
dissociation}
\label{deltay}
{\it Contributed by: L. Frankfurt, M. Strikman and M. Zhalov}

\subsubsection{Introduction} \bigskip

An important feature of small $x$ processes is the nontrivial
interplay between evolution in $\ln(x_0/x)$ and $\ln(Q^2/Q_0^2)$ on the
perturbative ladder.  Large $t$  processes accompanied by a large rapidity gap 
ensure that QCD evolution is suppressed as a function of $Q^2$ at small 
coupling.  As a result, it is possible to investigate $\ln (x_0/x)$ and 
$\ln(Q^2/Q_0^2)$ evolution separately.  Such phenomena include the transition 
from color transparency to color opacity in nuclei. 
Though color transparency is experimentally established \cite{E791,Sokoloff}, 
further studies are necessary to determine the range of energies and 
virtualities at which the phenomenon occurs.  There are a number of indirect 
indications for the color opacity regime although direct evidence is limited
\cite{FS06}. The rapidity gap processes we discuss here will provide 
additional means of addressing these questions.

A number of small $x$ processes  
which originate due to elastic parton scattering with small color-singlet 
$q\overline q$ dipoles (referred to as dipoles in the remainder of this
section) 
at large momentum transfer and at high energies have been suggested including
hard diffraction in $pp\to pX$ at large $t$.  Jet studies include
two jet production accompanied by a rapidity gap, the 
`coherent Pomeron' \cite{FS89}, and enhanced 
production of back-to-back dijets separated by a large 
rapidity gap \cite{Bjorken} relative to the dijet rate in the same kinematics 
without a gap \cite{MT,DT}.  Dijet production accompanied by a gap was 
studied at the Tevatron \cite{Abe:1998ip}. 
In addition, high $t$ vector meson photo- and electroproduction with a 
rapidity gap has also been proposed \cite{AFS,FR95,Bartels}. Over the last 
decade, theoretical and experimental vector meson studies were
focused on interactions with protons.  
HERA measured the relevant cross sections 
\cite{Derrick:1995pb,Adloff:2002em,Chekanov:2002rm,Aktas:2003zi,Aktas:2006qs}
in the $\gamma p$ center-of-mass range $20 \leq W_{\gamma p}\leq 200$ GeV.
The HERA data agree well with most predictions of QCD-motivated  
models \cite{Aktas:2006qs}, several of which use the LO BFKL approximation 
\cite{BFKL}.

It would clearly be beneficial to extend these studies to 
higher $W_{\gamma p}$ and over a larger range of rapidity gap, $\Delta y$, to 
investigate the $s_{(q \overline q)j}$ and $t$ dependencies 
of dipole-parton scattering where $j$ is the interacting parton.  
Here we summarize feasibility studies \cite{FSZ06_01,FSZ06} for probing these 
processes in UPCs at the LHC.

We focus on $\rho^0$
photoproduction at large $t$ with a rapidity gap, $\Delta y$, between the 
$\rho^0$ and the hadronic system $X$ produced in ultraperipheral $pA$ and 
$AA$ collisions,
\begin{equation}
\gamma + p(A) \to \rho^0 + \Delta y + X \, \, .
\label{eqvm}
\end{equation}
We consider the kinematics where $\Delta y \geq 4$, sufficiently large to 
suppress the fragmentation contribution.  Related investigations include
diffractive charm or dijet production where the hard final state is separated 
from the nucleon fragmentation region by large $\Delta y$.
For example, studies of the $A$ dependence of dijet production in {\it e.g.} 
$\gamma A\to ({\rm jet} + M_1) + \Delta y +({\rm jet} +M_2)$ can probe color 
transparency effects on gap survival in hard-photon induced processes 
\cite{Frankfurt:1996bh}.  CMS and ATLAS are 
well suited for such observations since they cover large rapidity intervals. 

The main variables are the mass of the system produced in the proton 
dissociation, $M_X$, the square of momentum transfer $-t\equiv Q^2= 
-(p_{\gamma}-p_{V})^2$, and the square of the $q\overline q$-parton elastic 
scattering energy
\begin{equation}
s_{(q \overline q)j}=xW_{\gamma p}^2 = x s_{\gamma p} \, \, .
\end{equation} 
Here 
\begin{equation}
x=\frac {-t} {(-t+M_X^2-m_N^2)} \, \, 
\label{xmin}
\end{equation}
is the minimum fraction of the proton momentum carried by the 
elastically-scattered parton for a given $M_X$ and $t$.  
At large $t$ and $W_{\gamma p}$, the gap, $\Delta y$, between the rapidity of
the produced vector meson and the final-state parton, at the leading edge of 
the rapidity range of the hadronic system $X$, is
\begin{eqnarray} 
\Delta y = \ln \frac {xW_{\gamma p}^2} {\sqrt{(-t)(M_V^2-t)}} \, \, .
\label{gapdef}
\end{eqnarray}
It is rather difficult to measure $M_X$ or $x$ 
directly.  However, they can be adequately determined by studying
the leading hadrons close to the rapidity gap; full reconstruction
is not required.

Generally, large $t$ scattering with a rapidity gap can be described as an 
incoherent sum of terms describing elastic quark and gluon scattering.
Each term is the product of the quasi-elastic large $t$ cross section of
$p(A) j \to V j$ and the density of parton $j$ in the target 
\cite{FS89,AFS,FR95}.  Large $t$ ensures two important simplifications: the 
parton ladder mediating quasi-elastic scattering 
is attached to the  projectile  via two gluons while the attachment of the 
dipole ladder to more than one target parton is strongly suppressed.  The 
gluon elastic-scattering cross section is enhanced by 81/16 relative to 
quark scattering.  Gluon scattering dominates over a wide $x$ range, 
constituting $\sim 80$\% (70\%) of the cross section at $x=0.1$ (0.3).  The 
$t$ dependence can 
be parametrized as a power law where the power is twice the number of
constituents in the hadron vertex,
$1/t^6$ for three quarks \cite{FS89} and 
$1/t^4$ for the $q \overline q$ system \cite{FR95}.
 
The cross section for vector meson photoproduction with target dissociation in
the range $-t \gg 1/r_V^2 >  \Lambda_{\rm QCD}^2$ where $r_V$ is the vector
meson radius; $W_{\gamma p}^2\gg M_X^2\gg m_N^2$; and fixed $x$ ($x< 1$) is 
\cite{FR95}
\begin{eqnarray}
\frac {d\sigma_{\gamma p\to V X}} {dt dx}=
\frac {d\sigma_{\gamma q \to V q}} {dt} \biggl [{81\over 16} g_{p}(x,t) 
+\sum_i [q_{p}^{i}(x,t)+{\overline q}_{p}^{i}(x,t)] \biggr ] \, \, .
\label{DGLAPBFKL}
\end{eqnarray}
Here $g_{p}(x,t)$, $q_{p}^{i}(x,t)$ and $\overline q_{p}^{i}(x,t)$ are the 
gluon, quark and antiquark distributions in the proton. The 
$\gamma q \rightarrow V q$ amplitude, $f_q (s_{(q \overline q)q} ,t)$, 
is dominated by quasi-elastic scattering of the small $q \overline q$ dipole 
configuration of the photon that transitions into the final-state vector meson.

Diffractive vector meson photoproduction from hadron and nuclear targets is a 
special case where evolution in $x$ is separated from evolution in the hard 
scale, see Ref.~\cite{FS89,MT,AFS,FR95}.  Since
$t$ is the same on all rungs of the ladder mediating quasi-elastic scattering, 
the amplitude $f_{q}(s_{(q \overline q)j},Q^2=-t)$ probes evolution in 
$\ln(1/x)$ at fixed $Q^2$. Because the momentum transfer is shared between 
two gluons, the
characteristic virtuality of $t$-channel gluons on the ladder is 
$\approx Q^2/4$ while the hard scale in the target parton density is 
$\approx Q^2$.

To lowest order in $\ln(1/x)$, the amplitude, 
$f_{q}(s_{(q \overline q)j},t)$, is independent of $W_{\gamma p}$ for fixed 
$t$.  Higher order terms in $\ln (1/x)$ were incorporated
in the leading and next-to-leading log approximations, including both $\ln Q^2$
and $\ln x$ effects so that $f_q$ increases with energy as a power of 
$\exp(\Delta y)$ in Eq.~(\ref{gapdef}) with a weak $t$ dependence,
\begin{eqnarray}
f_q(s_{(q \overline q)j},t)\propto 
\biggl (\frac {s_{(q \overline q)j}} {|t|}\biggr )^{\delta(t)} \, \, 
\label{dipamp}
\end{eqnarray}
for $|t| \gg M_V^2$.
Within NLO BFKL this dependence is not obvious since the 
solution may be given by a 
different saddle point at higher $Q^2$ \cite{Lipatov,Colferai:1999em}.
The value of $\delta(0)$ changes significantly between LO and NLO BFKL,
$\delta(0) \sim 0.6$ at LO and $\delta(0) \sim 0.1$ at NLO.  The difference
between NLO and resummed BFKL is smaller since
$\delta(0)\sim 0.2 - 0.25$ in resummed BFKL over a wide range of $Q^2$
\cite{Salam:2005yp}.  Hence we treat $\delta(t)$ as a free parameter
and generally assume it weakly depends on $t$.

Similar small $t$ processes could be described by the triple 
Pomeron approximation of the amplitude,
\begin{eqnarray}
f_q \propto \bigg( \frac{W_{\gamma p}^2}{M_X^2} \bigg)^{\alpha_{\Pomeron}^{\rm 
soft}(t)-1} \, \, ,
\label{triplepom}
\end{eqnarray}
where the soft Pomeron trajectory is
\begin{eqnarray}
\alpha^{\rm soft}_{\Pomeron}(t) = \alpha_0 + \alpha^\prime t
\label{softpomslope}
\end{eqnarray}
with $\alpha_0 \sim 1.08$ and $\alpha^\prime \sim 0.25$ GeV$^{-2}$.  
The amplitude decreases with energy for $-t \ge 0.4$ GeV$^2$.

We first use a simple parametrization of the HERA data based on a hard 
reaction mechanism \cite{FSZ06_01} to estimate the 
$\gamma p \to \rho^0 + \, \Delta y +\, X$ rates in $pA$ collisions 
at the LHC.  We find that it will be possible to extend the HERA energy range 
by a factor of 10.  We then analyze the $A$ dependence of the process and
show that it provides a critical test of the interplay between hard and 
soft dynamics as well as probes the onset of the hard black disk regime. 
We find that it will be possible to study this process to $W_{\gamma p}\sim 1$ 
TeV and study hard dynamics up to $xs_{\gamma p}/Q^2 \sim 10^{5}$, 
corresponding to an rapidity interval of $\sim 12$ units for gluon emission.
Hence the emission of several gluons in the ladder kinematics (when the 
rapidity interval between two gluons on the ladder is larger than one) is
possible. 

\subsubsection{Rapidity gap processes from $ep$ at HERA to $pA$ at the LHC} 
\label{sec2gap} \bigskip

The HERA experiments report cross sections integrated over
$M_X$ from $M_X=m_N$ up to an experimentally-fixed upper limit, ${\widehat M}$.
At fixed $t$, this corresponds to
the cross section integrated over $x$ from the $x_{\rm min}$ 
determined in Eq.~(\ref{xmin}) at $M_{X}={\widehat M}$ to $x=1$.

We described these data 
using the following expression,
based on Eq.~(\ref{DGLAPBFKL}), 
\begin{equation}
{\frac {d\sigma_{\gamma p\to \rho^0 X}} {dt}}=
\frac {C} { ({1-t/t_0} )^{4}}
\biggl(\frac {s_{\gamma p}} {{M_V}^2 -t} \biggr)^{2\delta(t)}I(x_{\rm min},t)
\, \, ,
\label{basic}
\end{equation}
where $|t_0| = 1$ GeV$^2$.  The cross section, $d\sigma_{\gamma q \rightarrow
V q}/dt$ is factorized into a component accounting for the $\gamma \rightarrow
V$ transition, $C/(1-t/t_0)^4$, and the dipole-parton scattering amplitude,
$f_q$.  The amplitude has been modified to account for the virtuality of the
recoiling parton, on the order of the soft scale, $|t_0|$
The factor $I(x_{\rm min}, t)$, is obtained by integrating over the parton
densities,
\begin{equation}
I(x_{\rm min},t) =
\int \limits_{x_{\rm min}}^{1} dx \, x^{2\delta(t)}
 \left[\frac {81} {16} g_{p}(x,t)+
\sum _{i}[q_{p}^i (x,t)+{\overline q}_{p}^i (x,t)] \right] \, \, 
\label{intx}
\end{equation}
where the CTEQ6M PDFs \cite{cteq6} have been employed.
The function $\delta (t)$ is parametrized as 
$\delta (t)=\delta_{0}+{\delta}^{\prime}t$.
The values of $\delta_0$, $\delta^{\prime}$ and $C$ were adjusted to provide a 
reasonable  description of the HERA $\rho^0$ data\footnote{There is a 
relatively small rapidity interval available for gluon emission in the color 
singlet ladder, $\ln (xs_{\gamma p}/Q^2) \leq 5$, 
in the HERA data.  Since only single 
gluon emission is allowed in the ladder kinematics, it is very difficult to 
apply a BFKL-type approximation.}.  The $t$ dependence was measured by H1 and 
ZEUS for  different $M_X$ cuts over a rather narrow interval of 
$W_{\gamma p}$.  As a result, these data cannot unambiguously fix the energy 
dependence of the dipole-parton amplitude in $\delta(t)$. We 
obtain a reasonable description of the data assuming both a relatively
weak energy dependence, $\delta(t)=0.1$ ($C=40$),
and a stronger energy dependence, $\delta(t)=0.2$ ($C=14$), for hard 
processes.  These 
values of $\delta(t)$ are significantly larger than those resulting
from extrapolation 
of the soft Pomeron trajectory in Eq.~(\ref{softpomslope}) to higher $t$,
even if a nonlinear term is introduced in the trajectory 
\cite{Erhan:1999gs}.
This can be seen by equating the exponents in Eqs.~(\ref{dipamp}) and 
(\ref{triplepom}) at $-t \geq 0.4$ 
GeV$^2$, $\delta(t)=\alpha^{\rm soft}_{\Pomeron}(t)-1\approx  0.08
+ 0.25 \, t$.  Our results are consistent with a rather weak $t$ dependence
of $\delta(t)$, hence we take $\delta^{\prime}=0$.  A very small negative 
value, $\delta^{\prime}=-0.01$ GeV$^{-2}$, improves agreement with the H1 
data at $-t>5$ GeV$^2$ with $\delta_{0}=0.2$.

As mentioned previously, in the hard regime the energy dependence
of the amplitude should be a weak function of $t$.
In $\rho^0$ photoproduction with a rapidity gap, large $t$ is necessary for
the hard mechanism to dominate.  However, for exclusive quarkonium 
photo/electroproduction or light vector meson electroproduction at large 
$Q^2$, the hard mechanism is expected to dominate at $t\sim 0$. 
Hence $\delta(t)$ should be similar to the energy 
dependence of the exclusive $\gamma^* p \to V p$ amplitude. At HERA,
the highest virtualities are reached in exclusive
$J/\psi$ electroproduction and  correspond to $\delta \sim 0.2$ for $t=0$
and $\delta \sim 0.1$ for $t\sim 1$ GeV$^2$ \cite{Levy:2005ap}.  
The observation that a
similar value of $\delta$ can describe the large-$t$ rapidity-gap data 
supports the interpretation of the data as due to hard 
elastic $q \overline q$ dipole-parton scattering.

\begin{figure}
\begin{center}
\centering
 \includegraphics[totalheight=0.4\textheight,width=0.45\textwidth]{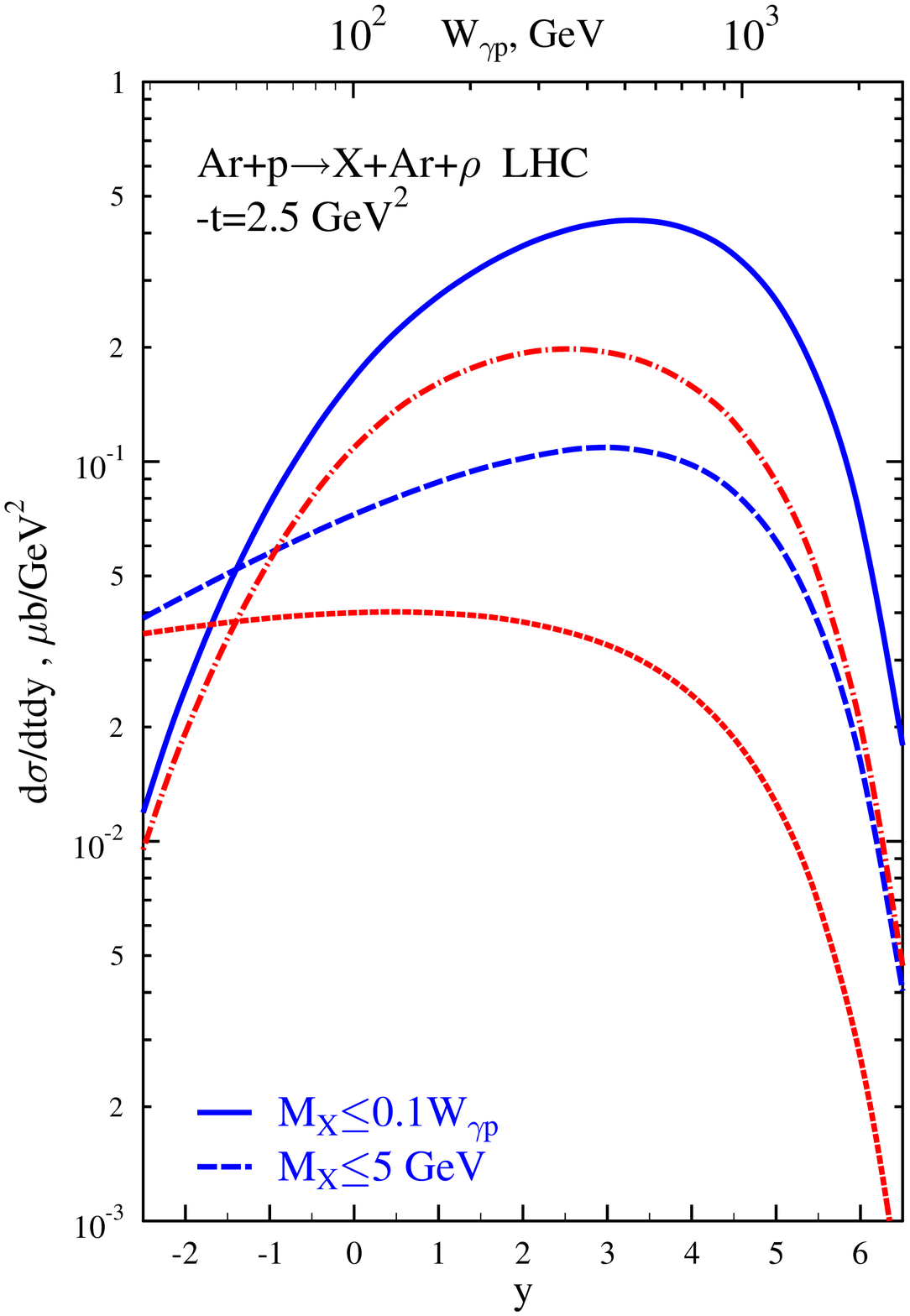}
 \includegraphics[totalheight=0.4\textheight,width=0.45\textwidth]{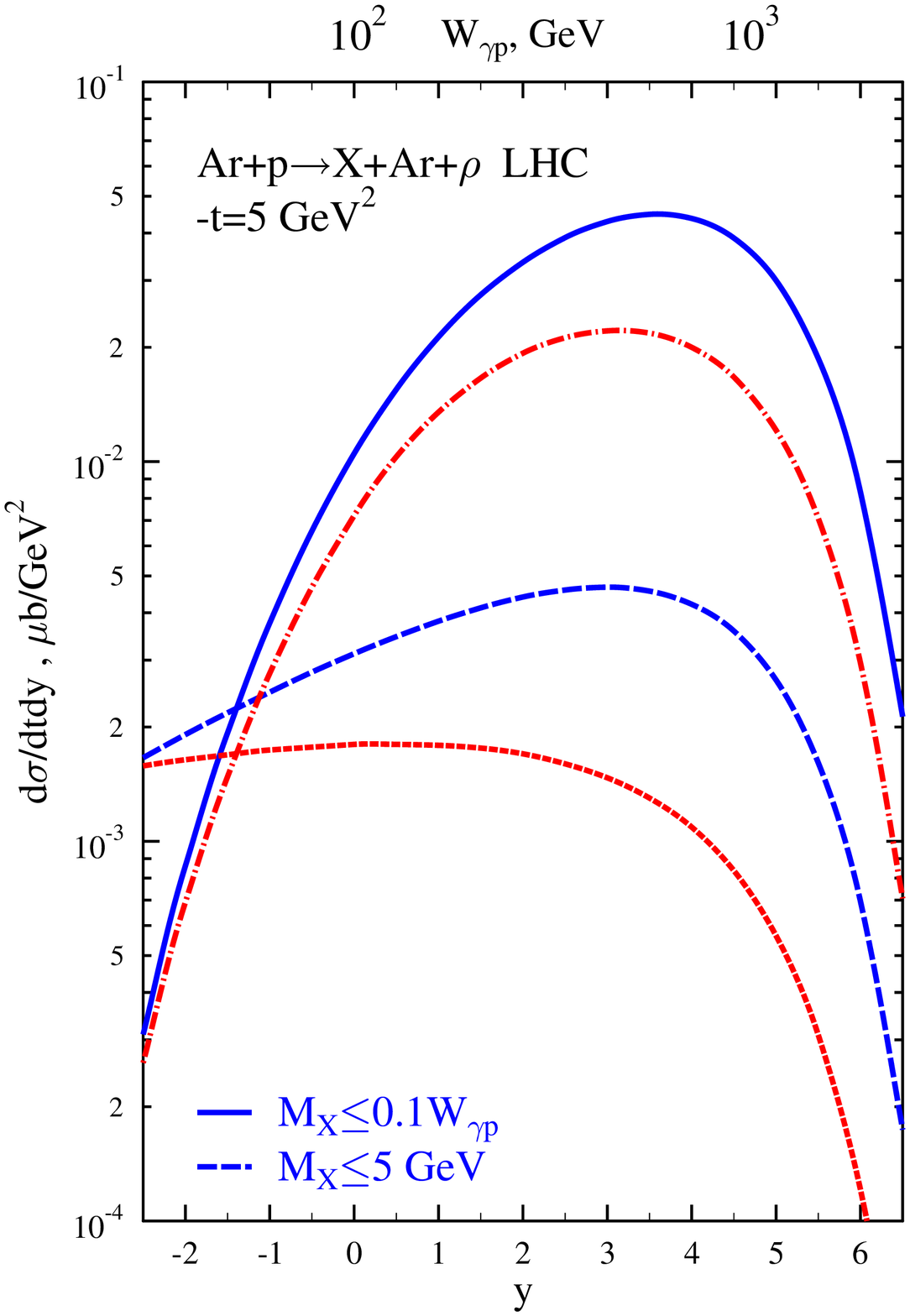}
 \caption[]{The $\rho^0$ rapidity distributions in ultraperipheral 
$p$Ar collisions at the LHC for two different $M_X$ cuts at the indicated
values of $t$ \protect\cite{FSZ06_01}.  The solid and dashed lines 
are calculations
with $\delta=0.2$ while the dot-dashed and short-dashed curves employ 
$\delta=0.1$.  
}
\label{arp}
\end{center}
\end{figure}

\begin{figure}
\begin{center}
\centering
 \includegraphics[totalheight=0.4\textheight,width=0.45\textwidth]{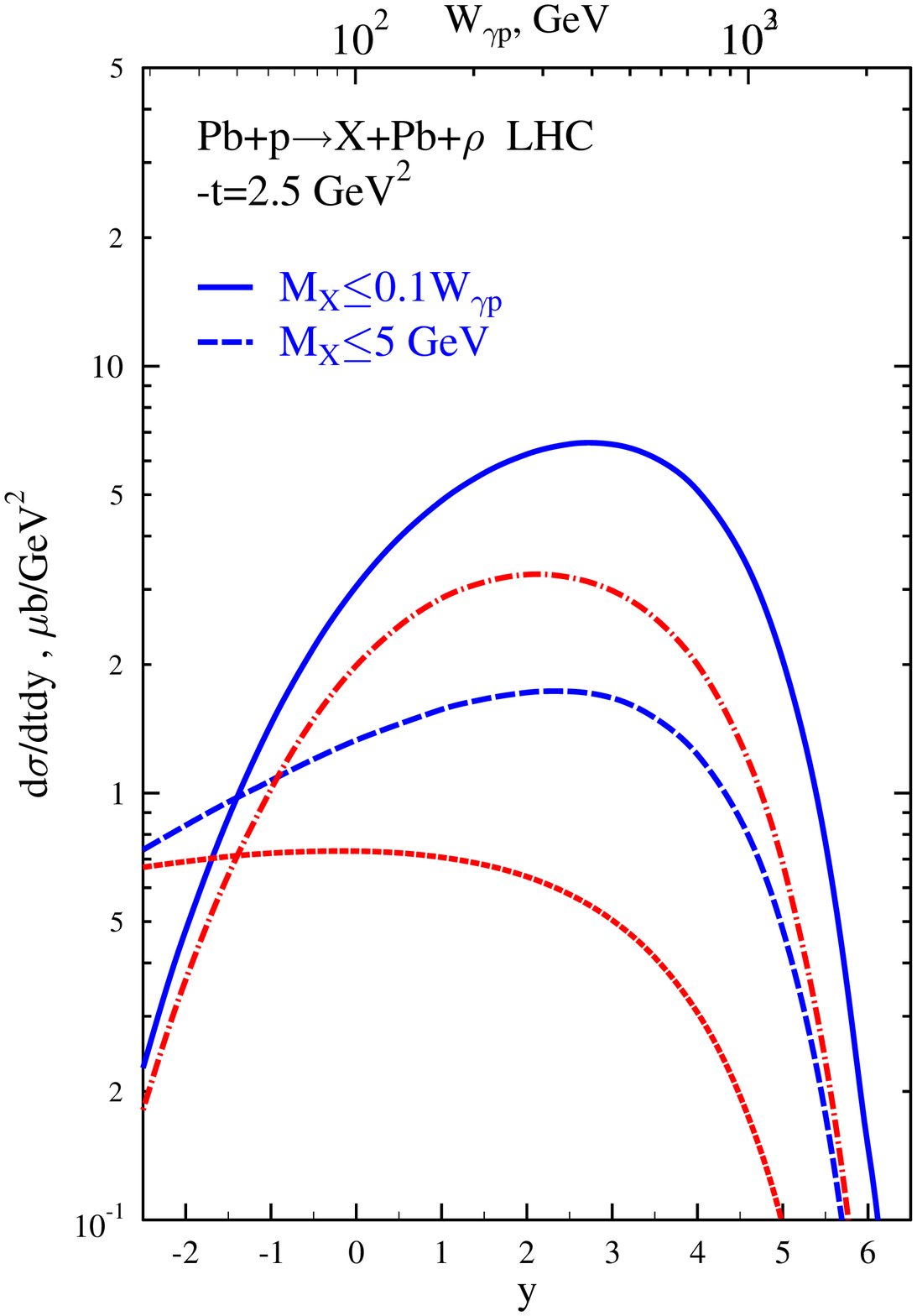}
 \includegraphics[totalheight=0.4\textheight,width=0.45\textwidth]{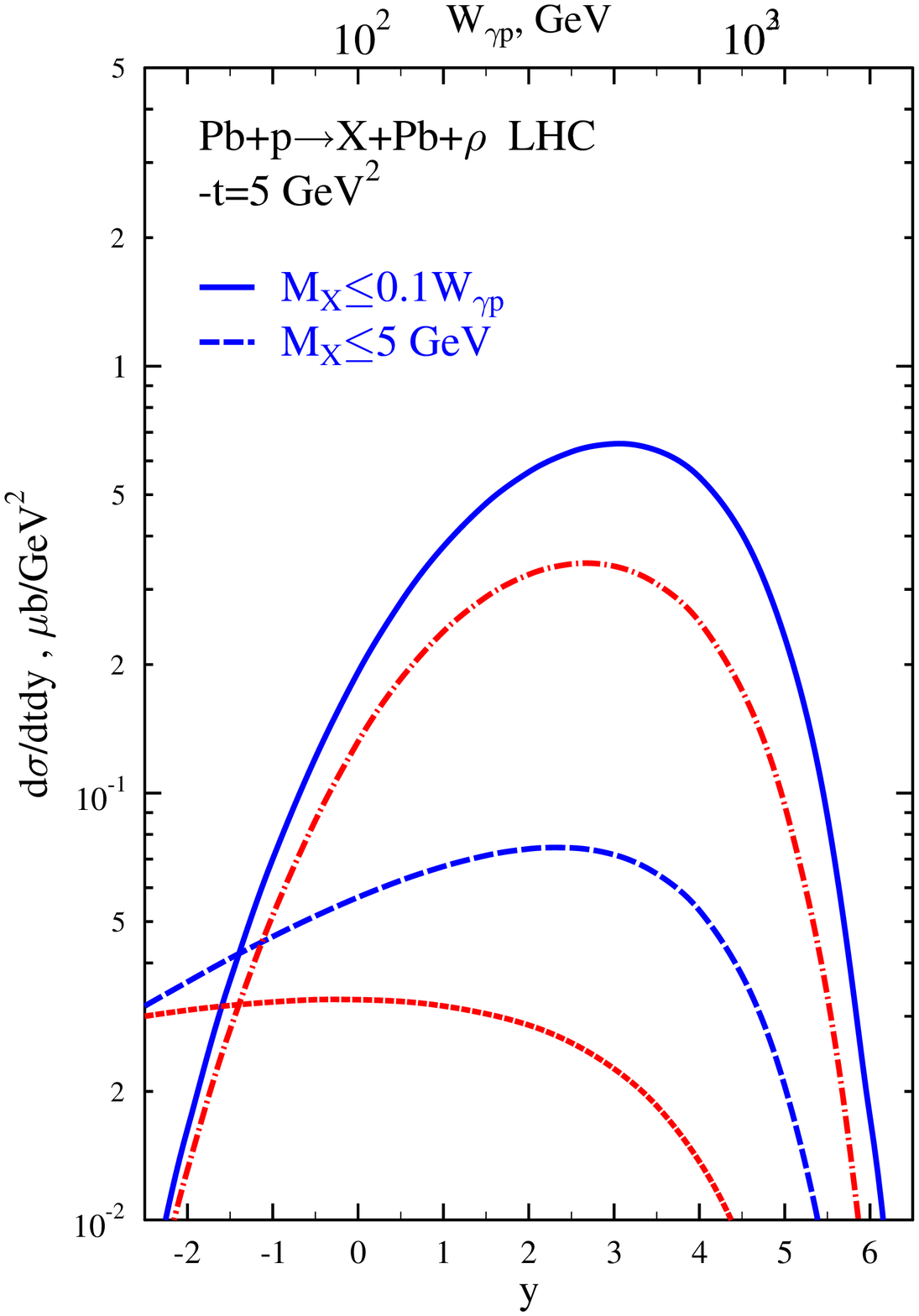}
\caption[]{The same as for Fig.~\protect\ref{arp} for $p$Pb collisions
\protect\cite{FSZ06_01}.
}
\label{pbp}
\end{center}
\end{figure}

It is unlikely  that further HERA studies will cover a sufficiently wide 
range of $W_{\gamma p}$ and $\Delta y$ to study the 
energy dependence of the large-$t$ elastic dipole-parton 
scattering amplitude. On the other hand, at the LHC, CMS and ATLAS will have 
sufficient rapidity coverage to study the process in Eq.~(\ref{eqvm}) 
in ultraperipheral $pA$ collisions. Hence
we use the parametrization of the $\gamma p\to X \rho^0$ cross section in
Eqs.~(\ref{basic}) and (\ref{intx}) to estimate the large-$t$ rapidity-gap 
$\rho^0$ cross section in ultraperipheral $pA$ and, later in $AA$
collisions at the LHC.
We do not address the $pA$ contribution from $\gamma A\to \rho^0 
X $ since it is very small
and can easily be separated experimentally.
The large-$t$ nucleon-dissociation cross section is then
\begin{equation}
{\frac {d\sigma_{pA\to \rho^0 X A}} {dt dy}} = \frac{dN_{\gamma}^{Z}(y)}{dk}
{\frac {d\sigma_{\gamma N\to \rho^0 X}(y)}{dt}}\, \, 
\end{equation}
where $dN_{\gamma}^{Z}(y)/dk$ is the photon flux generated by the ion
with energy $k = (M_{\rho^0}/2)\exp(y)$.
We consider intermediate and large 
momentum transfer in UPCs at the LHC, analogous to those
studied at HERA.

The cross section can be studied at fixed $t$ as a function of the $\rho^0$
rapidity with the restriction $M_X\leq 5 $ GeV to determine the energy 
dependence of the dipole-parton amplitude and thus $\delta(t)$. 
In this case, $x_{\rm min}$ does not depend on $W_{\gamma p}$
and the dipole-parton elastic scattering amplitude
varies with $W_{\gamma p}$ due to the increase of the rapidity gap with $y$.

We also study the cross section when $M_X \propto W_{\gamma p}$, specifically
${M}_{X} \leq 0.1\, W_{\gamma p}$. 
This cut corresponds to fixing $\Delta y$ and changing $x_{\rm min}$. 
Such studies could test the parton distribution 
functions and the reaction mechanism by extracting $I(x_{\rm min},t)$ from the
data in different $x_{\rm min}$ and $t$ bins.

We do not consider $W_{\gamma p} < 20$ GeV where
our HERA-based parametrization, Eqs.~(\ref{basic}) and (\ref{intx}), are
unreliable, particularly for $M_{X}\leq 5$ GeV.  In any case, the 
data indicate that the cross section is very small if $M_X\leq 2$ GeV.

The rapidity distribution of diffractive $\rho^0$ photoproduction accompanied
by a rapidity gap between the $\rho^0$ and 
the system $X$ produced by the target proton
break up is shown in Figs.~\ref{arp} and \ref{pbp} for $p$Ar and $p$Pb 
collisions respectively. The distributions are shown for two fixed values of
$t$: $-t = 2.5$ and 5 GeV$^2$.  We  use the same sets of cuts as those employed
in the HERA experiments.  The cut $M_X \leq 5$ GeV corresponds to a fixed
rapidity interval occupied by the hadrons in system $X$.  The energy-based
cut, $M_X \leq 0.1 W_{\gamma p}$, corresponds to the same minimum $\Delta y$ 
between the vector meson and the produced hadrons.

The choice $M_X \leq 5$ GeV gives a flatter
and broader rapidity distribution since $x_{\rm min}$ is independent of
$W_{\gamma p}$ and not very small.  When $M_X \leq 0.1 W_{\gamma p}$, smaller
values of $x_{\rm min}$ are reached for the same $-t$, giving a larger
cross section over most of the rapidity range, particularly for $-t = 5$ 
GeV$^2$.  The two choices exhibit the same behavior at large
forward rapidity due to the steep decrease of the photon flux.  
Results are also shown for two assumptions of $\delta(t)$: 0.2 and 0.1.  
The assumption $\delta(t) = 0.1$ narrows
the rapidity distribution, as does going to a higher $-t$.  The rates,
which can be obtained by multiplying the cross sections by luminosities of 6
$\mu$b$^{-1}$ for $p$Ar and $p$Pb respectively, are high.

The $t$-dependence of the cross section in Eq.~(\ref{basic}) 
should decrease more slowly than the asymptotic behavior of
the $(q \overline q) + j \rightarrow V + j$ cross section,
$\propto 1/t^4$.  As a result, the rate for $|t| > |t_{\rm min}| \ge 
2.5$~GeV$^2$ drop rather slowly with $t_{\rm min}$ (more slowly than
$1/t_{\rm min}^3$).
With the expected LHC $pA$ luminosities, the rates remain high up to
rather large $t$.  The rates for $-t > 10$ GeV$^2$ are only a 
factor of 10 smaller than for $-t> 5$ GeV$^2$.  The $J/\psi $ production rates 
would also be significant.  Although the rates are smaller than for $\rho^0$ 
production at fixed $t$, it would be possible to use $-t\ge 1$ GeV$^{2}$ in 
the analysis where the rates are larger than for the exclusive diffractive 
reaction, $\gamma p\to J/\psi p$.

Most events in these kinematics correspond to 
$x\ge 0.01$.  Thus we can primarily infer the energy dependence of the elastic 
$(q \overline q) j$ amplitude at different $Q^2$. Some events will also probe
as low as $x\sim 10^{-3}$.  However, it will be probably very difficult
to reach the $x$ range where quark scattering is larger than gluon scattering,
$x\ge 0.4$.  Overall, the energy range, $s_{\rm max}/s_{\rm min} \ge 
4\times 10^3$, is large enough for precision 
measurements of the energy dependence of the amplitude.
If $\delta(t) \approx 0.2$, the elastic cross section should increase by a 
factor of $\sim 30$ in the energy range.
 
\subsubsection{$A$ dependence of rapidity gap production in $AA$ collisions}
 \label{sec3gap} \bigskip
 
Since large $t$ rapidity gap processes,  $\gamma(\gamma^*) N \to V X$, 
are dominated by elastic $q\overline q$-parton scattering, these processes
provide a novel way to investigate small dipole interactions in the nuclear 
medium.
 
Ultraperipheral $AA$ collisions at the LHC will provide the first opportunity 
to investigate the new QCD regime of strong interactions at small coupling as
well as large target thickness.  Further studies will be possible at a future
$eA$ collider.  Ultraperipheral $AA$ collisions differ 
from $pA$ collisions since vector mesons can be produced by photons 
emitted from either nucleus. The cross section is the sum of the two 
contributions,
\begin{equation}
{\frac {d\sigma_{AA\to \rho^0 X AA^{\prime}}} {dydt}}= 
\frac{dN_{\gamma}^{Z}(y)}{dk} {\frac {d\sigma_{\gamma A\to \rho^0
X A^{\prime}}(y)}{dt}} + \frac{N_{\gamma}^{Z}(-y)}{dk}
{\frac {d\sigma_{\gamma  A\to \rho^0  X A^{\prime}}(-y)} {dt}} \, \, .
\end{equation}
Here $\sigma_{\gamma A\to \rho^0 X A^{\prime}}$ is the $\rho^0$ photoproduction
cross section with dissociation, the system $X$ results from 
diffractive dissociation of a
nucleon and $A^{\prime}$ is the residual nucleus. Several neutrons will be 
produced in the electromagnetic excitation of $A'$ 
by the photon-emitting nucleus, $A$, in Eq.~(\ref{eqvm}).

The system $X$ should be similar to that produced in nuclear DIS 
at similar $x$ and $Q^2\sim -t$ except that here the system can
be produced by both quark and gluon scattering.
The hadron spectrum is obtained from quark and gluon fragmentation in the
proportion of parton production
given by Eq.~(\ref{DGLAPBFKL}).  These hadrons should balance the 
vector meson transverse momentum.  The leading hadron momenta in the nuclear 
rest frame are $\sim -t/(2m_N x)$.
Hence, based on EMC measurements \cite{EMC}, we expect that, at large $t$ and 
$x \leq  0.05$, leading hadron absorption is small.
Nevertheless, a few neutrons will be produced in the nuclear
fragmentation region by final-state hadronic interactions 
\cite{Strikman:1998cc}. 
Therefore, either one or both ZDCs will detect several neutrons. 
Detecting the hadrons in $X$ can determine which nucleus emitted the photon, 
leading to the determination of the invariant $\gamma A$ energy. 

Studies of the $A$ dependence of 
$\gamma A\to \rho^0 X A^{\prime}$ at large $t$ can reveal the dynamics of 
the $(q \overline q) A$ interaction.  
Before discussing the predicted $A$ dependence in these kinematics,
we estimate the $A$ dependence at small $t$. 
At high energies, the photon is in an average configuration 
which interacts inelastically with a strength comparable to that of the
$\rho^0$. In this case, fluctuations in the interaction strength 
are rather small and the photoproduction cross section
can be calculated in the Gribov-Glauber approximation
for high-energy incoherent processes,
\begin{equation}
{\frac {d\sigma_{\gamma A\to \rho^0 X A}} {dt}}=
A_{\rm eff}{\frac {d\sigma_{\gamma p\to \rho^0 X}} {dt}} \, \, .
\label{iabr}
\end{equation}
The effective number of nucleons, $A_{\rm eff}$, determines the 
rapidity gap survival probability,
\begin{equation}
{\frac {A_{\rm eff}} {A}}= 
{1\over A} \int d^2b  \, T_A(b) \exp[-\sigma_{\rm in}^{\rho^0 N} T_A(b)] \,\, .
\label{aeffgapav}
\end{equation}
In the high energy regime, the growth of $\sigma_{\rm in}^{\rho^0 N}$ 
is significant.  Thus the suppression 
becomes quite large,  $A_{\rm eff}/A\sim A^{-{2\over 3}}$,
emphasizing the peripheral nature of the process.
 
At large $t$, the dominant component of the photon wavefunction responsible 
for vector meson photoproduction with nucleon dissociation is a $q \overline q$
dipole characterized by size $d\propto 1/\sqrt{|t|}$.  Leading and higher-twist
nuclear shadowing should decrease with $t$ due to color transparency. 
The contribution of planar (eikonal/Glauber rescattering) 
diagrams to the high-energy amplitude is canceled in a quantum field
theory \cite{Mandelstam63,Gribovcom}.  This result has recently been 
generalized to pQCD for the interaction of a small dipole with a large color
singlet dipole by $gg$ ladder exchanges: either of two color octet ladders 
\cite{BLV} or of multiple color singlet ladders \cite{Blok:2006ns}.
The primary distinction between a quantum-mechanical description of scattering
and a quantum field theory like QCD is that a field theory allows fluctuations
in the number of constituents in a given dipole configuration, all of which
can scatter in the target \cite{annual,Blok:2006ns}, while quantum
mechanics involves the interaction of systems with fixed number of 
constituents.  Each
constituent in a particular configuration can interact only once with a target
parton through a $t$ channel amplitude with vacuum quantum numbers.  Multiple
scattering thus arises when the interaction partners are viewed as collections
of partons, leading to a Gribov-Glauber type picture with causality and
energy-momentum conservation.

In the case of dipole-nucleus scattering, the first rescattering is given by 
the pQCD cross section for the interaction of the $q \overline q$ dipole of
transverse size $d$.  At leading order, the cross section can be written as
\cite{sigma,Frankfurt:1993it,Radyush},
\begin{equation}
\sigma_{\rm in}^{(q\overline q)N} (\tilde x, d^2) = 
\frac{\pi^2}{4}  C_F  d^2  \alpha_s (Q^2_{\rm eff}) 
\tilde x g (\tilde x, Q^2_{\rm eff}) \, \, .
\label{sigma_d_DGLAP2}
\end{equation}
where, similar to Eq.~(\ref{sigma_d_DGLAP}), $C_F = 4/3$, 
$d$ is the transverse size of the dipole, $Q^2_{\rm eff}
\propto 1/d^2$ 
is the effective dipole virtuality, $\tilde x=Q^2_{\rm eff}/W_{\gamma 
p}^2$ and $g(\tilde x, Q^2_{\rm eff})$ is the inclusive gluon density of the
target.  Since the dipole size scales as $1/\sqrt{|t|}$, at sufficiently large
$t$ and fixed $W_{\gamma p}$, 
$\sigma_{\rm in}^{(q\overline q)N}$ becomes small 
enough for interactions with more than three nucleons to be negligible. 
The rapidity gap survival probability then simplifies to
\begin{equation}
\frac{A_{\rm eff}}{A} = 1 - {\sigma_{\rm in}^{(q\overline q)N} \over A} 
\int d^2b \, T_A^2(b) \, \, .
\label{first_order}
\end{equation}
At fixed $t$, $\sigma_{\rm in}^{(q \overline q)N}$ increases with 
$W_{\gamma p}$ due to the growth
of the small $x$ gluon density, $\tilde x g_T(\tilde x, Q_{\rm eff}^2)
\propto ({W_{\gamma p}^2/Q_{\rm 
eff}^2})^{n}, n\ge 0.2$.  At large $W_{\gamma p}$, Eq.~(\ref{first_order})
breaks down and higher-order rescatterings involving the interaction of more 
than three nucleons with configurations containing three 
or more partons ($q\overline q g$ or higher) must be taken into account.  
The cross sections for such configurations should be larger than 
$\sigma_{\rm in}^{(q\overline q)N}$ in Eq.~(\ref{sigma_d_DGLAP2})
because the projectile has a non-negligible probability to consist
of several dipoles with sizes comparable to the initial dipole.  Therefore, 
in the following, we refer instead to an effective cross section,
$\sigma_{\rm eff}$, a parameter to model
the average dipole-nucleon interaction strength.  Although the eikonal-type
expansion in the number of rescatterings,
based on the average interaction strength, will somewhat overestimate
the absorption, it is still reasonable to use the eikonal approximation to 
estimate the suppression.  

\begin{figure}
\begin{center}
\epsfig{file=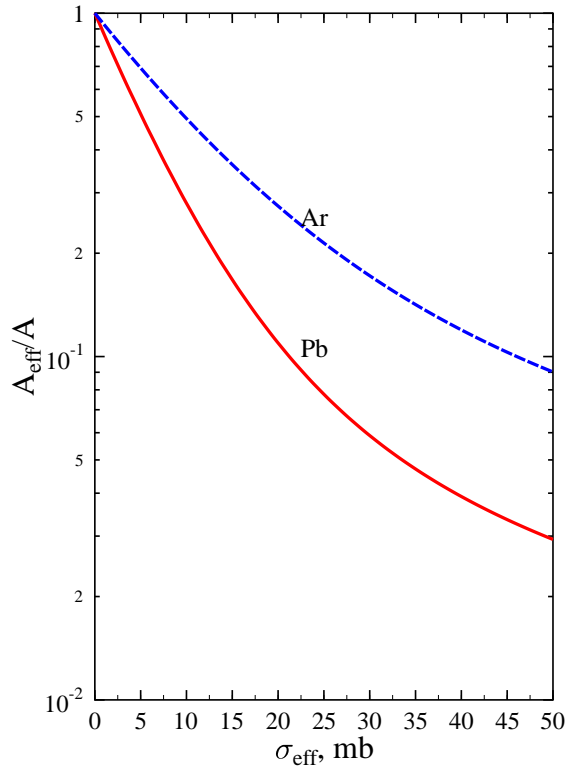, height=4in}
 \caption{The rapidity gap survival probability as a function of 
$\sigma_{\rm eff}$ \protect\cite{FSZ06_01}.}
 \label{gapsup}
\end{center}
\end{figure}

Figure~\ref{gapsup} shows $A_{\rm eff}/A$, calculated using 
Eq.~(\ref{aeffgapav}), as a 
function of $\sigma_{\rm eff}$.  The accuracy of the
calculated $A_{\rm eff}/A$ should increase both in the limit of small 
$\sigma_{\rm eff}$ where more than two scatterings is a small correction,
$\sigma_{\rm eff} \leq 3$ mb for $A\sim 200$, color transparency, and large 
$\sigma_{\rm eff}$, close to the color opacity or black disk regime.

Increasing $t$ at fixed $W_{\gamma p}$ leads to $A_{\rm eff}/A\to 1$, 
the onset of color transparency. When $W_{\gamma p} \sim 100$ GeV, a typical 
energy for UPCs at the LHC and the upper range of HERA energies, a $d=0.2$ fm 
dipole results in $\sigma_{\rm eff} \approx 5$ mb.
However, $A_{\rm eff}/A$ is considerably less than unity even for such a
relatively small value of $\sigma_{\rm eff}$, see Fig.~\ref{gapsup}. 
At these values of $\sigma_{\rm eff}$, the difference
between Eqs.~(\ref{aeffgapav}) and Eq.~(\ref{first_order}) is substantial.
since with $\sigma_{\rm eff} \approx 5$ mb and $A=200$, $A_{\rm eff}/A$ 
calculated with Eq.~(\ref{first_order}) is a factor of 1.6 
smaller than that of Eq.~(\ref{aeffgapav}).  The difference increases
with $\sigma_{\rm eff}$.  Hence either larger $t$ or
smaller $W_{\gamma p}$ is needed for complete color transparency as
described in Eq.~(\ref{ct}).

Thus increasing $W_{\gamma p}$ at fixed $t$ is expected to lead to the onset 
of the BDR for dipole interactions with the 
nuclear medium.  Vector meson photoproduction would then be strongly 
suppressed at central impact parameters so that the peripheral process
dominates with a cross section proportional to
$ A^{1/3}$. The suppression of the $\rho^0$ yield would then be comparable to 
the soft regime estimated employing Eq.~(\ref{aeffgapav}).
 
Higher-twist effects in these kinematics would also be manifested 
in the structure of the final state.  Since the higher-twist mechanism is 
more peripheral, a large suppression in the nuclear medium
would be combined with the emission of fewer neutrons.  The suppression could 
be determined by neutron multiplicity studies in the ZDC. 

In the leading-twist approximation,
the cross section is given by Eq.~(\ref{DGLAPBFKL}) where the nucleon parton 
distributions are replaced by the nuclear parton 
densities $g_A$, $q_A$ and $\overline{q}_A$,
 \begin{equation}
{\frac{d\sigma_{\gamma A\to V X A^{\prime}}} {dt dx}} =
\frac {d\sigma_{\gamma q\to V q}} {dt}
\bigg[ {81\over 16} xg_A(x,t) +\sum_i (xq_{A}(x,t)+x{\overline q_A}(x,t)) 
\bigg] \,\, .
\label{ct}
\end{equation}
The quark distributions 
do not deviate more than 10\% from a linear $A$ dependence for $0.05< x< 0.5$.
Current models of the nuclear gluon density, which dominates Eq.~(\ref{ct}),  
predict an enhancement of up to 20\%  for  $x\sim 0.1$ with perhaps some 
suppression at $x\ge 0.4$.  Hence the leading-twist approximation, 
Eq.~(\ref{ct}), predicts the onset of color transparency 
with increasing $t$, characterized by strong suppression of the dipole
interaction with the nuclear medium. 
The upper limit of
the photoproduction cross section in the impulse
approximation is
\begin{equation}
{\frac {d\sigma_{\gamma A\to \rho^0 X A}} {dt}}=
A{\frac {d\sigma_{\gamma p\to \rho^0 X}} {dt}} \, \, .
\label{ia}
\end{equation}

The reasonable agreement of the predicted behavior with the major features 
of large-$t$ rapidity-gap processes at HERA in the kinematics corresponding
to dipole-parton scattering at
$x\ge 0.05$ suggests that it is possible to trigger on high-energy
small $q \overline q$ dipole scattering without requiring small $x$. 
If the kinematics where $M_X$ corresponds to $x\leq 0.01$ could be reached, 
where leading-twist gluon shadowing is important \cite{Frankfurt:2003zd},
a further decrease of $A_{\rm eff}/A$ is possible. On the other hand,
elastic quarkonium photo/electroproduction is naturally at
small $x$. Thus $\rho^0$ production with nucleon dissociation provides a 
complementary, clean way to study interactions with the nuclear 
medium.  Hence, when $x \ll 10^{-2}$,
both leading and higher-twist effects in the dipole-parton and dipole-nucleus 
interactions are addressed.

Numerical estimates were made for
two scenarios at the LHC: the impulse approximation (IA) in Eq.~(\ref{ia}) 
where the cross section is proportional to $A$ and strong screening due to 
Glauber-Gribov multiple scattering (GA), implemented using 
Eq.~(\ref{aeffgapav}).  The GA result gives a lower limit on the rate 
while the IA is an upper limit.
We assume that absorption cross section for a small dipole should not be 
larger than the cross section for a hadron with the same valence quarks.  Thus 
$\sigma_{\rm in}^{\rho^0 N}$ in Eq.~(\ref{aeffgapav}) is based on an elastic 
$\rho^0 p$ scattering fit \cite{LD} and the vector dominance model.

\begin{figure}[t]
\begin{center}
\centering
 \includegraphics[totalheight=0.4\textheight,width=0.45\textwidth]{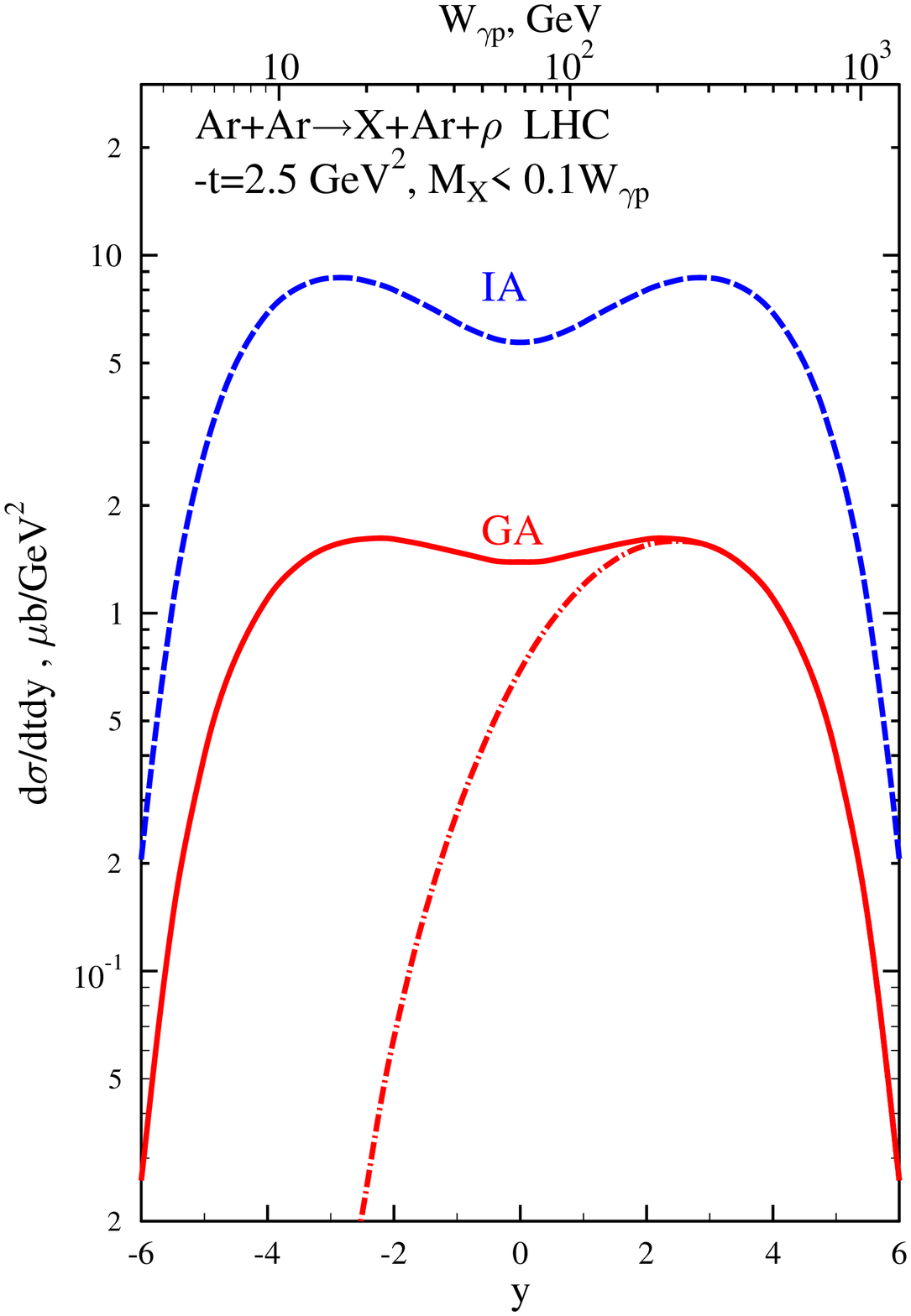}
 \includegraphics[totalheight=0.4\textheight,width=0.45\textwidth]{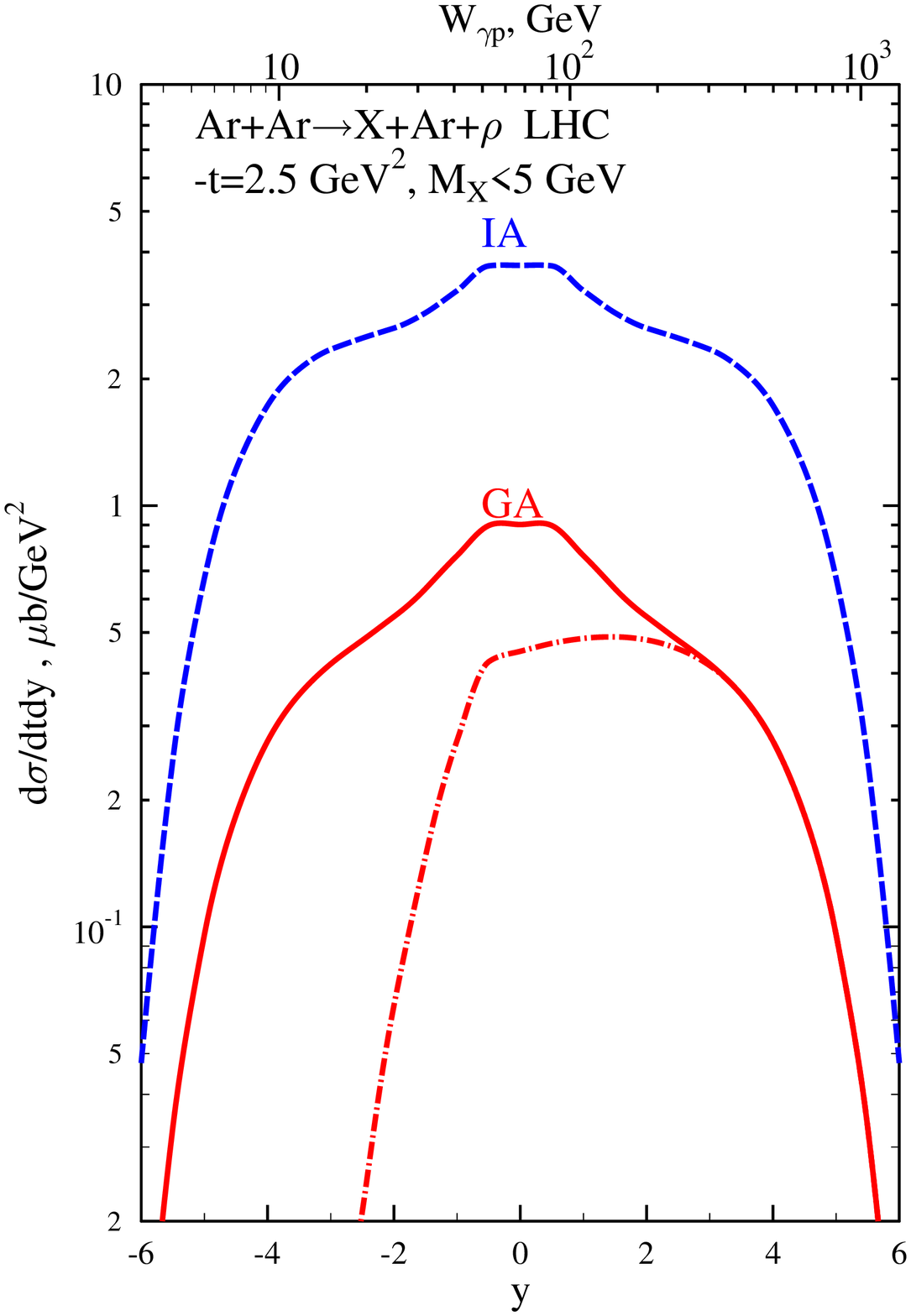}
\caption{The rapidity distribution of $\rho^0$ production with nucleon 
dissociation in Ar+Ar collisions at $-t=2.5$ GeV$^2$ \protect\cite{FSZ06_01}. 
The left-hand  
figure takes $M_{X}\leq 0.1W_{\gamma p}$ while the upper limit in the 
right-hand figure is fixed by restriction $M_{X}\leq 5$ GeV.  The dashed
curves are the impulse approximation while the solid curves include 
Glauber-Gribov screening, neglecting the small nuclear shadowing
correction.  The lower dashed curves show the contribution from a single
nucleus only.  
}
\label{art2i5}
\end{center}
\end{figure}

\begin{figure}[t]
\begin{center}
\centering
 \includegraphics[totalheight=0.4\textheight,width=0.45\textwidth]{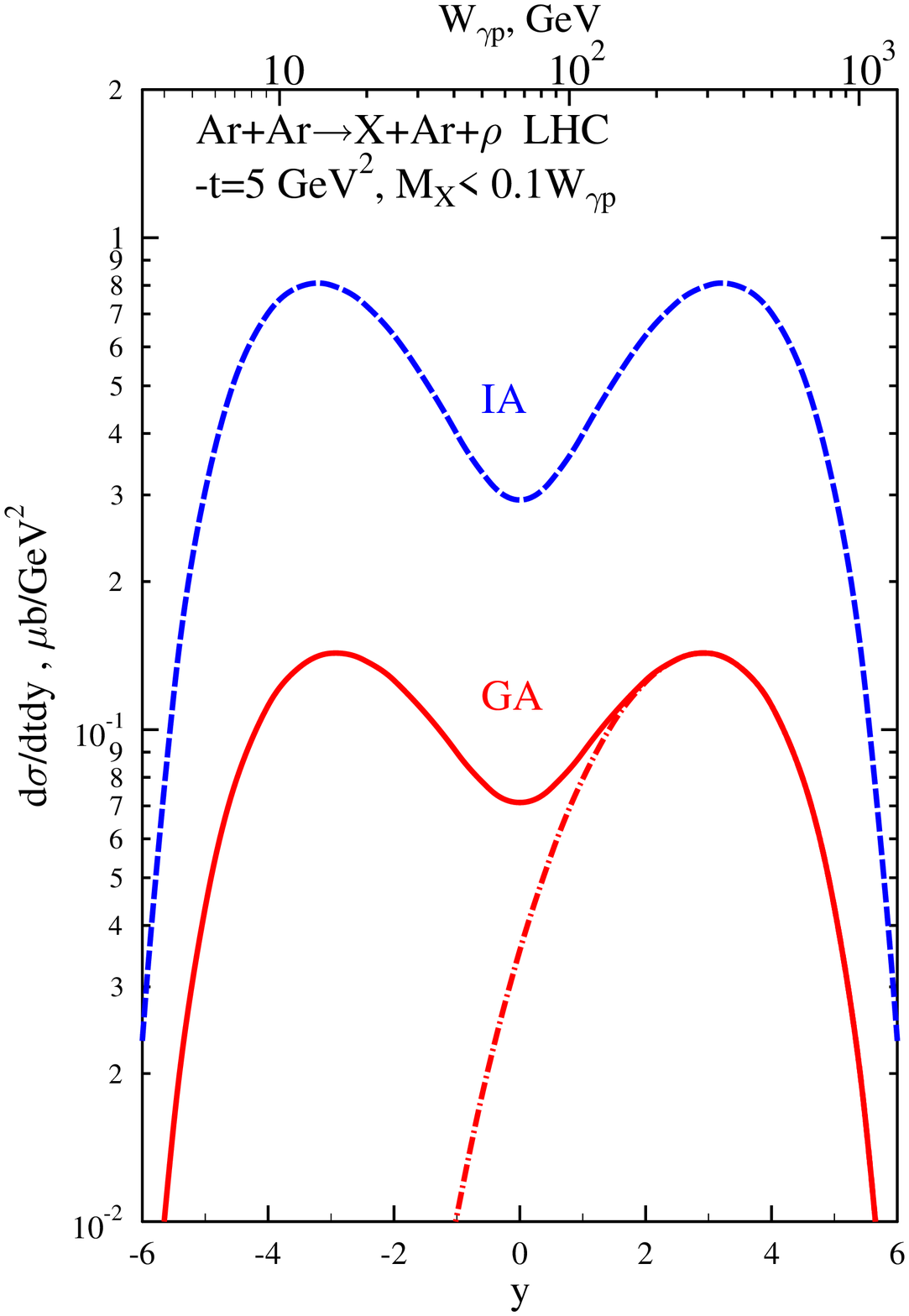}
 \includegraphics[totalheight=0.4\textheight,width=0.45\textwidth]{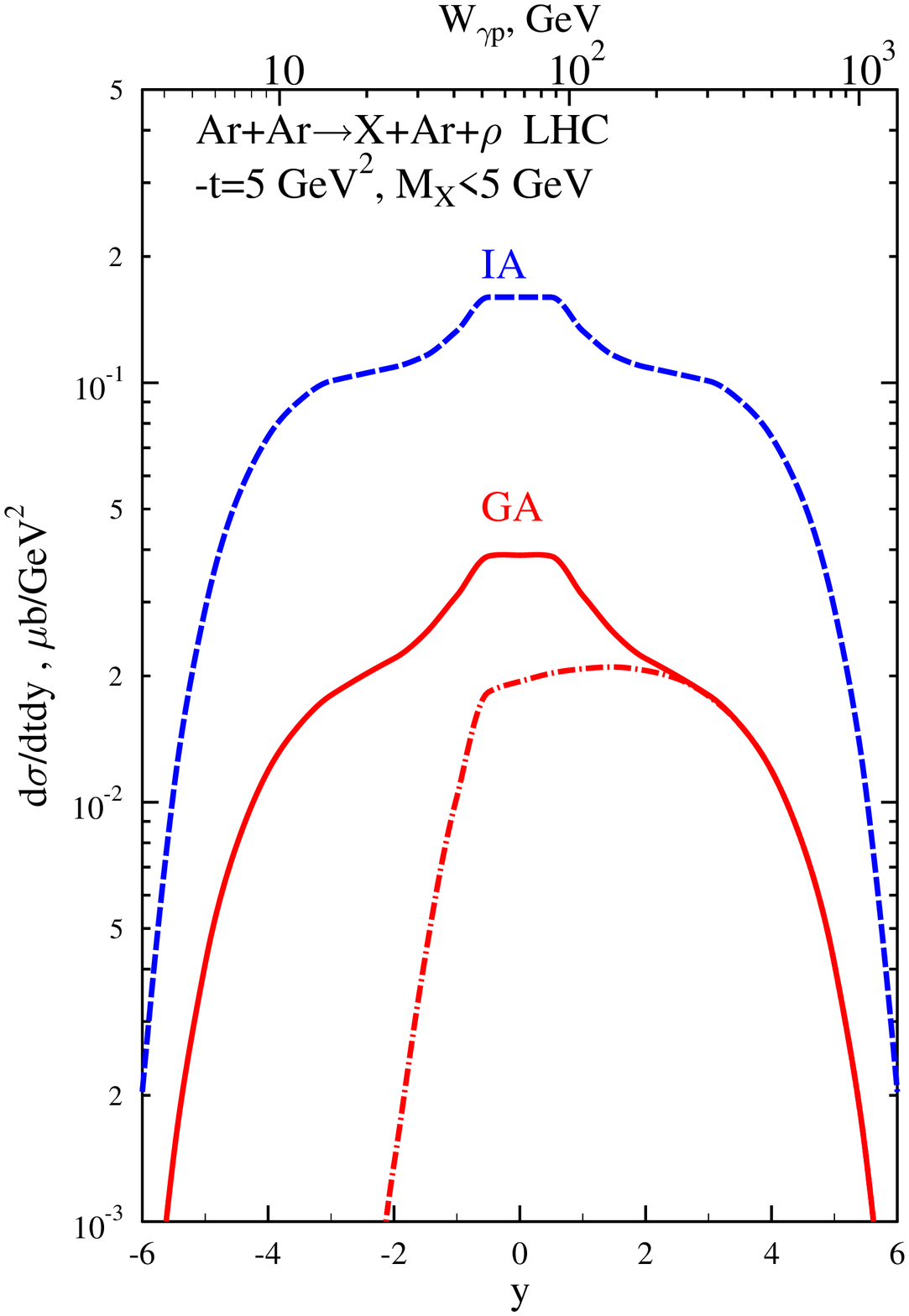}
 \caption{The same as Fig.~\protect\ref{art2i5} 
at $-t=5$ GeV$^2$ \protect\cite{FSZ06_01}.}
 \label{art5}
\end{center}
\end{figure}

Since the photon
that produces the $\rho^0$ can come from either nucleus, the $M_X$ cuts
described for $pA$ must be modified.  We again use the cut 
$M_X < 0.1 W_{\gamma p}$ but changes are needed for a fixed upper 
limit on $M_X$.  In $pA$ interactions, our parametrization is reasonable
for both cuts as long as $W_{\gamma p} > 30$ GeV. In lower energy $AA$ 
collisions, a large scattering energy in one nucleus corresponds to low
energy in the second nucleus.  The region $W_{\gamma p} < 20$ 
GeV is then reached and the fit becomes inapplicable for $M_X < 5$ GeV.  Thus, 
instead of a fixed upper limit of $M_X \leq 5$ GeV for all $W_{\gamma p}$, at
$W_{\gamma p} < 50$ GeV, we change from the fixed upper limit to a $W_{\gamma
p}$-dependent cut, $M_X \leq 0.1 W_{\gamma p}$.

\begin{figure}[t]
\begin{center}
\centering
 \includegraphics[totalheight=0.4\textheight,width=0.45\textwidth]{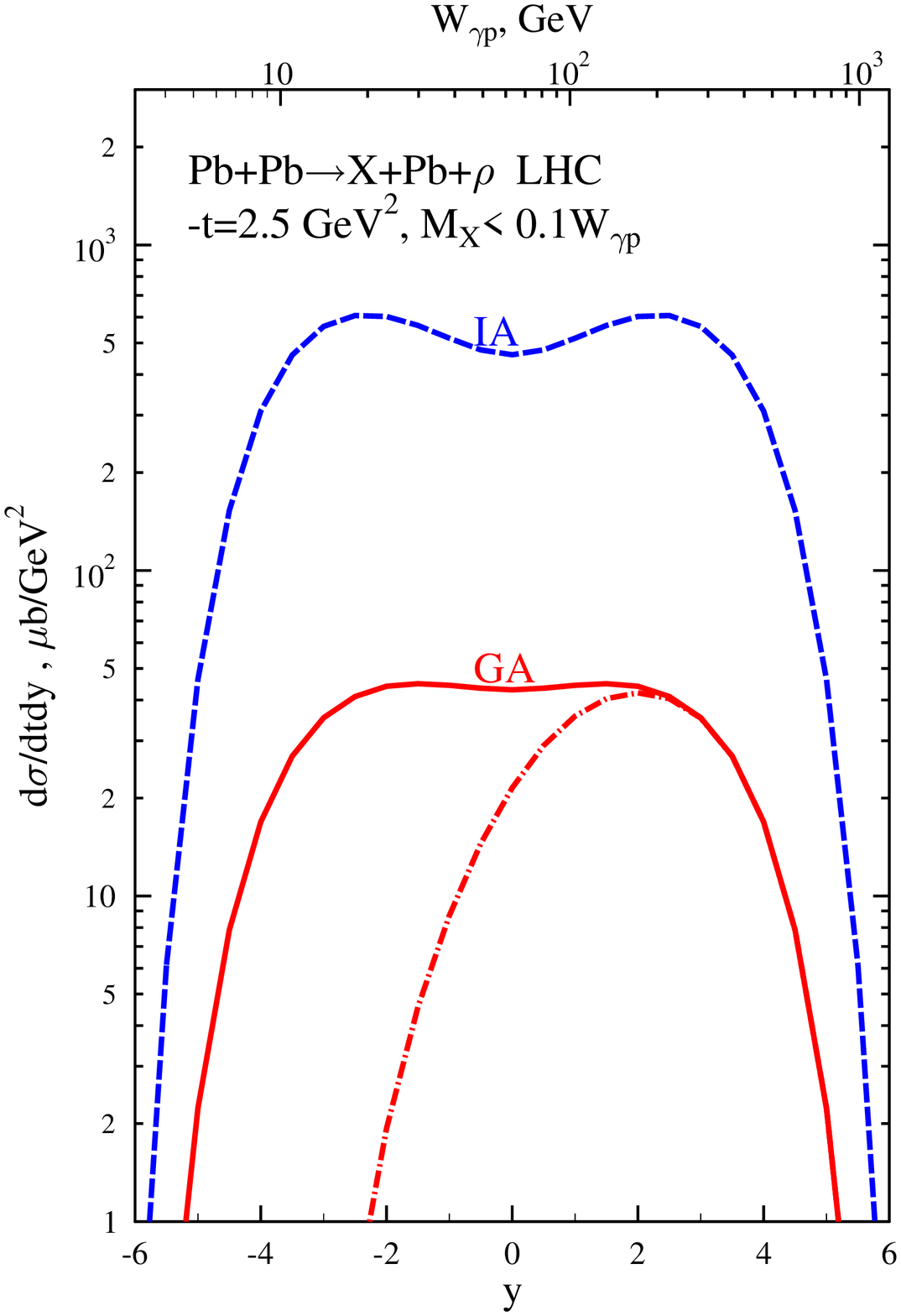}
 \includegraphics[totalheight=0.4\textheight,width=0.45\textwidth]{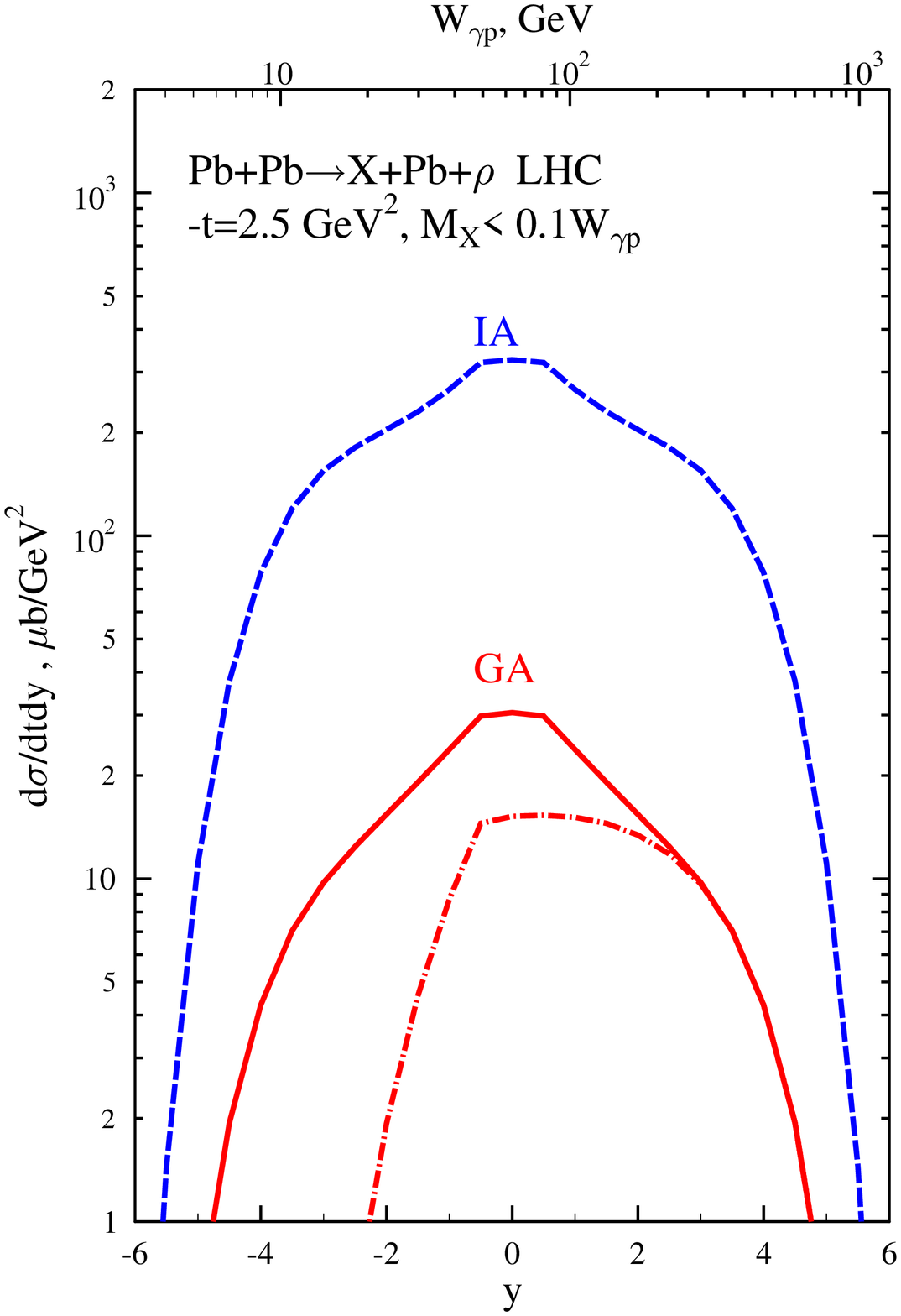}
 \caption{
The same as Fig.~\protect\ref{art2i5} for Pb+Pb collisions.  The rates can 
be estimated using the expected Pb+Pb luminosity, $L_{\rm Pb Pb}
=10^{-3}$ $\mu$b$^{-1}$ s$^{-1}$. \protect\cite{FSZ06_01}
}
 \label{pbt2i5}
\end{center}
\end{figure}

\begin{figure}[t]
\begin{center}
\centering
 \includegraphics[totalheight=0.4\textheight,width=0.45\textwidth]{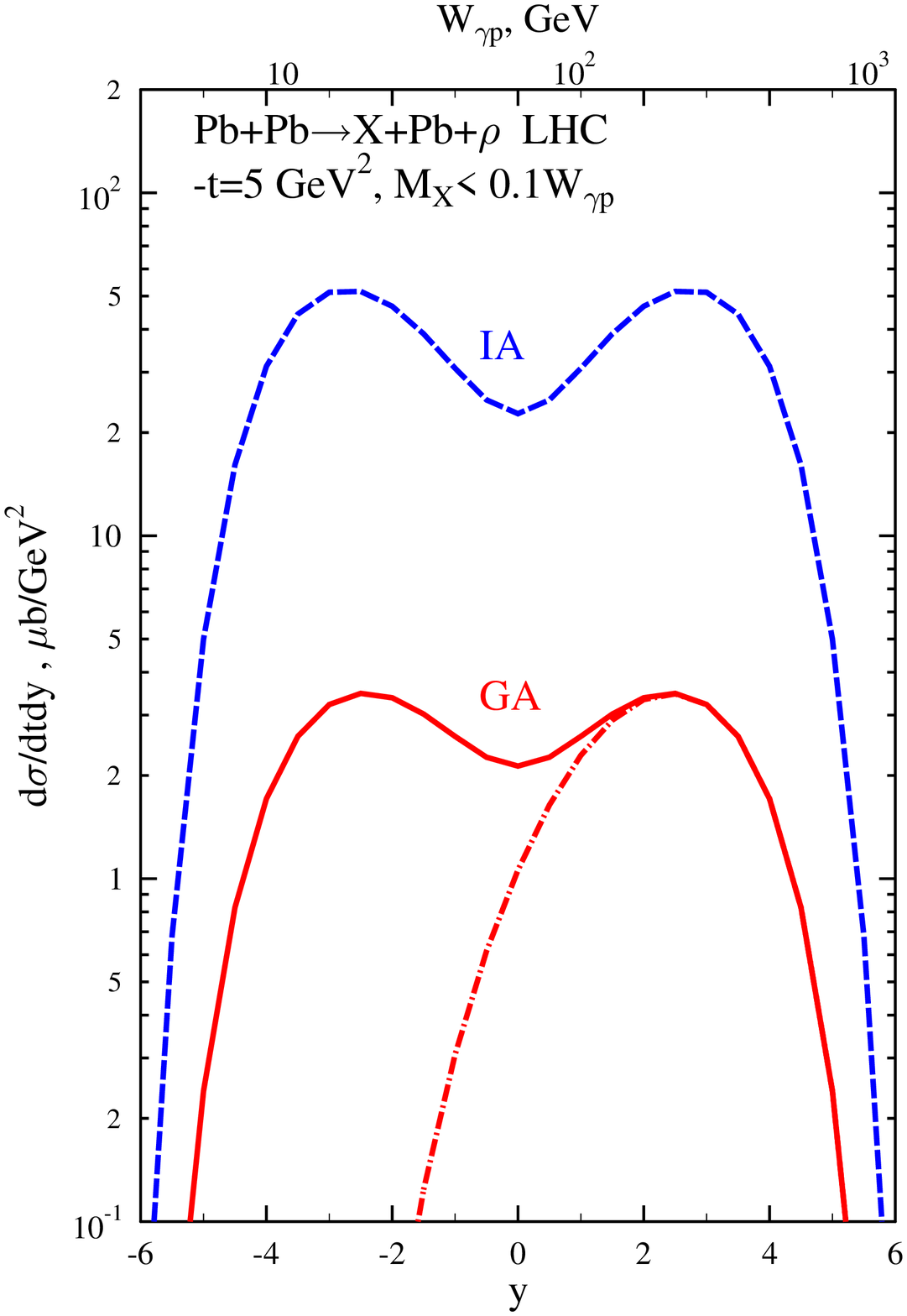}
 \includegraphics[totalheight=0.4\textheight,width=0.45\textwidth]{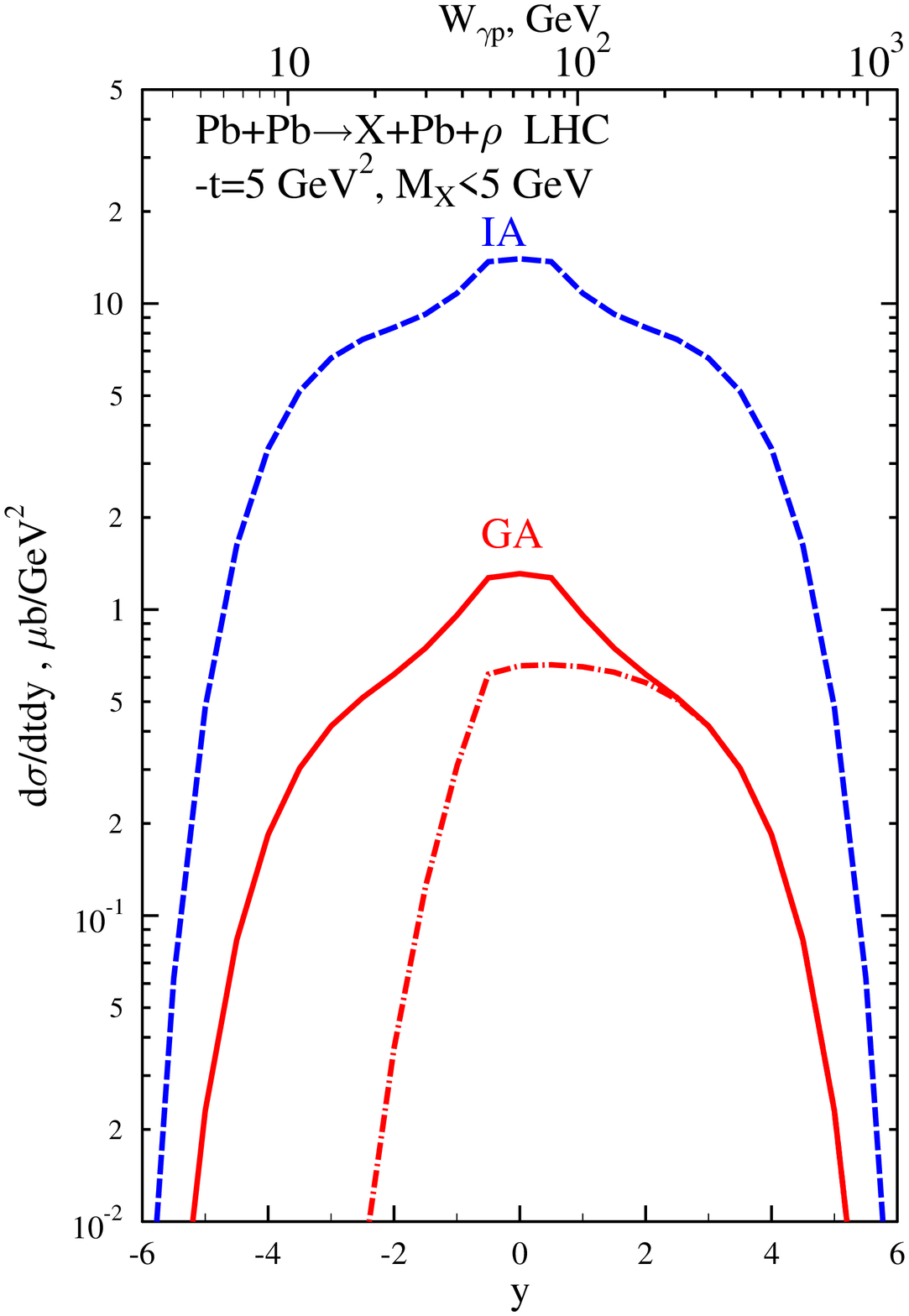}
 \caption{The same as in Fig.~\protect\ref{pbt2i5} at $-t=5$ GeV$^2$ 
\protect\cite{FSZ06_01}.}
 \label{pbt5}
\end{center}
\end{figure}

The $\rho^0$ rapidity distributions with nuclear breakup for Ar+Ar and 
Pb+Pb collisions are shown in Figs.~\ref{art2i5}-\ref{pbt5}.  
Figures~\ref{art2i5} and \ref{pbt2i5} show results for
$-t = 2.5$ GeV$^2$ while Figs.~\ref{art5} and \ref{pbt5} are for 
$-t = 5$ GeV$^2$.  The two $M_X$
cuts are shown for each value of $-t$ with the energy dependent cut, reaching
lower $x$, in the right part of each figure.  Recall that the two cuts become 
equivalent for $W_{\gamma p} \leq 50$ GeV.  Each figure shows two $AA$ curves
for each cut.  The upper limit on the cross section, obtained in the impulse 
approximation, see Eq.~(\ref{ia}), is shown in the dashed curves.  The results
obtained with Glauber-Gribov screening employing $\sigma_{\rm eff} = 
\sigma_{\rm in}^{\rho^0 N}(W_{\gamma N})$,
an effective lower limit, are shown in the solid curves.  Recall, however, that
the survival probability for the rapidity gap, shown in Fig.~\ref{gapsup}, 
is a strong
function of $\sigma_{\rm eff}$ and is thus sensitive to higher-twist effects.

The curves corresponding to a single nuclear target with the same energy and 
$\sigma_{\rm eff}$ are shown in the dot-dashed curves for one side of the 
collision.  These single-side curves are not exactly equivalent to
the $pA$ curves in Figs.~\ref{arp} and \ref{pbp} since the $AA$ energy 
is lower than the $pA$
energy, narrowing the rapidity distributions.  The behavior of the
single side distribution near midrapidity explains the shape of the $AA$
results.  The smooth decrease of the single-side result for $M_X < 0.1
W_{\gamma p}$ at $y<0$ leads to an $AA$ result that is either flat at
midrapidity ($-t = 2.5$ GeV$^2$) or has a dip in the middle ($-t = 5$ GeV$^2$).
On the other hand, the flatter single side behavior with the fixed upper limit
of $M_X$, corresponding to fixed $x_{\rm min}$, makes the $AA$ result increase
at midrapidity.

\begin{figure}
\begin{center}
\centering
 \includegraphics[totalheight=0.4\textheight,width=0.45\textwidth]{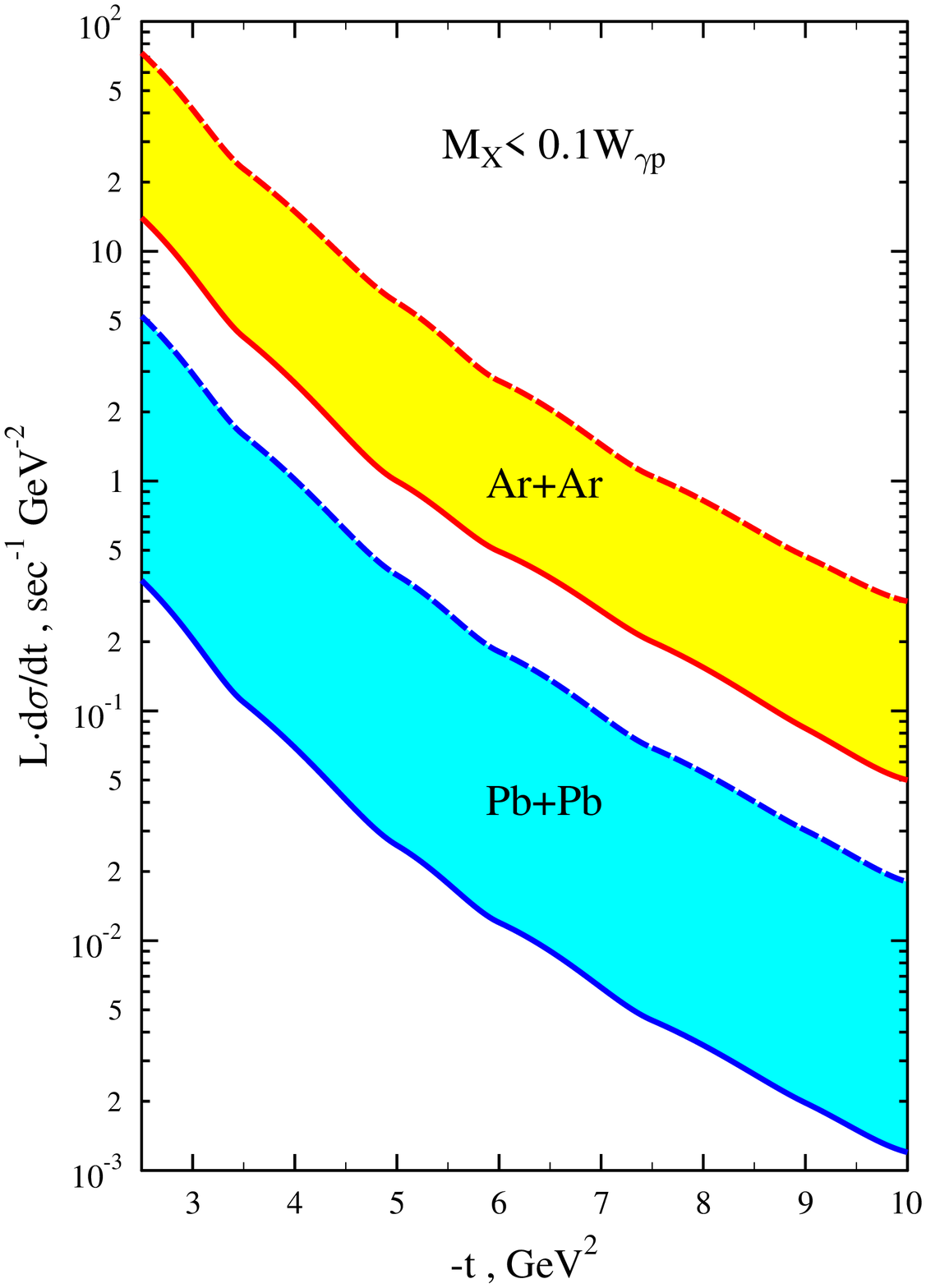}
 \includegraphics[totalheight=0.4\textheight,width=0.45\textwidth]{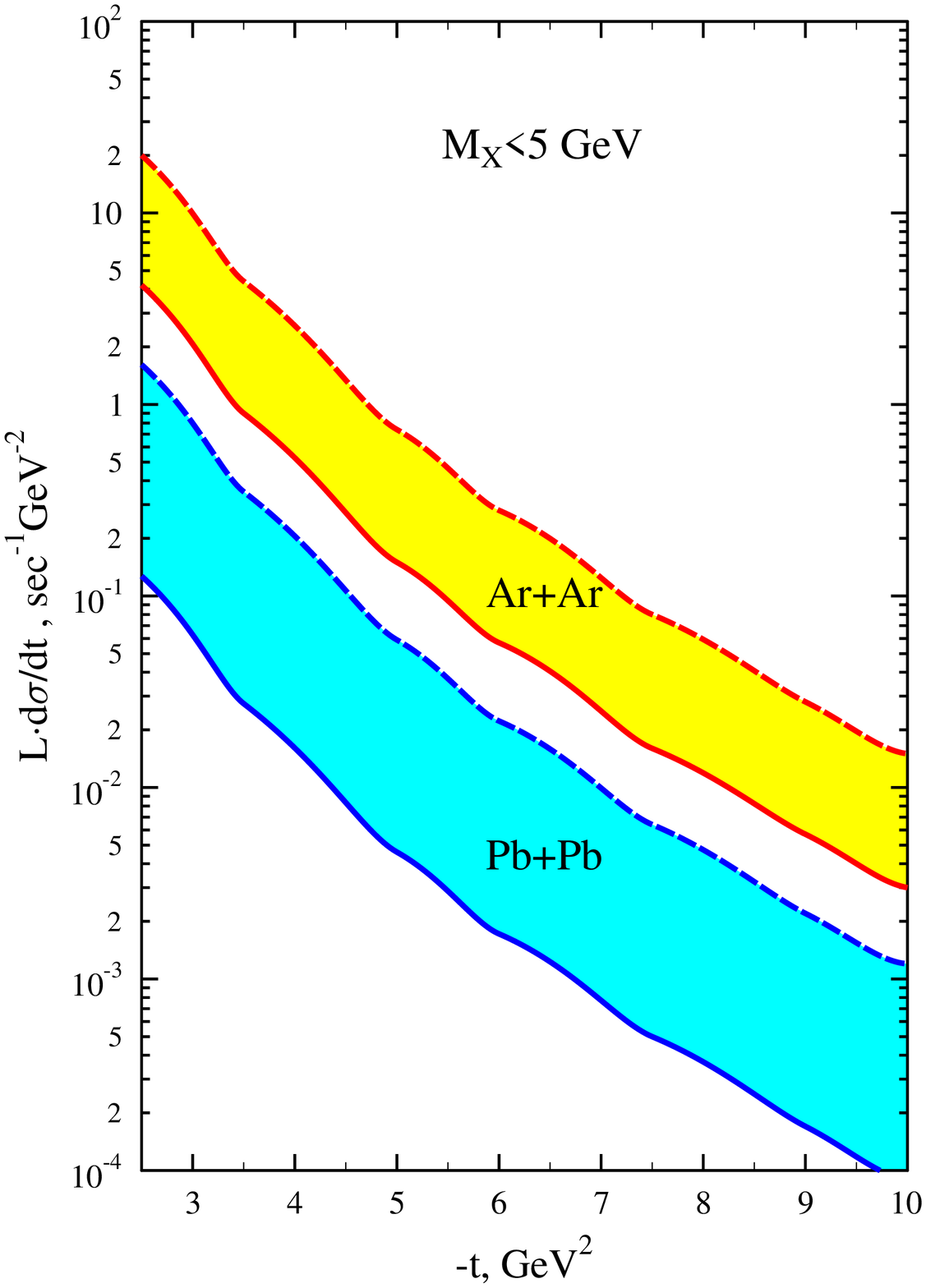}
\caption{The rapidity-integrated rates for $\rho^0$ photoproduction with a
rapidity gap in Ar+Ar and Pb+Pb UPCs as a function
of $-t$ \protect\cite{FSZ06_01}. The lower bound of the 
bands correspond to the Gribov-Glauber 
approach while the upper bound is the result in the impulse approximation.}
 \label{t10}
\end{center}
\end{figure}

The rapidity-integrated rates are shown in Fig.~\ref{t10}.  The rates 
decrease more rapidly for $M_X$ independent of energy.
This is not surprising since the average momentum fraction is larger.  The
shaded bands indicate the uncertainty between the IA (dashed
curves) and GA calculations with $\sigma_{\rm eff} = 
\sigma_{\rm in}^{\rho^0 N}(W_{\gamma N})$
(solid curves).  The larger suppression for Pb is demonstrated by the broader
band.   When the run time is taken into account, it is clear that the rates
will be sufficiently high for meaningful measurements out to $-t = 10$ GeV$^2$.

\subsubsection{Conclusions} \bigskip
 
Studies of rapidity gap processes in UPCs at the LHC will directly measure 
the energy dependence of the large-$t$ elastic amplitude of dipole-parton 
scattering.  The $\rho^0$ measurements will investigate the evolution of the
$A$ dependence over the transition between several QCD regimes: from soft 
physics to color transparency with increasing $t$ for fixed $W_{\gamma p}$ and
from color transparency to color opacity for fixed $t$ and increasing 
$W_{\gamma p}$.  These measurements will also study the
interplay of leading and higher-twist effects, a nontrivial 
function of $\Delta y$.
Altogether, these studies provide a new, powerful tool for studying  
small dipole interactions with the medium. 

\section{Determining the nuclear parton distributions}
{\it Contributed by: R. Vogt}
\label{section-pdf}
\subsection{Introduction}

Here we discuss three possible avenues for measuring the nuclear parton 
distributions through ultraperipheral heavy-ion collisions: heavy quark, dijet
and $\gamma +$jet photoproduction.  
Photoproduction occurs by ``direct'' and
``resolved'' production.  We will discuss both processes and compare the heavy
quark, dijet and $\gamma +$jet production rates from each one. 

``Direct'' photoproduction occurs when a photon emitted from one
nucleus interacts with a parton from the other nucleus, forming the final 
state.  There is only one leading order direct $Q \overline Q$ production
process, $\gamma g \rightarrow Q \overline Q$.  Thus $Q \overline Q$ production
is a rather clean probe of the nuclear gluon distribution.  Dijet production
also proceeds via an initial-state gluon, $\gamma g \rightarrow q \overline
q$.  However, there is an additional dijet production process, $\gamma q
\rightarrow g q$, the QCD Compton process.  In the case of massive quarks,
the heavy quark mass, $m_Q$, makes the $p_T$ distribution finite as $p_T
\rightarrow 0$.  Since the final state partons are massless in jet production,
$p_T$ is the only scale.  Thus some minimum $p_T$, $p_{T_{\rm min}}$, is chosen
to regulate the cross sections.  Finally, $\gamma +$jet production proceeds
via Compton scattering, $\gamma q \rightarrow \gamma q$.  Thus $\gamma +$jet
production is a direct probe of the nuclear quark and antiquark distributions.
This high $Q^2$ probe complements the nuclear deep-inelastic
scattering measurements of the charged parton distributions in the nucleus made
at lower $Q^2$.

A generic direct photoproduction cross section for ultraperipheral $AA$
collisions is obtained by 
convoluting the partonic photoproduction cross section, 
$d^2 \sigma_{\gamma i}/dt_1 du_1$,
with the photon flux from one nucleus, $dN_\gamma/dk$, 
and the parton distribution in the 
opposite nucleus, $F_i^A(x_2,Q^2)$,
\begin{eqnarray}
s_{_{NN}}^2\frac{d^2\sigma_{\rm dir}^{\gamma A}}{dt_{1\, _{NN}} 
du_{1\, _{NN}}} & = &
2 \int_{k_{\rm min}}^\infty dk {dN_\gamma \over 
dk} \int_{x_{2_{\rm min}}}^1 
\frac{dx_2}{x_2}  \nonumber \\
&  & \mbox{} \times \left[ \sum_{i=q,\overline q,g} F_i^A(x_2,Q^2)  
s^2 \frac{d^2 \sigma_{\gamma i}}{dt_1 du_1} \right] \, \, .
\label{maindir}
\end{eqnarray}
When the final state is a $Q \overline Q$ pair, $i = g$.  For dijet production,
$i = g$, $q$ and $\overline q$.  Finally, in the Compton scattering process,
$i = q$ and $\overline q$.
The partonic and hadronic Mandelstam invariants are $s$, $t_1$, $u_1$ and
$s_{_{NN}}$, $t_{1\, _{NN}}$, $u_{1\, _{NN}}$ respectively, defined later. The 
fractional momentum of the nucleon
carried by the gluon is $x_2$.  The minimum possible $x_2$, determined from
the nucleon-nucleon invariants using four-momentum conservation, is $x_{2_{\rm
min}} = -u_{1\, _{NN}}/(s_{_{NN}} + t_{1\, _{NN}})$. 
The photon momentum is denoted by $k$.  The
minimum photon momentum needed to produce the final state is $k_{\rm min}$. 
The spatial coordinates are $b$, the impact parameter,
and $z$, the longitudinal coordinate.  The factor of
two in Eq.~(\ref{maindir}) arises because both nuclei emit photons and
thus serve as targets.  For $pA$ collisions, this factor is
not included.  The incoherence of heavy quark and jet production
eliminates interference between the two production
sources \cite{Klein:1999gv}.  

The photon can also fluctuate into states with multiple $q\overline
q$ pairs and gluons, {\it i.e.}\ $|n(q \overline q)m(g)\rangle$, $n$ $q
\overline q$ pairs and $m$ gluons, the combination of which remains a color
singlet with zero flavor and baryon number. One
of these photon components can interact with a quark or gluon from the
target nucleus (``resolved'' production) \cite{Witten:ju}.  
The photon components are
described by parton densities similar to those used for protons except
that no useful momentum sum rule applies to the photon \cite{Sjostrand:1996wz}.
The quark and gluon constituents of the photon open up more channels for heavy
quark and jet photoproduction and could, in principle, lead to larger
rates for resolved production in certain regions of phase space.

The generic cross section for resolved photoproduction is
\begin{eqnarray}
\lefteqn{s_{_{NN}}^2\frac{d^2\sigma_{\rm res}^{\gamma A}}{dt_{1\, _{NN}} 
du_{1\, _{NN}}} 
= 2 \int_{k_{\rm min}}^\infty 
\frac{dk}{k} {dN_\gamma\over dk} \int_{k_{\rm min}/k}^1 \frac{dx}{x}
\int_{x_{2_{\rm min}}}^1 \frac{dx_2}{x_2}} \nonumber \\
&  & \mbox{} \times \left[ \sum_{i,j=q,\overline q, g} \!\! \left\{
F_i^\gamma (x,Q^2) F_j^A(x_2,Q^2) + F_j^\gamma (x,Q^2)
F_i^A(x_2,Q^2) \right\}
\hat{s}^2 \frac{d^2 \sigma_{ij}}{d\hat{t}_1 
d\hat{u}_1} \right] \, \, .
\label{mainres}
\end{eqnarray}
Since $k$ is typically
larger in resolved than direct photoproduction, the average photon
flux is lower in the resolved contribution.  In heavy quark production, $ij = q
\overline q$ and $gg$. In dijet production, 
$ij = qq$, $q q'$, $q \overline q$, $qg$, $gg$ $\cdots$.  
Finally, in $\gamma +$jet production, $ij = q
\overline q$, $qg$ and $\overline q g$.  Since 
the photon has no valence quarks,
the $q$ and $\overline q$ distributions in the photon are identical.  Again,
the factor of two accounts for the possibility of photon emission from each
nucleus.

The total photoproduction cross section is the sum of the direct and resolved
contributions \cite{Frixione:1995qc},
\begin{eqnarray}
s_{_{NN}}^2\frac{d^2\sigma_{\rm tot}^{\gamma A}}{dt_{1\, _{NN}} 
du_{1\, _{NN}}} = s_{_{NN}}^2\frac{d^2\sigma_{\rm dir}^{\gamma 
A}}{dt_{1\, _{NN}} 
du_{1\, _{NN}}} + s_{_{NN}}^2\frac{d^2\sigma_{\rm res}^{\gamma 
A}}{dt_{1\, _{NN}} 
du_{1\, _{NN}}} \, \, .
\label{phottot}
\end{eqnarray}

In the remainder of this introduction, we will discuss the common ingredients
of these calculations.  We first discuss the calculation of
the photon flux and the relevant kinematics.  We then turn to the expected
modifications of the nuclear parton distributions relative to those of the free
proton.  Finally, we present the photon parton distribution functions.  The
next two subsections deal with heavy quark and jet photoproduction. 

The photon flux is calculated using Eqs.~(\ref{wwr}).
The maximum center-of-mass energy, 
$\sqrt{s_{\gamma N}} \approx \sqrt{2E_{\rm max}m_p}$, for single
photon interactions with protons, $\gamma p \rightarrow Q \overline Q$
\cite{Fritzsch:1977zq}, at the LHC is given in Table~\ref{gamfacs}.  
At the LHC, the energies are high enough 
for $t \overline t$ photoproduction \cite{Klein:2000dk}.  
The total photon flux striking the target nucleus 
must be calculated
numerically.   The numerical
calculations are used for $AA$ interactions but the analytic flux 
in Eq.~(\ref{analflux}) is used for $pA$
interactions.  The difference between the numerical and analytic
expressions is typically less than 15\%, except for photon energies
near the cutoff. 

The nuclear parton densities $F_i^A(x,Q^2)$ in 
Eqs.~(\ref{maindir}) and (\ref{mainres}) can be
factorized into nucleon parton
densities, $f_i^N(x, Q^2)$, and a shadowing function $S^i(A,x,Q^2)$ that
describes the modification of the nuclear parton distributions in
position and momentum space
\begin{eqnarray}
F_i^A(x,Q^2) & = & S^i(A,x,Q^2)
f_i^N(x,Q^2) \label{nucglu} \, \, 
\end{eqnarray}
where $f^N_i(x,Q^2)$ is the parton density in the nucleon.  
In the absence of nuclear
modifications, $S^i \equiv1$.  While we have previously treated the spatial
dependence of shadowing,
\cite{Emel'yanov:1997pu,Emel'yanov:1998df,Emel'yanov:1998sy,Emel'yanov:1999bn,Vogt:2000hp}, we do not include it here.
We use the MRST LO parton distributions \cite{Martin:1998np}.  For 
$Q \overline Q$
production, we evaluate the nucleon parton densities at $Q^2= (a m_T)^2$
where $m_T^2 = p_T^2 + m_Q^2$, $a = 2$ for charm and 1 for bottom.  
The appropriate scale for jet production is $Q^2 = (ap_T)^2$ 
where we take $a = 1$.

\begin{figure}[htbp] 
\setlength{\epsfxsize=0.95\textwidth}
\setlength{\epsfysize=0.5\textheight}
\centerline{\epsffile{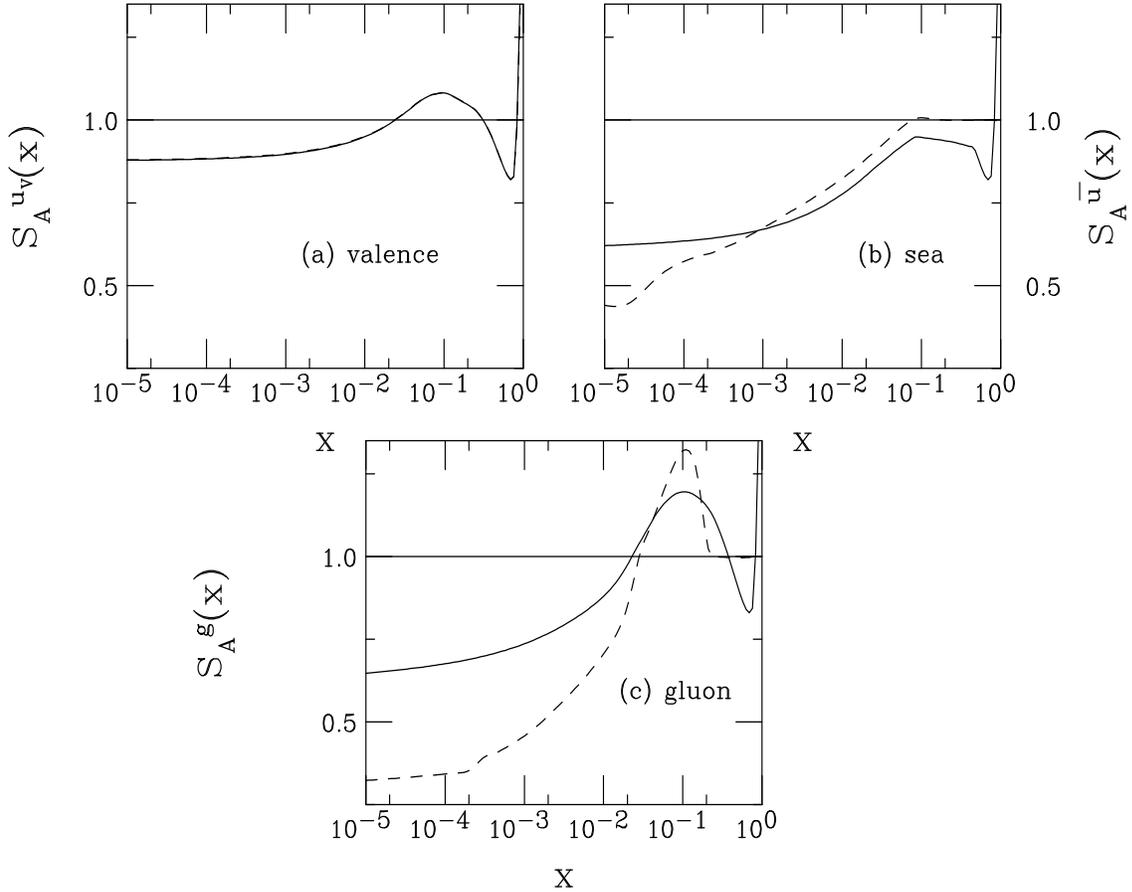}}
\caption[]{ The EKS98 and FGS shadowing parameterizations are compared at
the scale $Q = 2m_c = 2.4$ GeV.  The solid curves are the EKS98
parametrization, the dashed, FGS.
}
\label{shadcomp}
\end{figure}

We have chosen two recent parameterizations of the nuclear shadowing
effect which cover extremes of gluon shadowing at low $x$.  The Eskola
{\it et al.} parametrization, EKS98,  \cite{Eskola:1998iy,Eskola:1998df}
is based on the GRV LO
\cite{Gluck:1991ng} parton densities.  At the minimum scale, $Q_0$, valence 
quark shadowing is identical for
$u$ and $d$ quarks.  Likewise, the shadowing of $\overline u$ and
$\overline d$ quarks are identical at $Q_0$. Although the light quark
shadowing ratios are not constrained to be equal at higher scales, the 
differences between them are small.  Shadowing of the heavier flavor
sea, $\overline s$ and higher, is calculated separately at $Q_0$.  The
shadowing ratios for each parton type are evolved to LO for $1.5 < Q <
100$ GeV and are valid for $x \geq 10^{-6}$ \cite{Eskola:1998iy,Eskola:1998df}.
Interpolation in nuclear mass number allows results to be obtained for
any input $A$.  The parametrization by Frankfurt, Guzey and
Strikman, denoted FGS here, combines Gribov theory with hard
diffraction \cite{Frankfurt:2003zd}.  It is based on the CTEQ5M 
\cite{Lai:1999wy} parton
densities and evolves each parton species separately to NLO for $2 < Q
< 100$ GeV.  Although the given $x$ range is $10^{-5} < x < 0.95$, the
sea quark and gluon ratios are unity for $x > 0.2$.  The EKS98 valence
quark shadowing ratios are used as input since Gribov theory does not
predict valence shadowing.  The parametrization is available for $A = 16$, 
40, 110 and 206.  Figure~\ref{shadcomp} compares
the two parameterizations for $A \approx 200$ and $Q = 2m_c = 2.4$ GeV.
We take the EKS98 parametrization \cite{Eskola:1998iy,Eskola:1998df}, 
as a default but we also compare it to the FGS \cite{Frankfurt:2003zd}
results in some cases.

We now turn to the photon parton distributions.  There are
a few photon parton distributions
available~\cite{Gluck:1991jc,Gluck:1991ee,Drees:1984cx,Abramowicz:1991yb,Hagiwara:1994ag,Schuler:1995fk,Schuler:1996fc}.
The data \cite{PDFLIB,Bartel:1984cg} cannot definitively rule out any of 
these parton densities.  
As expected, $F_q^\gamma(x,Q^2) = F_{\overline
q}^\gamma (x,Q^2)$ flavor by flavor because there are 
no ``valence'' quarks in
the photon.  The gluon distribution in the photon is less well known.
We compare results with the GRV-G LO set \cite{Gluck:1991jc,Gluck:1991ee},
with a gluon distribution is similar to most of the other available sets 
\cite{Drees:1984cx,Hagiwara:1994ag,Schuler:1995fk,Schuler:1996fc}, to the
LAC1 set \cite{Abramowicz:1991yb} where the low $x$ gluon density is up 
to an order of magnitude higher.  The differences in the two photon parton
densities are most important for heavy quark production.

\begin{figure}[htb]
\setlength{\epsfxsize=0.95\textwidth}
\setlength{\epsfysize=0.4\textheight}
\centerline{\epsffile{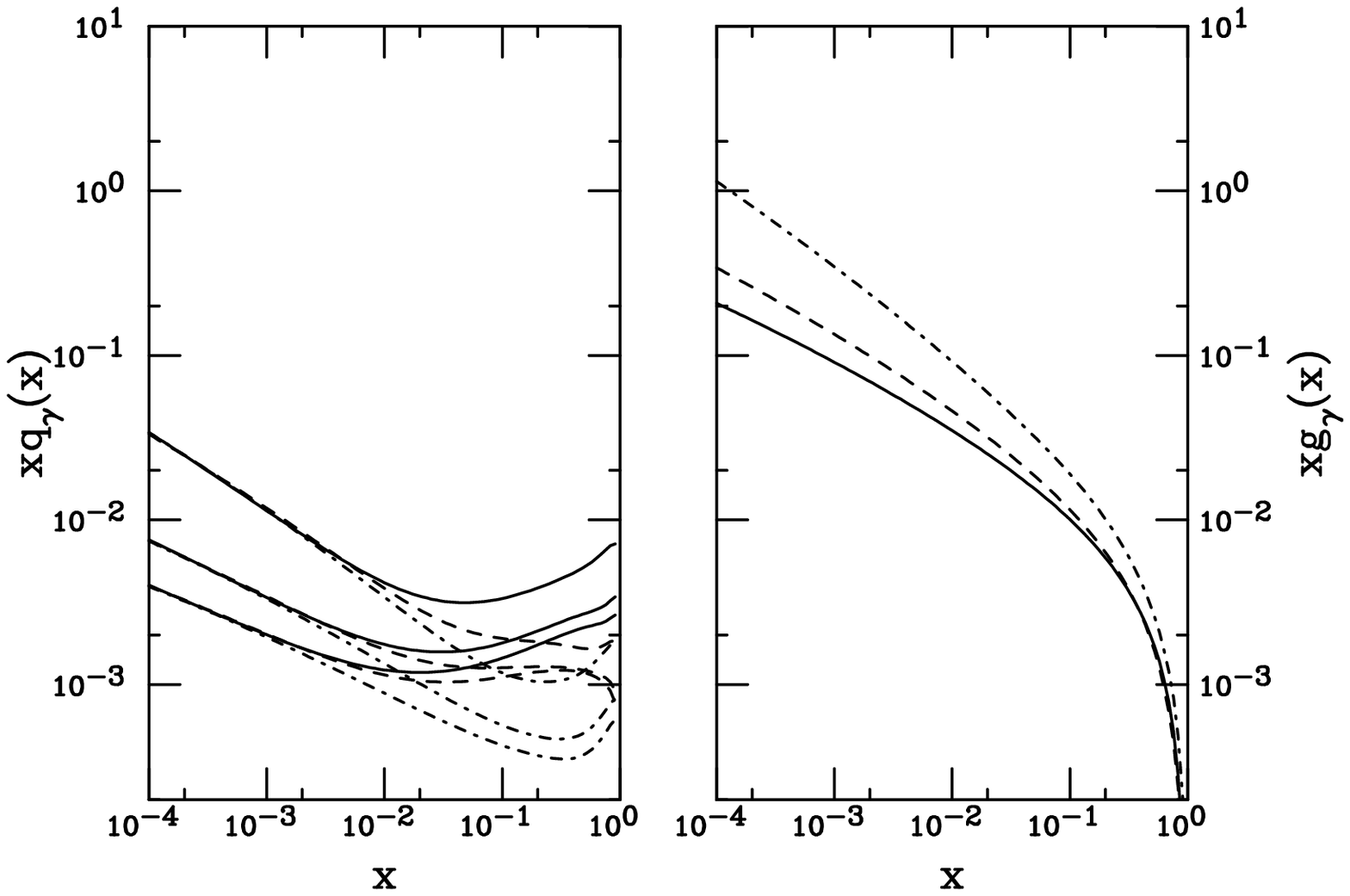}}
\caption[] {The GRV-G LO quark (a) and gluon (b) 
distributions of the photon.  In (a) the up (solid),
down (dashed) and strange (dot-dashed) distributions are evaluated at $2m_c$
(lower curves), $2m_b$ (middle curves) and $2m_t$ (upper curves).
In (b) the gluon
distributions are shown at $2m_c$ (solid), $2m_b$ (dashed) and $2m_t$
(dot-dashed).} 
\label{pdf_gamma_grv}
\end{figure}

The GRV-G LO photon parton densities are shown in Fig.~\ref{pdf_gamma_grv} 
for scales equal to $2m_c$, $m_b$ and $m_t$ where $m_c = 1.2$ GeV, $m_b = 4.75$
GeV and $m_t = 175$ GeV.  
This set has a minimum $x$ of $10^{-5}$
and $0.25 \leq Q^2 \leq 10^6$ GeV$^2$.  At low $x$, the $u$, $d$ and $s$
distributions are identical.  They diverge around $x \sim 10^{-3}$ with the
$u$ and $d$ distributions increasing with $x$ while the $s$ distribution
decreases until $x > 0.1$ where it turns up again.  As $x \rightarrow 1$ the
quark distributions become larger than the gluon distributions.

\begin{figure}[htb]
\setlength{\epsfxsize=0.95\textwidth}
\setlength{\epsfysize=0.4\textheight}
\centerline{\epsffile{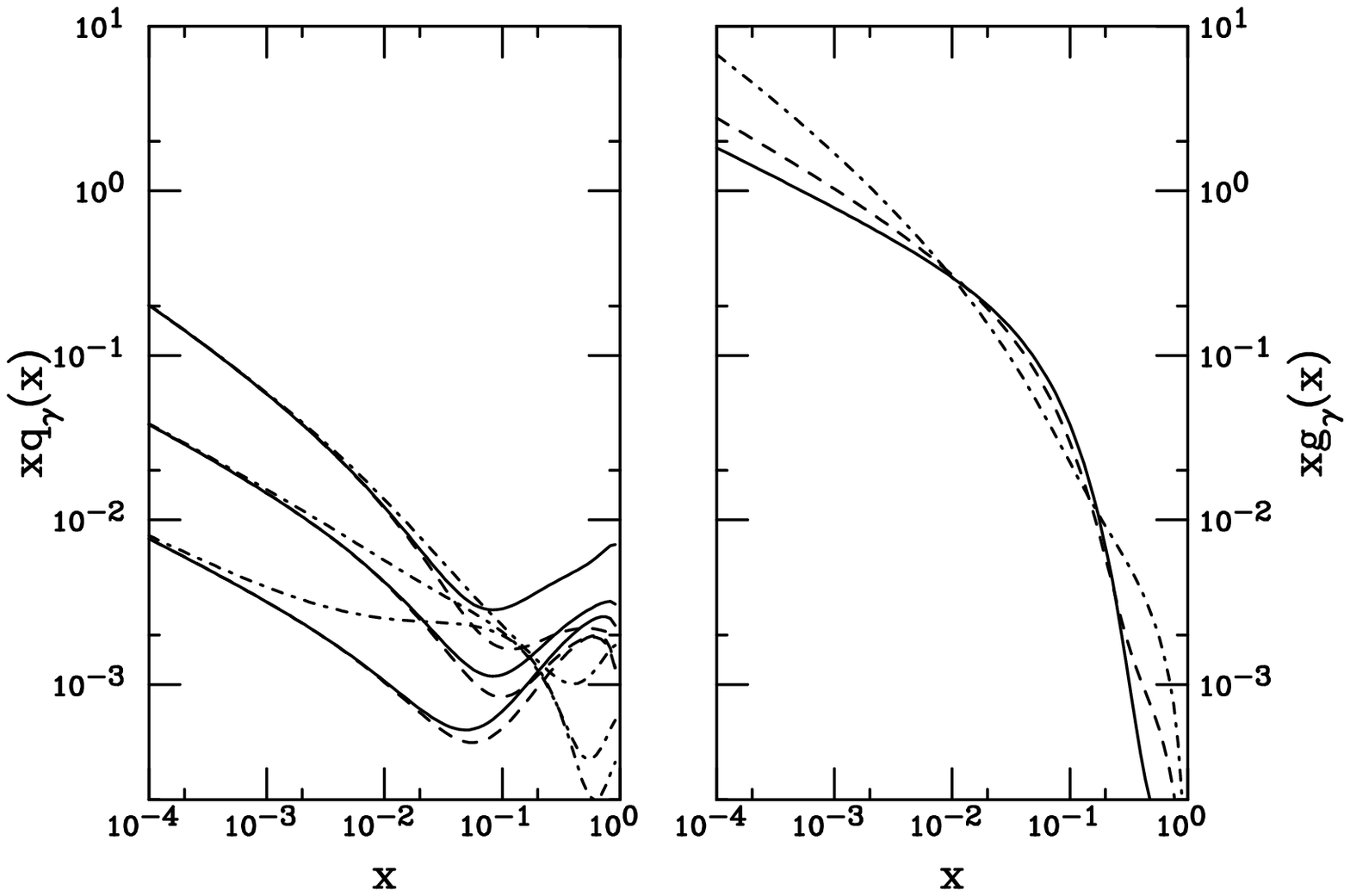}}
\caption[] {The LAC1 LO quark (a) and gluon (b) 
distributions of the photon.  In (a) the up (solid),
down (dashed) and strange (dot-dashed) distributions are evaluated at $2m_c$
(lower curves), $2m_b$ (middle curves) and $2m_t$ (upper curves).
In (b) the gluon
distributions are shown at $2m_c$ (solid), $2m_b$ (dashed) and $2m_t$
(dot-dashed).} 
\label{pdf_gamma_lac1}
\end{figure}

The LAC1 LO photon parton densities are shown in Fig.~\ref{pdf_gamma_lac1}
for the same scales.  
This set has a minimum $x$ of $10^{-4}$ and covers the
range $4 \leq Q^2 \leq 10^5$.
All the densities are somewhat higher than those of
GRV LO but they are less regular in shape, particularly the $s$ distribution
when $Q^2 = 4m_c^2$, possibly 
because this scale is rather close to $Q_0$.  
The gluon distributions are also rather irregular,
particularly at high $x$.  

The LAC1 densities are generally higher at low $x$
but the GRV-G gluon density is higher at $x>0.1$.  The LAC1 and GRV-G quark
distributions are also similar in this $x$ region.  Thus if relatively low $x$
values are reached, the LAC1 resolved results will be larger.  In the high $x$
region, the two densities will give either similar results or the GRV-G
densities may give a larger resolved component.  In any case, it is clear 
that, in certain kinematic regions, the
difference in the resolved yields due to the choice of photon parton density
could be significant.

With these ingredients, we turn to the specific final-state processes under
consideration.  We first discuss heavy quark photoproduction in
Section~\ref{hvq}.  Here the rates 
are high and the nuclear gluon distribution should be rather directly
accessible.  We then show expected results for direct and resolved jet
photoproduction in Section~\ref{jet}.  The additional channels for resolved
jet photoproduction could potentially enhance this contribution over the direct
contribution, obscuring the nuclear gluon distribution.  However, as we will
discuss, it might then be possible to examine the nuclear quark distribution.
Finally, we discuss how to distinguish between photoproduction and
hadroproduction at the LHC in Section~\ref{summary}.

\subsection{Heavy quark photoproduction}
{\it Contributed by: S. R. Klein, J. Nystrand and R. Vogt}
\label{hvq}

In this subsection we discuss photoproduction of massive $Q \overline Q$ pairs
at the LHC \cite{Klein:2002wm}.  We also discuss the dependence of the
resolved results on the photon parton density, comparing results from
with the GRV-G set \cite{Gluck:1991jc,Gluck:1991ee} (Ref.~\cite{Klein:2002wm})
to those with the 
LAC1 set \cite{Abramowicz:1991yb}.  We work to leading order in the strong
coupling constant $\alpha_s$.

We include all $Q \overline Q$ pairs in the total
cross sections and rates even though some of these pairs have masses below
the $H \overline H$ threshold where $H \overline H \equiv D \overline D$ and
$B \overline B$ for $c$ and $b$ quarks respectively.  No such distinctions
exist for top since it decays before hadronization.
Photoproduction is an inclusive process;
accompanying particles can combine with the $Q$ and $\overline Q$,
allowing the pairs with $M<2m_H$ to hadronize.  We assume the
hadronization process does not affect the rate.  

Direct $Q\overline Q$ pairs are produced in the reaction $\gamma(k) +
N(P_2) \rightarrow Q(p_1) + \overline Q(p_2) + X$ where $k$ is the
four momentum of the photon emitted from the virtual photon field of
the projectile nucleus, $P_2$ is the four momentum of the interacting
nucleon $N$ in ion $A$, and $p_1$ and $p_2$ are the four momenta of
the produced $Q$ and $ \overline Q$.  

On the parton level, the photon-gluon fusion reaction is $\gamma(k) +
g(x_2 P_2) \rightarrow Q(p_1) + \overline Q(p_2)$ where $x_2$ is the
fraction of the target momentum carried by the gluon.  The LO $Q
\overline Q$ photoproduction cross section for quarks with mass $m_Q$
is \cite{Jones:1977wx}
\begin{equation}
s^2 \frac{d^2 \sigma_{\gamma g}}{dt_1 du_1} = \pi \alpha_s(Q^2) \alpha e_Q^2 
B_{\rm QED} (s,t_1,u_1) \delta(s + t_1 + u_1)
\label{gamApart}
\end{equation}
where
\begin{equation}
B_{\rm QED} (s,t_1,u_1) = 
\frac{t_1}{u_1} + \frac{u_1}{t_1} + \frac{4m_Q^2 s}{t_1 u_1} \left[ 1 - 
\frac{m_Q^2 s}{t_1 u_1} \right] \, \, .
\label{bqeddef}
\end{equation}
At leading order (LO), the partonic cross section of the direct
contribution is proportional to $\alpha \alpha_s(Q^2) e_Q^2$, where
$\alpha_s(Q^2)$ is the strong coupling constant, $\alpha=e^2/\hbar c$ is the
electromagnetic coupling constant, and $e_Q$ is the quark charge, 
$e_c = e_t = 2/3$ and $e_b = -1/3$. 
Here $\alpha_s(Q^2)$ is evaluated to one loop at scale $Q^2$.  
The partonic invariants, $s$,
$t_1$, and $u_1$, are defined as $s = (k + x_2 P_2)^2$, $t_1 = (k -
p_1)^2 - m_Q^2 = (x_2 P_2 - p_2)^2 - m_Q^2$, and $u_1 = (x_2 P_2 -
p_1)^2 - m_Q^2 = (k - p_2)^2 - m_Q^2$.  In this case, $s = 4k \gamma_L
x_2m_p$ where $\gamma_L$ is the Lorentz boost of a single beam and
$m_p$ is the proton mass.  Since $k$ can be a continuum of
energies up to $E_{\rm beam} = \gamma_L m_p$, we define $x_1 = k/P_1$ 
analogous to the parton momentum fraction where $P_1$ is the nucleon
four momentum. For a detected quark in a nucleon-nucleon collision, the
hadronic invariants are then $s_{_{NN}} = (P_1 + P_2)^2$, $t_{1\, _{NN}} = 
(P_2 - p_1)^2 - m_Q^2$, and $u_{1\, _{NN}} = (P_1 - p_1)^2 - m_Q^2$.

We label the quark rapidity as $y_1$ and the antiquark rapidity as $y_2$.
The quark rapidity is related to the invariant $t_{1\, _{NN}}$ by 
$t_{1\, _{NN}} = -\sqrt{s_{_{NN}}} m_T e^{-y_1}$.  The
invariant mass of the pair can be determined if both the $Q$ and
$\overline Q$ are detected.  The square of the invariant mass, $M^2 =
s = 2m_T^2 (1 + \cosh(y_1 - y_2))$, is the partonic center-of-mass
energy squared.  For $Q \overline Q$ pair production, 
$k_{\rm min} = M^2/4\gamma_L m_p$.  At LO, $x_1
= (m_T/\sqrt{s_{_{NN}}})(e^{y_1} + e^{y_2})$ and $x_2 =
(m_T/\sqrt{s_{_{NN}}})(e^{-y_1} + e^{-y_2})$.  We calculate $x_1$ and $x_2$ as
in an $NN$ collision and then determine the flux in the lab frame for
$k = x_1 \gamma_L m_p$, equivalent to the center-of-mass frame in a
collider.  The photon flux is exponentially suppressed for $k>\gamma_L
\hbar c/R_A$, corresponding to a momentum fraction $x_1 > \hbar
c/m_pR_A$.  The maximum $\gamma N$ center-of-mass energy,
$\sqrt{s_{\gamma N}}$, is much lower than the hadronic $\sqrt{s_{_{NN}}}$.  

The cross section for direct photon-nucleon heavy quark
photoproduction is obtained by inserting Eq.~(\ref{gamApart}) into
Eq.~(\ref{maindir}). 
The equivalent hadronic invariants can be defined for photon four
momentum $k$ as $s_{\gamma N} = (k + P_2)^2$, $t_{1,\gamma N} = (P_2 -
p_1)^2 - m_Q^2$, and $u_{1, \gamma N} = (k - p_1)^2 - m_Q^2$
\cite{Smith:1991pw}.  The partonic and equivalent hadronic invariants for fixed
$k$ are related by $s = x_2s_{\gamma N}$, $t_1 = u_{1, \gamma N}$, and
$u_1 = x_2 t_{1, \gamma N}$.

\begin{figure}[htb]
\setlength{\epsfxsize=0.95\textwidth}
\setlength{\epsfysize=0.5\textheight}
\centerline{\epsffile{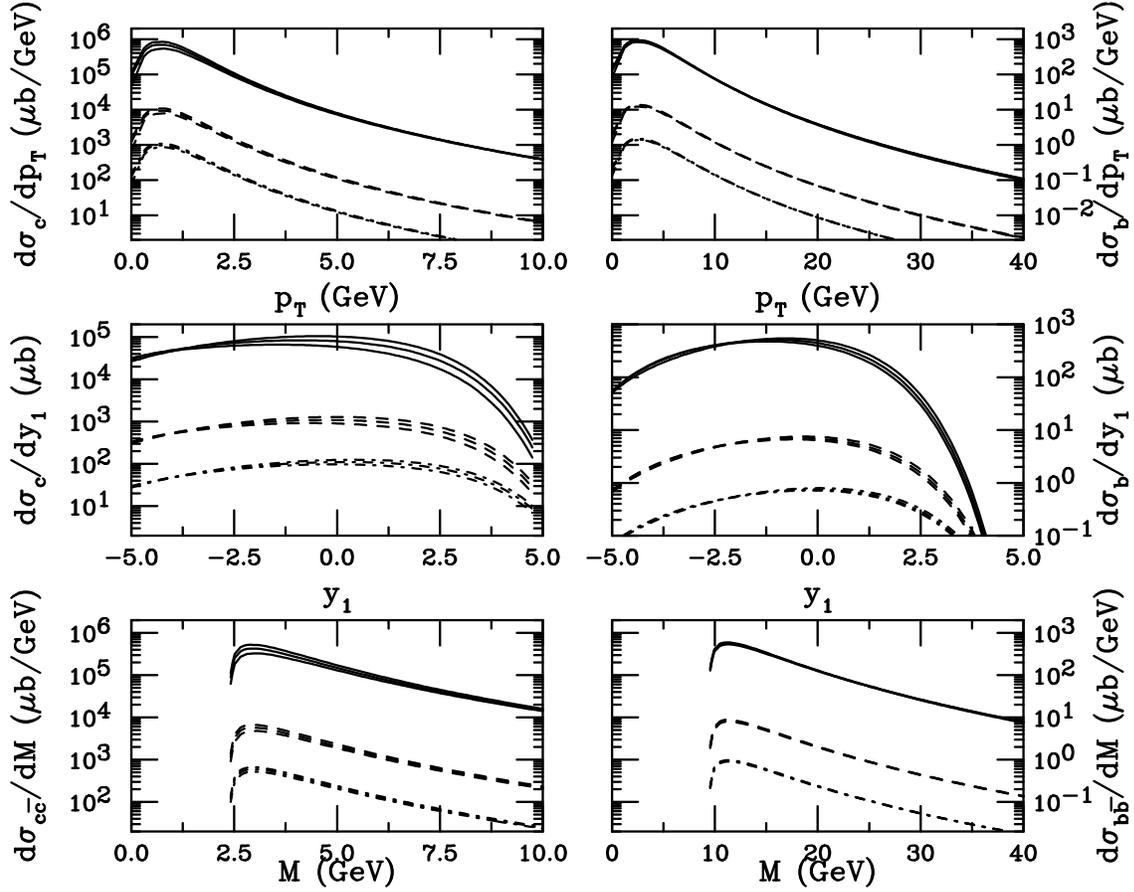}}
\caption[]{Direct $Q \overline Q$ photoproduction in peripheral 
$AA$ collisions.  The left-hand side is for charm while the
right-hand side is for bottom.  The single $Q$ $p_T$ (upper)
and rapidity (middle) distributions are shown along with the $Q \overline Q$ 
pair invariant mass distributions (lower).  The O+O (dot-dashed), 
Ar+Ar (dashed) and Pb+Pb (solid) results are given.  There are three curves
for each contribution: no shadowing, EKS98 and FGS.  At $y_1 > 0$, the highest
curve is without shadowing, the middle curve with EKS98 and the lower curve
with FGS.  The photon is coming from the left.}
\label{distcbdir}
\end{figure}

\begin{figure}[htb]
\setlength{\epsfxsize=0.95\textwidth}
\setlength{\epsfysize=0.5\textheight}
\centerline{\epsffile{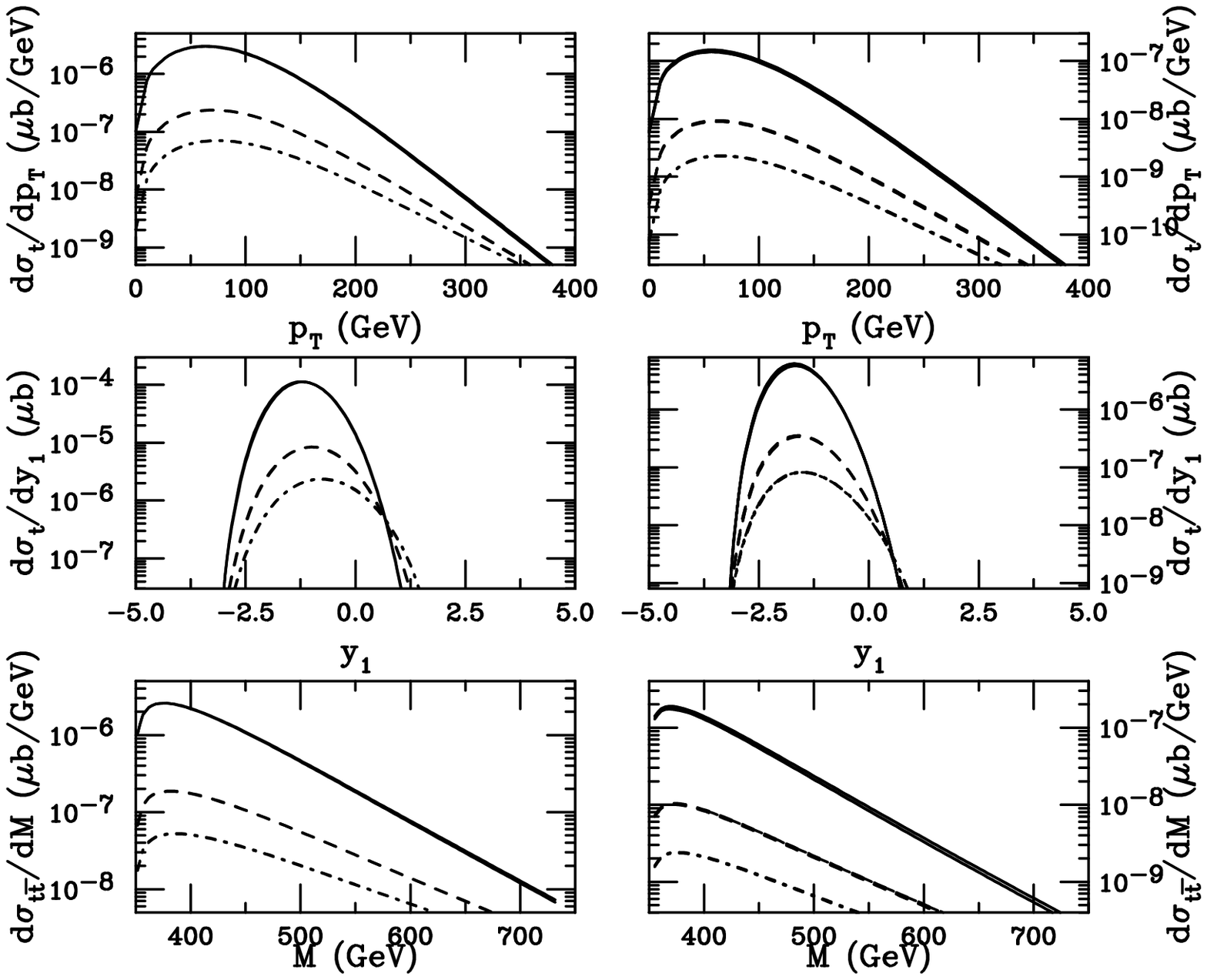}}
\caption[]{Direct (left) and resolved (right) $t \overline t$ 
photoproduction in peripheral 
$AA$ collisions.  Note the different scales on the $y$-axes for the two
production mechanisms.  The single $t$ $p_T$ (upper)
and rapidity (middle) distributions are shown along with the $t \overline t$ 
pair invariant mass distributions (lower).  The O+O (dot-dashed), 
Ar+Ar (dashed) and Pb+Pb (solid) results are given. 
The photon is coming from the left.}
\label{disttdirres}
\end{figure}

The charm and bottom photoproduction distributions are shown in
Fig.~\ref{distcbdir} for Pb+Pb, Ar+Ar and O+O collisions.  
The direct top photoproduction distributions for these three systems are 
given on the left-hand side of Fig.~\ref{disttdirres}.
There are three curves for each contribution, one
without shadowing and two with homogeneous nuclear
shadowing employing the EKS98 and FGS parameterizations.  
The photon comes from the left.  Then
$y_1<0$ corresponds to $k<\gamma_L x_2m_p$ in the center-of-mass (lab)
frame.  If the photon emitter and target nucleus are interchanged, the
resulting unshadowed rapidity distribution, $S^i=1$, is the mirror
image of these distributions around $y_1=0$. The
$Q$ and $\overline Q$ distributions are asymmetric around $y_1=0$.
The total heavy quark rapidity
distributions are then the sum of the displayed results with their mirror
images when both nuclei emit photons.  This factor of two, shown in
Eq.~(\ref{maindir}), is included
in the transverse momentum and invariant mass distributions.  Note that the
peak in the rapidity distributions moves towards more negative $y_1$ and the
distribution narrows as the
quark mass increases.  The $y_1$ phase space for a single top quark is
$\approx 3.7$, a decrease of more than a factor of two relative to charm.

\begin{table}[htbp]
\begin{center}
\caption[]{Direct $Q\overline Q$ photoproduction
cross sections integrated over $b>2R_A$ in peripheral $AA$ collisions.}
\label{gamAqqdir}
\vspace{0.4cm}
\begin{tabular}{|c|c|c|c|} \hline
& \multicolumn{3}{c|}{$\sigma^{\rm dir}$ (mb)} \\ \hline
$AA$ & no shad & EKS98 & FGS \\ \hline
\multicolumn{4}{|c|}{$c \overline c$} \\ \hline
O+O   & 1.66 & 1.50 & 1.35  \\
Ar+Ar & 16.3 & 14.3 & 12.3  \\
Pb+Pb & 1246 & 1051 & 850   \\ \hline
\multicolumn{4}{|c|}{$b \overline b$} \\ \hline
O+O   & 0.0081 & 0.0078 & 0.0075 \\
Ar+Ar & 0.073  & 0.070  & 0.066  \\
Pb+Pb & 4.89   & 4.71   & 4.42   \\ \hline
\multicolumn{4}{|c|}{$t \overline t$} \\ \hline
O+O   & $9.13 \times 10^{-9}$ & $9.27 \times 10^{-9}$ & $9.31 \times 10^{-9}$
\\
Ar+Ar & $2.86 \times 10^{-8}$ & $2.88 \times 10^{-8}$ & $2.87 \times 10^{-8}$
\\
Pb+Pb & $3.29 \times 10^{-7}$ & $3.21 \times 10^{-7}$ & $3.22 \times 10^{-7}$
\\ \hline
\end{tabular}
\end{center}
\end{table}

Since the distributions are shown on a logarithmic scale, shadowing appears to
be a rather small effect over most of phase space.  It is  
most prominent in the rapidity
distributions and are otherwise is only distinguishable for charm production at
low $p_T$ and low invariant mass.  Shadowing is largest at forward rapidities 
where low momentum fractions in the
nucleus are reached.  

The total cross sections for direct $Q \overline Q$
photoproduction are given in Table~\ref{gamAqqdir}\footnote{A typo in the
direct cross section code caused the cross sections in
Refs.~\cite{Klein:2002wm,Klein:2000dk} to be somewhat overestimated.  The
results given here are correct.}.  The EKS98 shadowing parametrization
has a $10-20$\% effect on the total $c \overline c$ cross section.  The effect
is smallest for O+O collisions, due to the small $A$, even though the energy
is higher and the effective $x$ values probed are smaller.  The stronger
shadowing of the FGS parametrization gives a $23-46$\% reduction of the $c
\overline c$ cross sections.  Both the $x$ and $Q$ values probed increase for
$b \overline b$ production.  Each of these increases reduces 
the overall effect of shadowing.  The EKS98 parametrization
results in only a 4\% reduction of the $b \overline b$ 
total cross sections, independent of $A$, while the FGS parametrization gives
an $8-10$\% effect.  Although we include the $t \overline t$ cross sections
with shadowing in Table~\ref{gamAqqdir}, 
we note that $m_t$ is larger than the maximum $Q$
for which the parameterizations may be considered reliable.  

\begin{figure}[htb]
\setlength{\epsfxsize=0.95\textwidth}
\setlength{\epsfysize=0.5\textheight}
\centerline{\epsffile{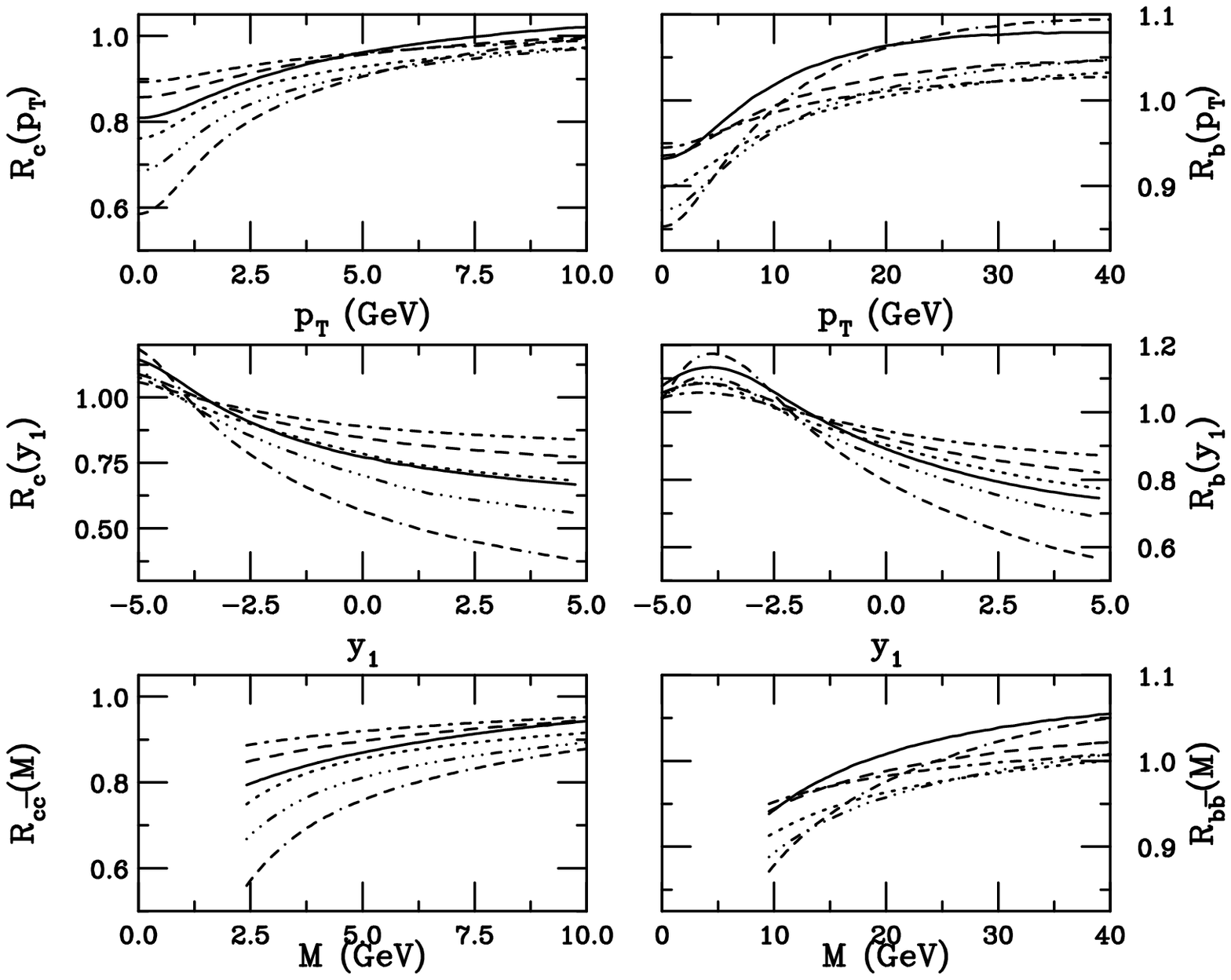}}
\caption[]{Shadowing in direct $Q \overline Q$ photoproduction in peripheral 
$AA$ collisions.  The left-hand side shows the results for charm while the
right-hand side gives the results for bottom.  The single $Q$ $p_T$ (upper)
and rapidity (middle) ratios are shown along with the $Q \overline Q$ 
pair invariant mass ratios (lower).   The results for the EKS98
(O+O (dot-dashed), Ar+Ar (dashed) and Pb+Pb (solid)) and FGS (O+O (dotted),
Ar+Ar (dot-dot-dot-dashed) and Pb+Pb (dash-dash-dash-dotted)) shadowing
parameterizations are given. The photon is coming from the left.}
\label{shadcbdir}
\end{figure}

The integrated cross sections provide incomplete information
about shadowing effects.  To provide a more complete picture of
the effects of shadowing on the distributions,
in Fig.~\ref{shadcbdir} we plot the ratio of the distributions with shadowing
included to those without shadowing, denoted $R_Q$ for the single quark
distributions and $R_{Q \overline Q}$ for the pair invariant mass
distributions.  The charm ratios are given on the left-hand side of
Fig.~\ref{shadcbdir} while the bottom ratios are on the right-hand side.
The ratios are given for both the EKS98 (solid, dashed and dot-dashed curves 
for Pb+Pb, Ar+Ar, and O+O respectively) and the FGS (dash-dash-dash-dotted,
dot-dot-dot-dashed and dotted curves for Pb+Pb, Ar+Ar and O+O respectively) 
shadowing parameterizations.  The distributions employing 
the FGS parametrization are more strongly
affected---the charm rapidity ratio for Pb+Pb with EKS98 is 
similar to the O+O ratio
with FGS.  Since the rapidity distributions are integrated over $p_T$, the
largest weight goes to low $p_T$ where the $Q^2$ evolution of the 
shadowing parameterizations is still small, producing the largest shadowing
effect at $Q \approx a\langle m_T \rangle$ where $a = 1$ for bottom and 2 
for charm and $\langle m_T \rangle \approx \sqrt{m_Q^2 + \langle p_T^2
\rangle}$. 

The lowest $x$ values occur at the highest forward rapidities when the
photon comes from the left, where the rapidity distribution begins to drop
off.  The lowest value of
$R_c(p_T = 0)$ corresponds to $R_c(y_1 = y_{1\, {\rm max}})$ where $y_{1 \,
{\rm max}}$ is the position of the peak of the rapidity distribution.
At midrapidity, just forward of the peak in the distribution, $R_c(y_1 =
0) \sim 0.75$ for EKS98 and 0.55 for FGS with the Pb beam.  The large
difference between these midrapidity values and from $R_c(y_1) = 1$ suggests 
that shadowing is measurable in these interactions.  While shadowing is reduced
at the larger $x$ and $Q$ for bottom production, it is still
significant enough for measurements to be feasible.  The top cross section is
too small for high statistics measurements.

\begin{figure}[htb]
\setlength{\epsfxsize=0.95\textwidth}
\setlength{\epsfysize=0.5\textheight}
\centerline{\epsffile{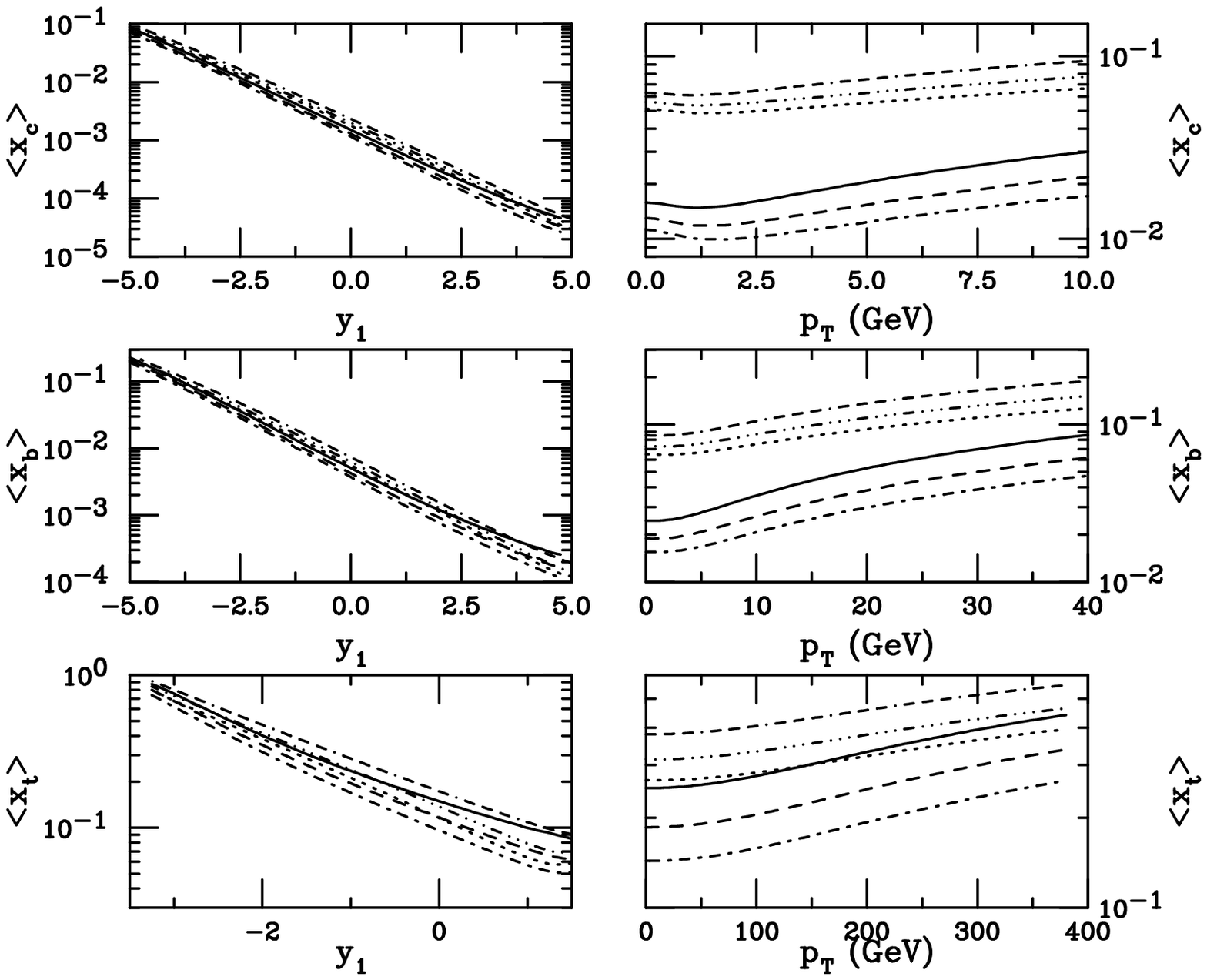}}
\caption[]{The average value of the nucleon parton momentum fraction $x$ as a
function of quark rapidity (left-hand side) and transverse momentum (right-hand
side).  The results are given for charm (upper), bottom (middle) and 
top (lower).  The direct values are given for
O+O (dot-dashed), Ar+Ar (dashed) and Pb+Pb (solid) while the resolved values
are given for O+O (dotted),
Ar+Ar (dot-dot-dot-dashed) and Pb+Pb (dash-dash-dash-dotted). (Resolved
production is calculated with the GRV-G photon parton distributions.)
The photon is coming from the left.}
\label{avexcbt}
\end{figure}

The typical nucleon $x$ ranges for charm, bottom and top production are shown 
in Fig.~\ref{avexcbt} as a function of quark rapidity (left-hand side) and 
transverse momentum (right-hand side).  It is then clear how the rapidity and
shadowing distributions in Fig.~\ref{shadcomp} map each other.  At large
negative rapidity, $\langle x_c \rangle \sim 0.1$ for charm, decreasing to
$\langle x_c \rangle \sim 10^{-5}$ at $y_1 \sim 5$.  The average $x$ for
bottom, $\langle x_b \rangle$,
increases by $m_b/m_c$ relative to charm.  For charm and bottom production,
there is not much difference between curves at different values of
$\sqrt{s_{_{NN}}}$.  Charm production is predominantly in the shadowing region
over all $y_1$ while, at large negative rapidity, bottom production reaches the
anti shadowing region.  On the other hand, top production is in the `EMC
region', $\langle x_t \rangle > 0.2$ for $y_1 < 0$.  

Figure~\ref{avexcbt} also illustrates how, as a function of quark $p_T$, the 
average $x$ corresponds to the peak of the rapidity distribution.  The average
value of $x$ changes slowly with $p_T$.  Some of this increase is due to the
growing value of $\langle m_T \rangle$ entering in the calculation of $x$.
However, the width of the rapidity distribution decreases with 
increasing $p_T$, an important effect, particularly for heavier flavors where
phase space considerations are important.

We now turn to the resolved (hadronic) contribution to the
photoproduction cross section.  The hadronic reaction, $\gamma N
\rightarrow Q \overline Q X$, is unchanged, but now, prior to the
interaction with the nucleon, the photon splits into a color singlet
state with some number of $q \overline q$ pairs and gluons.  
On the parton level, the resolved LO reactions are $g(xk) +
g(x_2 P_2) \rightarrow Q(p_1) + \overline Q(p_2)$ and $q (xk) +
\overline q(x_2 P_2) \rightarrow Q(p_1) + \overline Q(p_2)$ where $x$ is the
fraction of the photon momentum carried by the parton.  The LO diagrams for
resolved photoproduction are the same as
for hadroproduction except that one parton source is a photon
rather than a nucleon.
The LO partonic cross sections are 
\cite{Brock:1993sz}
\begin{eqnarray}
\hat{s}^2 \frac{d^2 \sigma_{q \overline q}}{d\hat{t}_1 d\hat{u}_1} & = 
& \pi \alpha_s^2(Q^2)
\frac{4}{9} \left( \frac{\hat{t}_1^2 + \hat{u}_1^2}{\hat{s}^2} 
+ \frac{2m_Q^2}{\hat{s}} \right) 
\delta(\hat{s} + \hat{t}_1 + \hat{u}_1) \, \, , \label{qqpartres} \\
\hat{s}^2 \frac{d^2 \sigma_{gg}}{d\hat{t}_1 d\hat{u}_1} & = & \frac{\pi 
\alpha_s^2(Q^2)}{16}
B_{\rm QED} (\hat{s},\hat{t}_1,\hat{u}_1) \nonumber \\
& & \mbox{} \times \left[ 3 \left( 1 - 
\frac{2\hat{t}_1 \hat{u}_1}{\hat{s}^2} \right)
- \frac{1}{3} \right] \delta(\hat{s} + \hat{t}_1 + \hat{u}_1)\, \, ,
\label{ggpartres}
\end{eqnarray}
where $\hat{s} = (xk + x_2P_2)^2$, $\hat{t}_1 = (xk - p_1)^2 - m_Q^2$,
and $\hat{u}_1 = (x_2P_2 - p_1)^2 - m_Q^2$.  The $gg$ partonic cross
section, Eq.~(\ref{ggpartres}), is proportional to the photon-gluon 
fusion cross section, Eq.~(\ref{gamApart}), with an additional factor 
for the non-Abelian
three-gluon vertex.  The $q \overline q$ annihilation cross section
has a different structure because it is a $\hat s$-channel process with
gluon exchange between the $q \overline q$ and $Q \overline Q$
vertices.  Modulo the additional factor in the $gg$ cross
section, the resolved partonic photoproduction cross sections are a
factor $\alpha_s(Q^2)/\alpha e_Q^2$ larger than the direct, $\gamma
g$, partonic photoproduction cross sections.  
The cross section for resolved $Q \overline Q$ photoproduction, using
Eq.~(\ref{mainres}) with the $q \overline q$ and $gg$ channels, is
\begin{eqnarray}
\lefteqn{s_{_{NN}}^2\frac{d^2\sigma^{\rm res}_{\gamma A \rightarrow 
Q \overline Q
X}}{dt_{1\, _{NN}} du_{1\, _{NN}}} = 2 \int_{k_{\rm min}}^\infty 
\frac{dk}{k} {dN_\gamma\over dk} \int_{k_{\rm min}/k}^1 \frac{dx}{x}
\int_{x_{2_{\rm min}}}^1 \frac{dx_2}{x_2}} \nonumber \\
&  & \mbox{} \times \left[ F_g^\gamma (x,Q^2) 
F_g^A(x_2,Q^2)  \hat{s}^2 
\frac{d^2 \sigma_{gg}}{d\hat{t}_1 d\hat{u}_1} \right. \nonumber  \\
&  & \mbox{} + \left. \sum_{q=u,d,s} F_q^\gamma (x,Q^2) 
\left\{
F_q^A(x_2,Q^2) + F_{\overline q}^A(x_2,Q^2) \right\}
\hat{s}^2 \frac{d^2 \sigma_{q \overline q}}{d\hat{t}_1 
d\hat{u}_1} \right] \, \, .
\label{qqbres}
\end{eqnarray}

\begin{figure}[htb]
\setlength{\epsfxsize=0.95\textwidth}
\setlength{\epsfysize=0.5\textheight}
\centerline{\epsffile{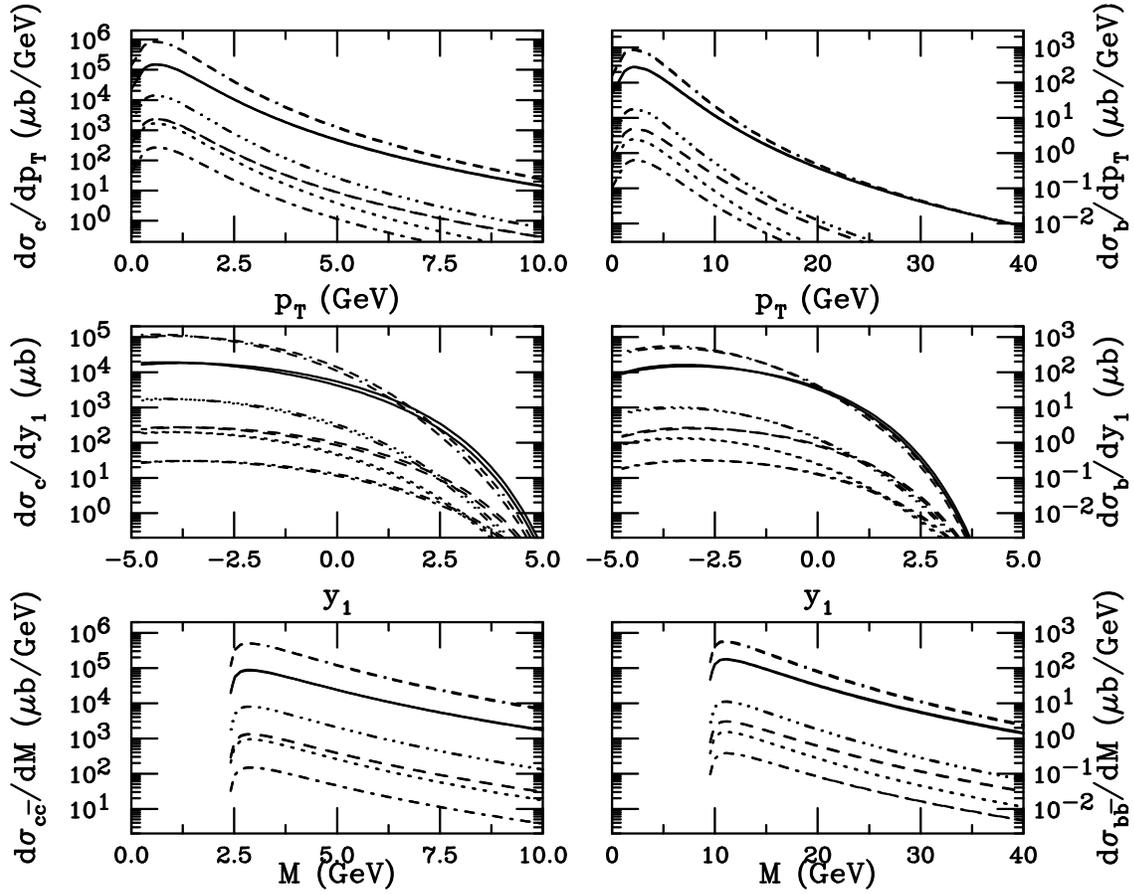}}
\caption[]{Resolved $Q \overline Q$ photoproduction in peripheral 
$AA$ collisions.  The left-hand side shows the results for charm while the
right-hand side gives the results for bottom.  The single $Q$ $p_T$ (upper)
and rapidity (middle) distributions are shown along with the $Q \overline Q$ 
pair invariant mass distributions (lower).  The results for the GRV-G 
(O+O (dot-dashed), Ar+Ar (dashed) and Pb+Pb (solid)) and LAC1 (O+O (dotted),
Ar+Ar (dot-dot-dot-dashed) and Pb+Pb (dash-dash-dash-dotted)) photon parton
densities are given.  There are two curves
for each contribution: no shadowing and EKS98.  At $y_1 > 0$, the highest
curve is without shadowing.  The photon is coming from the left.}
\label{distcbres}
\end{figure}

Figure~\ref{distcbres} shows the charm and bottom resolved photoproduction 
distributions in Pb+Pb, Ar+Ar and O+O collisions.  
The resolved top photoproduction distributions for these three systems are 
given on the right-hand side of Fig.~\ref{disttdirres}.
There are four curves for each contribution, two with the GRV-G set of
photon parton distribution functions and two with the LAC1 set.  One the two
curves for each set of photon parton distributions
is without shadowing while the other employs the EKS98 parametrization.

The difference between the two photon parton densities is quite large at
negative rapidities where the parton $x$ entering the photon parton
distribution is small.  The LAC1 resolved cross sections are largest here.
The GRV-G result is slightly larger at
forward rapidities although the results for the two sets are similar.
This crossover point occurs at larger forward rapidities for lighter nuclei
where the energy is higher.  For $c \overline c$ production, it occurs at
$y_1 \approx 1.75$ for Pb+Pb, 2.5 for Ar+Ar and 3 for O+O.  The larger $x$ of
$b \overline b$ production moves the crossover point backwards to $y_1 \approx
0.25$ for Pb+Pb, 1 for Ar+Ar and 1.5 for O+O, a shift of around 1.5 units
between charm and bottom production.  
At high $p_T$ and $M$, the GRV-G and LAC1 resolved distributions in Pb+Pb 
collisions approach each
other, showing that the differences in the two sets are reduced at high
scales.  The approach is more gradual for the higher energy light
ion collisions.  The same trend is seen in the mass distributions.
The GRV-G and LAC1 results are
indistinguishable for resolved top production.

The large difference in the resolved results will strongly influence whether
the direct or resolved contribution is greater.  This, in turn, directly
affects the capability to clearly measure the nuclear gluon distribution.
Thus Table~\ref{gamAqqres} compares the total $Q \overline Q$ resolved 
cross sections.  The LAC1 $c \overline c$ resolved cross sections are
$5-6$ times higher than the GRV-G cross sections while the difference for $b
\overline b$ production is a factor of $2.8-3.6$.  In both cases, the smallest
difference is for the heaviest ions, hence for the lowest energy and highest
$x$ values.  The
difference in the $t \overline t$ resolved cross sections is on the few per
cent level and is therefore negligible.

We may now compare these resolved cross sections to the direct cross sections
in Table~\ref{gamAqqdir}.  With the GRV-G set,
the resolved contributions are
$\sim 15$ and 20\% of the total charm and bottom photoproduction cross
sections respectively, comparable to the shadowing effect on direct production.
However, with the LAC1 set, the resolved contribution is equivalent
to or larger than the direct.  A measurement of the photon parton distributions
at low $x$ and $Q$ 
will thus be important for a precision measurement of gluon shadowing.

\begin{table}[htbp]
\begin{center}
\caption[]{Resolved $Q\overline Q$ photoproduction
cross sections integrated over $b>2R_A$ in peripheral $AA$ collisions.}
\label{gamAqqres}
\vspace{0.4cm}
\begin{tabular}{|c|c|c|c|c|c|} \hline
& \multicolumn{5}{c|}{$\sigma^{\rm res}$ (mb)} \\ \hline
& \multicolumn{3}{c|}{GRV-G} & \multicolumn{2}{c|}{LAC1} \\ \hline
$AA$ & no shad & EKS98 & FGS & no shad & EKS98 
\\ \hline
\multicolumn{6}{|c|}{$c \overline c$} \\ \hline
O+O   & 0.351   & 0.346 & 0.331 & 2.04 & 2.02 \\
Ar+Ar & 3.00    & 2.93  & 2.77  & 16.6 & 16.6 \\
Pb+Pb & 190   & 187 & 174 & 987  & 1007 \\ \hline
\multicolumn{6}{|c|}{$b \overline b$} \\ \hline
O+O   & 0.0029 & 0.0029 & 0.0029 & 0.0105 & 0.0106 \\
Ar+Ar & 0.0222 & 0.0226 & 0.0224 & 0.073  & 0.075 \\
Pb+Pb & 1.21   & 1.26   & 1.25   & 3.41   & 3.66 \\ \hline
\multicolumn{6}{|c|}{$t \overline t$} \\ \hline
O+O   & $2.81 \times 10^{-10}$ & $2.76 \times 10^{-10}$ & $-$ & $2.92 \times
10^{-10}$ & $2.88 \times 10^{-10}$ \\
Ar+Ar & $1.08 \times 10^{-9}$ & $1.04 \times 10^{-9}$ & $-$ & $1.09 \times 
10^{-9}$ & $1.05 \times 10^{-9}$ \\
Pb+Pb & $1.60 \times 10^{-8}$ & $1.48 \times 10^{-8}$ & $-$ & $1.62 \times 
10^{-8}$ & $1.49 \times 10^{-8}$ \\ \hline
\end{tabular}
\end{center}
\end{table}

\begin{figure}[htb]
\setlength{\epsfxsize=0.95\textwidth}
\setlength{\epsfysize=0.5\textheight}
\centerline{\epsffile{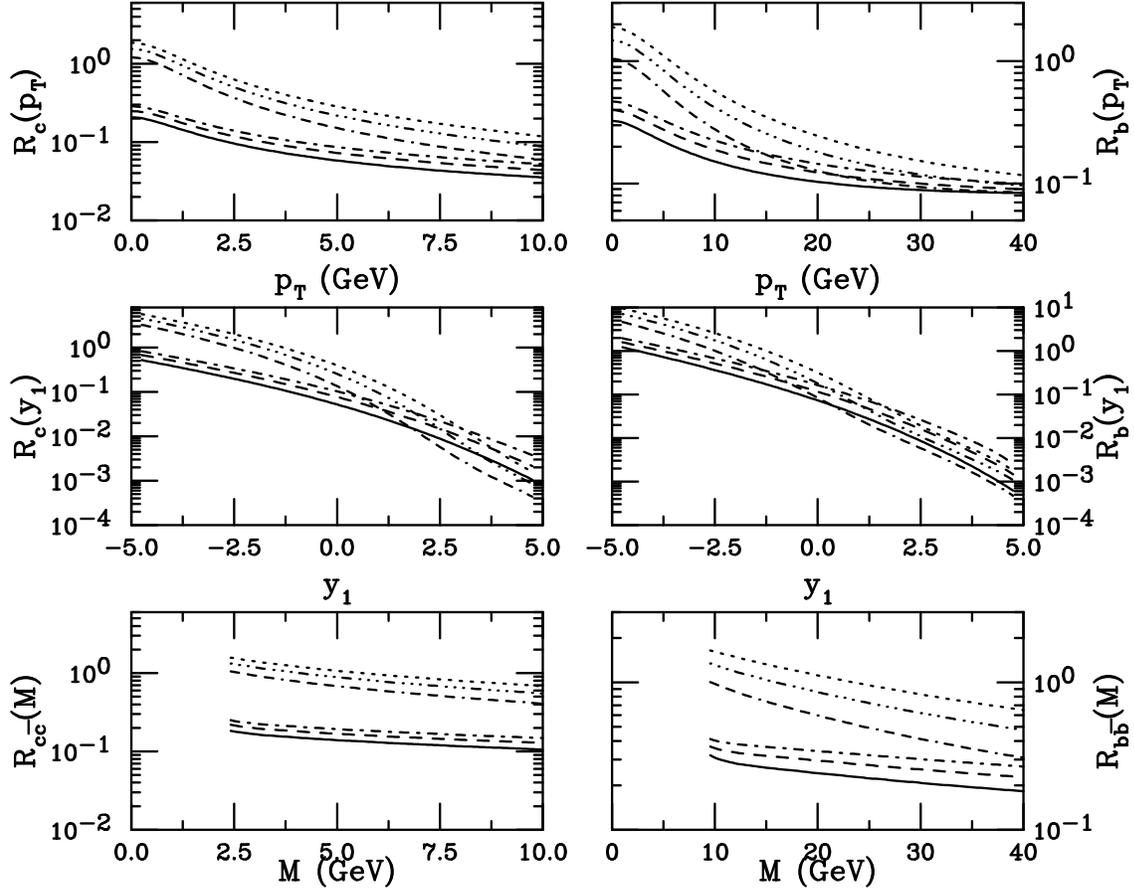}}
\caption[]{Resolved to direct $Q \overline Q$ photoproduction ratio
in peripheral 
$AA$ collisions.  The left-hand side shows the results for charm while the
right-hand side gives the results for bottom.  The EKS98 shadowing
parametrization is used in both cases.  The single $Q$ $p_T$ (upper)
and rapidity (middle) ratios are shown along with the $Q \overline Q$ 
pair invariant mass ratios (lower).   The results for the GRV-G
(O+O (dot-dashed), Ar+Ar (dashed) and Pb+Pb (solid)) and LAC1 (O+O (dotted),
Ar+Ar (dot-dot-dot-dashed) and Pb+Pb (dash-dash-dash-dotted)) photon parton
distributions are given. The photon is coming from the left.}
\label{cbresvsdir}
\end{figure}

Figure~\ref{cbresvsdir} shows the ratio of the resolved to direct charm 
and bottom production cross sections, $R_Q$, for the LAC1 and GRV-G 
photon parton densities.
If $R_Q <1$, direct production dominates and the most
direct information on the gluon distribution in the nucleus can be obtained.
The middle plots show that even the GRV-G resolved
contribution is equivalent to the direct one at large negative rapidity.
However, direct production dominates at $y_1 > 0$ for both photon parton
densities.  In this region, the resolved rapidity distribution drops rather
steeply with respect to the direct.  Recall that the direct rapidity
distribution is still rather broad at forward rapidities, see
Fig.~\ref{distcbdir}, particularly for charm.  This is also the region where
the nuclear gluon $x$ values are smallest.  In addition, at
sufficiently large $p_T$, direct production dominates.  The upper
plots of Fig.~\ref{cbresvsdir} show that $p_T \approx 2.5$ GeV should be
sufficient for charm while 10 is needed for bottom if the LAC1 set is employed.
There is no restriction on the $p_T$ range if the GRV-G set is more correct.
Better measurements of the photon parton densities should help settle this
issue.

Nuclear shadowing can also be studied in the resolved contribution although the
contribution from the gluon is now only a portion of the total.  In resolved
photoproduction, the $q \overline q$ channel gives a larger contribution than
in hadroproduction because the photon quark and antiquark distributions peak
at large $x$.  Indeed, the peak of the photon quark distribution is at higher
$x$ than the valence quark distribution in the nucleon.  Thus the $q \overline
q$ contribution increases close to threshold.  Although $c \overline c$ and $b
\overline b$ resolved photoproduction is not very near threshold, the effective
center-of-mass energies are reduced relative to $\sqrt{S_{\gamma N}}$ in
Table~\ref{gamfacs}. 

\begin{figure}[htb]
\setlength{\epsfxsize=0.95\textwidth}
\setlength{\epsfysize=0.5\textheight}
\centerline{\epsffile{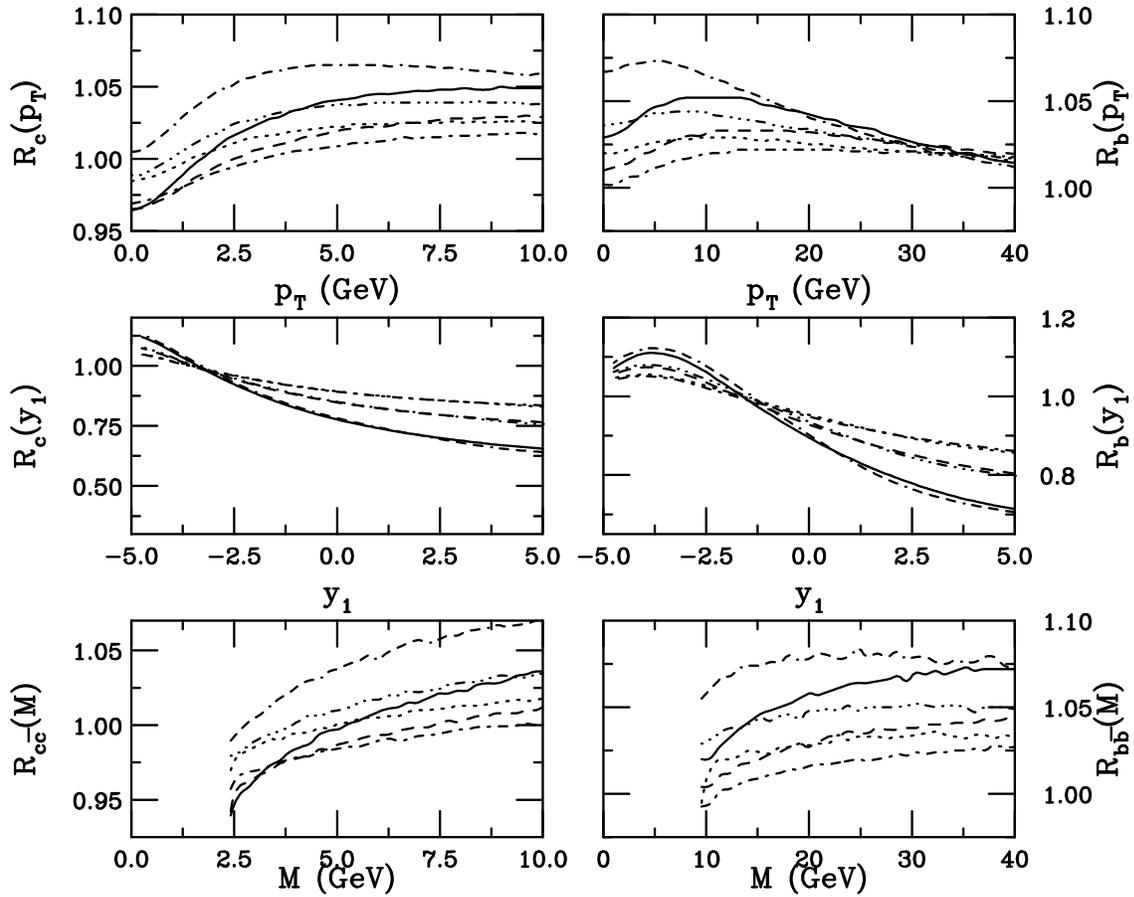}}
\caption[]{Shadowing in resolved $Q \overline Q$ photoproduction in peripheral 
$AA$ collisions.  The left-hand side shows the results for charm while the
right-hand side gives the results for bottom.  The EKS98 shadowing
parametrization is used in both cases.  The single $Q$ $p_T$ (upper)
and rapidity (middle) ratios are shown along with the $Q \overline Q$ 
pair invariant mass ratios (lower).   The results for the GRV-G
(O+O (dot-dashed), Ar+Ar (dashed) and Pb+Pb (solid)) and LAC1 (O+O (dotted),
Ar+Ar (dot-dot-dot-dashed) and Pb+Pb (dash-dash-dash-dotted)) photon parton
distributions are given. The photon is coming from the left.}
\label{shadcbres}
\end{figure}

Shadowing effects on the resolved contributions are shown in
Fig.~\ref{shadcbres} for the EKS98 parametrization.  The direct and
resolved ratios as a function of rapidity are remarkably similar, especially
for charm, as seen in a comparison of the
middle plots of Fig.~\ref{shadcbres} to those of Fig.~\ref{shadcbdir}.  The
similarity of the shadowing ratios
may be expected since the rapidity distributions best reflect the $x$ values of
the nuclear parton densities probed.  The additional $q \overline q$
contribution to resolved $b \overline b$ production is larger than for charm,
large enough to cause the small difference in $b \overline b$ shadowing between
direct and resolved production.  The basic kinematics of the partonic
interactions are the same for the nuclear partons but the momentum
entering the photon flux
is effectively changed by the ratio $k/k_{\rm min}$.  Thus
a higher momentum photon is needed to produce the effective $x$ entering the
photon parton momentum distributions.  As seen in Fig.~\ref{avexcbt}, there is
little difference in $\langle x_Q \rangle$ of the nucleon
between direct and resolved
photoproduction. 

The quark transverse momentum and pair
mass distributions are, however, more strongly affected since they are 
integrated over all rapidity.  The shift of the peak of the
resolved rapidity distributions to more negative rapidities increases the
average $x$ probed by the $p_T$ and $M$ distributions, decreasing the effects
of shadowing on these distributions.  This increase in $\langle x_Q \rangle$
as a function of $p_T$
due to the peak of the rapidity distribution is shown in Fig.~\ref{avexcbt}
for the GRV-G photon parton distribution.
Since the peak of the rapidity
distribution with the LAC1 set is at even larger negative rapidity than the
GRV-G, corresponding to larger nuclear momentum fractions,
the shadowing ratios in this case are larger as well.

The total rates are given in Table~\ref{gamAqqrates} for all three collision
systems, assuming a $10^6$ s run for each system.  
We have included the rates for both the GRV-G and LAC1
photon parton distribution sets.  The difference in rate can be up to a factor
of two when the LAC1 resolved cross sections are used.  The top rates,
particularly for the lighter nuclei, are lower than in
Ref.~\cite{Klein:2000dk}, due to a revision in the expected LHC light ion 
luminosities.  Thus using top quarks to measure the nuclear gluon distribution
at high $Q^2$ seems unlikely.

\begin{table}[htbp]
\begin{center}
\caption[]{Total $Q\overline Q$ photoproduction rates, integrated over 
$b>2R_A$ in peripheral $AA$ collisions. The rates are based on 
Tables~{\protect \ref{gamAqqdir}} and {\protect \ref{gamAqqres}} for a $10^6$ s
run.}
\label{gamAqqrates}
\vspace{0.4cm}
\begin{tabular}{|c|c|c|c|c|c|} \hline
\multicolumn{6}{|c|}{$N_{Q \overline Q}$} \\ \hline
& \multicolumn{3}{c|}{GRV-G} & \multicolumn{2}{c|}{LAC1} \\ \hline
$AA$ & no shad & EKS98 & FGS & no shad & EKS98 
\\ \hline
\multicolumn{6}{|c|}{$c \overline c$} \\ \hline
O+O   & $3.20 \times 10^8$ & $2.96 \times 10^8$ & $2.69 \times 10^8$ 
& $5.92 \times 10^8$ & $5.63 \times 10^8$ \\
Ar+Ar & $8.30 \times 10^8$ & $7.40 \times 10^8$ & $6.49 \times 10^8$
& $1.41 \times 10^9$ & $1.33 \times 10^9$ \\
Pb+Pb & $6.03 \times 10^8$ & $5.20 \times 10^8$ & $4.30 \times 10^8$
& $9.38 \times 10^8$ & $8.64 \times 10^8$ \\ \hline
\multicolumn{6}{|c|}{$b \overline b$} \\ \hline
O+O   & $1.76 \times 10^6$ & $1.71 \times 10^6$ & $1.66 \times 10^6$ 
& $2.98 \times 10^6$ & $2.94 \times 10^6$ \\
Ar+Ar & $4.09 \times 10^6$ & $3.98 \times 10^6$ & $3.81 \times 10^6$
& $6.28 \times 10^6$ & $6.24 \times 10^6$ \\
Pb+Pb & $2.56 \times 10^6$ & $2.51 \times 10^6$ & $2.38 \times 10^6$
& $3.07 \times 10^6$ & $3.52 \times 10^6$ \\ \hline
\multicolumn{6}{|c|}{$t \overline t$} \\ \hline
O+O   & 1.53 & 1.53 & - & 1.53 & 1.53 \\
Ar+Ar & 1.28 & 1.28 & - & 1.28 & 1.28 \\
Pb+Pb & 0.14 & 0.14 & - & 0.14 & 0.14 \\ \hline
\end{tabular}
\end{center}
\end{table}

These rates are based on total cross
sections, without any acceptance cuts or detector requirements included.  The
largest difference between the GRV-G and LAC1 resolved cross
sections lies at $y_1 < -2.5$.  Thus, for central detectors, the difference
in the rates would be reduced.  However, if heavy quark photoproduction is
also studied through its leptonic decays, the forward muon arm of ALICE, $2.4 <
y < 4$, would be sensitive to the photon parton densities when the photon comes
from the right-hand side, the mirror image of the plots.

The high $c \overline c$ and $b \overline b$ rates provide sufficient
statistics to distinguish between shadowing parameterizations and perhaps the 
photon parton distributions as well, even with finite acceptance.  The
rates for the different systems are all of the same order of magnitude.  The
lower photon flux for light, smaller $Z$ ions is compensated by
their higher luminosities.  Unfortunately
however, the top rates are discouragingly low, even for light ions.

There are a number of theoretical uncertainties in the
calculations shown here aside from the obvious ones in the nuclear and photon
parton densities and the photon flux.  
First, the calculation is to leading order only.
Higher order corrections can be significant, with large 
theoretical $K$ factors, and can affect the shape of the
distributions, particularly at large $p_T$, see Ref.~\cite{Smith:1991pw} for
a discussion of NLO photoproduction and Ref.~\cite{Vogt:2002eu} 
for discussions of
the $K$ factors in heavy quark hadroproduction.  There is also some 
uncertainty in the heavy quark mass and scale parameters.  Variations in the
mass generally cause larger changes in the cross sections than scale
variations.  Reference~\cite{Klein:2002wm} has 
a more detailed discussion of the uncertainties. 

One way to avoid some of these calculational uncertainties is
to compare the $pA$ and $AA$ photoproduction
cross sections at equal photon energies.  The parameter dependence
cancels in the direct $Q \overline Q$ production
ratio $\sigma(AA)/\sigma(pA)$.  In the equal speed
system, equal photon energies correspond to the same final-state
rapidities.  In $pA$ collisions, the photon almost always comes from
the nucleus due to its stronger photon field.  Thus the $pA$ rates 
depend on the free proton gluon
distribution.  The photon fluxes are different for $pA$ and $AA$
because the minimum radii used to determine $\omega_R$ are different:
$2R_A$ in $AA$ rather than $R_A+r_p$ in $pA$.  We use $r_p \approx 0.6$ fm
for the proton radius \cite{Breitweg:1997ed} but our results do not depend
strongly on $r_p$.

The $pA$ results are calculated in the equal-speed frame.  
At the LHC, the proton and nuclear beams must have the
same magnetic rigidity and, hence, different velocities and
per-nucleon energies.  Thus the equal-speed frame is boosted with
respect to the
laboratory frame and the maximum $pA$ energy per nucleon 
is larger than the $AA$ energy.  The $\gamma_L$ and $\sqrt{s_{_{NN}}}$ 
for $pA$ at the LHC in Table~\ref{gamfacs} are those of the equal-speed
system.  The $pA$ total cross sections for $Q \overline Q$ production 
are given in Table~\ref{gampccbb}.

\begin{table}[htbp]
\begin{center}
\caption[]{Direct and resolved $c\overline c$ and $b \overline b$ 
photoproduction cross sections integrated over $b > r_p + R_A$ 
in $pA$ collisions at the LHC.  }
\label{gampccbb}
\vspace{0.4cm}
\begin{tabular}{|c|c|c|c|} \hline
$pA$ & $\sigma^{\rm dir}$ ($\mu$b) & $\sigma^{\rm res}({\rm GRV})$ ($\mu$b) 
& $\sigma^{\rm res}({\rm LAC1})$ ($\mu$b) \\ \hline
\multicolumn{4}{|c|}{$c \overline c$} \\ \hline
$p$O  & 75.5  & 19.7 & 120.6 \\
$p$Ar & 335   & 81.1 & 486.3 \\
$p$Pb & 5492  & 1160 & 6371  \\ \hline
\multicolumn{4}{|c|}{$b \overline b$} \\ \hline
$p$O  & 0.419 & 0.190 & 0.773 \\
$p$Ar & 1.775 & 0.739 & 2.886 \\
$p$Pb & 26.83 & 9.60  & 34.68 \\ \hline
\multicolumn{4}{|c|}{$t \overline t$} \\ \hline
$p$O  & $1.54 \times 10^{-6}$ & $4.00 \times 10^{-8}$ & $4.00 \times 10^{-8}$
\\ 
$p$Ar & $4.40 \times 10^{-6}$ & $1.23 \times 10^{-7}$ & $1.24 \times 10^{-7}$
\\
$p$Pb & $3.00 \times 10^{-5}$ & $9.74 \times 10^{-7}$ & $9.86 \times 10^{-7}$ \
\\ \hline
\end{tabular}
\end{center}
\end{table}

The total $pA$ cross sections are generally smaller than the $AA$ cross
sections in Tables~\ref{gamAqqdir} and \ref{gamAqqres}.  Here the
cross sections are given in $\mu$b while the $AA$ cross sections are in units
of mb.  In hadroproduction, without shadowing, at the same energies, the
$AA/pA$ cross section ratio is $A$.  In photoproduction, the corresponding
ratio would be $2A$ since there is only a single photon source in $pA$.
Since the $AA$ and $pA$ results are given at different center-of-mass
energies, the comparison is not so straightforward.  However, even if the
$pA$ energy is lowered to that of the $AA$ collisions, the $AA/pA$ ratio is
larger than $2A$.  The difference in flux due to the change in the minimum
impact parameter from $2R_A$ in $AA$ to $r_p + R_A$ in $pA$ accounts for most
of the difference, especially for lighter nuclei where the difference $R_A -
r_p$ is smallest.  Reducing the minimum impact parameter increases the photon
flux \cite{Klein:2002wm}.

\begin{table}[htbp]
\begin{center}
\caption[]{Total $Q\overline Q$ photoproduction
rates in peripheral $pA$ collisions over a $10^6$ s run at the LHC.  
The rates are based on Table~{\protect \ref{gampccbb}}.} 
\label{gampqqrates}
\vspace{0.4cm}
\begin{tabular}{|c|c|c|c|c|c|c|} \hline
& \multicolumn{2}{c|}{$N_{c \overline c}$} & \multicolumn{2}{c|}{$N_{b 
\overline b}$} & \multicolumn{2}{c|}{$N_{t \overline t}$} \\ \hline
$pA$ & GRV-G & LAC1 & GRV-G & LAC1 & GRV-G & LAC1 \\ \hline
$p$O  & $9.52\times 10^8$ & $1.96\times 10^9$ & $6.09\times 10^6$ 
& $1.19\times 10^7$ & 15.8 & 15.8 \\
$p$Ar & $2.41\times 10^9$ & $4.76\times 10^9$ & $1.46\times 10^7$ 
& $2.70\times 10^7$ & 33.4 & 33.4 \\
$p$Pb & $4.92\times 10^9$ & $8.78\times 10^9$ & $2.79\times 10^7$ 
& $4.55\times 10^7$ & 23 & 23 \\ \hline
\end{tabular}
\end{center}
\end{table}

Clearly a $pA$ run at the same energy as $AA$ would reduce the uncertainties
due to the energy difference.  The same $x$ range of the proton and nuclear
gluon distributions would then be probed.  Such runs are possible but
with a loss of luminosity, leading to a reduction in the $pA$ rates.  However,
the $pA$ rates shown in Table~\ref{gampqqrates} are high--even higher 
than the $AA$ rates in the
same $10^6$ s interval thanks to the higher maximum $pA$ luminosities.  
Including the
reduced rates due to the lower energy, the $pA$ luminosity could be
significantly reduced without lowering the $c \overline c$ and $b \overline b$
statistics.  Then the only major
uncertainty would be the difference in photon flux between $AA$ and $pA$
interactions.  The uncertainties in the photon flux could be reduced by
measurements of other baseline processes, allowing a cleaner comparison.

The relatively higher $t \overline t$ rate in $pA$ collisions suggests that
the top quark charge of $2/3$ could be confirmed.  While less than 100 
$t \overline t$
pairs are produced in a $10^6$ s $pA$, run, essentially all the pairs fall 
into the $|y| \leq 2.5$ region.  Thus, ideally, a difference in rate by a 
factor of four due to $e_t^2$ could be detected in a single $pA$ run although
the combined results of several years of $pA$ runs would be better.

\subsection{Dijet photoproduction}
\label{jet}
{\it Contributed by: R. Vogt}

We now consider jet photoproduction in peripheral $AA$ and $pA$
interactions. In central collisions at RHIC, leading particles in jets are 
easier to detect above the high charged particle multiplicity background than
the jets themselves since these high $p_T$
particles can be tracked through the detector \cite{Adler:2002tq,Chiu:2002ma}.
In peripheral collisions, especially at LHC energies, jets should be easier
to isolate and may be observed directly using standard high energy jet 
detection
techniques \cite{Blyth}.  
We thus discuss the leading order $p_T$ distributions of jets
as well as leading particles.  We work at LO to avoid any ambiguities such as
jet reconstruction and cone size.  Note, however that the $p_T$
distributions are likely harder at NLO even though the $K$ factor appears to 
be relatively constant at high $p_T$ \cite{Eskola:cj}.
We also discuss the fragmentation of
jets and present the leading particle transverse momentum distributions, 
specifically charged pions, kaons and protons.

The hadronic reaction we study is $\gamma(k) + N(P_2) \rightarrow {\rm
jet}(p_1) \, 
+ \, {\rm jet}(p_2) \, + \, X$ where $k$ and $P_2$ are the photon and 
nucleon four-momenta.  The two parton-level
contributions to the jet yield in direct photoproduction 
are $\gamma(k) + g(x_2P_2) \rightarrow q(p_1) + \overline q(p_2)$ 
and $\gamma(k) + q(x_2P_2) \rightarrow g(p_1) + q(p_2)$ where also $q
\rightarrow \overline q$.  The produced partons are 
massless, requiring a minimum $p_T$ to keep the cross section finite.  
At LO, the jet yield is equivalent to the high $p_T$ parton yield.
The jet $p_T$ distribution is modified for photoproduction from {\it e.g.} 
Refs.~\cite{Emel'yanov:1999bn,Eskola:1988yh,Eskola:1996ce},  
\begin{eqnarray}
\label{jetdireq} s_{_{NN}}^2
\frac{d^2\sigma_{\gamma A \rightarrow {\rm jet} \, + \, {\rm jet}+ X}^{\rm 
dir}}{dt_{_{NN}} du_{_{NN}}} & = & 
2 \int_{k_{\rm min}}^\infty dk \frac{dN_\gamma}{dk}
\int_{x_{2_{\rm min}}}^1 \frac{dx_2}{x_2} \nonumber \\ 
& & \mbox{} \times  \bigg[ \sum_{i,j,l =
q, \overline q, g} F_i^A(x_2,Q^2) s^2 \frac{d^2\sigma_{\gamma i 
\rightarrow jl}}{dt du} \bigg]
\, \,  ,
\end{eqnarray}
where $x_2$ is the fraction of the initial hadron momentum carried by the 
interacting parton and $Q$ is the momentum scale of the interaction.
The extra factor of two on the right-hand side of
Eq.~(\ref{jetdireq}) arises because 
both nuclei can serve as photon sources in $AA$ collisions.  The partonic
cross sections are
\begin{eqnarray}
s^2 \frac{d^2 \sigma_{\gamma g \rightarrow q \overline q}}{dt du} & = 
& \pi \alpha_s(Q^2) \alpha e_Q^2  
\bigg[\frac{t^2 +u^2}{t u}  \bigg] \delta(s + t + u) \, \, , \label{gamApartj}
\\ 
s^2 \frac{d^2\sigma_{\gamma q \rightarrow gq}}{dt du} & = & - \frac{8}{3} \pi 
\alpha_s(Q^2)
\alpha e_Q^2 \bigg[ \frac{s^2 + t^2}{s t}\bigg] \delta(s + t + u) \, \, .
\label{QCDcompt}
\end{eqnarray}
The first is the photon-gluon fusion cross section, the only contribution to
massive
$Q \overline Q$ photoproduction \cite{Klein:2002wm}, while the second is the
QCD Compton process.
At LO, the partonic cross section is 
proportional to $\alpha \alpha_s(Q^2) e_Q^2$, where
$\alpha_s(Q^2)$ is the strong coupling constant to one loop, 
$\alpha=e^2/\hbar c$ is the
electromagnetic coupling constant, and $e_Q$ is the quark charge, $e_u = 
e_c = 2/3$
and $e_d = e_s = -1/3$. The partonic invariants, $s$,
$t$, and $u$, are defined as $s = (k + x_2 P_2)^2$, $t = (k -
p_1)^2 = (x_2 P_2 - p_2)^2$, and $u = (x_2 P_2 -
p_1)^2 = (k - p_2)^2$.  In this case, $s = 4k \gamma_L
x_2m_p$ where $\gamma_L$ is the Lorentz boost of a single beam and
$m_p$ is the proton mass.  Since $k$ can be a continuum of
energies up to $E_{\rm beam} = \gamma_L m_p$, we define $x_1 = k/P_1$,
analogous to the parton momentum fraction in the nucleon
where $P_1$ is the nucleon
four momentum. For a detected parton in a nucleon-nucleon collision, the
hadronic invariants are then $s_{_{NN}} = (P_1 + P_2)^2$, $T = (P_2 - p_1)^2$,
and $U = (P_1 - p_1)^2$.

The produced parton rapidities are $y_1$ and $y_2$.
The parton rapidity is related to the invariant $t_{NN}$ by $t_{NN} = -
\sqrt{s_{_{NN}}} p_T e^{-y_1}$.  At LO, $x_1
= (p_T/\sqrt{s_{_{NN}}})(e^{y_1} + e^{y_2})$ and $x_2 =
(p_T/\sqrt{s_{_{NN}}})(e^{-y_1} + e^{-y_2})$.  We calculate $x_1$ and $x_2$ as
in an $NN$ collision and then determine the flux in the lab frame for
$k = x_1 \gamma_L m_p$, equivalent to the center-of-mass frame in a
collider.  The photon flux is exponentially suppressed for $k>\gamma_L
\hbar c/R_A$, corresponding to a momentum fraction $x_1 > \hbar
c/m_pR_A$.  The maximum $\gamma N$ center-of-mass energy,
$\sqrt{S_{\gamma N}}$, is much lower than the hadronic $\sqrt{s_{_{NN}}}$.  
The equivalent hadronic invariants can be defined for photon four
momentum $k$ as $s_{\gamma N} = (k + P_2)^2$, $t_{\gamma N} = (P_2 -
p_1)^2$, and $u_{\gamma N} = (k - p_1)^2$
\cite{Smith:1991pw}.  The partonic and equivalent hadronic invariants for fixed
$k$ are related by $s = x_2s_{\gamma N}$, $t = u_{\gamma N}$, and
$u = x_2 t_{\gamma N}$.

\begin{figure}[htb]
\setlength{\epsfxsize=0.95\textwidth}
\setlength{\epsfysize=0.45\textheight}
\centerline{\epsffile{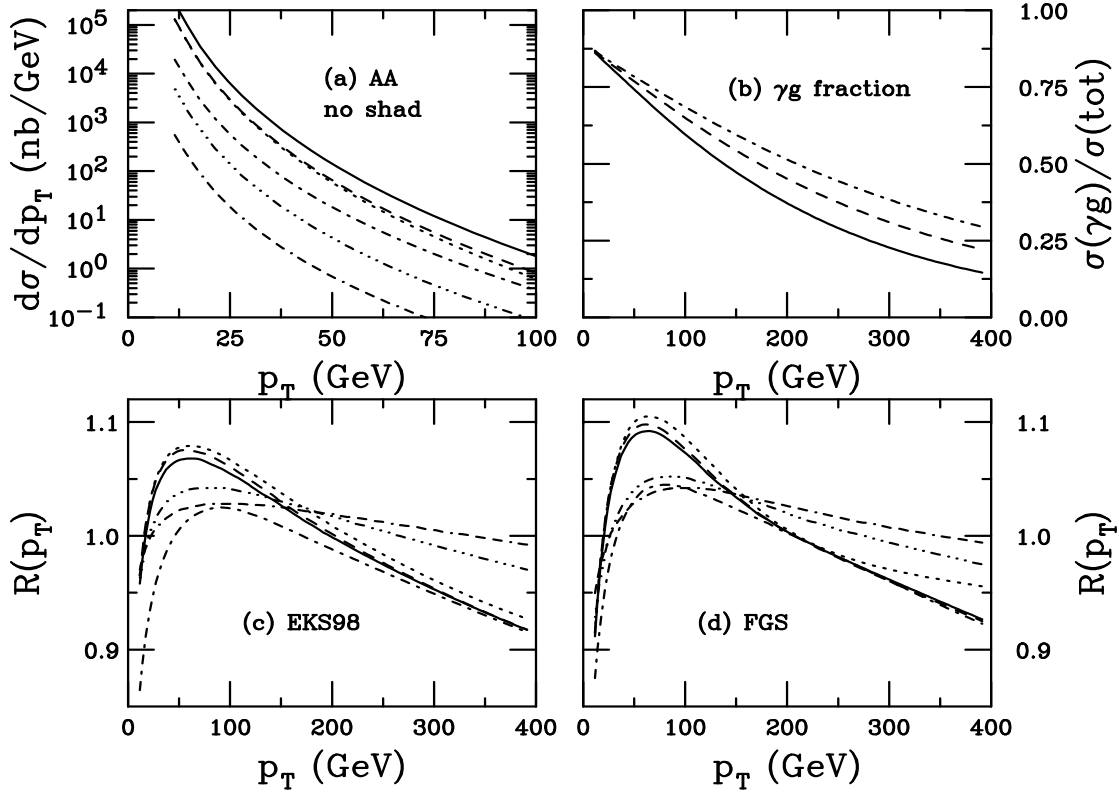}}
\caption[] {Direct dijet photoproduction in peripheral collisions. (a)
The $p_T$ distributions for $|y_1| \leq 1$ are shown for $AA$ 
collisions.  The solid curves is the total for Pb ions while the produced 
quarks (dashed), antiquarks (dotted) and gluons (dot-dashed) are shown
separately.  The total production for Ar (dot-dot-dot-dashed) and O
(dot-dash-dash-dashed) ions are also shown. (b) The fraction of gluon-initiated
jets as a function of $p_T$ for Pb+Pb (solid), Ar+Ar (dashed) and O+O
(dot-dashed) interactions. (c) The EKS98 shadowing ratios for produced partons.
The solid curve is the total for Pb ions while the ratios for produced 
quarks (dashed), antiquarks (dotted) and gluons (dot-dashed) are shown
separately.  The total ratios for Ar (dot-dot-dot-dashed) and O
(dot-dash-dash-dashed) ions are also shown. (d) The same as (c) for FGS.}
\label{jetdir}
\end{figure}

\begin{figure}[htb]
\setlength{\epsfxsize=0.95\textwidth}
\setlength{\epsfysize=0.55\textheight}
\centerline{\epsffile{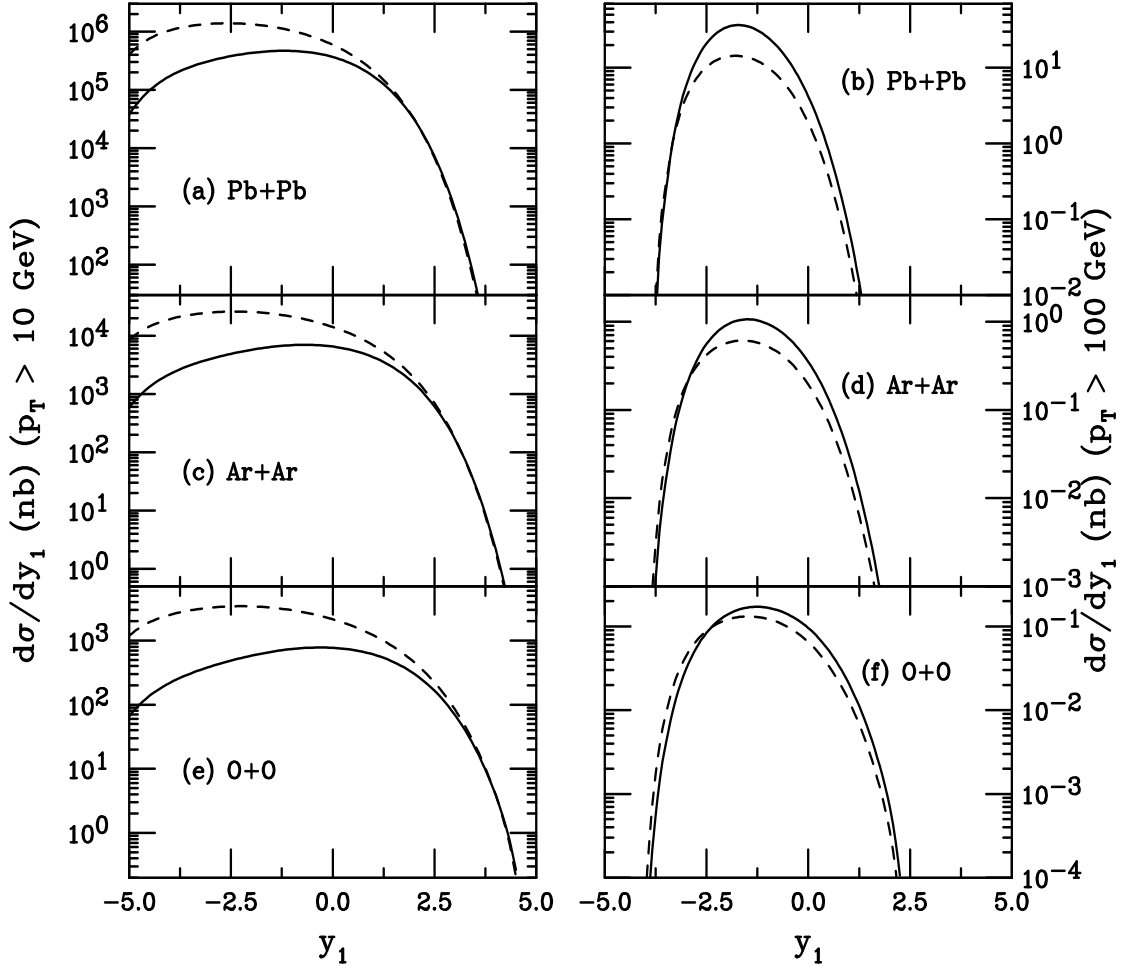}}
\caption[] {We compare the rapidity distributions of direct and resolved dijet
production without shadowing
in peripheral collisions. The left-hand side shows the results
for $p_T > 10$ GeV for (a) Pb+Pb, (c) Ar+Ar and (e) O+O collisions
while the right-hand side is for $p_T > 100$ GeV for (b) Pb+Pb, (d) Ar+Ar
and (f) O+O collisions.  The solid curves are 
the direct results while the dashed curves 
show the resolved results. The photon is coming from the left.}
\label{jet_dir_rap}
\end{figure}

The direct jet photoproduction $p_T$ distributions 
are given in Fig.~\ref{jetdir} for $AA$
interactions at the LHC.  For Pb+Pb collisions at $\sqrt{s_{_{NN}}} = 5.5$ 
TeV, we show the $p_T$ distributions of the produced quarks, 
antiquarks and gluons along with their sum.  For Ar+Ar collisions at
$\sqrt{s_{_{NN}}}=6.3$ TeV and O+O collisions 
at $\sqrt{s_{_{NN}}}=7$ TeV, we show only the total $p_T$ distributions.  
All the results are shown in the rapidity interval $|y_1|\leq 1$.
Extended rapidity
coverage, corresponding to  {\it e.g.} $|y_1| \leq 2.4$ for the CMS barrel and 
endcap systems, could increase the rates by a factor of $\approx 2$.  (The
increase in rate with rapidity acceptance is not linear in
$|y_1|$ because the rapidity distributions are asymmetric around $y_1 = 0$
and increasing $p_T$ narrows the rapidity distribution.  The effect of changing
the $y_1$ cut is closer to linear at low $p_T$ and larger at high $p_T$
because the peak is at $y_1 < -1$ for large $p_T$, as seen in the $y_1$ 
distributions on the right-hand side of Fig.~\ref{jet_dir_rap}.) 
At $p_T \approx 
100$ GeV, the cross sections are small.

There is a difference of $\approx 500$ in the Pb+Pb and O+O cross sections at
$p_T \approx 10$ GeV, decreasing to less than a factor of four at $\approx 400$
GeV.  The difference decreases with $p_T$ due to the larger phase space
available at high $p_T$ for the higher $\sqrt{s_{_{NN}}}$ systems.
The rates are nearly the same for all systems because the higher 
luminosities and higher $\sqrt{s_{_{NN}}}$ compensate
for the lower $A$ in lighter systems. 

The dijet jet hadroproduction cross sections are much higher because 
hard processes increase with the
number of binary collisions in $AA$ collisions, a factor of $\approx A^2$ 
for the minimum bias cross section.  (The relation is not exact
due to shadowing.)  Integration over all impact parameters leads
to $\approx A^2$ scaling while there is only a factor of $A$ in dijet
photoproduction since the photon flux is already integrated over
impact parameter for $b > 2R_A$.  This, combined with the lower effective
energy and fewer channels considerably reduces the photoproduction rates 
relative to hadroproduction. 

Quarks and antiquarks are produced in greatest abundance, with only a small
difference at high $p_T$.  Photon-gluon fusion alone
produces equal numbers of quarks and antiquarks.  The quark excess arises from
the QCD Compton diagram which also produces the small final-state 
gluon contribution.  The $\gamma (q + \overline q)$ 
contribution grows with $p_T$ since the valence quark distributions
eventually dominate production, 
as shown in Fig.~\ref{jetdir}(b) where the $\gamma g$
contribution is compared to the total.  At low $p_T$, the
$\gamma g$ contribution is $\approx 90$\% of the total, dropping to $10-30$\%
at $p_T \approx 400$ GeV.  At the large values of $x$ needed for high $p_T$ 
jet production, $f_p^{u_V} > f_p^g$.  
Thus the QCD Compton process eventually dominates dijet production,
albeit in a region of very low statistics.  The $\gamma g$ contribution is
larger for the lighter nuclei since the higher energies reduce the average $x$
values.  

The direct dijet photoproduction cross section
is significantly lower than the
hadroproduction cross section.  Some of this reduction is due to the different
couplings.  The photoproduction rate is reduced by a factor of $\alpha
e_Q^2/\alpha_s$, $\approx 100$.  There are also fewer 
diagrams for jet photoproduction relative to all $2
\rightarrow 2$ scatterings in hadroproduction.   In addition, the 
$gg \rightarrow gg$ hadroproduction process, with its high parton luminosity,
has no direct photoproduction equivalent.

Since the typical scales of jet production are large, 
the effects of shadowing, reflected in $R(p_T) = 
(d\sigma[S_A^i]/dp_T)/(d\sigma[S_A^i=1]/dp_T)$, are rather small, 
see Fig.~\ref{jetdir}(c) 
and (d), because the average $x$ is high.  The differences
between the two shadowing parameterizations are on the few percent level.
At low $p_T$, the produced quarks and antiquarks are mainly from gluons.
The produced gluons only come from quarks.  The peak for the produced quarks 
and antiquarks in $R(p_T)$
between $50 \leq p_T \leq 100$ GeV is
due to gluon anti-shadowing.
The total $R(p_T)$ for all produced partons in Pb+Pb collisions is dominated by
the $\gamma g$ contribution.  The maximum value of $S^i_A$ in the
antishadowing region is $\approx 1.07$ for
EKS98 and $\approx 1.1$ for FGS, reflecting the high $Q^2$ behavior of the
shadowing parameterizations.  

The EKS98 ratios for the produced quarks and
antiquarks in Fig.~\ref{jetdir}(c)
follow $R(p_T)$ for the total rather closely
over all $p_T$.  The quark and antiquark ratios 
are slightly above the total at low $p_T$ due to the
small $\gamma q$ contribution.  They continue to follow the total at high
$p_T$ since all the EKS98 ratios exhibit similar behavior at large $x$.  The
produced gluon ratio follows the quark ratios for
$p_T > 200$ GeV.  The
large $p_T$ contribution arises from the valence quarks.  Some antishadowing
is observed at low $p_T$, due to the valence quark contribution.  The total
ratios for the lighter ions are closer to unity for all $p_T$ due to their
smaller $A$.  

The results are similar for FGS, shown in Fig.~\ref{jetdir}(d), 
but with some subtle
differences.  The ratio $R(p_T)$ for produced gluons, arising 
from the $\gamma q$ contribution, exhibits a larger antishadowing effect on 
$R(p_T)$ because $S^{\overline q}_A$ 
is higher for this parametrization, 
see Fig.~\ref{shadcomp}.   The FGS antiquark shadowing
ratio goes to unity for $x>0.2$, causing the flattening of $R(p_T)$ for
antiquarks (dotted curve) due to the contribution from $\gamma \overline q
\rightarrow g \overline q$.  The FGS valence quark ratio is taken from
EKS98, resulting in the similarity of $R(p_T)$ in Figs.~\ref{jetdir}(c) and (d)
at high $p_T$.

Some care must be taken when
applying these parameterizations to high $p_T$ since the upper limit of their
fit range is 100 GeV.  While no extraordinary effects are seen in their 
behavior beyond this scale, 
the results should be taken as indicative only.  Finally, we
remark that we have only considered the range $|y_1|\leq 1$.  Including
contributions from all rapidities would increase the effect of shadowing since
large $|y_1|$ corresponds to smaller $x$ values.  

Figure~\ref{jet_dir_rap} shows the rapidity distributions with two 
different values of the minimum $p_T$, $p_T > 10$ GeV on the 
left-hand side and $p_T > 100$ GeV on the right-hand side.  
The results, given by the
solid curves, are shown without nuclear shadowing effects.  In
this case, the photon comes from the left.  There is a symmetric case where
the photon comes from the right, the factor of two on the $p_T$ distribution
in Eq.~(\ref{jetdireq}).  In this case, the $y_1$ distribution is reflected
around $y_1 = 0$.  With the 10 GeV cut, the distributions are rather broad
at negative $y_1$ where the photon has small momentum and, hence, large flux.
At large $y_1>0$, corresponding to small $x$ for the nucleon momentum fractions
and high photon momentum, the distributions fall rapidly since at high photon
momenta, the photon flux is cut off as $k \rightarrow k_{\rm max}$.  The
distributions with the 100 GeV cutoff are narrower because the edge of phase 
space is reached at lower values of $y_1$.  The rapidity distributions are 
broader in general for the lighter systems due to the higher 
$\sqrt{s_{_{NN}}}$.

\begin{figure}[htb]
\setlength{\epsfxsize=0.95\textwidth}
\setlength{\epsfysize=0.55\textheight}
\centerline{\epsffile{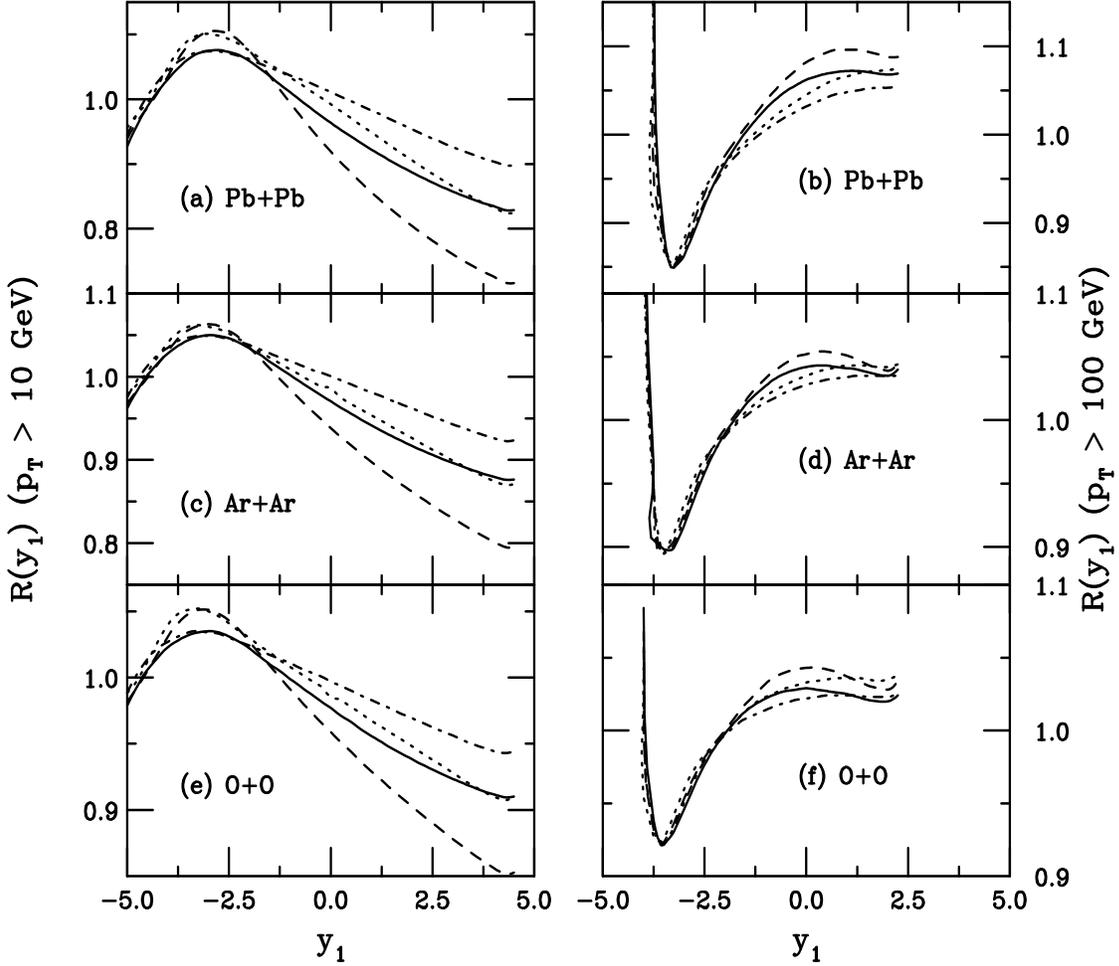}}
\caption[] {We compare shadowing ratios in direct and resolved dijet
production in peripheral collisions. The left-hand side shows the results
for $p_T > 10$ GeV for (a) Pb+Pb, (c) Ar+Ar and (e) O+O collisions
while the right-hand side is for $p_T > 100$ GeV for (b) Pb+Pb, (d) Ar+Ar
and (f) O+O collisions.  The solid and dashed curves give 
the direct ratios for the EKS98 and FGS parameterizations
respectively.  The dot-dashed and dotted curves 
show the resolved ratios for the EKS98 and FGS
parameterizations respectively.  The photon comes from the left.  Note the
difference in the $y$-axis scales here.}
\label{jet_dir_shad}
\end{figure}

Figure~\ref{jet_dir_shad} gives the ratio $R(y_1) = (d\sigma[S_A^i]/dy_1)/
(d\sigma[S_A^i = 1]/dy_1)$ for the two $p_T$ cuts.  The ratios reflect the 
direction of the photon, showing an antishadowing peak at $y_1 \sim -3$, an
EMC region at $y_1 < -4$ and a shadowing region for $y_1 > -0.5$ for $p_T >
10$ GeV, the left-hand side of Fig.~\ref{jet_dir_shad}.  The shadowing
effect is not large, $(20-25)$\% at $y_1
\sim 4$ for Pb+Pb collisions and decreasing with $A$.  The antishadowing peak
is higher for FGS while its shadowing effect is larger at positive $y_1$, as
also noted in the $p_T$-dependent ratios in Fig.~\ref{jetdir}.  A comparison
of the average effect around $|y_1| \leq 1$ with the $p_T$ ratios 
shown in Fig.~\ref{jetdir}, are in good agreement.  Even though $x_2$
is smaller for the lighter systems, the shadowing effect is also reduced.
Since shadowing also decreases with $Q^2$, the effect is even smaller for
$p_T > 100$ GeV, shown on the right-hand side of Fig.~\ref{jet_dir_shad}.  Here
the rise at $y_1 < -3.5$ is the Fermi motion as $x_2 \rightarrow 1$.  At 
$y_1 > -1.2$, the antishadowing region is reached.  The effect is rather small
here, only $\sim 5$\% at $y_1 \geq 0$.

We now turn to final-state particle production in the
hadronization of jets.  The particle with the highest $p_T$ is called the
``leading'' particle. 
The corresponding leading particle $p_T$ distribution
is \cite{Field:uq}
\begin{eqnarray}
\label{haddireq}
\frac{d\sigma_{\gamma A \rightarrow hX}^{\rm dir}}{dp_T} & = &
4 p_T \int_{\theta_{\rm min}}^{\theta_{\rm max}} 
\frac{d\theta_{\rm cm}}{\sin
\theta_{\rm cm}} \int dk \frac{dN_\gamma}{dk}
\int \frac{dx_2}{x_2}  \nonumber \\  & & \mbox{} \times \bigg[ \sum_{i,l =
q, \overline q, g} F_i^A(x_2,Q^2) 
\frac{d\sigma_{\gamma i \rightarrow lX'}}{dt} 
\frac{D_{h/l}(z_c,Q^2)}{z_c} \bigg] \, \, 
\end{eqnarray}
where the $X$ on the left-hand side includes all final-state hadrons in
addition to $h$ but
$X'$ on the right-hand side denotes the unobserved final-state parton.
The subprocess cross sections, $d\sigma/dt$, 
are related to 
$s^2 d\sigma/dt du$ in Eq.~(\ref{jetdireq})
through the momentum-conserving 
delta function $\delta(s + t + u)$ and division by $s^2$.
The integrals over rapidity have been replaced by an integral over
center-of-mass scattering angle, $\theta_{\rm min} \leq 
\theta_{\rm cm} \leq \theta_{\rm max}$, corresponding to a given
rapidity cut.  Here
$\theta_{\rm min} =0$ and $\theta_{\rm max} = \pi$ covers the full rapidity
range while $\theta_{\rm min} = \pi/4$ and $\theta_{\rm max} = 3\pi/4$ roughly
corresponds to $|y_1| \leq 1$. 
The fraction of the final
hadron momentum relative to that of the produced parton, $z_c$, appears 
in the fragmentation function, $D_{h/l}(z_c,Q^2)$, the probability to 
produce hadron
$h$ from parton $l$.  The fragmentation functions are assumed to be universal,
independent of the initial state.

The produced partons are fragmented into charged pions, charged kaons and
protons/antiprotons using LO fragmentation functions fit to $e^+ e^-$ data
\cite{Kniehl:2000fe}.  The final-state hadrons are assumed to be produced
pairwise so that $\pi \equiv (\pi^+ + \pi^-)/2$, $K \equiv (K^+ +
K^-)/2$, and $p \equiv (p + \overline p)/2$.  The equality of $p$ and
$\overline p$ production obviously does not describe low energy hadroproduction
well.  As energy increases, this approximation may become more
reasonable.  The produced hadrons follow the parent parton
direction.  
We have used the LO KKP fragmentation functions \cite{Kniehl:2000fe}.  
The KKP scale evolution is modeled using $e^+e^-$ data at
several different energies and
compared to $p \overline p$, $\gamma p$ and $\gamma\gamma$ data.
After some slight scale modifications \cite{Kniehl:2000hk} all the $h^-$ data
could be fit.  However, there are significant
uncertainties in fragmentation when the leading hadron takes most of
the parton momentum \cite{Zhang:2002py}, as is the case here.

We assume the same scale in the parton densities 
and the fragmentation functions, $Q^2 =
p_T^2$.  A larger scale, $p_T^2/z_c^2$, has sometimes been used in the parton
densities.  At high
$p_T$, where $z_c$ is large, the difference in the results for the two scales
is small.  We have not included any intrinsic transverse momentum broadening
in our calculations \cite{Gyulassy:2001nm,Vitev:2002vr}.  
This ``Cronin'' effect can be important when $p_T$ is small
but becomes negligible for transverse momenta larger than a few GeV.

\begin{figure}[htb]
\setlength{\epsfxsize=0.95\textwidth}
\setlength{\epsfysize=0.45\textheight}
\centerline{\epsffile{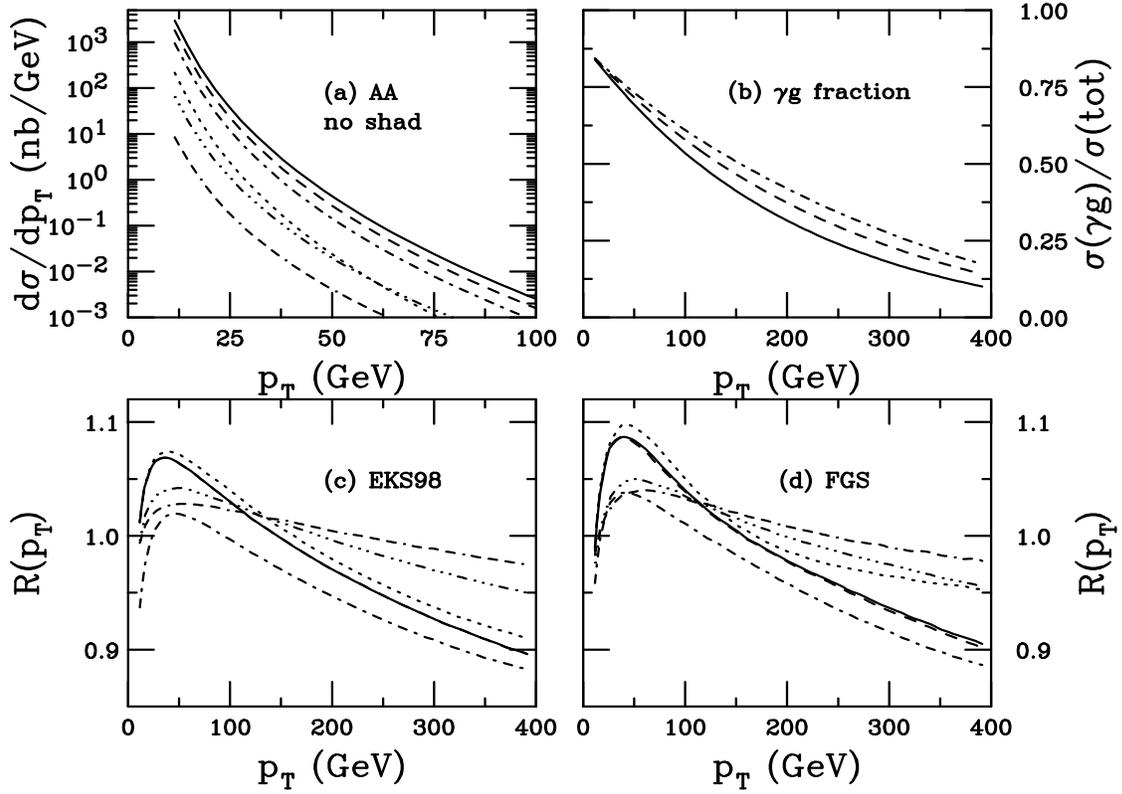}}
\caption[] {Direct photoproduction of leading hadrons 
in peripheral collisions. (a)
The $p_T$ distributions for $|y_1| \leq 1$ are shown for $AA$ 
collisions.  The solid curve is the total for Pb+Pb while the produced 
pions (dashed), kaons (dot-dashed) and protons (dotted) are shown
separately.  The total production for Ar+Ar (dot-dot-dot-dashed) and O+O
(dot-dash-dash-dashed) are also shown. (b) The fraction of gluon-initiated
hadrons as a function of $p_T$.  The curves are the same as in (a). 
(c) The EKS98 shadowing ratios for produced pions.
The solid curve is the total for Pb+Pb while the ratios for pions produced by
quarks (dashed), antiquarks (dotted) and gluons (dot-dashed) are shown
separately.  The total ratios for Ar+Ar (dot-dot-dot-dashed) and O+O
(dot-dash-dash-dashed) are also shown. (d) The same as (c) for FGS.}
\label{jethaddir}
\end{figure}

The corresponding hadron distributions from direct jet photoproduction
are shown in Fig.~\ref{jethaddir}(a) for $AA$ collisions.
The largest contribution to the total final-state charged particle production
is charged pions, followed by kaons and protons.  
Note that the leading hadron cross sections are lower than the partonic
jet cross sections, compare 
Figs.~\ref{jetdir}(a) and \ref{jethaddir}(a).  Several 
factors can account for this.  The maximum $\sqrt{s_{\gamma N}}$ is a factor 
of five or more less than $\sqrt{s_{_{NN}}}$ for $AA$ collisions.  The reduced 
number of processes available for direct dijet 
photoproduction is a significant contribution to the decrease.
Note also that the $p_T$ distribution is steeper for leading hadrons than for
the jets, as may be expected since the effective $p_T$ of the hadron
is higher than than of the produced parton.

The average $z_c$ for direct photoproduction of high $p_T$ particles 
is $\approx 0.4$
for particles with $p_T \approx 10$ GeV, increasing to $\langle z_c 
\rangle > 0.45-0.55$ for $p_T > 100$ GeV.  The lower $z_c$ values correspond to
lighter ion collisions.  In this $z_c$ region, the fragmentation
functions are not very well known.  As pointed 
out in Ref.~\cite{Zhang:2002py},
a small change in the fragmentation function fits can produce significant
changes at large $z_c$.  This region is not well constrained by
the $e^+ e^-$ data used in the fits.

The effect of fragmentation on the production channels is shown in
Fig.~\ref{jethaddir}(b) where we present the fraction of leading hadron 
production
from the $\gamma g$ channel for all charged hadrons.
The ratios are rather similar to those of the partonic jets although the
$\gamma g$ fraction is somewhat smaller due to the larger average $x$ of hadron
production with respect to jets, as we discuss later.

The shadowing ratios for charged pions produced in Pb+Pb collisions by quarks, 
antiquarks, gluons and the total from all partons,
are shown for the EKS98 and FGS parameterizations
in Fig.~\ref{jethaddir}(c) and (d).  The ratios for pion production in Ar+Ar
and O+O collisions are also shown.  The high $p_T$ FGS antiquark
ratios flatten out relative to the EKS98 ratio
because the $\gamma \overline q$ channel dominates gluon production at
high $p_T$.  The flattening behavior sets in earlier here because the $x$
for hadron production is higher than that for the jets.
The ratio of pions arising from produced gluons follows the valence ratio, as
expected. 
The ratios decrease with increasing $p_T$ due to 
the EMC effect for $x > 0.2$ when $p_T > 100$ GeV.  

\begin{figure}[htb]
\setlength{\epsfxsize=0.95\textwidth}
\setlength{\epsfysize=0.45\textheight}
\centerline{\epsffile{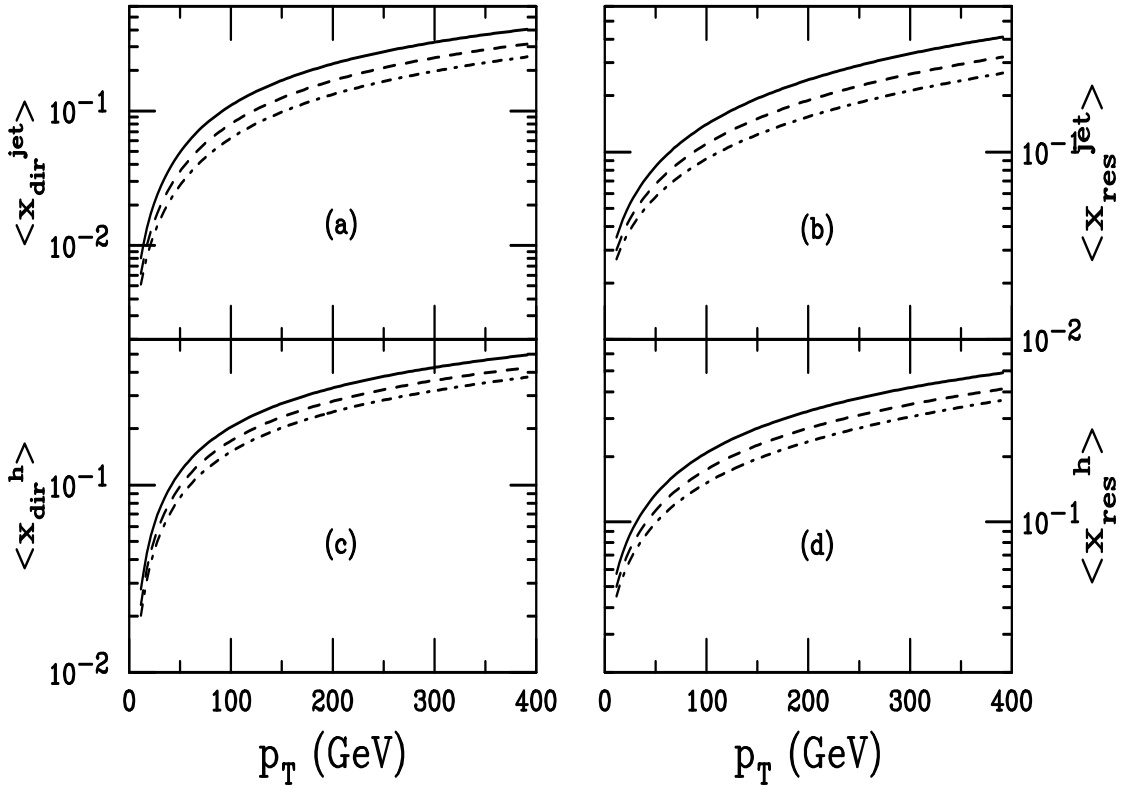}}
\caption[]{The average value of the nucleon parton momentum fraction $x$ as a
function of transverse momentum.  Results are given for (a) direct and (b) 
resolved gluon jet production and for (c) direct and (d) resolved pion 
production by gluons.  The results are given for
O+O (dot-dashed), Ar+Ar (dashed) and Pb+Pb (solid) interactions.}
\label{avexjet}
\end{figure}

We now discuss the relative values of the nucleon momentum fraction, $x$ for
parton and hadron production.  
On the left-hand side of Fig.~\ref{avexjet}, we compare the average $x$ values
for produced gluon jets (upper plot) and for pions produced by these gluons 
(lower plot).
We have chosen to compute the results for produced gluons alone to better
compare with resolved jet photoproduction, discussed next.  
Since we are interested in produced gluons, we only
consider the QCD Compton contribution, $\gamma q \rightarrow gq$.  
This channel is biased toward larger
momentum fractions than $\gamma g \rightarrow q \overline q$
since the gluon distribution is largest at 
small $x$ while the valence quark
distribution in the proton is peaked at $x \sim 0.2$. 
The average $x$ for 
a gluon jet is $\sim 0.005-0.008$ at $p_T\approx 10$ GeV, increasing to
$\sim 0.03-0.05$ at 50 GeV.
The smallest $x$ is from the highest energy O+O collisions.  The average $x$
increases with $p_T$, to $\sim 0.25-0.4$ at $p_T \sim 400$ GeV.
When final state pions are considered, in the lower left-hand plot, at low
$p_T$, the average $x$ is larger than for gluon jets.  At $p_T \approx 10$ GeV,
$\langle x \rangle \approx 0.02-0.03$ while at 50 GeV, 
$\langle x \rangle \approx 0.09 - 0.12$.  At high $p_T$, however, the
average $x$ becomes similar for jet and hadron production as
$\langle z_c \rangle$ increases to $\approx 0.6-0.7$ at $p_T \sim 400$ GeV.

We now turn to resolved production.  
The hadronic reaction, $\gamma N
\rightarrow {\rm jet} \, + {\rm jet}\, + X$, is unchanged, but in this case, 
prior to the
interaction with the nucleon, the photon splits into a color singlet
state of $q \overline q$ pairs and gluons.  
On the parton level, the resolved LO reactions are {\it e.g.} $g(xk) +
g(x_2 P_2) \rightarrow g(p_1) + g(p_2)$ where $x$ is the
fraction of the photon momentum carried by the parton.  The LO processes for
resolved photoproduction
are the same as those for LO $2 \rightarrow 2$
hadroproduction except that one parton source is a photon
rather than a nucleon.
The resolved jet photoproduction cross
section for partons of flavor $f$ in the
subprocess $ij\rightarrow kl$ in $AB$ collisions is, modified from 
Refs.~\cite{Emel'yanov:1999bn,Eskola:1988yh,Eskola:1996ce},
\begin{eqnarray} s_{_{NN}}^2
\frac{d\sigma^{\rm res}_{\gamma A \rightarrow {\rm jet} \, + \, {\rm 
jet}}}{dt_{_{NN}} du_{_{NN}}} & = & 2 \int_{k_{\rm min}}^\infty 
\frac{dk}{k} {dN_\gamma\over dk} \int_{k_{\rm min}/k}^1 \frac{dx}{x}
\int_{x_{2_{\rm min}}}^1 \frac{dx_2}{x_2} \label{sigjetres}  \\
&  & \mbox{} \times \sum_{{ij=}\atop{\langle kl \rangle}} \left\{
F_i^\gamma (x,Q^2) F_j^A(x_2,Q^2) + F_j^\gamma (x,Q^2)
F_i^A(x_2,Q^2) \right\} \nonumber \\ 
&  & \mbox{} \times  \frac{1}{1 + \delta_{kl}}
\left[\delta_{fk} \hat{s}^2\frac{d\sigma^{ij\rightarrow kl}}{d\hat t d\hat u}
(\hat t, \hat u) 
+ \delta_{fl} \hat{s}^2 \frac{d\sigma^{ij\rightarrow kl}}{ d\hat t d\hat u}
(\hat u, \hat t)
\right] \, \, 
\nonumber
\end{eqnarray}
where $\hat{s} = (xk + x_2P_2)^2$, $\hat{t} = (xk - p_1)^2$,
and $\hat{u} = (x_2P_2 - p_1)^2$.  
The $2 \rightarrow 2$ minijet subprocess cross sections, $d\sigma/d\hat t$, 
given in the Ref.~\cite{Owens:1986mp}, are related to 
$d\sigma/d\hat t d\hat u$ through the momentum-conserving 
delta function $\delta(\hat s + \hat t 
+ \hat u)$.
The sum over initial states includes all combinations of two parton
species with three flavors while the final state includes all pairs
without a mutual exchange and four flavors (including charm).  The
factor $1/(1 + \delta_{kl})$ accounts for identical particles in
the final state.

\begin{figure}[htb]
\setlength{\epsfxsize=0.95\textwidth}
\setlength{\epsfysize=0.45\textheight}
\centerline{\epsffile{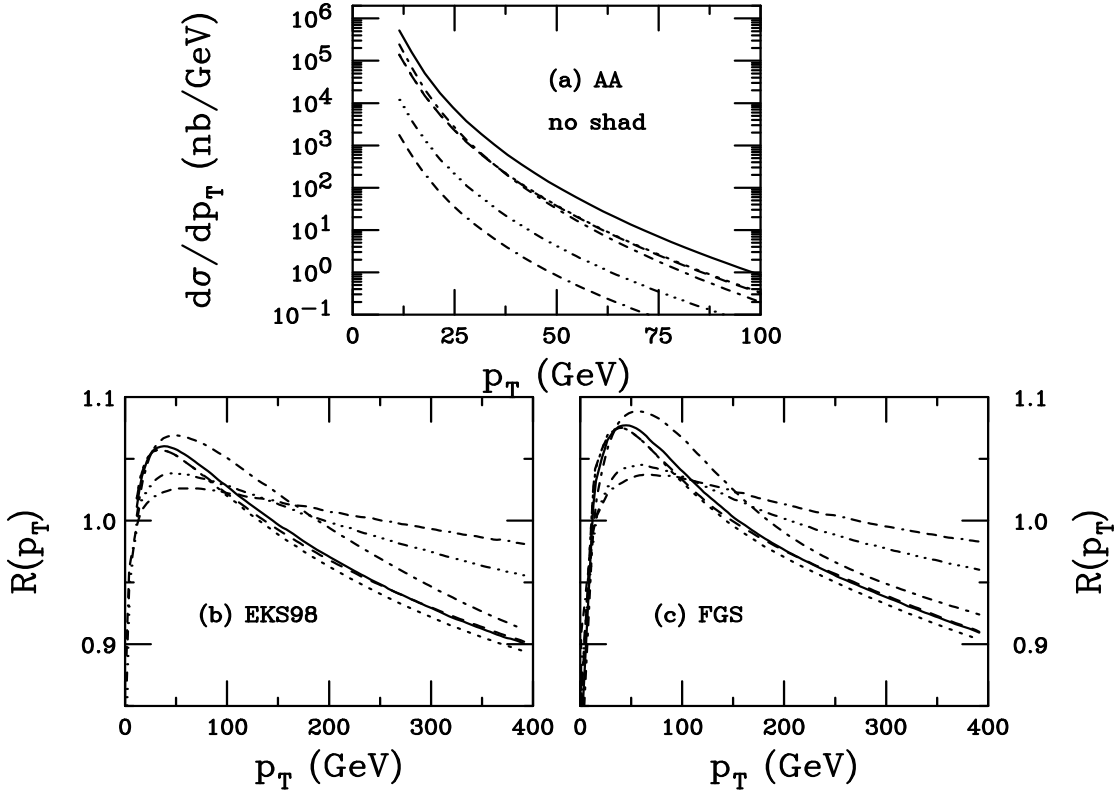}}
\caption[] {Resolved dijet photoproduction in peripheral $AA$ collisions. (a)
The Pb+Pb jet $p_T$ distributions with $|y_1| \leq 1$ are shown for quarks 
(dashed),
antiquarks (dotted), gluons (dot-dashed) and the total (solid).  We also show
the total jet $p_T$ distributions in Ar+Ar (dot-dot-dot-dashed) and 
O+O (dash-dash-dash-dotted) collisions. (b) The relative EKS98 shadowing 
contributions 
from quarks (dashed), antiquarks (dotted) and gluons (dot-dashed) as well
as the total (solid) are shown for Pb+Pb collisions.  The totals are also shown
for Ar+Ar (dot-dot-dot-dashed) and O+O (dash-dash-dash-dotted) interactions.
(c) The same as (b) for FGS.}
\label{pjetres}
\end{figure}

The resolved jet results, shown in Fig.~\ref{pjetres},
are independent of the photon parton densities for $p_T>10$ GeV.  
Along with the total quark, antiquark and gluon cross sections 
in Pb+Pb collisions, we also show the
individual partonic contributions to the jet $p_T$ distributions.  The 
produced gluon contribution dominates for $p_T < 25$ GeV but, by 50 GeV, 
quark and antiquark production becomes larger due to the increase of the $qg
\rightarrow qg$ channel relative to the $gg \rightarrow gg$ channel.  
We also show the total $p_T$ distributions
for Ar+Ar and O+O collisions.
For lighter nuclei, the crossover between gluon and quark/antiquark dominance
occurs at higher $p_T$ due to the higher collision energy.

The resolved dijet photoproduction contribution is two to three times larger 
than the direct for $p_T<50$ GeV, despite the
lower effective center-of-mass energy of resolved production.
The largest increase is for
the lightest nuclei since the lowest $x$ values are probed.  However, with
increasing $p_T$, the phase space is reduced.  The average
photon momentum is increased and, at large photon momentum, the flux drops
faster.
The average momentum fractions probed by the nuclear parton densities grows
large and only valence quarks contribute.  The lower effective energy of
resolved relative to direct photoproduction reduces the high $p_T$ phase space
for resolved production.  Thus, at the highest $p_T$, the resolved
rate is reduced relative to the direct by a factor of $4-9$.  The smallest 
decrease is for the
lightest system due to the higher effective $\sqrt{s_{_{NN}}}$.
Since resolved production has a narrower rapidity distribution than direct
production, increasing the rapidity coverage would not increase the rate
as much as for direct photoproduction.
  
Resolved production opens up many more channels through the parton components
of the photon.  Indeed, now all the $2 \rightarrow 2$ channels for LO jet
hadroproduction are available.  In addition, the quark and antiquark
distributions in the photon are the same.  These distributions are large 
at high momentum fractions, higher 
than the quark and antiquark distributions in the proton.  Thus
the quark and antiquark channels are enhanced relative to hadroproduction.
The largest difference between the quark and antiquark production rates 
is due to the difference between the valence and sea quark distributions in the
nucleus.  Where the valence and sea 
quark contributions are similar, as for $|y_1|\leq 1$, the difference is rather
small.  If all rapidities were included, the relative quark and antiquark
rates could differ more.

The direct and resolved rapidity distributions are compared in
Fig.~\ref{jet_dir_rap} for the two $p_T$ cuts, 10 and 100 GeV.  While the
$|y_1|\leq 1$ resolved contribution is a factor of two to three
larger than the direct at $p_T < 50$ GeV, a comparison of the $y_1$ 
distributions over all rapidities shows that the resolved contribution can be
considerably larger, a factor of $\sim 5-10$ at $y_1 < -3.5$ for $p_T
> 10$ GeV.  At $p_T > 100$ GeV, the resolved contribution is still equivalent
to or slightly larger than the direct at $y_1 < -3$ but drops below at
larger rapidities.  Thus, going to higher $p_T$ can separate direct from
resolved production, especially at forward rapidities.  
Recall that the produced gluons 
dominate resolved production at $p_T < 25$ GeV while they are only a small
contribution to direct production.  The largest gluon production channels are
typically $gg \rightarrow gg$ and $gq \rightarrow gq$.  As $y_1$ becomes large
and negative, the photon $x$ decreases while $x_2$ of the nucleon decreases,
leading to the dominance of the $gq$ channel.  The photon gluon distribution
is largest as $x$ decreases.  The valence quark distribution of the proton
is also important at high $p_T$, causing the resolved to direct ratio to
flatten for $y_1 > -2.5$ when $p_T >100$ GeV.

In Fig.~\ref{jet_dir_shad}, we compare the direct and resolved shadowing ratios
$R(y_1)$.  Shadowing is smaller for the resolved component
due to the larger $x_2$ for resolved production.  The difference
in the direct and resolved shadowing ratios is reduced for larger $p_T$.

\begin{figure}[htb]
\setlength{\epsfxsize=0.95\textwidth}
\setlength{\epsfysize=0.3\textheight}
\centerline{\epsffile{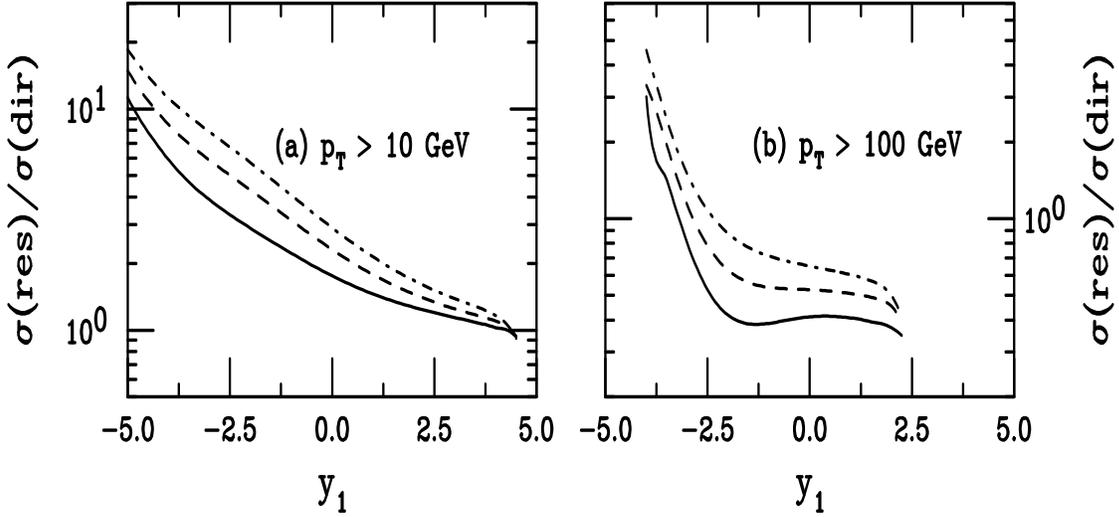}}
\caption[] {We present the resolved/direct dijet production ratios as a 
function of rapidity. In (a) we show the results
for $p_T > 10$ GeV 
while in (b) we show $p_T > 100$ GeV.  
The curves are Pb+Pb (solid), Ar+Ar (dashed) and O+O (dot-dashed).  The photon
comes from the left.}
\label{jet_dir_rat}
\end{figure}

To directly measure the nuclear parton densities, direct production should 
be dominant.  However, Fig.~\ref{jet_dir_rat} shows that a
$p_T$ cut is not very
effective for dijet production, even at forward rapidity.  
The resolved to direct production ratios
are all larger than unity for $p_T > 10$ GeV, even for large, positive $y_1$.
While the ratio is less than 1 for $y_1 > -2.5$ and $p_T >
100$ GeV, it is only $\sim 0.5$ for Pb+Pb, increasing to 0.8 for O+O.

Thus, although clean separation is possible at $p_T > 100$ GeV, precision
parton density measurements are not possible at these $p_T$'s due to the low
rate.  Other means of separation must then be found.  Resolved processes will
not be as clean as direct in the direction of the photon due to the breakup of
the partonic state of the photon.  The multiplicity in the photon fragmentation
region will be higher than in direct production where the nucleus should remain
intact.  A cut on multiplicity in the photon direction may help separate the
two so that, although there should be a rapidity gap for both direct and 
resolved photoproduction, the gap may be less prominent for resolved 
production.

Figure~\ref{pjetres}(b) shows the individual partonic 
shadowing ratios for Pb+Pb collisions with
the EKS98 parametrization.  The quark and antiquark shadowing ratios are 
very similar although the quark ratio becomes
larger for higher $p_T$ (higher $x$) values of $x$ due to the valence 
distribution.  Now the gluon ratio shows larger antishadowing since gluon 
production is now through the $gg$ and $qg$ channels rather than $\gamma q$ in
direct production, compare Fig.~\ref{jetdir}.  The FGS parametrization
gives similar results, Fig.~\ref{pjetres}(c).  However, since the small
FGS gluon antishadowing is stronger, $R(p_T)$ is larger for $p_T < 150$ GeV.

\begin{figure}[htb]
\setlength{\epsfxsize=0.95\textwidth}
\setlength{\epsfysize=0.45\textheight}
\centerline{\epsffile{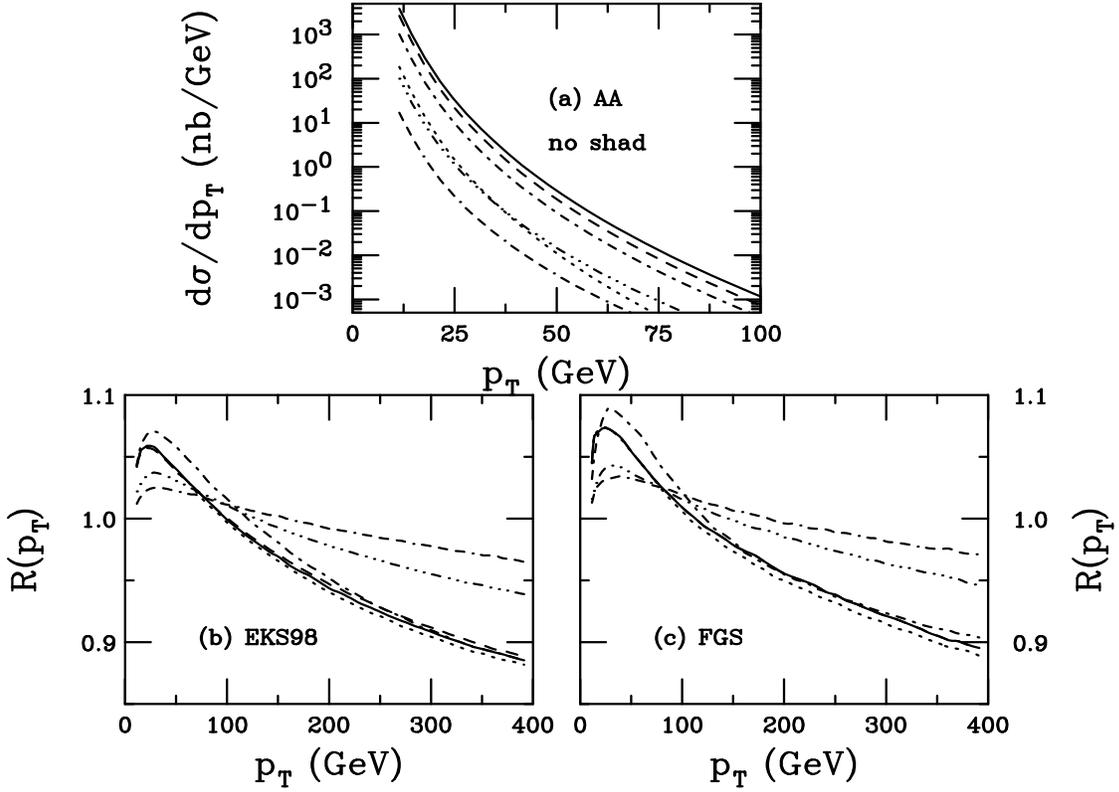}}
\caption[] {Resolved leading hadrons from dijet
photoproduction in peripheral collisions.
(a) The $p_T$ distributions for $|y_1| \leq 1$ are shown for $AA$
collisions.  The Pb+Pb results are shown for charged pions (dashed), kaons 
(dot-dashed), protons (dotted) and the sum of all charged hadrons (solid).   
The charged hadron $p_T$ distributions are also shown for
Ar+Ar (dot-dot-dot-dashed) and O+O (dot-dash-dash-dashed) collisions.
(b) The EKS98 shadowing ratios for produced pions.  For Pb+Pb collisions,
we show the ratios for pions produced by quarks (dashed), antiquarks 
(dotted), gluons (dot-dashed) and the total (solid) separately.  The ratios 
for pions produced by all partons are also shown for Ar+Ar (dot-dot-dot-dashed)
and O+O (dot-dash-dash-dashed) collisions. (c) The same as (b) for FGS.
}
\label{jethadres}
\end{figure}

The leading particle $p_T$ distributions from resolved dijet photoproduction 
are
\begin{eqnarray}
\frac{d\sigma^{\rm res}_{\gamma A \rightarrow hX}}{dp_T} & = & 4p_T   
\int_{\theta_{\rm min}}^{\theta_{\rm max}}
\frac{d\theta_{\rm cm}}{\sin \theta_{\rm cm}}
\int_{k_{\rm min}}^\infty 
\frac{dk}{k} {dN_\gamma\over dk} \int_{k_{\rm min}/k}^1 \frac{dx}{x}
\int_{x_{2_{\rm min}}}^1 \frac{dx_2}{x_2} \label{sighadres}  \\
&  & \mbox{} \times \sum_{{ij=}\atop{\langle kl \rangle}} \left\{
F_i^\gamma (x,Q^2) F_j^A(x_2,Q^2) + F_j^\gamma (x,Q^2)
F_i^A(x_2,Q^2) \right\} \nonumber \\ 
&  & \mbox{} \times  \frac{1}{1 + \delta_{kl}}
\left[\delta_{fk} \frac{d\sigma^{ij\rightarrow kl}}{d\hat t}(\hat t, \hat
u) 
+ \delta_{fl} \frac{d\sigma^{ij\rightarrow kl}}{d\hat t}(\hat u, \hat t)
\right] \frac{D_{h/k}(z_c,Q^2)}{z_c} \, \, .
\nonumber
\end{eqnarray}
The subprocess cross sections, $d\sigma/d\hat t$, 
are related to 
$\hat s^2 d\sigma/d\hat t d\hat u$ in Eq.~(\ref{sigjetres})
through the momentum-conserving 
delta function $\delta(\hat s + \hat t + \hat u)$ and division by $\hat s^2$.
The drop in rate between jets and high $p_T$ hadrons is similar to that in
direct photoproduction, as can be seen by comparison of Figs.~\ref{pjetres}
and \ref{jethadres} relative to Figs.~\ref{jetdir} and \ref{jethaddir}.  
Now that gluon fragmentation is also possible, the relative
pion contribution is larger than in direct photoproduction while
the relative proton contribution is significantly reduced.
The smaller effective center-of-mass energy for resolved
photoproduction lowers the phase space available for fragmentation. 
Baryon production is then reduced compared to light mesons.

The reduction in phase space for leading hadrons relative to fast
partons can be seen in the comparison of the average $x$ values for resolved 
photoproduction of jets and leading hadrons, shown on the right-hand side of 
Fig.~\ref{avexjet} for gluons and pions from gluons respectively.  At low 
$p_T$, the average $x$ of the gluon jet is $0.03 - 0.04$, increasing to $0.16 
- 0.24$ at $p_T \approx 200$ GeV, higher than for direct photoproduction, as
expected.  The $x$ values for hadron production are larger still,
$\approx 0.06$ at low $p_T$ while $x \approx 0.23-0.33$ at $p_T \approx 200$
GeV.

The shadowing ratios in Fig.~\ref{jethadres} also reflect the increasing $x$.
Now the antishadowing peak is shifted to $p_T \approx 30$ GeV since 
the average $x$ values are in the EMC region,
even at low $p_T$.  The values of $R(p_T)$ at high $p_T$ are somewhat lower
than for direct production due to the higher $x$.
The average $z_c$ of the fragmentation functions is also somewhat larger for
resolved production, $0.7 - 0.8$ at $p_T \approx 400$ GeV.  

Since the resolved jet cross section is larger
than the direct at low $p_T$, it is more difficult to make clean
measurements of the nuclear gluon distribution unless the two contributions 
can be separated by other methods.  Instead, the
large valence quark contribution at high $p_T$ suggests that jet 
photoproduction can probe the nuclear valence quark distributions
at larger $Q^2$ than previously possible.  At $p_T> 100$ GeV, more than 
half of direct jet production is through the $\gamma q$ channel.  
Unfortunately, the rates are low here, making high precision 
measurements unlikely.  However, the events should be very clean.

\subsection{$\gamma +$jet production}
\label{compt}
{\it Contributed by: R. Vogt}

A clean method of determining the quark distribution in the nucleus
at lower $p_T$ is the
process where a jet is produced opposite a photon in the final state, Compton
scattering in direct production.  The cross sections are reduced relative to
the jet$+$jet process since the coupling is $\alpha^2 e_Q^4$ in
the coupling rather than $\alpha \alpha_s e_Q^2$, 
as in dijet production.  In addition, the 
quark distributions are lower than the gluon, also reducing the rate.

We now discuss the jet and leading particle distributions
for direct and resolved $\gamma +$jet photoproduction.
Now, the hadronic process is $\gamma(k) + N(P_2) \rightarrow
\gamma(p_1) + {\rm jet}(p_2) \, + X$.
The only partonic contribution to the $\gamma+$jet yield in direct 
photoproduction is
$\gamma(k) + q(x_2P_2)  \rightarrow \gamma(p_1) +
q(p_2)$ (or $q \rightarrow \overline q$) where the 
produced quark is massless.  We now have
\begin{eqnarray}
\label{compdireq} s_{_{NN}}^2
\frac{d^2\sigma_{\gamma A \rightarrow \gamma + \, {\rm jet} + X}^{\rm 
dir}}{dt_{_{NN}} du_{_{NN}}} & = & 
2 \int dz \int_{k_{\rm min}}^\infty dk \frac{dN_\gamma}{dk}
\int_{x_{2_{\rm min}}}^1 \frac{dx_2}{x_2} \nonumber \\
& & \mbox{} \times \bigg[ \sum_{i=q, \overline q} 
F_i^A(x_2,Q^2) s^2 \frac{d^2\sigma_{\gamma i\rightarrow \gamma
i}}{dtdu} \bigg] 
\, .
\end{eqnarray}
The partonic cross section for the Compton process is
\begin{eqnarray}
\label{sigcompt}
s^2 \frac{d^2\sigma_{\gamma q\rightarrow \gamma q}}{dt du} = - \frac{2}{3} 
\pi \alpha^2 e_Q^4 \bigg[ \frac{s^2 + u^2}{s u}\bigg] \delta(s + t + u) \, \, .
\end{eqnarray}
The extra factor of two on the right-hand side of
Eq.~(\ref{compdireq}) again arises because 
both nuclei can serve as photon sources in $AA$ collisions.  
The kinematics are the same as in jet$+$jet
photoproduction,
described in the previous section.

\begin{figure}[htb]
\setlength{\epsfxsize=0.95\textwidth}
\setlength{\epsfysize=0.3\textheight}
\centerline{\epsffile{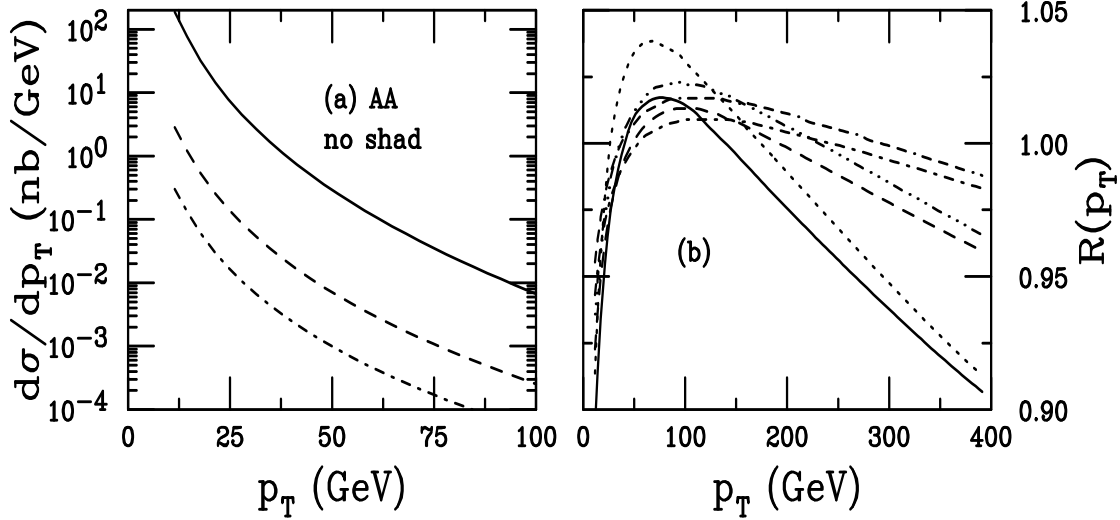}}
\caption[] {Direct $\gamma+$jet photoproduction in peripheral collisions. (a)
The $p_T$ distributions for $|y_1| \leq 1$ are shown for Pb+Pb (solid), Ar+Ar
(dashed) and O+O (dot-dashed) collisions. (b) The EKS98 shadowing ratios are 
shown for Pb+Pb (solid), Ar+Ar (dashed) and O+O
(dot-dashed) while the corresponding FGS ratios are shown for
Pb+Pb (dotted), Ar+Ar
(dot-dot-dot-dashed) and O+O (dot-dash-dash-dashed) collisions.  
The photon comes from the left.
}
\label{compdir}
\end{figure}

The direct $\gamma +$jet photoproduction results are given in 
Fig.~\ref{compdir}  
for $AA$ interactions at the LHC.  We
show the transverse momentum, $p_T$, distributions for all produced 
quarks and antiquarks in Pb+Pb, Ar+Ar and O+O collisions for $|y_1|\leq 1$.
The cross sections are lower than
those for $\gamma +$jet hadroproduction.  Direct $\gamma +$jet
photoproduction proceeds through fewer channels than
hadroproduction where the LO channels are $g q \rightarrow \gamma q$ and
$q \overline q \rightarrow g \gamma$, the same diagrams for resolved
$\gamma+$jet photoproduction.  This, along with the lower effective energy and
correspondingly higher $x$, reduces the photoproduction cross sections relative
to hadroproduction.  The lower $A$ scaling for photoproduction also restricts
the high $p_T$ photoproduction rate.

There is a drop of nearly three orders of magnitude between the dijet
cross sections in Fig.~\ref{jetdir} and the $\gamma+$jet cross sections 
in Fig.~\ref{compdir}.  Most of this
difference comes from the relative couplings, reduced by $\alpha_s/\alpha
e_Q^2$ relative to dijet photoproduction.  The
rest is due to the reduced number of channels available for direct $\gamma+$jet
production since more than half of all directly produced are gluon-initiated
for $p_T < 100$ GeV, see Fig.~\ref{jetdir}(b).

We have not distinguished between the quark and antiquark initiated jets.  
However, the quark-initiated jet
rate will always be somewhat higher due to the valence contribution. 
When $p_T < 100$ GeV, the quark and antiquark jet rates are very
similar since $x$ is still relatively low.  At higher $p_T$, the valence
contribution increases so that when $p_T = 400$ GeV, the quark rate is 1.5
times the antiquark rate.  Since the initial kinematics are the same for
$\gamma +$jet and jet+jet final states, the average momentum fractions
for $\gamma +$jet production are similar to those shown for
the $\gamma q \rightarrow gq$ channel in 
Fig.~\ref{avexjet}.

The shadowing ratios shown in Fig.~\ref{compdir}(b) are dominated by valence
quarks for $p_T > 100$ GeV.  The FGS ratio is slightly higher because the EKS98
parametrization includes sea quark shadowing.  The effect is similar to the
produced gluon ratios, at the same values of $x$ in Fig.~\ref{jetdir}(c) and
(d), since the final-state gluons can only come from quark and antiquark
induced processes.  

\begin{figure}[htb]
\setlength{\epsfxsize=0.95\textwidth}
\setlength{\epsfysize=0.55\textheight}
\centerline{\epsffile{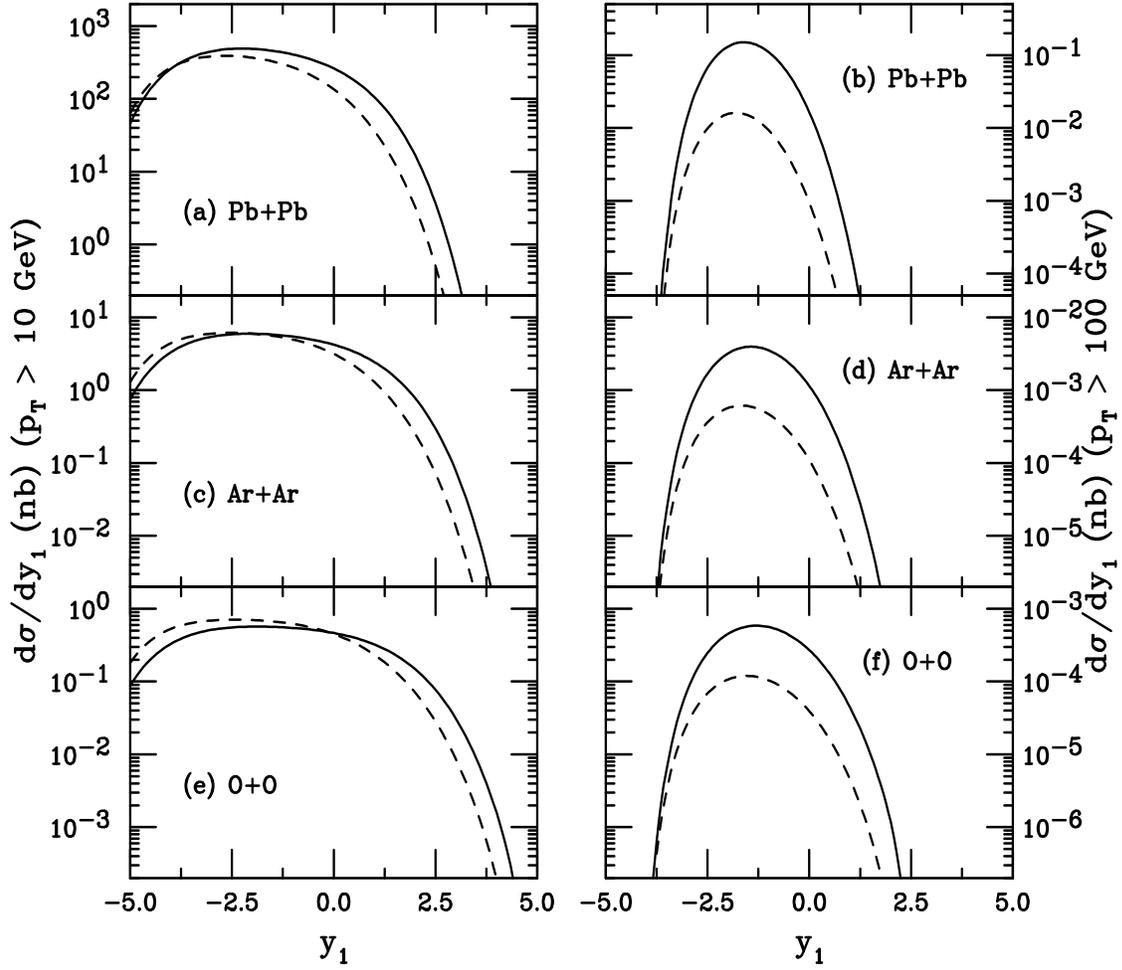}}
\caption[] {The rapidity distributions of direct and resolved
$\gamma+$jet 
photoproduction in peripheral collisions.  The left-hand side shows the results
for $p_T > 10$ GeV for (a) Pb+Pb, (c) Ar+Ar and (e) O+O collisions
while the right-hand side is for $p_T > 100$ GeV for (b) Pb+Pb, (d) Ar+Ar
and (f) O+O collisions.  The solid curves are 
the direct results while the dashed curves 
show the resolved results.  The photon comes from the left.  Note the different
scales on the $y$-axes.}
\label{compt_rap}
\end{figure}

We next present the rapidity distributions for the same two $p_T$ cuts used for
dijet photoproduction in Fig.~\ref{compt_rap}.  Note that the rapidity 
distribution for $p_T > 10$ GeV is broader at negative $y_1$ than the dijet
distributions in Fig.~\ref{jet_dir_rap} because direct dijet
production is dominated by $\gamma g \rightarrow q \overline q$ at these
$p_T$ while the valence distribution entering the
$\gamma q \rightarrow \gamma q$ does not drop
as rapidly at large $x_2$ as the gluon distribution.  When the turnover at
large negative $y_1$ occurs,
it is steeper than for the dijets.  However, it drops even more steeply at 
forward $y_1$ because the quark distribution is
smaller than the gluon at low $x_2$.  When $p_T > 100$ GeV, the $\gamma
+$jet $y_1$ distribution is narrower than the dijets since the
quark distributions drop faster with increasing $x_2$ at high $p_T$.

\begin{figure}[htb]
\setlength{\epsfxsize=0.95\textwidth}
\setlength{\epsfysize=0.55\textheight}
\centerline{\epsffile{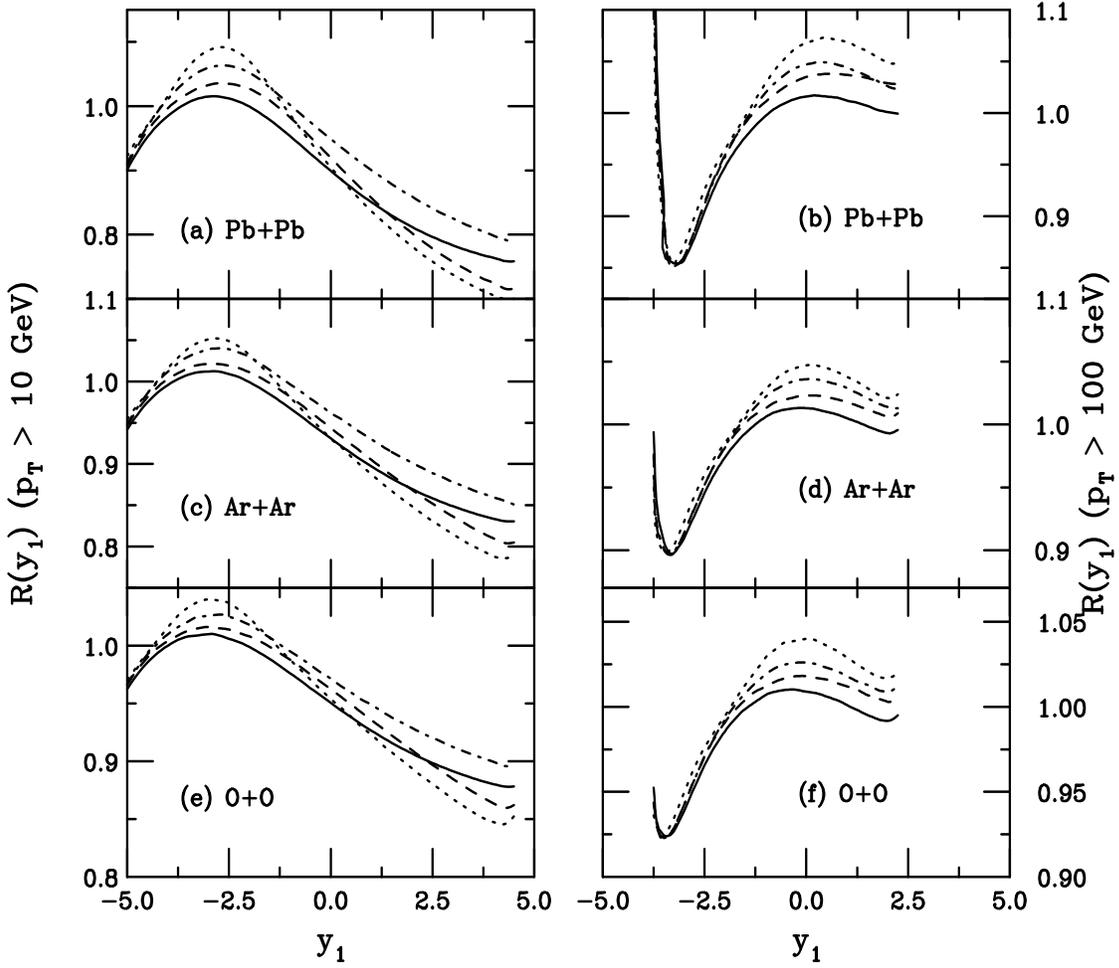}}
\caption[] {We compare shadowing ratios in direct and resolved $\gamma+$jet
production in peripheral collisions. The left-hand side shows the results
for $p_T > 10$ GeV for (a) Pb+Pb, (c) Ar+Ar and (e) O+O collisions
while the right-hand side is for $p_T > 100$ GeV for (b) Pb+Pb, (d) Ar+Ar
and (f) O+O collisions.  The solid and dashed curves give 
the direct ratios for the EKS98 and FGS parameterizations
respectively.  The dot-dashed and dotted curves 
show the resolved ratios for the EKS98 and FGS
parameterizations respectively.}
\label{compt_shad}
\end{figure}

The shadowing ratios as a function of $y_1$ are shown in Fig.~\ref{compt_shad}.
They exhibit some interesting differences from their dijet counterparts in
Fig.~\ref{jet_dir_shad} because of the different production channels.  At
$p_T > 10$ GeV, the antishadowing peak is lower at $y_1 \sim -2.5$ and the
shadowing is larger at $y_1 > 0$.  Although this may seem counter-intuitive, 
comparing the valence and sea quark shadowing ratios in Fig.~\ref{shadcomp}
can explain this effect.  Valence antishadowing, the same for EKS98 and FGS,
is lower than that of the gluon.  The sea quarks have either no antishadowing
(EKS98) or a smaller effect than the valence ratios (FGS).  Thus antishadowing
is reduced for direct $\gamma+$jet production.  At large $y_1$, the $x_2$ 
values, while smaller than those shown in Fig.~\ref{jet_dir_shad}(a) for $|y_1|
\leq 1$, are still moderate.  Since the evolution of the gluon distribution is
faster with $Q^2$, sea quark shadowing is actually stronger than gluon
shadowing at $p_T > 10$ GeV and low $x_2$.  When $p_T > 100$ GeV, the Fermi
momentum peak is not as prominent because the sharp increase in the valence
and sea shadowing ratios appears at higher $x_2$ than for the gluons, muting
the effect, particularly for the lighter systems.

\begin{figure}[htb]
\setlength{\epsfxsize=0.95\textwidth}
\setlength{\epsfysize=0.45\textheight}
\centerline{\epsffile{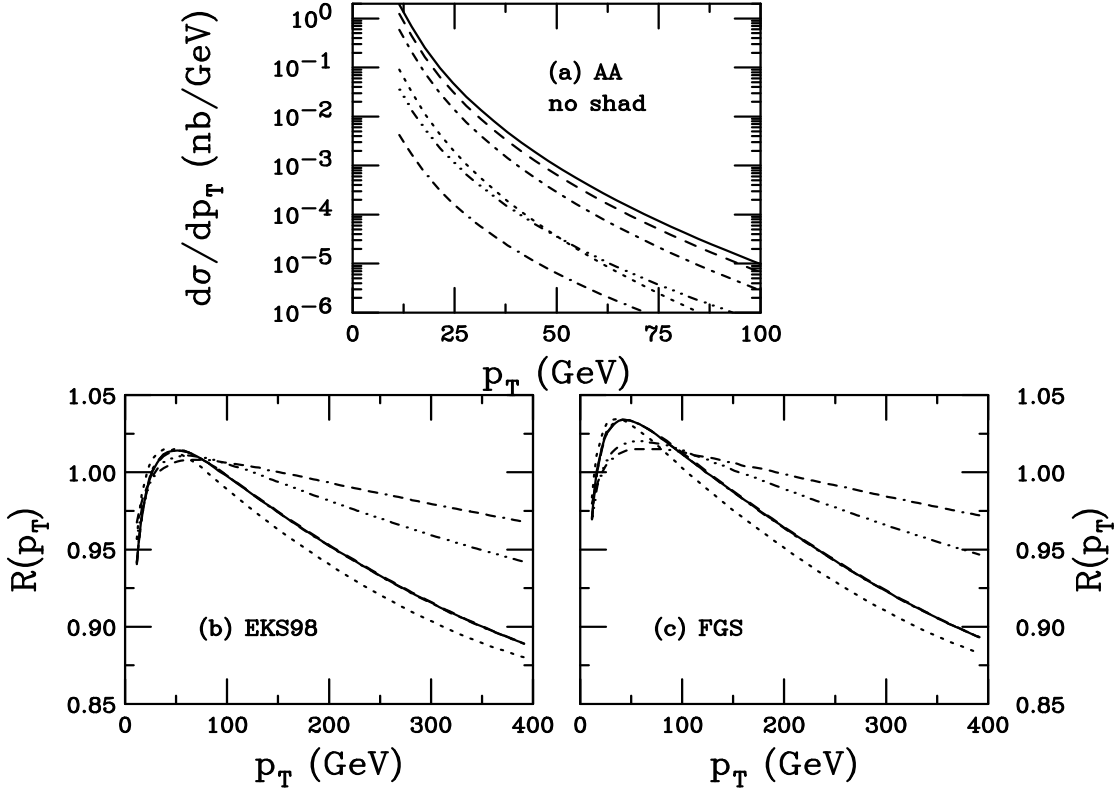}}
\caption[] {Direct leading hadrons from $\gamma+$jet photoproduction 
in peripheral collisions.
(a) The $p_T$ distributions for $|y_1| \leq 1$ are shown for $AA$
collisions.  The Pb+Pb results are shown for charged pions (dashed), kaons 
(dot-dashed), protons (dotted) and the sum of all charged hadrons (solid).   
The charged hadron $p_T$ distributions are also shown for
Ar+Ar (dot-dot-dot-dashed) and O+O (dot-dash-dash-dashed) collisions.
(b) The EKS98 shadowing ratios 
for produced hadrons. The results for pions, kaons and the charged hadron total
(solid) are nearly identical.  The proton result (dotted) is lower.  The total 
charged hadron ratios for Ar+Ar (dot-dot-dot-dashed) and O+O
(dot-dash-dash-dashed) collisions are also shown. (c) The same as (b) for FGS.}
\label{comphaddir}
\end{figure}

We now turn to a description of final-state hadron production opposite a
photon.  The leading particle $p_T$ distribution
is \cite{Field:uq}
\begin{eqnarray}
\label{haddircomp}
\frac{d\sigma_{\gamma A \rightarrow h X}^{\rm dir}}{dp_T} & = &
4 p_T \int_{\theta_{\rm min}}^{\theta_{\rm max}} 
\frac{d\theta_{\rm cm}}{\sin
\theta_{\rm cm}} \int dk \frac{dN_\gamma}{dk}
\int \frac{dx_2}{x_2} \\ &  & \mbox{} \times \bigg[ \sum_{i=q, \overline q} 
F_i^A(x_2,Q^2) 
\frac{d\sigma_{\gamma i \rightarrow \gamma i}}{dt} 
\frac{D_{h/i}(z_c,Q^2)}{z_c} \bigg] \, \,   \nonumber
\end{eqnarray}
where $X$ on the left-hand side includes the final-state gluon.
On the partonic level, 
both the initial and final state partons are identical so that parton $i$
fragments into hadron $h$ according to the 
fragmentation function, $D_{h/i}(z_c,Q^2)$.
The subprocess cross sections, $d\sigma/dt$, 
are related to 
$s^2 d\sigma/dt du$ in Eq.~(\ref{compdireq})
through the momentum-conserving 
delta function $\delta(s + t + u)$ and division by $s^2$.
Our results, shown in Fig.~\ref{comphaddir}, 
are presented in the interval $|y_1| \leq 1$.

The cross section for $\gamma+$hadron 
production are, again, several orders of magnitude
lower than the dijet calculations shown in Fig.~\ref{jethaddir}(a).
At the values of $z_c$ and $x$ important for dijet production, the final state
is dominated by quarks and antiquarks which fragment more frequently into
charged hadrons than do gluons.  While $\gamma g \rightarrow q \overline q$
produces quarks and antiquarks with identical distributions, the contribution 
from the $\gamma q
\rightarrow q g$ channel makes {\it e.g.}\ pion production by quarks and
antiquarks asymmetric.  For $p_T < 100$ GeV, 60\%
of the dijet
final state particles are pions, $\approx 33$\% kaons and $\approx 7$\%
protons.  As $p_T$ increases, the pion and proton contributions decrease
slightly while the kaon fraction increases.  In the case of $\gamma +$hadron 
final
states, there is no initial state gluon channel.  Thus the valence quarks 
dominate hadron production and the relative fraction of produced pions 
increases to 66\%.  The kaon and proton fractions are subsequently
decreased to $\approx 28$\% and $\approx 6$\% respectively.

The shadowing ratios, shown in Fig.~\ref{comphaddir}(b) and (c) for produced
pions, kaons and protons separately for Pb+Pb as well as the total ratios
for Ar+Ar and O+O collisions, reflect the quark-initiated processes.  We show
the results for all charged hadrons here since we do not differentiate between
quark and antiquark production. The ratios are almost identical for 
produced pions, kaons and charged
hadrons, are quite different from the ratios shown for pion production
by quarks and antiquarks in
Fig.~\ref{jethaddir}(c) and (d) since these pions originate 
from initial-state gluons and thus exhibit
antishadowing. The results are similar to pions from
gluon jets in Fig.~\ref{jethaddir}.  However, in this case 
the
ratios are slightly higher due to the relative couplings.  The proton ratios
are lower than those for pions and kaons due to the nuclear
isospin.  The dominance of $d$ valence quarks in nuclei reduces the proton
production rate since $d$ quarks are only half as effective at producing
protons as $u$ quarks in the KKP fragmentation scheme \cite{Kniehl:2000fe}.
This lower weight in the final state reduces the effectiveness of proton
production by the initial state, decreasing the produced proton shadowing 
ratios relative to pions and kaons.  
Valence quarks dominate the observed final state 
shadowing at these larger values of $x$, as in Fig.~\ref{avexjet}.

Now we turn to resolved production of $\gamma +$jet final states.  
The resolved jet photoproduction cross
section for partons of flavor $f$ in the
subprocess $ij\rightarrow k\gamma$ in $AB$ collisions is modified from 
Eq.~(\ref{sigjetres}) so that
\begin{eqnarray} s_{_{NN}}^2
\frac{d\sigma^{\rm res}_{\gamma A \rightarrow \gamma \, + \, {\rm 
jet}X}}{dt_{_{NN}} du_{_{NN}}} & = & 2 \int dz \int_{k_{\rm min}}^\infty 
\frac{dk}{k} {dN_\gamma\over dk} \int_{k_{\rm min}/k}^1 \frac{dx}{x}
\int_{x_{2_{\rm min}}}^1 \frac{dx_2}{x_2} \nonumber \\
&  & \mbox{} \times \sum_{{ij=}\atop{\langle kl \rangle}} \left\{
F_i^\gamma (x,Q^2) F_j^A(x_2,Q^2) + F_j^\gamma (x,Q^2)
F_i^A(x_2,Q^2) \right\} \nonumber \\ 
&  & \mbox{} \times \delta_{fk}
\left[\hat{s}^2\frac{d\sigma^{ij\rightarrow k\gamma}}{d\hat t d\hat u}(\hat t,
\hat u) 
+ \hat{s}^2 \frac{d\sigma^{ij\rightarrow k\gamma}}{d\hat t d\hat u}(\hat u, 
\hat t)
\right] \, \, .
\label{sigcompres} 
\end{eqnarray}
The resolved diagrams are those for hadroproduction of direct photons, $q g
\rightarrow q \gamma$ and $q \overline q \rightarrow qg$.
The $2 \rightarrow 2$ minijet subprocess cross sections are 
\cite{Owens:1986mp}
\begin{eqnarray}
\label{qgtogamq}
\hat s^2 \frac{d^2\sigma_{qg}}{d\hat t d\hat u} & = & - \frac{1}{3} 
\pi \alpha_s \alpha e_Q^2 \bigg[ \frac{\hat s^2 + \hat u^2}{\hat s \hat u}
\bigg] \delta(
\hat s + \hat t + \hat u) \, \, \\
\label{qqbtogamg}
\hat s^2 \frac{d^2\sigma_{q \overline q}}{d\hat t d\hat u} & = & 
\frac{8}{9} \pi \alpha_s \alpha e_Q^2 \bigg[ \frac{\hat t^2 + \hat u^2}{\hat t
\hat u} \bigg] \delta(\hat s + \hat t + \hat u) \, \, .
\end{eqnarray}
Note that
there is no factor $1/(1 + \delta_{kl})$, as in Eq.~(\ref{sigjetres}), since
there are no identical particles in
the final state.
\begin{figure}[htb]
\setlength{\epsfxsize=0.95\textwidth}
\setlength{\epsfysize=0.45\textheight}
\centerline{\epsffile{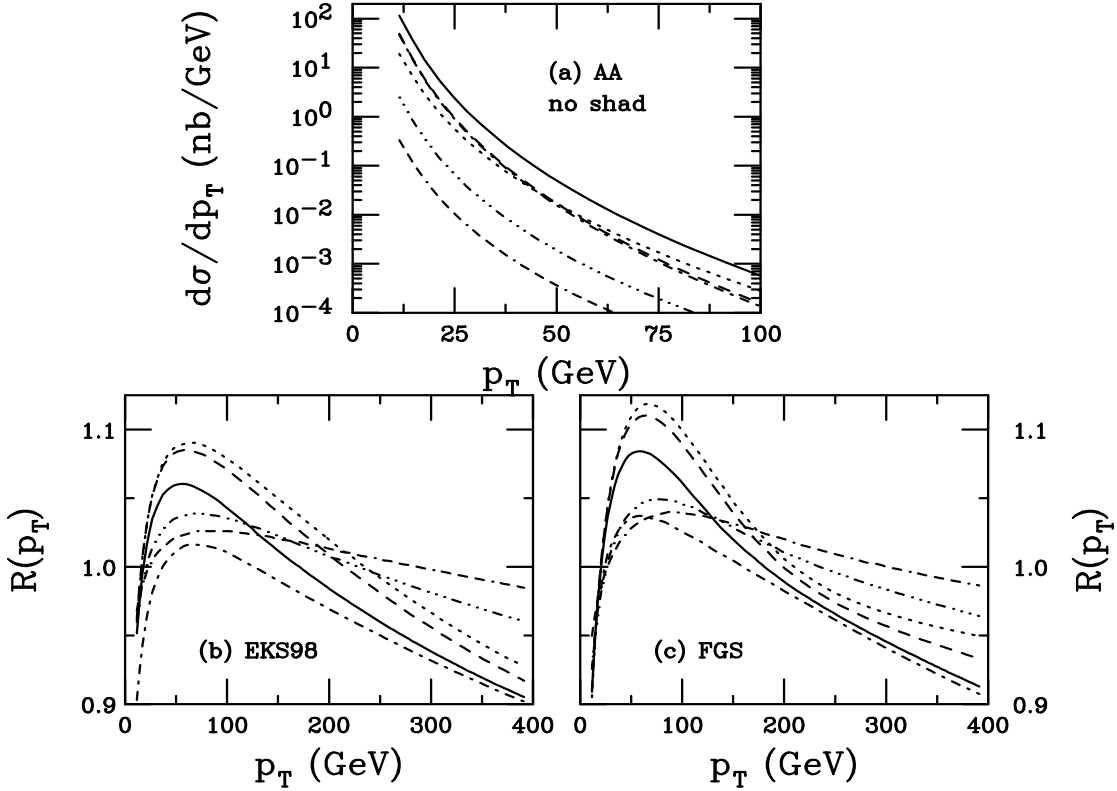}}
\caption[] {Resolved $\gamma +$jet photoproduction in peripheral $AA$ 
collisions. (a)
The Pb+Pb jet $p_T$ distributions with $|y_1| \leq 1$ are shown for quarks 
(dashed), antiquarks (dot-dashed), gluons (dotted) and the total (solid).  
We also show
the total jet $p_T$ distributions in Ar+Ar (dot-dot-dot-dashed) and 
O+O (dash-dash-dash-dotted) collisions. (b) The relative EKS98 shadowing 
contributions 
from quarks (dashed), antiquarks (dotted) and gluons (dot-dashed) as well
as the total (solid) are shown for Pb+Pb collisions.  The totals are also shown
for Ar+Ar (dot-dot-dot-dashed) and O+O (dash-dash-dash-dotted) interactions.}
\label{compjetres}
\end{figure}

The resolved jet results are shown in Fig.~\ref{compjetres} 
using the GRV LO photon parton densities.  Along with the
total partonic rates in Pb+Pb collisions, we also show the
individual partonic contributions to the jet $p_T$ distributions in
Fig.~\ref{compjetres}(a).  The total yields 
are slightly higher for the resolved
than the direct contribution where only one channel is
open and the coupling is smaller.  
Quark and antiquark production by the $qg$ process
is dominant for $p_T < 40$ GeV but, at higher $p_T$, gluon production dominates
from the $q \overline q$ channel.  The large values of $x$ again
makes the valence
quark contribution dominant at higher $p_T$.
The total $p_T$ distributions
for Ar+Ar and O+O collisions are also shown.

The strong antishadowing in the produced quark and antiquark ratios 
in Fig.~\ref{compjetres}(b) and (c) is due to the $qg$
channel.  The antiquark ratio is higher because the $qg$ parton luminosity
peaks at higher $x$ than the $\overline q g$ luminosity and at lower $x$ the
gluon antishadowing ratio is larger.  The difference between the
quark and antiquark ratios increases with $p_T$ since the average $x$ 
and thus the valence quark
contribution also grow with $p_T$.  At high $p_T$, the
flattening of the FGS quark and antiquark ratios is due to the
flattening of the gluon parametrization at $x > 0.2$.  

The final-state gluon ratio shows little antishadowing since it
arises from the $q \overline q$ channel.  The antishadowing in the EKS98 ratio
is due to the valence quarks while the higher ratio for FGS reflects the fact
that the antiquark ratios also show antishadowing for $x < 0.2$.  The ratio
for the total is essentially the average of the three 
contributions at low $p_T$, where they are similar, while at high $p_T$, 
where the $q \overline q$
channel dominates, the total ratio approximates the produced gluon
ratio in both cases.

The resolved rapidity distributions are also shown in Fig.~\ref{compt_rap} for
the two $p_T$ cuts.  The resolved distribution is not as broad at negative
$y_1$ as that of the dijet process in Fig.~\ref{jet_dir_rap} due to the 
smaller relative gluon contribution and the reduced number of channels 
available for the $\gamma+$jet process.  Note that the relative resolved to 
direct production is reduced here and the direct process is actually dominant
at positive $y_1 >0$ for $p_T > 10$ GeV and for all $y_1$ at $p_T > 100$ GeV.
The antishadowing peak is higher for resolved production, shown in 
Fig.~\ref{compt_shad}, thanks to the gluon contribution to resolved production.
\begin{figure}[htb]
\setlength{\epsfxsize=0.95\textwidth}
\setlength{\epsfysize=0.3\textheight}
\centerline{\epsffile{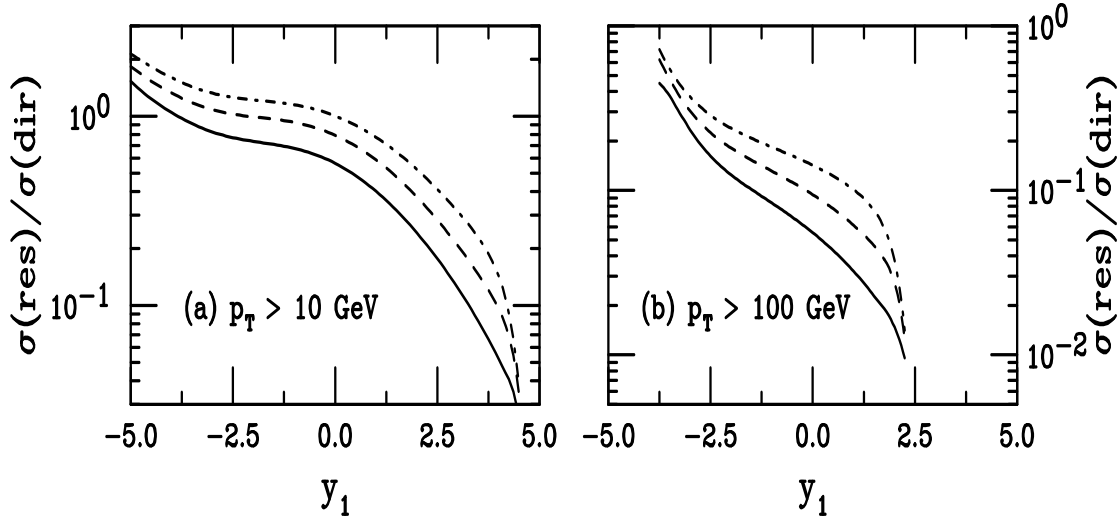}}
\caption[] {The resolved/direct $\gamma+$jet production ratios as a 
function of rapidity. The left-hand side shows the results
for $p_T > 10$ GeV
while the right-hand side is for $p_T > 100$ GeV.  
The curves are Pb+Pb (solid), Ar+Ar (dashed) and O+O (dot-dashed).  The photon
comes from the left.}
\label{compt_rat}
\end{figure}

Finally, we show the resolved to direct ratio in Fig.~\ref{compt_rat}.  The
direct rate alone should be observable at $y_1 > -4$ for Pb+Pb, $y_1 \sim -2.5$
for Ar+Ar and 0 for O+O and $p_T > 10$ GeV.  Direct production dominates over
all $y_1$ by a large factor when $p_T > 100$ GeV.  Although the rates are 
lower than the dijet, the dominance of direct $\gamma+$jet production
implies than the nuclear quark distribution can be cleanly studied.

The leading particle $p_T$ distributions of jets from $\gamma+$jet production
are
\begin{eqnarray}
\frac{d\sigma^{\rm res}_{\gamma A \rightarrow\gamma+ hX}}{dp_T} & = & 4p_T 
\int_{\theta_{\rm min}}^{\theta_{\rm max}}
\frac{d\theta_{\rm cm}}{\sin \theta_{\rm cm}}
\int_{k_{\rm min}}^\infty 
\frac{dk}{k} {dN_\gamma\over dk} \int_{k_{\rm min}/k}^1 \frac{dx}{x}
\int_{x_{2_{\rm min}}}^1 \frac{dx_2}{x_2} \nonumber \\
&  & \mbox{} \times \sum_{{ij=}\atop{\langle kl \rangle}} \left\{
F_i^\gamma (x,Q^2) F_j^A(x_2,Q^2) + F_j^\gamma (x,Q^2)
F_i^A(x_2,Q^2) \right\} \nonumber \\ 
&  & \mbox{} \times \delta_{fk}
\left[ {d\sigma\over d\hat t}^{ij\rightarrow k\gamma}(\hat t, \hat
u) 
+ {d\sigma\over d\hat t}^{ij\rightarrow k\gamma}
(\hat u, \hat t)
\right] \frac{D_{h/k}(z_c,Q^2)}{z_c} \, \, .
\label{sigcomphadres} 
\end{eqnarray}
The subprocess cross sections, $d\sigma/d\hat t$, 
are related to 
$\hat s^2 d\sigma/d\hat t d\hat u$ in Eq.~(\ref{sigcompres})
through the momentum-conserving 
delta function $\delta(\hat s + \hat t + \hat u)$ and division by $\hat s^2$.

\begin{figure}[htb]
\setlength{\epsfxsize=0.95\textwidth}
\setlength{\epsfysize=0.45\textheight}
\centerline{\epsffile{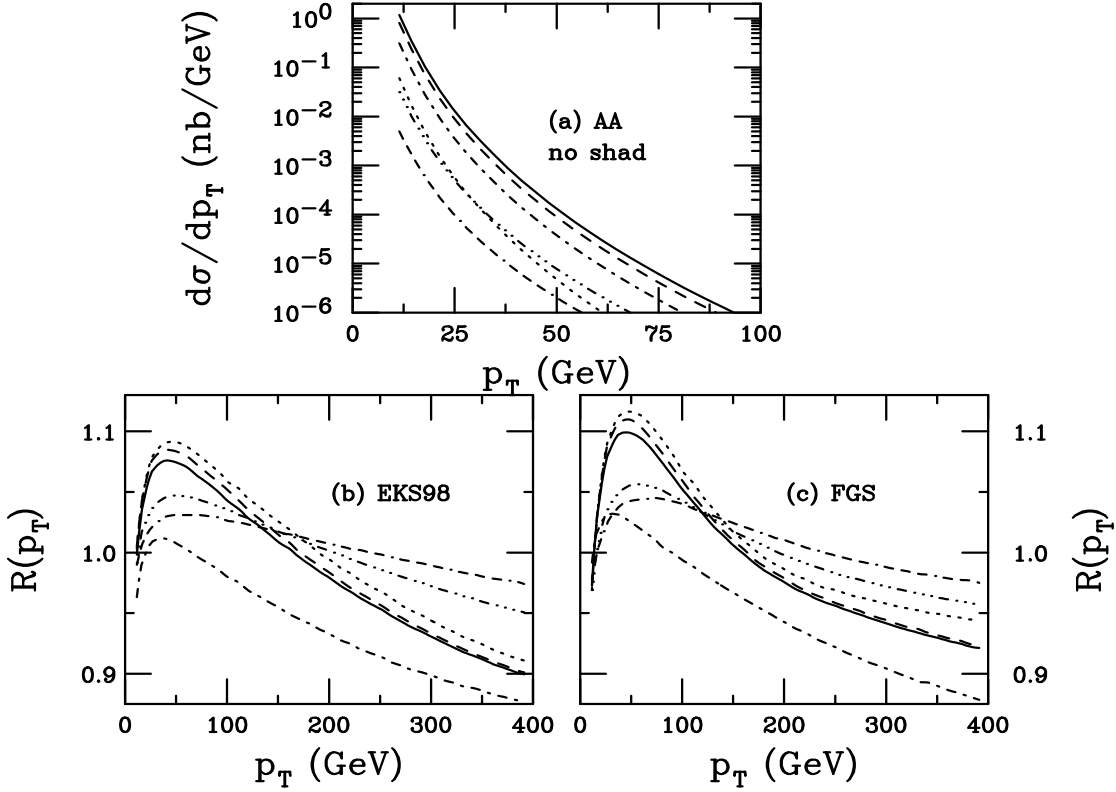}}
\caption[] {Resolved leading hadrons from $\gamma +$jet
photoproduction in peripheral collisions.
(a) The $p_T$ distributions for $|y_1| \leq 1$ are shown for $AA$
collisions.  The Pb+Pb results are shown for charged pions (dashed), kaons 
(dot-dashed), protons (dotted) and the sum of all charged hadrons (solid).   
The charged hadron $p_T$ distributions are also shown for
Ar+Ar (dot-dot-dot-dashed) and O+O (dot-dash-dash-dashed) collisions.
(b) The EKS98 shadowing ratios for produced pions.  For Pb+Pb collisions,
we show the ratios for pions produced by quarks (dashed), antiquarks 
(dotted), gluons (dot-dashed) and the total (solid) separately.  The ratios 
for pions produced by all partons are also shown for Ar+Ar (dot-dot-dot-dashed)
and O+O (dot-dash-dash-dashed) collisions. (c) The same as (b) for FGS.
}
\label{comphadres}
\end{figure}

The resolved $p_T$ distributions for hadrons
are shown in Fig.~\ref{comphadres}(a).  
Note that the resolved cross section for leading hadron production is
similar to direct production, shown in Fig.~\ref{comphaddir}(a).
The same effect is seen for dijet production in Figs.~\ref{jethadres} and
\ref{jethaddir}.  

The shadowing ratios are shown in Fig.~\ref{comphadres}.  
The difference between the shadowing ratios
for pions produced by quarks and antiquarks is rather large and reflects both
gluon antishadowing at low $p_T$ as well as the relative valence to sea
contributions for quark and antiquark production through $q(\overline q) g
\rightarrow q(\overline q) \gamma$.  In the FGS calculations, the
antiquark ratio reflects the flattening of the antiquark and gluon ratios
at $x > 0.2$.  Since pions produced by gluons come from the $q \overline q
\rightarrow \gamma g$ channel alone, only a small effect is seen, primarily in
the EMC region.  Now the total pion rates follow those for quark and antiquark
producing final-state pions than gluon.

Although our $p_T$-dependent calculations have focused on the midrapidity
region of $|y_1|\leq 1$, we have shown that extending the rapidity coverage 
could lead to greater sensitivity to the small $x_2$ region and larger
contributions from direct photoproduction, especially at low $p_T$.  

Thus $\gamma +$jet production is a good way to measure the nuclear
quark distribution functions.  Direct photoproduction is dominant 
at central rapidities for moderate values of $p_T$.  
Final-state hadron production is somewhat larger for 
direct production so that, even
if the rates are low, the results will be relatively clean.

\subsection{Uncertainties}
\label{summary}

There are a number of uncertainties in our results.  All our
calculations are at leading order so that there is some uncertainty in the
total rate, see Refs.~\cite{Klein:2002wm,Vogt:2002eu}.  
Some uncertainty also arises from the
scale dependence, both in the parton densities and in the fragmentation 
functions.  The fragmentation functions at large $z_c$ also introduce
uncontrollable uncertainties in the rates.  Hopefully more data will bring
the parton densities in the photon, proton and nucleus under better control
before the LHC begins operation.  The data from RHIC also promises to bring the
fragmentation functions under better control in the near future.

While the photon flux is also an uncertainty, it can be determined
experimentally.  The hadronic interaction
probability near the minimum radius depends on the matter distribution
in the nucleus.  Our calculations use Woods-Saxon distributions with
parameters fit to electron scattering data.  This data is quite
accurate.  However, electron scattering is only sensitive to the
charge distribution in the nucleus.  Recent measurements indicate that
the neutron and proton distributions differ in nuclei \cite{Trzcinska:sy}.
This uncertainty in the matter distribution is likely to limit the
photon flux determination.

The uncertainty in the photon flux can be reduced by
calibrating it with other measurements such as vector meson production,
$\gamma A \rightarrow VA$.  
Studies of well known two-photon processes, like lepton production,
can also help refine the determination of the photon flux.  With such
checks, it should be possible to understand the photon flux in $pA$
relative to $AA$ to better than 10\%, good enough for a useful shadowing
measurement.

\section{Small $x$ physics in UPCs}

\subsection{Identification of the QCD black disc regime}
{\it Contributed by: L. Frankfurt and M. Strikman}

\subsubsection{The black disk regime of nuclear scattering} \bigskip

A number of new, challenging QCD phenomena are related to the 
rapid increase of the gluon densities with decreasing $x$. 
As a result, the total inelastic cross section of the interaction of a small 
color singlet dipole with the target, given by Eq.~(\ref{sigma_d_DGLAP}) 
for LO pQCD, 
rapidly increases with incident energy. The increase in gluon density 
was directly observed in $J/\psi$ photo/electroproduction at HERA which found,
as predicted by pQCD, $\sigma_{\rm in} ^{q\overline q N}(s_{(q \overline q)N}, 
d \sim 0.3 \, {\rm fm}) 
\propto s_{(q \overline q)N}^{0.2}$ where 
$d \sim 0.3$ fm is a typical dipole size for $J/\psi$
production.  In addition, the proton structure function, evaluated to NLO
in a resummed series in $\alpha_s \ln(x_0/x)$ where $x_0$ is the starting
point for evolution in $x$, increases
similar to NLO DGLAP evolution for the energies studied so far
\cite{Salam:2005yp}. Thus pQCD predicts that the hard cross 
section should increase rapidly with energy.     

The increase in $\sigma_{\rm in}$ must be reduced at sufficiently high energies
to prevent the elastic cross section, proportional to 
$\sigma_{\rm tot}^2/R_A^2$, from exceeding $\sigma_{\rm tot}$ \cite{FKS,AFS}.  
Quantitative analyses show that the BDR should be reached in the ladder 
kinematics where the rapidity interval between gluon rungs on the ladder is
large (multi-Regge kinematics) so that NLL calculations are sufficient in 
pQCD. The relatively rapid onset of the BDR follows primarily from the large 
input hadron (nucleus) gluon distribution at the non-perturbative starting 
scale for QCD evolution, a consequence of spontaneously broken chiral symmetry 
and confinement. The predicted increase of the cross section with energy 
leads to complete absorption of the $q \overline q$ components of the photon
wavefunction at small impact parameters.  The components of the photon
wavefunction with $b > R_A$ produce a diffractive final state, calculable
in the strongly-absorptive, small-coupling QCD regime.  The absolute values 
and forms of these amplitudes naturally follow from the complete absorption 
in the BDR.

A variety of experimental observables with unambiguous predictions in the BDR 
\cite{Frankfurt:2001nt} will be discussed below. One example is the structure 
functions in the limit $x\to 0$: $F_2^h(x,Q^2)= c\, Q^2 \ln^3(x_0/x)$ where $c$
should be identical for hadrons and nuclei \cite{Frankfurt:2004fm}.  Note that 
BDR contribution is parametrically larger at high $Q^2$ than both the  
non-perturbative QCD result and the regime where pQCD evolution is valid with
$F_2^p \propto Q$. The dominance of the BDR contribution explains why it is
possible to evaluate the structure functions in the BDR without a quantitative 
understanding of the non-perturbative contributions at $Q \sim \Lambda_{\rm
QCD}$. At realistic energies, the universality of the structure functions may
only be achieved at small impact parameters.  

Another BDR prediction is the increase of the photo-absorption cross section
with energy as $c \ln^3(s_{(q \overline q)N}/s_0)$ where $c$ is calculable 
in QCD. Thus QCD predicts a stronger energy dependence of the photo-absorption
cross section than that of the Froissart bound for hadronic interactions.
Other model-independent phenomena in the BDR kinematics, such as
diffractive electroproduction of vector mesons and dijets on a nuclear target,
will be discussed below.  The theory of the BDR onset for 
high $p_T$ and hard phenomena with scales exceeding the BDR scale has been
described in the context of a number of models \cite{CGC}. 

The requirement of probability conservation (unitarity of the time-evolution 
operator of the quark-gluon wave packet) determines the kinematic region  
where the BDR may be accessible in hard interactions.  The simplest approach
is to consider the elastic-scattering dipole amplitude, $\Gamma(s_{(q 
\overline q)N},b)$, in 
the impact parameter representation.  The total, elastic and inelastic 
dipole-hadron cross 
sections can be written as 
\begin{equation}
\left. \begin{array}{l} 
\sigma_{\rm tot}(s_{(q \overline q)N}) \\[1ex]
\sigma_{\rm el}(s_{(q \overline q)N}) \\[1ex]
\sigma_{\rm in}(s_{(q \overline q)N})
\end{array}
\right\}
\;\; = \;\; \int d^2 b \; 
\left\{ \begin{array}{l} 
2 \, {\rm Re}\, \Gamma (s_{(q \overline q)N},b) \\[1ex]
|\Gamma (s_{(q \overline q)N},b)|^2 \\[1ex]
1 - |1-\Gamma (s_{(q \overline q)N},b)|^2 \, \, .
\end{array}
\right.
\label{unitarity}
\end{equation}
When elastic scattering is the non-absorptive complement of inelastic
scattering, the amplitude at a given impact parameter
is restricted such that $|\Gamma(s_{(q \overline q)N}, b)| \leq 1$ where 
$\Gamma(s_{(q \overline q)N},b) = 1$ corresponds to complete absorption, 
the BDR. 

The proximity of $\Gamma (s_{(q \overline q)N},b)$ to unity is an important 
measure of the dipole-nucleon interaction strength.
When $\Gamma(s_{(q \overline q)N},b)\ge 0.5$, the probability for an inelastic
dipole interaction, $|1- \Gamma (s_{(q \overline q)N},b)|^2$, exceeds 0.75,
close to unity.

\begin{figure}[htb]
\setlength{\epsfxsize=0.6\textwidth}
\setlength{\epsfysize=0.4\textheight}
\centerline{\epsffile{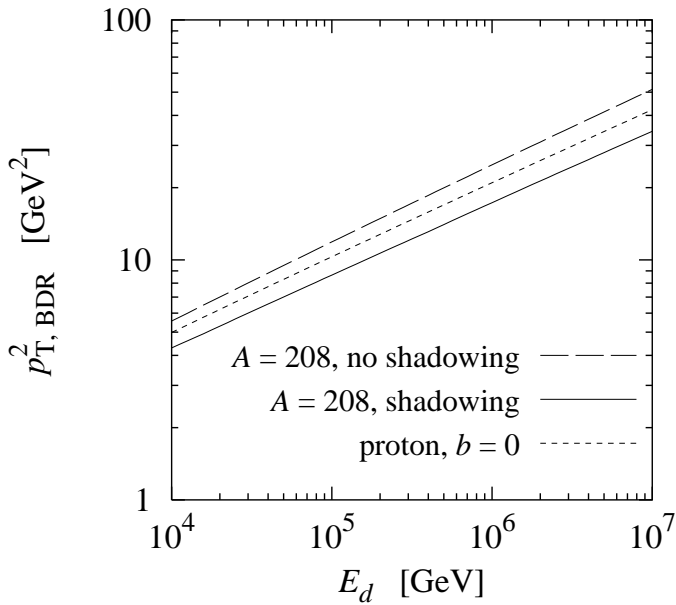}}
\caption[]{The dependence of $[p_T^{\rm BDR}]^2$ for gluon interactions 
with a proton at $b \sim 0$ (dotted line) and a lead nucleus without (dashed
line) and with (solid) leading-twist nuclear shadowing as a function of 
the incident gluon dipole energy in the 
rest frame of the target. Note that for an incident quark, $[p_T^{\rm BDR}]^2$ 
is a factor of two smaller.
}
\label{pbdr}
\end{figure}

Assuming that  the growth of 
$\Gamma(s_{(q \overline q)N},b)$ is proportional to
the nuclear thickness function, $T_A(b)$, given 
by pQCD for $\Gamma(s_{(q \overline q)N},b) 
\leq 1/2$, 
it is straightforward to estimate the highest $p_T$ at which the BDR remains
valid, $p_T^{\rm BDR}$ \cite{annual,FSW03}. 
Figure~\ref{pbdr} shows $[p_T^{\rm BDR}(s_{(q \overline q)N},b=0)]^2$ 
for gluon interactions with both a proton and a nucleus with $A \sim 208$.
The value of $p_T^{\rm BDR}$ is determined by the $p_T$ at which a single 
gluon would
be completely absorbed by the target like a colorless dipole of size
$d = \pi/Q \sim \pi/(2p_T)$. At $x \sim 10^{-4}$, the interaction scale for
which a colorless gluon dipole at the edge of the BDR is $Q^2 \sim 4p_T^{\rm
BDR} \sim $ few GeV$^2$, corresponding to $1-3$ gluon rungs on the ladder in
multi-Regge kinematics.
$b\sim 0$.  
This same kinematic region will also be 
covered in UPCs at the LHC. The $Q^2$ at which the BDR is reached for $q 
\overline q$ dipoles is about a factor of two smaller than for gluons at the
same energy.  This new strongly-interacting, small-coupling QCD regime is thus
fundamentally different from the leading-twist approximation in NLO pQCD.

Here we outline the basic features of hard production in the BDR which can 
distinguish it from competing phenomena.  

\subsubsection{Manifestations of the BDR for inclusive phenomena}

\subsection*{Nuclear structure functions and parton densities}

One distinct feature of the QCD Lagrangian is its conformal invariance in the 
limit where the bare quark masses can be neglected.  Conformal invariance is 
violated in QCD by spontaneously broken chiral symmetry.  Since the quark 
masses are typically neglected in hard scattering amplitudes, these amplitudes 
are conformally invariant except for effects due to the running of the 
coupling constant.  Conformal invariance of the moments of the structure 
functions leads to approximate Bjorken scaling up to corrections due to the 
$Q^2$ evolution.  It is often assumed that $p_T$ diffusion is unimportant 
after NLO effects are included in the BFKL approximation. This 
assumption is supported by numerical analysis of NLO BFKL approximation
\cite{Ciafaloni:2005xf}.

In contrast,  at sufficiently small $x$ where the BDR is reached 
and the pQCD series diverges, conformal invariance is grossly violated: 
approximate Bjorken scaling disappears. 
At the small $x$ values in the BDR probed at the LHC the structure function 
of a heavy nucleus with $R_A=1.2A^{1/3}$ fm has the form
\begin{equation}  
F_2^A(x,Q^2)= \sum_q \frac{e_q^2}{12\pi^2} 2\pi R_A^2 Q^2 \bigg[
\frac{1}{3} \ln A + \lambda \ln \bigg(\frac{x_0}{x}\bigg)\bigg] 
\theta(x_0-x) \, \, 
\label{f2abbdr}
\end{equation}
where $x_0$ does not depend on $A$.
The sum is over the number of active flavors with charge $e_q$. 
Since the DIS cross section is $\propto F_2/Q^2$, in this limit
the cross section becomes independent of $Q^2$.
The parameter $\lambda\approx 0.2$ characterizes the 
increase of the hard amplitudes with energy for moderate $Q^2$.
The first term in Eq.~(\ref{f2abbdr}) is overestimated since LT nuclear 
shadowing is neglected.  The result follows from the calculation of the 
nuclear structure function in terms of the
polarization operator of the photon \cite{Gribov} 
modified to include color transparency.
The structure function increase should change at asymptotically large energies
where the interaction radius significantly exceeds $R_A$,
\begin{equation}  
F_2^A(x,Q^2)= \sum_q \frac{e_q^2}{12\pi^2} 2\pi R_A^2 Q^2 \ln 
\bigg(\frac{x_0}{x}\bigg) \theta(x_0-x) \, \, .  
\end{equation}
  
The nuclear gluon density in the BDR, where the LT approximation breaks
down, can be defined from the Higgs-hadron scattering cross section because
the Higgs locally couples to two gluons. 
In the kinematics where the gluon-dipole interaction is in the BDR,
at moderately small $x$, the gluon distribution is
\cite{AFS}
\begin{equation}
xg_A(x,Q^2) = \frac{1}{12\pi^2} 2\pi R_A^2 Q^2 \bigg[\frac{1}{3} \ln A +
\lambda \ln\bigg(\frac{x_0}{x}\bigg)\bigg] \theta(x_0-x) \,\, .
\label{glubdr}
\end{equation}
Although the effect of color transparency was not taken into account in
Ref.~\cite{AFS}, it is included in Eq.~(\ref{glubdr}).

\subsection*{Nucleon structure functions and 
the total $\gamma N$ cross section}

In a nucleon, the onset of the BDR is accompanied by a fast increase of the 
interaction radius due to the steep decrease of nucleon density with impact 
parameter. As a result, 
\begin{eqnarray}
R_{N}^2({\rm eff})= R_N^2 +c \ln^2 (x_0/x) \, \, , 
\label{rneff}
\end{eqnarray}
leading to
\begin{equation}
F_2^N \propto \ln^3 \bigg(\frac{x_0}{x}\bigg) \,\, , \,\,\,\,\,\,\,\,\,\,
\sigma_{\gamma N} \propto \ln^3\bigg(\frac{s_{\gamma N}}{s_0}\bigg) \,\, .
\end{equation}
A similar phenomenon occurs only at extremely high energies in nuclei. The 
calculation of $c$ in Eq.~(\ref{rneff}) remains model dependent
except at ultrahigh energies where $R_A$ and $R_N$ are
determined by pion exchange.  Hence the total
$\gamma N$ cross section should grow faster with energy 
than the $NN$ cross section. The same is true for nuclei. Since the energy
increase is faster for a nucleon target, the ratio 
$\sigma_{\rm tot}^{\gamma A}/ A\sigma_{\rm tot}^{\gamma N}$, characterizing 
nuclear shadowing, should decrease with energy.  The fraction of the cross 
section due to heavy flavor production should then increase, asymptotically
reaching the SU(4)/SU(5) limit.
 
\subsection*{Inclusive jet and hadron production}

Since partons with $p_T \leq p_T^{\rm BDR}$ cannot propagate through nuclei 
without inelastic interactions, losing a significant fraction of their initial
energy and broadening the $p_T$ distribution \cite{FS06}, we 
expect leading-hadron suppression, similar to that 
observed in d+Au interactions at RHIC \cite{BRAHMS}.  The suppression 
strongly enhances scattering off the nuclear edge, resulting in back-to-back
correlations between high $p_T$ particles at central and forward rapidities. 
To study the $b$ dependence of this correlation, a centrality trigger is
necessary, along with the inclusive asymmetry observables defined in
Ref.~\cite{FS06}.  The suppression of the correlation is small if the
rapidity difference between the two jets is large \cite{FS06}.
It is also possible to study similar effects for leading charm production 
since $p_T^{\rm BDR} \geq m_c$.

The rise of the dijet cross section is expected to slow for $p_T \leq
p_T^{\rm BDR}$.  A similar decrease should be observed for back-to-back pions.
As shown in Ref.~\cite{strikman06} and in Section~\ref{section-pdf}, 
such studies will be feasible for 
$5 \times 10^{-5}\leq x\leq 10^{-2}$.

\subsubsection{Diffractive phenomena}

\subsection*{Inclusive diffraction}

Diffraction in the BDR emerges from the complementary components of the 
photon wavefunction that are not fully absorbed at $b\leq R_A$.  Thus it
directly reflects the photon wavefunction at the BDR resolution 
scale. The diffractive cross section should constitute about half the 
total cross section.  The difference from this limit is due to nuclear edge
effects.  Gribov's orthogonality argument for the derivation of the total 
cross section can be used to derive Eq.~(\ref{ccsb}), the BDR expression for 
the real photon cross section as a function of invariant mass $M$
\cite{Frankfurt:2001nt},
qualitatively different from pQCD. 

Dijet production dominates 
diffraction in the BDR.  Corrections arise from three jet production as in 
$e^+e^- \to q \overline q g$.  Dijet production is also strongly 
suppressed within the LT approximation where the cross section is proportional
to $1/p_T^8$.  Within the BDR, the jet cross section is proportional to
$A^{2/3}$ and decreases as $1/(-t) = 1/p_T^2$, as shown in 
Eq.~(\ref{ccsb}).  

\subsection*{Vector meson production}

The same approach gives the vector meson production cross section in the BDR,
corresponding to diagonal vector meson dominance with a total cross section 
of $2\pi R_A^2$,
 \begin{equation}
{{d\sigma_{\gamma A\to V A}}
\over dt} = \frac{3 \Gamma_{V\to e^+e^-}}{\alpha M_V}{(2\pi 
R_A^2)^2\over 16\pi} {4\left|J_1(\sqrt{-t}R_A)\right|^2
\over -t R_A^2} \, \, 
\label{vmccsb}
\end{equation}
where the first factor is equivalent to $|C_V|^2$ in Eq.~(\ref{csubv}).
The vector meson cross section in Eq.~(\ref{vmccsb}) decreases as
$1/M_V^4$ with $-t \sim M_V^2$ since $\Gamma_{V\to e^+e^-} \sim 1/M_V$ while in
the DGLAP regime the cross section decreases more rapidly as $M_V^{-8}$.
The different $M_V$ dependencies are reminiscent of the change in the $Q^2$ 
dependence of coherent vector meson production
from $\sigma_L \propto 1/Q^6$, $\sigma_T\propto 1/Q^8$ to $\sigma_L \propto 
1/Q^2$, $\sigma_T\propto 1/Q^4$  in the BDR \cite{Frankfurt:2001nt}.  
The $A$ dependence of 
the $t$-integrated cross section also changes from $A^{4/3}$ to $A^{2/3}$,
see Eq.~(\ref{vmccsb}).

As discussed in Section~\ref{neutron}, 
it will be difficult to push measurements of the 
coherent vector meson cross sections in $AA$ collisions to $s_{\gamma N}
\ge 2 E_N M_{V}$ at $y=0$ because it is impossible to distinguish
which nucleus emitted the photon. However, in $pA$ interactions, the $\gamma p$
contribution is much bigger than the $\gamma A$, making identification simpler,
see Section~\ref{paonium}.
 
There are two other ways to study the interaction of small dipoles up to 
$W_{\gamma N} \sim 1$ TeV in the BDR. One is vector meson production in 
incoherent diffraction which should change from $\sigma \propto A$ to
$\sigma \propto A^{1/3}$.
Another is high $t$ vector meson production in rapidity-gap events where a 
transition from the linear $A$ dependence of color transparency 
to the $A^{1/3}$ dependence in the BDR is expected.
The slope of the $t$ dependence of hard diffractive production by nucleons 
should rapidly increase with energy, $B = B_0 +c\ln^2(1/x)$ in the BDR
kinematics. 

\subsection{Testing saturation physics in ultraperipheral collisions}
{\it Contributed by: F. Gelis and U. A. Wiedemann}
\label{sec:cgc-desc}

Parton saturation is a phenomenon generically expected in hadronic 
collisions at sufficiently high center-of-mass energy. Within perturbative 
QCD, the linear evolution equation derived by Balitsky, Fadin, Kuraev and 
Lipatov \cite{KuraeLF1,BalitL1} describes the growth of the unintegrated
gluon distribution in a hadron as it is boosted towards higher
rapidities. This BFKL evolution formalizes the picture that large-$x$ 
partons in a hadronic wavefunction are sources for small-$x$ partons.  In the 
BFKL evolution, these small-$x$ contributions are generated by splitting 
processes such as $g\to gg$ which radiate into the phase space region newly 
opened up by the boost. This linear evolution leads to untamed growth of 
the parton density with $\log x$. It also leads to 
a power-like growth of hadronic cross sections with $\sqrt{s}$,
known to violate unitarity at ultra-high $\sqrt{s}$.

As first noted by Gribov, Levin and Ryskin \cite{GriboLR1}, at 
sufficiently high parton density,
nonlinear recombination processes such as $gg\to g$ cannot be neglected. These 
processes tame further growth of the parton distributions: a saturation 
mechanism of some kind must set in.  Treating
the partons as ordinary particles, it is possible to make a crude estimate 
of the onset of saturation from a simple mean-free path argument. The
recombination cross section for a gluon with transverse momentum $Q $ is
\begin{equation}
\sigma \sim \frac{\alpha_s(Q^2)}{Q^2}\; 
\end{equation}
while the number of gluons per unit transverse
area is given by
\begin{equation}\label{rho}
\rho \sim \frac{xg(x,Q^2)}{\pi r_h^2}\; ,
\end{equation}
where $r_h$ is the radius of the hadron and $x$ the momentum fraction of
the gluons. Saturation sets in when $\rho\sigma\sim 1$, or
equivalently for:
\begin{equation}\label{Qsaturation}
Q^2= Q_s^2 \sim \alpha_s(Q_s^2) \rho
\sim \alpha_s(Q_s^2)\frac{xg(x,Q_s^2)}{\pi R_A^2}\; .
\end{equation}
The momentum scale that characterizes this new regime, $Q_s$, is
called the saturation momentum \cite{Muell3}. Partons with transverse
momentum $Q> Q_s$ are in a dilute regime; those with $Q<Q_s$ are in
the saturated regime. Most generally, $Q_s$ characterizes
the scale at which nonlinear QCD effects become important. 
In the high energy limit, contributions from different nucleons in a nucleus
act coherently. For large nuclei, one thus expects 
$Q_s^2\propto\alpha_s(Q_s^2) A^{1/3}$. Another important parametric
characterization of the saturated region is obtained by estimating the 
number of 
partons occupying a small disk of radius $1/Q_s$ in the transverse plane. 
Combining Eqs.~(\ref{rho}) and (\ref{Qsaturation}) shows that the number is 
proportional to $1/\alpha_s$.  This is the parametrically
large occupation number of a classical field, supporting the idea that 
classical background field methods become relevant for describing nuclear 
wavefunctions at small $x$. 

Within the last two decades, the qualitative arguments given above have been
significantly substantiated. A more refined argument for the onset of 
saturation was given in Ref.~\cite{MuellQ1} where recombination is associated 
with a higher-twist correction to the DGLAP equation.  Early estimates of 
$Q_s$ in nucleus-nucleus collisions \cite{BlaizM1} do not 
differ much from more modern ones \cite{Muell8}. Finally, over the last decade,
nonlinear equations have been obtained 
which follow the evolution of the partonic systems from the dilute regime to
the dense, saturated, regime. These take different, equivalent, forms,
generically referred to as the JIMWLK equation. The resulting calculational
framework is also referred to as the color glass condensate (CGC) formalism. 

\subsubsection{The JIMWLK equation} \bigskip

In the original McLerran and Venugopalan model
\cite{McLerV1,McLerV2,McLerran:1994vd}, the fast partons are frozen, 
Lorentz-contracted color sources flying along the light-cone, constituting
a color charge density $\rho(\x_T)$. Conversely, the low $x$
partons are described by classical gauge fields, $A^\mu(x)$, determined
by solving the Yang-Mills equations with the source given by the
frozen partonic configuration.  An average over all acceptable
configurations must be performed.

The weight of a given configuration is a functional $W_{x_0}[\rho]$ of
the density $\rho$ which depends on the separation scale $x_0$ between
the modes which are described as frozen sources and the modes which
are described as dynamical fields. As one lowers this separation
scale, more and more modes are included among the frozen sources.
Therefore the functional $W_{x_0}$ evolves with $x_0$ according to a
renormalization group equation
\cite{JalilKLW1,JalilKLW2,Balit1,Kovch1,Kovch3,JalilKMW1,IancuLM1,IancuLM2,Weige1,FerreILM1}.

The evolution equation for $W_{x_0}[\rho]$, the so-called JIMWLK
equation, derived in Refs.~\cite{JalilKLW1,JalilKLW2,Balit1,%
Kovch1,Kovch3,JalilKMW1,IancuLM1,IancuLM2,Weige1,FerreILM1}, is
\begin{equation}
\frac{\partial {W_{x_0}[\rho]}}{\partial\ln(1/x_0)}
=\frac{1}{2}\int d^2x_T d^2y_T
\frac{\delta}{\delta {\rho_a(\vec x_T)}}\bigg[
{\chi_{ab}(\vec x_T,\vec y_T)}
\frac{\delta W_{x_0}[\rho]}{\delta\rho_b(\vec y_T)} \bigg]
 \, \, .
\label{eq:JIMWLK}
\end{equation}
The kernel, ${\chi_{ab}(\vec x_T,\vec y_T)}$, only depends
on $\rho$ via Wilson lines,
\begin{equation}
U(\vec x_T)\equiv {\cal P} \exp\left[ -ig \int_{-\infty}^{+\infty} dz^-
A^+(z^-,\vec x_T)\right]\; 
\end{equation}
where ${\cal P}$ denotes path ordering along the $x^-$ axis and $A^+$ is
the classical color field of the  hadron moving close to the speed
of light in the $+z$ direction. The field  $A^+$ depends implicitly on the
frozen sources, i.e.\ on $\rho(\vec x_T)$.

The JIMWLK equation can be rewritten as an infinite
hierarchy of equations for $\rho$, or equivalently $U$ correlation functions.
For example, the correlator ${\rm
Tr}\big<U^\dagger(\vec x_T)U(\vec y_T)\big>$ of two Wilson lines has
an evolution equation that involves a correlator of four Wilson
lines. If this 4-point correlator is assumed to be factorisable
into the product of two 2-point functions, a closed equation
for the 2-point function, the Balitsky-Kovchegov (BK)
\cite{Balit1,Kovch3} equation, is obtained,
\begin{eqnarray}
&&\frac{\partial {\rm Tr}\big<U^\dagger(\vec x_T)U(\vec y_T)\big>_{x_0}}
{\partial\ln(1/x_0)}
=
-\frac{\alpha_s}{2\pi^2}\int
\frac{d^2z_T \, (\vec x_T-\vec y_T)^2}{(\vec x_T-\vec z_T)^2(\vec y_T-\vec
z_T)^2}
\label{eq:BK} \\ \!\!\!
&&\times\Big[
N_c {\rm Tr}\big<U^\dagger(\vec x_T)U(\vec y_T)\big>_{x_0}
-{\rm Tr}\big<U^\dagger(\vec x_T)U(\vec z_T)\big>_{x_0}
{\rm Tr}\big<U^\dagger(\vec z_T)U(\vec y_T)\big>_{x_0}
\Big]\; .
\nonumber
\end{eqnarray}
The traces in Eq.~(\ref{eq:BK}) are performed over color indices.

When the color charge density is small, the
Wilson line, $U$, can be expanded in powers of $\rho$.  Equation~(\ref{eq:BK}) 
then becomes a linear evolution equation for the correlator
$\big<\rho(\vec x_T)\rho(\vec y_T)\big>_{x_0}$ or, equivalently, for
the unintegrated gluon density, the BFKL
equation. The same is true of Eq.~(\ref{eq:JIMWLK}) because, in
this limit, the kernel $\chi_{ab}$ becomes quadratic in $\rho$.

Similar to the BFKL or DGLAP evolution equations, the initial condition
is a non-perturbative input which can, in
principle, be modeled, adjusting the parameters
to fit experimental data. A simple input is the McLerran
and Venugopalan (MV) model with a local Gaussian for the initial 
$W_{x_0}[\rho]$,
\begin{equation}
W_{x_0}[\rho]=\exp\Big[ -\int \, d^2x_T
\frac{\rho(\vec x_T)\rho(\vec x_T)}{\mu^2}
\Big]\; .
\label{eq:MV-init-cond}
\end{equation}
Here, we stress that testing the predictions of the
CGC requires testing both the evolution with rapidity 
and the initial conditions.

The MV model requires an infrared cutoff at the scale
$\Lambda_{_{\rm QCD}}$ because assuming a local Gaussian
distribution ignores the fact that color neutralization occurs on
distance scales smaller than the nucleon size ($\sim
\Lambda_{_{\rm QCD}}^{-1}$): two color densities can 
only be uncorrelated if they are transversely separated by at 
least the distance scale
of color neutralization.  Note that the sensitivity to this infrared
cutoff gradually disappears as one lowers the separation scale $x_0$
in the JIMWLK equation. Indeed, in the saturated regime, color
neutralization occurs on distance scales of the order of
$Q_s^{-1}(x_0)$ \cite{IancuIM2}, the physical origin of the
universality of the saturated regime.

\subsubsection{Saturation in photon-nucleus collisions} \bigskip

High parton density effects can be tested in photo-nuclear UPCs.
Quite generically, the cross section for the process $AA\to F X$, where $F$
denotes a specific produced final state and $X$ unidentified debris from 
the nucleus, is
\begin{equation}
\sigma_{AA\to F X}(\sqrt{s_{_{NN}}})
=
\int_{k_{\rm min}}^{+\infty}dk \,
\frac{dN_\gamma}{dk}\;
\sigma_{\gamma A\to F X}(s_{\gamma N} = 2k \sqrt{s_{_{NN}}})\; .
\label{eq:gamma-A}
\end{equation}
In this formula, $s_{\gamma N} =2k\sqrt{S_{_{NN}}}$ is the square of the
center-of-mass energy of the $\gamma N$ system. The
minimum photon energy for production of $F$, $k_{\rm min}$, is 
determined from the invariant mass squared, $M^2$, of $F$,
\begin{equation}
k_{\rm min}=\frac{M^2}{2\sqrt{s_{_{NN}}}}\; .
\label{eq:wmin}
\end{equation}
In Eq.~(\ref{eq:gamma-A}), gluon saturation effects are included in the 
$\gamma A$ cross section in the
integral. In the next subsection, we discuss the effects of gluon
saturation on open $Q \overline Q$ production (detected as $D$ or $B$ mesons).

\subsubsection{Heavy quark production} \bigskip

Heavy quark production has been proposed as a UPC observable sensitive to 
saturation effects. Calculations which support
this statement treat the nucleus as a collection of classical color 
sources that acts via its color field.  These sources
produce a color field with which the $Q\overline{Q}$ pair
interacts. For a nucleus moving in the $+z$ direction, this color
field, expressed here in the Lorenz gauge, $\partial_\mu A^\mu=0$, is
\begin{equation}
  A^\mu(x) = -g\delta^{\mu +} \delta(x^-) 
\frac{1}{{\mathbf\nabla}_T^2} \rho(\vec x_T)\; 
\label{eq:A}
\end{equation}
where $\rho(\vec x_T)$ is the number density of color charges as a
function of the transverse position in the nucleus. The
scattering matrix for a quark traveling through this color field is
\begin{equation}
T(p,q)=2\pi\gamma^-\delta(p^--q^-)\epsilon\int d^2x_T 
e^{i(\vec q_T-\vec p_T)\cdot \vec x_T}\Big[U^\epsilon(\vec x_T)-1\Big]\; 
\label{eq:Qscatt}
\end{equation}
where $p$ ($q$) is the incoming (outgoing) four-momentum of the
quark and $\epsilon\equiv{\rm sign}(p^-)$\footnote{When $\epsilon = +1$, 
$U^\epsilon = U$ and when $\epsilon = -1$, $U^\epsilon = U^\dagger$.}. 
The Wilson line in the fundamental representation of SU$(3)$ that resums
all multiple scatterings of the quark on the color field of
Eq.~(\ref{eq:A}) is defined as
\begin{equation}
U(\vec x_T)\equiv T_-\exp\Big[
ig\int_{-\infty}^{+\infty} dz^- A^+_a(z^-,\vec x_T)t^a
\Big]\; 
\end{equation}
where $T_-$ denotes ordering in the variable $z^-$ with the fields with the
largest value of $z^-$ placed on the left.

\begin{figure}[htb]
\begin{center}
\resizebox*{9cm}{!}{\includegraphics{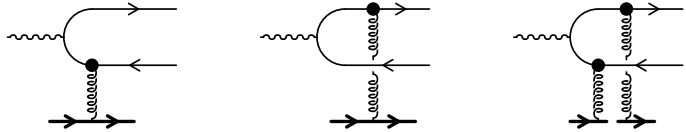}}
\end{center}
\caption[]{The three diagrams that contribute to the
  production of a $Q\overline{Q}$ pair in the interaction of a photon
  with the color field of the nucleus.}
\label{fig:01}
\end{figure}

 From this starting point, the cross section
for $\gamma A\to Q\overline{Q}X$ can be derived~\cite{GelisP1}. 
At leading order in
electromagnetic interactions, the three diagrams
in Fig.~\ref{fig:01} must be evaluated.  The black dot represents
the scattering matrix defined in Eq.~(\ref{eq:Qscatt}).
After summing these three diagrams, we obtain the amplitude
\begin{eqnarray}
&&{\cal M}^\mu(\vec k|\vec q,\vec p)
=\frac{i e_q}{2}\int\frac{d^2\vec l_T}{(2\pi)^2}
\int d^2x_{1T}d^2x_{2T}
\nonumber\\
&&\qquad\;
\times e^{i\vec l_T\cdot\vec x_{1T}}
e^{i(\vec p_T+\vec q_T-\vec k_T-\vec l_T)\cdot\vec x_{2T}}
\left(U(\vec x_{1T})U^\dagger(\vec x_{2T})-1\right)\;
\overline{u}(\vec q)\,
\Gamma^\mu\,v(\vec p)
\; 
\label{eq:DIS2}
\end{eqnarray}
where $l$ is the four-momentum transfer between the quark and the nucleus,
$e_q$ is the electric charge of the produced quark and
\begin{eqnarray}
\Gamma^\mu\equiv
\frac{\gamma^-(\slq-\slL+m)\gamma^\mu(\slq-\slK-\slL+m)\gamma^-}%
{p^-[(\vec q_T\!-\!\vec l_T)^2+m^2\!-\!2q^-k^+]
+q^-[(\vec q_T \!-\!\vec k_T\!-\!\vec l_T)^2+m^2]}
\; .
\label{eq:commonspinors}
\end{eqnarray}
Here $\gamma^-$ is the $-$ component of the Dirac matrices.
In these formulas, $k$, $q$ and $p$ are the four-momenta of the photon, quark 
and antiquark respectively. The cross section
is obtained from this amplitude by
\begin{eqnarray}
  &&
  d\sigma_{\gamma A\to Q\overline{Q}X}
  =\frac{d^3q}{(2\pi)^22q_0}\frac{d^3p}{(2\pi)^32p_0}
  \frac{1}{2k^-} 
  2\pi\delta(k^--p^--q^-)\nonumber\\
  &&\qquad\qquad\times
\left<{\cal M}^\mu(\vec k|\vec q,\vec p){\cal M}^{\nu *}(\vec k|\vec q,\vec p)
  \right>_\rho
  \epsilon_\mu(k)\epsilon_\nu^*(k)\; ,
\label{eq:cs1}
\end{eqnarray}
where $\epsilon_\mu(k)$ is the polarization vector of the photon (all 
possible polarizations should in principle be summed) and 
$\big<\cdots\big>_\rho$ denotes the average over all the possible
configurations of the distribution of color sources in the nucleus,
$\rho(\vec x_T)$, weighted by the functional $W[\rho]$ defined in
Section~\ref{sec:cgc-desc}.

\subsection*{Inclusive cross section}

After integrating over the phase space of the produced quark and antiquark,
the total cross section is \cite{GelisP1,GoncaM7} 
\begin{eqnarray}
\sigma_{\gamma A\to Q\overline{Q}X}
&=&\frac{\alpha e_Q^2}{2\pi^2} \int dl_T^2
\Big[\pi R_{_A}^2 C(x,l_T)\Big]
\nonumber\\
&&\qquad\times
\Big[1+\frac{4(l_T^2-m^2)}{l_T\sqrt{l_T^2+4m^2}}
{\rm arcth}\, \frac{l_T}{\sqrt{l_T^2+4m^2}}
\Big]\; ,
\label{eq:cs2}
\end{eqnarray}
with
\begin{equation}
C(x,l_T)\equiv \int d^2x_T\; e^{i\vec l_T\cdot \vec x_T}\;
\left<U(\vec x_T)U^\dagger(0)\right>_\rho\; .
\end{equation}
Note that Eq.~(\ref{eq:cs2}) depends on the modulus of $|\vec l_T|$.
The momentum fraction $x$ is given by $x=k_{\rm min}/k =
4m^2/s_{\gamma N}$, Eqs.~(\ref{eq:gamma-A}) and (\ref{eq:wmin}) with $M = 2m$. 
We emphasize that the $x$
dependence of $C(x,l_T)$ comes entirely from the $x$-evolution 
 of the distribution $W[\rho]$ of the classical color sources in
the nucleus. Therefore, the $x$ dependence of this cross section tests 
some predictions of the $W[\rho]$ evolution equations, Eq.~(\ref{eq:JIMWLK}), 
or the simpler BK equation for the evolution of the
correlator $\left<U(\vec x_T)U^\dagger(0)\right>_\rho$, Eq.~(\ref{eq:BK}).

After manipulation of Eq.~(\ref{eq:cs1}),
$\sigma_{\gamma A\to Q\overline{Q}X}$ can alternatively be
expressed in terms of the
dipole cross section \cite{GelisJ3}
\begin{equation}
\sigma_{\gamma A\to Q\overline{Q}X}=\int_0^1 dz\int d^2r_T
\left|\Psi(k|z,\vec r_T)\right|^2
\sigma_{\rm dip}(x,\vec r_T)\; .
\label{eq:cs3}
\end{equation}
In this formula, the ``photon
wavefunction'', $\Psi(k|z,\vec r_T)$, 
denotes the $Q\overline{Q}$ Fock component 
of the virtual photon light-cone wavefunction that corresponds to a 
quark-antiquark dipole of transverse size $\vec r_T$.
The square of the wavefunction is
\begin{eqnarray}
&&\left|\Psi(k|z,\vec r_T)\right|^2\equiv
\frac{N_c\,\epsilon_\mu(k)\epsilon_\nu^*(k)}{64\pi k_-^2 z(1-z)}
\int \frac{d^2l_T}{(2\pi)^2}\frac{d^2l^\prime_T}{(2\pi)^2}
e^{i(\vec l_T-\vec l^\prime_T)\cdot\vec r_T}
\nonumber\\
&&\qquad\qquad\qquad\times
{\rm Tr}_d\left((\slq+m)
\Gamma^\mu(\slp-m)
\Gamma^{\nu\prime\dagger}
\right)\; 
\end{eqnarray}
where Tr$_d$ indicates a trace over Dirac indices rather than a color trace.
The longitudinal momentum fraction, $z$, is defined as
$z=q^-/k^-$. The dipole cross section, an important quantity
in saturation physics, can be defined in terms of a Wilson
line correlator,
\begin{equation}
\sigma_{\rm dip}(\vec r_T)
=
2\int d^2b \;[1-S(\vec b,\vec r_T)]\; ,
\label{eq:dipole-cs}
\end{equation}
with
\begin{equation}
S(\vec b,\vec r_T)\equiv \frac{1}{N_c}
{\rm Tr} \,\left<U(\vec b+\frac{\vec r_T}{2})
U^\dagger(\vec b-\frac{\vec r_T}{2})\right>_\rho \; .
\end{equation}
The above expressions are valid for both $\gamma A$ and $\gamma p$ 
interactions.  The only difference is that the averages are performed over
the color field of a nucleus or a proton respectively. 

Several models of the dipole cross section have been used 
to fit the HERA $\gamma p$ data. Golec-Biernat and W\"usthoff
used a very simple parametrization \cite{GolecW1,GolecW2},
\begin{equation}
\sigma_{\rm dip}(x,\vec r_T)=\sigma_0 \left[
1-e^{-\frac{1}{4}Q_s^2(x)r_T^2}
\right]\; ,
\label{eq:GBW}
\end{equation}
which shows good agreement with the data at $x<10^{-2}$
and moderate $Q^2$. In this formula, the scale $Q_s(x)$ has the $x$-dependent
form
\begin{equation}
Q_s^2(x)=Q_0^2 \left(\frac{x_0}{x}\right)^\lambda\; .
\label{eq25}
\end{equation}
A fit of HERA $F_2$ data suggests $\lambda\approx 0.29$. The
parameter $Q_0$ is set to 1~GeV with $x_0\approx 3\times 10^{-4}$ for a 
proton.  In the nucleus $Q_0^2$ must be scaled by $A^{1/3}$.  However,
this model fails at large $Q^2$. The high $Q^2$ behavior was improved in 
Ref.~\cite{BarteGK1} where the dipole cross section was parametrized to
reproduce pQCD for small dipoles. Even if these approaches are inspired by 
saturation physics, they do not derive the dipole
cross section from first principles. Recently, Iancu, Itakura and
Munier \cite{IancuIM3} derived an expression of the dipole
cross section from the color glass condensate framework and obtained
a good fit of the HERA data with $\sigma_0$, $\lambda$ and $Q_0$ ($(x_0)$)
as free parameters. An
equally good fit was obtained by Gotsman, Levin, Lublinsky and
Maor who derived the $x$ dependence of the dipole cross section by
numerically solving the BK equation, including DGLAP corrections
\cite{GotsmLLM1}.

\subsection*{Diffractive cross section}

Starting from the dipole cross section, Eq.~(\ref{eq:dipole-cs}), 
the elastic dipole cross section is \cite{Muell9}
\begin{equation}
\sigma^{\rm elastic}_{\rm dip}(\vec r_T)
=
\int d^2b\; [1-S(\vec b,\vec r_T)]^2\; .
\end{equation}
If diffractive $Q\overline{Q}$ production is viewed as a sum
of elastic dipole scatterings \cite{GelisP2,GoncaM3},
\begin{eqnarray}
\sigma^{\rm diff}_{\gamma A\to Q\overline{Q}X}
&=&
\int d^2b\int_0^1 dz\int d^2r_T
\left|\Psi(k|z,\vec r_T)\right|^2\nonumber\\
&&\qquad\qquad\times
\Big[
1-\frac{1}{N_c}
{\rm Tr} \,\left<U(\vec b+\frac{\vec r_T}{2})
U^\dagger(\vec b-\frac{\vec r_T}{2})\right>_\rho
\Big]^2\; .
\end{eqnarray}
Therefore, to simultaneously predict the inclusive and
diffractive cross sections, a description of the source
distribution, $W[\rho]$, that contains some information about
the transverse profile of the nucleus is needed.  If only the inclusive 
cross section is calculated, a model of the impact-parameter integrated
total dipole cross section is sufficient.

\subsection*{Example results}

Several models, with various assumptions and
degrees of sophistication, exist in the literature
\cite{GoncaM7,GoncaM3,GoncaM4}. The dipole cross section on a proton is
calculated with Eqs.~(\ref{eq:GBW}) and (\ref{eq25})
employing $\lambda\approx 0.29$, $x_0\approx 3.04\times 10^{-4}$ and
$\sigma_0\approx 23.03$ mb \cite{GolecW1,GolecW2}. The dipole 
cross section for a nucleus is 
obtained using Glauber scattering, 
\begin{equation}
\sigma_{\rm dip}(x,\vec r_T)=
2\int d^2b \;\Big[1-\exp \left(-\frac{1}{2} T_A(b)
\sigma_{\rm dip}^{p}(x,\vec r_T)\right)\Big]\; . 
\end{equation}
Although the average 
dipole size decreases with $x$ \cite{GoncaM3}, the effect
is not very significant. More importantly, if all other parameters are
kept fixed, the dipole is larger in the diffractive cross section than in
the inclusive cross section ~\cite{GoncaM3}.

\begin{figure}[htb]
  \begin{center}
    \resizebox*{8cm}{!}{\includegraphics{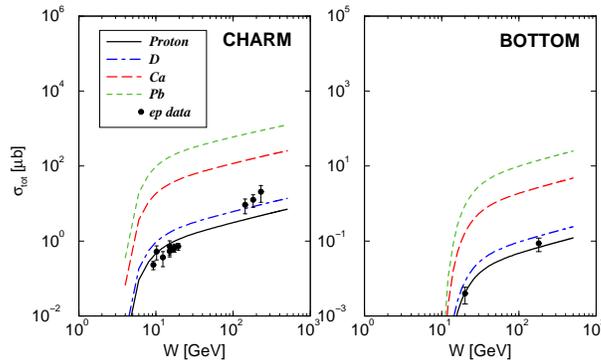}}
  \end{center}
  \caption[]{The inclusive $\gamma A\to Q\overline{Q}X$
    cross section for (left-hand side) charm and (right-hand side) bottom
    as a function of $W$, the center-of-mass energy in the
    $\gamma A$ system.  Reprinted from Ref.~\protect\cite{GoncaM3} with
    permission from Springer-Verlag.}
\label{fig:04}
\end{figure}

\begin{figure}[htb]
  \begin{center}
    \resizebox*{8cm}{!}{\includegraphics{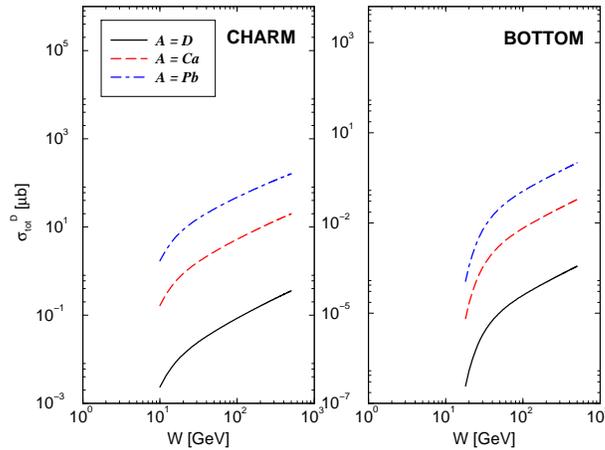}}
  \end{center}
  \caption[]{The diffractive $\gamma A\to Q\overline{Q}X$
    cross section for (left-hand side) charm and (right-hand side) bottom
    as a function of $W$, as a function of the center-of-mass energy in the
    $\gamma A$ system, for various nuclei.  Reprinted from 
    Ref.~\protect\cite{GoncaM3} with
    permission from Springer-Verlag.}
\label{fig:05}
\end{figure}

Figure \ref{fig:04} shows the results of Ref.~\cite{GoncaM3} for
inclusive $c\overline{c}$ and $b\overline{b}$ production. The
cross sections are given for protons and deuterons as well as
calcium and lead nuclei.  The proton case is
compared to $ep$ data.
Similar results for the diffractive cross section are
displayed in Fig.~\ref{fig:05} for deuterons, calcium and lead.
The diffractive cross section is about a factor of ten lower 
than the inclusive cross section.

\section{Two-photon physics at the LHC}

Since photons couple to all charged particles, two-photon 
processes involve a wide range of reactions.  The large ion
charge leads to high two-photon rates.  In this section, we
discuss some of the available physics processes that can be 
studied at the LHC due to these high rates.  We begin with
lepton-pair production in strong fields in Section~\ref{pureqed}
and then consider hadronic final states in Section~\ref{ggpotential}.
Finally, Section~\ref{alice-gammagamma} discusses the observation
of two-photon proceses at the LHC.

\subsection{Pure QED processes}
\label{pureqed}

The lepton pair production cross section in heavy-ion collisions
is extremely high, about $2\times 10^5$ b for $e^+e^-$ production
by lead beams at the LHC.  The large coupling, proportional to
$Z^4 \alpha$ invites discussion of nonperturbative effects.  One particularly
interesting process is the production of antihydrogen via positron
capture by an antiproton \cite{newref3}.
Multiple lepton pair production by a single heavy-ion event is
also interesting, as discussed in Section~\ref{strong}.  In
addition, Section~\ref{bfpp} considers a special case of
pair production, bound-free pairs, where an $e^+ e^-$ pair 
is produced so that the $e^-$ is bound to one of the incident
ions.  Although the bound-free pair production cross section is smaller
than the total lepton-pair cross section, it has a number of important 
implications for the LHC.  Pair production has been studied at RHIC
\cite{Adams:2004rz} and the Fermilab Tevatron \cite{CDF}, in addition 
to fixed-target experiments \cite{newref2,secx:vane92,secx:vane94}.

\subsubsection{Bound-free pair production}
\label{bfpp} \bigskip
{\it Contributed by: S. R. Klein} \bigskip

Bound-free pair production (BFPP) occurs when an $e^+e^-$ pair is produced 
and the $e^-$ becomes bound to one of the incident nuclei.  This process has 
a large cross section, about 280 b for lead at the LHC \cite{BFPP}.
BFPP is important because it limits the LHC luminosity with heavy-ion beams.  
The reaction changes the charge, $Z$, of a beam ion while leaving its 
momentum essentially unchanged.  In essence, the process produces a beam of 
single-electron ions.  Since the ions have an altered charge to
mass ratio, they follow a different trajectory from the bare ion beam,
eventually striking the LHC beam pipe.  With lead beams, the change in charge 
is $1/Z\approx 0.012$ and the ion strikes the beam pipe
several hundred meters downstream from the interaction point.  At the maximum 
design luminosity for lead, $10^{27}$ cm$^{-2}$s$^{-1}$, the single-electron 
beam carries about 280,000 particles/s, dumping about 25 Watts of power into
a relatively small section of the beam pipe, enough to overwhelm the LHC 
magnet cooling systems and causing the struck magnet to quench 
\cite{LHC450,klein01,Jowett:2003my,pac2005}.  It is 
necessary to keep the LHC luminosity low enough to prevent this from
happening.

Several different BFPP calculations have been made.  
One such calculation for capture into a $K$-shell orbital \cite{BFPPHencken},
\begin{equation}
\sigma_{A B \rightarrow Ae^- + B + e^+} = 
Z_A^5Z_B^2\bigg[a\log(\gamma_{L \, {\rm CM}}) + b^\prime \bigg] \, \, ,
\end{equation}
shows the scaling with beam species and energy.  Here,
$Z_A$ is the charge of nucleus $A$ that the electron is bound to, 
$Z_B$ is the charge of $B$, $\gamma_{L \, {\rm CM}}$ the Lorentz boost of a 
single beam in the center-of-mass frame
and $a$ and $b^\prime$ are
constants fit to lower energy data.  This approach was used to obtain the 280 
b cross section given above.  
Capture to higher $s$ orbitals decreases by a factor of $\sim 1/n^3$, a factor
of 8 reduction for the $L$-shell.  
The net effect of all the higher orbitals is to increase 
the cross section by about 20\%.  
Some earlier calculations have obtained BFPP cross section of $\sim 100$ b
but the value of 280 b is also in agreement with other calculations 
\cite{Agger}. In addition, BFPP was measured at the CERN SPS for 158 
GeV/nucleon lead beams on a number of fixed targets \cite{CERNSPS1,CERNSPS2}.  
The data are in reasonable agreement with the calculations of 
Ref.~\cite{BFPPHencken}.

Bound-free pair production has been observed during the 2005 RHIC run
with copper beams \cite{BFPP}.
Although the cross section for bound-free pair production of $^{+28}$Cu is 
small, only 150 mb, the change in $Z/A$
due to electron capture is larger than for heavier ions.   
The single-electron $^{+28}$Cu ions struck the beam pipe about 136 m 
downstream from the interaction point, producing
hadronic showers in the beam pipe and accelerator magnets.  The 
ionization caused by the hadronic showers, with a rate of about 10 Hz,
consistent with theory predictions, was
detected by small PIN diodes.  

\subsubsection{Strong field effects in lepton pair production: 
Coulomb corrections and multiple pair production}
\label{strong} \bigskip
{\it Contributed by: A. J. Baltz, K. Hencken, and G. Baur} \bigskip

In this section we discuss the strong photon-ion
coupling constant and how the nonperturbative QED effects arising from its
strength might be observed in lepton pair production at the LHC.
While the role of higher-order QED in electromagnetic heavy-ion
reactions is interesting in itself, it is also useful as a simpler
model for investigating aspects of nonperturbative QCD.  Though the primary
heavy-ion program involves the technically more challenging quantitative
understanding of nonperturbative QCD, the more tractable theoretical treatment 
of higher-order QED should be experimentally verified.
At present, both the experimental and theoretical state-of-the art of 
higher-order QED is unsatisfactory in UPCs.  As we discuss, although
some theoretical questions remain, more experimental data is greatly needed.
Although there was sufficient interest during the planning stages of RHIC
\cite{frbt}, no definitive
experimental tests of higher-order QED have yet been performed.
Here we review the theoretical and experimental situation and discuss some
experimental probes of nonperturbative QED effects at the LHC.

Leading-order calculations of charged-particle induced lepton pair production
date back to the work of Landau and Lifshitz \cite{Landau:1934zj} and Racah 
\cite{Racah:1937XX}.  The 1937 Racah formula for the total cross section 
is remarkably accurate when compared with more recent Monte Carlo calculations
\cite{Bottcher:1989ia,Hencken:1998hf,Alscher:1996gn} of $e^+ e^-$ production.
However, for lead or gold beams, $Z\alpha \sim 0.6$ is
not small.  Higher-order effects may then be non-negligible. In addition,
strong-field QED effects are expected to be more
pronounced at small impact parameters which can be well defined in heavy-ion 
collisions, making it possible to test this expectation.
Calculations have suggested large nonperturbative enhancements relative to
perturbative results at low energies  on one hand  and, on the other, 
significant reductions at
ultrarelativistic energies.  Neither effect has yet been verified
experimentally.

Coupled-channel calculations of $e^+ e^-$ production have been performed 
at low kinetic energies, 1-2 GeV per nucleon
\cite{Rumrich:1991xs,Momberger:1991XX,Rumrich:1992kp}.
A significant increase over perturbation
theory was found.  An enhancement relative to perturbation theory was
also obtained in coupled-channel calculations of $b=0$ fixed-target Pb+Pb 
interactions at 200 GeV/nucleon \cite{Thiel:1995XX}.  

These calculated enhancements were obtained from large cancellations of 
positive and negative time contributions to the pair 
creation probability, with some contributions orders of magnitude larger than
the signal.  The coupled-channel basis is necessarily incomplete.
This limitation, combined with other approximations, may render the method 
impractical.  For example, a factor of 50 enhancement was found in a 
calculation of bound-free pair production in central Pb+Pb fixed-target 
interactions at 1.2 GeV/nucleon ($\gamma_L = 2.3$) 
\cite{Rumrich:1991xs,Rumrich:1992kp}.  When the basis was expanded to include
a 70\% larger pertubative cross section \cite{Baltz:1994XX}, the
higher-order result decreased the enhancement to a factor of nine even though
the basis increased.

There are two interesting higher-order strong field effects: Coulomb 
corrections and multi-photon
exchanges from either one or both ions.
In this treatment, only one $e^+ e^-$ pair is
assumed to be present at any intermediate time step. 
Retarded propagators can then be utilized 
to calculate higher-order Coulomb effects on the total cross section and 
uncorrelated final electron or positron states 
\cite{Baur:1990za,Best:1991uw,Hencken:1994hp,Baltz:2001dp,Aste:2001te}.

The exact solution of the Dirac equation for an
electron in the field of the two nuclei has been studied in the limit
$\gamma_L \rightarrow\infty$.
An all-order summation can be
made in the high-energy limit in the related problem of bound-free pair
production \cite{Baltz:1996hr}.
The summation can be done analytically for free pair 
production \cite{secx:SegevW98,Baltz:1998zb}.  After integration over $b$, the
total cross section is identical to the leading-order result 
\cite{Baltz:1998zb,Segev:1998ur}.  The CERN SPS pair production
data \cite{secx:vane97} also showed perturbative scaling \cite{Segev:1998ur}.  
These data, obtained 
from 160 GeV/nucleon Pb and 200 GeV/nucleon S beams on C, Al, Pa, and Au 
targets, are the only 
available ultrarelativistic $e^+ e^-$ data spanning
a large part of the total cross section.
They showed that the cross sections
scale as the product of the squares of the projectile
and target charges, $(Z_A Z_B)^2$ \cite{secx:vane97}, in
contrast to predictions of
$e^+ e^-$ photoproduction on a heavy target, which shows a negative
(Coulomb) correction, proportional to $Z^2$, well described by
Bethe-Maximon theory \cite{secx:betheM54,secx:daviesBM54}.

Subsequently, it was argued \cite{secx:ivanovSS99,Lee:1999ey,secx:leeM01} 
that a more careful regularization of the propagator than that of
Refs.~\cite{Baltz:1998zb,Segev:1998ur} was needed.  Negative Coulomb
corrections then reappeared, in agreement with Bethe-Maximon
theory. This result was confirmed by numerical 
calculations with a properly regularized
propagator.  The exact semi-classical total cross section 
for $e^+ e^-$ production with $A \sim 200$
is reduced by 28\% at the SPS, 17\% at RHIC and
11\% at the LHC \cite{Baltz:2004dz}.
These calculations are in apparent disagreement with the SPS data.
However, the coupled-channel treatment of the same basic
reaction \cite{Thiel:1995XX} finds an enhancement of the pair production
probability at $b \sim 0$.  The difficulties in the method have been 
previously noted.

At RHIC, the first experimental observation of $e^+ e^-$ pairs
accompanied by nuclear
dissociation was made by STAR \cite{Adams:2004rz}.
As discussed in Section~\ref{section-vm-interference} and 
Ref.~\cite{Baur:2003ar},
this corresponds to pair production with $\langle b \rangle \sim 20$ fm.
Comparison with perturbative QED calculations set
a limit on higher-order corrections of
$-0.5 \sigma_{\rm QED} < \Delta \sigma < 0.2 \sigma_{\rm QED}$, 
at a 90\% confidence
level. Detailed leading-order QED calculations are carried out in 
Ref.~\cite{HBT2}.
The electromagnetic excitation of both ions is included in the semi-classical
approach according to Ref.~\cite{Baur:2003ar}. 

A comparison to calculations without dissociation
in the STAR acceptance gives an indication
of the relative difference between the perturbative and higher-order
results. Within the STAR acceptance, the calculated exact result is 
17\% lower than the perturbative one \cite{Baltz:2004dz}, $\Delta \sigma 
= -0.17 \sigma_{\rm QED}$, not excluded by STAR. On the other hand, the small
impact parameter should enhance higher-order processes.
 
A sample numerical calculation has been performed using the
same method for $e^+ e^-$ production by Pb+Pb ions with cuts in a possible
detector setup in the forward region \cite{Bocian04} 
at the LHC.  For electron and positron energy $E$ and angle $\theta$ such that
$3 < E < 20$ GeV and $0.00223 < \theta < 0.00817$ radians, the
perturbative cross section of 2.88 b without a form factor is reduced 
by 18\%, to 2.36 b, in an exact numerical calculation.  If forward $e^+ e^-$ 
pairs are employed for luminosity measurements at LHC, it seems 
necessary to consider the Coulomb corrections to the predicted cross sections.

Section~\ref{tonyref} discusses $\mu^+ \mu^-$ pairs
as a $\gamma \gamma$ luminosity monitor.  While it is
straightforward to calculate the perturbative $\mu^+ \mu^-$ pair
production rate in heavy-ion collisions, the importance of Coulomb 
corrections is somewhat less clear than for $e^+ e^-$.  Analytic arguments 
suggest that Coulomb corrections are
small for $\mu^+ \mu^-$ production \cite{HKS2,Hencken:2006ir}.  On the other 
hand, numerical calculations of the total $\mu^+ \mu^-$ cross sections, 
employing the same method as the exact $e^+ e^-$ calculations for RHIC and LHC 
mentioned previously, find larger relative reductions 
with respect to perturbation 
theory, 22\% for RHIC and 14\% for LHC.

A second higher-order effect is multiple pair production in a single collision
which restores unitarity, violated at leading order if only single pair
production is assumed. The leading order single pair production probability is
interpreted as the average number of pairs produced in a single collision.
Integration of this probability over impact parameter gives the total 
multiple-pair production cross section.   The matrix element for multiple-pair 
production can be factorized into an antisymmetrized product of pair production
amplitudes.  Calculating the total multiple-pair production probability,
neglecting the antisymmetrization of the amplitude, recovers the Poisson
distribution \cite{Baur:2001jj,BaurHT98}.  
There are also multi-particle
corrections to single pair production which contribute up to 5\% of the 
probability \cite{Baur:2001jj}.
Studies of multiple-pair production for ALICE \cite{HBT2} 
found that about 10\% of the produced pairs detected in the inner tracker come
from multiple-pair production.

Lighter ion runs at the same $\gamma_L$ and with the same lepton
pair acceptance could provide experimental verification of the predicted
Coulomb corrections, observable through deviations from the 
predicted $Z^4$ scaling for $A=B$, so far unobserved at RHIC or 
the SPS. Asymmetric collisions, with $Z_A \neq Z_B$, could
also help separate higher-order corrections from multi-photon
exchange with only one or with both ions.

\subsection{Physics potential of two-photon and electroweak processes}
\label{ggpotential}
{\it Contributed by: K. Hencken, S. R. Klein, M. Strikman and R. Vogt}

This section briefly describes some processes accessible through two-photon
interactions, including vector meson pair production and heavy flavor meson
spectroscopy.  We also briefly discuss tagging two-photon processes through
forward proton scattering as a way to enhance searches for electroweak final
states.  Finally, we mention the possibility of using $\gamma \gamma
\rightarrow e^+e^-$ as a luminosity monitor at colliders.

{\it Double vector meson production, $\gamma\gamma \rightarrow V V$}:
Double vector meson production in $pA$ and $AA$ collisions is hadronically
forbidden for kinematics where both the rapidity difference between one 
vector meson and the initial hadron and the rapidity difference between the two
vector mesons is large.  The negative $C$-parity of the vector mesons forbids
the process to proceed via vacuum exchange.  Accordingly, two-photon processes 
are the dominant contribution.  Studies of final states where
one of the vector mesons is heavy,  
such as $J/\psi  \rho^0$, can measure the
two-gluon form factor of the vector meson for the first time to determine the
transverse size of the gluon in the vector meson. It is expected that
the $t$-dependence of $\gamma \gamma \rightarrow J/\psi V$ is 
very broad with more than 30\% of the events at $p_T \ge 1$ GeV$/c$. 
In the case when both pairs are heavy, {\it e.g.}
$J/\psi J/\psi$, the BFKL regime can be probed.
On the other hand, if both mesons are light, {\it e.g.} $\rho^0 \rho^0$ or 
$\rho^0 \phi$, the Gribov-Pomeranchuk factorization theorem can be tested 
in a novel way.  For a more extensive discussion and rate estimates
in $AA$ collisions, similar to those in $pA$, see Ref.~\cite{FELIX}. 

{\it Heavy flavor meson spectroscopy}:
While single vector meson production is forbidden in two-photon
processes, it is possible to study heavy $Q \overline Q$ pair production. 
The $\gamma\gamma \rightarrow Q \overline Q$ production rate is 
directly proportional to the two-photon decay width,
$\Gamma_{\gamma\gamma}$. The two-photon luminosity is about three orders 
of magnitude larger than that at LEP. Thus it may be possible to
distinguish between quark and gluon-dominated resonances, ``glueballs''.
For the production rates in $AA$ collisions, see 
Refs.~\cite{Baur:2001jj,FELIX,TPHIC,Nystrand:1998hw,vidovic2}. 

One important background to meson production in two-photon processes is 
vector meson photoproduction followed by radiative decay.  For 
example, in ultraperipheral Pb+Pb collisions at the LHC, the $J/\psi$ 
photoproduction rate followed by the decay
$J/\psi\rightarrow\gamma\eta_c$ is about 2.5 per minute, much higher 
than the $\gamma\gamma\rightarrow\eta_c$ rate \cite{BKN}. The two channels
have similar kinematics, complicating any measurement of the two-photon 
coupling.

{\it Two-photon tagging and electroweak processes}:
Tagging two-photon interactions would enhance the detection capability of
electroweak processes such as $W^+ W^-$ pairs, $H^0$
and $t \overline t$ final states.
Detection of far-forward scattered protons has been routinely used
at $pp$ and $ep$ colliders to select 
diffractive events. It not only suppresses backgrounds, allowing
more efficient event selection, but also
improves event reconstruction by employing the measured proton
momentum. 

At the LHC, forward proton
detectors can also be used to measure photoproduction 
\cite{piotrzk,CMS-TOTEM}. The
acceptance of the detectors recently proposed by TOTEM and ATLAS at about 
$\pm 220$ m from the interaction point is determined primarily by the strong
dipole fields of the LHC beam line. Protons can be then detected if their
fractional energy loss, $\xi$, as a result of photon or Pomeron exchange, is
significant. For high luminosity running, the $\xi$ acceptance at $\pm 220$ m
is $0.01 < \xi < 0.1$. Unfortunately,
this acceptance does not match the typical fractional energy loss of ions in
UPCs. However, newly proposed detectors at $\approx \pm 420$ m from the 
interaction point will extend the acceptance down to $\xi \sim 2\times
10^{-3}$ \cite{loi}. With such detectors, a fraction of two-photon 
interactions in $pA$ and $AA$ collisions
employing light ions such as Ar or Ca can be double tagged: 
both the forward scattered proton and ion (or both ions)
are detected. Of course, detectors at both $\pm 220$ m and $\pm 420$ m 
can be used to tag diffractive scattering since larger values 
of $\xi$ are usually involved. 

Thus forward proton detectors provide unique and powerful
capabilities for tagging UPCs at the LHC.  This tagging would
allow selection and measurement of electroweak processes with
small cross sections.  For example, in a one month $p$Pb run,
10 $\gamma \gamma \rightarrow W^+ W^-$ events are expected. These
$W^+ W^-$ pairs are sensitive to the quartic gauge couplings and
would be characterized by a small pair $p_T$.  Single $W$ bosons
will be produced at high $p_T$ in $\gamma A$ and $\gamma p$
interactions with much higher statistics, similar to previous
studies at HERA \cite{Baur,Diener}.  In addition, photoproduction
of $t \overline t$ pairs could provide a measure of the top quark
charge \cite{Klein:2000dk}.  In all these examples, measuring the forward proton
improves background suppression as well as reconstruction of the
event kinematics.  Forward proton measurements can also be used to extend and
cross-check other techniques, such as large rapidity gap signatures, which
will be exclusively used in heavy-ion collisions.

Coherent $W^+$ photoproduction, $\gamma p \rightarrow W^+ n$,
is a way to measure the $W^+$ electromagnetic coupling \cite{Baltz:2007hw}. 
Detection of a single neutron in the ZDCs without additional hadrons
could serve as a trigger. Coherent photoproduction was studied in both $pp$ 
and $pA$ collisions, along with incoherent production in $pp$ collisions,
accompanied by proton breakup.  The coherent and incoherent $pp$ production 
rates were found to be comparable with a few $W^+$ produced per year.


\subsection{Photon-photon processes with ALICE}
{\it Contributed by: Yu. Kharlov and S. Sadovsky}
\label{alice-gammagamma}

\subsubsection{$e^+ e^-$ pairs in the ALICE forward detectors}

Multiple $e^+e^-$ pair production in ultraperipheral
heavy-ion collisions mainly affects the inner and forward ALICE detectors,
located a short radial distance away from the beam axis. These detectors are
the Inner Tracking System (ITS), the T0 and V0 detectors and the
Forward Multiplicity Detector (FMD), see Section~\ref{section-alice} 
for details. The
load of the T0 and V0 detectors is of particular importance because
these detectors provide the Level-0 ALICE trigger signals.

\begin{figure}[htbp]
  \begin{center}
    \includegraphics[width=14cm]{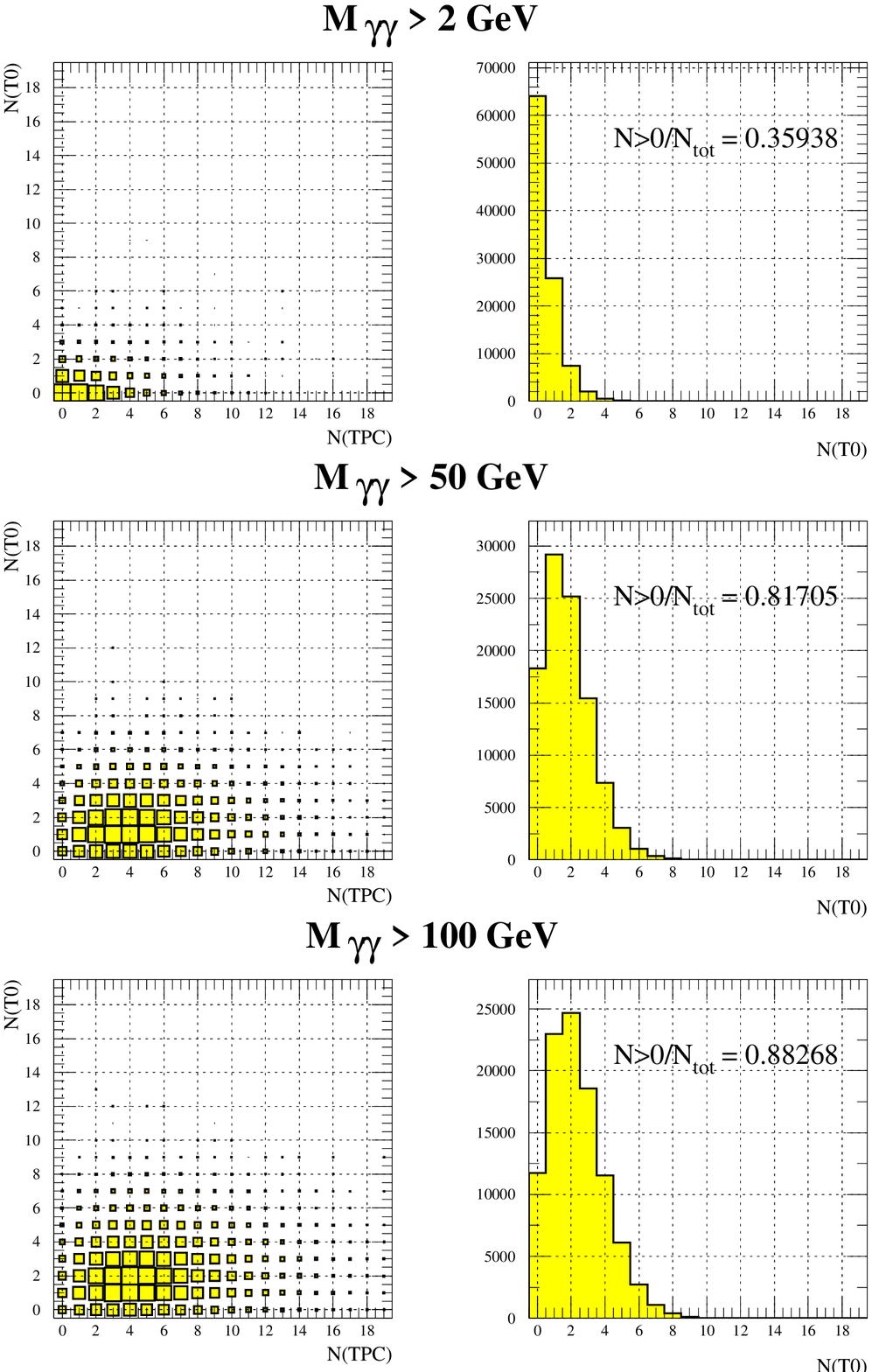}
  \end{center}
  \caption[]{Charged track multiplicities in the TPC and T0 detectors in
minimum bias events.}
  \label{fig:TPC-T0}
\end{figure}

A full simulation of an electron or position track from $e^+e^-$ pair 
production through the T0, V0 and FMD detectors in \PbPb\ collisions 
was performed.  The software package for
ALICE simulation and reconstruction, {\sc aliroot} \cite{ALIROOT}, was
used. An event generator for $e^+e^-$ pair production, 
{\sc epemgen}~\cite{ALICE-INT-2002-27}, was incorporated into {\sc aliroot}. 
This generator simulates the lepton $p_T$ and $y$ distributions according to 
the five-dimensional differential cross section
$d^5\sigma/dp_{T+}dy_+dp_{T-}dy_-d\phi_{+-}$ calculated in
Ref.~\cite{Alscher:1996gn}.

Only the ITS, T0, V0, and FMD detectors and the beam pipe were taken into
account. Three magnetic fields, $B = 0.2$, 0.4 and 0.5~T, were simulated.  
The cross sections and detection rates for at least one $e^\pm$ in
the detectors in \PbPb\ collisions at $L =
10^{6}~\mbox{kb}^{-1}\mbox{s}^{-1}$ are shown in Table~\ref{tab:xsec-aliroot}. 

The single electron cross sections rapidity distribution in 
multiple $e^+ e^-$ pair production
is very flat over a wide rapidity range, giving relatively large cross
sections even at forward rapidity.  However, the rapidity acceptance
is not the only factor determining the cross sections in
Table~\ref{tab:xsec-aliroot}.  They also strongly depend on the inner radii of
the detectors, representing an effective low $p_T$ cut.  The left and right
rapidity coverage of the T0 detectors are very similar and the inner radii are
the same, resulting in nearly the same rates on the left and right-hand sides.
On the other hand, while the right V0 detector has larger $\eta$ coverage,
its larger inner radius reduces the cross section so that the left V0 detector
has a higher cross section.  The right-hand FMD covers twice the $\eta$ range
of the left-hand detector.  Since the two detectors have the same inner radii,
the right-hand cross section is twice as large.
\begin{table}[htbp]
\begin{center}
  \caption[]{The electron cross sections and detection rates in the T0, V0 and 
FMD detectors in {\sc aliroot} for $B=0.2$, 0.4 and 0.5~T.}
\vspace{0.4cm}
  \begin{tabular}{|c|c|c|c|c|c|}
    \hline
Detector & $B$ (T) & \multicolumn{2}{c|}{Right} &
                      \multicolumn{2}{c|}{Left} \\ \cline{3-6}
         &          & $\sigma$ (kb) & Rate (MHz) &
                      $\sigma$ (kb) & Rate (MHz) \\
         \hline
T0                     & 0.2 & $1.7$ & $1.7$ & $1.9$ & $1.9$ \\ 
Right: $-5<\eta <-4.5$ & 0.4 & $0.7$ & $0.7$ & $0.7$ & $0.7$ \\ 
Left: $2.9 < \eta<3.3$ & 0.5 & $0.4$ & $0.4$ & $0.5$ & $0.5$ \\ \hline
V0                      & 0.2 & $1.7$ & $1.7$ & $3.8$ & $3.8$ \\ 
Right: $-5.1<\eta<-2.5$ & 0.4 & $0.6$ & $0.6$ & $2.0$ & $2.0$ \\ 
Left: $1.7 <\eta<3.8$   & 0.5 & $0.4$ & $0.4$ & $1.2$ & $1.2$ \\ \hline
FMD                     & 0.2 & $7.9$ & $7.9$ & $3.8$ & $3.8$ \\ 
Right: $-5.1<\eta<-1.7$ & 0.4 & $3.1$ & $3.1$ & $1.8$ & $1.8$ \\ 
Left: $1.7<\eta< 3.4$   & 0.5 & $2.2$ & $2.2$ & $1.1$ & $1.1$ \\ \hline
  \end{tabular}
  \label{tab:xsec-aliroot}
\end{center}
\end{table}
The forward detector load due to $e^+e^-$ pair production is 
sufficiently high to be an important background, especially 
for $B=0.2$~T. The load should be
compared to the maximum L0 trigger rate, $\sim 200$~kHz.


\subsubsection{Detection of $\GG$ processes in ALICE}

\subsection*{$\GG\to X$}

We now consider detection of $\GG\to X$ in the ALICE
central detectors. Charged particles with $p_T$ larger than 100 MeV
which pass the SPD2 or SSD2 trigger can be detected in the TPC with full
azimuthal coverage and $|\eta| < 0.9$ \cite{TPC-TDR}. Photons and
electrons with energies greater than 100~MeV in $|\eta| < 0.12$ and 
$\Delta\phi = 100^\circ$ can be detected by PHOS \cite{PHOS-TDR}.

To detect two-photon minimum bias events in ALICE, it is
important to have hits in the T0 detector since T0 defines the event
timing and starts a pre-trigger \cite{ALICE-INT-2001-38}.   
Figure~\ref{fig:TPC-T0} shows
the correlation between charged track multiplicities in T0 and the TPC
in $\gamma\gamma \to X$ events for three $\gamma\gamma$ invariant
mass ranges: $M_{\gamma\gamma} > 2$, 50 and 100 GeV. These
correlations demonstrate that the detection efficiency for low invariant
mass $\gamma\gamma$ pairs in minimum bias events is small but the cross
section is high.  On the other hand, the small cross section at higher
$\gamma\gamma$ invariant mass is compensated by higher detection
efficiency.

\begin{figure}[htb]
  \begin{center}
    \includegraphics[width=8cm]{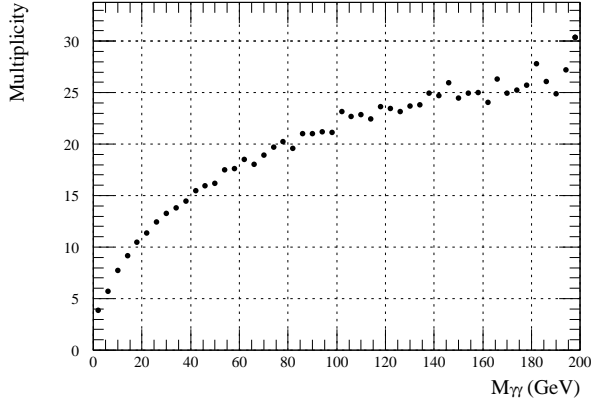}
  \end{center}
  \caption[]{The average charged particle multiplicity in
    $\GG$ interactions as a function of the $\GG$ invariant mass predicted by
    {\sc pythia} \protect{\cite{Pythia}.}  Reprinted from 
    Ref.~\protect\cite{Baur:2001jj} with permission from Elsevier.}
  \label{fig:gg-mult}
\end{figure}

The charged track multiplicity in $\GG$ collisions is similar to that
in hadronic collisions at the same center-of-mass energy because the
main contribution to $\GG$ interactions comes from strong
interactions of vector mesons \cite{SchulerS97}.  At high
multiplicities, the $\GG$ events cannot be exclusively detected in a 
restricted acceptance like that of the TPC.  The charged particle multiplicity 
in the TPC pseudorapidity range predicted by {\sc pythia} \cite{Pythia} 
is shown in Fig.~\ref{fig:gg-mult} as a function of
$M_{\gamma \gamma}$.

\begin{figure}[htb]
  \begin{center}
    \includegraphics[width=8cm]{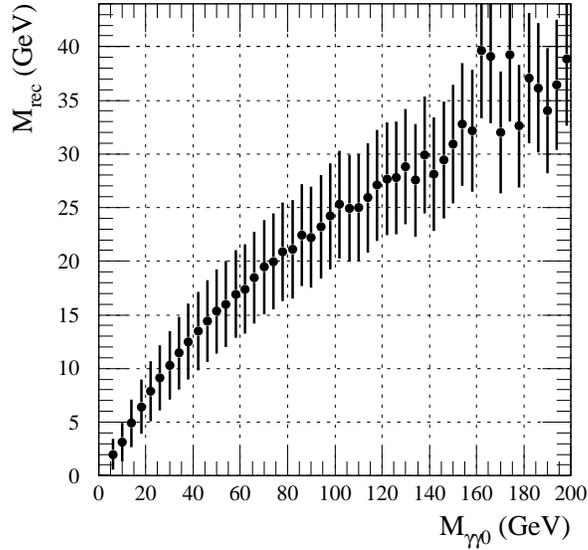}
  \end{center}
  \caption[]{The correlation between the reconstructed invariant mass and the 
 true $\gamma \gamma$ invariant mass.}
  \label{fig:alice-rec_vs_gen}
\end{figure}

Because particles escape in the forward region, the detected invariant
mass is less than the true $M_{\gamma \gamma}$. In
Fig.~\ref{fig:alice-rec_vs_gen} the correlation between the invariant
mass of the reconstructed event and the true invariant mass predicted by the 
event generator {\sc tphic} \cite{TPHIC} 
are shown. The vertical error bars show the width of the
measured mass distribution.
Up to 80\% of the total $\gamma \gamma$ mass can be lost. 
A mass-unfolding procedure to 
reconstruct the true collision energy, similar to that used 
by the L3 collaboration to measure the total $\GG$ interaction cross section  
\cite{L3:01}, can be applied. 

\begin{figure}[htb]
  \begin{center}
    \includegraphics[width=14cm]{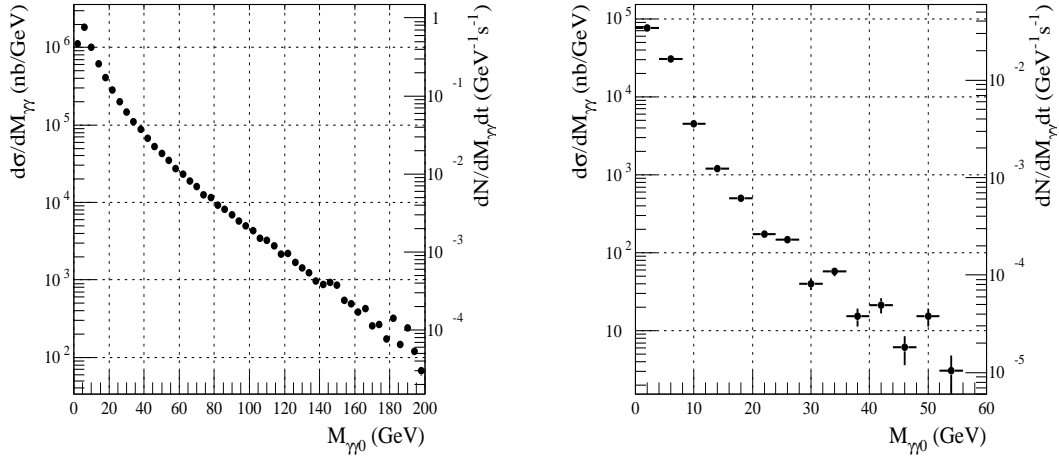}
  \end{center}
  \caption[]{The differential cross section (left axis) and event rate (right 
axis) for reconstructed $\gamma \gamma$ invariant mass from $\gamma\gamma
\rightarrow X$ (left-hand side) and $\gamma \gamma \rightarrow \mu^+ \mu^-$
(right-hand side) in Pb+Pb collisions.}
  \label{fig:xs-gg->X}
\end{figure}

On the left-hand side of
Fig.~\ref{fig:xs-gg->X}, the detected $\GG\to X$ cross sections (left axis)
and event rates (right axis) in \PbPb\ collisions are shown. The event rate
is calculated for an average luminosity of $0.42~\mbox{mb}^{-1}\mbox{s}^{-1}$ 
\cite{Morsch2001}. An event is
assumed to be detected if it is selected by the SPD2 multiplicity trigger
and if secondary particles come into the acceptance of
the TPC and PHOS.
A total $\GG \rightarrow X$ cross section of 52 mb for at $M_{\GG} >
2.3$~GeV in \PbPb\ collisions at $\sqrt{s_{_{NN}}} = 5.5$~TeV is used. The
reconstructed $\gamma \gamma$ cross section is 25~mb with an integrated
rate of 10 Hz.

\subsection*{$\GG\to \mu^+\mu^-$}
\label{tonyref}

Muon pair production must be measured exclusively
since both muons have to be detected. Therefore the SPD2 trigger cannot 
select these events since at least three charged particles are required
in SPD2. An SSD2 trigger can select $\GG\to \mu^+\mu^-$. The right-hand side
of 
Fig.~\ref{fig:xs-gg->X} shows the cross section for events selected by
an SSD2 multiplicity
trigger. The Pb+Pb event
rate for $L = 0.42~\mbox{mb}^{-1} \mbox{s}^{-1}$ is also shown. The
geometric efficiency is
about $4\%$. The low efficiency is due to the dependence of the lepton pair
production cross section on scattering angle, $\theta$, 
in the $\GG$ center-of-mass frame \cite{Barklow90},
$$
  \frac{d\sigma_{\gamma\gamma \rightarrow l^+ l^-}}{d\cos\theta} \propto
  \frac{1+\cos^2 \theta}{1 - \beta^2 \cos^2 \theta} \, \, .
$$
These lepton pairs can be used as a $\gamma \gamma$ 
luminosity monitor since they are easy to detect and simple to
calculate.  Since they are detected exclusively, the background is very small
\cite{ShamovT98Alice,BudnevGMS72}.

\subsection*{Quarkonium production}

Two-photon collisions can be used to study $C$-even charmonium and
bottomonium states ($\eta_c$, $\chi_{c0}$, $\chi_{c2}$, $\eta_b$,
$\chi_{b0}$, $\chi_{b2}$). 
The two-photon widths of $C$-even charmonia were determined from $\gamma
\gamma$ processes in $e^+ e^-$ collisions at LEP \cite{Abdallah:2003xk},,
BELLE \cite{Nakazawa:2004gu} and CLEO \cite{Dobbs:2005yk}.  
The corresponding properties
of the bottomonium states still remain unknown.  Predictions for the unknown
quarkonia two-photon widths are given in Ref.~\cite{Baur:2001jj}, following
Refs.~\cite{Kwong88,Bodwin94}.
The production cross sections are high enough for millions of $\eta_c$ and
$\chi_c$ states to be produced in a $10^6$ s \CaCa\ run while $\sim 1000$  
bottomonium states can be produced. The production cross sections and
rates are shown in
Table~\ref{tab:xs-charmonia}.
\begin{table}[htbp]
\begin{center}
\caption[]{Cross sections and quarkonia production rates in 
$10^6$~s \PbPb\ and \CaCa\ LHC runs.}
\vspace{0.4cm}
\begin{tabular}{|c|c|c|c|c|c|c|} \hline
 &  &  &
     \multicolumn{2}{c|}{$\sigma(AA \to AA R)$ ($\mu$b)} &
     \multicolumn{2}{|c|}{Rate (per $10^6$ s)} \\ \cline{4-7}
State ($R$) & $M$ (GeV) & $\Gamma_{\gamma\gamma}$ (keV) &
Pb+Pb & Ca+Ca & Pb+Pb & Ca+Ca   \\ \hline
$\eta_c$    & 2.979 & $7.4$ & 540 & 3.7 & $5.4 \times 10^4$ 
& $1.6 \times 10^7$ \\
$\chi_{c0}$ & 3.415 & $4.0$ & 170 & 1.2 & $1.7 \times 10^4$ 
& $4.8 \times 10^6$ \\
$\chi_{c2}$ & 3.556 & $0.46$ & 85 & 0.59 & $8.5 \times 10^4$ 
& $2.4 \times 10^6$ \\
$\eta_b$    & 9.366 & $0.43$ & 0.32 & 0.0028 & 32 & $1.1 \times 10^3$ \\
$\chi_{b0}$ & 9.860 & $2.5 \times 10^{-2}$ & 0.015 & $1.5 \times 10^{-4}$ 
& 1.5 & 600 \\
$\chi_{b2}$ & 9.913 & $6.7 \times 10^{-3}$ & 0.020 & $1.8 \times 10^{-4}$ 
& 2.0 & 720 \\ \hline
\end{tabular}
\label{tab:xs-charmonia}
\end{center}
\end{table}

The quarkonium states need to be detected exclusively. As an example, we 
discuss charmonium measurements in Ca+Ca interactions.  We restrict the event
to 2, 4 or 6 charged tracks with the sum of the charges in
the TPC equal to zero 
and not more than 2 photons in PHOS.  Since no particle
identification is assumed for the charged tracks, all charged
particles are assigned to be pions. The rates in Table~\ref{tab:xs-charmonia}
were simulated the {\sc tphic} generator \cite{TPHIC}. Further selection
criteria were applied to restrict the sum of the transverse momenta of the
detected particles to $\sum p_T < 50$~MeV/$c$. The main source of simulated
background, $\GG\to X$, had to pass the
same selection criteria as the charmonium signal. The $\GG \to X$ 
cross section in \CaCa\ collisions with $W_{\GG}
> 2.3$~GeV is $0.38$~mb, corresponding to $1.5\times 10^9$ events
in a $10^6$~s run. 

\begin{figure}[htb]
\parbox[t]{0.95\hsize}{
  \centerline{\resizebox{\hsize}{!}{\includegraphics{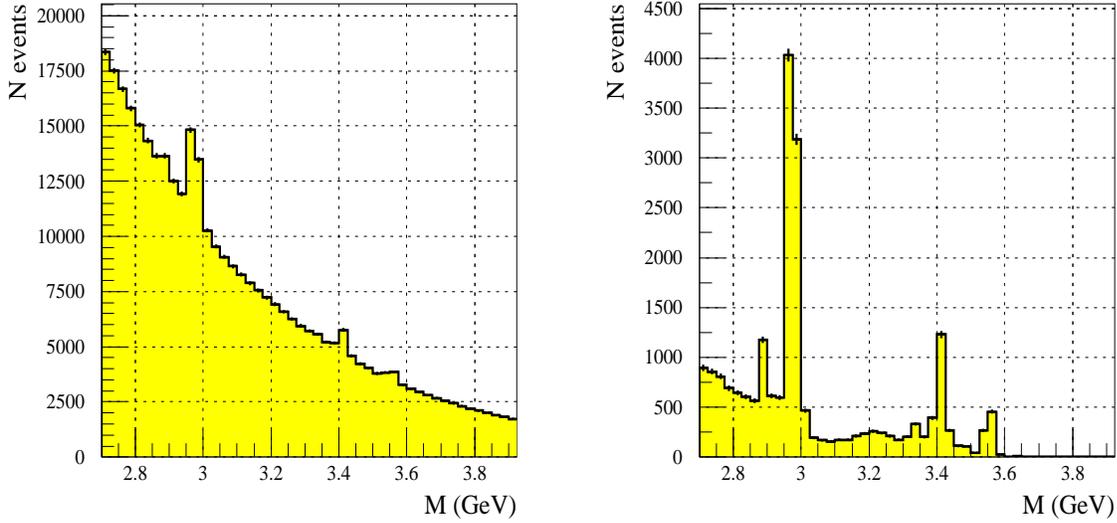}}}
  \caption[]{The $\GG$ invariant mass distribution with $\eta_c$, $\chi_{c0}$ 
and $\chi_{c2}$ peaks in a $10^6$ s Ca+Ca run before (left) and after (right)
background subtraction.  Reprinted from Ref.~\protect\cite{ALICE-PPRvolII}
with permission from Institute of Physics.}
  \label{fig:charmonium+minb}
}
\end{figure}

Figure~\ref{fig:charmonium+minb} shows the number of
events as a function of invariant mass in the central ALICE
detectors with (left-hand side) and without (right-hand side) the background.
The peaks at the $\eta_c$, $\chi_{c0}$ and $\chi_{c2}$ masses are
visible. The mass spectrum after background subtraction, fitted
by an exponential, is shown on the right-hand side of
Fig.~\ref{fig:charmonium+minb}. During 
one run, $\sim 7000$ $\eta_c$, 1200 $\chi_{c0}$, and 700
$\chi_{c2}$ can be detected. The non-resonant
background, as well as additional peaks to the left of the charmonium
states are explained by misidentification of charged tracks which
spreads the invariant mass of the detected system and shifts it to lower 
masses.  Note that the background from $\gamma \Pomeron \rightarrow J/\psi
\rightarrow \gamma \eta_c$, larger than the $\gamma \gamma \rightarrow 
\eta_c$ rate, has not been included.  These events may swamp the signal if the 
soft photon is not identified.

The bottomonium states are much harder to detect. The main decay channel
for $C$-even bottomonium is to two gluons. Due to the high mass, 
the number of hadrons produced by gluon fragmentation
is rather large. The average multiplicity of the $\eta_b$ decay
products is predicted to be 18. There are many $\pi^0$s and strange mesons
among the final-state particles. Due to the restricted ALICE acceptance,
especially the small aperture of PHOS, the probability of detecting all
the bottomonium decay products is very low. We simulated $10^5$ $\eta_b$
events,  $\sim 100$ times higher than the production rate, and reconstructed
none of them. Therefore $\eta_b$ and $\chi_b$ detection
in ALICE remains an open question.

\subsection*{Expected rates in the central barrel}

The expected lepton pair yields in two-photon interactions 
were estimated from the geometrical acceptance of the ALICE central barrel 
and muon arm. Events were generated based on
Refs.~\cite{Klein:1999qj,starlight,Baltz:2002pp,Nystrand:1998hw}. 
The rates were calculated for a Pb+Pb luminosity  
of $5 \times 10^{26}$ cm$^{-2}$s$^{-1}$.

The geometrical acceptance of the ALICE central barrel is defined as 
$|\eta| < 0.9$ and $p_T > 0.15$ GeV/$c$ while, for the muon arm, 
$2.5 \leq \eta \leq 4.0$ and $p_T > 1$ GeV/$c$ is used. Both track are
required to be within the acceptance cuts for the event to be reconstructed. 
In the TRD, a trigger cut of $p_T > 3.0$ GeV/$c$ will be necessary in 
central collisions.  It is not clear if this is also necessary for 
ultraperipheral events. The rates for $e^+e^-$ pairs are calculated for both 
$p_T > 0.15$ GeV/$c$ and $p_T > 3$ GeV/$c$. 

\begin{figure}[htb]
\centering\mbox{\epsfig{file=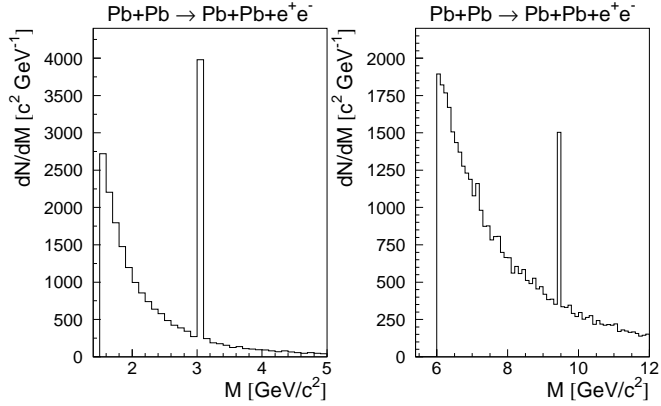,width=.6\textwidth}}
\caption[]{Invariant mass distributions for $\gamma \gamma \to e^+e^-$.
Both leptons are within the geometrical acceptance of the central barrel. The 
left-hand side shows the expected yield for $M>1.5$ GeV/$c^2$ in 
$2 \times 10^4$ s at design luminosity (an integrated Pb+Pb luminosity of 
10~$\mu$b$^{-1}$). The right-hand side shows the yield for $M>6$ GeV/$c^2$ in
$2 \times 10^6$ s (1~nb$^{-1}$). Only the natural
widths of the $1 S$ vector mesons have been included.} 
\label{UPC:minv}
\end{figure}

The lepton pair rates for pairs with $M>1.5$ GeV/$c^2$ are given in 
Table~\ref{UPC:llrates}.
The expected $e^+e^-$ yields in the central barrel are shown in 
Fig.~\ref{UPC:minv} for $M > 1.5$ GeV/$c^2$ in a $2 \times 10^4$~s run 
(left) and for $M > 6.0$ GeV/$c^2$ in a $2 \times 10^6$~s run (right). 
The approximate quarkonium $1S$ rates are also shown.  Higher-lying $S$ states 
are not included

\begin{table}[htbp] 
\begin{center} 
\caption[]{Expected lepton pair yields for $M>1.5$ GeV within the ALICE 
geometrical acceptance.}
\vspace{0.4cm}
\label{UPC:llrates}
\begin{tabular} { | c | c | c | } \hline
Selection  &  Geometrical Acceptance   & Rate (per $10^6$~s) \\ \hline 
\multicolumn{3}{|c|}{$e^+e^-$} \\ \hline
All         &  100\%                    & 7 $\times 10^7$    \\
$|\eta|<0.9$, $p_T > 0.15$ GeV$/c$   &  1.0\%        & 7 $\times 10^5$    \\
$|\eta|<0.9$, $p_T >3$ GeV$/c$ &  0.02\% & $1.4\times 10^4$  \\ \hline 
\multicolumn{3}{|c|}{$\mu^+\mu^-$} \\ \hline
All         &  100\%                    & 2.2 $\times 10^7$    \\
$2.2\leq\eta\leq 4.0$, $p_T >1.0$ GeV$/c$ & 0.26\%  & $6\times 10^4$ \\ \hline 
\end{tabular}
\end{center} 
\end{table}

\section{Measuring beam luminosity with UPCs}
\label{beam-luminosity}

{\it Contributed by: A.~J.~Baltz, S.~N.~White and I.A.~Pshenichnov}

\subsection{Introduction}
\label{beam-lum-intro}

The determination of the absolute luminosity at the LHC is the responsibility
of the individual experiments. The usual procedure is to select a physical 
process, a luminosity monitor, that is very stable with respect to luminosity. 
The yield of the luminosity monitor is taken with the rest of the data 
and, at some point, the cross section of the monitor is calibrated and 
used for absolute normalization of the data.

For the purposes of accelerator operation, it is sufficient to have a 
stable luminosity monitor which can be used for commissioning and 
optimizing the setup of the machine. The monitor can typically be 
calibrated to an accuracy of 10\% based on
accelerator instrumentation which determines the intensity and 
distribution of the stored beams~\cite{Adler:2003pb}.

In the same way, luminosity monitors used by experiments can be 
calibrated to this $\sim 10$\%  accuracy by comparing counting rates to
delivered luminosity. Since the desired luminosity uncertainty 
is typically of order 2\%, the accelerator-based 
calibration alone is insufficient. Achieving the higher 
necessary precision requires accurate knowledge of the monitor cross section, 
if it is calculable, or direct comparison with another cross section.

There are 
electromagnetic processes which can be calculated to the required accuracy
both for heavy-ion and proton beams at the LHC. 
Since lepton pair production depends primarily on the ion charge 
and only weakly on its internal structure, it may be a good ion monitor
process.
The new ion luminosity monitoring technique,  
described in the following sections, is usable during normal beam 
conditions and is a by-product of heavy-ion data taking.  

Luminosity measurements in $pp$ collisions are more problematic than
in the Pb+Pb scenario described below since there are no large cross
sections which are both calculable and free of detector modeling.
Instead both ATLAS \cite{DetectorTDRs} and CMS/TOTEM \cite{CMS-TOTEM}
plan to measure small angle $pp$
elastic scattering during runs with special optics at relatively
low luminosity.  Elastic scattering data can yield an absolute
luminosity measurement if it can be extended into the calculable
pure Coulomb regime, as proposed by ATLAS.  Alternatively, TOTEM
has proposed using a luminosity-independent method for deriving
the total cross section to normalize the elastic scattering data.
In both cases, the expected uncertainty in the luminosity determination
is roughly the desired 2\%.

In order to make effective use of the precision luminosity measurements
in the elastic scattering runs, a stable monitor of the relative luminosity
which can be employed both during the special runs and high luminosity
runs is needed.  It is critical that this monitor be relatively
insensitive to machine-related background processes since the machine
luminosities differ by a large factor.  Although ATLAS has a system of counters
designed for this purpose (LUCID), both ATLAS and CMS can use the ZDCs
developed for heavy-ion runs in $pp$ monitoring.  The ZDCs are stable
monitors but need to be calibrated.  The ZDC cross section 
is predicted to be $\sim 9$\% of the inelastic cross section.
Thus $pp$ elastic scattering can then be used to calibrate the ZDC so that
the ZDCs can be used to calibrate the accelerator-based measurements
and calculate the luminosity in $pp$ and heavy-ion runs.

\subsection{Luminosity monitoring at RHIC and LHC}
\label{beam-lum-rhic-lhc}

In spite of the significant differences between the four RHIC experiments, 
all experiments incorporated an identical minimum bias interaction 
trigger which served as a standard luminosity monitor.
The Zero Degree Calorimeters, ZDCs, trigger 
events in which at least one neutron is emitted in each 
beam direction. The calorimeters planned for LHC, like those at 
RHIC, will be sensitive to beam neutrons with transverse momentum 
$p_T\leq 200$ MeV/$c$. 
Measurements at SPS and RHIC confirmed that, over the full range 
of centralities, hadronic interactions of Pb or Au ions always result 
in neutron emission within the ZDC acceptance.

In addition to these collisions, the ZDC trigger is sensitive to 
ultraperipheral interactions resulting in 
mutual electromagnetic dissociation (MED). The MED calculation, used for 
absolute luminosity determination, is 
discussed below.  At RHIC, data taken with the ZDC trigger were analyzed 
to determine the fraction of electromagnetic events based on event topology 
and particle multiplicity.  The total cross section, including both hadronic 
and electromagnetic contributions, was calculated to 5\% accuracy with
the ZDC trigger.
  
The RHIC ZDC cross section, $\sigma_{\rm tot}$, is 10.8 b for Au+Au
collisions at $\sqrt{s_{_{NN}}} = 130$ GeV. A similar calculation predicts 
14.8 b for Pb+Pb collisions at $\sqrt{s_{_{NN}}} = 5.5$ TeV.
Further measurements at RHIC, which will improve the accuracy of the ZDC 
cross section, are expected to reduce the uncertainty in the 
LHC prediction to $\sim 2\%$.

\subsection{Mutual electromagnetic dissociation as a luminosity monitor}
\label{RHIC&Notations}

A method to measure and monitor beam luminosity in heavy-ion
colliders was proposed in Ref.~\cite{Baltz:1998ex}.
According to this method, the rate of mutual electromagnetic dissociation 
events, $R^{\rm MED}$, measured by the ZDCs
provides the luminosity,
\begin{equation}
L = \frac{R^{\rm MED}}{\sigma^{\rm MED}} \, \, ,
\end{equation}
if the mutual electromagnetic dissociation cross section, $\sigma^{\rm MED}$, 
is computed with sufficient accuracy. 
Simultaneous forward-backward single neutron emission from each beam is a
clear signature of mutual 
electromagnetic dissociation which 
proceeds by mutual virtual photon absorption.
The excitation and subsequent decay of the Giant Dipole 
Resonances (GDR) in both nuclei is 
responsible for the bulk of this process.
In heavy nuclei, such as gold or lead, single neutron emission, $1n$, 
is the main mechanism of GDR decay. 

Measurements of neutron emission in mutual dissociation of gold nuclei
recently performed at RHIC give some confidence in the ZDC
technique~\cite{Adler:fq}, including the theoretical interpretation 
of the data necessary for the luminosity measurements~\cite{Chiu:2001ij}.

Table~\ref{tab:ratio21}, from Ref.~\cite{Chiu:2001ij},
presents the measured ratios of the ZDC
hadronic cross section, $\sigma_{\rm geom}$, to the total ZDC cross section,
$\sigma_{\rm tot}$, including mutual electromagnetic dissociation.  This 
ratio agrees well with both Weizs\"{a}cker-Williams calculations 
employing measured
photodissociation cross sections as input~\cite{Baltz:1998ex} and
with {\sc reldis} calculations~\cite{Pshenichnov:2001qd}.

\begin{table}[htbp]
\begin{center}
\caption[]{Experimental and theoretical ratios of mutual dissociation 
cross sections \protect\cite{Chiu:2001ij}.  
See the text for an explanation of the notation.  Copyright 2002 by the
American Physical Society (http://link.aps.org/abstract/PRL/v89/e021302).}
\vspace{0.4cm}
\begin{tabular}{|c|c|c|c|c|c|}
\hline
& PHENIX & PHOBOS & BRAHMS & Ref.~\protect\cite{Baltz:1998ex} & 
Ref.~\protect\cite{Pshenichnov:2001qd} \\ 
\hline
$\sigma_{\rm tot}$ (b) & -- & -- & -- & 10.8$\pm$0.5 & 11.2  \\
 & & & & & \\ 
$\sigma_{\rm geom}$ (b)&  --  & -- & -- & 7.1  & 7.3 \\
 & & & & & \\ 
$ \frac{\sigma_{\rm geom}}{\sigma_{\rm tot}}$
 & 0.661$\pm$0.014 &  0.658$\pm$0.028 & 0.68$\pm$0.06 &  0.67 & 0.659 \\
 & & & & & \\ 
 $\frac{\sigma(1nX| {\cal D})}{ \sigma_{\rm tot}} $
 & 0.117$\pm$0.004 & 0.123$\pm$0.011 & 0.121$\pm$0.009 & 0.125 & 0.139  \\
 & & & & & \\ 
 $\frac{\sigma(1nX| 1nY)}{ \sigma(1nX| {\cal D})} $
 & 0.345 $\pm$ 0.012  & 0.341 $\pm$ 0.015 & 0.36 $\pm$0.02 & 0.329&  -- \\
 & & & & & \\ 
$\frac{\sigma(2nX| {\cal D})}{ \sigma(1nX| {\cal D})}$ & 0.345$\pm$0.014
 & 0.337 $\pm$ 0.015 & 0.35$\pm$0.03 & -- & 0.327 \\
 & & & & & \\ 
 $\frac{\sigma(1nX| 1nY)}{ \sigma_{\rm tot}} $
& 0.040$\pm$0.002 & 0.042$\pm$0.003 & 0.044$\pm$0.004 & 0.041$\pm$0.002 & -  \\
& & & & & \\ 
\hline
\end{tabular}
\label{tab:ratio21}
\end{center}
\end{table}

\begin{figure}[htb]
\begin{centering}
   {\includegraphics[width=8.5cm]{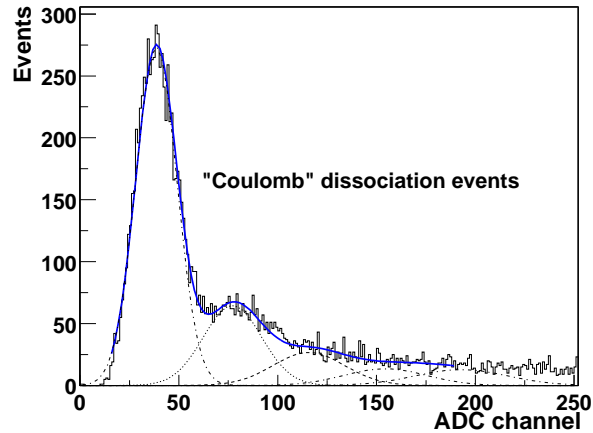}}
    \caption[]{
Energy spectrum for the left-hand ZDC with ``Coulomb'' selection
 for events with a single neutron in the right-hand ZDC 
\protect\cite{Chiu:2001ij}.  Copyright 2002 by the
American Physical Society (http://link.aps.org/abstract/PRL/v89/e012302).}
\label{fig:rawdist}
 \end{centering}
 \end{figure}

Figure~\ref{fig:rawdist} shows the energy spectrum obtained in one ZDC when
the opposite ZDC measures only one neutron.  
Requiring only one neutron in one of
the ZDCs provides ``Coulomb'' event selection.  The total number of events
in the spectrum of Fig.~\ref{fig:rawdist} after background subtraction
corresponds to the cross section $\sigma(1nX| {\cal D})$ for
the $(1nX|{\cal D})$ topology.  Here $1n$ signifies one neutron,
$X$ denotes the undetected particles emitted along with the single neutron 
and ${\cal D}$ denotes an arbitrary 
dissociation mode for the other nucleus.

The decay topology $(1nX| 1nY)$ corresponds to
exactly one neutron in each ZDC accompanied by undetected
particles $X$ and $Y$ respectively and gives rise to the highest peak in
the energy spectrum shown in Fig.~\ref{fig:rawdist}.  The topology trigger 
with a single neutron in each ZDC is about 35\% of the total $(1nX|{\cal 
D})$ topology, as shown in Table~\ref{tab:ratio21}.  The $(2nX|1nY)$ topology,
with two neutrons in the left-hand ZDC, gives rise to the second peak in
Fig.~\ref{fig:rawdist}.   Emission of a second neutron in the $(2nX|{\cal D})$
topology is about 35\% of the single neutron topology $(1nX|{\cal D})$,
see Table~\ref{tab:ratio21}.  The table also shows the ratios of 
$\sigma(1nX| 1nY)$ and $\sigma(1nX| {\cal D})$ to the total ZDC cross section, 
$\sigma_{\rm tot}$.

The $(1nX| 1nY)$ topology is useful for luminosity measurements because 
contamination from hadronic events is small and the dependence on the nuclear
radius is weak, as shown in 
Ref.~\cite{Baltz:1998ex}.  Table~\ref{tab:ratio21} shows
that the ZDC cross section ratios measured by three RHIC experiments (PHENIX,
PHOBOS and BRAHMS) 
agree well with each other and with the calculations.  The ZDC has also
successfully tagged UPC events with $\rho^0$ production 
by virtual photons~\cite{Adler:2002sc,Baltz:2002pp}.

In addition to employing the $(1NN| 1nY)$ topology, it is also possible
\cite{Pshenichnov:ALICE} to use the sum 
$\sigma(1nX| 1nY)+\sigma(1nX| 2nY)+\sigma(2nX| 1nY)+\sigma(2nX| 2nY)$
as a luminosity monitor, as explained below.

At the LHC, the advantage of the proposed 
methods~\cite{Baltz:1998ex,Pshenichnov:ALICE}
is the use of the ZDCs, intended for centrality measurements in 
hadronic heavy-ion collisions~\cite{ZDC-ALICE99,ZDC-ALICE02,ZDC-CMS}.
Therefore, no additional instrumentation is needed 
for luminosity measurements. However, a key ingredient is an accurate
calculation of the neutron emission cross sections in mutual electromagnetic 
dissociation $\sigma^{\rm MED}$, the subject of this chapter.

References~\cite{Korot1,Korot2} show
that, due to the excitation of discrete nuclear states, there will
be high energy photons in the forward direction which could be a 
signature of UPCs. It seems however, that there is presently
no practical experimental means to use these high energy photons in the LHC
experiments.
A related process has been discussed in cosmic ray physics \cite{Anchor}
where TeV gamma rays originate from the excitation and de-excitation 
of cosmic ray nuclei in the cosmic microwave background (CMB) radiation.
Thus ultrarelativistic nuclei may be viewed as
`relativistic mirrors' which boost low energy photons 
from the CMB and the equivalent photon spectrum, respectively, to very high 
energies.

\subsection{Unique characteristics of mutual electromagnetic 
dissociation of heavy ions}

Since the first pioneering studies of the electromagnetic 
dissociation~\cite{Heckman,Olson}, the process has commonly been
defined as disintegration of one of the nuclei involved in a UPC even though
their nuclear densities do not overlap. Recent
experiments~\cite{Scheidenberger:tr,Hill:ie} have measured  
projectile or target dissociation,
respectively.   

Both at RHIC and the LHC, single electromagnetic dissociation, when only one
of the nuclei is excited and dissociates, far exceeds the geometric cross 
section, $\sigma_{\rm geom}$, due to direct
nuclear overlap~\cite{Baur:2001jj,Baltz:as,Pshenichnov:2001qd,Krauss:1997vr}. 
As a result, electromagnetic dissociation and $e^+e^-$ pair production
(when followed by electron capture)  
reduce the beam lifetime in colliders~\cite{Baltz:as}.

Both nuclei may be disintegrated in one 
event by the corresponding Coulomb fields of their collision partners
\cite{Pshenichnov:2001qd}. Here we focus on the mutual dissociation
of lead ions at the LHC to monitor and measure luminosity by detection of
forward-backward neutrons.  Details can be found in 
Ref.~\cite{Pshenichnov:2001qd,Pshenichnov:ALICE}.

\subsection{Leading order mutual electromagnetic Pb+Pb dissociation}

The Weizs\"{a}cker-Williams 
method~\cite{Baur:2001jj,Krauss:1997vr} treats 
the impact of the Lorentz-boosted Coulomb field of nucleus $A$ 
as the absorption of equivalent photons by nucleus $B$.     

\begin{figure}[htb]
\unitlength=1.0cm 
\begin{center}
\begin{picture}(7.0,7.0)(3.,-1.)
\Text(2.5,3.8)[]{\Large {\bf LO}}  
\SetScale{0.8} 
\SetWidth{1.0}
\ArrowLine(0,100)(60,100)  
\Text(0,3.12)[]{$A$} 
\SetWidth{0.5}  
\GCirc(60,100){4}{1.} 
\SetWidth{1.0}
\ArrowLine(64,100)(120,100)  
\Text(2.48,3.12)[]{$A$}   
\SetWidth{0.5}
\Photon(60,97)(60,36){3}{4} 
\Text(1.36,1.76)[]{$E_1$} 
\SetWidth{1.0}
\ArrowLine(0,33)(60,33)  
\Text(0,0.64)[]{$B$} 
\SetWidth{0.5}  
\GCirc(60,33){4}{0} 
\SetWidth{2.0}
\ArrowLine(60,33)(120,33)  
\Text(2.48,0.64)[]{$B^\star$}
\SetWidth{0.5}  
\GCirc(120,100){4}{0}
\SetWidth{2.0} 
\ArrowLine(120,33)(180,33)  
\Text(4.72,0.64)[]{$B^\star$}
\SetWidth{0.5}  
\GCirc(120,33){4}{1.} 
\Photon(120,97)(120,36){3}{4} 
\Text(3.05,1.76)[]{$E_2$}
\SetWidth{2.0}
\ArrowLine(120,100)(180,100)  
\Text(4.72,3.12)[]{$A^\star$}
\SetOffset(6.8,2.2)
\Text(2.5,3.5)[]{\large {\bf NLO$_{12}$}}    
\SetWidth{1.0}
\ArrowLine(0,100)(60,100)  
\Text(0.,3.05)[]{$A$} 
\SetWidth{0.5}  
\GCirc(60,100){4}{1.} 
\SetWidth{1.0}
\ArrowLine(64,100)(120,100)  
\Text(2.50,3.05)[]{$A$}   
\SetWidth{0.5}
\Photon(60,97)(60,36){3}{4} 
\Text(1.36,1.76)[]{$E_1$} 
\SetWidth{1.0}
\ArrowLine(0,33)(60,33)  
\Text(0,0.64)[]{$B$} 
\SetWidth{0.5}  
\GCirc(60,33){4}{0} 
\SetWidth{2.0}
\ArrowLine(60,33)(120,33)  
\Text(2.48,0.64)[]{$B^\star$}
\SetWidth{0.5}  
\GCirc(120,100){4}{0}
\SetWidth{2.0} 
\ArrowLine(120,33)(220,33)  
\Text(5.52,0.64)[]{$B^\star$}
\SetWidth{0.5}  
\GCirc(120,33){4}{1.} 
\Photon(120,97)(120,36){3}{4} 
\Text(3.05,1.76)[]{$E_2$}
\GCirc(140,100){4}{0}
\Photon(140,97)(140,36){3}{4} 
\Text(4.24,1.76)[]{$E_3$}
\GCirc(140,33){4}{1.}
\SetWidth{2.0}
\ArrowLine(120,100)(220,100)  
\Text(5.52,3.05)[]{$A^\star$}
\SetOffset(6.8,-1.8)
\Text(2.5,3.5)[]{\large {\bf NLO$_{22}$}}    
\SetWidth{1.0}
\ArrowLine(-20,100)(60,100)  
\Text(0,3.05)[]{$A$} 
\SetWidth{0.5}  
\GCirc(60,100){4}{1.} 
\SetWidth{1.0}
\ArrowLine(64,100)(120,100)  
\Text(2.50,3.05)[]{$A$}   
\SetWidth{0.5}
\Photon(60,97)(60,36){3}{4}
\Text(0.88,1.76)[]{$E_1$} 
\Photon(40,97)(40,36){3}{4} 
\GCirc(40,100){4}{1.}
\GCirc(40,33){4}{0}
\Text(2.0,1.76)[]{$E_2$} 
\SetWidth{1.0}
\ArrowLine(-20,33)(60,33)  
\Text(0,0.64)[]{$B$} 
\SetWidth{0.5}  
\GCirc(60,33){4}{0} 
\SetWidth{2.0}
\ArrowLine(60,33)(120,33)
\Line(40,33)(60,33)  
\Text(2.48,0.64)[]{$B^\star$}
\SetWidth{0.5}  
\GCirc(120,100){4}{0}
\SetWidth{2.0} 
\ArrowLine(120,33)(220,33)  
\Text(5.52,0.64)[]{$B^\star$}
\SetWidth{0.5}  
\GCirc(120,33){4}{1.} 
\Photon(120,97)(120,36){3}{4} 
\Text(3.12,1.76)[]{$E_3$}
\GCirc(140,100){4}{0}
\Photon(140,97)(140,36){3}{4} 
\Text(4.24,1.76)[]{$E_4$}
\GCirc(140,33){4}{1.}
\SetWidth{2.0}
\ArrowLine(120,100)(220,100)  
\Text(5.52,3.05)[]{$A^\star$}   
\end{picture} 
\end{center} 
\caption[]{The electromagnetic excitation and mutual dissociation of 
relativistic nuclei. Open and closed circles denote elastic and 
inelastic vertices, respectively. The
LO contribution is shown on the left-hand side. The NLO contributions with 
single and double photon exchange, ${\rm NLO_{12}}$, and with two 
double-photon exchange, ${\rm NLO_{22}}$, are shown on the right-hand side. 
} 
\label{fig:6:EMsecond}  
\end{figure}
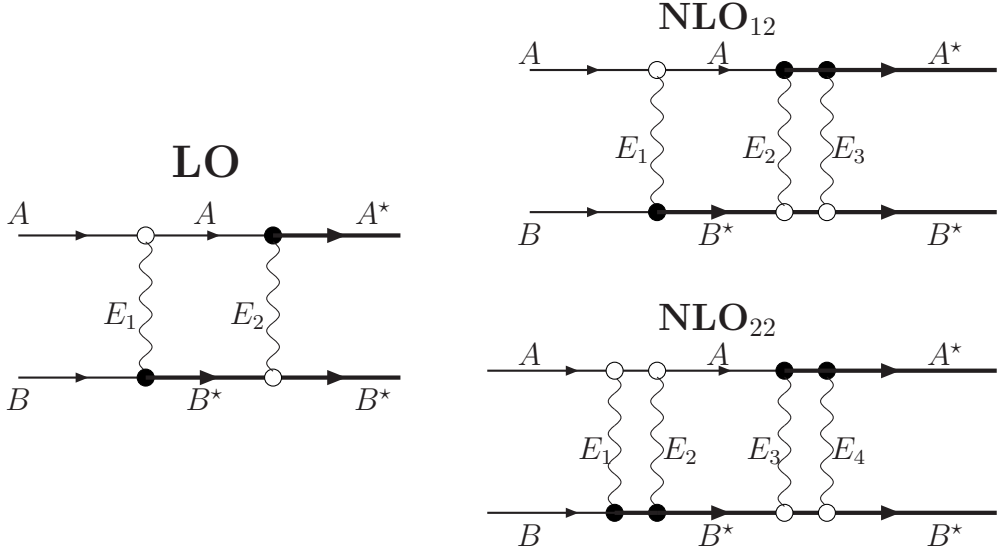

Figure~\ref{fig:6:EMsecond} shows the leading and next-to-leading order
processes contributing to mutual electromagnetic dissociation, with each
order treated independently (see Refs.~\cite{Baur:2003ar,Pshenichnov:2001qd}
for details).  The open and closed circles on the diagrams denote elastic
and inelastic vertices, respectively.  Thus, at LO, a photon with energy
$E_1$ is exchanged between $A$ and $B$, leaving $B$ in excited state
$B^*$ after absorption of the photon.  A photon with energy $E_2$ is 
exchanged between $B^*$ and $A$ and absorbed by $A$,
exciting it to $A^*$.  Both excited nuclei dissociate.  There is no
time ordering and, for calculational purposes, the photon emission spectrum
does not depend on whether the nuclei are excited or not.  The photon exchange 
between ground-state nuclei is the primary
exchange while the photon exchange between an excited nucleus and a 
ground-state nucleus is a secondary photon exchange.  There is a
complementary diagram to NLO$_{12}$, NLO$_{21}$, where nucleus $B$ is
excited by double photon absorption while $A$ is excited by single photon
absorption.  

The cross section
for dissociation of $A$ and/or $B$  
to final states $i$ 
and $j$ (single and mutual dissociation) respectively, is
\begin{equation}
\sigma^{\rm (S,M)ED}(i| j) = 2\pi\int\limits_{b_{c}}^{\infty}  db \, b \, 
P_{A}^i(b) \, P_{B}^j(b) \, \, 
\label{eq:6:SM}
\end{equation}
where SED stands for dissociation of only one of the nuclei 
(single electromagnetic dissociation) and MED is for 
mutual electromagnetic dissociation.  
When only one nucleus is dissociated, the cross section includes only one
probability factor, {\it i.e.} either $P_{A}^i(b)$ or $P_{B}^j(b)$ is
unity.
The lower limit of integration, $b_{c}$, is a sharp cutoff, approximately 
given by the sum of the nuclear radii, $b_{c}\approx R_{A}+ R_{B}$, 
to separate the nuclear and electromagnetic interaction domains. 
The choice of the lower limit is discussed further in 
Sec.~\ref{sec:6:HADRONIC}.
In Eq.~(\ref{eq:6:SM}), probability for dissociation of $B$ 
at impact parameter $b$ is defined as
\begin{equation}
P_{B}^j(b)=e^{-m_{B}(b)}\int\limits_{E_{\rm min}}^{E_{\rm max}} 
dE_1 \, \frac{d^3N_{\gamma A}}{dE_1 d^2b} \, 
\sigma_{B}(E_1)\, f_{B}(E_1,j)
\label{eq:6:P1}
\end{equation}
where $m_{B}(b)$ is the mean number of photons absorbed by nucleus 
$B$,
\begin{equation}
m_{B}(b)=
\int\limits_{E_{\rm min}}^{E_{\rm max}} dE_1 \,
\frac{dN^3_{\gamma A}}{dE_1 d^2b} \, 
\sigma_{B}(E_1) \, \, .
\label{mb}
\end{equation}
Here $dN_{\gamma A}/dE_1 d^2b$ 
is the virtual photon spectra from nucleus $A$ at $b$ from Eq.~(\ref{wwr}),
$\sigma_{B}(E_1)$ and $f_{B}(E_1,j)$ are the total photo-absorption cross 
section and the branching ratio for dissociation into final state $j$ due to
absorption of a photon with energy $E_1$ by $B$ \cite{Pshenichnov:2001qd}. 
The expression for $P_{A}^i(b)$ is obtained by exchanging
subscripts and taking $j \rightarrow i$. The neutron emission threshold is 
used for $E_{\rm min}$ while
$E_{\rm max}\approx\gamma_L/R_{A,B}$.  In the case of a collider, the Lorentz 
factor of the heavy-ion 
beam is boosted to the rest frame of the collision partner,
$\gamma_L^{\rm rest} = 2\gamma^2_L -1$.
At the LHC, the Coulomb fields of the ions are extremely Lorentz-contracted
with $\gamma_L \sim 1.7\times 10^7$. 
 
The total LO cross section for mutual electromagnetic dissociation by 
two-photon exchange, as shown in Fig.~\ref{fig:6:EMsecond}, is
\begin{eqnarray}
\sigma^{\rm MED}_{\rm LO} & = & 2\pi\int\limits_{b_{c}}^{\infty} db \, b \,
[m_{A}(b) e^{-m_{A}(b)}][m_{B}(b) e^{-m_{B}(b)}] \nonumber \\
& = & 2\pi 
\int\limits_{b_{c}}^{\infty} db \, b \, m_A^2(b) e^{-2m_A(b)} \, \, .
\label{eq:6:LO}
\end{eqnarray}
The last equality assumes $A=B$ and $Z_A=Z_B$. In the case of 
single dissociation, SED, only one factor of $m_A(b) \exp[-m_A(b)]$ is
included.
Note that we have taken $f_B(E,j) = f_A(E,i) \equiv
1$ in Eq.~(\ref{eq:6:P1}) since no final state is specified and the sum over 
branching ratios is by definition unity at each photon energy.

\subsection{Next-to-leading-order mutual electromagnetic dissociation}
\label{sec:6:NLO-SECT}

In addition to the LO mutual dissociation process,
a set of NLO processes with three or four photon exchanges 
can be considered.  
The total MED cross section for the three photon process, ${\rm NLO_{12}}$,
shown in Fig.~\ref{fig:6:EMsecond} is
\begin{eqnarray}  
\sigma^{\rm MED}_{\rm NLO_{12}} & = &
2\pi\int\limits_{b_{c}}^{\infty} db \, b \,
[m_{A}(b) \, e^{-m_{A}(b)}] \bigg[\frac{m_{B}^2(b)}{2} 
\, e^{-m_{B}(b)} \bigg] \nonumber \\ & = &
2\pi\int\limits_{b_{c}}^{\infty} db \, b\,
\frac{m_A^3(b)}{2} \, e^{-2m_A(b)} \, \, 
\label{eq:6:NLO12}
\end{eqnarray}
where again $A = B$ is assumed in the last equality.
The complementary process, ${\rm NLO_{21}}$, with the excitation 
of $B$ via double photon absorption is equally possible 
and has the same cross section.  Likewise the total SED cross section for
breakup of one nucleus by exchange of two photons is
\begin{equation}  
\sigma^{\rm SED}_{\rm NLO_2} = 2\pi\int\limits_{b_{c}}^{\infty} db \, b \,
\frac{m_A^2(b)}{2} \, e^{-m_A(b)} \, \, .
\label{eq:6:NLO2}
\end{equation}

The MED cross section for four photon exchange, denoted ${\rm NLO_{22}}$ in 
Fig.~\ref{fig:6:EMsecond}, is
\begin{eqnarray}  
\sigma^{\rm MED}_{\rm NLO_{22}} & = & 
2\pi\int\limits_{b_{c}}^{\infty} db \, b\,
\bigg[\frac{m_{A}^2(b)}{2} \, e^{-m_{A}(b)} \bigg]\bigg[
\frac{m_{B}^2(b)}{2} \, e^{-m_{B}(b)} \bigg] \nonumber \\ & = & 
2\pi\int\limits_{b_{c}}^{\infty} db \, b \,
\frac{m_A^4(b)}{4} \, e^{-2m_A(b)} \, \, 
\label{eq:6:NLO22}
\end{eqnarray}
when $A = B$ in the last equality.
The SED cross section for exchange of three or more photons is the sum over
the series $[m_A^n(b)/n!]$ for $n\geq 3$,
\begin{equation}  
\sigma^{\rm SED}_{\rm NLO_{3+}} = 2\pi\int\limits_{b_{c}}^{\infty} db \, b\,
e^{-m_A(b)} \sum_{n=3}^\infty
\frac{m_A^n(b)}{n!} \, \, .
\label{eq:6:NLO3p}
\end{equation}
In MED, the exchange of at least three photons on one side is referred to as
triple excitation or NLO$_{\rm TR}$.

One can calculate the sum of all contributions to single and mutual 
electromagnetic exchange using
the prescription of Ref.~\cite{Baltz:1998ex}:
\begin{equation}
\sigma^{\rm S(M)ED}_{\rm tot} = 2\pi\int\limits_{b_{c}}^{\infty} db \, b \,
[1-e^{-m_A(b)}]^E \, \, 
\label{SMTOT}
\end{equation}
where $E = 1$ for SED and 2 for MED.
Since the collision probability for each ion
without photon exchange is equal to $\exp[-m_A(b)]$,
Eq.~(\ref{SMTOT}) is evident.

In a more detailed treatment, the character of
the intermediate state would be considered since a given 
photon energy $E_1$ leads 
to a specific set of intermediate states.  The excitation cross sections of
the intermediate states differ somewhat relative to the unexcited nuclei 
introducing correlations between the photon energies which are not considered 
here.

\subsection{Hadronic nuclear dissociation in grazing collisions}
\label{sec:6:HADRONIC}

At grazing impact parameters, $b\sim R_{A} + R_{B}$, 
nuclei are partly transparent 
to each other. Hadronic nucleon-nucleon collisions may be absent in peripheral 
events with a weak overlap of diffuse nuclear surfaces
while electromagnetic interactions
may lead to electromagnetic 
dissociation. Both hadronic and electromagnetic interactions may occur in
the same event. For example, 
a neutron-neutron collision in the density overlap 
zone may lead to neutron ejection accompanied by photon emission
and absorption in electromagnetic interactions.  

Therefore, a smooth transition from purely 
nuclear collisions at $b< R_{A} + R_{B}$ to electromagnetic 
collisions at $b> R_{A} + R_{B}$ takes place. Such a transition region
was considered in the ``soft-sphere'' model of Ref.~\cite{Aumann:1994ev}. 
A similar approach was adopted in Ref.~\cite{Baltz:1998ex}, where the cross 
section for nuclear or electromagnetic dissociation alone or for both together
was written in an unexponentiated form as
\begin{equation}
\sigma = 2\pi \int^{\infty}_{0} db \, b \,
\biggl({\cal P}^{\rm nuc}(b)+{\cal P}^{\rm ED}(b)-{\cal P}^{\rm nuc}(b)
{\cal P}^{\rm ED}(b) \biggr) \, \, 
\end{equation}
where ${\cal P}^{\rm nuc}(b)$ and ${\cal P}^{\rm ED}(b)$ are
the probabilities 
of nuclear and electromagnetic dissociation at $b$.  Including the limits of
integration for each 
term separately, we have
\begin{eqnarray}
\sigma & = & 2\pi \int^{b^{\rm nuc}_{c}}_{0} db \, b \, {\cal P}^{\rm nuc}(b)
\, + \, 2\pi \int^{\infty}_{b^{\rm ED}_{c}} db \, b \, {\cal P}^{\rm ED}(b)
\nonumber \\ & & \mbox{} - \, 2\pi \int^{b^{\rm nuc}_{c}}_{b^{\rm ED}_c} db 
\, b \, {\cal P}^{\rm nuc}(b) {\cal P}^{\rm ED}(b) \, \, .
\label{FULLP}
\end{eqnarray}
Here individual impact parameter cutoff values, $b^{\rm nuc}_{c}$ and  
$b^{\rm ED}_{c}$, were used for the nuclear and electromagnetic interactions. 
However,  the simpler expression,
\begin{equation}
\sigma = \sigma^{\rm nuc}+\sigma^{\rm ED} =
2\pi\int^{b_{c}}_{0} db \, b \, {\cal P}^{\rm nuc}(b)+
2\pi\int^{\infty}_{b_{c}} db \, b \, {\cal P}^{\rm ED}(b),
\label{TRUNCATEDP}
\end{equation}
is widely used with a single cutoff, $b_c$, chosen so that 
$b^{\rm ED}_c< b_c < b^{\rm nuc}_c$.  Using a single cutoff allows the 
first and second terms of Eq.~(\ref{FULLP}) to be simplified while 
the third term is eliminated.  Numerical results based on Eqs.~(\ref{FULLP})
and (\ref{TRUNCATEDP}) are similar, as
shown for the ``sharp-cutoff'' and ``soft-sphere''
models of Ref.~\cite{Aumann:1994ev}.  In the case of heavy nuclei,
the difference between realistic values of
$b^{\rm ED}_c$, $b^{\rm nuc}_c$ and $b_c$ is less than 1 fm.
As a result, the third term in Eq.~(\ref{FULLP}) turns out to be small. 
Finally, the nuclear and electromagnetic contributions can be studied
separately using Eq.~(\ref{TRUNCATEDP}).  This separation is important
for understanding nuclear and electromagnetic
dissociation at ultrarelativistic  
colliders where the nuclear and 
electromagnetic interaction products populate different, 
non-overlapping rapidity regions. In the widely-used BCV 
parametrization \cite{BenCook}, $b_c$ 
is 
\begin{equation}
b_c= R_{\rm BCV}\bigl( A^{1/3}+B^{1/3}-X_{\rm BCV}(A^{-1/3}
+B^{-1/3})\bigr) \, \, .
\label{BCV}
\end{equation}
The parameters $R_{\rm BCV}=1.34$ fm and $X_{\rm BCV}=0.75$ 
are obtained by fitting Glauber-type calculations of the total nuclear 
reaction cross Sections~\cite{BenCook}.  The fragment angular
distributions, very sensitive to $b_c$, can be described by the BCV 
parametrization \cite{Grunschloss:ds}.

Even when the nuclear densities partly overlap and only a few 
nucleon-nucleon collisions occur, intense hadron production 
is expected at LHC energies. These secondary hadrons will
be produced at midrapidity while neutrons from electromagnetic dissociation
are emitted close to beam rapidity. This difference can be used to disentangle 
hadronic and electromagnetic dissociation. 

The cross section for the removal (abrasion) of $a_1$ nucleons from 
projectile $A$ by interaction with target $B$ may be derived
from Glauber multiple-scattering theory~\cite{Hufner}
\begin{equation}
\sigma^{\rm nuc}(a_1)={A\choose{a_1}} 2\pi
\int\limits_{0}^{\infty} db \, b \,
[1-P(b)]^{a_1} [P(b)]^{A-a_1} \, \, .
\label{abr_main}
\end{equation}
Here $P(b)$ is the overlap of the projectile, 
$T_{A}(s)$, and target, $T_{B}(|\vec b - \vec s|)$,
thickness functions at impact parameter $b$
\begin{equation}
P(b)= \frac{1}{A}\int d^2s T_{A}(s)
\exp[-\sigma_{NN}T_{B}(|\vec{b} - \vec{s}|)] \, \, .
\end{equation}
The nuclear densities are parametrized by Woods-Saxon distributions 
for heavy nuclei.

More details and numerical results for hadron-nuclear dissociation in Pb+Pb 
collisions at the LHC can be found 
in Ref.~\cite{Pshenichnov:2001qd}. 
Here we only give the expression for the dissociation cross section when no
nucleons are ejected, due to partial transparency of surface nucleons,
\begin{equation}
\sigma^{\rm nuc}(0)= 2\pi
\int\limits_{0}^{\infty} db \, b \, [P(b)]^{A} \, \, .
\label{abr_0}
\end{equation}

\subsection{Mutual electromagnetic excitation as a filter for close 
collisions}

\begin{figure}[htb]
\begin{centering}
{\includegraphics[width=1.08\textwidth]{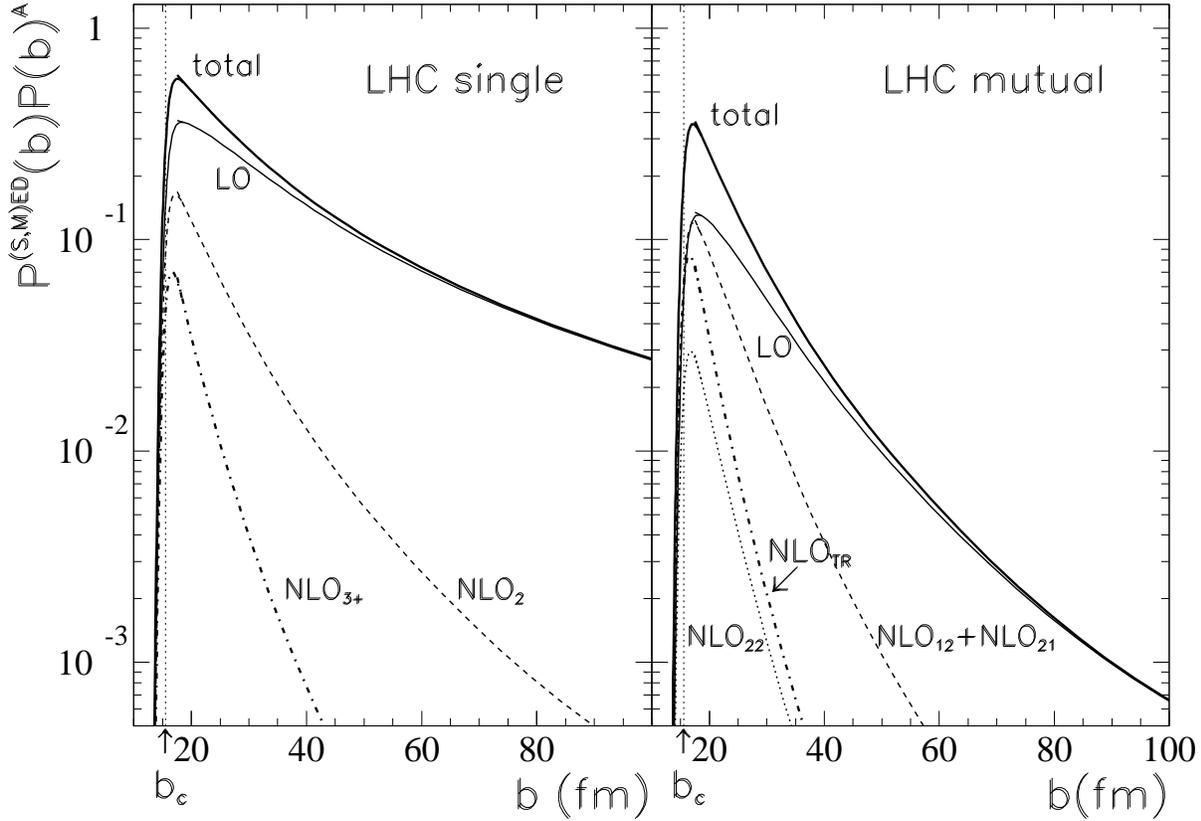}}
\caption[]{The probability of single (left panel) and mutual (right panel) 
electromagnetic dissociation to LO and NLO as a function of $b$
in $2.75A + 2.75A$ TeV Pb+Pb collisions at the LHC predicted
by the {\sc reldis} and soft-sphere models.  The thick solid lines
give the sum of the LO and NLO contributions. The value $b_c=15.54$~fm is 
indicated.}
\label{fig:6:prob}
\end{centering}
\end{figure}

In this section, we compare the probability distributions for single and mutual
electromagnetic dissociation.  In single electromagnetic dissociation, only
one of the two nuclei dissociate.  The cross section for the two processes can
be written as
\begin{equation}
\sigma^{\rm (S,M)ED} = 2 \pi\int\limits_{b_{c}}^{\infty}  db \, b \,
{\cal P}^{\rm (S,M)ED}(b) \label{sandm}
\end{equation}
where ${\cal P}^{\rm SED}(b)$ and ${\cal P}^{\rm MED}(b)$ are the probabilities
for single and mutual electromagnetic dissociation.
The expressions for ${\cal P}^{\rm MED}(b)$ can be taken 
from Eqs.~(\ref{eq:6:LO}), (\ref{eq:6:NLO12}) and  (\ref{eq:6:NLO22}).
The expressions for ${\cal P}^{\rm SED}$ follow from Eqs.~(\ref{eq:6:NLO2})
and (\ref{eq:6:NLO3p}). 
Equation~(\ref{sandm}) corresponds to the sharp-cutoff approximation.  
For comparison, in the soft-sphere model, the cross section
without the sharp cutoff, $b_c$, can be written as
\begin{equation}
\sigma^{\rm (S,M)ED} = 2\pi\int\limits_{0}^{\infty}  db \, b \, 
[ P(b)]^{A} \, {\cal P}^{\rm (S,M)ED}(b) \, \, ,
\end{equation}
as follows from Eq.~(\ref{abr_0}) when no nucleons are ejected.
The product $[P(b)]^{A} {\cal P}^{\rm (S,M)ED}(b)$
is presented in Fig.~\ref{fig:6:prob} for single 
and mutual electromagnetic dissociation  for each of the LO and NLO processes
discussed previously.  Using the sharp-cutoff for heavy nuclei gives a few 
percent error on $\sigma^{\rm (S,M)ED}$, within the uncertainty introduced by
the photonuclear cross section data used in calculations.
 
The largest contributions to the MED cross section comes from `close'
collisions with $b\sim b_c$, where the probability to absorb a virtual 
photon is large, and two or more photons can be absorbed
by each nucleus. The MED probabilities decrease faster with $b$ than SED.  
Thus mutual dissociation can be used as a filter for selecting collisions
with $b \sim b_c$.

The relative NLO contributions to MED are enhanced compared to
SED, as shown in Fig.~\ref{fig:6:prob}. 
The sum NLO$_{12} \, + \, {\rm NLO}_{21}$ is similar to the LO contribution
when $b\sim b_c$.
In this region, the probability of triple excitations, NLO$_{\rm TR}$, is also 
comparable to the LO contribution. 
However, all the NLO contributions decrease faster with $b$ than the LO 
contribution.  Thus the NLO cross sections in Table~\ref{tab:6:TLO-NLO} 
are lower than the LO cross sections. 
\begin{table}[htb]
\begin{center}
\caption[]{The total LO, individual NLO corrections and summed MED cross 
sections for $2.75A + 2.75 A$ TeV Pb+Pb
collisions at the LHC \protect\cite{Pshenichnov:2001qd}.  
}
\vspace{0.4cm}
\begin{tabular}{|c|c|c|c|c|}
\hline
$\sigma^{\rm MED}_{\rm LO}$ (b) & $\sigma^{\rm MED}_{\rm NLO_{12}} +
\sigma^{\rm MED}_{\rm NLO_{21}}$ (b) & $\sigma^{\rm MED}_{\rm NLO_{22}}$ (b) & 
$\sigma^{\rm MED}_{\rm NLO_{\rm TR}}$ (b) & $\sigma^{\rm MED}_{\rm tot}$ (b)
\\ \hline
3.92 & 1.50 & 0.23 & 0.56 & 6.21 \\ \hline
\end{tabular}
\label{tab:6:TLO-NLO}
\end{center}
\end{table}

The MED cross sections presented here were calculated using the
modified {\sc reldis} code~\cite{Pshenichnov:2001qd} employing a special 
simulation mode for MED events.
Table~\ref{tab:6:TLO-NLO} gives the inclusive LO, NLO$_{12}$, NLO$_{22}$,
NLO$_{\rm TR}$ contributions and total cross sections with
$f(E,i) = f(E,j) = 1$ in $\sigma^{\rm MED}(i|j)$.  The LO contribution is
$\sim 63$\% of $\sigma^{\rm MED}({\rm tot})$ 
at LHC energies. The sum of the NLO contributions gives an additional
$\sim 28$\%. Therefore, the remaining contribution, $\sim 9$\% of the total 
MED cross section, is due to exotic triple nuclear excitations with three or 
more photons absorbed by at least one nucleus.

Electromagnetic heavy-ion excitation is widely used to study nuclear structure,
as demonstrated by fixed-target experiments at intermediate
energies~\cite{Emling:1994gu,Aumann-Emling}.  Experimental studies of MED 
at the LHC can provide valuable information about double and triple nuclear 
excitations in electromagnetic interactions, particularly for multi-phonon 
resonances. Triple excitation data is very important for
triple giant resonance excitations since there are currently 
no data on such extreme excitations.  The first theoretical predictions for
the energies and widths of such states are given in 
Ref.~\cite{dePassos:2001dc}.

The number of forward neutrons emitted in the dissociation process and
detected in the ZDCs can be used to study multiple excitations,
even when the ZDC resolution cannot determine the exact number of neutrons 
in one of the ZDCs or for poor statistics.  
To demonstrate the utility of the dissociation process, we assume that the
dissociation channel of one of the nuclei in MED is unknown.  Then the
inclusive MED cross sections, $\sigma^{\rm MED}(1nX| {\cal D})$, 
$\sigma^{\rm MED}(2nX| {\cal D})$ and $\sigma^{\rm MED}(3nX| 
{\cal D})$ for emission of one, two and three neutrons, 
by one of the nuclei, respectively, can be considered.   The $X$ 
indicates that neutron emission can be
accompanied by some number of undetected particles.  
In the notation of Section~\ref{RHIC&Notations}, 
${\cal D}$ denotes an arbitrary 
dissociation mode for the other nucleus so that $f(E,i)
\equiv 1$.

\begin{table}[htb]
\begin{center}
\caption[]{The MED cross sections for $2.75A+2.75A$ TeV
Pb+Pb collisions at the LHC where $X$ and $Y$ denote other particles emitted
from the nucleus with the neutrons and ${\cal D}$ is an arbitrary dissociation 
channel for the other nucleus ($f(E,i)=1$).  Results are given at LO alone
and with the sum of the NLO contributions included 
\protect\cite{Pshenichnov:2001qd}.
}
\vspace{0.4cm}
\begin{tabular}{|c|c|c|}
\hline
 Final State & $\sigma_{\rm LO}$ (mb) & $\sigma_{\rm LO} + 
\sigma_{{\rm NLO}_{12}} + \sigma_{{\rm NLO}_{21}} + \sigma_{{\rm NLO}_{22}}$
(mb) \\ \hline
$(1nX| 1nY)$      & 750  & 805  \\
$(1nX| {\cal D})$ & 1698 & 2107 \\
$(2nX| {\cal D})$ & 443  & 654 \\
$(3nX| {\cal D})$ & 241  & 465 \\ \hline
\end{tabular}
\label{tab:6:TLONLO}
\end{center}
\end{table}

The LHC cross sections for several MED channels are given in 
Table~\ref{tab:6:TLONLO}.  The specific branching ratios for
the final-state channels, $f(E,1nX)$, $f(E,2nX)$ and
$f(E,3nX)$, are calculating by simulating neutron emission from a
lead nucleus following the absorption of photons with energy $E$.
The probability $P_A$ in Eq.~(\ref{eq:6:P1}) is modified by the 
branching ratio in the integral over $E$ while the factor 
$\exp[-m_A(b)]$ remains the same for both nuclei when $A=B$.
If one final-state neutron is required for both nuclei, the cross 
section $\sigma^{\rm MED}(1nX| 1nY)$ is reduced relative to 
$\sigma^{\rm MED}(1nX| {\cal D})$ since both branching ratios are
included in the probabilities. 

The relative NLO contributions are very different for one, two and
three neutron emission. 
The NLO contribution to $\sigma^{\rm MED}(1nX| 1nY)$ 
is small, $\sim 7$\%.  On the other hand, the NLO correction to 
$\sigma^{\rm MED}(3nX|{\cal D})$ is almost
a factor of two.  This large increase is because
the NLO processes shown in Fig.~\ref{fig:6:EMsecond} include
nuclear excitation due to double photon absorption, particularly
double GDR excitation. Since the average GDR energy for gold and lead
nuclei is $\sim 13-14$ MeV, double GDR introduces, on average, 
$26-28$ MeV excitation, above the three neutron emission threshold.
Thus the $1n$ and $2n$ emission cross sections have smaller NLO corrections
than the $3n$ cross sections. 
Measurements of the forward $3n$ emission rates in ALICE
may detect multiple GDR excitations in 
lead. 

\subsection{Reliability of the {\sc reldis} predictions}

The reliability of the {\sc reldis} code was studied in 
Ref.~\cite{Pshenichnov:2001qd} by examining its sensitivity to
variations in the input data and parameters.  A good description of the
existing SED data on lead and gold nuclei at the CERN SPS 
Refs.~\cite{Golubeva,Scheidenberger:tr,Hill:ie,Dekhissi:1998ke} was obtained. 
As shown in Sec.~\ref{RHIC&Notations}, good agreement with the
first RHIC data on mutual dissociation~\cite{Chiu:2001ij} was also found.

The photonuclear cross sections for specific neutron
emission channels ($f(E,i) \neq 1$, $f(E,j)\neq 1$)
were calculated by two different models of
photonuclear reactions, {\sc gnash}~\cite{Gandini:1998} 
and {\sc reldis}, see Table~\ref{tab:6:TPdir} and 
Ref.~\cite{Pshenichnov:2001qd} for details.  In addition, two different values 
for the probability of direct neutron emission in the $1n$ channel, 
$P^{\rm dir}_n = 0$ and 0.26, were used in the {\sc reldis} code.

\begin{table}[htbp]
\begin{center}
\caption[]{The LO and NLO MED cross section 
are presented for the {\sc gnash} and {\sc reldis} codes in 
$2.75A + 2.75A$ TeV Pb+Pb 
collisions\protect\cite{Pshenichnov:2001qd}.  
The sensitivity of the MED cross sections in selected channels
to the photon energy range, $E_\gamma$,
the probability of direct single neutron emission, $P^{\rm dir}_n$, and
the input photonuclear cross sections is illustrated.   
The recommended values are shown in boldface.  For comparison, the predicted
value of $\sigma^{\rm MED}_{\rm LO}(1n|1n)$ in the GDR region $(E_\gamma \leq
24$ MeV) calculated in Ref.~\protect\cite{Baltz:1998ex}
is 533 mb.
}
\vspace{0.4cm}
\begin{tabular}{|c|c|c|c|c|c|}
\hline
& GDR region &\multicolumn{2}{|c|}{quasi-deuteron region} &
\multicolumn{2}{|c|}{all $E_\gamma$ }\\ \cline{5-6}
& $E_\gamma\leq 24$ MeV &\multicolumn{2}{|c|}{$E_\gamma\leq 140$ MeV }&
 \multicolumn{2}{|c|}{$\sigma_{\rm LO}^{\rm MED} + \sigma_{\rm NLO_{12}}^{\rm
MED} + $ }\\ \cline{2-4}
& $\sigma_{\rm LO}^{\rm MED}$ (mb) &\multicolumn{2}{|c|}{$\sigma_{\rm LO}^{\rm 
MED}$ (mb)} & \multicolumn{2}{|c|}{$\sigma_{\rm NLO_{21}}^{\rm
MED} +  \sigma_{\rm NLO_{22}}^{\rm MED}$ (mb)}\\ \cline{2-6}
 & {\sc reldis} & {\sc gnash} & {\sc reldis} & {\sc reldis} & {\sc reldis} \\
Channel & $P^{\rm dir}_n=0$ & & $P^{\rm dir}_n=0$ & $P^{\rm dir}_n=0$ & 
$P^{\rm dir}_n=0.26$ \\  \hline\hline
$(1nX|1nY)$ & 519 & 488 &  544 & 727 & {\bf 805} \\ 
$(1nX|2nY) + (2nX|1nY)$ & 154 & 220 & 217 
& 525 & {\bf 496} \\ 
$(2nX| 2nY)$ & 11 & 24 & 22 &  96 & {\bf 77} \\ 
LMN &  684 &  732 & 783 & 1348 & {\bf 1378} \\ \hline
\end{tabular}
\label{tab:6:TPdir}
\end{center}
\end{table}

At the LHC, secondary nuclei are produced by electromagnetic 
dissociation of beam nuclei induced by interactions with residual gas and 
collimator material.  These nuclear fragments diverge from the primary beam
because of their scattering angle and their different $Z/A$ relative
to the primary beam.  Since these fragments do not fall within the 
acceptance of the
collimation system, they induce a significant heat load in the superconducting 
magnets when they hit the magnet vacuum chamber. The yields of specific 
nuclear fragments from SED, MED and fragmentation of beam nuclei were
calculated using {\sc reldis} and abrasion-ablation models to estimate the 
heat load at the LHC \cite{Assmann:2004km,Jowett:2004rd}.

The cross sections for one or two neutron emission are given in 
Table~\ref{tab:6:TPdir} for different maximum values of the photon energy,
$E_\gamma \leq E_{\rm max}$,
the upper limit in the energy integrals in Eqs.~(\ref{eq:6:P1}) and (\ref{mb}).
Results are shown for the GDR region, $E_\gamma \leq 24$ MeV, energies up to 
the quasi-deuteron absorption region, $E_\gamma \leq 140$ MeV, and the full
range.  In addition to the specified one and two neutron emission channels,
a cumulative value, the Low Multiplicity Neutron (LMN) emission 
cross section,
\begin{eqnarray}
\sigma^{\rm MED}({\rm LMN}) & = & \sigma^{\rm MED}(1nX| 1nY) +
\sigma^{\rm MED}(1nX| 2nY) \nonumber \\ & & \mbox{} + 
\sigma^{\rm MED}(2nX| 1nY) + \sigma^{\rm MED}(2nX| 2nY) \, \, , 
\end{eqnarray}
is also shown.

Table~\ref{tab:6:TPdir} shows that there is a $\sim 10$\% ambiguity
in $\sigma (1nX| 1nY)$, mainly due to uncertainties in the 
photo-neutron cross sections measured in experiments with real photons. 
However, when the sum of the one and two neutron 
emission channels, $\sigma^{\rm MED}({\rm LMN})$, is considered, the 
uncertainty is reduced to $\sim 2$\%. 
The sum, $\sigma^{\rm MED}({\rm LMN})$, is also more stable with respect to 
other parameters relative to the other cross sections in 
Table~\ref{tab:6:TPdir}, as discussed in Ref.~\cite{Pshenichnov:2001qd}.
Therefore, $\sigma^{\rm MED}({\rm LMN})$ serves as a cumulative neutron 
emission rate useful for luminosity measurements at heavy-ion colliders.

\begin{figure}[htbp]
\begin{centering}
{\includegraphics[height=0.9\textheight]{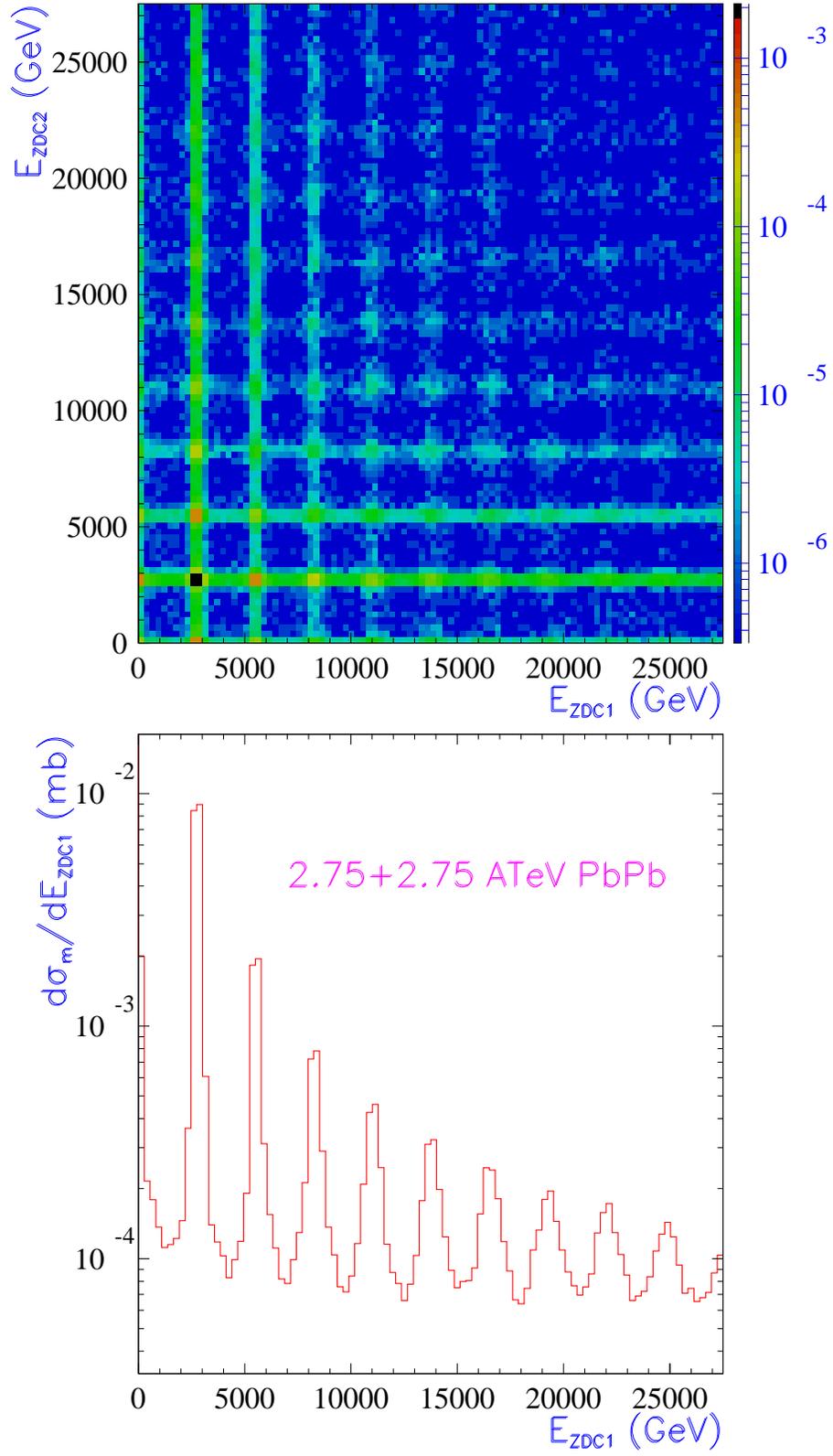}}
\caption[]{Top panel: The distribution of the total forward-backward neutron
energy emitted in MED in \PbPb\ collisions at the LHC.
Bottom panel: The energy distribution in one ZDC obtained
by projection of the top plot.  The results are given for the LO process 
without including the ZDC energy resolution.  From 
Ref.~\protect\cite{Pshenichnov:2001qd}.}
\label{fig:6:ezdc}
\end{centering}
\end{figure}

At collider energies, neutron emission 
in mutual electromagnetic dissociation is not entirely exhausted
by the simultaneous excitation and giant resonance decays in
both of the colliding nuclei.  In addition to mutual GDR excitation,
asymmetric processes, such as GDR excitation of one nucleus accompanied
by a photonuclear reaction in the other nucleus, are very likely.
The presence of such asymmetric dissociations is clear in 
Fig.~\ref{fig:6:ezdc} which shows the forward neutron energy distributions.

The ALICE ZDC has several advantages relative to the RHIC ZDCs.
The forward neutron energy resolution is expected to be $\sim 10$\% at the
LHC \cite{ZDC-ALICE99,ZDC-ALICE02} while it is $\sim 20$\% at RHIC
\cite{Adler:fq}.  As a result, the $3n$ and $4n$ emission
channels can be unambiguously identified by the ALICE ZDC, making it possible
to study multiple GDR excitations.

\section{Hard photoproduction at HERA}
{\it Contributed by: M. Klasen}

\subsection{Introduction}
\label{heraintro}

In view of possible photoproduction studies in ultraperipheral heavy-ion
collisions at the LHC, we briefly review the present theoretical
understanding of photons and hard photoproduction processes at HERA,
discussing the production of jets, light and heavy hadrons, quarkonia, and
prompt photons. We address in particular the extraction of the strong
coupling constant from photon structure function and inclusive jet
measurements, the infrared safety and computing time of jet definitions, the
sensitivity of dijet cross sections on the parton densities in the photon,
factorization breaking in diffractive dijet production, the treatment of the
heavy-quark mass in charm production, the relevance of the color-octet
mechanism for quarkonium production, and isolation criteria for prompt
photons.

\begin{figure}
 \centering
 \includegraphics[width=.66\columnwidth]{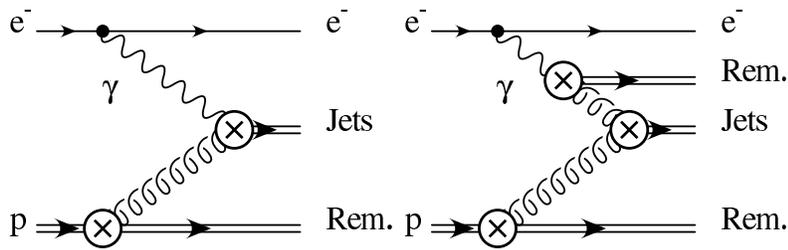}
 \caption[]{Factorization of direct (left) and resolved (right)
 photoproduction in the QCD-improved parton model \protect\cite{kla02}.  
Here Rem.\ indicates the proton and photon remnants.  Copyright 2002 by the
American Physical Society (http://link.aps.org/abstract/RPM/v74/p1221).}
\label{fig:1}
\end{figure}

Electron-proton scattering at HERA is dominated by the exchange of
low-virtuality (almost real) photons \cite{kla02}. If the electron is
anti-tagged or tagged at small angles, the photon flux from the electron can
be calculated in the Weizs\"acker-Williams approximation, where the energy
spectrum of the exchanged photons is given by
\bea
 \hspace*{-10mm}
 f_{\gamma/e}^{\rm brems}(x)&=&\frac{\alpha}{2\pi}\left[
 \frac{1+(1-x)^2}{x}\ln\frac{Q^2_{\max}(1-x)}{m_e^2 x^2}
 + 2 m_e^2 x\left(\frac{1}{Q^2_{\max}}-\frac{1-x}{m_e^2 x^2}\right)
 \right]\hspace*{10mm}
\eea
and the subleading non-logarithmic terms modify the cross section typically
by 5\% \cite{kes75}. In the QCD-improved parton model, valid for hard
scatterings, the photons can then interact either directly with the partons
in the proton (Fig.\ \ref{fig:1}, left) or resolve into a hadronic
structure, so that
their own partonic constituents interact with the partons in the proton
(Fig.\ \ref{fig:1}, right). While this separation is valid at leading order
(LO) in QCD perturbation theory, the two processes are intimately linked at
next-to-leading order (NLO) through the mandatory factorization of a
collinear singularity that arises from the splitting of the photon into a
quark-antiquark pair and induces a mutual logarithmic factorization scale
dependence in both processes. In close analogy to deep-inelastic
electron-proton scattering, one can define a photon structure function
\bea
 \hspace*{-5mm}
 F_2^{\gamma}(Q^2)&=&\sum_q 2 x e_q^2\lg f_{q/\gamma}(Q^2) \right.
\nonumber \\ & & \mbox{} \left. +\frac{\alpha_s(Q^2)}
 {2\pi}\le C_{q}\otimes f_{q/\gamma}(Q^2)
 + C_{g}\otimes f_{g/\gamma}(Q^2)\re
 +\frac{\alpha}{2\pi}e_q^2C_{\gamma}\rg
\eea
that is related to the parton densities in the photon and has been measured
in electron-positron collisions at LEP. Even the strong coupling constant
$\alpha_s$ that appears in the expression above can be determined rather
precisely in fits to these data \cite{alb02}. A convenient modification of
the $\ms$ factorization scheme consists in absorbing the point-like Wilson
coefficient
\bea
 \hspace*{-10mm}
 C_{\gamma}(x)&=&2N_C\, C_g(x)=3\, \left[ \left( x^2+(1-x)^2\right)
   \,\ln \, \frac{1-x}{x}
 + 8x(1-x)-1 \right] \hspace*{10mm}
\eea
in the Altarelli-Parisi splitting function $P_{q\leftarrow\gamma}^{\dis} =
P_{q\leftarrow \gamma}^{\ms} - e_q^2\, P_{q\leftarrow q}\otimes C_{\gamma}$
\cite{Gluck:1991ee}.

\subsection{Inclusive and diffractive jet production}

While at LO hadronic jets are directly identified as final-state partons,
their definition becomes subtle at higher orders, when several partons (or
hadrons) can be combined to form a jet. According to the standardization of
the 1990 Snowmass meeting, particles $i$ are added to a jet cone $J$ with
radius $R$, if they are a distance 
$R_i = \sqrt{(\eta_i-\eta_J)^2+(\phi_i-\phi_J)^2} < R$
from the cone center. However, these broad combined jets are difficult to
find experimentally, so that several modifications (mid-points, additional
seeds, iterations) have been successively applied by the various
experiments. The deficiencies of the cone algorithm are remedied in the
longitudinally invariant $k_T$-clustering algorithm, where one uses only the
combination criterion $R_{ij} < 1$ for any pair of particles $i$ and $j$.
Unfortunately, this algorithm scales numerically with the cubic power of the
number, $N$, of particles involved. Only recently a faster version has
been developed making use of geometrical arguments and diagrammatic methods
known from computational science \cite{cac06}. The publicly available {\tt
FastJet} code scales only with $N\ln N$ and is now rapidly adopted, in
particular for the LHC, where the particle multiplicity is high.

\begin{figure}
 \centering
 \includegraphics[width=.5\columnwidth]{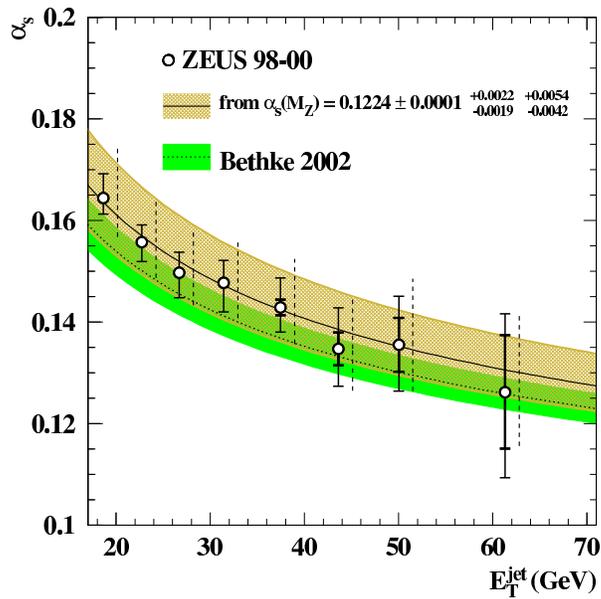}
 \caption[]{Strong coupling constant as measured from scaling
 violations in inclusive single-jet production at ZEUS.
Reprinted from Ref.~\protect\cite{alphas} with permission from Elsevier. }
\label{fig:2}
\end{figure}

Single (inclusive) jets benefit from high statistics and the presence of
a single (transverse) energy scale $E_T$, which makes them easily accessible
experimentally and their prediction theoretically stable. The
$E_T$-distribution of the single-jet cross section can then be used to
determine e.g.\ the strong coupling constant from scaling violations, as
shown in Fig.\ \ref{fig:2}.
However, the single-jet cross section,
\bea
 \hspace*{-10mm}
  \frac{d^2\sigma}{dE_T d\eta}
  &=& \sum_{a,b} \int_{x_{a,\min}}^1 dx_a \, x_a 
  f_{a/A}(x_a,M_a^2) \, x_b f_{b/B}(x_b,M_b^2)
  \frac{4E_AE_T}{2x_aE_A-E_Te^{\eta}}
  \frac{d\sigma}{dt} \, \, ,
 \hspace*{10mm}
\eea
includes a convolution over one of the longitudinal momentum fractions
of the partons so that parton densities cannot be uniquely determined.

\begin{figure}
 \centering
 \includegraphics[width=0.9\columnwidth]{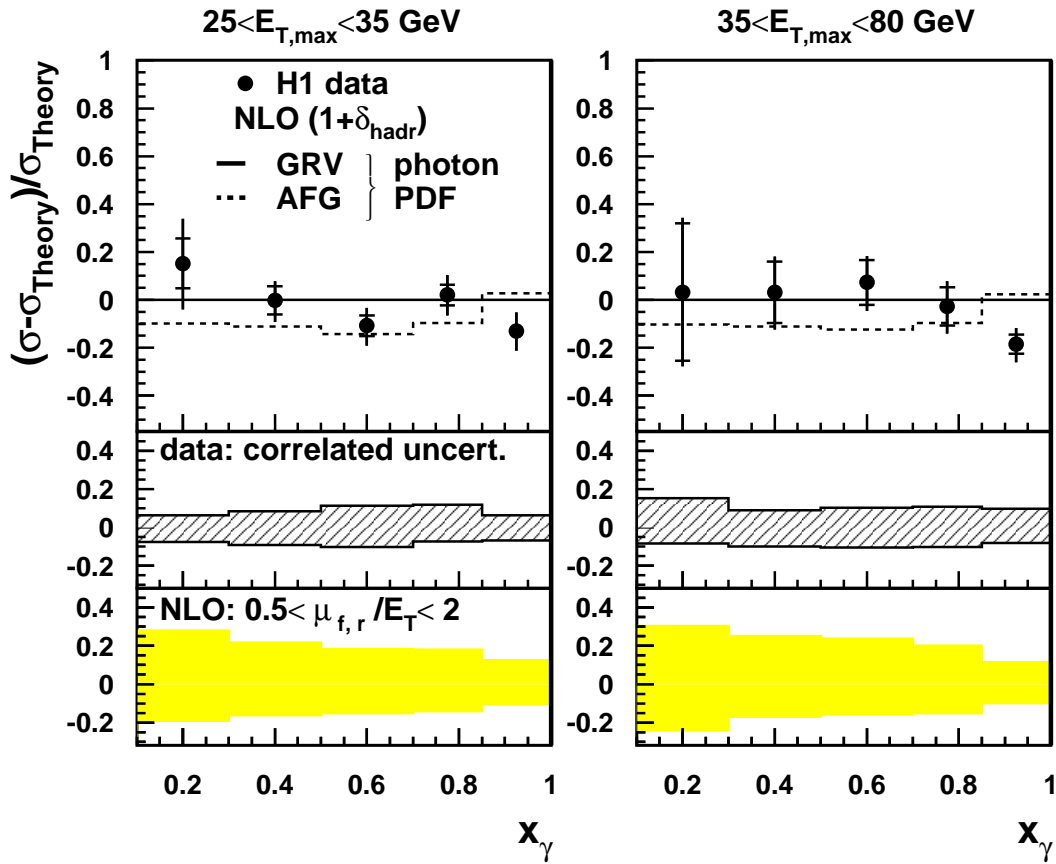}
 \caption[]{Sensitivity of the dijet photoproduction cross
 section as measured by H1 on the GRV and AFG parameterizations of the
 parton densities in the photon.
Reprinted from Ref.~\protect\cite{dijets} with permission from 
Springer-Verlag. }
\label{fig:3}
\end{figure}
In addition to the transverse energy $E_T$ and pseudorapidity $\eta_1$ of
the first jet, the inclusive dijet cross section
\beq
  \frac{d^3\sigma}{dE_T^2d\eta_1d\eta_2}
  = \sum_{a,b} x_a f_{a/A}(x_a,M_a^2) x_b f_{b/B}(x_b,M_b^2)
  \frac{d\sigma}{dt}
\eeq
depends on the pseudorapidity of the second jet $\eta_2$. In LO only two
jets with equal transverse energies can be produced and the observed
momentum fractions of the partons in the initial electrons or hadrons
$x_{a,b}^{\rm obs} = \sum_{i=1}^{2} E_{T_i}e^{\pm\eta_i} / (2E_{A,B})$
equal the true momentum fractions $x_{a,b}$. If the energy transfer
$y=E_\gamma/E_e$ is known, momentum fractions for the partons in photons
$x_\gamma^{\rm obs}=x_{a,b}^{ \rm obs}/y$ can be deduced. In NLO, where a
third jet can be present, the observed momentum fractions are defined by the
sums over the two jets with highest $E_T$ and they match the true momentum
fractions only approximately. Furthermore, the transverse energies of the
two hardest jets no longer need to be equal to each other. Even worse, for
equal $E_T$ cuts and maximal azimuthal distance, $\Delta\phi=\phi_1-\phi_2=
\pi$, the NLO prediction becomes sensitive to the method chosen for the
integration of soft and collinear singularities. The theoretical cross
section is then
strongly scale dependent and thus unreliable. This sensitivity also
propagates into the region of
large observed momentum fractions. It is thus preferable to cut on the
average $\overline{E}_T=(E_{T_1}+E_{T_2})/2$. The sensitivity of the dijet
photoproduction cross section as measured by H1 on the GRV and AFG
parameterizations of the  parton densities in the photon is shown in
Fig.\ \ref{fig:3}.

\begin{figure}
 \centering
 \includegraphics[width=0.66\columnwidth]{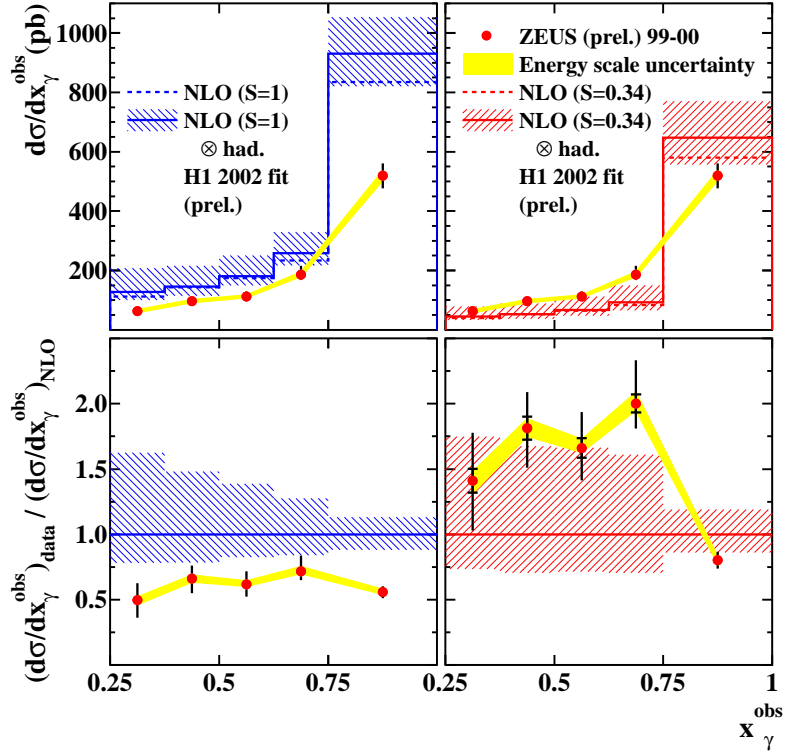}
 \caption[]{Dependence of the diffractive dijet cross section on
 the observed longitudinal momentum fraction of the scattered photon at
 ZEUS \protect\cite{diffr}.}
\label{fig:4}
\end{figure}
In diffractive processes with a large rapidity gap between a leading proton
\cite{kla04a}, neutron \cite{kla06} or some other low-mass hadronic state
and a hard central system, QCD factorization is expected to hold for
deep-inelastic scattering, so that diffractive parton densities can be
extracted from experiment, but
to break down for hadron-hadron scattering, where initial-state rescattering
can occur. In photoproduction, these two factorization
scenarios correspond to direct and
resolved processes, which are, however, closely related, as noted in 
Section~\ref{heraintro}.
It is thus interesting to investigate the breakdown of factorization in
kinematic regimes where direct or resolved processes dominate. This can
either be done by measuring the dependence on the photon virtuality $Q^2$
(transition from virtual to real photons) \cite{kla04b}, $E_T$ (direct
processes are harder than resolved photons), or $x_\gamma^{\rm obs}$ (unity for
direct processes at LO). The $x_\gamma^{\rm obs}$ distribution is confronted 
with the hypothesis of no (or global) factorization breaking
(left) and with a suppression factor $S$ of 0.34 \cite{kai03} applied to 
resolved processes only (right) in Fig.\
\ref{fig:4}. Note that the interdependence of direct
and resolved processes requires the definition of a new factorization scheme
with suppression of the scale-dependent logarithm also in the direct
contribution \cite{kla05}
\bea
 M(Q^2,S)_{\overline{\rm MS}} &=& \le-\frac{1}{2N_c} P_{q_i\leftarrow
 \gamma}(z)\ln\lrr\frac{M_{\gamma}^2 z}{p_T^{*2}(1-z)}\rrr+{Q_i^2\over2} \re S
 \nonumber \\
 && \ -\frac{1}{2N_c} P_{q_i\leftarrow\gamma}(z)
 \ln\lrr\frac{p_T^{*2}}{zQ^2+y_s s}\rrr.
\eea

\subsection{Light and heavy hadron production}

\begin{figure}
 \centering
 \includegraphics[width=\columnwidth]{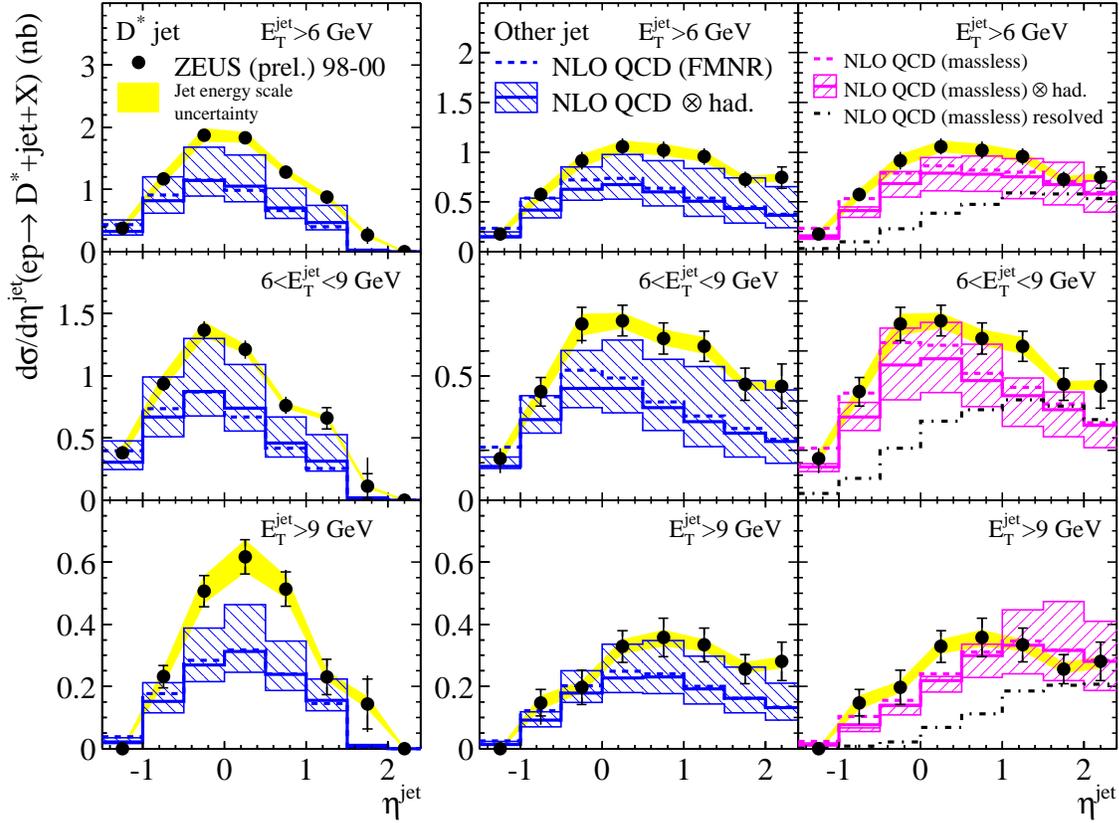}
 \caption[]{Rapidity distributions of $D^*$-mesons and associated
 jets as measured by ZEUS and compared to massive (fixed-order) and massless
 (variable flavor number scheme) calculations \protect\cite{charm}.}
\label{fig:5}
\end{figure}

If individual hadrons are experimentally identified, the cross sections
above have to be modified to include convolutions over fragmentation
functions $D(z)$. For light quarks and gluons, these non-perturbative, 
universal distributions must be fitted to $e^+e^-$ data, but then produce
successful predictions for HERA data at NLO. For heavy quarks, the
fragmentation functions can in principle be calculated perturbatively, if
the heavy-quark mass is kept finite (``fixed-order scheme''), although,
for large $E_T$, they must be evolved using renormalization group equations
(for example at ``next-to-leading logarithm'') \cite{cac01}. An alternative
method is to fit the fragmentation functions for $D$- and $B$-mesons again
to $e^+e^-$ data at large $E_T$ (``variable flavor number scheme''). If, in
addition, the finite mass terms are kept in the hard coefficient functions,
one moves from a ``zero-mass scheme'' to a ``general-mass scheme'' and can
achieve a smooth transition from large to small $E_T$ \cite{kni05}. A
comparison of both theoretical approaches to recent $D^*$+jet data from ZEUS
is shown in Fig.\ \ref{fig:5}. While the massive calculation with central
scale choice clearly underestimates the data, the variable flavor number
scheme allows not only for direct, but also for resolved-photon
contributions and tends to give a better description of the data over the
full rapidity range. Note that both predictions have been multiplied by
hadronization corrections modeled with Monte Carlo simulations. While
several calculations for inclusive single-hadron production with real
photons are available, a theoretical investigation of the transition region
to virtual photons and of the production of two hadrons, for example in the
forward region, is still needed.

\begin{figure}
 \centering
 \includegraphics[width=0.66\columnwidth]{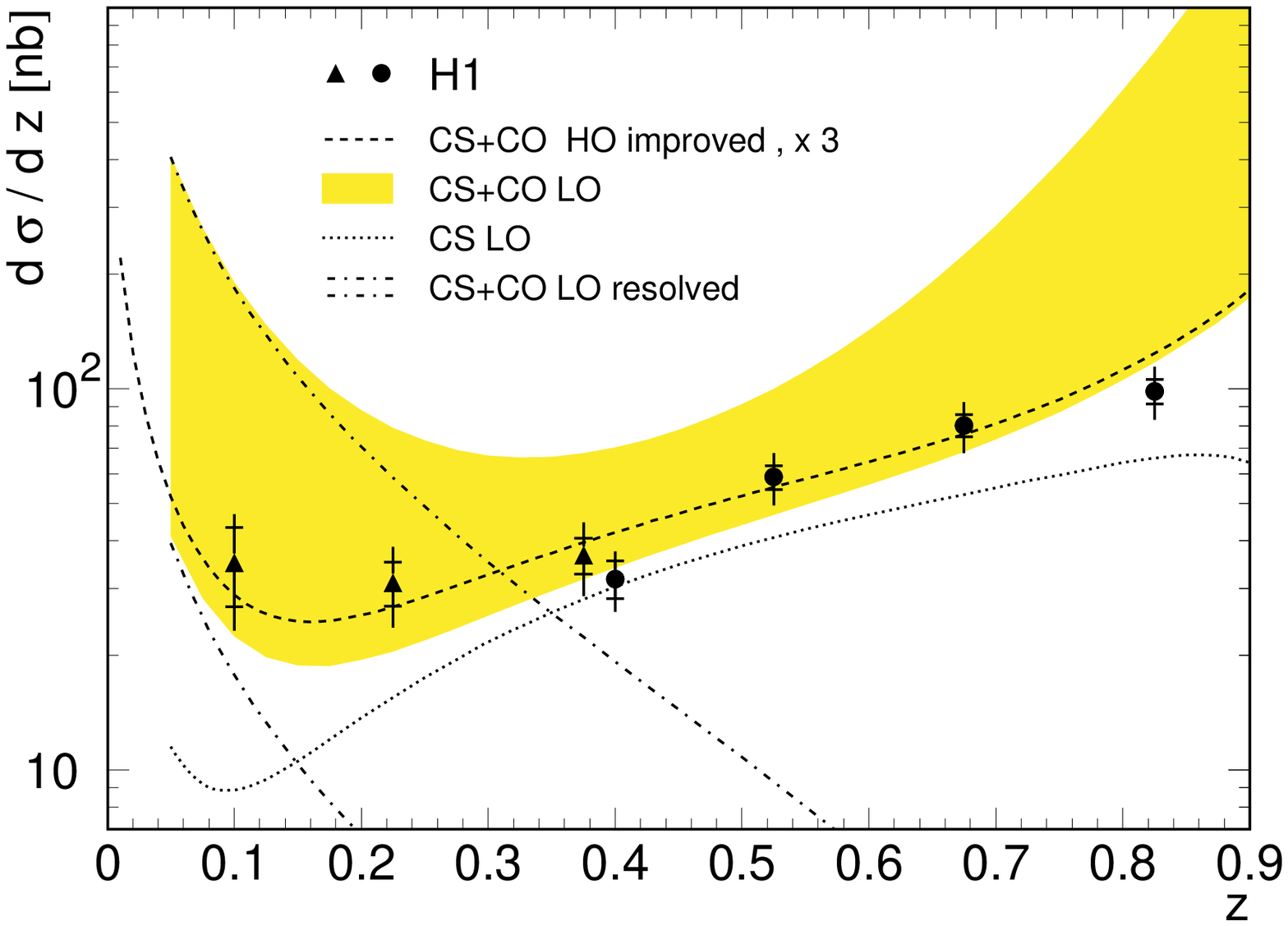}
 \caption[]{Direct and resolved contributions to the
 color-singlet and color-octet $J/\psi$ energy distribution in
 photoproduction at HERA \protect\cite{jpsi}.
Reprinted from Ref.~\protect\cite{alphas} with permission from 
Springer-Verlag. }
\label{fig:6}
\end{figure}
The production of heavy quark-antiquark bound states is still far from
being understood theoretically. While color-singlet (CS) states are to some
extent formed already during hard collisions, their contribution has been
shown to be both theoretically incomplete due to uncanceled infrared
singularities as well as phenomenologically insufficient due to an
order-of-magnitude discrepancy with the measured $J/\psi$ $p_T$-spectrum 
at the Tevatron. On the other hand, non-relativistic QCD
(NRQCD) allows for a systematic expansion of the QCD Lagrangian in the
relative quark-antiquark velocity and for additional color-octet (CO)
contributions with subsequent color neutralization through soft gluons.
Then $J/\psi$-production in photon-photon collisions at LEP can be
consistently described \cite{kla02b}, as can be the photoproduction data from
HERA in Fig.\ \ref{fig:6}. At HERA, the color-octet contribution becomes
important only at small momentum-transfer $z$ of the photon to the
$J/\psi$. Unfortunately, recent CDF data do not support the prediction
of transverse polarization of the produced $J/\psi$ at large $p_T$ as
predicted from the on-shell fragmentation of final-state gluons within
NRQCD.  Further experimental and theoretical studies are thus urgently
needed.

\subsection{Prompt photon production}

\begin{figure}
 \centering
 \includegraphics[width=0.66\columnwidth]{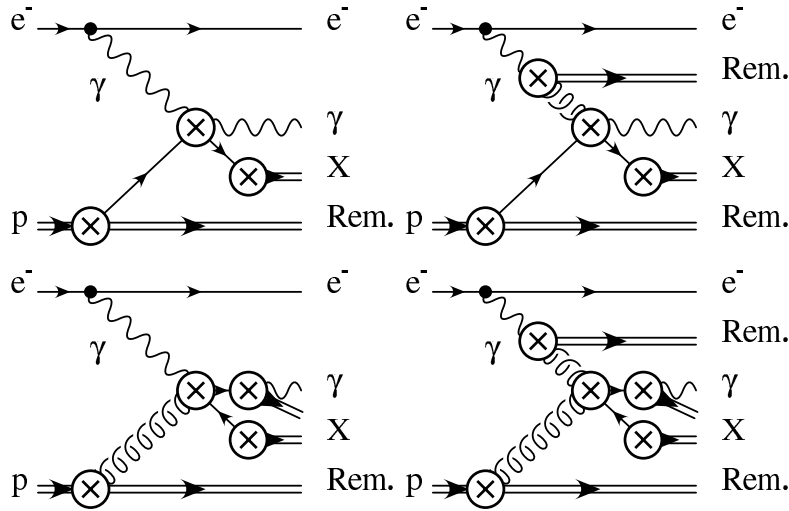}
 \caption{Factorization of prompt photon photoproduction \protect\cite{kla02}.
  Copyright 2002 by the American Physical Society  
(http://link.aps.org/abstract/RPM/v74/p1221).}
\label{fig:8}
\end{figure}
\begin{figure}
 \begin{center}
 \includegraphics[width=0.33\columnwidth,clip=]{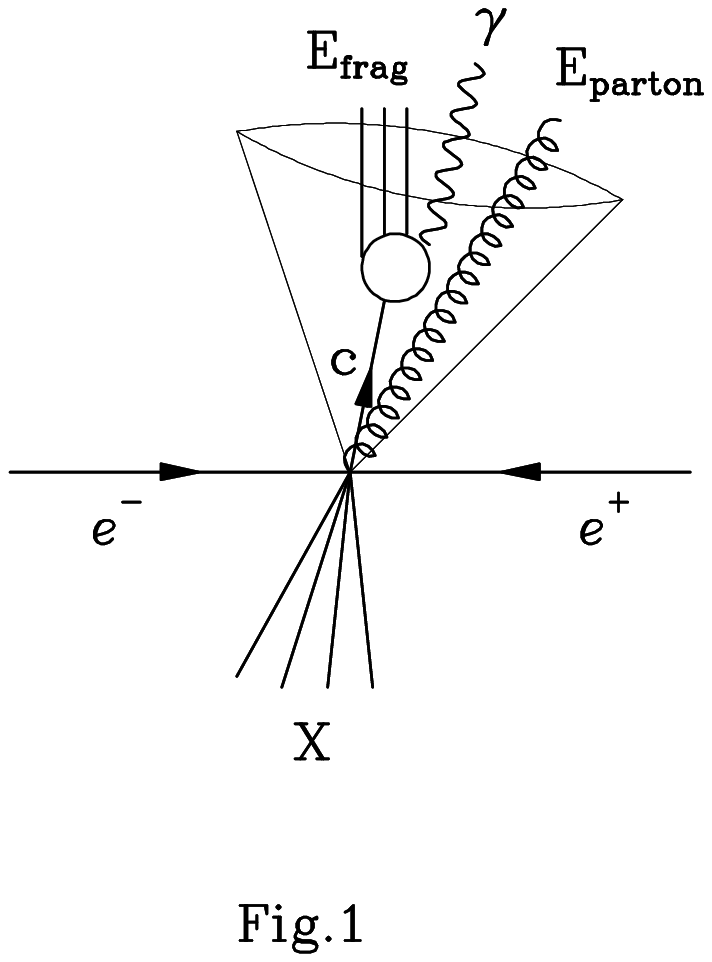}
 \caption[]{
 Illustration of an isolation cone containing a parton $c$ that fragments
 into a photon $\gamma$ plus hadronic energy $E_{\rm frag}$ 
 \protect\cite{kla02}. In addition,
 a gluon enters the cone and fragments giving hadronic energy
 $E_{\rm parton}$. Copyright 2002 by the American Physical Society
 (http://link.aps.org/abstract/RPM/v74/p1221).}
\label{fig:7}
 \end{center}
\end{figure}

The production of prompt photons in association with jets receives
contributions from direct and resolved initial photons as well as direct
and fragmentation contributions in the final state, as shown in Fig.\
\ref{fig:8}.
Photons produced via fragmentation usually lie inside hadronic jets while
directly produced photons tend to be isolated from the final state hadrons.
The theoretical uncertainty coming from the non-perturbative fragmentation
function can therefore be reduced if the photon is isolated in phase space.
At the same time the experimental uncertainty coming from photonic decays of
$\pi^0$, $\eta$, and $\omega$ mesons is considerably reduced.
Photon isolation can be achieved by limiting the (transverse)
hadronic energy $E_{(T)}^{\rm had}$ inside a cone of size $R$ around the
photon to
\beq
 E_{(T)}^{\rm had}<\epsilon_{(T)} E_{(T),\gamma} \, \, ,
\eeq
illustrated in Fig.\ \ref{fig:7}.
Recently an improved photon isolation criterion
\beq
 \sum_i E^{\rm had}_{(T),i} \theta(\delta-R_{i}) < \epsilon E_{(T),\gamma}
 \lrr{1-\cos\delta\over 1-\cos\delta_0}\rrr \, \, ,
\eeq
has been proposed, where $\delta\leq\delta_0$ and $\delta_0$ is now the
isolation cone \cite{isolation}. This procedure allows the fragmentation
contribution to vanish in an infrared safe way.

\begin{figure}
 \centering
 \includegraphics[width=0.66\columnwidth]{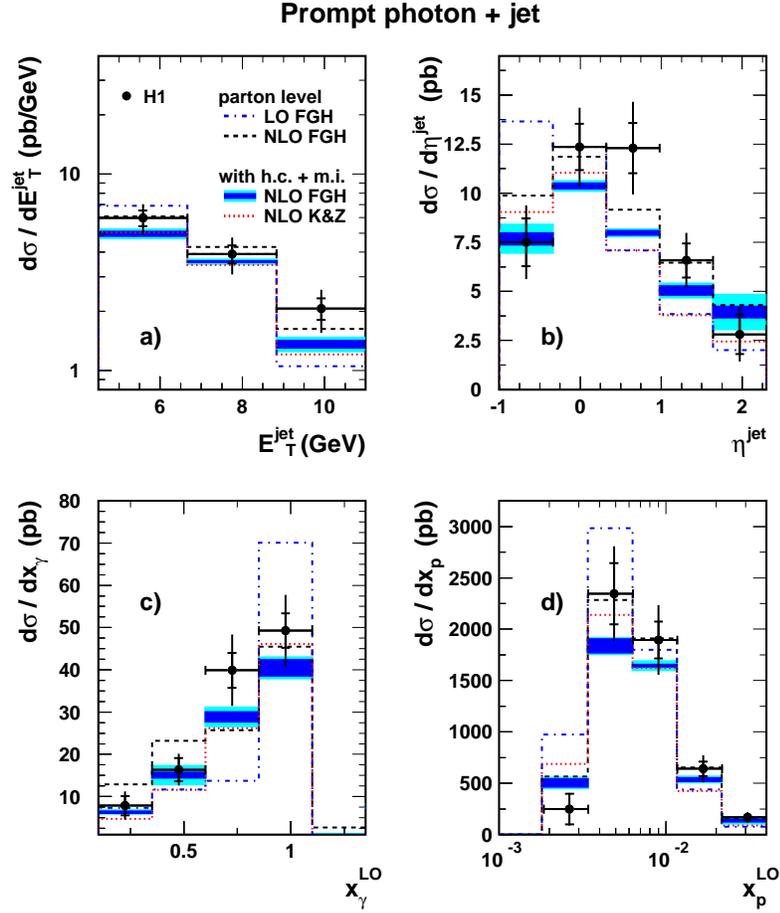}
 \caption[]{Various distributions of photoproduction of prompt
 photons in association with jets as measured by H1 and compared to two
 different QCD calculations.  Reprinted from Ref.~\protect\cite{phojet} 
with permission from Elsevier. }
\label{fig:9}
\end{figure}
Photoproduction of prompt photons and jets has been measured
by the H1 collaboration and compared with two QCD predictions, which differ
in their inclusion of NLO corrections to the resolved and fragmentation
contributions. Only after modeling hadronization corrections and multiple
interactions with Monte Carlo generators, the measured distributions shown
in Fig.\ \ref{fig:9} agree with the QCD predictions, showing the particular
sensitivity of photon final states to hadronic uncertainties.

\subsection{Summary}

Photoproduction processes have been abundantly measured at HERA and
stimulated many theoretical studies, ranging from the
investigation of the foundations of QCD as inscribed in its factorization
theorems, over the determination of its fundamental parameter, the strong
coupling constant, to improvements in our understanding of proton and photon
structure as well as light and heavy hadron formation.

With the shutdown of HERA on July 1, 2007, many questions, in particular in 
the diffractive and non-relativistic kinematic regimes, will remain
unanswered for quite some time until the eventual construction of
a new electron-hadron collider such as eRHIC or an International Linear
Collider. Photon-induced processes in ultraperipheral heavy-ion collisions
may offer a chance to continue investigations in this interesting
field, opening in addition a window to nuclear structure, if these processes
can be experimentally isolated.

\section{UPC lessons from RHIC}
{\it Contributed by: D. d'Enterria, S. R. Klein, J. Seger and S. N. White} 

\subsection{RHIC}
{\it Contributed by: J. Seger}

In the Relativistic Heavy Ion Collider,
counter-rotating beams of fully ionized nuclei collide head on at each of six
locations around the 2.4 mile ring. Particle species ranging from protons 
to gold can be accelerated, stored and collided at RHIC. RHIC can study 
both ``symmetric'' collisions of equal ion species, such as Au+Au, 
and ``asymmetric'' collisions of unequal ion species, such as d+Au. 
Collisions of polarized protons can also be studied. The top energy for 
heavy-ion beams is 100 GeV/nucleon while for protons it is
250 GeV. The design luminosity was $2 \times 10^{26}$
cm$^{-2}$s$^{-1}$~\cite{Harrison:2003sb}. During the initial run in the fall 
of 2000, Au+Au collisions at $\sqrt{s_{_{NN}}}= 130$ GeV were 
studied at the target luminosity of $2 \times 10^{25} $
cm$^{-2}$s$^{-1}$, 10\% design. In 2001, RHIC reached the design energy of 
$\sqrt{s_{_{NN}}} = 200$ GeV with a peak luminosity of $3.7 \times 10^{26}$ 
cm$^{-2}$s$^{-1}$. The 2003 run consisted of 9 weeks of polarized $pp$
collisions and 11 weeks of d+Au collisions while the 2004 run consisted of 12 
weeks of Au+Au collisions at $\sqrt{s_{_{NN}}} = 200$ GeV followed by 1 
week of Au+Au collisions at $\sqrt{s_{_{NN}}}= 62.4$ GeV and 5 weeks of 
polarized $pp$ collisions. The 2005 run included 9 weeks of Cu+Cu collisions
at $\sqrt{s_{_{NN}}} = 200$ GeV and 2 weeks at $\sqrt{s_{_{NN}}} = 62.4$ GeV.
The 2006 run was dedicated to $pp$ collisions.  The 2007 run is another 
Au+Au run.  While peak luminosities have risen
above design values, it was not until the 2004 run that the integrated 
luminosities finally outpaced the design projections, providing an integrated 
luminosity of 1270 $\mu$b$^{-1}$ to STAR in the 12 week Au+Au 
run~\cite{secx:rhic04}.

Studies of mutual Coulomb dissociation by photon-nucleus scattering were made 
with data from three RHIC experiments: PHENIX, PHOBOS and 
BRAHMS~\cite{Chiu:2001ij}.   The
cross section for mutual Coulomb dissociation was found to be comparable to the
geometric cross section, in good agreement with Ref.~\cite{Baltz:1998ex}.
This process may thus be useful for luminosity monitoring.

Ultraperipheral collisions occur with great frequency at RHIC. While the cross
section for coherent interaction is large, these events typically produce 
fewer than 10 charged particles (often only two).  This low multiplicity 
implies significant background, making triggering difficult. For RHIC, the
experimental challenge is to not throw events away: effective triggers
are critical for selecting a reasonable data set. The ability to
trigger on either very low multiplicity events or on nuclei only mildly 
affected by collisions is crucial to successful UPC studies at RHIC.

\subsection{STAR results}
{\it Contributed by: S. Klein and J. Seger}
\label{sec:star}

\begin{figure}[ht]
\centering
 \setlength{\unitlength}{1 cm}
  \includegraphics[totalheight=2 in]{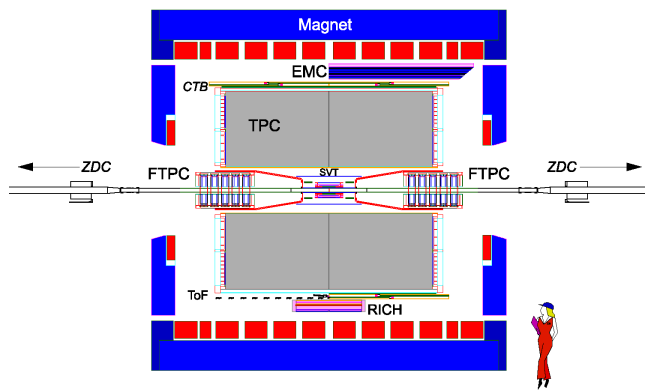}
   \caption[]{A schematic view of the STAR detector at RHIC. 
Reprinted from Ref.~\protect\cite{secx:ackermann03} with permission from
Elsevier.}
  \label{fig:STAR}
\end{figure}

While all four RHIC experiments have shown interest in UPCs,
most of the physics results published to date come from the Solenoidal 
Tracker at RHIC (STAR) experiment~\cite{secx:Ackermann99}. The STAR detector, 
shown in Fig.~\ref{fig:STAR}, tracks charged particles in a 4.2 m long Time 
Projection Chamber (TPC)~\cite{secx:wieman97} with an inner radius of 50 cm 
and an outer radius of 2 m. A
solenoidal magnet surrounds the TPC. In 2000, the TPC was operated in a 0.25 T
magnetic field. In subsequent runs, the magnetic field was operated primarily 
at the design value of 0.5 T with small data sets taken at 0.25 T.
Tracks in the TPC are reconstructed in the pseudorapidity range $|\eta| <
1.5$.  Tracks with $p_T > 100$ MeV/$c$ are reconstructed with high
efficiency. Tracks can be identified by their energy loss in the TPC. 
STAR also has two radial-drift forward TPCs, one on each side of the central
TPC, to extend tracking capabilities into the range of $2.5 < |\eta| < 4.0$.
In addition, STAR is installing a barrel electromagnetic calorimeter covering
$|\eta| < 1$, and an endcap electromagnetic calorimeter on the west pole tip
of the detector covering $1.086 < \eta < 2$.

Each RHIC detector includes two ZDCs, located at $z = \pm
18$ m from the interaction point to detect undeflected neutrons from nuclear
breakup. These calorimeters are sensitive to single neutrons and have an 
acceptance of close to 100\% for neutrons from nuclear
breakup~\cite{Chiu:2001ij,secx:adler01}.

Although each RHIC detector also includes beam-beam counters (BBCs), the 
implementation differs between the experiments.  For STAR, the BBCs are 
a hexagonal scintillator array structure $\pm 3.5$ m from the interaction 
point. They detect charged particles with $2 < |\eta| < 5$ with full coverage 
in $\phi$. Requiring coincidence between the two BBCs can reduce background 
contributions from beam-gas events.  The STAR BBCs were partially implemented 
for the 2001 run and fully implemented for the 2003
run~\cite{Surrow:2002zr}.

\subsubsection{Triggers} \bigskip

The primary STAR trigger detector is a cylindrical Central Trigger Barrel (CTB)
consisting of 240 scintillator slats surrounding the TPC. Each slat in the 
CTB covers a pseudorapidity interval of $\Delta\eta = 0.5$ and a range in 
azimuthal angle $\phi$ of $\Delta\phi = \pi/30$.  The scintillator is
sensitive to charged particles with $p_T > 130$ MeV/$c$ in the 0.25 T magnetic 
field. The acceptance for charged particles depends on where the particles 
are produced.  For a vertex at the center of
the TPC, the trigger barrel is sensitive only to charged particles with
$|\eta|< 1$. The finite acceptance of the CTB therefore limits its usefulness 
since many events of interest (such as $e^+e^-$ pairs) produce tracks at 
high $|\eta|$.

Some ultraperipheral events with zero tracks reaching the CTB can be extracted 
from the STAR minimum-bias data.  The STAR minimum-bias trigger requires 
coincident neutron
signals in the East and West ZDCs. It therefore triggers only on the subset of
ultraperipheral events that include mutual nuclear excitation.  
STAR found that the majority of such events deposited a single neutron in 
each ZDC.  Very few events deposited more than three neutrons into either ZDC.

For those events with tracks that do reach the CTB, STAR can select events 
with a particular topology or multiplicity.  STAR initially focused on 
selecting events with a two particle final state.  The CTB was divided in four
azimuthal quadrants for this `topology' trigger~\cite{secx:bieser03}. Single 
hits were required in the opposite side quadrants while the top and bottom 
quadrants acted as vetoes to suppress cosmic rays.
The topology trigger did not place any requirement on ZDC signals. Analysis 
of the ZDC signal in events selected by the topology trigger shows that, in 
almost all cases, both ZDC's are empty. Datasets with this trigger therefore 
consist primarily of events with no nuclear excitation. To extend the 
triggering capabilities to events with more than two
tracks in the final state requires modification of the trigger algorithms.

\subsubsection{$e^+e^-$ production} \bigskip

Exclusive $e^+e^-$ pair production has been observed in ultraperipheral 
collisions at RHIC~\cite{Adams:2004rz,secx:morozov03}.  The two tracks are 
approximately back-to-back in the transverse plane due to the small $p_T$ 
of the pair, $\sim$ 5 MeV/$c$.  The maximum cross section is at low invariant 
mass with a peak at small forward angles. Many of the tracks have such low 
transverse momentum that, even in a 0.25 T field, they do not reach the CTB.
Thus triggering is limited to the minimum bias
trigger: $e^+e^-$ pair production with mutual nuclear excitation.

The STAR data is compared to two different calculations of the pair production
probability, $P_{ee}(b)$. The first
calculation uses the equivalent photon approximation \cite{Baur:2001jj}. 
The photon flux is calculated from each nucleus using the 
Weizs\"{a}cker-Williams approach. The
photons are treated as if they were real~\cite{secx:baur91}.  Then $e^+e^-$ 
pair production is calculated to leading order~\cite{Budnev:1974de}. 
The $p_T$ spectrum of a photon with energy $\omega$ is given
by~\cite{Klein:2000aa,Vidovic:1992ik}
\begin{equation}\label{eq:photonpT}
\frac{dN_\gamma}{d^2p_T d\omega} = \frac{Z^2 \alpha^2 |p_T|^2}{\pi^2} 
\left[ \frac{F(p_T^2+\omega^2/\gamma^2)}{p_T^2+\omega^2/\gamma^2} 
\right]^2 
\end{equation}
where $\alpha$ is the electromagnetic coupling constant, $Z$ is the nuclear 
charge and $F$ is the nuclear form factor.  The calculation of the form factor
uses a Woods-Saxon nuclear density distribution
with $R_{\rm Au} = 6.38$ fm and skin thickness of 0.535 fm~\cite{Klein:1999qj}.

\begin{figure}[ht]
\vspace{0.2 in}
  \centering
  \includegraphics[totalheight=2 in, clip]{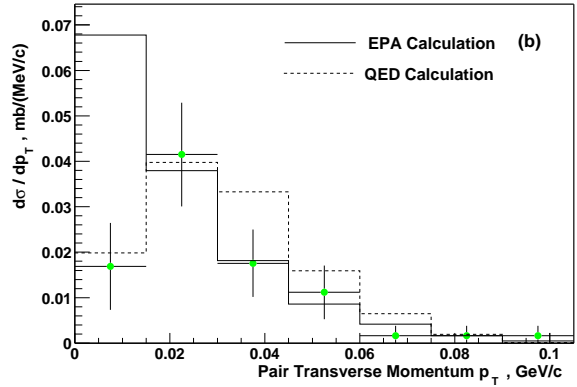}
  \caption[]{The $e^+ e^-$ pair $p_T$ distribution.  The data (points) are
  compared with the EPA (solid histogram) and LO QED (dashed histogram)
  predictions.  The error bars include both statistical and systematic 
  errors ~\protect\cite{Adams:2004rz}.  Copyright 2004 by the American Physical
  Society  (http://link.aps.org/abstract/PRC/v70/e301902).}
  \label{fig:ee_pt}
\end{figure}

The second calculation is a LO QED pair production calculation
\cite{Alscher:1996gn}.  This calculation includes the photon virtuality.  
Within the measured kinematic range, the results differ mainly in the pair
$p_T$ spectrum.  Figure~\ref{fig:ee_pt} compares the $p_T$ distributions
of the two calculations with the data.  The QED calculation is in much
better agreement.

In the ${\rm Au \, Au} \rightarrow {\rm Au}^* \, {\rm Au}^*
\, e^+e^-$ analysis, STAR identifies 52 UPC $e^+e^-$
pairs in an 800,000 event sample at $\sqrt{s_{_{NN}}} =200$ GeV with the
0.25 T magnetic field setting.  Within the limited kinematic range, STAR 
measures a cross section of $\sigma= 1.6 \pm 0.2 \pm 0.3$ 
mb~\cite{Adams:2004rz},
1.2$\sigma$ lower than the equivalent photon prediction of 2.1 mb and close to
the QED calculations $\sigma_{\rm QED}= 1.9$ mb.  The $e^+ e^-$ measurement 
can be used to put limits on changes in the cross section due to higher 
order corrections. At a 90\% confidence level, higher order corrections to 
the cross section must be within the range $-0.5\sigma_{\rm QED} 
< \Delta \sigma < 0.2\sigma_{\rm QED}$.

A study of $e^+e^-$ production in sulfur-induced fixed-target heavy-ion 
collisions at $\sqrt{s_{_{NN}}}= 20$ GeV found that the positrons had a higher 
average energy than the electrons~\cite{secx:vane92}.  This difference may 
be explained by Coulomb corrections~\cite{secx:vane94} since electrons are 
attracted to nuclei while positrons
are repelled.  Calculations show that interference between the leading
(two-photon) and the next-to-leading
\begin{figure}[ht]
\vspace{0.2 in}
  \centering
  \includegraphics[totalheight=2 in,clip]{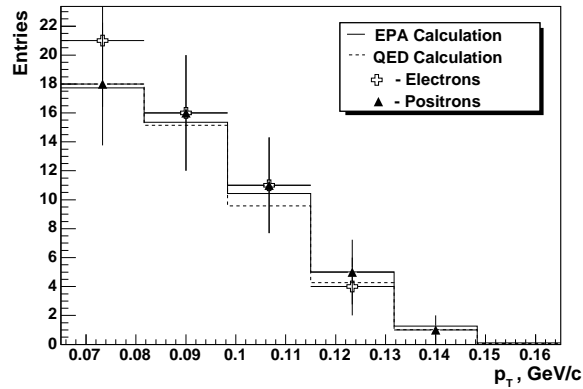}
  \caption[]{The $p_T$ spectra of produced electrons (open crosses) and 
   positrons (solid triangles).  The solid histogram shows the EPA 
  calculation~\protect\cite{Adams:2004rz}.  Copyright 2004 by the American
  Physical Society  (http://link.aps.org/abstract/PRC/v70/e301902).}
  \label{fig:e+e-}
\end{figure}
order (three-photon) channels can lead to 30-60\% asymmetries in some kinematic
variables~\cite{Brodsky:1968}.  Figure~\ref{fig:e+e-} 
compares the $p_T$ spectra of produced electrons and positrons at RHIC.  No 
large asymmetry is seen, indicating that it is not yet necessary to invoke 
higher-order terms.

\subsubsection{$\rho^0$ production} \bigskip

Photonuclear $\rho^0$ production has also been observed at
RHIC~\cite{Adler:2002sc}. Events with no nuclear excitation are observed in 
the STAR topology-triggered data.  Such events are labeled $(0n,0n)$ to 
indicate that there are no neutrons in either ZDC. Events involving nuclear 
excitation in addition to $\rho^0$ production can be observed in the STAR 
minimum bias data.  These events are
labeled $(Xn,Xn)$ to indicate that at least one neutron was deposited into 
each ZDC.

\begin{figure}[!ht]
\centering
 \includegraphics[width=6cm]{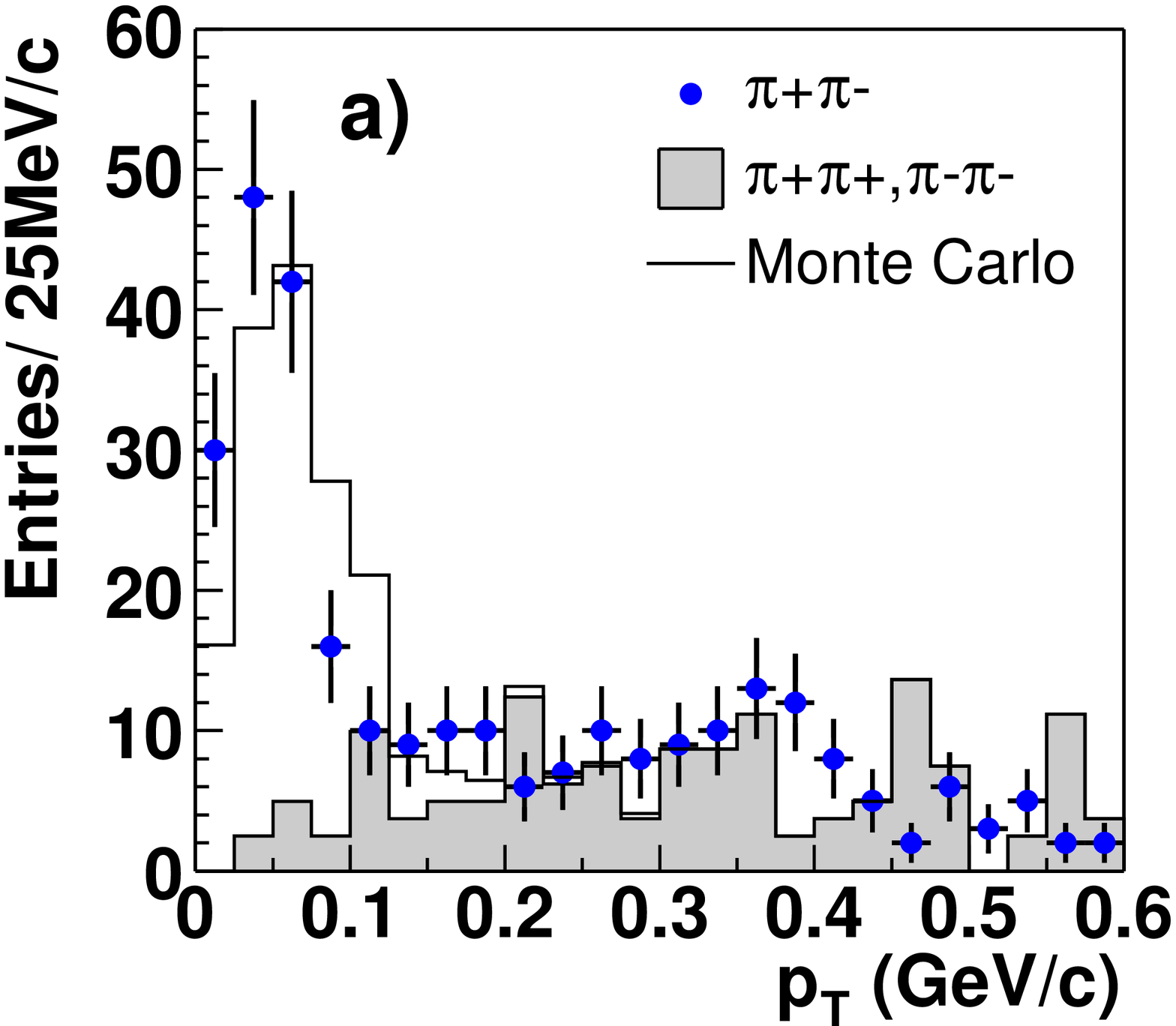}
 \includegraphics[width=6cm]{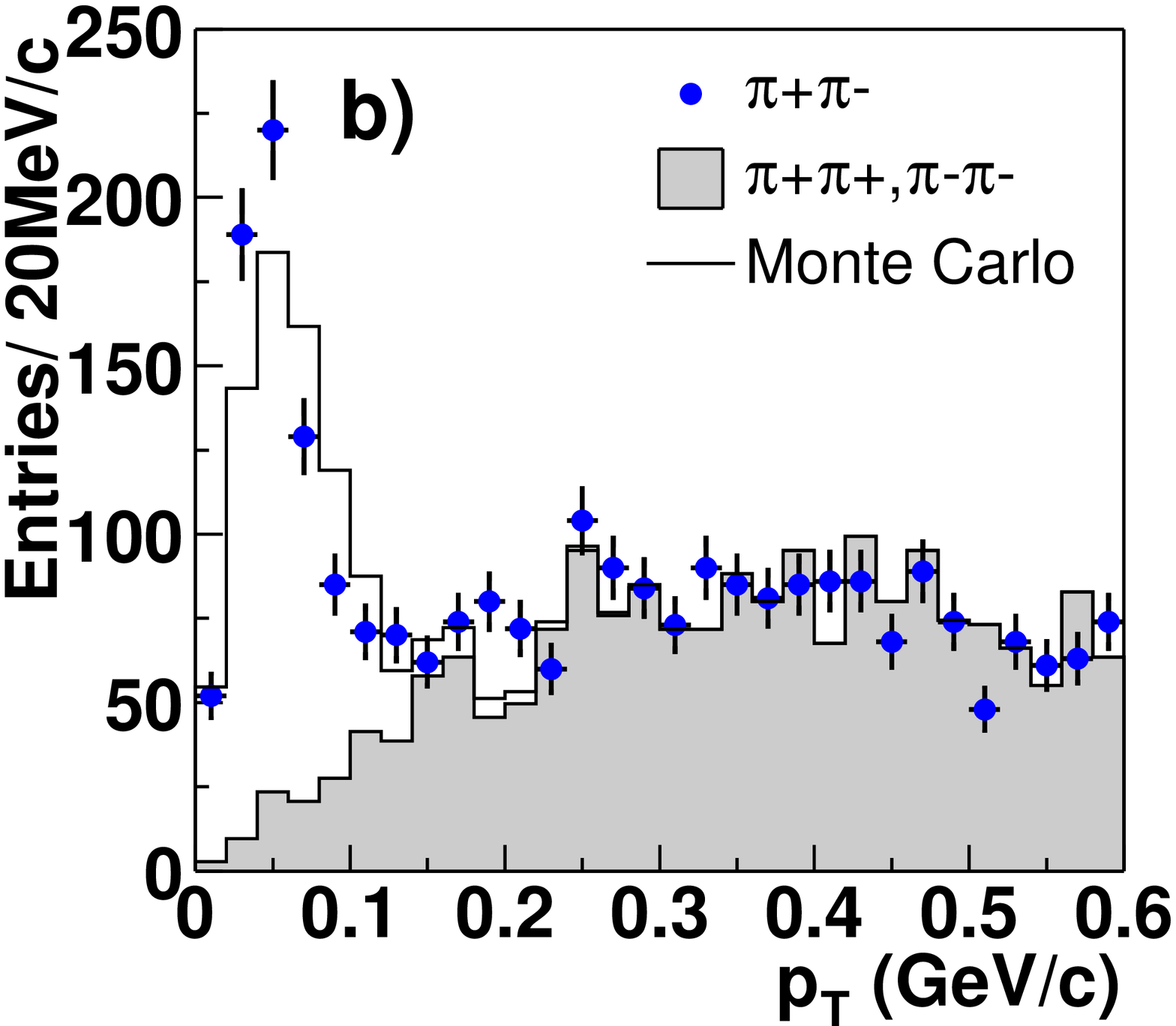}
\caption[]{ The $p_T$ spectra of pion pairs for two-track events selected by 
the STAR (a) topology ($0n,0n$) and (b) minimum bias
($Xn,Xn$) triggers at $\sqrt{s_{_{NN}}} = 130$ GeV. 
The points are $\pi^+ \pi^-$ pairs and the 
shaded histograms are the normalized like-sign combinatorial background.  
The open histograms are the simulated
$\rho^0$ spectra superimposed on the background~\protect\cite{Adler:2002sc}.
Copyright 2002 by the American Physical Society 
(http://link.aps.org/abstract/PRL/v89/e272302).}
\label{fig:rho_ptandm}
\end{figure}

Figure~\ref{fig:rho_ptandm} shows the transverse momentum spectrum of 
$\pi^+ \pi^-$ pairs (points) in Au+Au collisions at 
$\sqrt{s_{_{NN}}} = 130$ GeV.  A clear peak at $p_T < 150$ MeV$/c$, 
the signature for coherent coupling, can be 
observed.  The like-sign combinatorial background (shaded histogram),
normalized to the signal for $p_T > 250$ MeV/$c$, does not show such a peak. 
The open histogram is a Monte Carlo simulation~\cite{Klein:1999qj} for 
coherent $\rho^0$ production accompanied by nuclear breakup superimposed on 
the background. The simulation, including not only the nuclear form factor
but also the photon $p_T$ distribution and the interference of production 
amplitudes from both gold nuclei, matches the data reasonably well.

\begin{figure}[ht]
  \centering
  \includegraphics[totalheight=2 in]{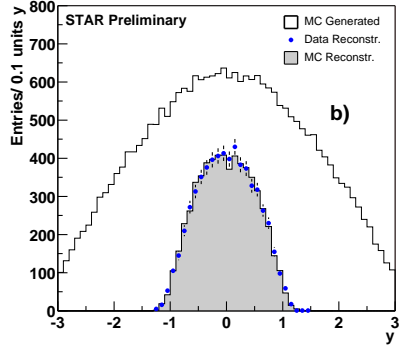}
  \caption[]{The STAR $\rho^0$ minimum bias ($Xn,Xn$) rapidity distribution 
data (points) compared to the normalized reconstructed (shaded histogram) and
generated (open histogram) simulated events
at $\sqrt{s_{_{NN}}} = 200$ GeV.  Reprinted from 
Ref.~\protect\cite{Meissner:2003me} with permission from Elsevier.}
  \label{fig:rho_rapidity}
\end{figure}

The Monte Carlo simulation also closely matches the observed rapidity 
distribution, shown in Fig.~\ref{fig:rho_rapidity} for the $\sqrt{s_{_{NN}}} =
200$ GeV Au+Au data.
The STAR exclusive $\rho^0$ acceptance is about $40\%$ for $| y_{\rho}| < 1$.
Above $| y_{\rho}| = 1$, the acceptance is small and this region is excluded
from the analysis.  The cross sections are extrapolated to the full $4\pi$ 
by Monte Carlo.  For coherent $\rho^0$ production at $\sqrt{s_{_{NN}}}
= 130$ GeV accompanied by mutual nuclear break-up ($Xn,Xn$), the measured
cross section is $\sigma({\rm Au Au} \rightarrow {\rm Au}^*_{Xn}{\rm Au}^*_{Xn}
\rho^0) = 28.3\pm2.0\pm6.3$ mb. By selecting single neutron signals in both 
ZDCs, STAR obtains $\sigma({\rm Au Au} \rightarrow {\rm Au}^*_{1n}
{\rm Au}^*_{1n}\rho^0) = 2.8\pm0.5\pm0.7$ mb.  These
cross sections are in agreement with calculations~
\cite{Klein:1999qj,Baltz:1998ex}.

\begin{figure}[ht]
  \centering
  \includegraphics[totalheight=2 in]{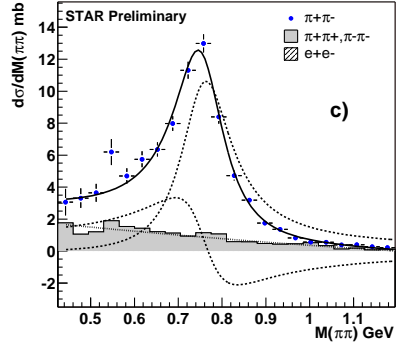}
  \caption[]{The $d\sigma_{{\rm Au \, Au} \rightarrow {\rm Au}^*
{\rm Au}^* \rho^0}/dM_{\pi\pi}$ spectrum for two-track ($Xn,Xn$) events with
$p_T  <$ 150 MeV/$c$ in the STAR minimum bias data.  The shaded
  histogram is the combinatorial background. The hatched histogram includes an
additional contribution from coherent $e^+e^-$ pairs.  The solid curve is the 
sum of a Breit-Wigner (dashed curve), a mass-independent contribution from 
direct $\pi^+\pi^-$ production and interference (dotted curve), 
and a second order polynomial for the residual background (dot-dashed 
curve).  Reprinted from 
Ref.~\protect\cite{Meissner:2003me} with permission from Elsevier.}
  \label{fig:rho_massfit}
\end{figure}

Figure~\ref{fig:rho_massfit} shows the 200 GeV $d\sigma/dM_{\pi\pi}$ 
spectrum for events with $p_T < 150$ MeV/$c$ (points).
The fit (solid curve) is the sum of a relativistic Breit-Wigner for 
$\rho^0$ production and a S\"{o}ding interference term for direct 
$\pi^+\pi^-$ production~\cite{secx:soding66}. A second-order polynomial 
(dash-dotted) describes the combinatorial
background (shaded histogram) from grazing nuclear collisions and incoherent
photon-nucleon interactions.The $\rho^0$ mass and width are consistent 
with accepted values~\cite{secx:groom00}.  Alternative parameterizations such as
a modified S\"{o}ding parametrization~\cite{Breitweg:1997ed} and a 
phenomenological Ross-Stodolsky parametrization~\cite{secx:ross66} yield 
similar results. Incoherent $\rho^0$ production, where a photon interacts with 
a single nucleon, yields high $p_T$ $\rho^0$.  The small number of $\rho_0$
that survive the low-$p_T$ cut are indistinguishable from
the coherent process. A coherent two-photon background, 
${\rm Au \, Au} \rightarrow {\rm Au}^*{\rm Au}^*l^+l^-$, contributes mainly 
at low invariant mass, $M_{\pi\pi}
< 0.5$ GeV/$c^2$.  A second order polynomial models these residual background
processes.

\begin{figure}[ht]
  \centering
  \includegraphics[totalheight=2 in]{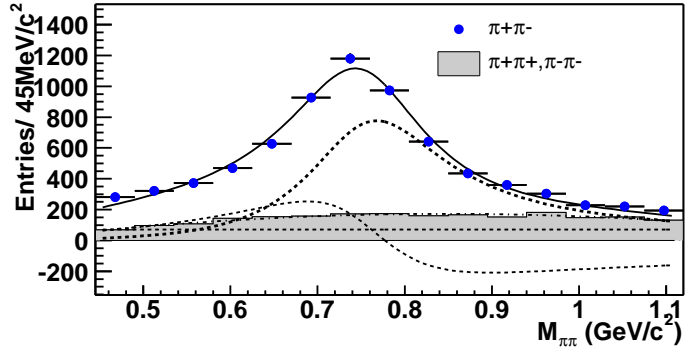}
  \caption[]{The $dN/dM_{\pi\pi}$ invariant mass distribution for two-track 
events in $\sqrt{s_{_{NN}}}=200$ GeV d+Au collisions at STAR
\protect\cite{Meissner:2003me}.  No $p_T$ cut is applied to the d+Au data.
Reprinted from 
Ref.~\protect\cite{Meissner:2003me} with permission from Elsevier.}
  \label{fig:dAu_massfit}
\end{figure}

The 2003 d+Au run also yielded $\rho^0$ events. These asymmetric collisions 
involve two distinct processes, depending on whether
the gold or the deuterium emits the photon: ${\rm Au} \rightarrow \gamma {\rm 
Au}$ followed by $\gamma {\rm d} \rightarrow {\rm d} \rho^0$ or
${\rm d } \rightarrow 
\gamma {\rm d}$ and $\gamma {\rm Au} \rightarrow {\rm Au} \rho^0$. Photon
emission is much more likely from the gold nucleus.  When a photon scatters 
from  the gold scatters from off the deuteron, it may scatter coherently, 
leaving the deuterium intact or incoherently,
dissociating the deuterium.  About 10$^6$ d+Au collisions were recorded 
using the STAR topology trigger with a subsample that required a neutron 
signal in the ZDC.  Requiring a neutron signal in the ZDC cleanly separates 
incoherent reactions in which the deuterium breaks up.  
Figure~\ref{fig:dAu_massfit} shows a clear $\rho^0$
signal in the preliminary $M_{\pi\pi}$ invariant mass spectrum. Further 
analysis is underway.

\begin{figure}[tb]
\centering
\includegraphics[width=6cm]{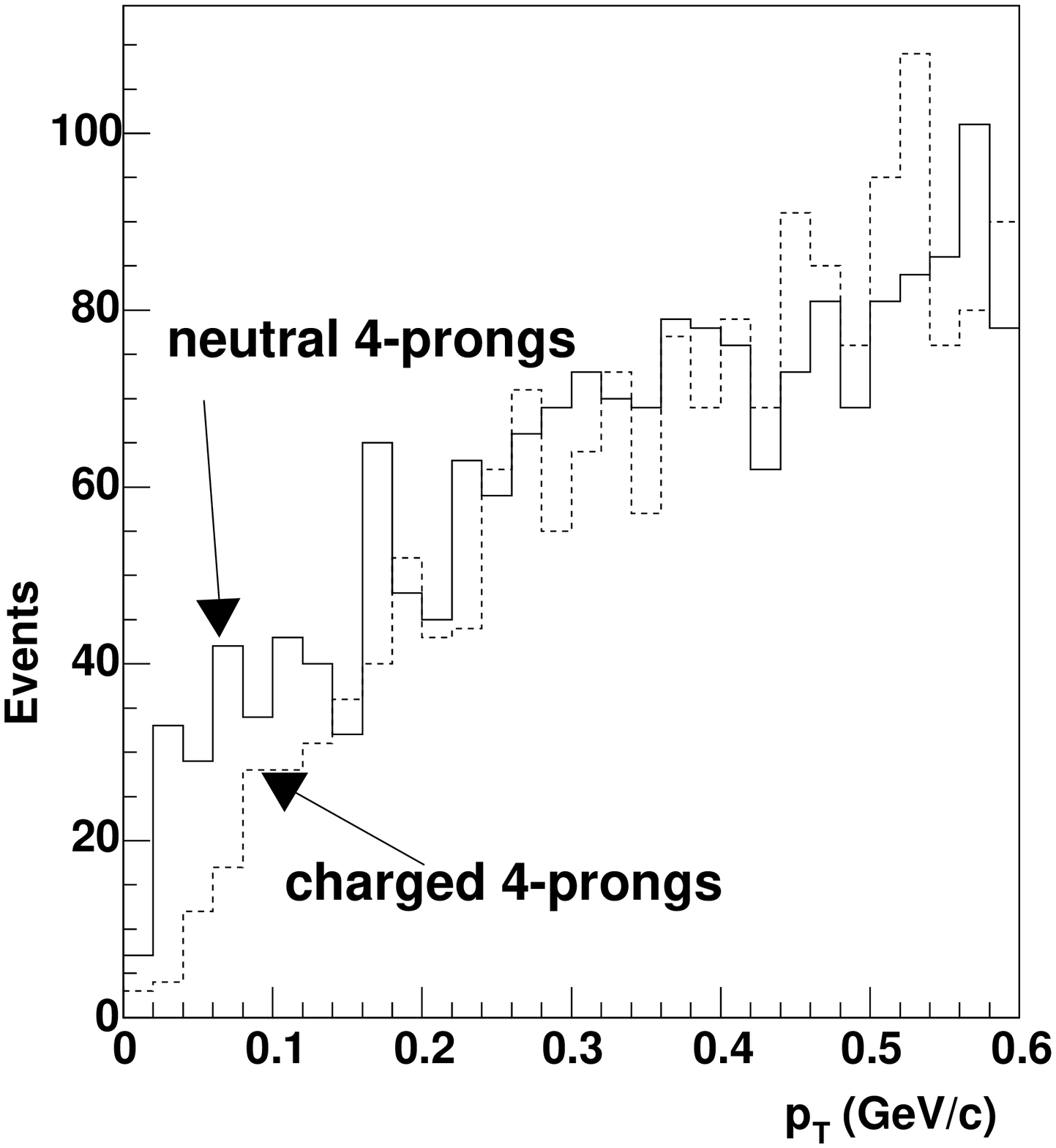}
\includegraphics[width=6cm]{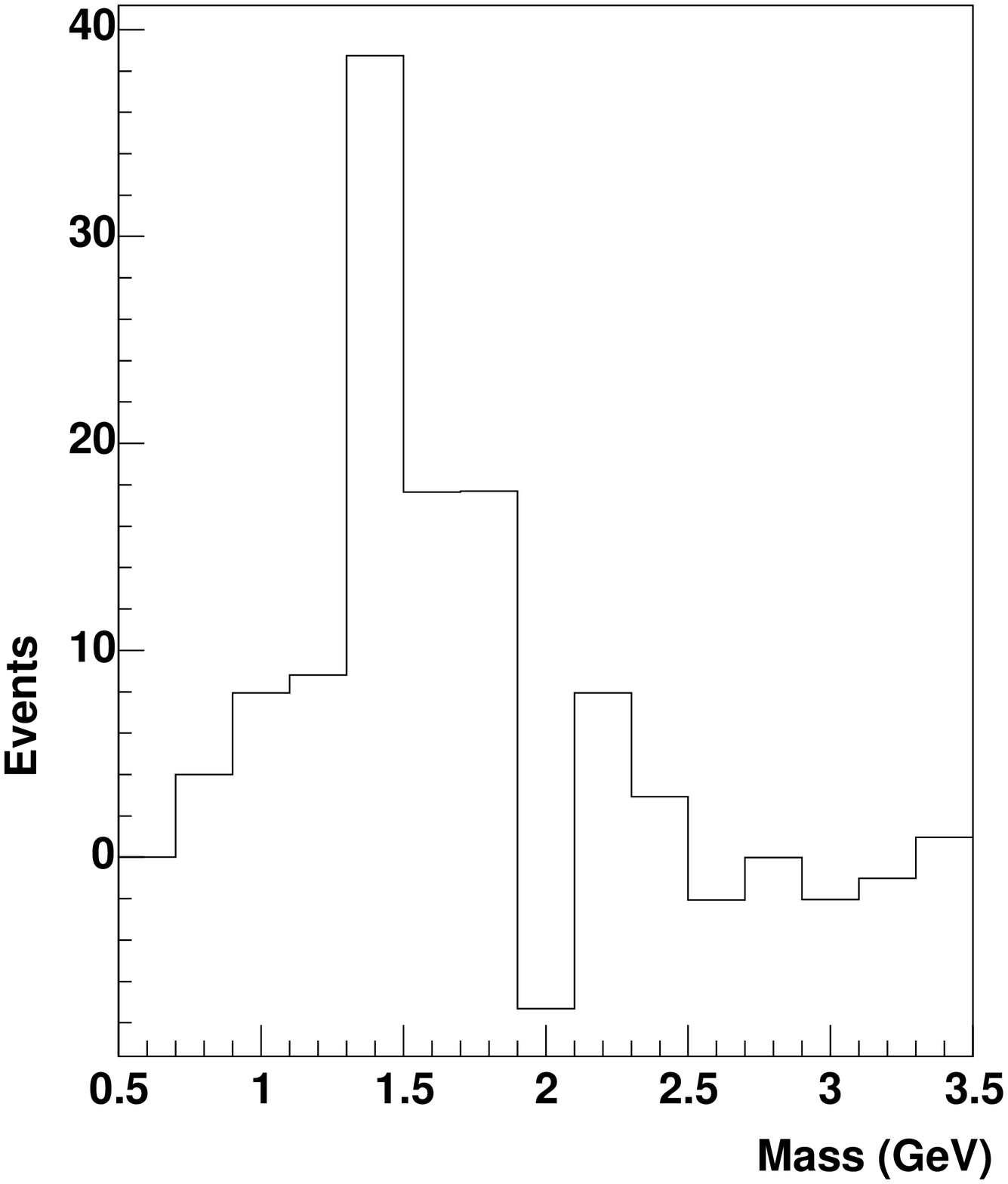}
\caption[]{Left-hand side: The $p_T$ spectrum for 4-prong UPC final states.  
The solid histogram is for neutral combinations ({\it e.g.} 
$\pi^+\pi^-\pi^+\pi^-$) while the dashed histogram indicates 
charged combinations ({\it e.g.} $\pi^+\pi^+\pi^+\pi^-$).  
The difference is the net coherent photoproduction
signal.  Right-hand side:  The background-subtracted mass spectrum of 4-prong 
coherent production, treating all charged particles as pions. From 
Ref.~\protect\cite{spendis}.  Copyright 2005 by the American Institute of
Physics.}
\label{fig:4prong}
\end{figure}

STAR has also observed  photoproduction of four-pion final states.
Fig.~\ref{fig:4prong} shows an excess of zero net charge four-prong final
states such as $\pi^+\pi^-\pi^+\pi^-$ at low $p_T$ in $\sqrt{s_{_{NN}}} = 200$
GeV Au+Au collisions, as expected for coherent 
photoproduction.  
Since no particle identification is applied the particles are assumed to be
pions.  No excess is seen for finite net-charged final states.
The four-pion mass is peaked around $\sim 1.5$ GeV$/c^2$, as also shown in 
Fig.~\ref{fig:4prong}, consistent with a
$\rho^{0 \prime}$ decaying to four pions.

\subsubsection{Interference}
\label{star-interference} \bigskip

Stringent event selection criteria were 
used to select a clean, low-background sample~\cite{Klein:2004kq} from the
STAR 200 GeV Au+Au data to study $\rho^0$ interference. 
The magnitude of the interference depends on the 
ratio of the amplitudes for $\rho^0$ production from the two nuclei.  Away 
from $y =0$, the amplitudes differ and the interference is reduced. Thus 
this analysis focuses on the midrapidity region. We neglect $\rho^0$ 
candidates with $|y| < 0.1$ to avoid possible cosmic ray
contamination. A Monte Carlo is used to calculate the expected
interference for different rapidity ranges ~\cite{Klein:1999qj,Klein:2000aa}.

\begin{figure}[ht]
  \centering
  \includegraphics[totalheight=2 in]{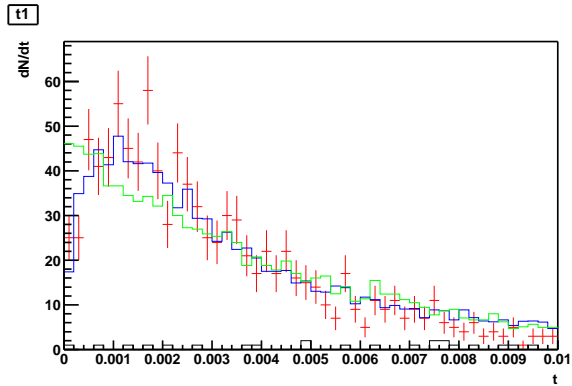}
  \caption[]{The raw (uncorrected) STAR $t_\bot$ spectrum for the $\rho^0$ 
topology sample in $0.1 < |y| <  0.5$.  The points are the data.  The dashed 
histogram is a simulation including interference while the dot-dashed 
histogram is a calculation without interference.  The solid histogram 
with $dN/dt \sim 0$ is the like-sign background \protect\cite{Klein:2004kq}.}
  \label{fig:interf_t}
\end{figure}

STAR studies the $p_T$ spectra using the variable $t_{\bot} = p_T^2$. At RHIC
energies, the longitudinal component of the 4-momentum transfer is small so 
that $t \approx t_{\bot}$. Without interference, $dN/dt\propto
\exp(-bt)$~\cite{Klein:2000aa,Alvensleben:1970aa}.  
Figure~\ref{fig:interf_t} compares
the uncorrected minimum bias data for $0.1 < | \eta | < 0.5$ with two
simulations, with and without interference. Both simulations include the 
detector response. The data has a significant downturn for $t < 0.001$ 
GeV$^2$, consistent with $\langle b \rangle =18$ fm expected for a 
$\rho^0$ accompanied by mutual excitation~\cite{Baur:2003ar}. These data 
match the calculation that includes
interference but not the one without interference.

\begin{figure}[!ht]
\hspace*{-0.2cm}
\centering
\begin{minipage}{12cm}
 \includegraphics[width=6cm,height=6cm]{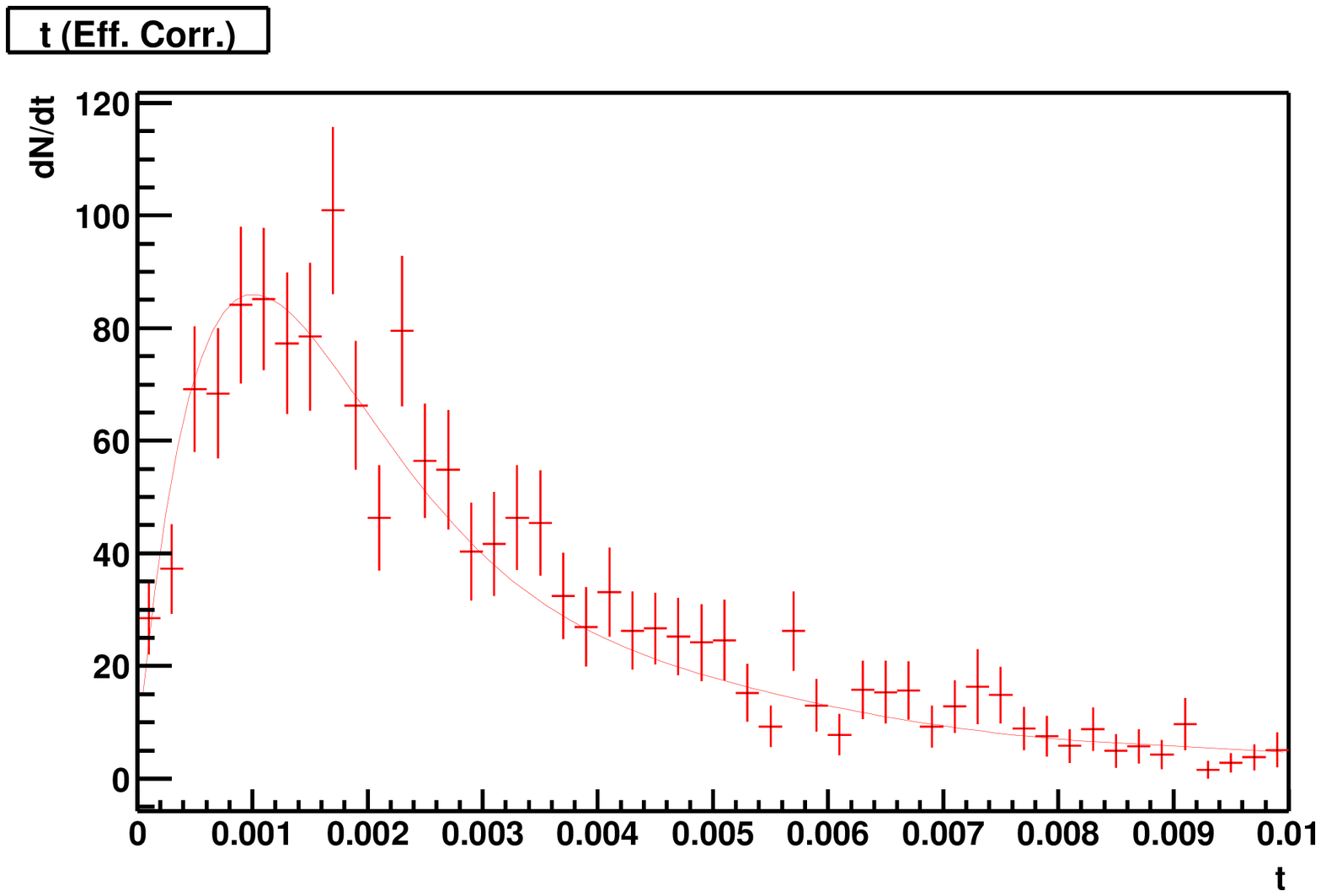}
\hspace*{-0.2cm}
 \includegraphics[width=6cm,height=6cm]{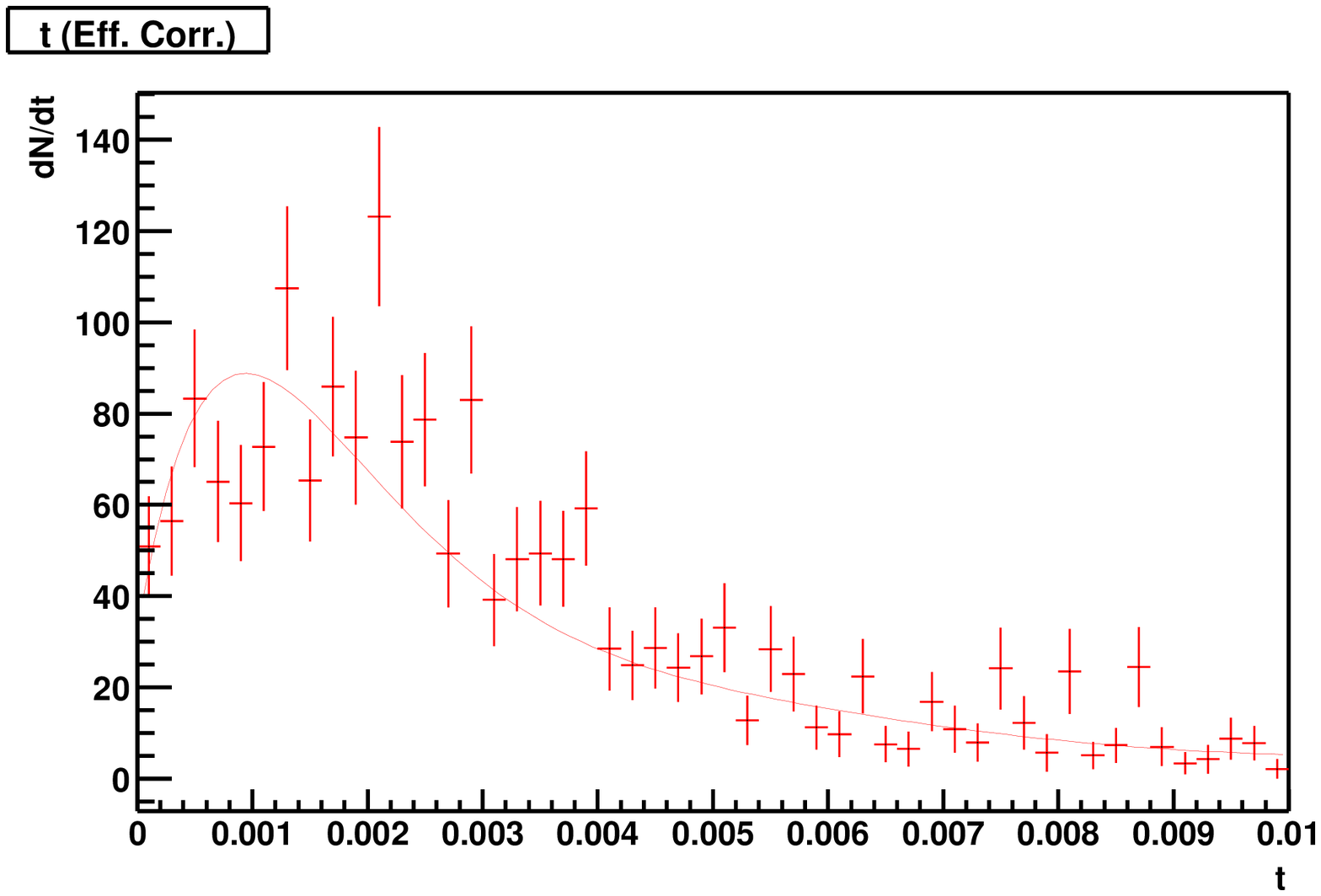}
 \includegraphics[width=6cm,height=6cm]{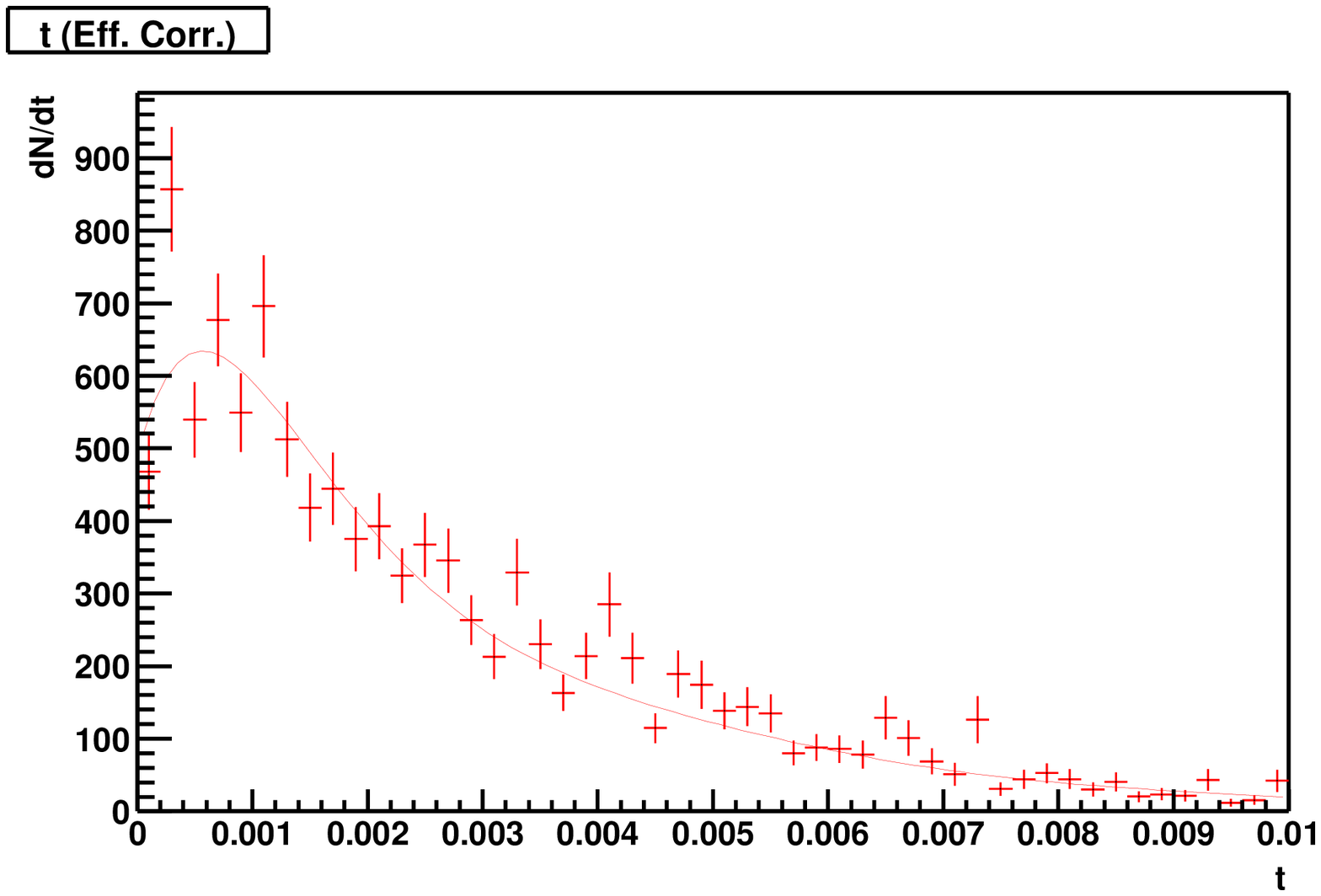}
  \includegraphics[width=6cm,height=6cm]{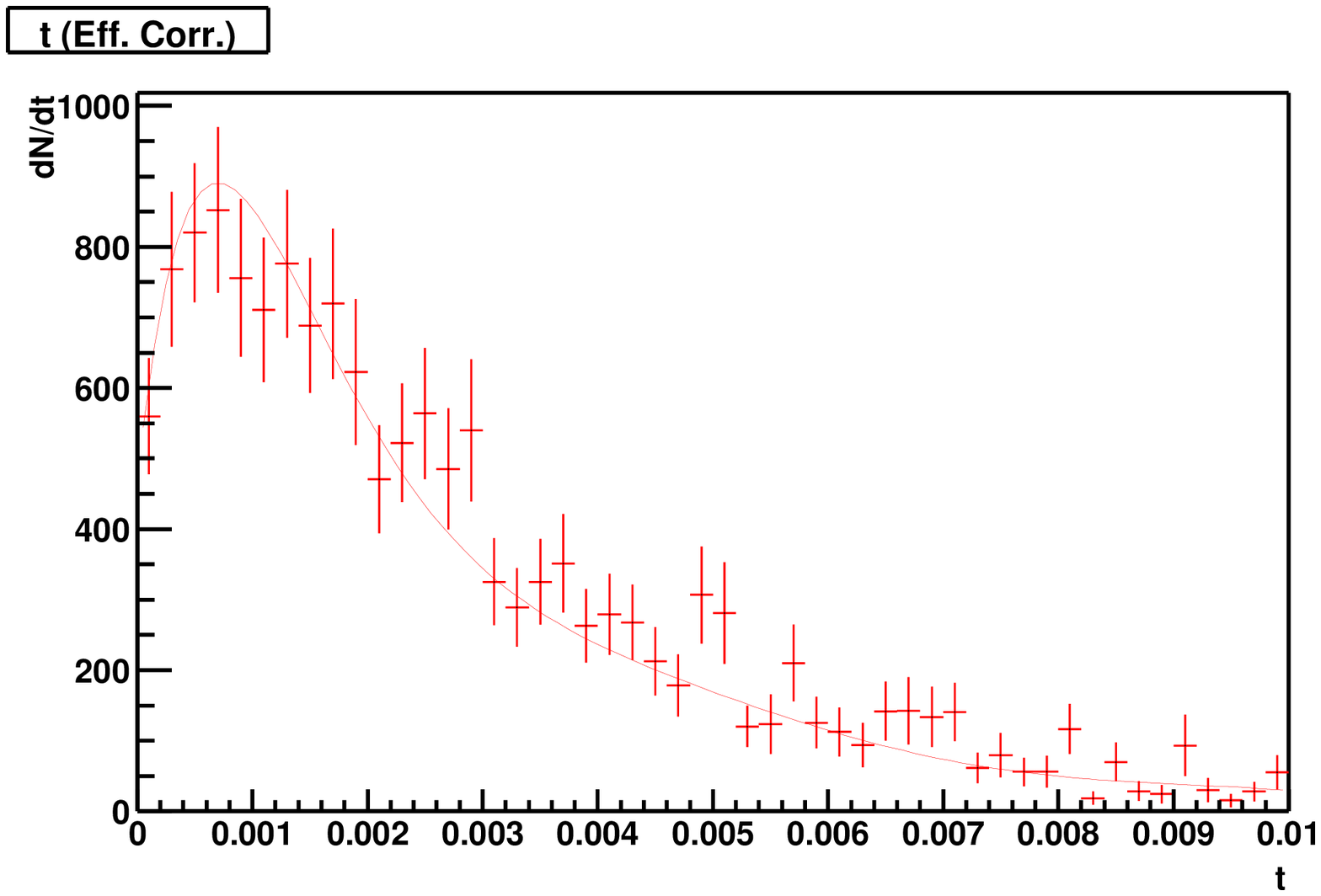}
  \end{minipage}
\setlength{\unitlength}{0.5 in}
  \begin{picture}(0,0)
  \put(-8,3){\shortstack[l]{(a)\\STAR Preliminary \\Min. Bias \\0.1 $< | y| <$ 0.5}}
  \put(-8,-2){\shortstack[l]{(c)\\STAR Preliminary \\Topology \\0.1 $< | y| <$ 0.5}}
  \put(-3,3){\shortstack[l]{(b)\\STAR Preliminary \\Min. Bias \\0.5 $< | y| <$ 1.0}}
  \put(-3,-2){\shortstack[l]{(d)\\STAR Preliminary \\Topology \\0.5 $< | y| <$ 1.0}}
  \end{picture}
\caption[]{The efficiency-corrected $\rho^0$ distributions as a function of
$t_{\bot}$ spectrum for $\rho^0$ from the minimum bias trigger with (a) $0.1 < 
| y| < 0.5$ and (b) $0.5 < | y| < 1.0$ and the topology trigger with (c) $0.1 
< | y| < 0.5$ and (d) $0.5 < | y| < 1.0$.  The points are the data while the 
curve is a fit to Eq.~(\protect\ref{eq:interf}) \protect\cite{Klein:2004kq}.}
 \label{fig:interf_data}
\end{figure}

The efficiency-corrected data are shown in
Fig.~\ref{fig:interf_data}~\cite{Klein:2004kq}.  The minimum bias and 
topology data are shown separately in two rapidity bins: $0.1 < | y | < 0.5$ 
and $0.5 < | y | < 1.0$.  The data is fit with three parameter,
\begin{equation}\label{eq:interf}
\centering
 \frac{dN}{dt} = a \exp(-Bt)[1 + c(R(t)- 1)]
\end{equation}
where $R(t)= {\rm Int}(t)/{\rm Noint}(t)$ is the ratio of the Monte Carlo $t$
distribution with and without interference. The factor $a$ provides an 
overall normalization, $B$ is the slope and $c$ quantifies the interference
effect: $c = 0$ corresponds to no interference, while
$c =1$ corresponds to the expected interference \cite{Klein:2000aa}. 
This functional form separates the interference effect $c$
from the nuclear form factor $B$.

Table~\ref{tab:interf_values} gives the fit results.
\begin{table}[htbp]
  \begin{center}
  \label{tab:interf_values}
  \caption[]{The fit results for the two different triggers in two rapidity 
bins \protect\cite{Klein:2004kq}.}
\vspace{0.4cm}
\begin{tabular}{|c|c|c|c|c|c|} \hline
 Trigger & $|y_{\rm min}|$ & $|y_{\rm max}|$ & $B$ (GeV$^{-2}$) & $c$ 
& $\chi^2$/DOF \\ \hline
 minimum bias & 0.1 & 0.5 & $301 \pm 14$ & $1.01 \pm 0.08$ & 50/47 \\
 minimum bias & 0.5 & 1.0 & $304 \pm 15$ & $0.78 \pm 0.13$ & 73/47 \\
 topology & 0.1 & 0.5 & $361 \pm 10$ & $0.71 \pm 0.16$ & 81/47 \\
 topology & 0.5 & 1.0 & $368 \pm 12$ & $1.22 \pm 0.21$ & 50/47 \\ \hline
\end{tabular}
\end{center}
  \end{table}
At small rapidities the amplitudes are similar and the
interference reduces the cross section at $p_T =0$ more than at larger 
rapidities.  In the minimum bias data, the interference extends to higher 
$p_T$ because it has a smaller average slope, $B$, than the topology
data.

The fit values of $B$ for the minimum bias and topology $\rho^0$ data differ 
by $\sim 20$\%. This difference may be attributed to the different impact 
parameter distributions caused by tagging the minimum bias data by nuclear 
breakup since the photon flux decreases as the inverse square of the impact
parameter, $1/b^2$. When $b$ is a few times $R_A$, the $\rho^0$s are more 
likely to be produced on the side of the target near the photon emitter 
than on the far side.  Thus $\rho^0$ production is concentrated on the 
near side, leading to a smaller effective production volume and a smaller 
$B$ in the minimum bias data since $B \propto 1/b^2$.

The four values of $c$ are consistent within errors.  The weighted average 
is $c =0.93\pm 0.06$.  The preliminary interference measured by STAR is
$93\pm6 \, ({\rm stat.})\pm 8\, ({\rm syst.})\pm 15 ({\rm theory})\%$ of the 
expected~\cite{Klein:2004kq}.

\subsubsection{Proton-nucleus and deuteron-nucleus results} \bigskip

In Au+Au collisions, coherent vector meson
photoproduction has a strong signature since most of the
signal is at $p_T <{\rm few} \, \hbar/R_A\approx 100$
MeV/$c$.

\begin{figure}[tbhp]
  \centering \includegraphics[height=4.5cm]{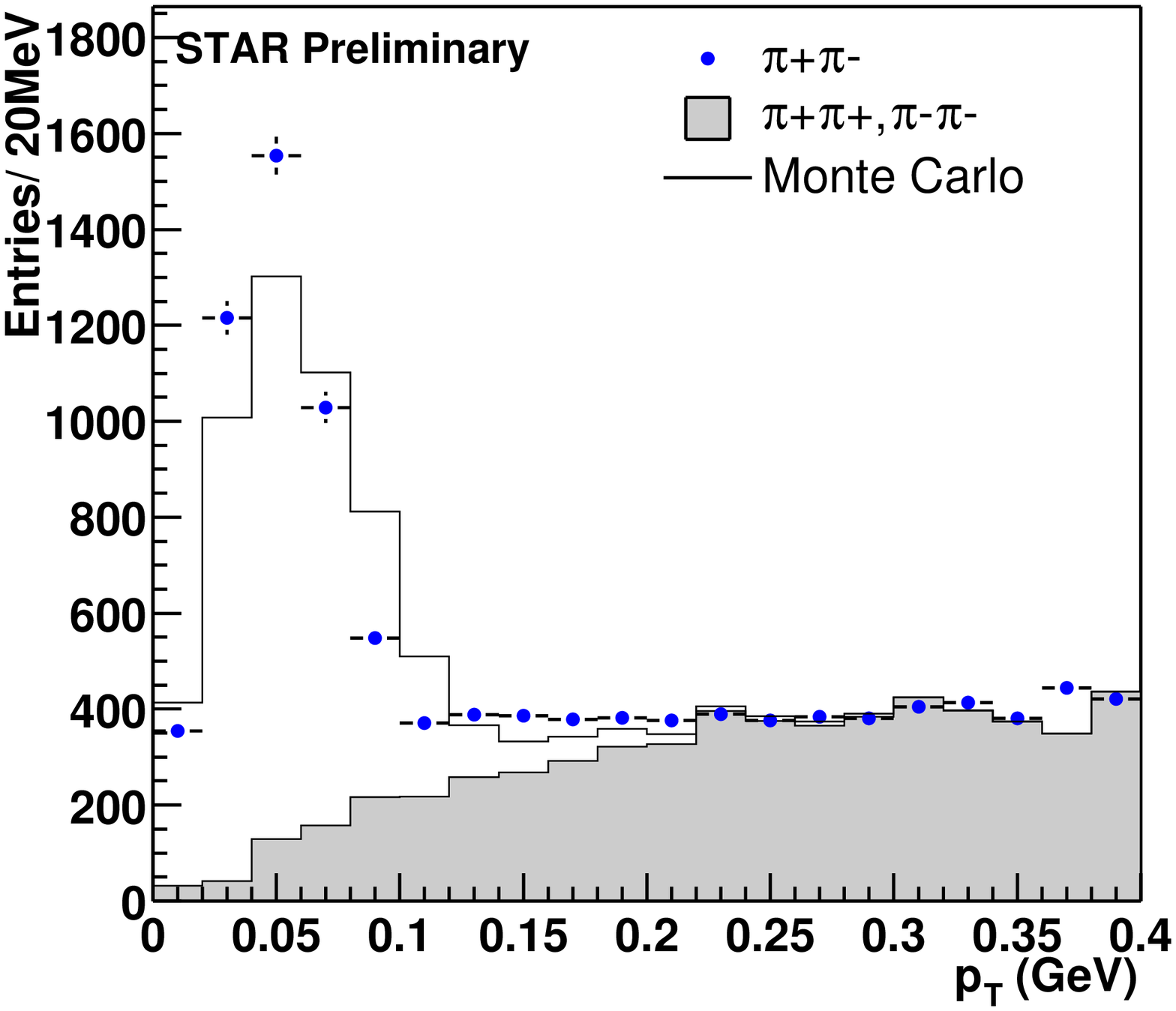}
  \includegraphics[height=4.5cm]{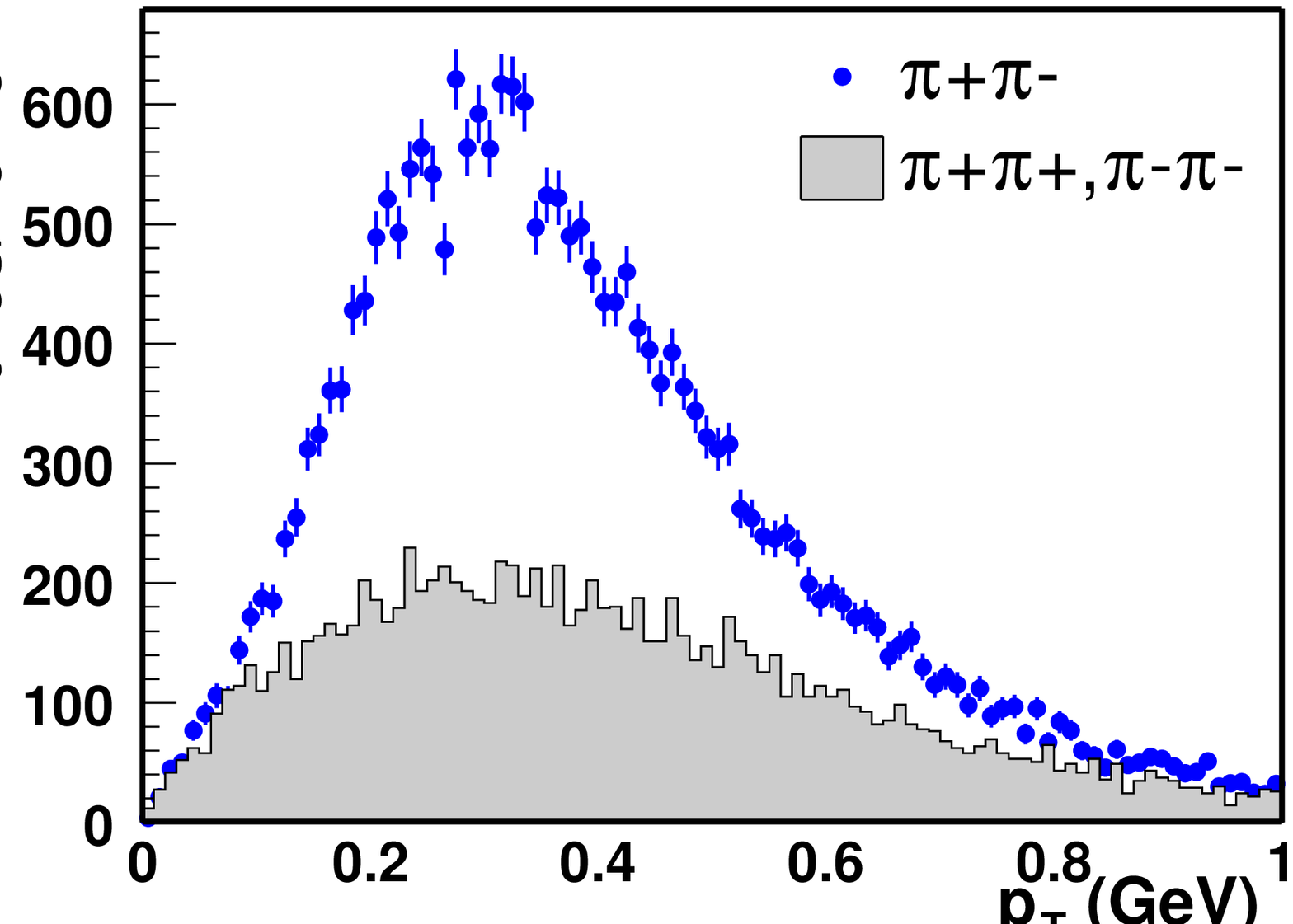} 
\caption[]{The $p_T$
  spectrum of $\rho^0$ photoproduction in Au+Au (left-hand side) relative 
  to d+Au (right-hand side) 
  collisions \protect\cite{STARda}.  The points are the data while the
  shaded histogram is the like-sign background.  On the left-hand side, the 
  like-sign background has been scaled to match the zero net-charge data for
  $p_T>150$ MeV/$c$ to include the background from incoherent $\rho^0$ 
  production as a background.}
\label{fig:STARda}
\end{figure}

In contrast, the deuteron is small and has only two nucleons.
Thus the coherent enhancement is limited and the $p_T$ constraint is
not useful. Nevertheless, by taking advantage of the large solid angle
coverage, a fairly clean $\rho^0$ sample can be isolated \cite{STARda}.  Figure
\ref{fig:STARda} compares the $\rho^0$ $p_T$ spectra from STAR
in Au+Au and d+Au collisions.  
These events were selected using a trigger that required a low charged
multiplicity in the region $|\eta| < 1$ in coincidence with a ZDC
signal that indicated deuteron
dissociation.  This trigger had a reasonably high selectivity but the
deuteron breakup requirement reduced the $\rho^0$ cross sections considerably.

Data was also taken with a trigger that did not require a neutron
signal from deuteron breakup.  This trigger was sensitive to reactions where
the deuteron remained intact and also where the neutron acquired a
large $p_T$ through deuteron dissociation. A $\rho^0$ signal is also
seen with this trigger.  In addition to the $\rho^0$ at large $p_T$, a
small peak is visible at very low $p_T$, consistent with photon
emission from the deuteron.  Both signals (with and without the
neutron) had similar $\pi\pi$ invariant mass spectra.
In addition to the $\rho^0$, direct $\pi^+\pi^-$ production
was measured.  The proportion of direct $\pi^+\pi^-$ to $\rho^0$ production
was comparable to that observed in both Au+Au collisions at RHIC and $ep$
collisions at HERA.  

\subsection{PHENIX Results}
{\it Contributed by: D. d'Enterria and S. N. White} 
\label{phenix_res}

\subsubsection{Introduction} \bigskip

During the 2003 RHIC d+Au run at $\sqrt{s_{_{NN}}}= 200$ GeV, 
PHENIX commissioned new detectors in the forward region.  PHENIX installed a 
proton calorimeter (fCAL) on either side of the outgoing beam pipe near the 
ZDC ($z=\pm 18$ m) and embedded a shower 
maximum position-sensitive hodoscope (SMD) between the first and second ZDC 
modules~\cite{pCal}. 
The fCAL~\cite{pCal} detects protons from deuteron breakup. Since beam-energy 
protons have a lower rigidity than the deuteron beam, protons from 
deuteron breakup are deflected into the fCAL by an accelerator dipole (DX) 
located at $z=\pm 11$ m. 
The SMD recorded the transverse momentum of neutrons interacting in the ZDC by 
measuring the centroid of the shower secondaries. 

The cross section for the deuteron dissociation reaction, d+Au$\rightarrow
n+p+$Au has been calculated \cite{Klein:2003bz} and found to be 1.38 b 
($\pm 5\%$) with 0.14 b due to hadronic diffraction dissociation 
\cite{Glauber} and the rest due to electromagnetic dissociation in the field 
of the gold nucleus. PHENIX measured 
this dissociation cross section with a trigger sample requiring 
$E_{\rm ZDC}\geq 10$ GeV in either ZDC (in the gold or deuteron direction).

During the 2004 RHIC Au+Au run at $\sqrt{s_{_{NN}}} = 200$ GeV, PHENIX 
also commissioned a trigger to study 
high mass $e^+e^-$ pair production in UPCs. Two sources of high 
mass $e^+e^-$ pairs are relevant for this measurement. The high mass 
continuum from $\gamma\gamma\rightarrow e^+e^-$ was measured for 
$M \geq 1.8$ GeV, significantly above the range explored by STAR.  
A $J/\psi\rightarrow e^+e^-$ sample was also observed in 
photoproduction off the Au target~\cite{dde_qm05}. Coherent and 
incoherent $J/\psi$ production cross sections have been 
calculated~\cite{Klein:1999qj,starlight,Baltz:2002pp,Strikman}.  
The contributions to the $J/\psi$
$p_T$ distribution from the two 
processes should be distinguishable due to their different shapes.

The high mass dilepton measurement is interesting because it 
demonstrates the feasibility of triggering on hard photoproduction processes 
with small cross section. Deuteron dissociation, the earliest calculation of 
diffractive dissociation to be found in the 
literature, has a large cross section and is both theoretically and 
experimentally clean. Thus PHENIX used this process to calibrate the cross
sections of other processes produced in d+Au interactions ~\cite{leitch}.
	
\subsubsection{Deuteron diffractive dissociation}

\subsection*{d+Au Cross Sections}
 
In addition to deuteron dissociation, the total d+Au inelastic cross section 
interesting for the RHIC program. The inelastic cross section is sampled in 
the experiments, the ``minimum-bias trigger", for use as a luminosity monitor. 
Once the minimum-bias cross section is known, the cross sections of other 
processes recorded during the same luminosity interval can also be calculated. 

There are two approaches to this cross section normalization. In the first, it 
is derived from known, elementary, $NN$ inelastic cross sections using the
Glauber model with a Woods-Saxon density distribution.
The second approach, adopted by PHENIX, is to directly determine the 
minimum-bias trigger cross section by comparing to the 
reliably-calculated~\cite{Klein:2003bz} 
deuteron dissociation process measured by PHENIX in 2003.

\subsection*{Instrumentation}

The four RHIC experiments have midrapidity spectrometers with different 
characteristics but all share identical ZDCs 
located at $z= \pm 18$ m. The ZDCs cover $\pm 5$ cm in $x$ and $y$ about 
the forward beam direction and have an energy resolution of $\sigma_E/E<21\%$
for 100 GeV neutrons within $x,y \leq 4.5$ cm~\cite{pCal}. Almost all 
non-interacting spectator neutrons are detected in the ZDCs while 
charged particles are generally swept out of the ZDC region by strong (16
Tm) accelerator dipoles at $z =\pm 11$ m.
These dipoles sweep spectator protons from deuteron dissociation beyond the 
outgoing beam trajectory since they have twice the deuteron charge-to-mass 
ratio. In PHENIX, the spectator protons are detected in the fCal~\cite{pCal}.
		
PHENIX used two additional hodoscopes (beam-beam counters or BBCs)~\cite{eBBC},
located at $z =\pm 1.5$ m and covering $3.0\leq |\eta| \leq 3.9$, as the 
main minimum-bias trigger. Events with one or more charged particles hitting 
both the $+z$ and $-z$ BBCs fired this trigger. Determining the BBC
cross section, $\sigma_{\rm BBC}$, is equivalent 
to determining the luminosity for the PHENIX data. All d+Au events 
occurring well within the $z$ interval between the BBCs fire this trigger 
with an $88\pm4\%$ efficiency \cite{eBBC}.  The efficiency decreases for 
$|z_{\rm vertex}|\geq 40$ cm. Thus $\sigma_{\rm BBC}$ was determined using 
only events within the interval $|z_{\rm vertex}| < 40$ cm. A correction was 
then applied to the fraction of all RHIC events within this interval. 
The $z$ distribution of the data can be determined using time-of-flight 
measurements between the ZDCs for events with a north-south coincidence of
the ZDCs (with single event resolution of $\sigma_z\sim 2$ cm).

\begin{figure}
\centering
    \epsfxsize=0.4\hsize
     \epsffile{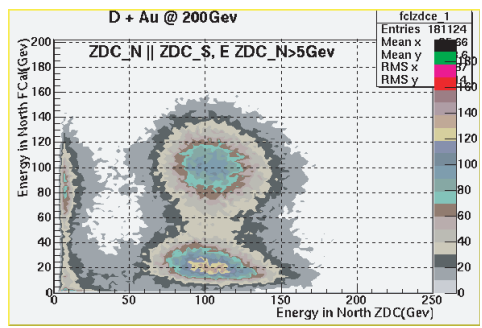}
\caption[]{The energy deposition in the proton calorimeter (fCal) as a 
function of energy deposition in the ZDC (neutron calorimeter) for events 
with some activity in the ZDC in the deuteron beam direction 
\protect\cite{ZDC}. This sample includes absorptive stripping as well as 
d$\rightarrow n p$.  Copyright 2005 by the American Institute of Physics.}
\label{fcalvszdc}
\end{figure}

\subsection*{Data analysis}

Typical event rates were several kHz with $L\sim (1-4)\times10^{28}$ 
cm$^{-2}$s$^{-1}$ 
for all processes considered here. Therefore the analysis was based on a 
representative data sample with 
\begin{eqnarray}
N^{\rm trig} (\mathtt{ BBC}) & = & 2.3 \times 10^5 \\
N^{\rm trig}(\mathtt{ ZDC_{Au}\, .OR. \,ZDC_d}) & = & 4.6 \times 10^5 
\end{eqnarray}
events where the energy deposited in the ZDC is greater than 10 GeV.
The subscripts Au and d represent the
Au or d directions. The second trigger is sensitive to deuteron dissociation, 
characterized by a 100 GeV neutron in ZDC$_{\rm d}$ and a 100 GeV proton 
in the fCal with no activity at midrapidity.
	
Additional data samples were recorded with one of the RHIC beams 
intentionally displaced by up to 1 mm to measure the fraction of triggers 
due to d+Au collisions relative to beam-gas background. The largest 
background was $\leq3\%$.  The quoted rates have been corrected for the 
measured background.  The BBC event yield, corrected for the accelerator 
interaction distribution is
\begin{equation}
N^{\rm corr} (\mathtt { BBC}) = 228634 \pm0.5\% \, \, .
\end{equation}

\subsection*{Deuteron dissociation analysis}

As stated previously, deuteron dissociation events have a clear signature in 
PHENIX. This is illustrated in Fig.~\ref{fcalvszdc} which shows the forward 
(deuteron direction) energy deposited in the neutron and proton calorimeters 
in ZDC-triggered events.
	
The SMD distribution confirmed that neutrons have a small 
angular divergence.  Consequently there is only a small correction for ZDC 
acceptance. Absorptive stripping events where one nucleon in the deuteron 
is absorbed in the target is the main potential background for the 
deuteron dissociation sample. PHENIX used an iterative procedure to extract 
the dissociation event yield, fitting the ZDC$_{\rm d}$ + fCal total energy to 
the sum of $100 +200$ GeV lineshapes and correcting for the calculated 
efficiency as successive cuts on activity in other detectors were applied. 
The first two iterations yield 
$N({\rm d} \rightarrow n+p) = 157149$ and 156951
events, showing that the procedure is clearly stable.

The final result is :
\begin{eqnarray}
\sigma_{\rm BBC} & = & \frac{N^{\rm corr}( \mathtt{BBC})}{N({\rm 
d}\rightarrow n p)} \, \sigma_{{\rm d}\rightarrow n p} 
= \frac{228634}{158761} \, 1.38 \, 
(\pm0.5\%) \, {\rm b} \nonumber \\
& = & 1.99\, (\pm1.6\%\pm5.0\%)\, \rm{b} \, \, ,
\end{eqnarray}
the quantity needed for the luminosity normalization.
In order to compare with Glauber calculations in the literature, a correction 
was applied for the 88\% BBC detector efficiency to obtain the inelastic
d+Au cross section,
\begin{equation}
\sigma_{\rm inel}^{\rm dAu} = \frac{\sigma_{\rm BBC}}{0.88} 
= 2.26 \, \pm \, 1.6\% \, \pm \, 5.0\%\, \pm\, 4.5\%\, \rm{b} \, \,.
\end{equation}
The last two errors reflect the theoretical uncertainty on $\sigma_{{\rm d}
\rightarrow np}$ and the BBC efficiency uncertainty.
A similar analysis yields the ZDC$_{\rm Au}$ cross section
for $E>10$ GeV, 
also used as a minimum bias trigger,
\begin{equation}
\sigma_{\rm ZDC_{Au}} = 2.06 \, \pm \, 1.7\% \, \pm \, 5.0\% \, \rm{b} \, \, .
\end{equation}

\subsubsection{$J/\psi$ and high mass $e^+e^-$ photoproduction} \bigskip

Ultraperipheral electromagnetic interactions of nuclei can be calculated using 
the equivalent photon approximation with $b\geq 2 R_{A}$. As long as the photon
squared momentum transfer is restricted 
to $Q^2<1/R_{A}^2$, the photon spectrum is dominated by 
coherent emission with a $Z^2$ enhancement in the equivalent 
$\gamma A$ luminosity. Because the coupling strength, 
$Z^2\alpha$, is close to unity, additional low energy photon exchanges 
occur with high probability, particularly at $b \sim 2R_A$ \cite{Baur:2003ar}. 
These low energy photons excite collective nuclear resonances, such as the
GDR, very effectively.  The excited nuclei return to their ground state 
predominantly by the emission of one or two neutrons. Neutron tagging 
is particularly useful for triggering UPC events and, when combined with a 
rapidity gap requirement on the side of the photon-emitting nucleus, is a 
powerful trigger selection criteria in heavy-ion collisions. The fraction 
of $J/\psi$ events with at least one neutron tag is calculated to be $60$\%.

Several calculations can be found in the 
literature \cite{Klein:1999qj,starlight,Baltz:2002pp} and an 
event generator, {\sc starlight}, is used at RHIC to simulate both coherent 
vector meson production and $\gamma\gamma\rightarrow e^+e^-$. Recently
\cite{Strikman} a calculation of incoherent $\gamma \, {\rm Au} \, \rightarrow
J/\psi X$ production was also presented. This calculation 
considers the same coherent photon flux but instead of quasi-elastic $J/\psi$ 
production off the entire nucleus it considers the corresponding production 
off individual target nucleons. Signatures of the incoherent process are a 
broader $J/\psi$ $p_T$ distribution and a higher neutron multiplicity due to 
interactions of the recoiling nucleon within the nucleus, see the discussion
in Section~\ref{vmabreakup}.

\subsection*{Trigger selection}

PHENIX has excellent capabilities for electron identification since it 
includes a high resolution electromagnetic calorimeter (EMCal) and Ring 
Imaging \v{C}erenkov (RICH) counters. The RICH and EMCal cover the same 
rapidity acceptance, $|\eta|\leq 0.35$, as the PHENIX tracking system in two 
approximately back-to-back spectrometer arms covering $\Delta\phi = \pi$. In 
addition to the tracking coverage near $\eta =0$, PHENIX used the BBC 
hodoscopes to trigger on inelastic heavy-ion collisions.  
The ZDCs measure the number of neutrons from beam 
dissociation and can be used to trigger on one or more neutrons in either 
beam direction. 

The PHENIX ultraperipheral dielectron trigger combined three of the trigger 
elements to select one or more beam dissociation neutrons in the ZDC, at 
least one electromagnetic cluster in the EMCal with energy and 
a rapidity gap signaled by no hits in one or the other of the BBC counters.
\begin{equation}
\mathtt {UPC \, Trigger = (EMCal \geq 0.8 \, GeV).AND.(ZDC_N.OR.
ZDC_S).AND.\overline{BBC}} \, \, .
\end{equation}
This very loose trigger yielded $8.5\times 10^6$ events out of 
$1.12\times 10^9$ recorded minimum 
bias interactions. Thus the UPC trigger comprised less than $0.5 \%$ 
of the inelastic cross section and a negligible part of the 
available trigger bandwidth.
		
\begin{figure}[tbh]
\centering
    \epsfxsize=0.9\hsize
     \epsffile{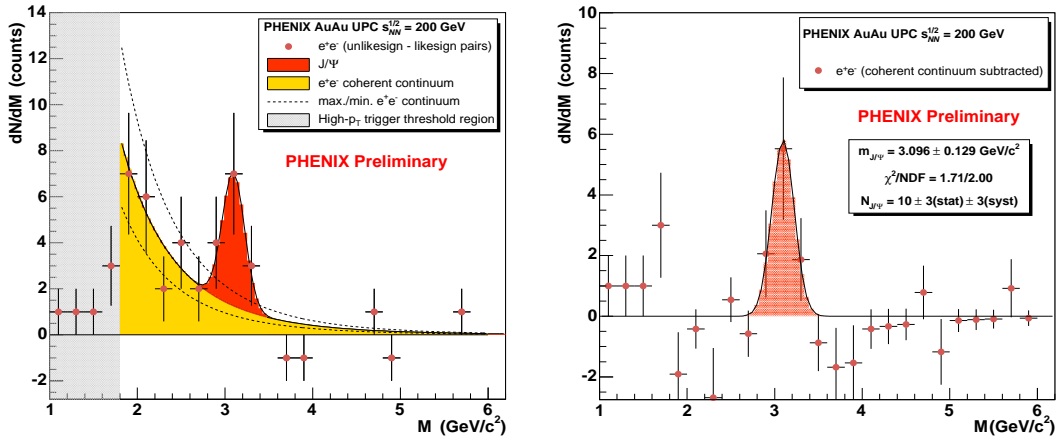}
\caption[]{Preliminary invariant mass distribution of $e^+e^-$ pairs 
measured by PHENIX in ultraperipheral Au+Au collisions 
at $\sqrt{s_{_{NN}}} = 200$ GeV. The left-hand plot shows a fit assuming 
an $e^+e^-$ continuum and a $J/\psi$ signal.
The two dashed curves indicate the continuum uncertainty. The right-hand
plot shows the signal after continuum subtraction.  From 
\protect\cite{dde_qm05}.}
\label{fig:phenix1}
\end{figure}

\subsection*{Event selection}

The main features of dilepton photoproduction are small pair
transverse momentum and low multiplicity tracks (both characteristic 
of diffractive processes).  Coherent $J/\psi$ production is primarily at 
midrapidity, $|y| \leq 1$. For a charged particle track to be reconstructed in 
the tracking detectors, $n_{\rm track}\leq 15$ and $|z_{\rm vertex}|\leq 30$ 
cm was required.
		
The integrated luminosity corresponding to 
this data sample was calculated after the vertex cut was applied and 21\% of 
the data with different running conditions was removed.  Using the number of 
minimum bias interaction triggers in the remaining sample and the $6.3 \pm
0.5$ b minimum bias Au+Au cross section~\cite{ppg014}, we find
$\int dt L = 120 \pm 10$ $\mu$b$^{-1}$.

The momentum of electron candidate tracks was measured using the deflection 
in the magnetic spectrometer. After defining electron candidate trajectories 
and momenta in the spectrometers, cuts consistent with electron response 
in the RICH and EMCal were imposed.  At least two photomultipliers were
required to have a \v{C}erenkov signal in the correct region of the RICH and
at least one electron was required to deposit greater than 1 GeV energy in the
EMCal.  Finally, electron candidates had to occupy different spectrometer 
arms since low $p_T$ $J/\psi$s decay to back-to-back electrons.

\begin{figure}[tbh]
\centering
    \epsfxsize=0.4\hsize
     \epsffile{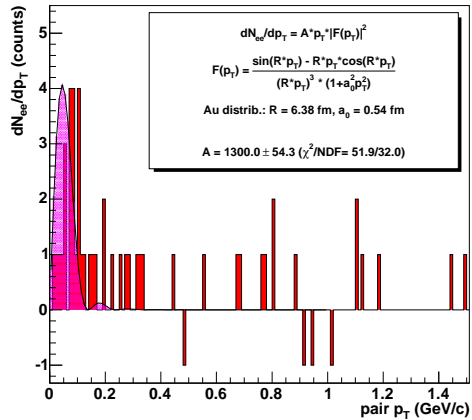}
\caption[]{The $J/\psi \rightarrow e^+ e^-$ $p_T$ distribution from 
ultraperipheral Au+Au collisions compared to a calculation of coherent 
photoproduction with a realistic nuclear form factor.  From 
\protect\cite{dde_qm05}.}
\label{fig:phenix2}
\end{figure}

\subsection*{Results}

The event selection cuts yielded 42 $e^+ e^-$ signal candidates and 7
$e^\pm e^\pm$ candidates with $M \geq 1.8$ GeV/$c^2$.  A like-sign subtraction
was performed to estimate the combinatorial background, resulting in
the signal spectrum shown in Fig.~\ref{fig:phenix1}. 
	
To extract the $J/\psi$ signal, the continuum spectrum was fit to 
a power law (with the power determined from a full simulation).  The number 
of events with $1.8<M<2.0$ GeV/$c^2$ was used to estimate the continuum, 
shown in Fig.~\ref{fig:phenix1}. The extracted signal is $10 \, \pm \, 3
\, ({\rm stat})\, \pm \, 3 \, ({\rm syst})$ 
events.  The continuum subtraction dominates the systematic error.
	
Inclusive hadronic $J/\psi$ production in heavy-ion collisions has a 
broad $p_T$ distribution with an average $p_T$ of $\sim 1.5$ 
GeV/$c$~\cite{leitch}. This should be compared to the measured coherent 
photoproduction $p_T$ distribution peaked at $p_T\approx 80$ MeV/$c$,
as expected when all pairs with $M > 1.8$ GeV/$c^2$ 
are included, see Fig.~\ref{fig:phenix2}.  Figure~\ref{fig:phenix2} also
shows the expected shape due to the Au form factor.
	
The $J/\psi$ photoproduction cross section was calculated, correcting for 
detector acceptance and cut efficiencies obtained by simulating $J/\psi$
production with the expected 
$p_T$ distribution.  The geometrical 
acceptance and efficiencies reduce the $J/\psi$ yield in $|y| \leq 0.5$ by 
5.0 \% and 56.4 \% respectively. The preliminary cross section at $y=0$ is
\begin{equation}
\frac{B d\sigma_{J/\psi}}{dy} = 48 \, \pm \, 14 \, ({\rm stat}) \,
\pm \, 16 \, ({\rm syst}) \, \mu{\rm b} \, \, ,
\end{equation}
in good agreement with the 58 $\mu$b {\sc starlight}
\cite{Klein:1999qj,starlight,Baltz:2002pp} 
prediction. In future Au+Au runs, PHENIX will see a 10-fold increase in 
event yield, making detailed studies of both coherent and quasi-elastic
$J/\psi$ photoproduction possible.  PHENIX will also commission a second 
trigger, sensitive to $J/\psi \rightarrow \mu^+\mu^-$ at large rapidity 
where the quasi-elastic signal will dominate~\cite{Strikman}. 
Nevertheless, the present low-statistics measurement clearly demonstrates 
the feasibility of small cross section diffractive measurements in heavy-ion
collisions at RHIC and the LHC.

\section{LHC detector overview of UPC potential}

\subsection{Introduction}
\label{section-overviewDetectors}
{\it Contributed by: P. Yepes}

The ALICE, ATLAS and CMS collaborations plan to take data at the LHC 
with heavy-ion beams.  ALICE was specifically designed for heavy-ion physics 
and intends to address both soft and hard physics. CMS and ATLAS were designed 
for hard physics and initially focused on proton-proton collisions.
However their potential for heavy-ion physics was soon pointed out.
In this chapter, a brief description of each detector is presented with 
special emphasis on those features most relevant for UPCs.

Table~\ref{table:summaryDetectors} shows the main features
of the LHC detectors.  All three detectors have 
complete azimuthal tracking coverage over different rapidity regions. ALICE is 
limited to $|\eta| \lesssim$ 1 while ATLAS and CMS extend their coverage 
to $|\eta| < 2.4$.  The latter two detectors have tracking systems that can be 
read out at every beam crossing. The ALICE TPC provides excellent resolution,
$\Delta p_T/p_T =1.5$\%, for low momentum particles, $0.05 <p_T< 2$ GeV/$c$. 
However, ALICE can only be read out with a rate on the order of kHz.
The ATLAS (CMS) momentum resolution is $\Delta p_T/p_T\approx 3$\% ($<2$\%). 
ATLAS can reconstruct low momentum particles down to $p_T =0.5$ GeV/$c$, 
while CMS measures tracks as low as $p_T =0.2$ 
GeV/$c$~\cite{ferenc07,CMS-HI-PTDR}.
\begin{sidewaystable}
\caption[]{Summary of the main characteristics of the ALICE, ATLAS and CMS 
detectors.}
\vspace{0.4cm}
{\footnotesize
\begin{tabularx}{\linewidth}{| X | X | X | X | X | X |} \hline
 Component &          &   & ALICE \cite{ALICE-TDR}  & ATLAS \cite{DetectorTDRs}
& CMS \cite{CMS-PTDR1,CMS-HI-PTDR}\\ \hline 
 Tracking  & Acceptance & $\eta$       & $|\eta|<0.9$ & $|\eta|<2.4$ 
& $|\eta|<2.4$ \\ \cline{3-6}
 && $\phi$      & $0<\phi<2\pi$ & $0<\phi<2\pi$ & $0<\phi<2\pi$ \\ \cline{3-6} 
 && $p_T$ & $p_T > 0.05$ GeV/$c$ & $p_T > 0.5$ GeV/$c$ & $p_T > 0.2$ GeV/$c$ 
\\  \cline{2-6} 
 & Resolution  & $\Delta p_T/ p_T$ &1.5\% ($p_T <2$ GeV/$c$) 
&$\approx 3$\% & $<2$\%, $p_T < 100$ GeV/$c$ \\ 
 &          &         & 9.0\%, $p_T = 100$ GeV/$c$ &      &  \\ \cline{1-6} 
 Muons                 & Acceptance & & $-4<\eta<-2.5$ & $|\eta|< 2.4$  
& $|\eta| < 1$, $p_T > 3.5$ GeV/$c$ \\ 
  &   & &     & $p_T> 4.5$ GeV/$c$ & $|\eta| > 1$, $p_T > 1.5$ GeV/$c$ 
\\ \cline{1-6}
 Particle ID & $\pi/K$ &    & $0.1<p< 3$ GeV/$c$ & TBD  & $0.2<p_T<1$ GeV/$c$ 
\\ \cline{2-6} 
     &  $K/p$    &       & $0.2<p< 5$ GeV/$c$ &  TBD  &  $0.4<p_T<2$ GeV/$c$ 
\\ \cline{2-6}
     & $e/\pi$ &     & $0.1<p< 25$ GeV/$c$ & $p > 2$ GeV/$c$ & $0.1<p<0.2$ 
GeV/$c$, $p> 2$ GeV/$c$\\ \cline{1-6} 
 Electromagnetic  & Acceptance   & $\eta$ & $|\eta|< 0.12$ & $|\eta|<3.1$ 
& $|\eta|<3$ \\ \cline{3-6}	     
Calorimetry & & $\phi$ & $1.22\pi<\phi<1.78\pi$ & $2\pi$ & $2\pi$\\ \cline{2-6}
& Segmentation & $\Delta\phi \times \Delta\eta$ & $0.0048 \times 0.0048$ 
& $0.025 \times 0.003$ & $0.0175 \times 0.0175$ \\ \cline{2-6} 
 &    & Longitudinal  &  No  &   Yes     &     No   \\ \cline{2-6}
 & Resolution ($E$ in GeV) & $\Delta E / E$ & $ 0.03/\sqrt{E} \oplus 0.03/E$ & 
$0.1/\sqrt{E} \oplus 0.005$ & $0.027/\sqrt{E} \oplus 0.0055$ \\ 
	     &              &       & $ \oplus \, 0.01$ &     & \\ \cline{2-6}
& Technology  &  & PbWO$_4$ crystals & Liquid Ar (LAr) & PbWO$_4$ crystals  
\\ \cline{1-6}
 Hadronic & Acceptance & $\eta$ & NA  & $|\eta|<3$ & $|\eta|<3$ \\ \cline{3-6} 
 Calorimetry &   & $\phi$ & NA    & $2\pi$ & $2\pi$ \\	     \cline{2-6}
& Segmentation & $\Delta\phi \times \Delta\eta$ & NA & $0.1 \times 0.1$ 
& $0.087 \times 0.087$ \\ \cline{2-6} 
&    & Longitudinal  &   NA     &   Yes   &     Yes  \\ \cline{2-6}
 & Resolution   & $\Delta E / E$ & NA   &    $0.5/\sqrt{E} \oplus 0.02$ &
$1.16/\sqrt{E} \oplus 0.05$ \\ \cline{2-6}
 & Technology   &  & NA  & Pb Scint(B) - LAr(F)  & Cu Scint   \\ \cline{1-6}
 Forward     & Acceptance & $\eta$   & NA & $3<|\eta|<4.9$  & $3<|\eta|<5$ 
\\ \cline{3-6}	     
 Calorimetry &  & $\phi$ & NA   & $2\pi$ & $2\pi$ \\	     \cline{2-6}
 & Segmentation & $\Delta\phi \times \Delta\eta$ & NA & $0.1 \times 0.1$ 
& $0.087 \times 0.087$     \\ \cline{2-6} 
 & Technology   &  & NA  & Cu/LAr - W/LAr  & Fe/quartz fibers\\ \cline{1-6}
 Very    & Acceptance & $\eta$ & NA  & NA  & $5.3<\eta<6.7$ \\ \cline{3-6}
 Forward &   & $\phi$& NA    & NA  & $2\pi$ \\	     \cline{2-6}
 Calorimetry & Segmentation & $\Delta\phi \times \Delta\eta$ & NA & NA 
& $\pi/8$ $(\pi/4)$  Had (EM) \\ \cline{1-6} 
 Forward  & Acceptance & $\eta$ & NA & NA & $5.3<\eta<6.7$ \\ \cline{3-6}
 Tracking (TOTEM) &    & $\phi$& NA           & NA  & $2\pi$ \\ \cline{1-6} 
 Zero-Degree & Acceptance & $|\eta|$ (neutrals) & $\gtrsim 8.6$ & $\gtrsim 8.3$
& $\gtrsim 8.3$ \\ \cline{3-6}
 Calorimeters (ZDC)  &    & $\phi$& $2\pi$ & $2\pi$  & $2\pi$ \\ \cline{1-6} 
\end{tabularx} 
}
\label{table:summaryDetectors}
\end{sidewaystable}

ALICE is equipped with a muon spectrometer 
with full azimuthal acceptance in the 
rapidity range $-4 < \eta < -2.5$.  ATLAS has large acceptance, 
$|\eta| < 2.4$ for muons with $p_T > 4.5$ GeV/$c$. In CMS, muons with 
$p_T > 3.5$ GeV/$c$ will be detected in the central region,
$|\eta| < 1$, while the forward muon detector, $1 < |\eta| < 2.4$, has muon
acceptance for $p_T > 1.5$ GeV/$c$.   

ALICE is best designed for particle identification. It can separate pions from 
kaons in the range $0.1 <p< 3$ GeV/$c$, kaons from protons over $0.2 <p< 5$ 
GeV/$c$, and electrons from $\pi^0$'s for $0.1<p<25$ GeV/$c$. Studies in 
CMS indicate good low $p_T$ capabilities using the three layers of the 
silicon pixel tracker to achieve $\pi$, $K$ and $p$ separation 
within $0.4 < p_T<$ 1 GeV/$c$ ~\cite{ferenc07,CMS-HI-PTDR}. In addition, a 
conservative range over which electrons can be separated from neutral 
pions is $2 < p_T <  20$ GeV/$c$. 
   
The features of the electromagnetic calorimeters are also given in 
Table~\ref{table:summaryDetectors}. ALICE has a PbWO$_4$ photon spectrometer
with excellent spatial and energy resolution, albeit small acceptance, and
a larger lead/scintillator electromagnetic calorimeter. ATLAS 
and CMS have hermetic calorimeters employing liquid argon and PbW0$_4$ 
crystals, respectively,
both covering $|\eta|\lesssim 3$. CMS has slightly better resolution
while ATLAS provides additional information on the longitudinal shower shape.

Both ATLAS and CMS are equipped with large coverage, $|\eta| <5$, hadron 
calorimetry.  The CMS copper-scintillator calorimeter has slightly
finer transverse segmentation than ATLAS.  However, ATLAS combines 
lead scintillators with liquid argon to achieve a factor of two better energy 
resolution than CMS. 
Both experiments also feature ZDCs ($|\eta|\gtrsim 8.5$ for neutrals),
a basic tool for neutron tagging in ultraperipheral heavy-ion interactions.
CMS has an additional electromagnetic/hadronic calorimeter, CASTOR ($5.3<
|\eta|<6.7$), and shares the interaction point with the TOTEM experiment, 
providing two additional trackers at very forward rapidities, T1 ($3.1 
<|\eta|< 4.7$) and T2 ($5.5 <|\eta|< 6.6$)~\cite{CMS-TOTEM}.

\subsection{ The ALICE detector}
{\it Contributed by: V. Nikulin, J. Nystrand, S. Sadovsky and E. Scapparone} 
\label{section-alice}

\begin{figure}[tbp]
\centering
  \includegraphics[width=0.8\textwidth]{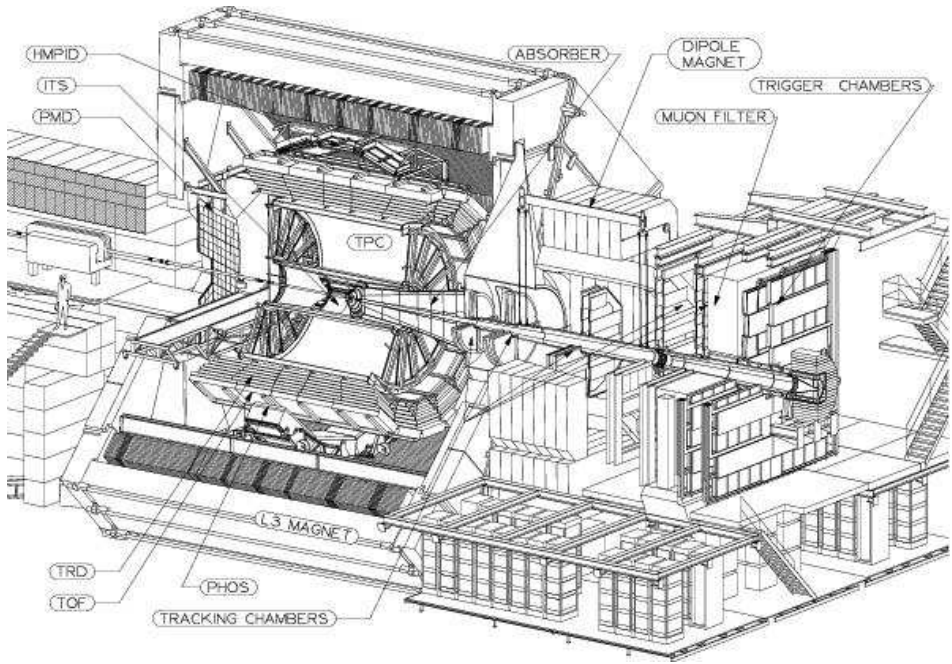}
  \caption{
    The ALICE Detector.  The ZDCs, positioned more than 100 m from
    the interaction point, are not shown.}
    \label{fig:ALICEdetector}
\end{figure}

The ALICE detector \cite{ALICE-TDR}, shown in Fig.~\ref{fig:ALICEdetector},
is designed to study
the physics of strongly interacting matter at extreme energy
densities where the formation of a new phase of matter, the
quark-gluon plasma, is expected. 
The detector  is designed to cope with up to 8000 particles per unit rapidity.
It consists of a central part which measures hadrons, electrons and photons, 
a forward muon spectrometer and two zero degree calorimeters
located up- and downstream from the detector \cite{ALICE-TDR}.

The central (barrel) part, which covers the full azimuth over $|\eta| < 0.9$ 
in pseudorapidity,  is embedded in a large solenoid magnet. 
The barrel detectors include a silicon inner tracking system (ITS),
a cylindrical time-projection chamber (TPC), three time-of-flight (TOF) 
arrays, ring-imaging Cerenkov (HMPID) and transition radiation (TRD) counters 
and a high resolution electromagnetic calorimeter (PHOS). 
Several smaller detectors (FMD, V0, T0) are located at small angles
forward and backward of midrapidity.  Note that the Forward Muon Spectrometer, 
shown on the right-hand side of Fig.~\ref{fig:ALICEdetector}, is actually at
backward rapidity according to the LHC frame convention.

A solenoidal magnetic field of $0.2 - 0.5$ T allows full 
tracking and particle identification down to $p_T \sim 100$ MeV/$c$.
The optimal field strength and volume is a compromise between momentum 
resolution, momentum acceptance and tracking efficiency. 

\subsubsection*{The inner tracking system}

The ITS consists of six 
layers of high-resolution detectors. 
The innermost four layers are equipped with  
silicon pixel and drift detectors designed to handle the high multiplicities
expected.  The outer layers, at a $\sim 50$ cm radius, are equipped with 
double-sided silicon micro-strip detectors. 
Four of the layers have analog readout for independent particle 
identification via $dE/dx$ in the non-relativistic region, which gives the 
ITS standalone capability as a low $p_T$ particle spectrometer.

The Level-0 (L0) trigger for the pixel detectors is under development.  The
$10^6$ fast logical {\tt OR} outputs are grouped into 1200 processor
inputs by serialization/de-serialization.  The processor enables
implementation of a very flexible decision algorithm.  The possibilities
of using this trigger for UPC events should be carefully studied.

\subsubsection*{The time projection chamber}

The inner radius of the TPC (90 cm) is fixed by the maximum acceptable hit 
density, 0.1 cm$^{-2}$.  The 250 cm outer radius is determined by the 
length required for the $dE/dx$ resolution to be better than 10\%. 
The large granularity, $\sim 5\times 10^8$ pixels, together with an 
optimized gas choice, ensures good two-track resolution.

\subsubsection*{Particle identification}

Good particle identification over a wide $p_T$ range is one of the strongest 
features of the ALICE detector.  Several detector systems are used for particle
identification:  the ring-imaging Cerenkov detector (HMPID), the transition
radiation detector (TRD) and the time-of-flight array (TOF).  A short 
description of the functioning of these detectors in UPCs is given here.

The HMPID, covering 15\% of the $(\eta,\phi)$
acceptance, is optimized for high $p_T$ particle
detection. The six-layer TRD will identify 
electrons with momenta above 1~GeV/$c$.  In order to reduce the large energy
dissipation by the TRD front-end electronics, it remains in stand-by mode
when nothing interesting is happening.  A pre-trigger signal is required to 
wake up the electronics within 100 ns of the interaction.  In hadronic
interactions, the fast logical {\tt OR} of the small angle T0 detector
is used as a pre-trigger.  Since the T0 detectors are only useful in a 
high-multiplicity environment and cannot be employed in UPCs, 
the {\tt OR} from the TOF can be used as an alternate TRD pre-trigger.

The large TOF detector covers a cylindrical surface with
polar angle acceptance of $45^\circ < \theta < 135^\circ$.  
The TOF consists of
90 modules with 18 $\phi$ sectors and 5 $z$ segments.  
The TOF modules are
made up of multi-gap resistive plate chambers (MRPCs) with an intrinsic timing 
resolution of $\sim 100$ ps. The 5 longitudinal
segments have different numbers of strips, according to their position.
There are 1674 MRPC strips in the TOF, each with 96 pads, for a total of
$\sim 1.6 \times 10^5$ readout pads.  The TOF L0 trigger can provide
the total hit multiplicity.  A more sophisticated Level-1 (L1) trigger could be
applied.  

\subsubsection*{Photon detectors}

The photon spectrometer (PHOS) is intended to measure direct photons 
and high $p_T$ neutral mesons. PHOS is a single arm, high-resolution 
electromagnetic calorimeter, positioned 4.6 m below the interaction vertex.
It covers 8 m$^2$, $\sim 10$\% of the barrel 
solid angle, with $\sim 17,000$ channels of scintillating 
PbWO$_4$ crystals.  Thus PHOS could be used as veto counter to select events
with ``abnormally'' low multiplicity.

The electromagnetic calorimeter (EMCAL) will be
an ALICE upgrade.  It will be a medium-resolution scintillation sandwich 
calorimeter.  While the EMCAL is primarily for jet studies, the L0 EMCAL
trigger could probably be adapted to UPC needs.
Additional studies are required.  

The photon multiplicity detector (PMD) is a pre-shower detector that measures
the ($\eta,\phi$) distribution of the photons in the region, $1.8 < 
\eta < 2.6$.  The PMD consists of two identical planes of  proportional 
honeycomb chambers with a 3 interaction length thick lead converter in between
the chambers.

\subsubsection*{Small angle detectors}

ALICE has a number of smaller detector systems (ZDC,  FMD,  V0, and T0) 
positioned at small angles for triggering.
Two ZDCs are located in the accelerator tunnels 100 m away from the 
interaction point. Their pseudorapidity acceptance is $8.6 < |\eta|$.
They measure the spectator nucleons (both neutrons and protons) in 
the collision. 

The other small detectors are located asymmetrically with respect to the
interaction point.  The right-hand arrays are at negative rapidity while the
left-hand detectors are at positive rapidity.
The forward multiplicity detector (FMD) is a silicon strip ring 
counter with about 25000 channels. 
It measures charge particle production in the pseudorapidity ranges 
$-3.4 < \eta < -1.7$ (right) and $1.7 < \eta < 5.1$ (left).
The T0 (beam-beam) detector, 12 Cerenkov 
radiators coupled to photomultiplier tubes, is located at
$2.9 < \eta < 3.3$ (left) and $-5.0 < \eta 
< -4.5$ (right).  It produces fast signals with 
good timing resolution, $\sigma \sim 50$ ps, allowing online reconstruction
of the main vertex.  The V0 detector, 72 plastic scintillators grouped in 
five rings, covers $-3.8 <  \eta < -1.7$ (right) and $2.8 < \eta < 5.1$ (left).
It provides a minimum bias trigger for the central detectors and can be used
for centrality determination in $AA$ collisions and to validate the trigger
signal in the muon spectrometer in $pp$ collisions.

\subsubsection*{Forward muon spectrometer}

The forward muon spectrometer, covering the backward region 
$-4.0 <  \eta  < -2.5$ in the LHC reference frame,
will study quarkonium decays to dimuons.
The expected mass resolution is $\sim 100$ MeV at 10 GeV, sufficient 
to distinguish between the $\Upsilon$ $S$ states.
The muon spectrometer consists of a composite absorber with 
high-$Z$ (small angles) and low-$Z$ (near the front) materials, 
located 90 cm behind the interaction point; 
a large dipole magnet with a 3 T m field integral; 10 planes of thin, 
high-granularity tracking chambers with $\sim 100$ $\mu$m spatial
resolution; a 1.2 m iron muon filter;
and four trigger chambers for the L0 trigger.  The muon filter sits
between the tracking and trigger chambers to reduce the trigger chamber
background.

\subsubsection*{Trigger and data acquisition}

The ALICE trigger is especially important because the very large event size 
causes severe data acquisition and storage problems. 
ALICE features a complex and flexible trigger. 
Several detectors provide input to the different trigger levels to select 
signals such as centrality, high $p_T$ electrons, muons, or photons.
Several trigger levels are foreseen.
\begin{itemize}
\item {\bf Level-0 (L0):} This is the fast, minimum-bias interaction trigger,
issued after $\sim 0.8$ $\mu$s, used as strobe for some electronics. 
It includes various decisions from the T0 and the muon spectrometer as well as
from other auxiliary sub-detectors such as the cosmic telescope. 
Recently, L0 triggers were developed for the ITS, PHOS and TOF detectors. 
The T0 decision is used as a TRD pre-trigger for hadronic interactions.
\item {\bf Level-1 (L1):} This trigger, with a latency of $\sim 6.5$ $\mu$s,
receives additional information from the ZDC, PHOS, TRD, FMD and  PMD.
\item {\bf Level-2 (L2):} Relatively slow detector decisions are included 
with a delay of $\sim 88$ $\mu$s so that it is possible to veto events
where a second high multiplicity event occurs just before or soon after the
trigger of interest (TPC past-future 
protection). 
\item {\bf High Level Trigger (HLT):} The HLT is an on-line computing farm 
with several hundred commodity processors providing further event selection 
and event compression.
\end{itemize}

Most UPC events are characterized by very low multiplicity.
Unfortunately, the low multiplicity background tends to be rather large. 
Therefore, L0 background suppression is necessary.
UPC events with nuclear dissociation can be selected by combining low
multiplicity with nuclear break-up using the ZDCs, greatly suppressing 
the background.  However, the ALICE ZDCs are located too far from the 
main detector to be used at L0.
Thus UPC events need to be triggered by the central barrel and/or the 
muon spectrometer.
 
UPC events are normally rejected by the standard ALICE L0 trigger.
However, recent developments employing the ITS pixel trigger are quite  
encouraging.  The trigger could apply $p_T$ cuts to the L0 signal, 
rejecting the low $p_T$ background due to $e^+e^-$ 
pair production, $\sigma_{\rm tot} \sim 200$ kb \cite{HKS},
to study UPCs in the barrel. 
Detailed studies of this option are underway.
The EMCAL, currently under construction, will be used as an
additional dijet trigger. 

The muon spectrometer trigger could be successfully used 
to detect UPCs with final-state muons.  The PHOS can be used to 
efficiently veto high multiplicity events at L0, considerably reducing 
the UPC background for muon events.
 
The offline analysis, including ZDC information and reconstruction
of the parent particle $p_T$ distributions, could select events of 
interest.  Further studies are presented in section 
\ref{alice-gammagamma}. 

\paragraph*{Barrel trigger strategies}

The `elastic' and `inelastic' UPC classes
require different triggers. The identification of exclusive vector meson 
production is based on reconstruction of the entire event (the two
tracks from the decay) and identifying coherent production
through the low lepton pair $p_T$.  On the other hand, $\gamma$-parton 
interactions must be
identified by a rapidity gap between the photon-emitting nucleus 
and the produced particles.

The very different topology of ultraperipheral interactions relative to 
central nucleus-nucleus collisions leads to different trigger
requirements.  
In the case of hadronic interactions, the forward detectors trigger
large multiplicity events.  This is not possible for
UPCs since they are characterized by voids of produced 
particles, rapidity gaps,
several units wide. To detect ultraperipheral events, it is necessary to 
have a low-level trigger sensitive to the production of a few charged 
particles around midrapidity \cite{ALICE-PPRvolII}.

In ALICE, the fast response, large pseudorapidity coverage, 
$| \eta | < 1$, and high segmentation, of the TOF make it 
well suited for a L0 trigger in the central region. 
Since the T0 detectors are not used in UPCs, the fast
{\tt OR} TOF signal can be used as a pre-trigger for the
TRD.  The pad signals from each of the 90 
modules are included in the TOF L0 trigger.  They can provide information on
event multiplicity and topology, important for the development of a UPC 
trigger.  
Hits in several TOF pads are required for the pre-trigger.

A possible trigger scheme for exclusive $\rho^0$, $J/\psi$
and $\Upsilon$ photoproduction is
described below.
\begin{itemize}
\item {\bf L0:} The TOF L0 multiplicity coupled with a suitable
topology cut can provide a trigger for exclusive events with exactly
two charged tracks in the central barrel.  The forward detectors are available
at L0 and can identify the presence of one more more rapidity gaps.  For
example, if only the V0 detectors (V0L, $2.8 < \eta < 5.1$, and V0R,
$-1.7 > \eta > 3.8$) are available at L0, then if there is no signal in V0L
but a signal in V0R, there is a rapidity gap of at least $-1.7 > \eta > 3.8$.
Thus the ALICE trigger logic
unit can carry the information on track multiplicity as well as  
rapidity gap.
\item {\bf L1:} The main trigger cut for $J/\psi \rightarrow e^+e^-$ and
$\Upsilon \rightarrow e^+e^-$ decays at this level will be identification 
of one electron and one positron in the TRD. If a more accurate measurement 
of the central barrel multiplicity is available, it could be used to
select events with exactly two charged tracks. Information from the
ZDCs may be used to select events with or without Coulomb breakup.
\item {\bf HLT:} The HLT may be used to require exactly two opposite-sign
tracks from the primary vertex in the TPC. Using the reconstructed momenta, 
a cut on the summed track $p_T$ can be applied.  Such a cut is highly
efficient for suppressing the incoherent background. Some
of the pions from coherent $\rho^0$ decays could be misidentified as 
electrons at L1 in the TRD. Due to the extremely high $\rho^0$ rate, it 
may be necessary to apply an invariant mass cut in the HLT to scale down
these events so that they do not occupy the full bandwidth.
\end{itemize}
A similar triggering scheme for other photonuclear events is given below. 
\begin{itemize}
\item {\bf L0:} At L0, there would be an asymmetric signal in the V0 
counters: low or intermediate multiplicity on one side and no signal 
from the opposite side, supplemented by a low-multiplicity trigger in
the central arm, such as from the TOF.
\item {\bf L1:} The ZDC on the same side as the rapidity gap
should be empty. The signal in the ZDC on the opposite side should
be low.
\item {\bf HLT:} The photonuclear event rate will be high but only a small 
fraction of these events will be interesting. In addition, the asymmetric 
signature at L0 and L1 are also caused by beam-gas interactions. The
HLT will be needed to reject beam-gas events and select the interesting 
photonuclear events such as open charm production.
\end{itemize}

The expected vector meson and lepton pair yields from two-photon
interactions were estimated using the geometrical acceptance of the
ALICE central barrel and muon arm. Events were generated from a Monte
Carlo model based on the calculations
in Refs.~\cite{Klein:1999qj,starlight,Nystrand:1998hw,Baltz:2002pp}.  The rates
were calculated assuming a Pb+Pb luminosity of $5 \times
10^{26}$ cm$^{-2}$s$^{-1}$.  The ALICE acceptance is defined as
$|\eta| < 0.9$ and $p_T > 0.15$~GeV/$c$ in the central barrel, 
$-4 \leq \eta \leq -2.5$ and $p_T > 1.0$ GeV/$c$ in the muon spectrometer. 
Both tracks are required to be within the acceptance for the event to be
reconstructed.  A trigger cut of $p_T > 3.0$ GeV/$c$
is necessary for central collisions in the TRD. The expected
vector meson and lepton pair rates are shown in
Table~\ref{tab:ALICE-rates}~\cite{ALICE-PPRvolII}.
\begin {table}[htb] 
\begin{center}
\caption[]{The expected yields from Pb+Pb UPCs for several final states
within the geometrical acceptance of the ALICE central barrel
\protect\cite{ALICE-PPRvolII}.}
\vspace{0.4cm}
\begin{tabular} {|c|c|c|} \hline 
Final State & Acceptance  &  Rate/$10^6$ s  \\ \hline
$\rho^0\to\pi^+ \pi^-$  & central barrel  & $2 \times 10^8$ \\
$J/\psi\to e^+e^-$      & central barrel  & $1.50 \times 10^5$   \\
$\Upsilon(1S)\to e^+e^-$& central barrel  & 400 -- 1400     \\
$e^+e^-, M>1.5$~GeV/$c^2$ & central barrel, $p_T>0.15$~GeV/$c$
            & $7 \times 10^5$ \\
$e^+e^-, M>1.5$~GeV/$c^2$ & central barrel, $p_T>3$~GeV/$c$  &
            $1.4\times 10^4$ \\
$\mu^+\mu^-, M>1.5$~GeV/$c^2$ & muon spectrometer, $p_T>1$~GeV/$c$  &
            $6 \times 10^4$ \\ \hline
\end{tabular}
\end{center}
\label{tab:ALICE-rates}
\end{table}

\paragraph*{ITS low multiplicity trigger}

A low level 
trigger for ultraperipheral processes in ALICE can
be based on the charged track multiplicity in the central rapidity
region. The ITS \cite{ITS-TDR}, useful 
for fast charged-multiplicity measurements, is considered as a trigger here.
The ITS consists of six coaxial
cylindrical detectors: two pixel detectors (SPD1 and SPD2);
two drift detectors (SDD1 and SDD2); and two strip detectors (SSD1
and SSD2). Table~\ref{tab:its} presents their main features.

\begin{table}[htb]
\begin{center}
\caption[]{The elements of the ALICE inner tracker, their type and their radial
distance from the beam pipe.}
\vspace{0.4cm}
\begin{tabular}{|c|c|c|} \hline
Detector & Type  & Radius (cm)  \\ \hline
SPD1  & Pixel &  3.9 \\ 
SPD2  & Pixel &  7.6  \\ 
SDD1  & Drift & 14.0  \\ 
SDD2  & Drift & 24.0  \\ 
SSD1  & Strip & 40.0  \\ 
SSD2  & Strip & 45.0  \\ \hline
\end{tabular}
\end{center}
\label{tab:its}
\end{table}

The ALICE detector will typically operate in a solenoidal magnetic
field of strength $0.2 < B < 0.5$~T. Although the nominal field value is 0.2 T 
\cite{ALICE-TDR}, ALICE is likely to run at the highest field value 
\cite{LHCC2003-049}. 
The magnetic field restricts the kinematic acceptance for
charged particles.  At 0.2~T, the minimum transverse
momenta, $p_T^{\rm min}$, necessary for charged particles to reach SPD1 and
SPD2, the inner detectors, are 1.2~MeV/$c$ and 2.3~MeV/$c$ respectively. 
The minimum $p_T$ needed to reach the outermost detector, SSD2,
is 13.5~MeV/$c$. (For higher field values, $p_T^{\rm min}$ increases 
linearly with $B$.  Thus low fields are best for studying soft physics.)
Particle absorption in the beam pipe and detector layers 
is not taken into account.
The detector load is defined as the cross section for having $N_e$ charged 
hits from electrons or positrons in the detector acceptance.
The load
\cite{ALICE-INT-2002-11} and the corresponding UPC trigger rate estimates
are presented in Table~\ref{tab:sigma(n')}. 

\begin{table}[htb]
\begin{center}
\caption[]{The cross sections and rates of $N_e$ $e^+$ or $e^-$ interacting
in the SPD1, SPD2 and SSD2 layers of the ITS detectors in Pb+Pb
collisions with $L_{\rm PbPb} = 
10^3 ~\mbox{b}^{-1}\mbox{s}^{-1}$ and $B = 0.2$ T.}
\vspace{0.4cm}
\begin{tabular}{|c|c|c|c|c|c|c|} \hline
 & \multicolumn{2}{c|}{SPD1}
 & \multicolumn{2}{c|}{SPD2} 
 & \multicolumn{2}{c|}{SSD2} \\
 & \multicolumn{2}{c|}{$p_T>1.2$ MeV/$c$, $|\eta|<1.5$}
 & \multicolumn{2}{c|}{$p_T>2.3$ MeV/$c$, $|\eta|<1.5$} 
 & \multicolumn{2}{c|}{$p_T>13.5$ MeV/$c$, $|\eta|<1$} \\ \cline{2-7}
	     $N_e$ & $\sigma_{N_e}$ (b) & Rate (Hz)
     & $\sigma_{N_e}$ (b) & Rate (Hz) 
     & $\sigma_{N_e}$ (b) & Rate (Hz)               \\ \hline
1 & 13000 & $1.3\times 10^7$ & 4600 & $4.6\times 10^6$ & 140 & 
$1.4\times 10^5$ \\
2 & 4400  & $4.4\times 10^6$ & 1200 & $1.2\times 10^6$ &  19 & 
$1.9\times 10^4$  \\
3 & 87    & $8.7\times 10^4$ &  8.4 & $8.4\times 10^3$ & 0.3 & 
$3.0\times 10^{2}$ \\
4 & 20    & $2.0\times 10^4$ &  1.7 & $1.7\times 10^3$ & 
$3.4\times 10^{-2}$ & $34$ \\
5 & 4.3   & $4.3\times 10^3$ & 0.36 & $3.6\times 10^2$ & 
$3.7\times 10^{-3}$ & $3.7$ \\
\hline
\end{tabular}
\end{center}
\label{tab:sigma(n')}
\end{table}
These rates have to be compared to the
hadronic collision rates.
The impact-parameter integrated Pb+Pb cross section is about 8~b.
Thus the ALICE L0 trigger rate should be at least comparable to this 
cross section, $\sim 8$~kHz, up to a limitation of 0.4 MHz due to the
LHC clock. Thus the SPD2 trigger is required to select events with at 
least three charged particles while $p_T^{\rm min}$ is high enough for
the SSD2 trigger rate to be able to function for $N_e \ge 1$. 
On the other hand, charge conservation in
$\gamma \gamma$ processes requires the restriction $N_e \ge 2$.
Note that at higher field values, the increased $p_T^{\rm min}$ would mean
reduced event rates in all the detectors.

The L0 pixel trigger is under development.

\paragraph*{TOF Trigger backgrounds}

The UPC signal is characterized by a few tracks in an otherwise empty detector.
The most important trigger background is the fake trigger rate (FTR)
due to spurious hits in the TOF.
The main source of FTR for the TOF L0 trigger 
is the MRPC noise, measured to be 0.5 Hz/cm$^2$. 
The MRPC noise is due to ionizing particles in
the chamber.  The fraction of MRPC noise from the front-end electronics is
just few percent, measured by switching off the MRPC high voltage.
The main source of background noise in the TOF during ALICE operation are 
beam-gas collisions, 
beam misinjection and neutrons from Pb+Pb interactions. 
To be conservative, we assume an MRPC noise level of 2.5 Hz/cm$^{2}$,
a factor 5 larger than the measured value.

\begin{figure}[htb]
\centering\mbox{\epsfig{file=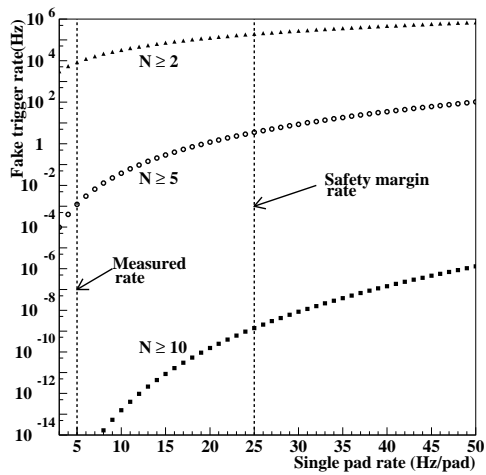,width=.45\textwidth}}
\caption{The L0 FTR as a function of the single pad rate. Reprinted from 
Ref.~\protect\cite{ALICE-PPRvolII}
with permission from Institute of Physics.} 
\label{UPC:ftr}
\end{figure}

\begin{figure}[!t]
\centering\mbox{\epsfig{file=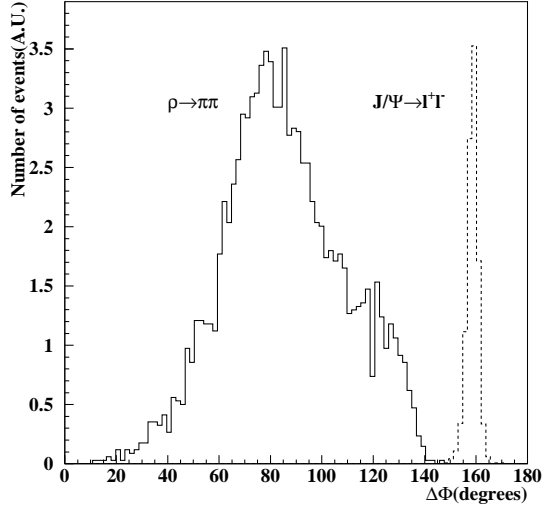,width=.5\textwidth}}
\caption{The azimuthal difference between the $J/\psi$ and $\rho^0$
decay products.  Reprinted from Ref.~\protect\cite{ALICE-PPRvolII}
with permission from Institute of Physics.}
\label{UPC:phi}
\end{figure}

Figure~\ref{UPC:ftr} shows the FTR as a function of the rate in a single
TOF pad.
The FTR is $\simeq 1$ Hz when five or more pads are fired, $N_{\rm pad}\geq 
5$ while the FTR for $N_{\rm pad} = 2$ is 200 kHz. Such high rate, 
unmanageable at L0, can be further reduced by using the vector meson 
decay topology. The
$J/\psi \rightarrow l^+ l-$ and $\rho^0 \rightarrow \pi\pi$
decays were simulated using 
{\sc starlight} \cite{Klein:1999qj,starlight,Baltz:2002pp}.  The decay products
were tracked through the TPC volume in a 0.5 T magnetic field without any
secondary interactions or multiple scattering effects on the track 
direction. Efficiencies for containing both decay products  
in the volume of $\epsilon_{\rm cont}^{J/\psi} =16.7$\% and 
$\epsilon_{\rm cont}^{\rho} =8.3$\% respectively were found.
Figure~\ref{UPC:phi} shows the distribution of the azimuthal opening between 
the two decay products, $\Delta \varphi$, in the plane orthogonal to the 
beam axis.  Although smeared by the magnetic field, a clear topology is still 
evident.  The FTR can be reduced by a factor $f_{\rm top}$ using the $\Delta
\varphi$ distribution, resulting in an additional ``efficiency'', 
$\epsilon_\phi$, for detecting the decay products.  
The $J/\psi$ decay FTR can be 
reduced by $f_{\rm top}
= 18$ by selecting only pairs of pads (one for each decay product) within a 
$150^{\circ} \leq \Delta\varphi \leq 170^{\circ}$ window with $\epsilon_\phi =
1$.  The $\rho^0$ FTR can be reduced by $f_{\rm top} = 9$ by selecting only 
pairs of pads in a
$70^{\circ} \leq \Delta\varphi \leq 110^{\circ}$ window with $\epsilon_\phi =
0.6$ for the $\rho^0$ signal.

Further FTR reduction can be obtained for both decays
by synchronizing the ITS readout with the beam bunches.  This is possible 
because the LHC accelerator timing, trigger and control (TTC) system
distributes fast timing signals from the RF 
generators and synchronous with the circulating beams to all experiments.
The bunch clock frequency, 40 MHz, is the same in proton and ion modes even 
though the ion bunch spacing is 125 ns.  
Since the ITS readout has a 20 ns duration, it can read out five times
between bunches.  By aligning the readout with the beam bunches and the bunch
clock while vetoing the next four readout pulses,
the noise can be reduced by an additional factor of five for each track,
a factor of 25 reduction in the combinatorial background for 
$N_{\rm cell} = 2$.
Then the L0 FTR is less than $200 / 25 f_{\rm top}$~kHz 
for the two vector meson decays so that
\begin{eqnarray} 
J/\psi \,\, {\rm FTR} & < & 440\,\, {\rm Hz} \\
\rho^0 \,\, {\rm FTR} & < & 880\,\, {\rm Hz} \, \, .
\end{eqnarray}
This L0 FTR should be compared to the $J/\psi$ and $\rho^0$ signal rates,
\begin{eqnarray} 
J/\psi \,\, {\rm Rate} & = & L_{\rm PbPb} \, B(J/\psi \rightarrow l^+ l^-) \, 
\sigma_{J/\psi} \, \epsilon_{\rm cont}^{J/\psi} \nonumber \\ 
& = & 0.5 \, {\rm mb}^{-1} \, {\rm Hz} \times
0.12 \times 32\, {\rm mb} \times 0.167 = 0.32 \, {\rm Hz} \\
\rho^0 \,\, {\rm Rate} & = & L_{\rm PbPb} \, \sigma_{\rho} \, 
\epsilon_{\rm cont}^{\rho}
\, \epsilon_{\phi} \nonumber \\ & = & 0.5 \, {\rm mb}^{-1} \, {\rm Hz} \times
5200\, {\rm mb} \times 0.083 \times 0.6 = 120\, {\rm Hz} \, \, .
\end{eqnarray}
Note that the $J/\psi \rightarrow l^+ l^-$ branching ratio in the $J/\psi$
rate is the sum of the branching ratios in the electron and muon decay 
channels.

Thus the TOF can tag vector meson decays at L0.  Detailed studies of
varies ultraperipheral processes are underway.

\paragraph*{TPC trigger backgrounds}

Experience from RHIC shows that coherent events can be identified with 
good signal to background ratios when the entire event is reconstructed and 
a cut is applied on the summed transverse momentum of the event. The incoherent
background can be estimated by reconstructing events with two same-sign
tracks, {\it e.g.}\ $\pi^+\pi^+$ or $\pi^-\pi^-$ for
$\rho^0 \rightarrow \pi^+\pi^-$ \cite{Adler:2002sc}. The main heavy vector 
meson background will most likely be lepton pairs produced in two-photon 
interactions.  Since these pairs are produced coherently, they are not 
rejected by a pair $p_T$ cut \cite{Klein:1999qj,starlight,Baltz:2002pp}.  

The following TPC background sources have been investigated: 
peripheral $AA$ interactions; incoherent $\gamma A$
interactions and cosmic ray muons. These same sources were also considered in 
a STAR study \cite{Nystrand:1998hw}. 

The trigger contribution from cosmic ray muons was not negligible in STAR 
since the scintillator counters used 
in the central trigger barrel surrounded the TPC and covered a large area. 
Measurements from L3$+$Cosmics, which also used
scintillators surrounding a large volume (the L3 magnet), observed a cosmic
ray muon rate five times
lower than that calculated for STAR because L3 is 
about 100~m underground. In STAR, the cosmic ray trigger rate was reduced 
by a topology cut on the zenith angle. 
If a silicon pixel detector is used for triggering, the 
area susceptible to cosmic ray muon triggers is greatly reduced since at least
one of the tracks will point to the vertex.

Peripheral $AA$ interactions have been studied using 5000 events with
$13 < b < 20$ fm generated by FRITIOF 7.02 \cite{Pi:1992ug}.
The inelastic Pb+Pb cross section for this 
range is 2.1 b, corresponding to about 25\% of the 8 b total inelastic cross 
section.  Of these 5000 events, 435 (9\%) had between one and five charged
tracks in the TPC.  A subset of these, 97 events (2\% of the original 5000) 
had two charged tracks in the TPC.  The
cross section for exactly two charged tracks in the TPC is then $0.02 \times
2.1 \, {\rm b} = 40$ mb, an order of magnitude lower than the $\rho^0$ 
photoproduction cross section.  In addition, the summed $p_T$ distribution 
for two charged tracks is peaked at higher values in the
background events than in the signal events.

Incoherent photonuclear interactions might be an important background at 
the trigger level, and at the analysis level for inclusive events. 
Direct $\gamma$-parton interactions are only a small fraction of the total 
$\gamma A$ cross section.  The bulk of the vector meson cross section can
be described by generalized vector meson 
dominance, see Section~\ref{vector}. Since the virtual photon energy 
spectrum has a peak much lower than the beam energy, these interactions
resemble interactions between the beam nucleus and a hadron 
nearly at rest. However, these photonuclear events have a much broader $p_T$
distribution than that of coherent $\rho^0$ and $J/\psi$ production.

The total photonuclear cross section can be calculated by integrating over the 
virtual photon energy spectrum,
\begin{equation}
\sigma_{\gamma A}^{\rm tot} = 2 \, \int_{k_{\rm min}}^{\infty} \, dk\, 
 \frac{dN_\gamma}{dk} \, \sigma_{\gamma A}(k) \, \, .
\end{equation}
The factor of two arises because each nucleus can act as both photon emitter 
and target. With a minimum photon energy cutoff of $k_{\rm min} = 10$~GeV 
in the rest frame of the target nucleus and assuming that $\sigma_{\gamma {\rm
Pb}}(k)$ is independent of photon energy, $\sigma_{\gamma{\rm Pb}} = 15$~mb
\cite{Engel:1996yb}, $\sigma_{\gamma A}^{\rm tot} = 44$ b. 
The DTUNUC 2.2 event generator \cite{Engel:1996yb} was used to simulate
50000 $\gamma$Pb events. Of these, 1595 (3\%) left one to five charged tracks
in the TPC, a cross section of $44 \, {\rm b} \times 0.03 = 1.4$ b, 
larger than the peripheral $AA$ cross
section fulfilling the same criteria.

Thus the TPC backgrounds appear to be under control and we conclude that 
$\rho^0$ and $J/\psi$ photoproduction can be triggered on without being
swamped by background.

\subsection{The ATLAS detector}
{\it Contributed by: S. N. White} 

\begin{figure}[htbp]
\begin{center}
\psfig{file=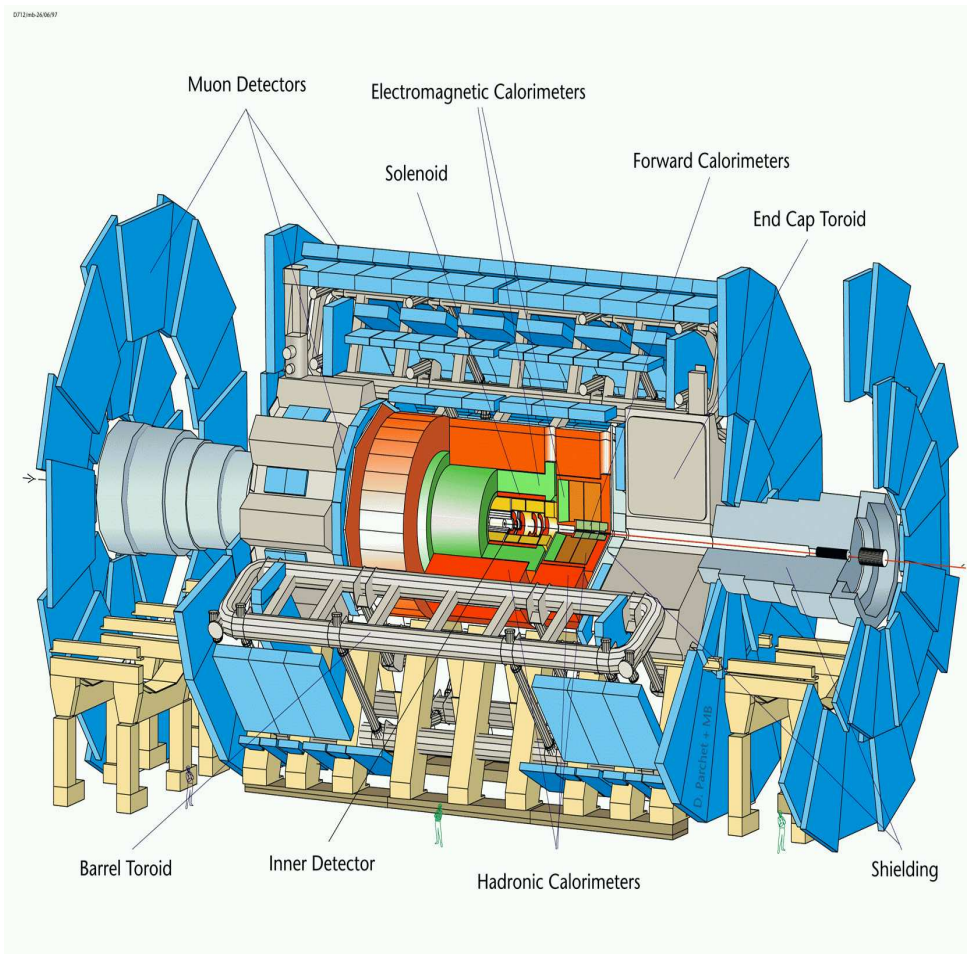,width=5in,height=5in}
\end{center}
\caption[]{The ATLAS Detector.}
\label{ATLAS_big} 
\end{figure}

The ATLAS detector is designed to study 14 TeV $pp$ collisions. 
The physics pursued by the
collaboration is vast and includes Higgs boson and SUSY searches and
other scenarios beyond the Standard Model. To achieve these goals at a full
machine luminosity of $10^{34}$ cm$^{-2}$s$^{-1}$, the calorimeter is
designed to be as hermetic 
as possible and has extremely fine segmentation. The detector, shown in
Fig.~\ref{ATLAS_big}, is a combination of three subsystems: the inner
tracking system, the 
electromagnetic and hadronic calorimeters and a full coverage muon detector.
The inner tracker is composed of a finely-segmented silicon pixel
detector; a semiconductor tracker (SCT) and the transition radiation
tracker (TRT). The segmentation is optimized for $pp$ collisions at
design luminosity.

The ATLAS calorimeters are divided into
electromagnetic and hadronic sections and cover
$|\eta| < 4.9$. The EM calorimeter is an accordion liquid argon
device, finely segmented longitudinally (lines of constant $\eta$) and 
transversely (in $\phi$) over
$|\eta| < 3.1$.  

The ATLAS electromagnetic calorimeter has three longitudinally-segmented 
sections.  The first is closest to the beam pipe while the third
is furthest away.  The first longitudinally-segmented section has granularity
$\Delta\eta \times \Delta\phi = 0.003\times 0.1$ in the barrel and is
slightly coarser in the endcaps.  Note that  $\Delta \phi$ is larger in
the first longitudinally-segmented section because the showers spread more
in $\phi$ here.  The second 
longitudinally-segmented section is composed of $\Delta\eta \times
\Delta\phi = 0.025 \times 0.025$ cells while the last segmented section has
$\Delta\eta \times \Delta\phi = 0.05 \times 0.05$ cells. In addition, a
finely segmented, $\Delta \eta \times \Delta \phi = 0.025 \times 0.1$, 
pre-sampler system is placed in front of the electromagnetic calorimeter. The 
overall energy resolution of the electromagnetic calorimeter is
$10\%/E^{1/2} +0.5\%$. The calorimeter also has good pointing 
resolution, $60 / E^{1/2}$ mrad, for photons and  better than 200 ps timing
resolution for showers with $E > 20$ GeV. 

The hadronic calorimeter is also segmented longitudinally and transversely.
The barrel calorimeter is a lead scintillator 
tile structure with a granularity of $\Delta\eta \times
\Delta\phi = 0.1 \times 0.1$.  In the endcaps, liquid argon technology is
used for radiation hardness.  The granularity in the endcaps is the same
as in the barrel. The energy resolution is $50\%/E^{1/2} +2\%$ for pions.  
                
The very forward region, up to $|\eta| < 4.9$, is covered by the
Forward Calorimeter, an axial-drift liquid argon
calorimeter \cite{TDR12}. 
The ATLAS muon spectrometer is located behind the calorimeters, 
shielded from hadronic showers. The spectrometer uses several tracking devices 
and a toroidal magnet system. Most of the
volume is covered by monitored drift tubes. In the forward region, where
the rate is high, cathode strip chambers are used. The standalone
muon momentum resolution is $\sim 2$\% for muons with 
$10 < p_T < 100$ GeV/$c$.  The performance of each subsystem is summarized 
in Ref.~\cite{DetectorTDRs}.

\subsection{The CMS detector}\label{appar}
{\it Contributed by: D. d'Enterria and P. Yepes}

\begin{figure}[hbtp]
\begin{center}
\psfig{file=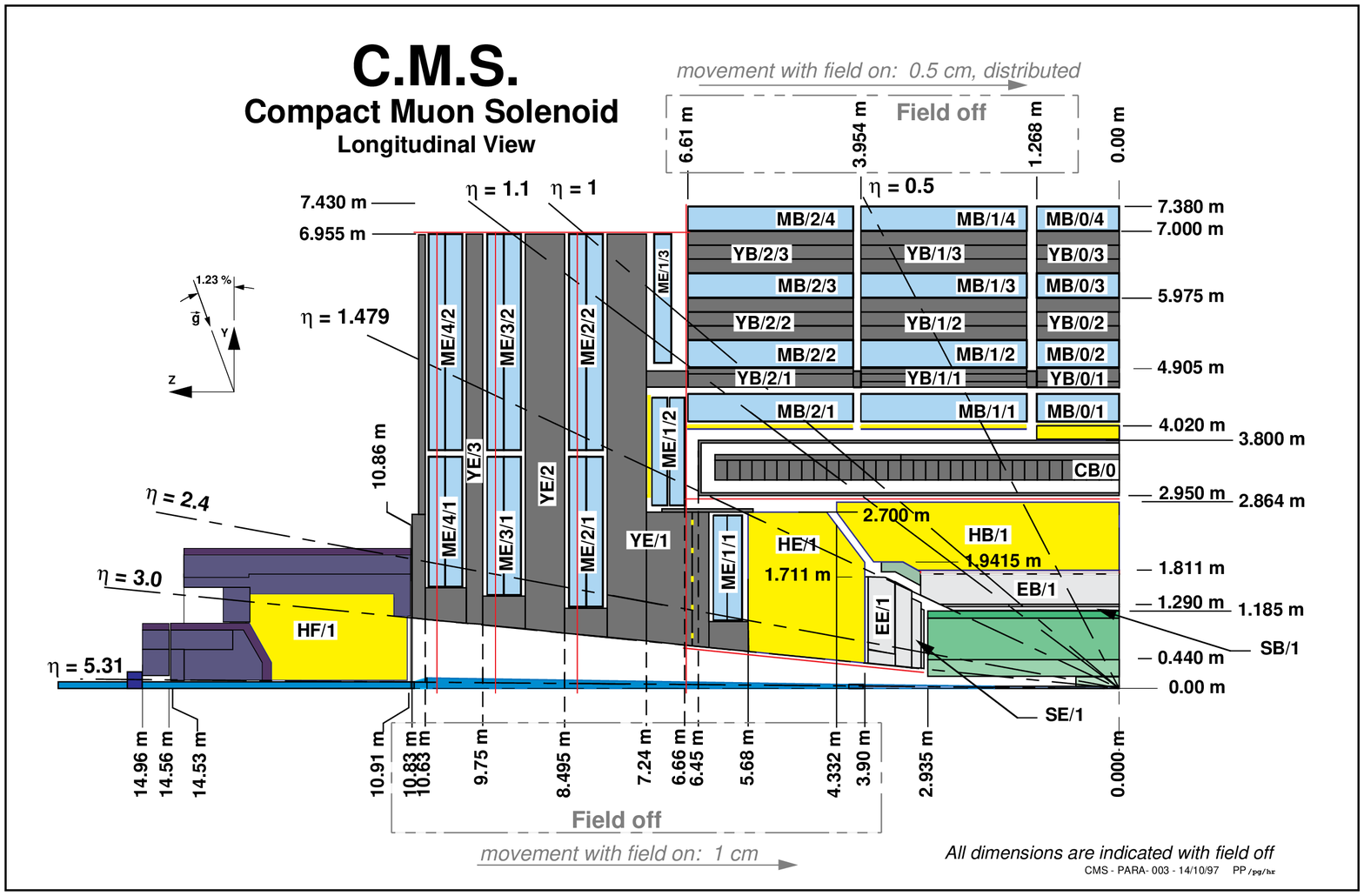,width=5in}
\caption { The CMS detector: a longitudinal view \protect\cite{CMS-PTDR1}.}
\label{geom}
\end{center}
\end{figure}

The CMS detector is designed to identify and precisely 
measure muons, electrons, photons and jets over a large energy and 
rapidity range.  A detailed description of detector elements can be found
in the Technical Design Reports~\cite{Htdr,Mtdr,Etdr,Ttdr,CMS-PTDR1}. 
An overall view of 
one quadrant of the detector is shown in Fig.~\ref{geom}. The central element 
of CMS is the magnet, a 13~m long, 6~m diameter, high-field solenoid with an 
internal radius of $\sim 3$ m, providing a uniform 4 T magnetic field. The 
tracking system, electromagnetic, and hadronic calorimeters are positioned 
inside the magnet, while the muon detector 
is outside. The tracker covers the pseudorapidity region $|\eta| < 2.4$ 
while the electromagnetic and hadronic calorimeters cover $|\eta| < 3$ and 
$|\eta| < 5$ respectively. The complete CMS geometry 
is included in the detailed {\sc geant}-4 based simulation package, CMSSW
\cite{CMS-PTDR1}.

\vspace*{0.7 cm}

\subsubsection*{Tracker} 
\label{trackingDetectors}

Starting from the beam axis, the tracker ($|\eta|< 2.4$) is 
composed of two different types of detectors: pixel layers and silicon strip 
counters. The pixel detector is composed of 3 barrel layers which are located 
at 4.5~cm, 7.5~cm and 10~cm from the beam 
axis and 2 endcap disks in each of the forward and backward directions 
with a possibility of a third set of disks to be
added later.
The barrel layers, covering rapidities up to $|\eta| = 2.1$, are made of 
more than $9.6 \times 10^6$, $1.6 \times 10^7$ and $2.24 \times 10^7$ 
pixels for the inner, middle and outer layers respectively, 
with pixel dimensions of $100 \times 150$ \micron$^2$. 
The inner Si strip counter consists of 4 cylindrical layers in the barrel 
and 3 mini-disks in the endcap. The detectors have 80 \micron\ pitch
and a strip length of 6.1~cm.

\subsubsection*{Electromagnetic Calorimeter}

The electromagnetic calorimeter, ECAL, is composed of $\sim 83000$ 
scintillating PbWO$_4$ crystals. The barrel
part of the ECAL covers the pseudorapidity range $|\eta| < 1.479$. 
The front face of the crystals is at 1.29~m radius. 
Each crystal has a square cross section of $22 \times 22$ mm$^2$ and a length 
of 230~mm, corresponding to $25.8 X_0$. 
The crystal cross section 
corresponds to $\Delta\eta \times \Delta\phi = 0.0175 \times 0.0175$ 
in the barrel region. The endcap crystal 
calorimeter covers the pseudorapidity range $1.48 < |\eta| < 3$. 
A pre-shower is located 
in front of the endcap 
crystal calorimeter covering the pseudorapidity range $1.5 < |\eta| < 2.5$. 

\subsubsection*{Hadronic Calorimeter}

The hadronic calorimeter has two parts: a central section, HB and HE, 
covering $|\eta| < 3$ and a forward/backward section, HF, covering $3 < |\eta|
< 5$.  The central calorimeter consists of the hadron barrel, HB, and hadron 
endcap, HE, both located inside the CMS magnet cryostat. It is a sampling 
calorimeter made of scintillator/copper plates. 
The forward calorimeter is located 6~m downstream of the HE endcaps.
The granularity is $\Delta\eta \times \Delta\phi \sim 0.087 \times 0.087$.

\subsubsection*{Muon system} 
The muon system is composed of drift tubes in the barrel region, MB, $|\eta|<
1.3$; cathode strip chambers in the endcap regions, ME, $0.9 < |\eta| < 2.4$ 
and resistive plate chambers in both 
barrel and endcap, covering $|\eta|< 2.1$, dedicated to triggering. 
All the muon chambers are positioned approximately perpendicular to the muon 
trajectories and cover the pseudorapidity range $|\eta| < 2.4$. 
 
\subsubsection*{Trigger and Data Acquisition}

The trigger and data acquisition system of a collider experiment plays an 
important role because the collision frequency and overall 
data rates are much larger than the rate at which events can be written
to mass storage. Events seen by the detector are inspected
online and only those with potentially interesting physics are selected for
further offline reconstruction.
 
In CMS, this online inspection of events is divided into two steps. 
The first step, the Level-1 Trigger (L1), brings the
data flow below 100 GBytes/s, while the second step or High-Level Trigger, 
HLT, further reduces the data rate to 100 MBytes/s.

The L1 trigger is on raw quantities in the calorimeters and muon detectors. 
It can select jets, electron, photons, and muons with transverse momenta
above certain thresholds. On the other hand, the HLT
has access to information from all detectors, identical to that 
available in the offline analysis.

Triggering in UPCs, characterized by very low particle multiplicities 
depositing a relatively small amount of energy in the central part of the 
detector, is done by requiring a neutron signal in one of the ZDCs in 
combination with at least one large rapidity gap between the produced system 
and the beam rapidity.  Such a trigger will select UPCs with nuclear breakup. 
Even if the rates are high due to backgrounds, they can be reduced in the HLT 
where the full event information is available for detailed analysis, see 
Section~\ref{sec:cms_upc_trigg}.

\section{Conclusions}

Ultraperipheral collisions at the LHC can provide a new means of studying
small $x$ physics, continuing along the
road pioneered at HERA in the last decade. The rates and 
collision energies for many inclusive and diffractive hard phenomena 
will be high enough to extend the HERA studies
to a factor of ten lower $x$ and, for the first time, explore hard phenomena
at small $x$ with nuclei in the same $x$ range as the proton.  This larger
reach in $x$ will explore a kinematic range  where nonlinear effects 
should be much larger than at HERA and the leading-twist approximation should 
break down. It would then be possible to test various theoretical 
predictions for the new high gluon density QCD regime such as 
parton  saturation and the physics of the black disk regime. Except for 
measurements of the parton distributions, none of the information accessible 
in UPCs will be available from the other studies at 
the LHC. UPC photons allow smaller 
virtualities to be probed than hadronic collisions so that larger nonlinear 
effects on the parton distributions could be measured at the same $x$.
In addition, photons are cleaner probes
than hadrons, simplifying the interpretation of UPC data compared to 
parton-parton interactions in hadron collisions.

UPC studies require good particle tracking over a large solid angle 
combined, for many analyses, with good particle
identification and selective triggers able to select few-particle final 
states with specified topologies.   All of the
LHC detectors are well suited to this task with large usable solid 
angles and various forms of particle identification. 
As amply demonstrated at RHIC, triggering is a bigger challenge.  
However, the particle-physics style, multi-level
triggers seem up to this challenge.  Triggers and analyses for several 
key benchmark processes were presented here,
showing that the detectors are able to collect and analyze UPC events. 

\section*{Acknowledgments}
We thank C. Bertulani, D. Brandt, U. Dreyer, J. M. Jowett,
V. Serbo, D. Trautmann and C. Weiss for helpful discussions.
This work was supported in part by the US Department of
Energy, Contract Numbers DE-AC02-05CH11231 (S.~R.~Klein and R.~Vogt);
W-7405-Eng-48 (R. Vogt); 
DE-FG02-93ER40771 (M.~Strikman); DE-AC02-98CH10886 (A.~J.~Baltz and
S.~N.~White); and DE-FG03-93ER40772 (P.~Yepes).
The work of R.~Vogt was also supported in part by the National Science 
Foundation Grant NSF PHY-0555660. D. d'Enterria acknowledges support
by the 6th EU Framework Programme contract MEIF-CT-2005-025073.
V. Nikulin and M. Zhalov would like to express their acknowledgment
for support by CERN-INTAS grant no 05-103-7484 .

\end{document}

%% file: abrev.tex
\def\GG{\gamma\gamma}

\def\BE{\begin{equation}}
\def\EE{\end{equation}}

\def\lsim{\mathrel{\raise.3ex\hbox{$<$\kern-.75em\lower1ex\hbox{$\sim$}}}}
\def\gsim{\mathrel{\raise.3ex\hbox{$>$\kern-.75em\lower1ex\hbox{$\sim$}}}}
\def\to{\rightarrow}

\def\fun#1#2{\lower3.6pt\vbox{\baselineskip0pt\lineskip.9pt}}

\makeatletter
\def\lesssim{\mathrel{\mathpalette\vereq<}}
\def\vereq#1#2{\lower3pt\vbox{\baselineskip1.5pt \lineskip1.5pt
\ialign{$\m@th#1\hfill##\hfil$\crcr#2\crcr\sim\crcr}}}
\def\gtrsim{\mathrel{\mathpalette\vereq>}}
\makeatother
\newcommand{\PbPb}{Pb+Pb}           
\newcommand{\CaCa}{Ca+Ca}           
\newcommand{\micron}{$\mu$m}

\def\x{{\mathbf x}}

\newcommand{\slL}{\raise.15ex\hbox{$/$}\kern-.53em\hbox{$l$}}
\newcommand{\slP}{\raise.15ex\hbox{$/$}\kern-.53em\hbox{$P$}}
\newcommand{\slp}{\raise.1ex\hbox{$/$}\kern-.63em\hbox{$p$}}
\newcommand{\slq}{\raise.1ex\hbox{$/$}\kern-.53em\hbox{$q$}}
\newcommand{\slv}{\raise.1ex\hbox{$/$}\kern-.63em\hbox{$v$}}
\newcommand{\slR}{\raise.15ex\hbox{$/$}\kern-.53em\hbox{$R$}}
\newcommand{\slQ}{\raise.15ex\hbox{$/$}\kern-.53em\hbox{$Q$}}
\newcommand{\slK}{\raise.15ex\hbox{$/$}\kern-.53em\hbox{$k$}}

\newcommand \Pomeron {I\!\!P}
\newcommand \jp {$J/\psi \,$}

\newcommand{\UPC}{ultraperipheral collisions }

\def\be{\begin{equation}}
\def\ee{\end{equation}}
\def\bea{\begin{eqnarray}}
\def\eea{\end{eqnarray}}

\def\mean#1{\ensuremath{\langle #1\rangle}}
\newcommand{\jps}{J/\psi}
\newcommand{\upsi}{\Upsilon}

\newcommand{\gaga}{\gamma\,\gamma}

\newcommand{\gA}{\gamma\,A}
\newcommand{\gpb}{\gamma\, {\rm Pb}}

\newcommand{\eepair}{e^+e^-}
\newcommand{\mumupair}{\mu^+\mu^-}
\newcommand{\lele}{l^{+}\,l^{-}}

\def\ttt#1{\texttt{\small #1}}

\def\ms {\overline{\rm MS}}
\def\dis{{\rm DIS}_{\gamma}}
\def\lrr{\left( }
\def\rrr{\right) }
\def\le{\left[ }
\def\re{\right] }
\def\lg{\left\{ }
\def\rg{\right\} }
\def\beq{\begin{equation}}
\def\eeq{\end{equation}}